\documentclass[fleqn, usenatbib]{mnras}
\usepackage{newtxtext, newtxmath}
\usepackage{float}
\usepackage{longtable}
\usepackage{lscape} 
\usepackage{multirow}
\usepackage{subfigure}
\usepackage{mhchem}
\usepackage[rightcaption]{sidecap}
\usepackage{graphicx}
\DeclareUnicodeCharacter{2212}{\ensuremath{-}}
\usepackage[T1]{fontenc}
\usepackage{graphicx}	
\usepackage{amsmath}
\usepackage[dvipsnames]{xcolor}
\usepackage{adjustbox}
\usepackage{lineno}

\DeclareRobustCommand{\VAN}[3]{#2}
\let\VANthebibliography\thebibliography
\def\thebibliography{\DeclareRobustCommand{\VAN}[3]{##3}\VANthebibliography}

\title[Ethanol and dimethyl ether in hot cores]{ATOMS: ALMA Three-millimeter Observations of Massive Star-forming regions -XXI. A Large-sample Observational Study of Ethanol and Dimethyl Ether in Hot Cores}

\author[Zhiping Kou et al.]{
Zhiping Kou,$^{1,2,3}$\thanks{E-mail: kouzprl@gmail.com}
Xiaohu Li,$^{1,4}$\thanks{E-mail: xiaohu.li@xao.ac.cn}
Sheng-Li Qin,$^{3}$\thanks{E-mail: qin@ynu.edu.cn}
Tie Liu,$^{5}$
E. Mannfors,$^{6}$
Xindi Tang,$^{1}$
Prasanta Gorai,$^{7,8}$
\newauthor
Guido Garay,$^{9, 10}$
Swagat R. Das,$^{9}$
Pablo Garc\'ia,$^{10, 9}$
Leonardo Bronfman,$^{9}$
M. Juvela,$^{6}$
Li Chen,$^{3}$
\newauthor
Xunchuan Liu,$^{5}$
Patricio Sanhueza,$^{11, 12, 13}$
Yaping Peng,$^{14}$
Long-Fei Chen,$^{15,16}$
Jiahang Zou,$^{3,5}$
\newauthor
Dongting Yang,$^{3}$
L. Viktor T\'{o}th,$^{17,18}$
Lokesh Dewangan, $^{19}$
Hong-Li Liu,$^{3}$
James O. Chibueze, $^{20, 21, 22}$
\newauthor
and Ziyang Li$^{3}$
\\
$^{1}$Xinjiang Astronomical Observatory, Chinese Academy of Sciences,
150 Science 1-Stree, Urumqi, Xinjiang 830011, People's Republic of China\\
$^{2}$University of Chinese Academy of Sciences, Beijing 100049, People's Republic of China\\
$^{3}$School of Physics and Astronomy, Yunnan University, Kunming 650091, People's Republic of China\\
$^{4}$Xinjiang Key Laboratory of Radio Astrophysics, 150 Science 1-Street, Urumqi, Xinjiang 830011, People's Republic of China\\
$^{5}$Shanghai Astronomical Observatory, Chinese Academy of Sciences, 80
Nandan Road, Shanghai 200030, People’s Republic of China\\
$^{6}$Department of Physics, University of Helsinki, PO Box 64, 00014 Helsinki, Finland\\
$^{7}$Rosseland Centre for Solar Physics, University of Oslo, PO Box 1029 Blindern, 0315 Oslo, Norway\\
$^{8}$Institute of Theoretical Astrophysics, University of Oslo, PO Box 1029 Blindern, 0315 Oslo, Norway\\
$^{9}$Departamento de Astronom\'{i}a, Universidad de Chile, Las Condes, 7591245 Santiago, Chile\\
$^{10}$Chinese Academy of Sciences South America Center for Astronomy, National Astronomical Observatories, CAS, Beijing 100101, China\\
$^{11}$Department of Earth and Planetary Sciences, Institute of Science Tokyo, Meguro, Tokyo, 152-8551, Japan\\
$^{12}$National Astronomical Observatory of Japan, National Institutes of Natural Sciences, 2-21-1 Osawa, Mitaka, Tokyo 181-8588, Japan\\
$^{13}$Department of Astronomical Science, SOKENDAI (The Graduate University for Advanced Studies), 2-21-1 Osawa, Mitaka, Tokyo 181-8588, Japan\\
$^{14}$Department of Physics, Faculty of Science, Kunming University of Science and Technology, Kunming 650500, People's Republic of China\\
$^{15}$School of Physics and Electronic Science, Guizhou Normal University, Guiyang 550025, China\\
$^{16}$Guizhou Provincial Key Laboratory of Radio Astronomy and Data Processing, Guiyang 550025, China\\
$^{17}$Department of Astronomy, E\"{o}tv\"{o}s Lor\'{a}nd University, P\'{a}zm\'{a}ny P\'{e}ter s\'{e}t\'{a}ny 1/A, H-1117, Budapest, Hungary\\
$^{18}$University of Debrecen, Faculty of Science and Technology, Egyetemt\'{e}r 1, H-4032 Debrecen, Hungary\\
$^{19}$Physical Research Laboratory, Navrangpura, Ahmedabad 380 009, India\\
$^{20}$Department of Mathematical Sciences, University of South Africa, Cnr Christian de Wet Rd and Pioneer Avenue, Florida Park, 1709 Roodepoort, South Africa\\
$^{21}$Centre for Space Research, North-West University, Potchefstroom Campus, Private Bag X6001, Potchefstroom 2520, South Africa\\
$^{22}$Department of Physics and Astronomy, Faculty of Physical Sciences, University of Nigeria, Carver Building, 1 University Road, Nsukka 410001, Nigeria\\
}

\date{Accepted 2025 March 7. Received 2025 March 7; in original form 2024 October 21}

\pubyear{\the\year{}}

\begin{document}
\label{firstpage}
\pagerange{\pageref{firstpage}--\pageref{lastpage}}
\maketitle

\begin{abstract}
Hot cores, as a stage of massive star formation, exhibit abundant line emissions of COMs. We present a deep line survey of two isomers of C$_2$H$_6$O: ethanol (C$_2$H$_5$OH; EA), and dimethyl ether (CH$_3$OCH$_3$; DE) as well as their possible precursor CH$_3$OH towards 60 hot cores by using the ALMA 3 mm line observations. EA is detected in 40 hot cores and DE is detected in 59 hot cores. Of these, EA and DE are simultaneously detected in 39 hot cores. We calculate rotation temperatures and column densities of EA and DE by using the XCLASS software. The average rotation temperature of EA is higher than that of DE, whereas the average column density of EA is lower than that of DE. Combined with previous studies of hot cores and hot corinos, we find strong column density correlations among EA and DE ($\rho$ = 0.92), EA and CH$_3$OH ($\rho$ = 0.82), as well as DE and CH$_3$OH ($\rho$ = 0.80). The column density ratios of EA/DE versus the column densities of CH$_3$OH remain nearly constant with values within \verb|~|1 order of magnitude. These strong correlations and the stable ratios, suggest that EA, DE, and CH$_3$OH could be chemically linked, with CH$_3$OH potentially serving as a precursor for EA and DE. Compared with chemical models, the three different warm-up timescale models result in the systematic overproduction of EA and the systematic underproduction of DE. Therefore, our large sample observations can provide crucial constraints on chemical models.
\end{abstract}


\section{Introduction} \label{sec:Intro.}
More than 320 molecules have been identified in the interstellar medium (ISM) or circumstellar shells\footnote{\url{https://cdms.astro.uni-koeln.de/classic/molecules/}}. Of various astronomical sources, the regions of star formation display rich and diverse chemistry. Through the analysis of molecular spectral lines, it is possible to investigate the physical processes at different stages of star formation, which in turn serves to study how physical conditions influence the formation and complexity of molecules. 
Compared to low-mass stars (M $<$ 2 M$_{\odot}$), the formation process of high-mass stars (M $>$ 8 M$_{\odot}$) remains less thoroughly understood. Nevertheless, the characterization of high-mass star formation processes can be simply classified through different observational signatures, including massive pre-stellar cores, high-mass protostellar objects, hot cores, hyper-, and ultra-compact H{\sc ii} regions. Among them, the hot cores, characterized by relatively high excitation temperatures ($>$100 K), high gas densities (n$_{\rm H_2}$ = 10$^5$ - 10$^8$ cm$^{-3}$) and compact sizes ($<$ 0.1 pc) \citep{2000prpl.conf..299Kurtz, cesaroni2005massive}, shows rich line emissions from complex organic molecules (COMs). COMs are defined as carbon-bearing molecules containing at least six atoms \citep{2009ARA&A..47..427H}. Up to now, more than 80 COMs have been detected \citep{McGuire_2022}. Line surveys conducted by radio telescopes have extensively detected numerous molecules, including COMs that contain oxygen and nitrogen atoms in hot cores \citep{2007A&A...470..639F, 2010ApJ...711..399Qin, 2013A&A...550A..46B, 2015ApJ...803...39Qin, 2017A&A...604A..60B, 2017ApJ...847...87S, 2019AA...632A..57Csengeri, 2020A&A...641A..54Coletta, 2020A&A...635A.198Belloche, 2021MNRAS.505.2801L,2024ApJ...962...13CLi}.

There is a special class of molecules, known as isomers, which are characterized by having the same number of atoms but distinct arrangements of atoms in space. The relative abundance of isomers can be attributed to diverse physical and chemical factors, including temperature, pressure, radiation sources, and the molecules involved in the formation reactions \citep{2018ApJ...852...70Bergan}. 
Therefore, isomers may be used to study the effects of physical and chemical conditions on the formation of molecules \citep{2007ApJ...660.1588B}. There are two isomers in the C$_2$H$_6$O group (Ethanol [C$_2$H$_5$OH], hereafter EA and Dimethyl Ether [CH$_3$OCH$_3$], hereafter DE). DE was first observed in 1974 towards the Orion Nebula molecular cloud \citep{1974ApJ...191L..79S}. EA was first observed in 1975 towards the Sagittarius B2 (Sgr B2) molecular cloud \citep{1975ApJ...196L..99Z}.  
Subsequently, EA and DE have been widely detected across various sources, including cold interstellar clouds, low-mass star-forming regions, high-mass star-forming regions, and protoplanetary disks \citep{2003ApJ...593L..51Cazaux, 2005ApJ...632..973Jorgensen, 2015A&A...582A..64LykkeL, 2015ApJ...804...81T, 2017A&A...604A..60B, 2017A&A...598A..59Rivilla, 2019AA...628A...2B, 2020A&A...635A..48Manigand, 2020A&A...635A.198Belloche,  2022A&A...659A..29Brunken, 2023A&A...673A..34Agundez,2023ApJ...958..174LiuM}. 
Previous smaller sample observations reported a possible chemical correlation between EA and DE in the low-mass star-forming region and high-mass star-forming region, respectively \citep{2007A&A...465..913B, 2020A&A...635A.198Belloche, 2022ApJ...939...84Baek, 2022MNRAS.512.4419P}. 

Experiments by \cite{2009A&A...504..891O} showed that EA and DE can be formed on CH$_3$OH-rich ice. A follow-up experimental study by \cite{2015PCCP...17.3081Maity} used methanol (CH$_3$OH) and methanol-carbon monoxide ice to study the formation of EA and DE when exposed to ionizing radiation. 
Both experiments indicated that CH$_3$OH ice and CH$_4$ ice can result in the formation of EA and DE.  
The models of \cite{2022ApJS..259....1G} have shown that DE is predominantly formed through both the ice and gas phases, while EA is almost formed in the early cold collapse stage on dust grains.
The chemical model presented by \cite{2022ApJS..259....1G} did not take into account the reaction pathways of EA and DE given by these experiments, i.e., the formation of EA and DE from photodissociation products of methanol.

To the best of our knowledge, large sample observations of EA and DE molecules with similar spatial resolutions and the same spectral coverage are still lacking.
It is still unknown to what extent the physical environment of the source influences the abundances of EA and DE. Therefore, instead of studying individual objects, we aim to provide a large sample analysis of EA and DE with the ALMA (ATOMS) survey towards the 60 hot cores targeted in this work, examining their abundance variations and gaining insights into their formation pathways.

The structure of the paper is as follows: Observations and hot core sample are described in Section \ref{sec:Obser.}, and the observational results and statistical analysis are reported in Section \ref{sec:Result and Statistical Analysis}. In Section \ref{sec:Dis.}, we compare the column densities and abundances of EA, DE, and CH$_3$OH with other sources and chemical models and discuss the column density correlations of EA, DE, and CH$_3$OH. Section \ref{sec:Con.} summarises the main results of this work and draws conclusions.

\section{Observations and Hot core sample} \label{sec:Obser.} 
\subsection{Observations} \label{sec:2.1} 
The observations were carried out with  ALMA band 3 towards 146 Infrared Astronomical Satellite (IRAS) clumps (2019.1.00685.S, PI: Tie Liu). The 146 massive clumps were selected from the CS J = 2 - 1 surveys \citep{bronfman1996cs}. Atacama Compact 7-m Array (ACA) and 12-m array (C43-2 or C43-3 configurations) observations were conducted from late September to mid-November 2019. Observational mode was set to single-point mode, with typical on-source times of 3 minutes for 12-m array and 8 minutes per source for ACA 7-m array. The observations employed the Band 3 receivers in dual polarization mode. Eight spectral windows (SPWs) were configured,  including 6 higher spectral resolutions windows (SPWs 1–6, bandwidth 58.59 MHz with spectral resolution \verb|~|0.2-0.4 km s$^{-1}$) and 2 broad bandwidth windows (SPWs 7-8, bandwidth 1875 MHz with spectral resolution \verb|~|1.6 km s$^{-1}$). The frequency ranges of SPWs 7 and 8 are 97536-99442 MHz and 99470-101390 MHz respectively. The data were calibrated and imaged using the Common Astronomy Software Applications package (CASA; \citep{2007ASPC..376..127McMullin}). The data were first presented by \cite{2020MNRAS.496.2790LiuT, 2020MNRAS.496.2821LiuT}, where more details about the spectral setups, flux and phase calibrators can be found. SPWs 7-8 were used to create continuum images from the line-free channels centred at  99.4 GHz (or 3 mm). 
Visibility data from the ALMA 12-m arrays are cleaned using the CASA 5.6 tclean task. This was done by applying the natural weighting, setting the gridder with the ‘pblimit’ parameter of 0.2, and a multiscale deconvolver.  All images are corrected for the primary beam.
In this study, we only employ 12 m array data as hot cores typically have smaller source sizes and the COM lines are less affected by missing flux issues, focusing on two broad bandwidth windows (SPWs 7-8) to identify EA and DE. The spectra of our sample were extracted from an area corresponding to the size of the beam, centred on the peak of the continuum. The angular resolution and maximum recovered angular scale for the 12-m array observations are \verb|~| 1.2$^{''}$ - 1.9$^{''}$ and \verb|~| 14.5$^{''}$ - 20.3$^{''}$, respectively. The mean 1$\sigma$ rms noise of  SPWs 7 and 8 are better than 10 mJy beam$^{-1}$ per channel for spectral lines.

\onecolumn
\begin{landscape}
\begin{longtable}{ccccccccccccc}
\caption{\large Physical parameters of hot cores} \label{tab:1} \\
\hline
\multicolumn{1}{c}{Source} & \multicolumn{1}{c}{Group} & \multicolumn{1}{c}{RA} & \multicolumn{1}{c}{DEC} & \multicolumn{4}{c}{$\text{C}_2\text{H}_5\text{OH}$ (CDMS)} & & \multicolumn{4}{c}{$\text{CH}_3\text{OCH}_3$ (CDMS)} \\
\cline{5-8}
\cline{10-13}
 & & \multicolumn{1}{c}{hh:mm:ss} & \multicolumn{1}{c}{o ${'}$ ${''}$} & \multicolumn{1}{c}{$T_{\text{rot}}$ (K)} & \multicolumn{1}{c}{$N$$\times$10$^{16}$ (cm$^{-2}$)} & \multicolumn{1}{c}{$E_{\text{up}}$(K)} & \multicolumn{1}{c}{$f_{\text{CH$_3$OH}}$} & & \multicolumn{1}{c}{$T_{\text{rot}}$ (K)} & \multicolumn{1}{c}{$N$$\times$10$^{16}$ (cm$^{-2}$)} & \multicolumn{1}{c}{$E_{\text{up}}$(K)} & \multicolumn{1}{c}{$f_{\text{CH$_3$OH}}$} \\
\hline
\endfirsthead

\multicolumn{13}{c}
{\tablename\ \thetable\ -- \textit{Continued from previous page}} \\
\hline
\multicolumn{1}{c}{Source} & \multicolumn{1}{c}{Group} & \multicolumn{1}{c}{RA} & \multicolumn{1}{c}{DEC} & \multicolumn{4}{c}{$\text{C}_2\text{H}_5\text{OH}$ (CDMS)} & & \multicolumn{4}{c}{$\text{CH}_3\text{OCH}_3$ (CDMS)} \\
\cline{5-8}
\cline{10-13}
 & & \multicolumn{1}{c}{hh:mm:ss} & \multicolumn{1}{c}{o ${'}$ ${''}$} & \multicolumn{1}{c}{$T_{\text{rot}}$ (K)} & \multicolumn{1}{c}{$N$$\times$10$^{16}$ (cm$^{-2}$)} & \multicolumn{1}{c}{$E_{\text{up}}$(K)} & \multicolumn{1}{c}{$f_{\text{CH$_3$OH}}$} & & \multicolumn{1}{c}{$T_{\text{rot}}$ (K)} & \multicolumn{1}{c}{$N$$\times$10$^{16}$ (cm$^{-2}$)} & \multicolumn{1}{c}{$E_{\text{up}}$(K)} & \multicolumn{1}{c}{$f_{\text{CH$_3$OH}}$} \\
\hline
\endhead

\hline \multicolumn{13}{r}{\textit{Continued on next page}} \\
\endfoot

\endlastfoot
\hline
I08303-4303 & $C$ & 08:32:08.68 & -43:13:45:78  & - &  - &  - & - & &  [150]$^{1}$ & 7.5 & 10-25 & $0.3\pm0.001$ \\
I08470-4323 & $A$ & 08:48:47.79 & -42:54:27.90  & $165\pm9$ & $5.0\pm0.2$ & 35-199 & $0.058\pm0.005$ & & $122\pm14$ & $8.3\pm0.5$ & 10-137 & $0.097\pm0.006$\\
I09018-4816 & $C$ & 09:03:33.46 & –48:28:01.69  & - & - & - & - & & [150] & 2.6 & 10-25 & $0.136\pm0.007$ \\
I11298-6155$^{*}$ & $D$ & 11:32:05.59 & –62:12:25.62  & - & - & - & - & & [150] & 1.2 & 10-25 & $0.075\pm0.001$ \\
I12326-6245$^{*}$  & $D$ & 12:35:35.09 & –63:02:31.91  & - & - & - & - & & [150] & 7.5 & 10-25 & $0.202\pm0.001$ \\
I13079–6218  & $A$ & 13:11:13.75 & –62:34:41.55 & $182\pm12$ & $5.7\pm0.8$ & 35-407 & $0.044\pm0.006$ & &  $129\pm5$ & $35.7\pm0.4$ & 10-347 & $0.271\pm0.029$ \\
I13134–6242    & $A$    & 13:16:43.20 & –62:58:32.30  & $196\pm10$ & $8.5\pm0.1$ & 35-617 & $0.077\pm0.001$ & & $108\pm7$ & $8.9\pm1.1$ & 10-196 & $0.081\pm0.010$ \\
I13140-6226   & $A$     & 13:17:15.49 & –62:42:24.42  & [150] & 0.65 & 35 & $0.055$  & & [150] & 1.2 & 10-25 & $0.093$ \\
I13471-6120$^{*}$ & $D$  & 13:50:41.81 & –61:35:10.67    & - & - & - & - & & $129\pm18$ & $27.3\pm6.4$ & 10-347 & $0.21\pm0.049$ \\
I13484-6100   & $C$     & 13:51:58.31 & –61:15:41.50     & - & - & - & - & & [150] & 6.0  & 10-137 & $0.545$ \\
I14498-5856   & $A$     & 14:53:42.68 & –59:08:52.89      & $127\pm24$ & $0.9\pm0.5$ & 35-127 & $0.021\pm0.008$  & & [150] & 4.2 & 10-25 & $0.058\pm0.005$ \\
I15254–5621$^{*}$ & $B$ & 15:29:19.39 & –56:31:22.34    & $152\pm15$ & $10.2\pm0.8$ & 35-406 & $0.057\pm0.001$  & & [150] & 8.5  & 10-25 & $0.045\pm0.003$ \\
I15437-5343   & $A$     & 15:47:32.73 & –53:52:38.80    & 125 & 1.55 & 35-142 & $0.021\pm0.001$  & & [150] & 5.0  & 10-25 & $0.04\pm0.002$ \\
I15520-5234$^{*}$ & $B$ & 15:55:48.47 & –52:43:06.75    & [150] &  1.2 & 126.707 & $0.020\pm0.001$ & & $109\pm39$ & $5.1\pm2.69$ & 10-101 & $0.116\pm0.058$  \\
I16060-5146$^{*}$ & $B$ & 16:09:52.64 & –51:54:54.49    & $127\pm14$ &  $2.1\pm1.1$ & 135-172 & $0.142\pm0.068$  & & [150] & 3.4  & 10-25 & $0.12$ \\
I16065-5158$^{*}$ & $B$ & 16:10:19.99 & –52:06:07.25    & $177\pm28$ &  $7.4\pm1.7$ & 35-406 & $0.056\pm0.008$ & & $127\pm11$ & $11.2\pm1.5$  & 10-347 & $0.086\pm0.005$ \\
I16071-5142$^{*}$ & $B$ & 16:10:59.59 & –51:50:23.37    & $168\pm23$ &  $3.8\pm1.9$ & 35-247 & $0.039\pm0.018$ & & $145\pm7$ & $20.3\pm0.2$  & 10-392 & $0.207\pm0.002$\\
I16076–5134   & $C$     & 16:11:26.59 & –51:41:57.84     & - & - & - & - & & [150] & 6.2  & 10-25 & $0.364\pm0.021$ \\
I16164-5046$^{*}$ & $B$ & 16:20:11.08  & –50:53:14.75    & $199\pm5$ & $5.0\pm0.1$ & 35-456 & $0.055\pm0.001$ & & $197\pm25$ & $11.5\pm2.9$  & 10-196 & $0.126\pm0.031$\\
I16172–5028$^{*}$ & $D$ & 16:21:02.97 & –50:35:12.60   & -   & - & - & - & & [150] & 5.0  & 10-25 &  $0.088\pm0.001$ \\
I16272-4837C1    & $A$   & 16:30:58.77  & –48:43:53.57    & $200\pm25$ & $15.1\pm0.5$ & 35-793 & $0.047\pm0.001$ & & $140\pm10$  & $56.2\pm5.6$  & 10-347 & $0.175\pm0.012$\\
I16272-4837C2  & $C$     & 16:30:57.29 & –48:43:39.87  & - & - & - & - & & [150]  & 7.2  & 10.21395 &  $0.342\pm0.001$ \\
I16318–4724$^*$ & $B$  & 16:35:33.96  & –47:31:11.59      & $174\pm5$ & $17.0\pm0.1$ & 35-617 & $0.061\pm0.004$ & & $156\pm7$  & $28.1\pm1.9$  & 10-257 & $0.1\pm0.001$ \\
I16344-4658   & $A$    & 16:38:09.49  & –47:04:59.73       & $134\pm19$  & $1.4\pm0.5$ &  35-159 & $0.029\pm0.009$ & & $138\pm10$   & $14.3\pm1.3$  & 10-266 & $0.298\pm0.019$ \\
I16348–4654$^{*}$ & $B$ & 16:38:29.65  & –47:00:35.67      & $160\pm5$   & $30.1\pm3.0$ & 35-430 & $0.020\pm0.001$ & & $126\pm7$   & $102.0\pm16.3$  & 10-347 & $0.068\pm0.006$ \\
I16351-4722$^{*}$ & $B$ & 16:38:50.50 & –47:28:00.68  & [150]   & 4.7 & 126.707 & $0.04$  & & [150]   & 9.2  & 10-25 & 0.074 \\
I16458-4512$^{*}$ & $D$ & 16:49:30.04 & –45:17:44.58  & - & -   & -  & - & & [150]    & 6.3  & 10-25 & $0.42\pm0.001$ \\
I16484-4603   & $A$   & 16:52:04.66  & –46:08:33.85       & $135\pm9$ & $3.4\pm0.1$ & 35-384 & $0.020\pm0.001$ & & $124\pm19$   & $10.8\pm2.0$  & 10-347 & $0.064\pm0.008$\\
I16547-4247 & $A$   & 16:58:17.18 & –42:52:07.57    & 135    & 1.35 &  35-172 & $0.034\pm0.001$ & & [150] &  1.2 & 10-25 &  $0.003\pm0.001$ \\
I17008-4040   & $A$    & 17:04:22.91  & –40:44:22.91       & $159\pm7$ & $17.7\pm0.4$ & 35-793 & $0.055\pm0.001$ & & $149\pm32$ & $26.4\pm0.8$ & 10-347 & $0.083\pm0.022$\\
I17016-4124C1   & $A$   & 17:05:10.97  & –41:29:06.95     & $158\pm5$  & $26.0\pm1.4$ & 35-455 & $0.056\pm0.002$ & & $140\pm5$  & $45.7\pm1.9$ & 10-347 & $0.099\pm0.001$\\
I17016-4124C2$^{*}$ & $D$   & 17:05:11.20  & –41:29:07.05  & - & -  & - & - & & [150]  & 20 & 10-25 &  $0.065\pm0.001$ \\
I17016-4124C3   & $C$     & 17:05:11.09  & –41:29:02.75 & -  & -  & - & - & & [150]   & 5.2 & 10-25 &  $0.065\pm0.001$ \\
I17158-3901C1   & $A$   & 17:19:20.43  & –39:03:51.58     & 120  & 1.2 & 79-157 & $0.042\pm0.003$ & & [150] & 3.0 & 10-25 & $0.09\pm0.005$  \\
I17158-3901C2  & $C$    & 17:19:20.47  & –39:03:49.20  & -   & -  & - & - & & [150]   & 1.5 & 10-25 & $0.236\pm0.001$ \\
I17175-3544C1$^{*}$ & $D$ & 17:20:53.46   & –35:47:02.16   & - & -  & - & - & & [150] & 2.6 & 10-25 &  $0.004\pm0.001$ \\
I17175-3544C2 & $A$ & 17:20:53.42  & –35:46:57.72  & $184\pm8$  & $125.0\pm15.0$ & 35-647 & $0.198\pm0.020$ & & $164\pm9$  & $85.1\pm5.1$ & 10-392 & $0.135\pm0.005$ \\
I17220-3609$^{*}$  & $B$ & 17:25:25.22  & –36:12:45.34      & $161\pm13$  & $10.2\pm2.2$ & 35-407 & $0.036\pm0.006$ & & $127\pm8$ & $32.3\pm3.2$ & 10-347 & $0.115\pm0.007$ \\
I17233-3606$^{*}$ & $D$ & 17:26:42.46   & –36:09:17.85  & -  & -  & - & - & & $144\pm14$  & $31.2\pm4.0$ & 10-347 & $0.325\pm0.028$ \\
I17441-2822$^{*}$  & $E$ & 17:47:20.17   & –28:23:04.74    & $156\pm12$  & $14.4\pm1.5$ & 35-431 & $0.1\pm0.003$  & & - & - & - & - \\
I18032–2032C1$^*$ & $D$ & 18:06:14.92   & –20:31:43.22   & - & - & - & - & & [150] &  9.6 & 10-25 &  $0.087\pm0.001$ \\
I18032–2032C2  & $A$   & 18:06:14.88   & –20:31:39.59    & $117\pm7$  & $1.9\pm0.2$ & 35-247 & $0.042\pm0.003$ & & $114\pm14$  & $4.8\pm1.6$ & 10-101 & $0.103\pm0.029$ \\
I18032–2032C3  & $C$   & 18:06:14.80   & –20:31:39.26  & -  & -  & - & - & & [150]  & 3.6 & 10-25 &  $0.2\pm0.001$ \\
I18032–2032C4  & $A$    & 18:06:14.66  & –20:31:31.57     & $159\pm5$  & $12.0\pm0.4$ & 35-445 & $0.057\pm0.001$ & & $105\pm7$  & $13.8\pm1.7$ & 10-137 & $0.065\pm0.004$ \\
I18056-1952   & $A$     & 18:08:38.23  & –19:51:50.31     & $215\pm29$   & $106.0\pm3.1$ & 35-588 & $0.035\pm0.001$ & & $183\pm6$  &  $223.6\pm5.6$ & 10-196 & $0.074\pm0.002$ \\ 
I18089-1732  & $A$     & 18:11:51.45  & –17:31:28.96     & $180\pm31$    & $9.8\pm0.1$ & 35-445 & $0.037\pm0.001$ & & $138\pm11$    & $38.9\pm3.8$ & 10-347 & $0.149\pm0.008$ \\
I18117–1753    & $A$   & 18:14:39.51  & –17:52:00.08      & $178\pm16$    & $4.0\pm0.2$ & 35-407 & $0.025\pm0.001$ & & $102\pm33$    & $14.8\pm5.9$ & 10-266 & $0.092\pm0.031$ \\
I18159–1648C1 & $A$    & 18:18:54.66  & –16:47:50.28       & $176\pm20$    & $3.4\pm0.2$ & 35-362 & $0.031\pm0.003$ & & $114\pm6$    & $15.4\pm1.1$ & 10-266 & $0.14\pm0.016$ \\
I18159–1648C2  & $A$   & 18:18:54.34  & –16:47:49.97        & $154\pm30$    & $1.9\pm0.2$ & 35-247 & $0.033\pm0.003$ & & $132\pm18$    & $7.4\pm1.6$ & 10-137 & $0.124\pm0.025$ \\
I18182–1433  & $C$    & 18:21:09.05    & –14:31:47.88 & - & - & - & - & & [150]  & 5.8 & 10-25 & 0.27 \\
I18236–1205  & $C$   & 18:26:25.79     & –12:03:53.08       & - & - & -  & - & & [150]  & 5.2 & 10-25 & $0.305\pm0.017$ \\
I18290–0924 & $C$    & 18:31:44.13      & –09:22:12.25  & - & -    & - & - & & [150]  & 9.7 & 10-25 & $0.17$  \\
I18316–0602  & $A$   & 18:34:20.91      & –05:59:42.00      & 115    & 2.8 & 35-185  & $0.032\pm0.001$ & & [150]  & 6.5 & 10-25 & $0.044\pm0.002$  \\
I18411–0338  & $A$      & 18:43:46.23   & –03:35:29.77  & [150]    & 4.2 & 126-198 & $0.046\pm0.003$ & & [150]  & 4.6 & 10-25 &  $0.051\pm0.001$ \\
I18469–0132$^{*}$ & $B$ & 18:49:33.05   & –01:29:03.34     & $127\pm14$    & $3.2\pm0.9$ & 79-157 & $0.023\pm0.005$ & & $119\pm9$ & $24.6\pm1.7$ & 10-266  & $0.175\pm0.001$ \\
I18507+0110$^{*}$  & $B$   & 18:53:18.56 & +01:14:58.23 & $160\pm7$ & $34.0\pm2.3$ & 35-533 & $0.031\pm0.001$ & & $153\pm7$ & $168.0\pm8.5$ & 10-490 & $0.147\pm0.005$ \\
I18507+0121  & $A$    & 18:53:18.01  & +01:25:25.56    & $157\pm11$     & $13.8\pm1.7$ & 35-533 & $0.032\pm0.003$ & & $140\pm20$ & $43.5\pm7.5$ & 10-490 & $0.103\pm0.015$  \\
I18517+0437  & $A$    & 18:54:14.24   & +04:41:40.65   & 130  & 2.2 & 79-282  & $0.023\pm0.001$  & & [150] & 3.6 & 10-25 & $0.021\pm0.001$  \\
I19078+0901$^*$ & $B$ & 19:10:13.16   & +09:06:12.49    & 130 & 2.5 & 79-157 & $0.028\pm0.001$  & & [150] & 3.8 & 10-25 & $0.024\pm0.001$  \\
I19095+0930$^{*}$ & $B$ & 19:11:53.99  & +09:35:50.27    & $91\pm9$ & $8.2\pm0.9$ & 35-230 & $0.019\pm0.002$  & & [150] & 20 & 10-25 & $0.028\pm0.001$ \\
\hline
\multicolumn{11}{l}{\normalsize {$*$ symbol represents this source associated with H{\sc ii} regions; $-$ symbol represents no transition line of the molecule has been detected.}}\\
\multicolumn{11}{l}{\normalsize {1: The T$_{\rm rot}$ in square brackets indicates that it was fixed during the fitting process. We use this value to fit the column density.}}\\
\end{longtable}
\end{landscape}
\twocolumn
\subsection{Hot core sample} \label{sec:2.2} 
The study is based on high-sensitivity, high-resolution ALMA observations as part of the ATOMS project. Typical nitrogen- and oxygen-containing COMs, namely C$_2$H$_5$CN, CH$_3$CHO, and CH$_3$OH were employed to identify hot cores, resulting in the identification of a total of 60 hot cores, with 45 being newly detected \citep{2022MNRAS.511.3463Qin}. This sample covers a broad range of luminosities ($\sim$10$^{1.2}$-10$^{7}$ L$_{\odot}$), distances from the Sun ($\sim$1-12 kpc), and distances from the Galactic center ($\sim$0.2-10 kpc) (see Table A1 in \citealp{2020MNRAS.496.2790LiuT}). Of the 60 hot cores, 24 are associated with Ultra-Compact H{\sc ii} regions (UCH{\sc ii} regions) \citep{2022MNRAS.511.3463Qin, 2023MNRAS.520.3245Z}. Hence, observations of EA and DE based on the 60 hot cores can be used to study the abundance variations of EA and DE in various physical environments, as well as to constrain the reaction pathways of EA and DE.

\section{Results and Statistical Analysis} \label{sec:Result and Statistical Analysis}

\subsection{Line identification} \label{sec:3.1}

\begin{figure*}
\centering
\includegraphics[height=8cm,width=16cm]{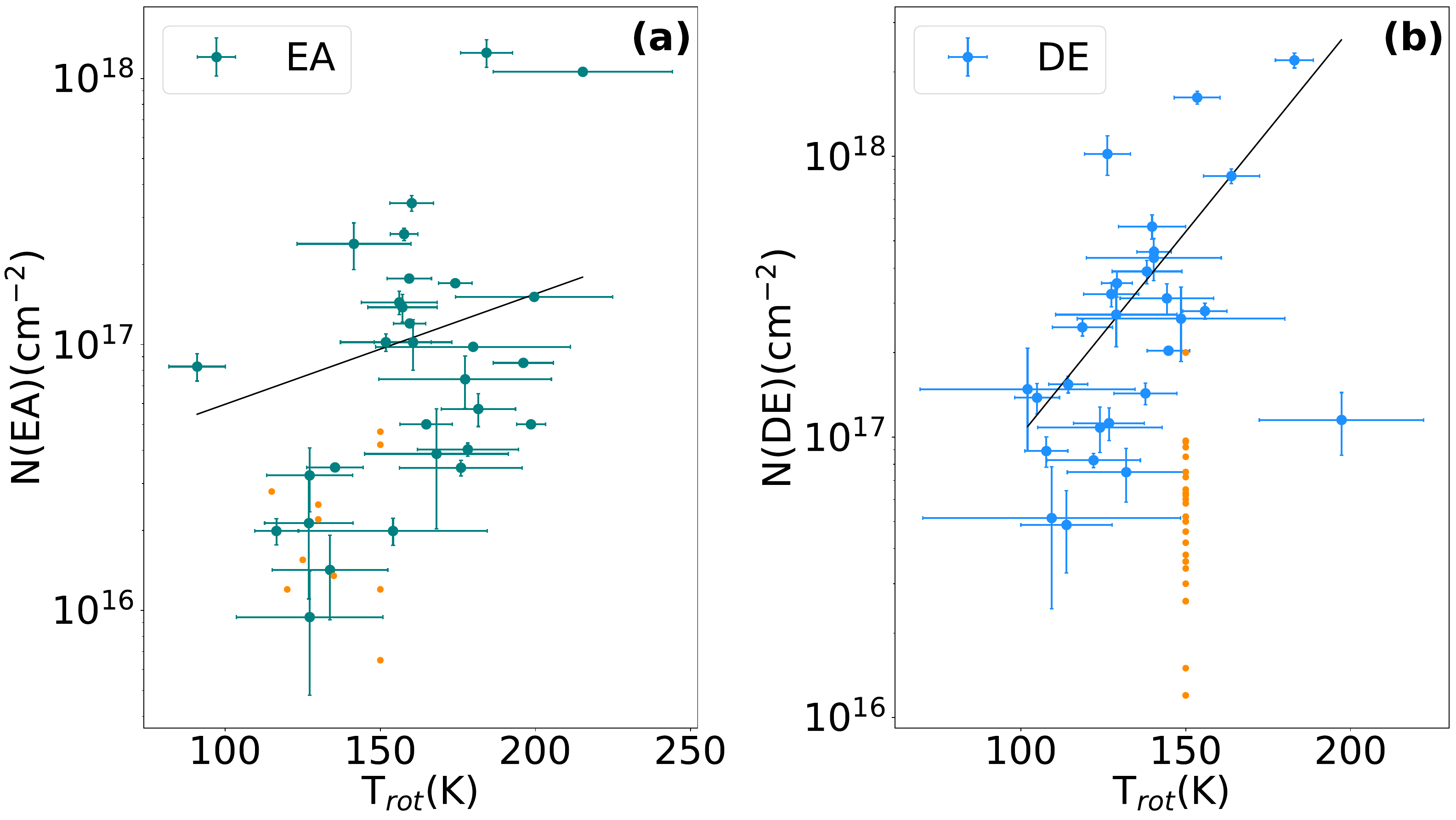}
\caption{\small The left panel shows the correlation between rotation temperatures and column densities for EA, and the right panel shows the correlation between rotation temperatures and column densities for DE. The orange points in the upper panels are sources for which the rotation temperature and column density were not optimized. These points were not included in the fitting. The weighted least squares fits are shown as black solid lines.}  
\label{fig:1}
\end{figure*}
The eXtended CASA Line Analysis Software Suite (XCLASS\footnote{\url{https://xclass.astro.uni-koeln.de}}; \citep{2017A&A...598A...7Moller}) is employed for the identification of DE and EA towards the 60 hot cores.
The XCLASS fitting program takes the source size, the line width, the velocity offset, the rotation temperature, and the column density as input parameters. Assuming molecular gas is in local thermodynamic equilibrium (LTE) and solving the radiative transfer equation, XCLASS produces theoretical spectra using spectroscopic data from the Cologne Database for Molecular Spectroscopy (CDMS\footnote{\url{https://cdms.astro.uni-koeln.de/classic/}}; \citep{2005JMoSt.742..215M}) and the Jet Propulsion Laboratory molecular databases (JPL\footnote{\url{http://spec.jpl.nasa.gov/}}; \citep{1998JQSRT..60..883Pickett}). We used the XCLASS version 1.2.5 for the line fitting, which  incorporates an embedded SQLite database file containing spectroscopic data from CDMS and JPL. We upgraded this database file by using the XCLASS task UpdateDatabase. In this work, we use the same procedure to identify and calculate parameters for EA and DE, as described in \cite{2022MNRAS.511.3463Qin}. 
We have estimated the continuum contribution to the line emission using Eq. (5) from the paper of \cite{2020A&A...635A..48Manigand}. The correction factors  range from 0.99-1.1. And then the continuum contribution to the line emission can be neglected.
To evaluate the dust attenuation, we used the dust opacity equation (see Eq.(4) of \cite{2021A&A...651A...9Moller} to calculate the dust optical depth. The dust optical depths in our sample are less than 0.08. The hydrogen column densities were obtained from \cite{2025A&A...694A.166C} (in their Tables C1-C4). Thus, we neglect continuum emission and dust attenuation in our spectral line fitting.  The molecular line identification and calculation are made in two steps:  

 - In the first step, we take deconvolved angular sizes of the continuum sources as source sizes which are given by \cite{2022MNRAS.511.3463Qin} in their Table 1. The velocity offsets are determined with reference to the rest frequency of the CH$_3$OH transition line, 13(-2,12)-12(-3,10) E, v$_{\mathrm{t}}$=0, at 100.6389 GHz. Then, three parameters: rotation temperature, column density, and line width are set to be free parameters for fitting the observed spectra using the XCLASS software. The rotation temperature, column density, and line width were manually adjusted to achieve a reasonable match between the synthetic and observed spectra. 
 
 - In the second step, the line width is set as a fixed parameter based on the first step. We then employ the Modeling and Analysis Generic Interface for eXternal numerical codes (MAGIX; \citep{2013A&A...549A..21Moller}) to obtain optimized rotation temperatures and column densities. In the calculation, the Levenberg-Marquardt algorithm (LM) and the Markov chain Monte Carlo algorithm (MCMC) are employed to estimate the errors in the derived rotation temperatures and column densities. 

Table \ref{tab:1} presents calculated results for EA and DE. 
In some sources, the MAGIX calculations do not converge, so we show the results from the first step. For sources where only one or two molecular lines of DE and EA were detected, the rotation temperature could not be constrained. We calculated the column densities using three different temperatures (100 K, 150 K, and 200 K), which showed a slight increase in column density, with the variation remaining within 20\%. We chose a median value of 150 K to derive their column densities. 
The identified transition lines of EA and DE above 3$\sigma$ are listed in Table \ref{tab:A.1} and \ref{tab:A.2}. We also marked blending problems of EA and DE in Table \ref{tab:A.1} and \ref{tab:A.2}. 
The whole observed spectra with modelled ones are shown in Fig. \ref{fig:B1} in Appendix \ref{app:B}. We also plot modelled spectra of three molecules (CH$_3$OH, C$_2$H$_5$CN, and CH$_3$OCHO) from \cite{2022MNRAS.511.3463Qin} in order to assess the potential blending issue. The blending transitions are also denoted in Fig. \ref{fig:B1}. 

\begin{figure*}
\begin{minipage}{0.15\linewidth}
\includegraphics[height=5.5cm,width=8.5cm]{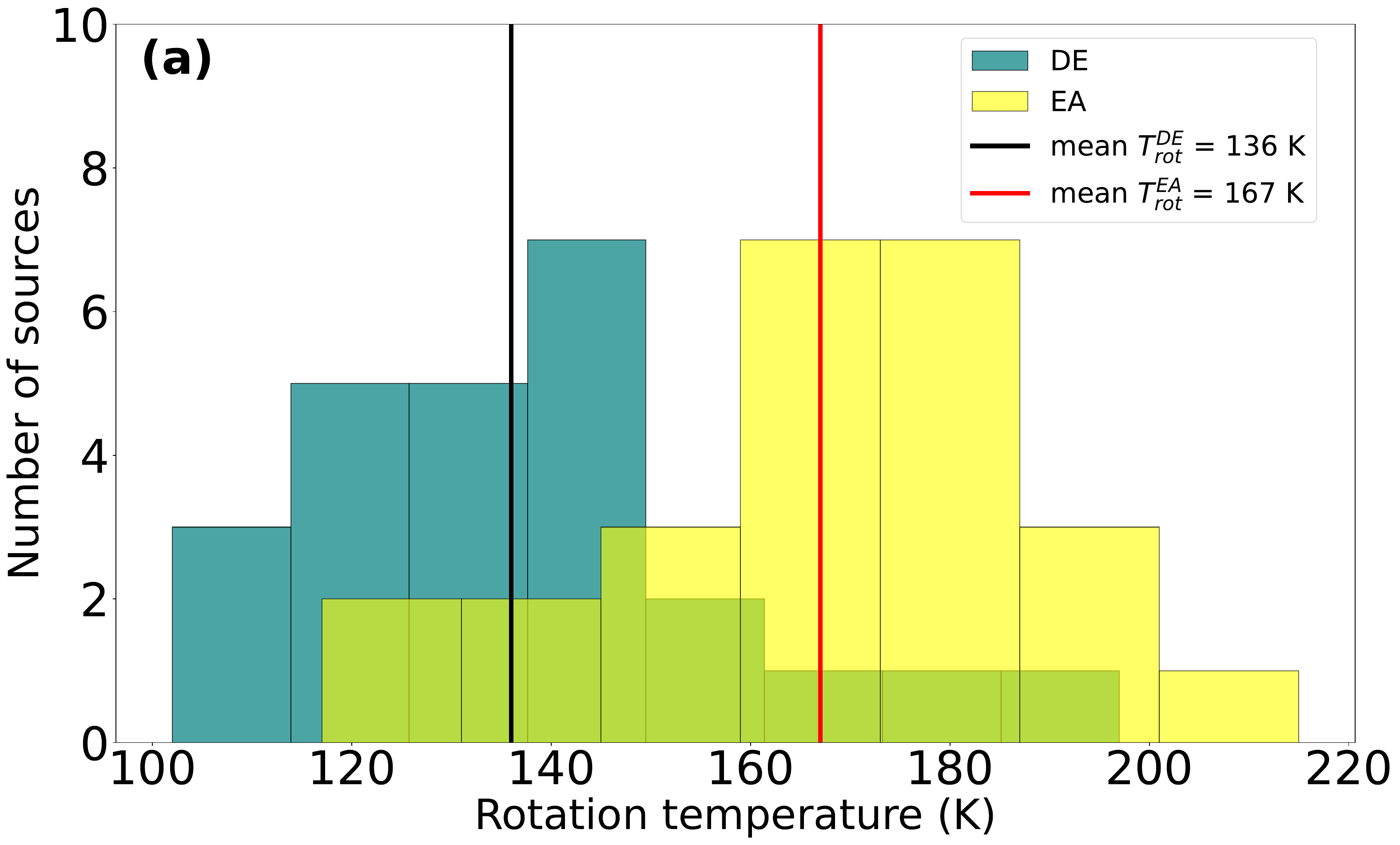}
\end{minipage}  
\hfill 
\begin{minipage}{0.5\linewidth}
\centering 
\includegraphics[height=5.5cm,width=8.5cm]{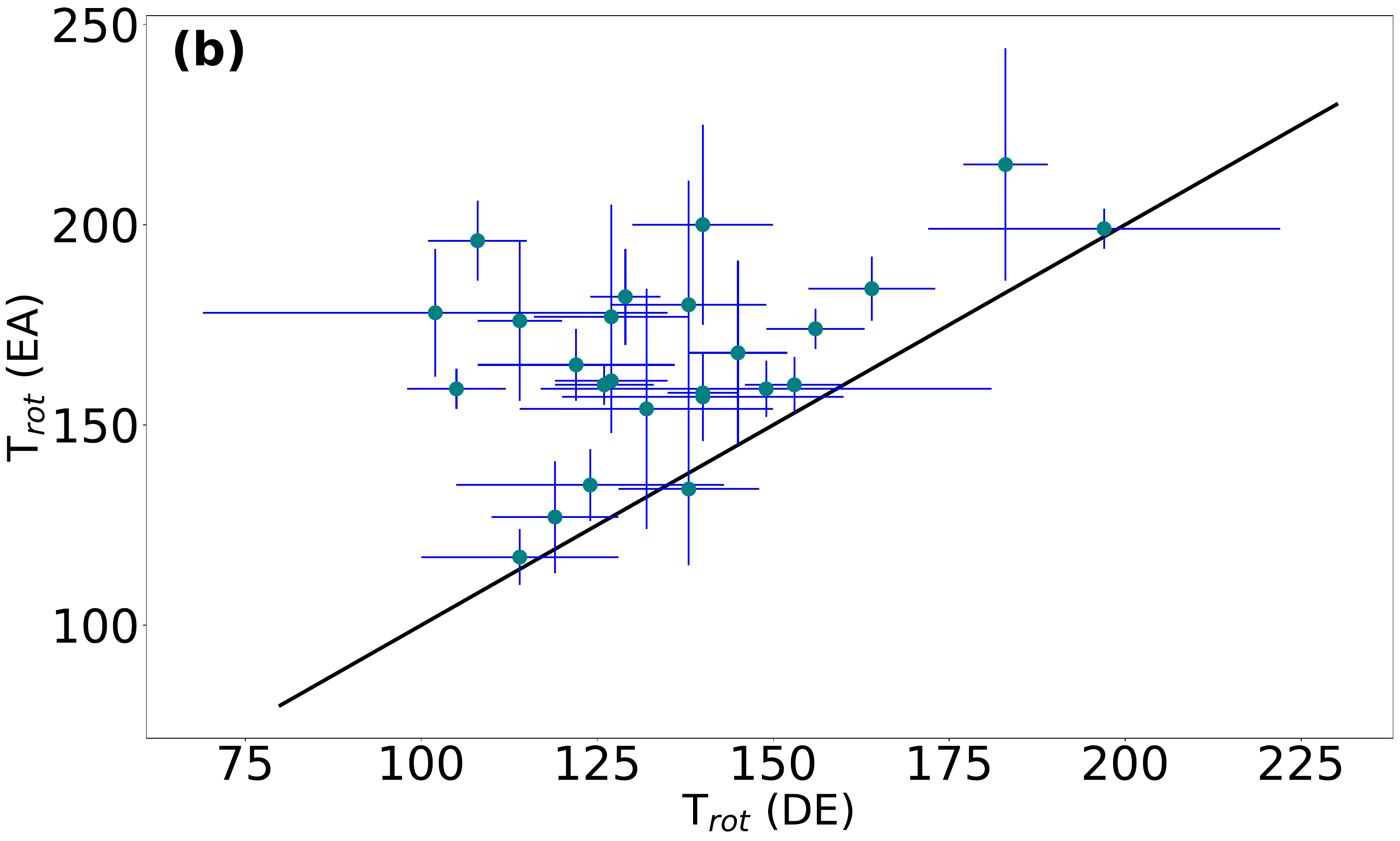}
\end{minipage} 
\caption{\small Left: The distribution of rotation temperature. The red line shows the mean rotation temperature of EA (167 K), and the black line shows the mean rotation temperature of DE (136 K). Right: The rotation temperatures of EA are plotted as a function of the rotation temperatures of DE, where the black line indicates that the rotational temperatures of EA are equal to DE.}
\label{fig:2}
\end{figure*}

The ATOMS project covers the H40$\alpha$ recombination line at 99022.952 MHz. This recombination line originates from ionized gas associated with H{\sc ii} regions.
Based on whether EA and DE are detected and whether hot cores are associated with H{\sc ii} regions, we classified them into 5 groups, as listed in Table \ref{tab:2}.
 
\begin{table}
\centering
\scriptsize
\caption{\normalsize Classification of sources based on EA, DE detection and hot core association with H{\sc ii} regions.}
\begin{tabular}{p{0.5cm}ccccc}
\hline
\hline
Group & EA detection & DE detection & Associated H{\sc ii} region & Number of sources \\
\hline
A & Yes & Yes & No & 25 \\
B & Yes & Yes & Yes & 14\\
C & No & Yes & No & 11\\
D & No & Yes & Yes & 9 \\
E & Yes & No & Yes & 1 \\
\hline
\end{tabular}
\label{tab:2}
\end{table}

Group A: 25 sources without associated H{\sc ii} regions detected both EA and DE.
We find that 6 sources (IRAS 13140-6226, IRAS 15437-5343, IRAS 16344-4658, IRAS 17158-3902C1, IRAS 18411-0338, and IRAS 18517+0437) have fewer than five emission lines of EA detected, with the maximum upper level energy being less than 200 K. The column densities of EA in these 6 sources range from 4.5 $\times$ 10$^{15}$ to 4.0 $\times$ 10$^{16}$ cm$^{-2}$. In IRAS 16547-4247, only two blended transition lines of DE are detected, thus we give an estimated value for the column density of DE assuming a rotation temperature of 150 K. 

Group B: 14 sources associated with H{\sc ii} regions have both EA and DE detections. 
IRAS 16348-4654, IRAS 18507+0110 and IRAS 17175-3544C2 exhibit higher column densities of DE and EA. The column densities of EA or DE in these three sources are at the level of $\sim$ 10$^{18}{\rm cm}^{-2}$. 

Group C: in 11 sources without H{\sc ii} regions, only DE was detected, with the maximum upper level energy being less than 138 K. 
Fewer than 3 transition lines for DE are detected in 9 sources, with upper level energies from 10 to 25 K. 

Group D: 9 sources associated with H{\sc ii} regions have only DE detected. In IRAS 13471-6120 and IRAS 17233-3606, more than two emission lines of DE are detected with the maximum upper level energy of 347 K. These two sources present higher rotation temperatures (\verb|~|136 K) and column densities (\verb|~|$2.6 \times 10^{17}$ cm$^{-2}$) compared with the other 7 sources. 
For the other 7 sources, only two emission lines of DE are detected with upper level energies of 10-25 K. 
The ATOMS spectral setup covers the two lowest upper level energy of EA transition lines, which are 6(1,6)-5(1,5) with E$_{\rm up}$ = 18 K and A$_{\rm ij}$ = 1.1 $\times$ 10$^{-8}$ s$^{-1}$, and 8(2,7)-8(1,8) with E$_{\rm up}$ = 35 K and A$_{\rm ij}$ = 5.2 $\times$ 10$^{-6}$ s$^{-1}$. 
The transition line of EA with upper level energy of 18 K is not detected in our sample.
Because we do not cover the transition line of EA with lower upper level energy (\textless 35 K) and higher Einstein A coefficient, observations at other wavelengths are needed to confirm the absence of EA in these sources, where only two emission lines of DE are detected with upper level energies ranging from 10 to 25 K.

Group E: 15 unblended transition lines of EA with upper level energies ranging from 35 to 432 K are detected in IRAS 17441-2822, but the transition lines of DE are not detected in this source.

The moment 0 maps of EA at 100990 MHz and DE at 99324 MHz are shown in Figs. \ref{fig:C1}. These two lines are not blended with other molecules. Overall, the molecular line emissions and the continuum emissions show very similar morphologies in most sources. It should be noted that in some sources, the molecular line emissions of EA and DE are not well associated with the continuum emissions. In such case, the adopted source sizes from the continuum may lead to an underestimation of molecular column densities if the molecular emission more compact than the continuum emissions and vice versa.

\subsection{Statistical analysis} \label{sec:3.2}

For EA, there are two types of amplitude motions: the torsions of the CH$_3$ and OH groups. In the ground vibrational state, the CH$_3$ torsion can be neglected. Due to the torsion of the OH group, EA consists of two distinguishable conformers: one is anti conformer with the lower lying and another one is gauche conformer with the doubly degeneracy. The gauche conformer can be divided into two distinguishable states: gauche$^+$ and gauche$^-$ \citep{2018A&A...620A.170J:, 2023ApJ...958..111Bhat}. 
From Fig. \ref{fig:B1}, out of 60 hot cores, more than 3 transitions of EA are detected in 37 hot cores. The upper level energies for EA range from 35 to 793 K (see Table \ref{tab:A.1}).
The single transition of EA was detected with intensities exceeding 3$\sigma$ in three sources: IRAS 15520-5234 and IRAS 16351-4722 (at 100358.9 MHz with E${\rm up}$ = 126 K), and IRAS 13140-6226 (at 100990.1 MHz with E${\rm up}$ = 35 K). The single transition detected in each of the three sources is identified as an emission line from EA, as it was not blended with other molecular lines in the 37 sources where more than three EA transitions were detected. 
Rotation temperatures towards 40 hot cores range from 91 to 215 K with an average value of 154 K. Column densities towards 40 hot cores range from 6.5 $\times$ 10$^{15}$ to 1.3 $\times$ 10$^{18}$ cm$^{-2}$ with an average value of 1.3 $\times$ 10$^{17}$ cm$^{-2}$.
The relationship between T$_{\rm EA}$ and N$_{\rm EA}$ is shown in Fig. \ref{fig:1}a. Pearson’s co-efficient for EA between rotation temperatures and column densities is 0.4, indicating that column densities of EA are not correlated with rotation temperatures.

DE, an asymmetric top molecule with two CH$_3$ groups, shows large amplitude motion along the CO bond, resulting in the splitting of a rotational level into four sub-states: AA, EE, AE, and EA \citep{2023ApJ...958..111Bhat}. In our sample, the EA and AE sub-states are overlapped, resulting in a triple peak spectral shape for DE (see Fig. \ref{fig:B1}). 
Out of 59 hot cores, more than 3 transitions of DE are detected in 29 hot cores. 29 sources have only two transition lines with upper level energies of 10 and 25 K. The upper energy levels for DE range from 10 to 490 K (see Table \ref{tab:A.2}).
The single transition of DE (at 99325.2 MHz with E$_{\rm up}$=10 K) was detected with intensities exceeding 3$\sigma$ in IRAS 16271-4837C2. This transition is identified as an emission line from DE, as it was not blended with other molecular lines in the 29 sources where more than three DE transitions were detected.
Rotation temperatures towards 59 hot cores range from 102 to 197 K, with an average value of 143 K. Column densities towards 59 hot cores range from 1.2 $\times$ 10$^{16}$ to 2.2 $\times$ 10$^{18}$ cm$^{-2}$ with the average value of 2.2 $\times$ 10$^{17}$ cm$^{-2}$. 
The relationship between T$_{\rm DE}$ and N$_{\rm DE}$ is shown in Fig. \ref{fig:1}b. Pearson’s co-efficient for DE between rotation temperatures and column densities is 0.5, indicating that column densities of DE are not correlated with rotation temperatures.

We present the distributions of rotation temperatures for EA and DE in Fig. \ref{fig:2}, which include 25 sources with well fitted rotation temperatures for groups A and B.
The rotation temperatures for EA range from 117 to 215 K with an average value of $167\pm14$ K, while the rotation temperatures for DE range from 102 to 197 K with an average value of $136\pm12$ K, as shown in Fig. \ref{fig:2}a. Rotation temperatures between EA and DE show no correlation, with Pearson’s co-efficient of 0.44. In the same sources, the rotation temperatures of most EA are larger than that of DE, as shown in Fig. \ref{fig:2}b.
The column densities of EA range from $1.4 \times$ 10$^{16}$ to $1.3 \times$ 10$^{18}$ cm$^{-2}$, with an average value of $1.9 \pm 0.1 \times$ 10$^{17}$ cm$^{-2}$. 
The column densities of DE range from  $4.8 \times$ 10$^{16}$  to $2.2 \times$ 10$^{18}$ cm$^{-2}$, with an average value of  $4.2 \pm 0.1 \times$ 10$^{17}$ cm$^{-2}$. The average column densities of EA and DE are at the same order of magnitude. We will discuss the correlation between the column densities of EA and DE in Section \ref{sec:4.1}. 
 

\section{Discussion} \label{sec:Dis.}
\begin{figure*}
\begin{minipage}{0.3\linewidth}
\includegraphics[height=4cm,width=5.5cm]{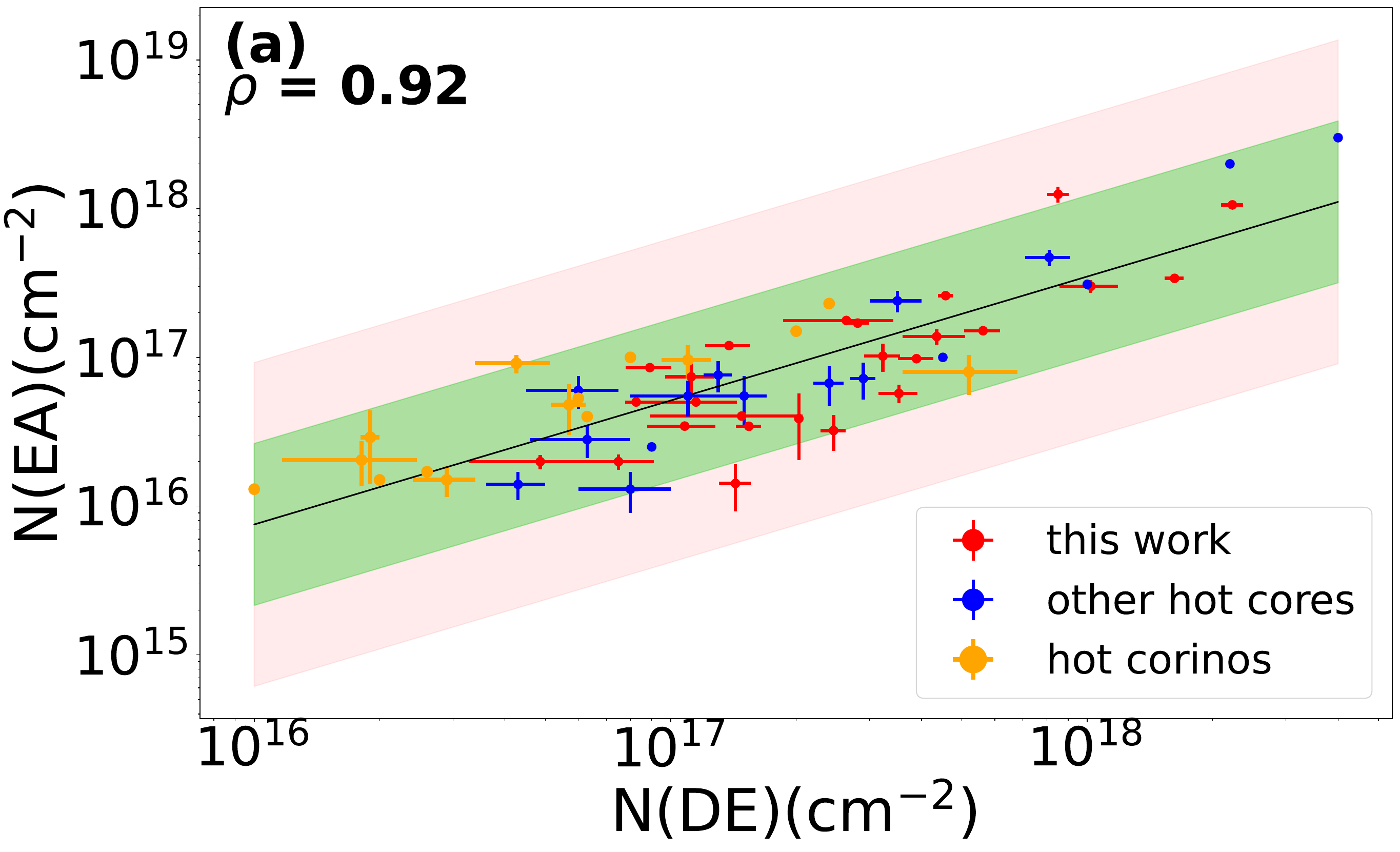}
\end{minipage}
\hfill 
\begin{minipage}{0.3\linewidth}
\centering 
\includegraphics[height=4cm,width=5.5cm]{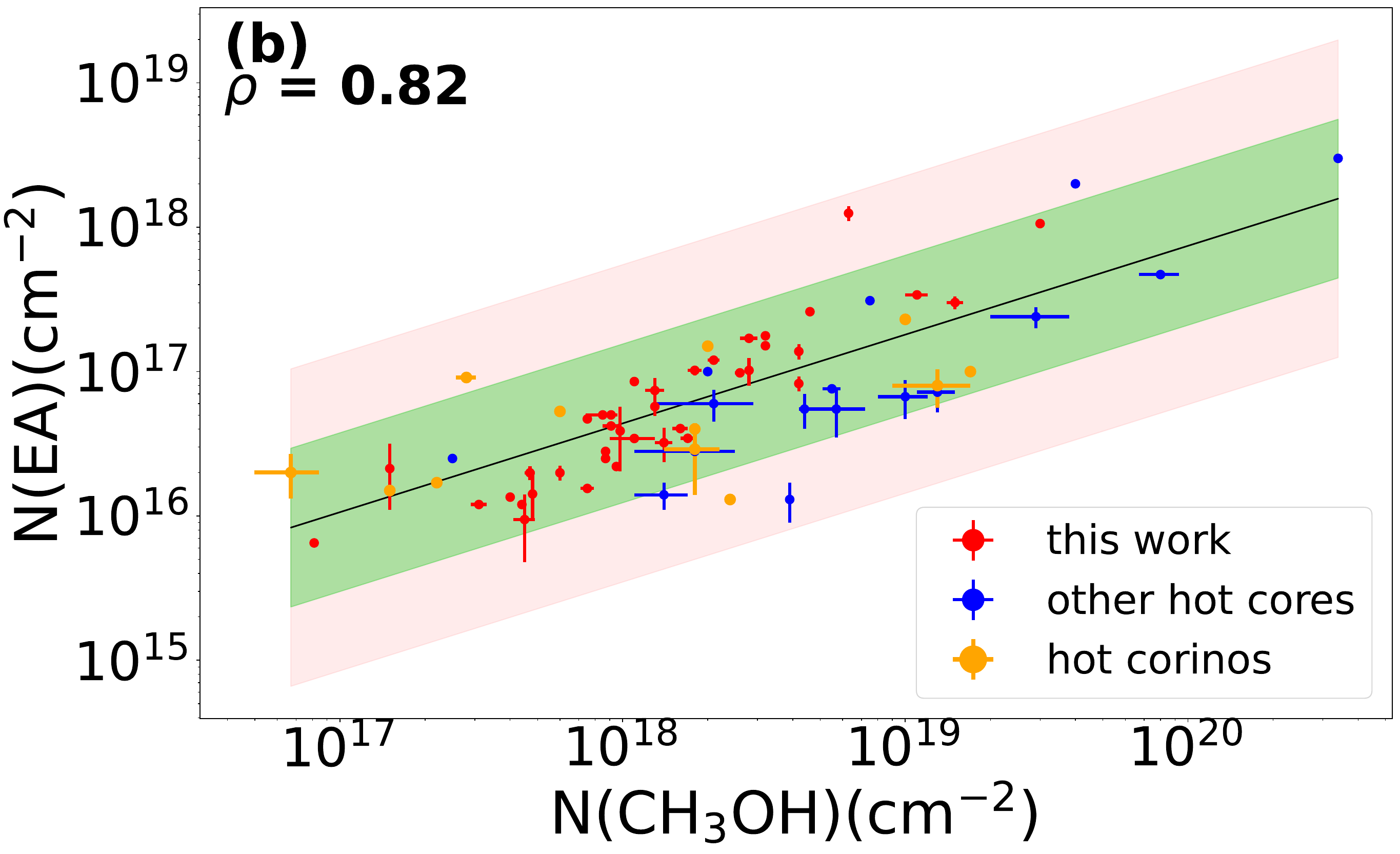}
\end{minipage} 
\hfill 
\begin{minipage}{0.3\linewidth} 
\includegraphics[height=4cm,width=5.5cm]{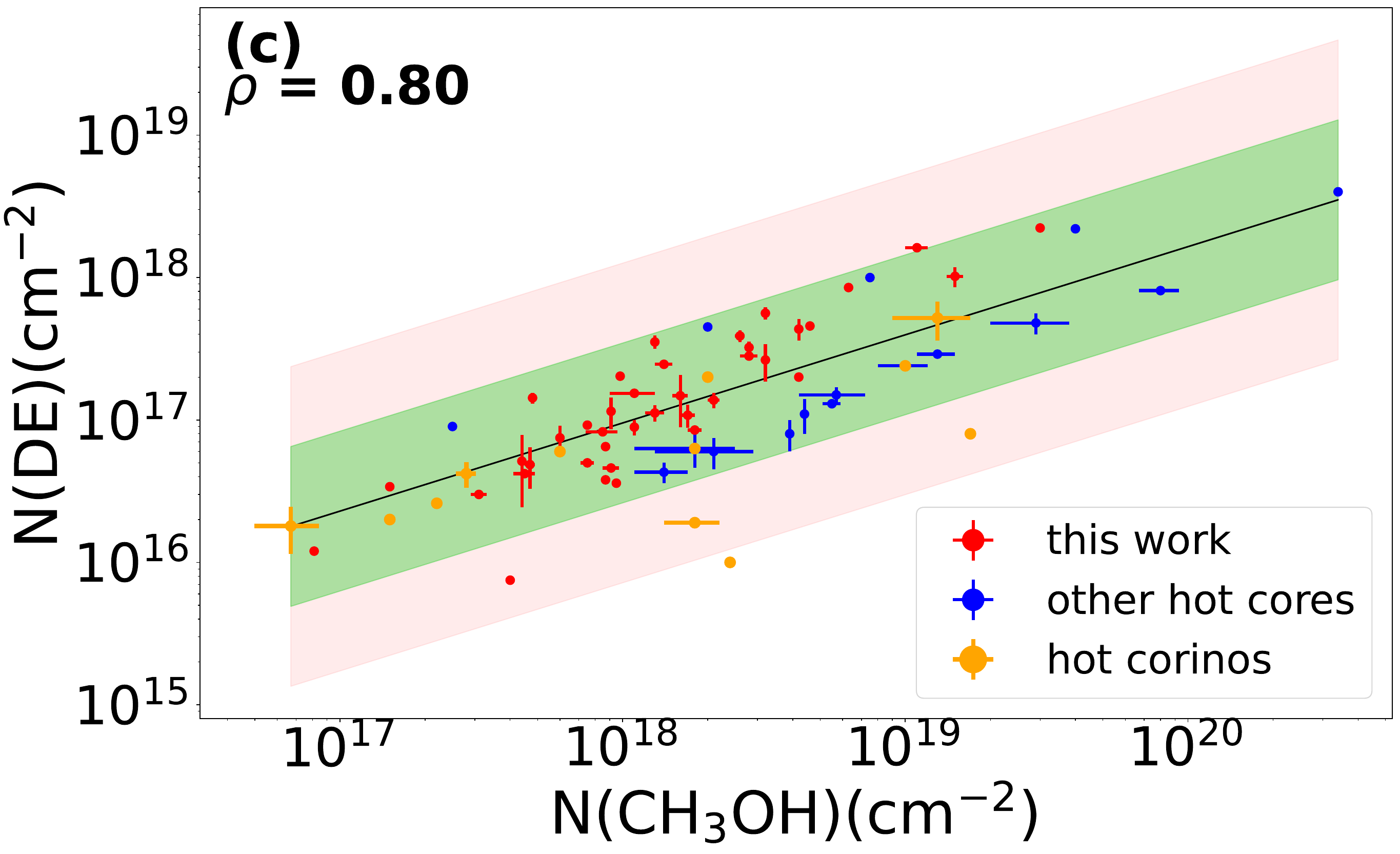}
\end{minipage} 
\caption{\small Left: the column densities of DE vs. the column densities of EA. Middle: the column densities of EA vs. the column densities of CH$_3$OH. Right: the column densities of DE vs. the column densities of CH$_3$OH. Pearson’s co-efficient is displayed in the top left corner of these figures. The linear least-squares fits for all dots are shown as black lines. The limegreen shaded and red shaded regions show a factor of one and two standard deviations from this fit, respectively. The data are derived from Tables \ref{tab:1} and \ref{tab:A.4}.}  
\label{fig:3}
\end{figure*}

\begin{figure*}
\centering
\includegraphics[height=8.77cm,width=14.33cm]{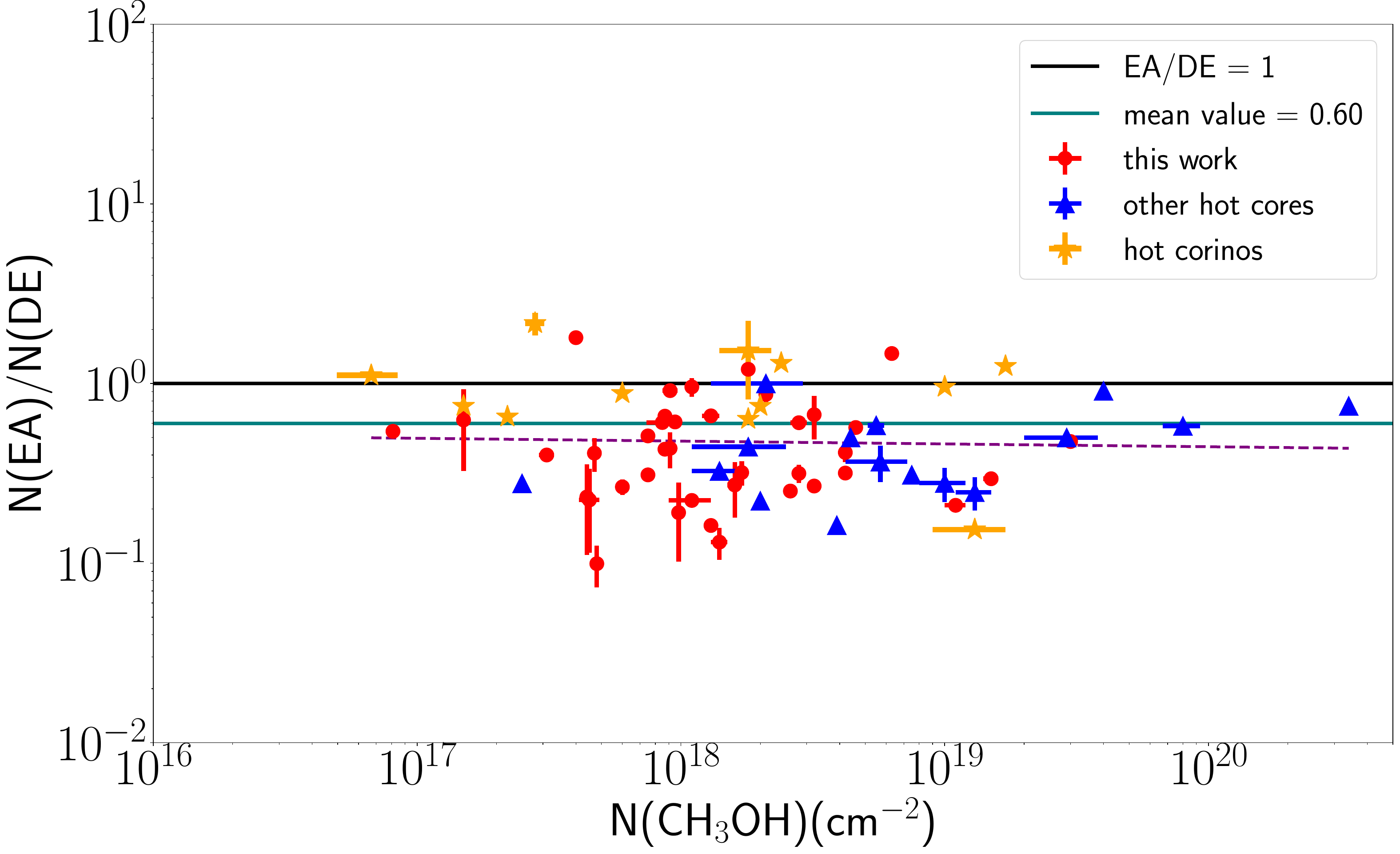} 
\caption{\small The relationship between column densities of CH$_3$OH and the ratios of EA/DE. The teal line shows the mean value of the column density ratios of EA/DE, 0.60. The linear least-squares fit for all dots is shown as a dashed purple line.}
\label{fig:4} 
\end{figure*}   


\begin{figure}
\centering
\begin{minipage}{\linewidth}
\centering
\includegraphics[height=5.5cm,width=8.5cm]{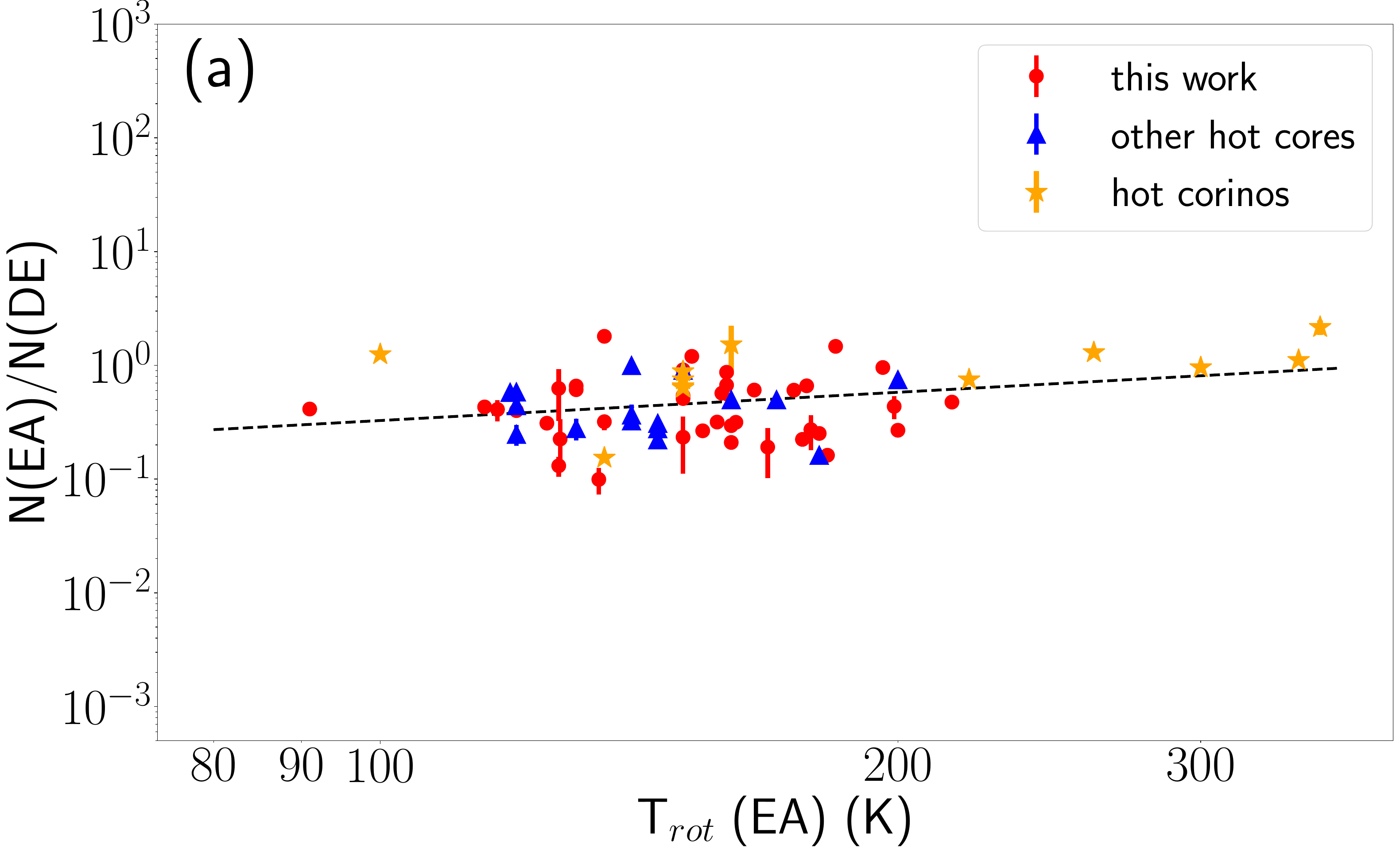}
\end{minipage} 
\vspace{0.5cm} 
\begin{minipage}{\linewidth}
\centering
\includegraphics[height=5.5cm,width=8.5cm]{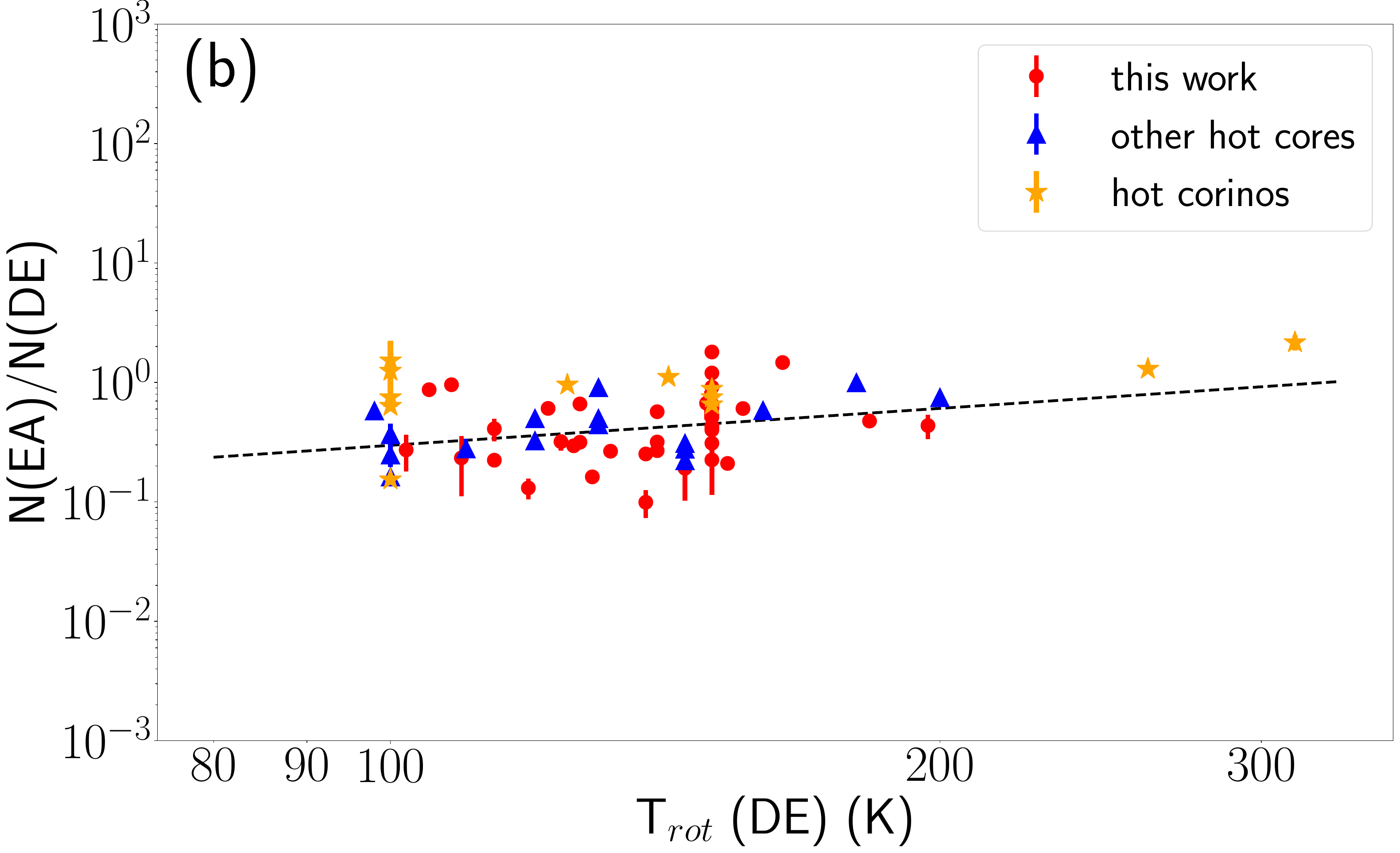}
\end{minipage} 
\caption{\small The relationship between the ratios of EA/DE and rotation temperatures of EA (upper panel), and rotation temperatures of DE (lower panel). The linear least-squares fits are shown as dashed black lines. }
\label{fig:6}
\end{figure}

\subsection{Comparison with other sources and parameter correlations} \label{sec:4.1} 
To further investigate the relationships among methanol, EA, and DE, we take the column densities of methanol in our sample from \cite{2022MNRAS.511.3463Qin} and gather data from other hot cores and hot corinos reported in the literature, where interferometric observations are conducted with reliable source sizes estimated. In Table \ref{tab:A.4}, we summarize the column densities of CH$_3$OH, EA, and DE from the literature. 
We compared the rotation temperature of EA and DE between our sample (25 sources) and the sources from the literature (28 sources).
The average rotation temperatures of EA and DE in our sample are $167\pm14$ K and $131\pm11$ K, while the values from the literature are $171\pm12$ K and $136\pm12$ K, respectively. The comparable rotation temperatures between ours and those in the literature indicate that the measured column densities are consistent. We computed the molecular abundances relative to methanol by f$_{\rm CH_3OH}$ = N(EA or DE)/N(CH$_3$OH), which are listed in Table \ref{tab:1} and Table \ref{tab:A.4}.
The relative abundances range from 0.02 to 0.19 for EA and from 0.06 to 0.29 for DE, respectively, which are in agreement
with previous results in hot cores (from 0.005 to 0.31 for EA and from 0.004 to 0.26 for DE, respectively) and hot corinos (from 0.003 to 0.1 for EA and from 0.01 to 0.36 for DE, respectively).

Exploring the correlations between relative molecular abundances could provide insights into the formation processes of COMs.
Fig.\ref{fig:3}a presents the result of linear least-squares fitting to column densities between EA and DE. 
A positive correlation was found between the column densities of EA and DE, with Pearson’s co-efficient of 0.92. 
97\verb|%| of the sources are within a factor of one deviation from this fit.
Previous observations of EA and DE also presented a strong linear relationship between abundances and column densities by \cite{2007A&A...465..913B} in 7 high-mass YSOs ($\rho$ \verb|>| 0.9), \cite{2020A&A...635A.198Belloche} in 6 solar-type protostars ($\rho$ = 0.99), and \cite{2022ApJ...939...84Baek} in 12 high-mass star-forming regions ($\rho$ = 0.89). 
In Figs.\ref{fig:3}b and \ref{fig:3}c, we perform a linear fitting to examine the relationship among CH$_3$OH, EA, and DE. Strong correlations between CH$_3$OH and EA, as well as CH$_3$OH and DE, are presented with Pearson’s co-efficient of 0.82 and 0.80, spanning 3-4 orders of magnitude in column density (\verb|~|10$^{15}$ - 10$^{18}$  cm$^{-2}$ for EA and DE, \verb|~|10$^{16}$ - 10$^{20}$ cm$^{-2}$ for CH$_3$OH). 
Figure \ref{fig:3}b and c show systematically higher values of DE and EA column densities compared to those of methanol in our work.  
Note that some other studies calculated column densities of methanol using isotope ratios, including $^{13}$CH$_3$OH and CH$_3$$^{18}$OH, whereas we used the main isotopologue, CH$_3$OH, through spectral fitting as discussed in \cite{2022MNRAS.511.3463Qin}.

Figure \ref{fig:4} shows the column density ratios of EA/DE as a function of the column densities of CH$_3$OH. 
In Fig. \ref{fig:4} and the subsequent analysis of Fig. \ref{fig:6} and Fig. \ref{fig:7}, we include all 39 sources from groups A and B, as well as 28 sources from the literature.
Of the 67 sources, we find that in most sources, the column densities of DE are higher than EA. However, there are 11 sources in which the column densities of EA are higher than DE, and one source in which the column densities of EA and DE are equal. 
The column density ratios of EA to DE are within one order of magnitude, ranging from 0.1 to 2.1 with an average value of 0.60.
Despite the column densities of CH$_3$OH spanning 4 orders of magnitude, the column densities of EA and DE exhibit similar column densities within the same source, regardless of hot cores or hot corinos.
The strong correlations we find among CH$_3$OH, EA, and DE ($> 0.8$), along with the nearly constant ratios may hint at a chemical link between EA and DE, which we discuss further in Sect. \ref{sec:4.2}. DE has also been found to be correlated with other O-bearing COMs, such as methyl formate \citep{2020A&A...641A..54Coletta} and ketene \citep{2024MNRAS.533.1583LiChuanshou}.

\subsection{Implications for the chemistry of EA and DE} \label{sec:4.2}
\begin{figure*}
\begin{minipage}{0.15\linewidth}
\includegraphics[height=5.5cm,width=8.5cm]{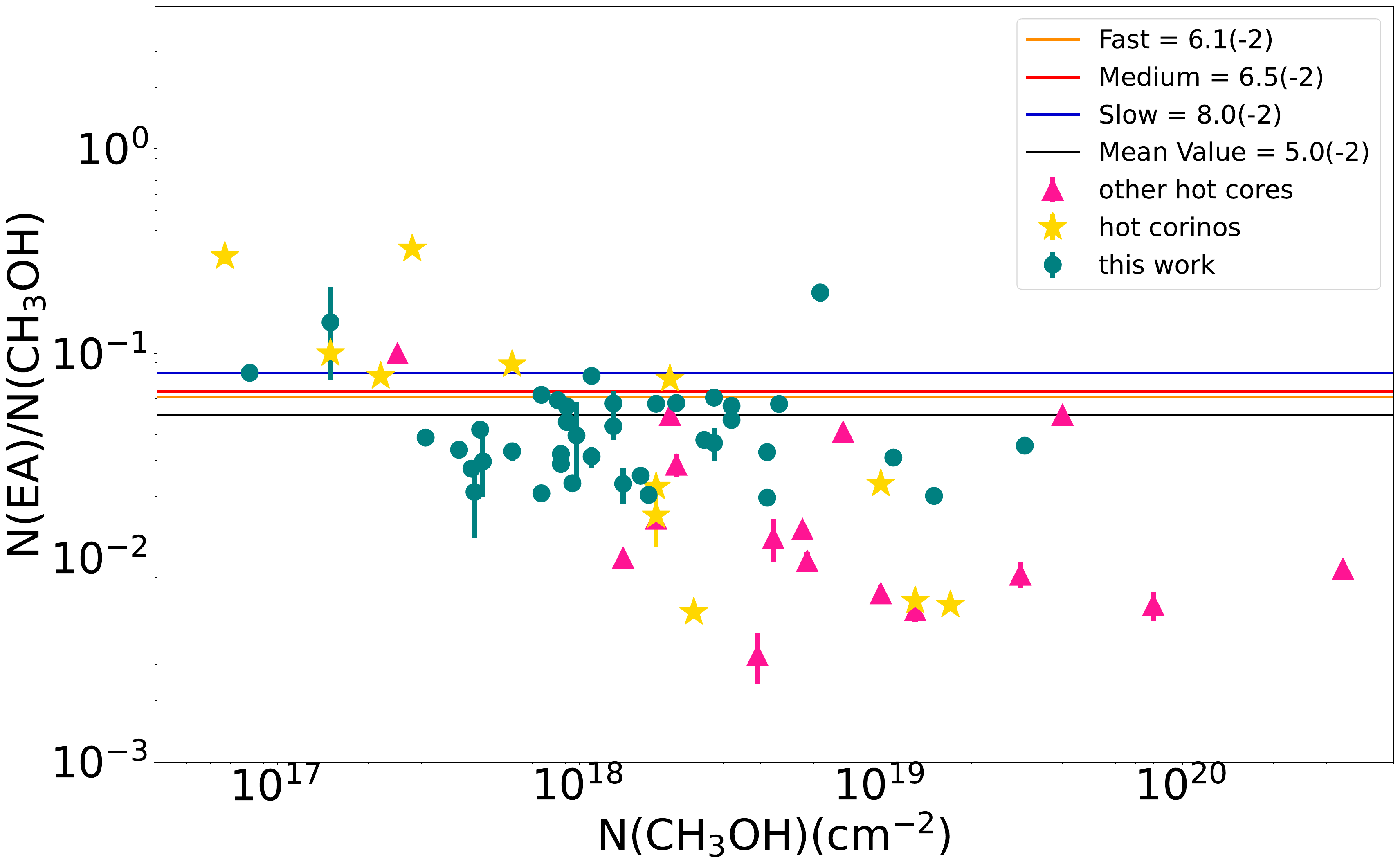}
\end{minipage}
\hfill 
\begin{minipage}{0.5\linewidth}
\centering 
\includegraphics[height=5.5cm,width=8.5cm]{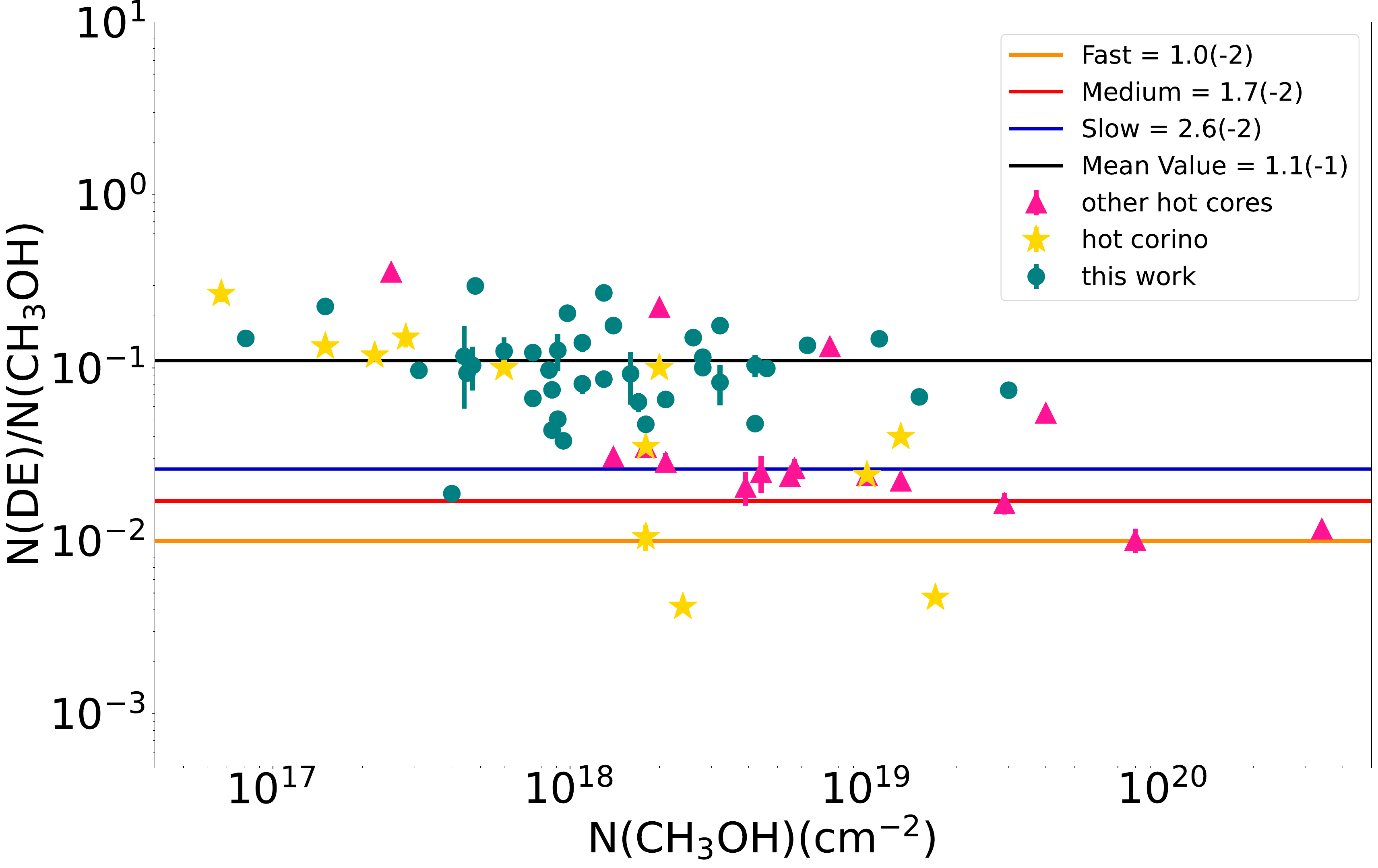}
\end{minipage}  
\caption{\small Left: The relationship between the column densities of CH$_3$OH and the column densities ratios of EA/CH$_3$OH. Right: The relationship between the column densities of CH$_3$OH and the column densities ratios of DE/CH$_3$OH. The black lines show the mean of the column density ratios of EA/CH$_3$OH and DE/CH$_3$OH. The darkorange, red, and blue represent the different warm-up timescales from simulations ~\protect\citep{2022ApJS..259....1G}. (Fast: $5 \times 10^4$ yr, Medium: $2 \times 10^5$ yr, Slow: $ 1\times 10^6$ yr.)}
\label{fig:7}
\end{figure*}
The strong linear correlations of the column densities of CH$_3$OH, EA and DE (Fig. \ref{fig:3}), along with the nearly constant ratios (Fig. \ref{fig:4}), may indicate chemical links among CH$_3$OH, EA and DE.
Experimental studies showed that EA and DE are formed via radical–radical recombination involving CH$_3$OH and its photodissociation products (hydroxymethyl radical (CH$_2$OH), methoxy radical (CH$_3$O), and methyl radical (CH$_3$)) \citep{2009A&A...504..891O, 2018ApJ...852...70Bergan, 2021ApJ...913...61Yocum}

\begin{equation}
\begin{aligned}
    \ce{CH3 + CH2OH -> CH3CH2OH, and}
\end{aligned}
\end{equation}
\begin{equation}
\begin{aligned}
    \ce{CH3 + CH3O -> CH3OCH3.}
\end{aligned}
\end{equation}

Although experimentally determined ratios depend on the ice composition, total UV fluence, heating rate, and other parameters, the ratios for radical–radical recombination yields relative abundances of EA and DE within the same order of magnitude. 
Notably, EA/DE abundance ratios remain consistent regardless of fluences, temperatures, and ice compositions when radical–radical recombination (CH$_3$O, CH$_2$OH, and CH$_3$) is considered \citep{2009A&A...504..891O}. 

Different types of sources hint at different physical environments. Nevertheless, the EA/DE ratios of our sample remain within one order of magnitude regardless of whether the sources are hot corinos or hot cores. In addition to Fig. \ref{fig:4} of Sect. \ref{sec:4.1} showing the relationship between the column density ratios of EA/DE and the column densities of CH$_3$OH, Fig. \ref{fig:6}a and b show the relationship between the column density of ratios of EA/DE and the rotation temperatures of both EA and DE. The ratios of EA/DE do not show an obvious upward or downward trend with the rotation temperatures of both EA and DE. 
The nearly stable ratios suggest that the EA/DE ratio does not depend on the physical environment. If reactions involving the radical products of CH$_3$OH dissociation are the dominant production mechanism for EA and DE in the hot cores and hot corinos, with other mechanisms that do not affect their abundance by more than an order of magnitude, the relatively stable EA/DE ratios observed in our sample can be explained.

Figures \ref{fig:7} a and b compare the column density ratios of EA and DE with respect to CH$_3$OH against chemical models \citep[hereafter G22]{2022ApJS..259....1G}. 
The G22 models include three phases: bulk ice mantle, ice surface, and gas phase. The evolution of the hot core in the G22 model has two stages: a cold collapse stage and a static (fixed density) warm-up stage. Three warm-up timescales are employed in the G22 models: 5 $\times$ 10$^4$ yr (fast), 2 $\times$ 10$^5$ yr (medium), and 1 $\times$ 10$^6$ yr (slow). When compared the observed EA and DE abundances relative to CH$_3$OH with the predicted values from the fast, medium and slow models, we find a systematic underproduction of DE and a slight overproduction of EA in the chemical models of G22. 
In the chemical models of G22, 60\verb|%| of DE mainly forms in association with CH$_2$ and CH$_3$OH in ices via the reactions:
\begin{equation}
    \ce{CH2 + CH3OH -> CH3OCH3}
\end{equation}
and
\begin{equation}
    \ce{CH2 + CH3O -> CH3OCH2}
\end{equation}
\begin{equation}
    \ce{H + CH3OCH2 -> CH3OCH3}
\end{equation}
CH$_3$O is formed by the hydrogenation of H$_2$CO or H abstraction from
CH$_3$OH. While 40\verb|%| of DE is formed from CH$_3$OH, protonated CH$_3$OH and ammonia in the gas phase  via the reactions:
\begin{equation}
    \ce{CH3OH + CH3OH2+ -> CH3OCH4+ + H2O}
\end{equation}
and
\begin{equation}
    \ce{CH3OCH4+ + NH3 -> CH3OCH3 + NH4+}.
\end{equation}

\cite{Jin_2020} suggested the formation of DE through the reaction of methanol with ground methylene (CH$_2$), given that the inclusion of non-diffusive grain-surface reaction can not reproduce the observed abundance of DE in prestellar cores.
However, when G22 models account for non-diffusive grain-surface chemical processes and reactions of methanol with ground-state methylene, these mechanisms remain insufficient to reproduce the observed gas-phase abundance of DE. We propose that DE formation in chemical models involving ground-state CH$_2$ reactions may not represent the main pathway.
There are two main production pathways for EA in the chemical models of G22: C$_2$H$_5$ radicals react with the photolysis product OH of water and CH$_2$ reacts with CH$_3$OH on dust grains in the early cold collapse stage via:
\begin{equation}
    \ce{C2H5 + OH -> C2H5OH}
\end{equation}
and 
\begin{equation}
    \ce{CH2 + CH3OH -> C2H5OH}.
\end{equation}
Thus, we can find a correlation between EA and DE by CH$_3$OH in the chemical model. 
The efficiency of the methylene reactions with methanol and H-atom abstraction or addition to hydrocarbons in the bulk ice is not well constrained in the models.
We suggest that reactions involving the radical products of CH$_3$OH dissociation should be taken into account in the formation of EA and DE within chemical models.

\section{Conclusions} \label{sec:Con.}
We present a large sample hot core survey focusing on dimethyl ether (CH$_3$OCH$_3$; DE) and ethanol (C$_2$H$_5$OH; EA) towards 60 hot cores from the ATOMS project. We derived the rotation temperatures and the column densities for the 60 sources. We analyse column density correlations and molecular ratios between EA and DE. Our conclusions are as follows:

1. EA is detected in 40 hot cores and DE is detected in 59 hot cores among 60 hot cores. Both EA and DE are detected in 39 hot cores.

2. The average rotation temperature of EA is higher than DE, with EA having a temperature of $167\pm14$ K compared to $136\pm12$ K of DE. The average column density of DE is slightly higher than EA, with DE having a column density of $4.2\pm0.1$ $\times$ 10$^{17}$ cm$^{-2}$ compared to $1.9\pm0.1$ $\times$ 10$^{17}$ cm$^{-2}$ of EA.

3. We find a very strong correlation ($\rho$ = 0.92) between the column densities of EA and DE, and a very strong correlation ($\rho$ \verb|>| 0.80) between the column densities of CH$_3$OH and EA, as well as CH$_3$OH and DE.
We also find that the column densities of EA and DE are within the same order of magnitude in the same source, with an average EA/DE ratio of 0.60.
These correlations and nearly constant ratio hint at a potential chemical link between EA and DE via CH$_3$OH, as suggested by previous experiments and models. 

4. The comparison with experiments suggests that reactions involving the radical products of CH$_3$OH dissociation may be the predominant mechanism for the production of EA and DE.

5. Compared with chemical models, the abundance of EA is systematically overproduced while the abundance of DE is systematically underproduced in chemical models. Our large hot core sample can provide crucial constraints for chemical models.

ALMA line surveys towards a large sample of hot cores provide insight into the formation of COMs. In the future, JWST mid-infrared observations of solid-phase COMs are needed to study their  transition from the ice phase to the gas phase to  constrain the reaction pathways.

\section*{Acknowledgements}
We would like to thank the anonymous referee for their constructive comments. This work has been supported by the National Key R\&D Program of China (No. 2022YFA1603100), the National Natural Science Foundation of China (grant Nos. 12033005, 12073061, 12122307, and 12473025), the Tianchi Talent Program of Xinjiang Uygur Autonomous Region, the Natural Science Foundation of Xinjiang Uygur Autonomous Region (2024D01E37), and Natural Science Foundation of Xinjiang Uygur Autonomous Region (2022D01B221).
P. Garc\'ia is sponsored by the Chinese Academy of Sciences (CAS), through a grant to the CAS South America Center for Astronomy (CASSACA).
SRD acknowledges support from the Fondecyt Postdoctoral fellowship (project code 3220162) and ANID BASAL project FB210003.
H.-L. Liu is supported  by Yunnan Fundamental Research Project (grant No.\,202301AT070118, 202401AS070121), and by Xingdian Talent Support Plan--Youth Project.
GG and LB gratefully acknowledge support by the ANID BASAL project FB210003.
This work is sponsored in part by the Chinese Academy of Sciences (CAS), through a grant to the CAS South America Center for Astronomy (CASSACA) in Santiago, Chile. This paper makes use of the following ALMA data: ADS/JAO.ALMA\#2019.1.00685.S. ALMA is a partnership of ESO (representing its member states), NSF (USA), and NINS (Japan), together with NRC (Canada), MOST and ASIAA (Taiwan), and KASI (Republic of Korea), in cooperation with the Republic of Chile. The Joint ALMA Observatory is operated by ESO, AUI/NRAO, and NAOJ.
%
\section*{Data Availability}
The raw data are available in ALMA archive. 
\bibliographystyle{mnras}
\bibliography{References} 
\appendix
\section{Additional tables} \label{app:A}
Table \ref{tab:A.1}-\ref{tab:A.2} lists the identified transitions of C$_2$H$_5$OH  and CH$_3$OCH$_3$ focused on in this paper. Table \ref{tab:A.4} gives source parameters from the literature used for comparison in Section \ref{sec:4.1}. 

\onecolumn
\begin{longtable}{cccccc}
\caption{\normalsize Identified transitions of Ethanol} \label{tab:A.1} \\
\hline
\multicolumn{1}{c}{Species} & \multicolumn{1}{c}{Transition} & \multicolumn{1}{c}{Frequency (MHz)} & \multicolumn{1}{c}{$E_{\rm up}$ (K)} & \multicolumn{1}{c}{Log$_{10}$($A_{\rm ij}$) (s$^{-1}$)} & Comments \\
\multicolumn{1}{c}{Database} &  &  &  & & \\
\hline
\endfirsthead

\multicolumn{5}{c}
{\tablename\ \thetable\ -- \textit{Continued from previous page}} \\
\hline
\multicolumn{1}{c}{Species} & \multicolumn{1}{c}{Transition} & \multicolumn{1}{c}{Frequency (MHz)} & \multicolumn{1}{c}{$E_{\rm up}$ (K)} & \multicolumn{1}{c}{Log$_{10}$($A_{\rm ij}$) (s$^{-1}$)} & Comments \\
\multicolumn{1}{c}{Database} &  &  &  & &  \\
\hline
\endhead

\hline \multicolumn{5}{r}{\textit{Continued on next page}} \\
\endfoot

\endlastfoot
C$_2$H$_5$OH  & 21(1,21)-21(0,21), g- & 97535.9816 & 246.22202 & -4.94163 \\
CDMS          & 26(0,26)-26(1,26), g- & 97549.7236 & 340.71828 & -4.93098 \\
              & 24(1,24)-24(0,24), g- & 97562.8607 & 300.59403 & -4.93403 \\
              & 20(1,20)-20(0,20), g- & 97574.0683 & 229.65032 & -4.94427 \\
              & 25(1,25)-25(0,25), g-$^{\star}$ & 97600.41610 & 320.26966 & -4.93155 & blended with CH$_3$OCHO, v=1 \\
              & 27(0,27)-27(1,27), g-$^{\star}$ & 97631.31940 & 361.94530 & -4.92798 & blended with SO$_2$ \\
              & 26(1,26)-26(0,26), g- & 97645.16580 & 340.72070 & -4.92976 \\
              & 19(1,19)-19(0,19), g-$^{\star}$ & 97649.54140 & 213.85535 & -4.94678 & blended with CH$_3$OCHO \\
              & 27(1,27)-27(0,27), g-$^{\star}$ & 97698.52110 & 361.94694 & -4.92708 & blended with SO$_2$ \\ 
              & 28(0,28)-28(1,28), g-$^{\star}$ & 97708.86540 & 383.94705 & -4.92580 & blended with CH$_3$OCHO, v=1 \\
              & 28(1,28)-28(0,28), g-$^{\star}$ & 97755.57550 & 383.94815 & -4.92519 & blended with CH$_3$OCHO, v=1 \\              
              & 18(1,18)-18(0,18), g- & 97774.34830 & 198.83725 & -4.94931 \\              
              & 29(0,29)-29(1,29), g- & 97784.08670 & 406.72349 & -4.92467 \\  
              & 29(1,29)-29(0,29), g- & 97815.93060 & 406.72430 & -4.92423 \\              
              & 30(1,29)-30(2,29), g-$^{\star}$ & 97828.95140 & 444.44702 & -5.01714 & blended with aGg-(CH$_2$OH)$_2$ \\
              & 30(0,30)-30(1,30), g-$^{\star}$ & 97857.41130  & 430.27447 & -4.94433 & blended with CH$_3$OCHO, v=1 \\     
              & 30(1,30)-30(0,30), g- & 97877.42120 & 430.27514 & -4.94404 \\              
              & 31(0,31)-31(1,31), g-$^{\star}$ & 97932.39970 & 454.59989 & -4.94980 & blended with CH$_3$OCHO, v=1 \\              
              & 17(1,17)-17(0,17), g- & 97962.87620 & 184.59670 & -4.95176 \\
              & 31(1,31)-31(0,31), g- $^{\star}$ & 97998.07440 & 454.60031 & -4.95361 & blended with CH$_3$OCH$_3$ \\ 
              & 4(3,1)-3(2,1), g+  & 98005.46540 & 76.16202 & -5.38642 \\  
              & 32(1,32)-32(0,32), g- & 98021.3226 & 479.69975 & -5.43074 \\ 
              & 32(0,32)-32(1,32), g- & 98025.1344 & 479.70051 & -5.44123 \\              
              & 33(0,33)-33(1,33), g- & 98056.16590 & 505.57311 & -4.96820 \\              
              & 31(1,30)-31(2,30), g- & 98060.67790 & 469.29102 & -5.01037 \\              
              & 33(1,33)-33(0,33), g- & 98091.86630 & 505.57338 & -4.96714 \\              
              & 12(2,10)-11(3,9), g- & 98157.69100 & 132.22948 & -7.63129 \\
              & 34(1,34)-34(0,34), g-$^{\star}$ & 98163.25800 & 532.22083 & -5.09402 & blended with $^{13}$CH$_3$CH$_2$CN \\   
              & 34(0,34)-34(1,34), g-$^{\star}$ & 98163.50680  & 532.22070 & -5.09412 & blended with $^{13}$CH$_3$CH$_2$CN \\
              & 4(3,2)-3(2,2), g- & 98173.82080 & 681.52515 & -5.07476 \\
              & 38(2,36)-38(3,36), g- & 98178.82080 & 681.52515 & -5.07476 \\
              & 16(1,16)-16(0,16), g- & 98230.35070 & 171.13372 & -4.95442 \\  
              & 35(0,35)-35(1,35), g-$^{\star}$ & 98234.35230 & 559.64175 & -5.01013 & blended with CH$_3$$^{13}$CH$_2$CN and CH$_3$COCH$_3$ \\
              & 35(1,35)-35(0,35), g-$^{\star}$ & 98237.47890 & 559.64190 & -5.01011 & blended with CH$_3$$^{13}$CH$_2$CN and CH$_3$COCH$_3$ \\                                 
              & 32(1,31)-32(2,31), g- & 98274.09660 & 494.90763 & -5.00682\\              
              & 36(1,36)-36(0,36), g-$^{\star}$ & 98312.02280 & 587.83648 & -4.93783 & blended with CH$_3$COCH$_3$ \\              
              & 37(0,37)-37(1,37), g-$^{\star}$ & 98386.49140 & 616.80412 & -4.91841 & blended with $^{33}$SO \\
              & 37(1,37)-37(0,37), g- & 98388.08630 & 616.80420 & -4.91840 \\
              & 33(1,32)-33(2,32), g- & 98440.56050 & 521.29706 & -4.99655 \\ 
              & 38(0,38)-38(1,38), g-$^{\star}$ & 98464.38620 & 646.54498 & -4.90874 & blended with CH$_3$COCH$_3$ \\ 
              & 38(1,38)-38(0,38), g-$^{\star}$ & 98465.57970 & 646.54504 & -4.90874 & blended with CH$_3$COCH$_3$ \\     
              & 15(1,15)-15(0,15), g- & 98585.09460 & 158.44813 & -4.95888 \\
              & 34(1,33)-34(2,33), g- & 98602.78900 & 548.45978 & -4.99259 \\
              & 40(0,40)-40(1,40), g- & 98634.20 & 708.377 & -5.1738 \\
              & 40(0,40)-40(1,40), g- & 98634.20 & 708.377 & -5.1738 \\
              & 41(1,41)-41(0,40), g- & 98638.54400 & 708.337 & -5.1738 \\
              & 30(2,29)-30(1,29), g- & 98755.43430 & 444.47033 & -5.00494 \\
              & 36(1,35)-36(2,35), g- & 98801.0158 & 605.10107 & -4.9967 \\              
              & 13(1,13)-13(0,13), g-$^{\star}$ & 98878.3058 & 135.37443 & -5.21771 & blended with CH$_3$OCHO and CH$_3$COCH$_3$ \\ 
              & 35(2,34)-35(1,34), g- & 98931.05140 & 576.39820 & -4.98832 \\
              & 28(2,27)-28(1,27), g- & 98946.2593 & 397.11968 & -5.01351 \\              
              & 28(2,27)-28(1,27), g- & 98946.2593  & 397.11968 & -5.01351 \\              
              & 14(1,14)-14(0,14), g- & 98983.55600 & 146.53742 & -4.97592 \\
              & 13(4,9)-12(5,7), g- & 99109.6943  & 156.94978 & -6.14826 \\              
              & 27(2,26)-27(1,26), g- & 99143.6938 & 374.60755 & -5.01752 \\              
              & 38(2,37)-38(1,37), g- & 99181.373  & 664.83514	& -5.02364 \\             
              & 39(2,38)-39(1,38), g- & 99268.001  & 695.85924 & -4.98139 \\              
              & 26(2,25)-26(1,25), g- & 99439.9777 & 352.87178 & -5.02102 \\            
              & 17(3,14)-17(2,15), anti & 99524.0608 & 141.33193 & -5.02148 \\ 
              & 42(1,41)-42(2,41), g- & 99544.0206 & 793.56436 & -5.03044 \\                          
              & 25(2,24)-25(1,24), g- & 99864.37230 & 331.91320 & -5.02382 \\
              & 18(3,15)-18(2,16), anti$^{\star}$ & 99975.86150 & 156.85020 & -5.00820 & blended with NH$_2$CN \\ 
              & 6(1,6)-5(1,5), g+ & 100194.32240 & 74.95922 & -5.07575 \\
              & 16(3,13)-16(2,14), anti & 100358.93690 & 126.70774 & -5.02380 \\           
              & 6(1,6)-5(1,5), g- & 100365.04230 & 79.63728 & 	-5.02354\\              
              & 24(2,23)-24(1,23), g- $^{\star}$ & 100452.03290 & 	311.73236 & -5.02563 & blended with CH$_3$OCHO, v=1 \\  
              & 15(2,14)-14(3,11), anti$^{\star}$ & 100660.16550 & 104.97351 & -5.91482 & blended with CH$_3$OCHO \\
              & 8(2,7)-8(1,8), anti$^{\star}$ & 100990.11310 & 35.17324 & -5.28474 & blended with CH$_3$OCHO, v=1 \\            
              & 23(2,22)-23(1,22), g-$^{\star}$ & 101243.61200  & 292.33033 &	-5.02624 & blended with CH$_3$CHO \\             
\hline 
\multicolumn{5}{l}{The transition marked with $\star$ represents the EA transition blended with other molecules.}\\
\end{longtable}

\begin{longtable}{cccccc}
\caption{\normalsize Identified transitions of Dimethyl ether} \label{tab:A.2} \\
\hline
\multicolumn{1}{c}{Species} & \multicolumn{1}{c}{Transition} & \multicolumn{1}{c}{Frequency (MHz)} & \multicolumn{1}{c}{$E_{\rm up}$ (K)} & \multicolumn{1}{c}{Log$_{10}$($A_{\rm ij}$) (s$^{-1}$)} & Comments \\
\multicolumn{1}{c}{Database} &  &  &  & &  \\
\hline 
\endfirsthead

\multicolumn{5}{c}%
{{\tablename\ \thetable{} -- continued from previous page}} \\
\hline
\multicolumn{1}{c}{Species} & \multicolumn{1}{c}{Transition} & \multicolumn{1}{c}{Frequency (MHz)} & \multicolumn{1}{c}{$E_{\rm up}$ (K)} & \multicolumn{1}{c}{Log$_{10}$($A_{\rm ij}$) (s$^{-1}$)} & Comments \\
\multicolumn{1}{c}{Database} &  &  &  & & \\
\hline 
\endhead

\hline \multicolumn{5}{r}{{Continued on next page}} \\
\endfoot

\endlastfoot
CH$_3$OCH$_3$ & 16(3,14)-15(4,11)AA$^{\star}$ & 97990.62900 & 136.60559 & -5.89847 & blended with SO$_2$ and CS  \\
CDMS          & 16(3,14)-15(4,11)EE$^{\star}$ & 97993.38200 & 136.60558 & -5.89837  & blended with SO$_2$ and CS  \\
              & 16(3,14)-15(4,11)EA $^{\star}$ & 97996.09800 & 136.60556 & -5.89840 & blended with C$_2$H$_5$OH  \\
              & 16(3,14)-15(4,11)AE$^{\star}$ & 97996.17400 & 136.60557 & -5.89839 & blended with C$_2$H$_5$OH  \\
              & 20(8,12)-21(9,13)AA$^{\star}$ & 98278.87100 & 304.93929 & -6.06688 & blended with CH$_3$OCHO \\              
              & 20(8,13)-21(9,13)AA$^{\star}$ & 98278.89000 & 304.93929 & -6.06713 & blended with CH$_3$OCHO \\ 
              & 20(8,12)-21(9,12)EE$^{\star}$ & 98278.89700 & 304.93929 & -6.06700 & blended with CH$_3$OCHO \\               
              & 20(8,13)-21(9,13)EA$^{\star}$ & 98278.91200 & 304.93929 & -6.06724 & blended with CH$_3$OCHO \\              
              & 20(8,13)-21(9,13)EE$^{\star}$ & 98281.10500 & 304.93925 & -6.06699 & blended with CH$_3$OCHO \\               
              & 20(8,12)-21(9,13)AE$^{\star}$ & 98281.11200 & 304.93925 & -6.06702 & blended with CH$_3$OCHO \\               
              & 20(8,13)-21(9,12)AE$^{\star}$ & 98281.13100 & 304.93925 & -6.06670 & blended with CH$_3$OCHO \\ 
              & 20(8,12)-21(9,12)EA$^{\star}$ & 98283.33000 & 304.93922 & -6.06662 & blended with CH$_3$OCHO \\     
              & 25(6,20)-24(7,17)AA & 98934.79400 & 346.86981 & -5.90165 \\              
              & 25(6,20)-24(7,17)EE & 98935.19400 & 346.86983 & -5.90452 \\              
              & 25(6,20)-24(7,17)EA & 98935.51900 & 346.86840 & -5.91226 \\
              & 25(6,20)-24(7,17)AE & 98935.67200 & 346.86841 & -5.90143 \\ 
              & 25(6,20)-24(7,18)EE & 98941.21600 & 346.86983 & -8.09800 \\              
              & 25(6,20)-24(7,18)EA & 98941.76800 & 346.86842 & -7.52693 \\              
              & 25(6,19)-24(7,17)EE & 99177.14200 & 346.88130 & -8.09585 \\              
              & 25(6,19)-24(7,17)EA & 99177.47200 & 346.88131 & -7.52489 \\              
              & 25(6,19)-24(7,18)AA & 99182.68200 & 346.88127 & -5.89851 \\
              & 25(6,19)-24(7,18)EE & 99183.16400 & 346.88130 & -5.90113 \\   
              & 25(6,19)-24(7,18)AE & 99183.56800 & 346.88132 & -5.89846 \\
              & 25(6,19)-24(7,18)EA & 99183.72100 & 346.88132 & -5.90886 \\  
              & 4(1,4)- 3(0,3)EA$^{\star}$ & 99324.36200 & 10.21395 & -5.25730 & blended with aGg-(CH$_2$OH)$_2$ \\
              & 4(1,4)- 3(0,3)AE & 99324.36400 & 10.21395 & -5.25729 \\             
              & 4(1,4)- 3(0,3)EE & 99325.21700 & 10.21255 & -5.25726\\              
              & 4(1,4)- 3(0,3)AA & 99326.07200 & 10.21245 & -5.25724 \\
              & 23(9,14)-24(10,15)AA & 99607.63600 & 391.66490 & -6.03791 \\              
              & 23(9,15)-24(10,14)AA & 99607.64000 & 391.66490 & -6.03796 \\
              & 23(9,14)-24(10,14)EE & 99608.49400 & 391.66494 & -6.03788 \\              
              & 23(9,15)-24(10,15)EA & 99609.35000 & 391.66498 & -6.03792 \\              
              & 23(9,15)-24(10,15)EE & 99609.61800 & 391.66500 & -6.03787 \\              
              & 23(9,14)-24(10,15)AE & 99610.47200 & 391.66504 & -6.03788 \\              
              & 23(9,15)-24(10,14)AE & 99610.47600 & 391.66504 & -6.03790 \\
              & 23(9,14)-24(10,14)EA & 99611.59800 & 391.66495 & -6.03791 \\ 
              & 14(2,13)-13(3,10)AA & 99833.66900 & 100.60712 & -5.93844 \\
              & 14(2,13)-13(3,10)EE & 99836.45200 & 100.60711 & -5.93834 \\             
              & 14(2,13)-13(3,10)EA & 99839.22800 & 100.60710 & -5.93827\\              
              & 14(2,13)-13( 3,10)AA & 99326.07200 & 100.60710 & -5.93836 \\  
              & 22(5,18)-21(6,15)AA $^{\star}$ & 100434.16900 & 265.87335 & -5.87452 & blended with CH$_3$COOH and CH$_3$COCH$_3$ \\     
              & 22(5,18)-21(6,15)EE $^{\star}$ & 100435.46400 & 265.87327 & -5.87493 & blended with CH$_3$COOH and CH$_3$COCH$_3$ \\                      
              & 22(5,18)-21(6,15)EA $^{\star}$ & 100436.70700 & 265.87333 & -5.87597& blended with CH$_3$COOH and CH$_3$COCH$_3$ \\                 
              & 22(5,18)-21(6,15)AE $^{\star}$ & 100436.81300 & 265.87334 & -5.87446 & blended with CH$_3$COOH and CH$_3$COCH$_3$ \\        
              & 6(2,5)- 6(1,6)EA & 100460.41300& 24.70684 & -5.36012 \\
              & 6(2,5)- 6(1,6)AE & 100460.43700 & 24.70684 & -5.36021 \\             
              & 6(2,5)- 6(1,6)EE$^{\star}$ & 100463.07500 & 24.70682 & -5.36007 & blended with HC$_3$N, v$_7$=1 \\               
              & 6(2,5)- 6(1 6)AA$^{\star}$ & 100465.72600 & 24.70680 & -5.36004 & blended with HC$_3$N, v$_7$=1 \\               
              & 26(10,16)-27(11,17)AA & 100918.26400 & 489.36427 & -6.01218 \\              
              & 26(10,17)-27(11,16)AA & 100918.26500 & 489.36427 & -6.01233 \\
              & 26(10,16)-27(11,16)EE & 100919.35700 & 489.36432 & 	-6.01219 \\              
              & 26(10,17)-27(11,17)EE & 100919.55700 & 489.36433 & 	-6.01219 \\              
              & 26(10,17)-27(11,17)EA & 100920.44900 & 489.36438 &	-6.01203 \\              
              & 26(10,16)-27(11,17)AE & 100920.64900 & 	489.36424 & -6.01232 \\              
              & 26(10,17)-27(11,16)AE & 100920.65000 & 	489.36424 & -6.01280 \\
              & 26(10,16)-27(11,16)EA & 100920.85000 & 	489.36425 & -6.01202 \\             
              & 19(4,16)-18(5,13)AA & 100946.83600 & 195.82029 & -5.86034 \\
              & 19(4,16)-18(5,13)EE & 100949.00300 & 195.82025 & -5.86030 \\             
              & 19(4,16)-18(5,13)EA & 100951.09100 & 195.82021 & -5.86063\\              
              & 19(4,16)-18(5,13)AE & 100951.24900 & 195.82021 & -5.86010\\   
\hline 
\multicolumn{5}{l}{The transition marked with $\star$ represents the DE transition blended with other molecules.}\\
\end{longtable}

\begin{landscape}
\begin{small}
\begin{longtable}{ccccccccccc}
\caption{\normalsize Other sources from literature of CH$_3$OH, DE and EA} \label{tab:A.4} \\
\hline
\multicolumn{1}{c}{Source} & \multicolumn{1}{c}{Source} & \multicolumn{1}{c}{$N_{\rm methanol}$ $\times10^{16}$ (cm$^{-2}$)} & T$_{\rm EA}$ (K) & \multicolumn{1}{c}{$N_{\rm EA}$ $\times10^{16}$ (cm$^{-2}$)}  &  $f_{\rm CH_3OH}$ & T$_{\rm DE}$ (K) & \multicolumn{1}{c}{$N_{\rm DE}$$\times10^{16}$(cm$^{-2}$)} &  $f_{\rm CH_3OH}$  & \multicolumn{1}{c}{EA/DE} & \multicolumn{1}{c}{Refs.} \\
\hline
\endfirsthead

\multicolumn{9}{c}
{{\tablename\ \thetable{} -- continued from previous page}} \\
\hline
\multicolumn{1}{c}{Source} & \multicolumn{1}{c}{Source} & \multicolumn{1}{c}{$N_{\rm methanol}$ $\times10^{16}$ (cm$^{-2}$)} & T$_{\rm EA}$ (K) & \multicolumn{1}{c}{$N_{\rm EA}$ $\times10^{16}$ (cm$^{-2}$)}  &  $f_{\rm CH_3OH}$ & T$_{\rm DE}$ (K) & \multicolumn{1}{c}{$N_{\rm DE}$$\times10^{16}$(cm$^{-2}$)} &  $f_{\rm CH_3OH}$  & \multicolumn{1}{c}{EA/DE} & \multicolumn{1}{c}{Refs.} \\
\hline
\endhead

\hline \multicolumn{9}{r}{{Continued on next page}} \\
\endfoot
 
\endlastfoot
Hot corino & B1-c & ($^{18}$O) $180\pm40$ & $160\pm60$ & $2.9\pm1.5$ & $0.016\pm0.004$ & $100\pm10$ & $1.9\pm0.1$ & $0.01\pm0.001$ & 1.05 & 1 \\
 & L 483 & (f) $1700$ & 100 & $10$ & $0.006$ & 100 & $8$ & $0.005$ & 1.25 & 2 \\
 & BRH 71 IRS1 & (f) $240$ & 260 & $1.3$ & $0.005$ & 260 & $1.0$ & $0.004$ & 1.3 &  3 \\
 & IRAS 16293A & (f) $1300\pm400$ & $135\pm27$ & $8.0\pm2.4$ & $0.006$ & $100\pm20$ & $52\pm16$ & $0.04$ & 0.47 & 4 \\
 & IRAS 16293B & ($^{18}$O) $1000$ & $300$ & $23$ & $0.023$ & $125$ & $24$ & $0.024$ & 0.95 & 5 \\
 & G211.47-19.27S & (f) $28\pm2.3$ & $352\pm4$ & $9.1\pm0.57$ & $0.325\pm0.006$ & $313\pm3$ & $4.2\pm0.87$ & $0.15\pm0.018$ & 2.14 & 6 \\
 & G208.68-19.20N1 & (f) $6.7\pm1.73$ & $342\pm6$ & $2.0\pm0.68$ & $0.298\pm0.024$ & $142\pm7$ & $1.8\pm0.645$ & $0.269\pm0.027$ & 1.12 & 6 \\
 & IRAS 2A1 & (f) $180$ & $150$ & $4.0$ & $0.222$ & $100$ & $6.3$ & $0.035$ & 0.63 & 7 \\
 & SVS 13A & (f) $200$ & $220$ & $15$ & $0.075$ & $100$ & $20$ & $0.1$ & 0.75 & 7 \\
 & IRAS 4A2 & (f) $60$ & $150$ & $5.3$ & $0.088$ & $150$ & $6.0$ & $0.1$ & 0.88 & 7 \\
& IRAS 4B & (f) $15$ & $150$ & $1.5$ & $0.1$ & $150$ & $2.0$ & $0.133$ & 0.75 & 7 \\
 & Serp-MM18a & (f) $22$ & $150$ & $1.7$ & $0.077$ & $150$ & $2.6$ & $0.118$ & 0.65 & 7 \\
\hline
Hot core & AFGL 4176 & ($^{13}$C) $550\pm40$ & $120\pm45$ & $7.6$ & $0.014\pm0.001$ & $160\pm15$ & $13$ & $0.023\pm0.002$ & 0.58 & 8 \\
 & G31.41+0.31 & ($^{18}$O) $8000\pm1300$ & $119\pm14$ & $47$ & $0.006\pm0.001$ & $98$ & $81$ & $0.01\pm0.014$ & 0.58 & 9 \\
 & G345.5+1.5 & ($^{18}$O) $140\pm30$ & $140\pm40$ & $1.4\pm0.3$ & $0.01$ & $120\pm20$ & $4.3\pm0.7$ & $0.03\pm0.001$ & 0.32 & 10 \\
 & G19.88-0.53 & ($^{18}$O) $440\pm13$ & $170\pm30$ &  $5.5\pm1.5$ & $0.013\pm0.003$ & $130\pm20$ & $11\pm3$ & $0.025\pm0.006$ & 0.5 & 10 \\
 & G23.21-0.37 & ($^{18}$O) $2900\pm900$ & $160\pm40$ & $24\pm4$ & $0.008\pm0.001$ & $120\pm10$ & $48\pm8$ & $0.017\pm0.002$ & 0.68 & 10 \\
 & G35.03+0.35 & ($^{18}$O) $180\pm70$ & $120\pm40$ & $2.8\pm0.7$ & $0.015\pm0.002$ & $130\pm20$ & $6.3\pm1.7$ & $0.035\pm0.004$ & 0.44 & 10 \\
 & IRAS 16547-4247 & ($^{18}$O) $210\pm80$ & $140\pm40$ & $6.0\pm1.5$ & $0.028\pm0.004$ & $180\pm40$ & $6.0\pm1.5$ & $0.028\pm0.004$ & 1.0 & 10 \\
 & NGC 6334 I(N)-SM2 & ($^{18}$O) $390\pm11$ & $180\pm20$ & $1.3\pm0.4$ & $0.003\pm0.001$ & $100\pm20$ & $8.0\pm2.0$ & $0.02\pm0.005$ & 0.16 & 10 \\
 & NGC 7129 FIRS 2 & ($^{13}$C) $34000$ & $200$ & $300$ & $0.009$ & $200$ & $400$ & $0.012$ & 0.75 & 11 \\
 & CygX-N30-1 & ($^{13}$C) $1300\pm200$ & $120\pm60$ & $7.2\pm2.0$ & $0.005\pm0.001$ & $100\pm20$ & $29\pm2$ & $0.022\pm0.002$ & 0.24 & 12 \\
 & CygX-N30-2 & ($^{13}$C) $1000\pm200$ & $130\pm60$ & $6.7\pm2.0$ & $0.007\pm0.001$ & $110\pm20$ & $24\pm2$ & $0.024\pm0.003$ & 0.27 & 12 \\
 & CygX-N30-3 & ($^{13}$C) $570\pm150$ & $140\pm60$ & $5.5\pm2.0$ & $0.01\pm0.001$ & $100\pm20$ & $15\pm2$ & $0.026\pm0.003$ & 0.36 & 12 \\
 & Sgr-N2 & (f) $4000$ & $150$ & $200$ & $0.05$ & $130$ & $220$ & $0.055$ & 0.90 & 13 \\
 & Sgr-N3 & (f) $750$ & $145$ & $31$ & $0.04$ & $145$ & $100$ & $0.133$ & 0.31 & 13 \\
 & Sgr-N4 & (f) $25$ & $145$ & $2.5$ & $0.1$ & $145$ & $9.0$ & $0.36$ & 0.27 & 13 \\
 & Sgr-N5 & (f) $200$ & $145$ & $10$ & $0.05$ & $145$ & $45$ & $0.225$ & 0.22 & 13 \\ 
\hline   
\multicolumn{9}{l}{$^{13}$C: The column density for CH$_3$OH was derived by multiplying the column density from the fit of $^{13}$CH$_3$OH.}\\
\multicolumn{9}{l}{$^{18}$O: The column density for CH$_3$OH was derived by multiplying the column density from the fit of CH$_3$$^{18}$OH.}\\
\multicolumn{9}{l}{f: The column density for CH$_3$OH was derived from the fit.}\\
\multicolumn{9}{l}{[1]\cite{2020AA...639A..87VanGelder} [2]\cite{2019AA...629A..29Jacobsen}  [3]\cite{2020ApJ...891...61Yang} [4] \cite{2020A&A...635A..48Manigand} [5]\cite{2018A&A...620A.170J:} [6]\cite{2022ApJ...927..218Hsu} [7]\cite{2020A&A...635A.198Belloche} }\\
\multicolumn{9}{l}{[8]\cite{2019AA...628A...2B} [9]\cite{2023AA...677A..15Mininni} [10]\cite{2023AA...678A.137Chen} [11]\cite{2014A&A...568A..65Fuente} [12]\cite{2021A&A...655A..86V} [13]\cite{2019A&A...628A..27Bonfand}}\\
\end{longtable}
\end{small}
\end{landscape}

\section{Additional figures}  \label{app:B}
Figure \ref{fig:B1} shows the best-fitting results for sources detected by CH$_3$OCH$_3$ and C$_2$H$_5$OH. The method for calculating the local noise for each window was adopted from \cite{2021NatAs...5..188Loomis}. 

\begin{figure}
\centering
{\includegraphics[height=4.5cm,width=15.93cm]{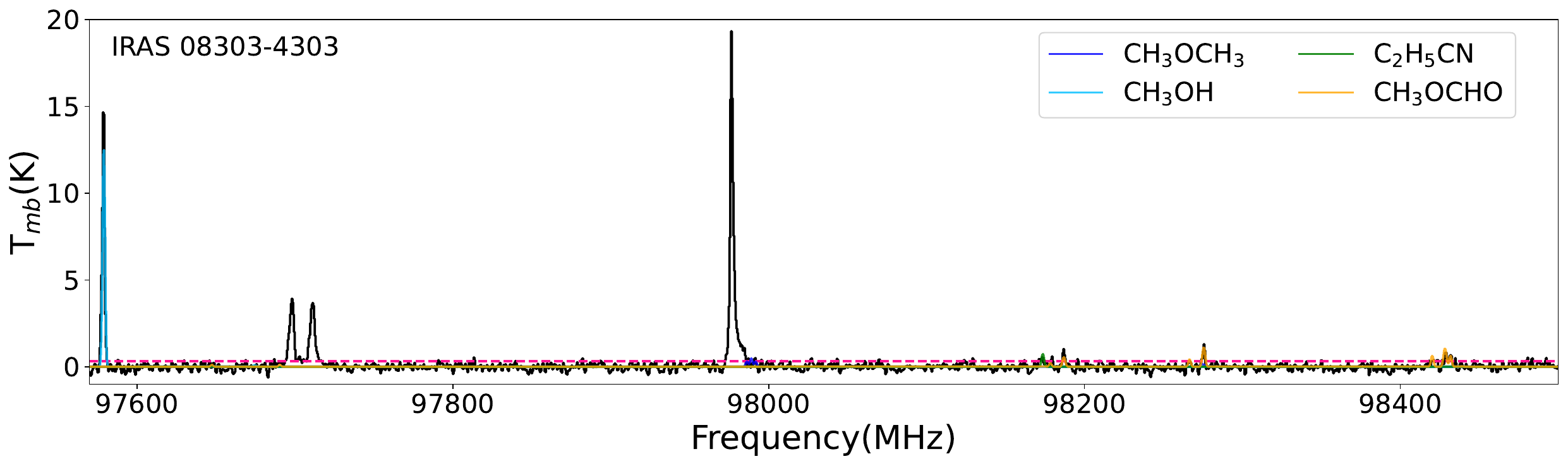}}
\quad
{\includegraphics[height=4.5cm,width=15.93cm]{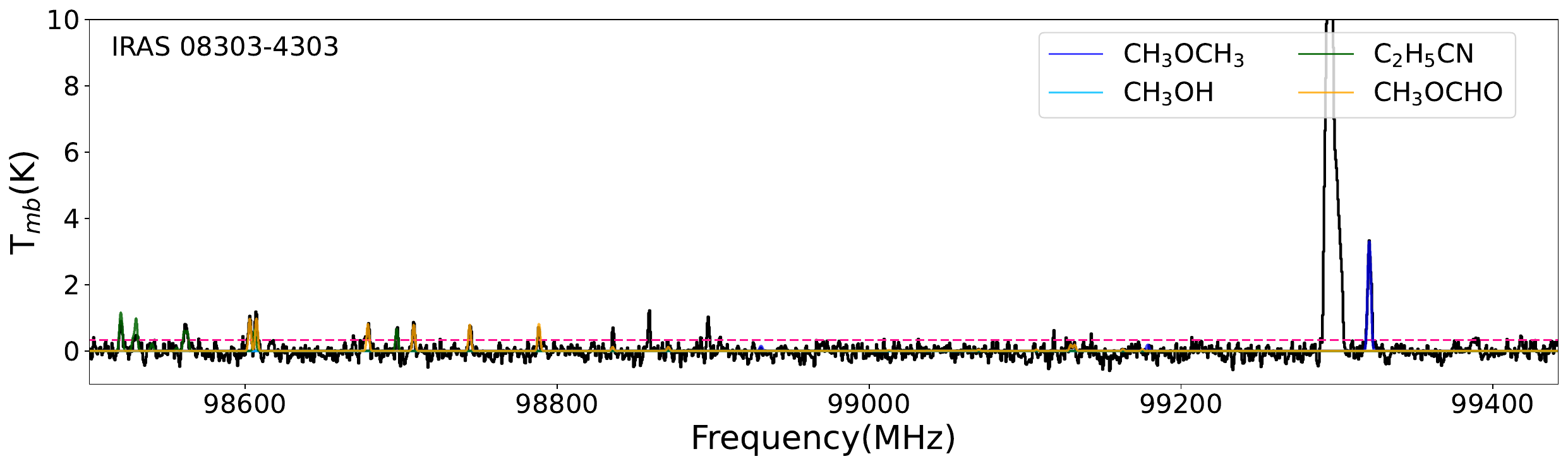}}
\quad
{\includegraphics[height=4.5cm,width=15.93cm]{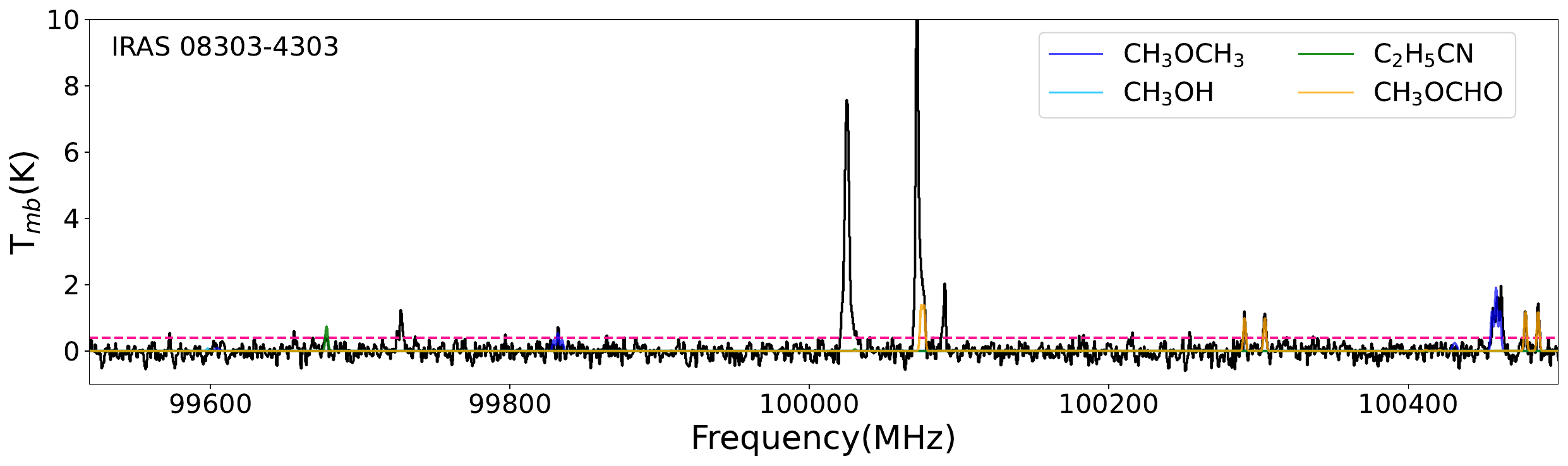}}
\quad
{\includegraphics[height=4.5cm,width=15.93cm]{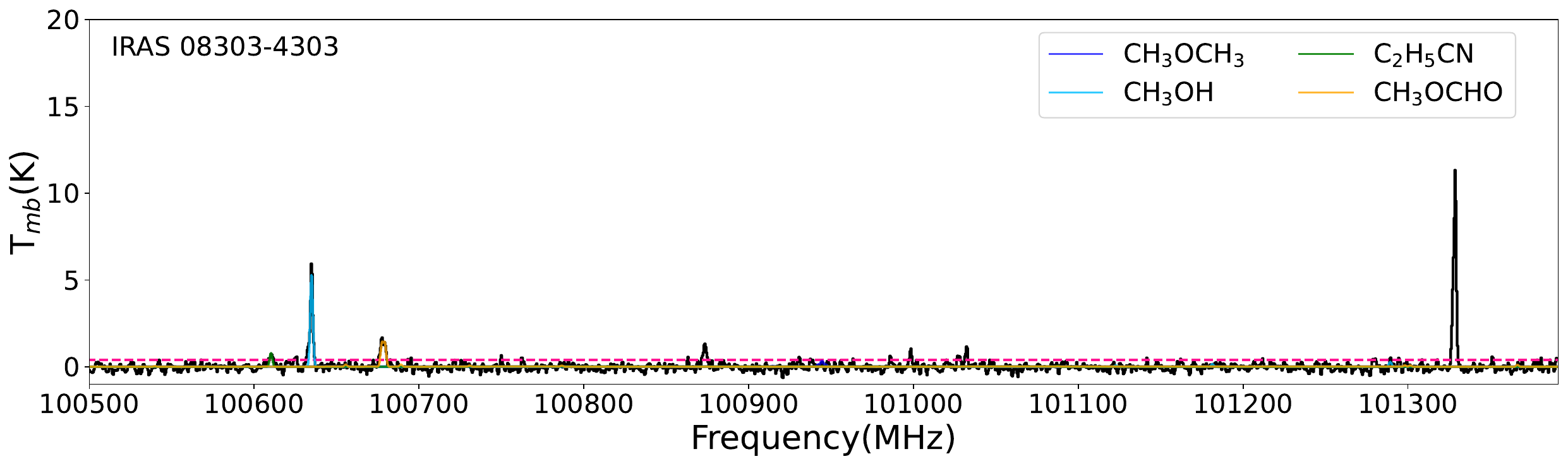}}
\quad
{\includegraphics[height=4.78cm,width=15.93cm]{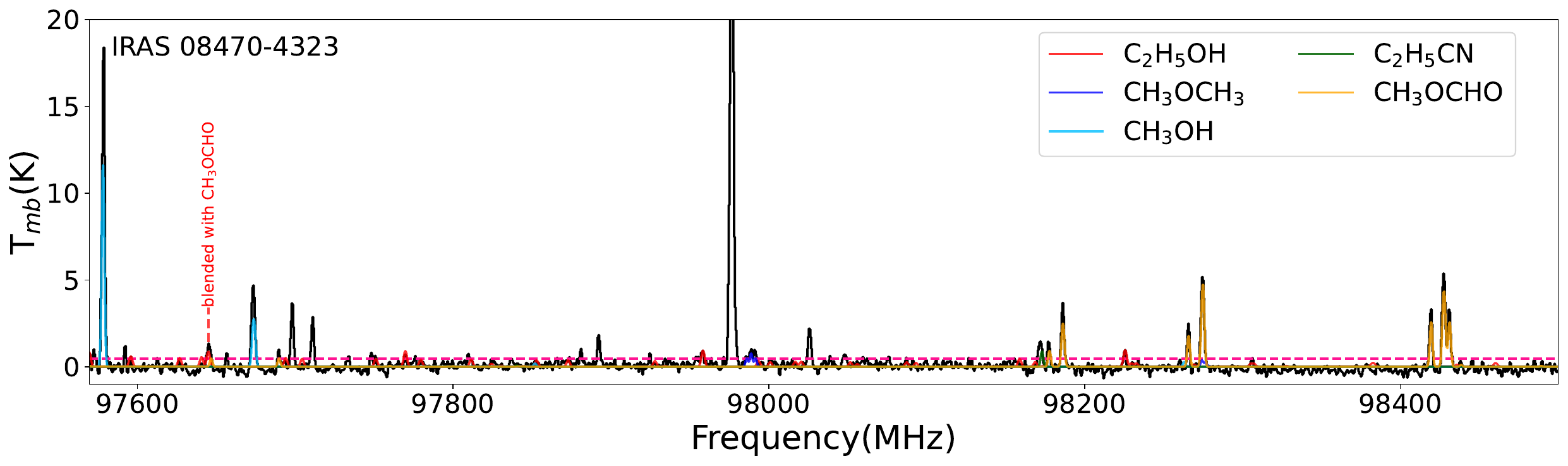}}
\caption{The observed spectrum is shown in black, with the synthetic spectrum for each species coded in colour, respectively. The horizontal pink dashed line indicates the 3 $\sigma$ noise level in each window. The blended lines of C$_2$H$_5$OH and CH$_3$OCH$_3$ are labeled in red and blue text, respectively.}
\label{fig:B1}
\end{figure}
\setcounter{figure}{\value{figure}-1}
\begin{figure}
\centering
{\includegraphics[height=4.78cm,width=15.93cm]{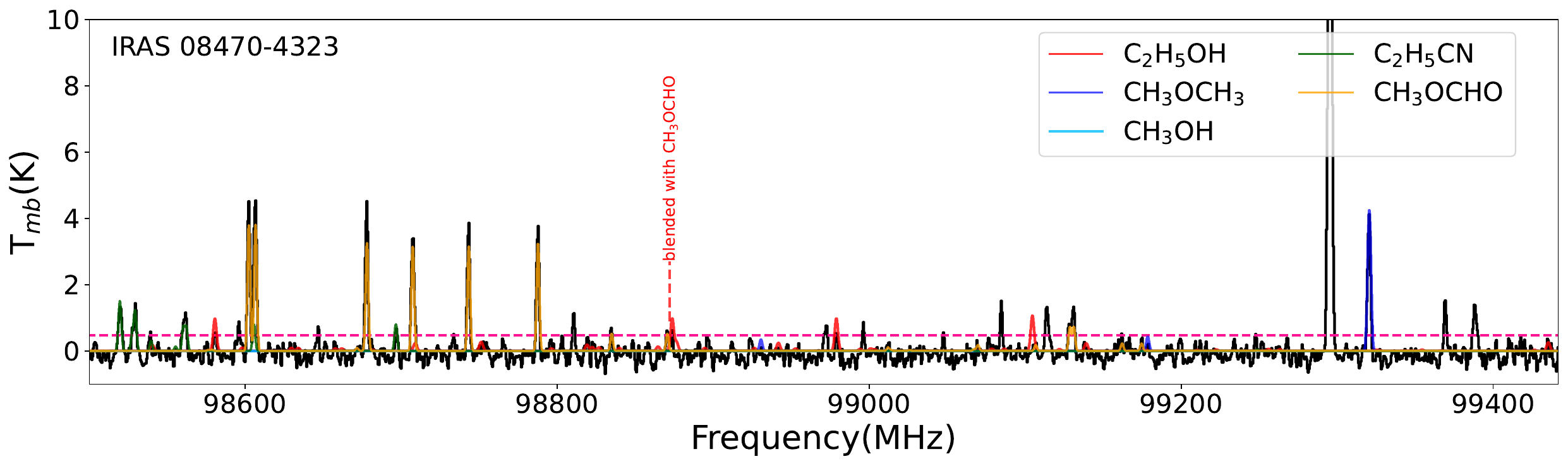}}
\quad
{\includegraphics[height=4.78cm,width=15.93cm]{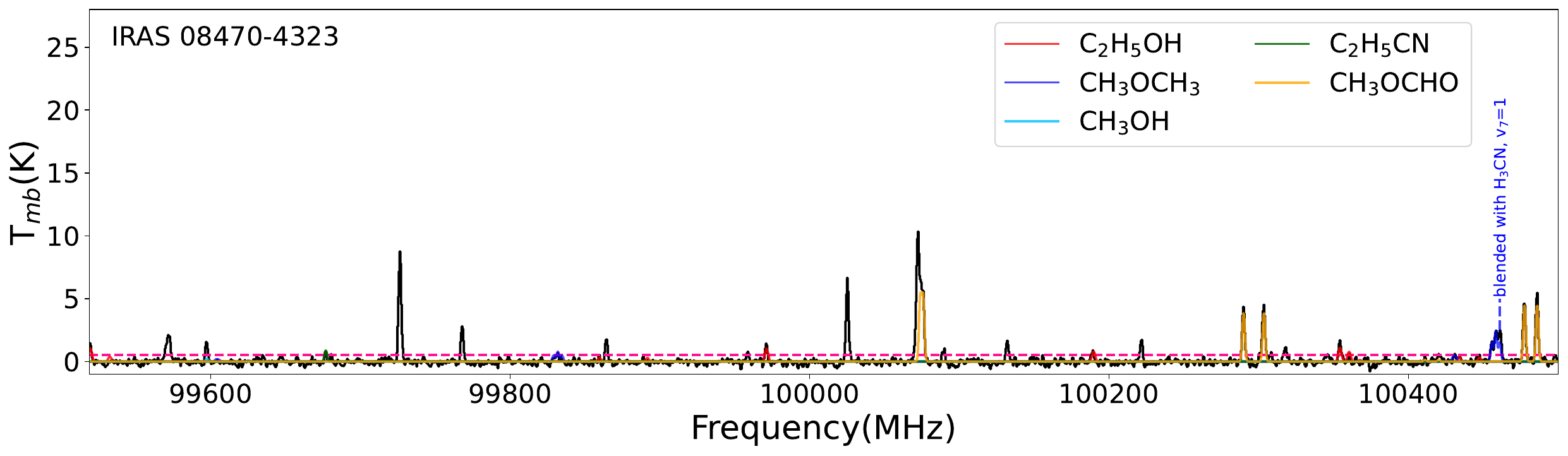}}
\quad
{\includegraphics[height=4.78cm,width=15.93cm]{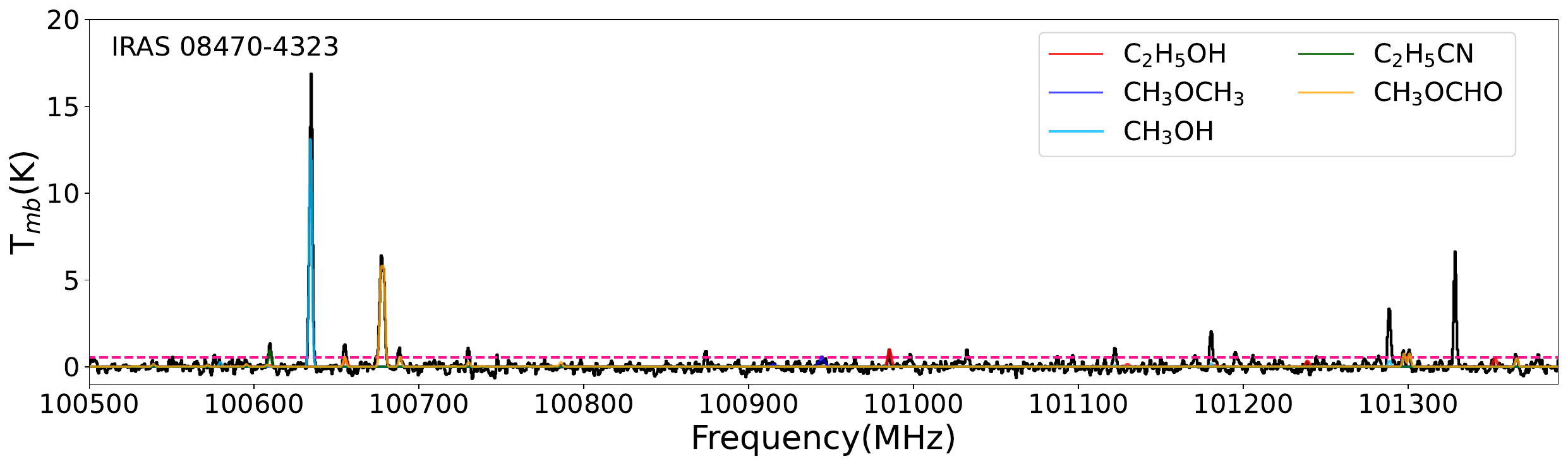}}
\quad
{\includegraphics[height=4.5cm,width=15.93cm]{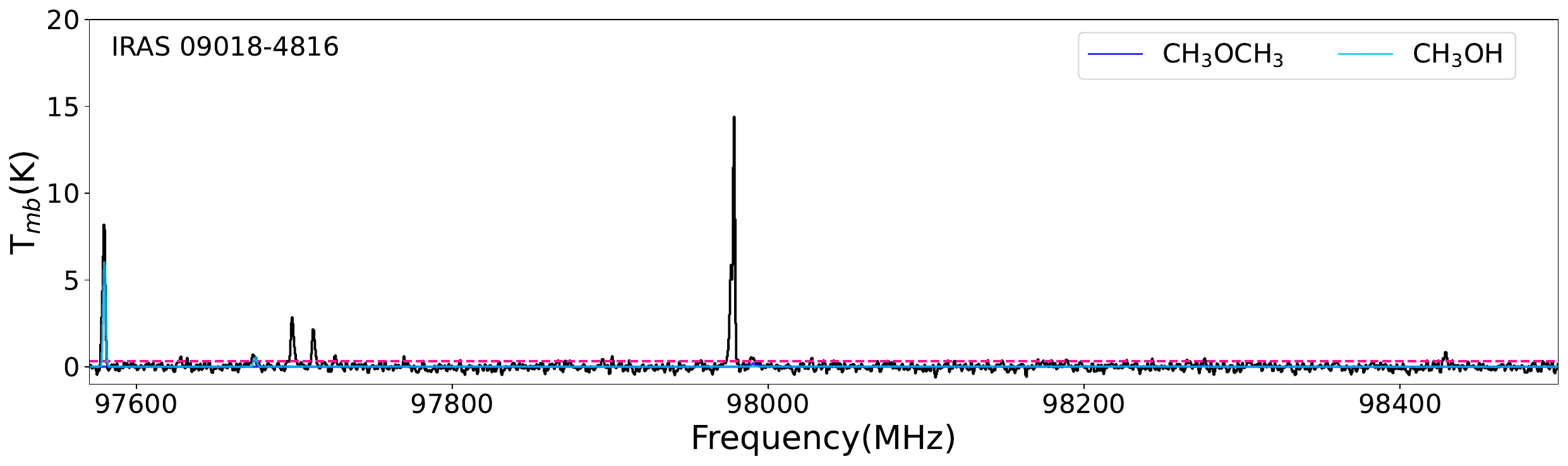}}
\quad
{\includegraphics[height=4.5cm,width=15.93cm]{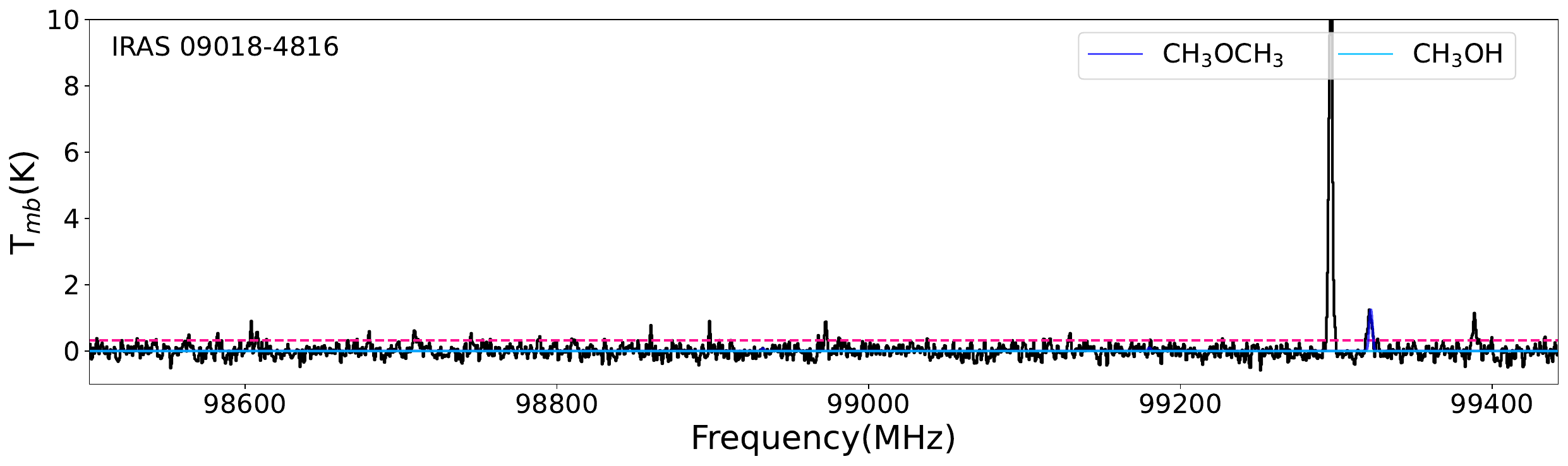}}
\caption{Continued.}
\end{figure}
\setcounter{figure}{\value{figure}-1}
\begin{figure}
\centering
{\includegraphics[height=4.5cm,width=15.93cm]{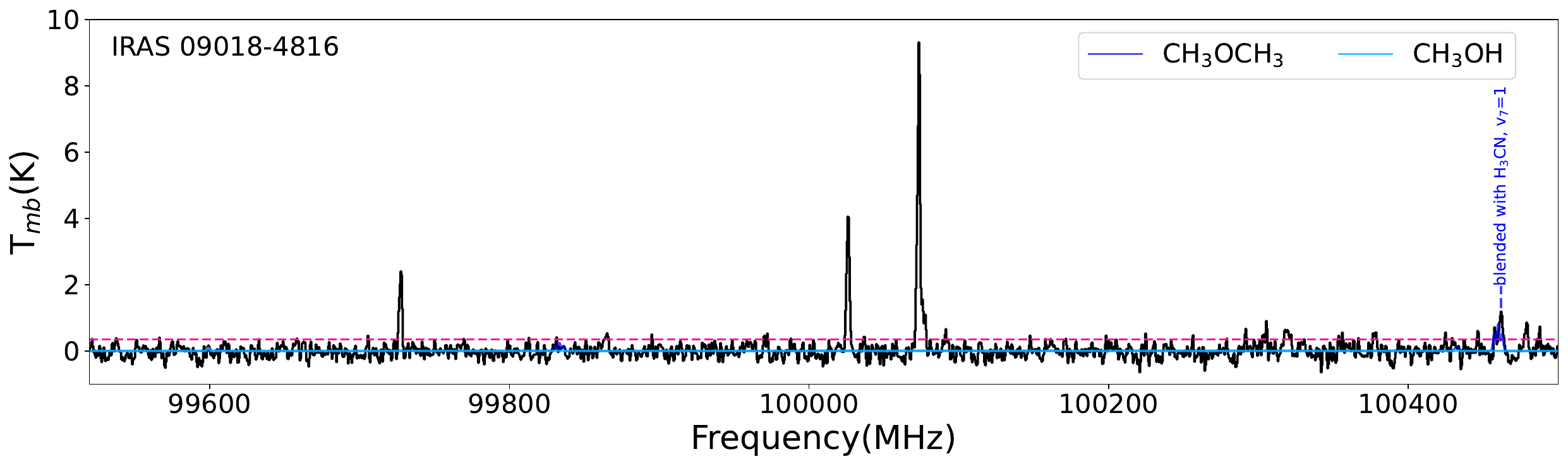}}
\quad
{\includegraphics[height=4.5cm,width=15.93cm]{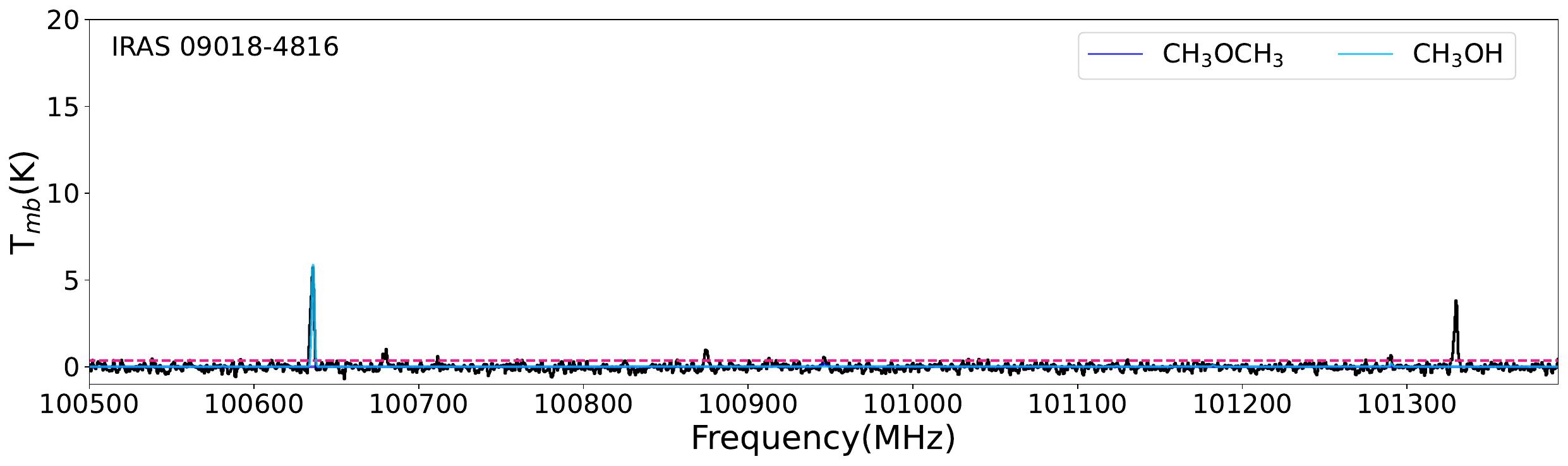}}
\quad
{\includegraphics[height=4.5cm,width=15.93cm]{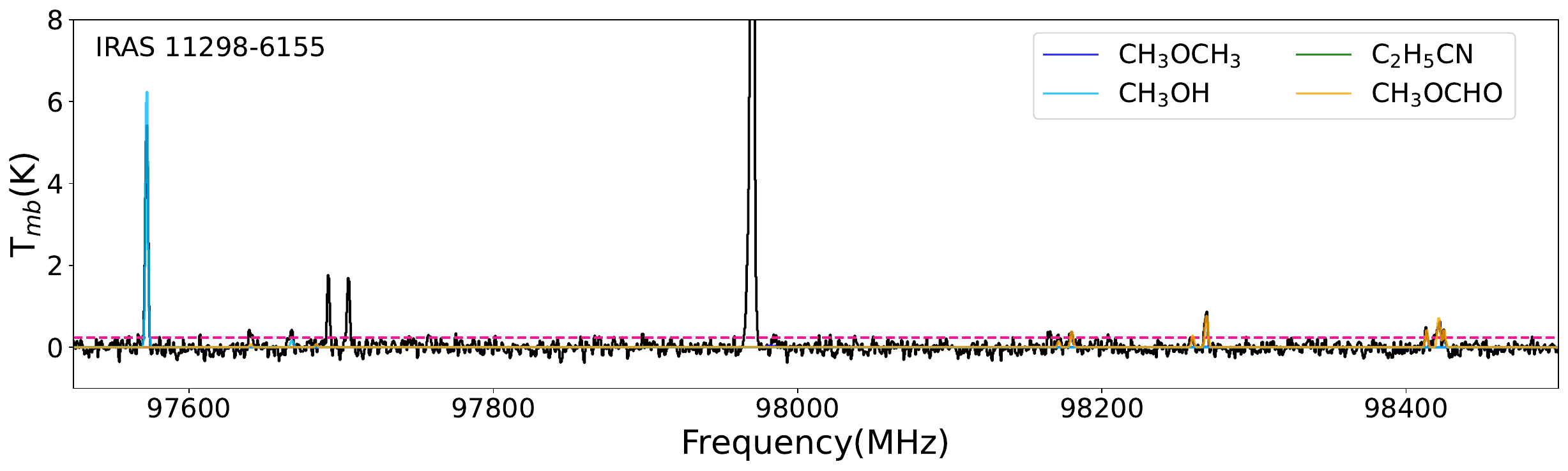}}
\quad
{\includegraphics[height=4.5cm,width=15.93cm]{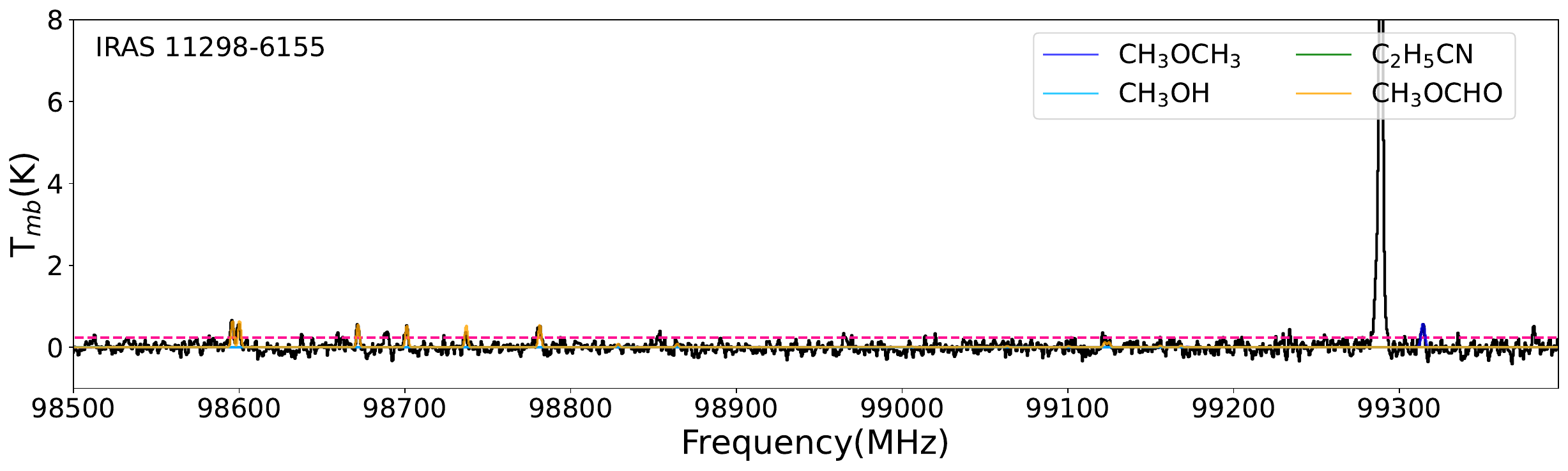}}
\quad
{\includegraphics[height=4.5cm,width=15.93cm]{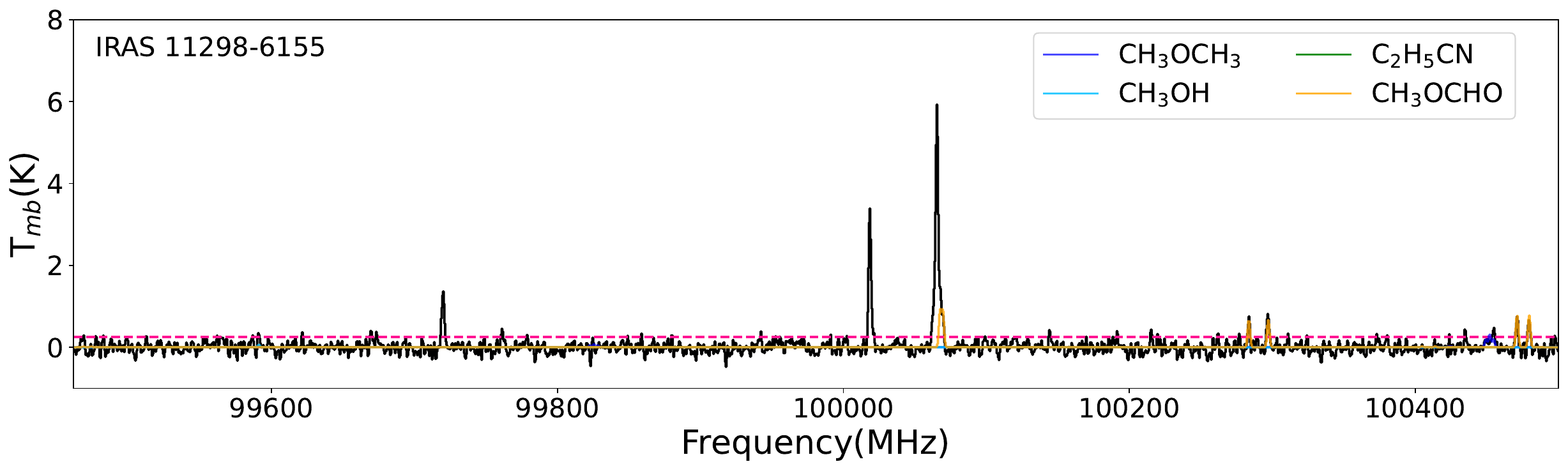}}
\caption{Continued.}
\end{figure}
\setcounter{figure}{\value{figure}-1}
\begin{figure}
  \centering 
{\includegraphics[height=4.5cm,width=15.93cm]{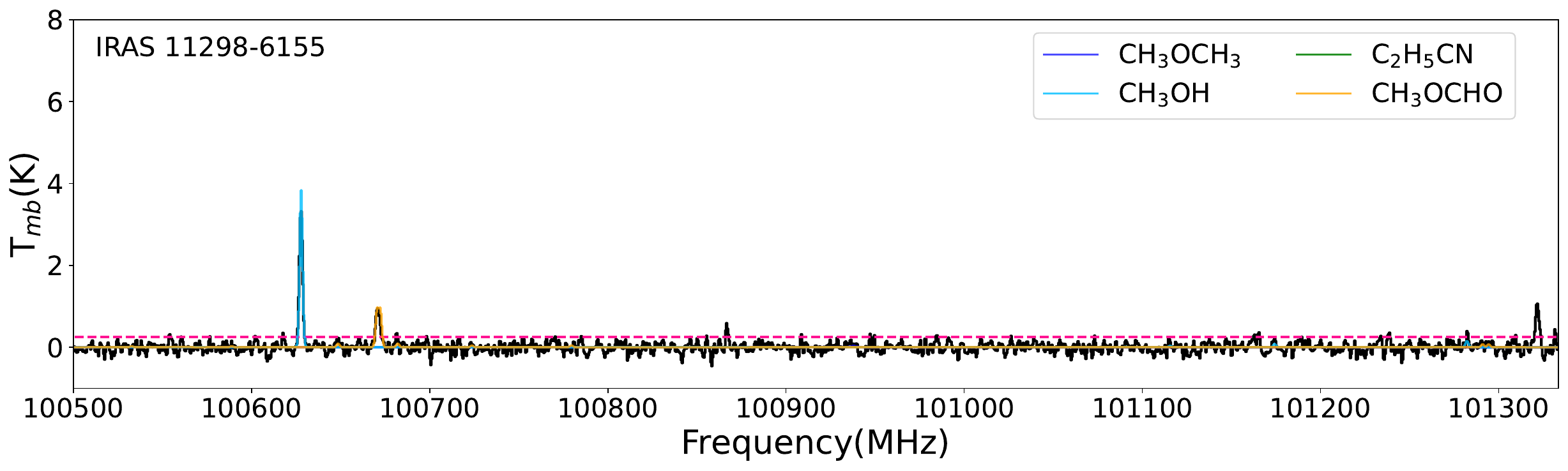}}
\quad
{\includegraphics[height=4.5cm,width=15.93cm]{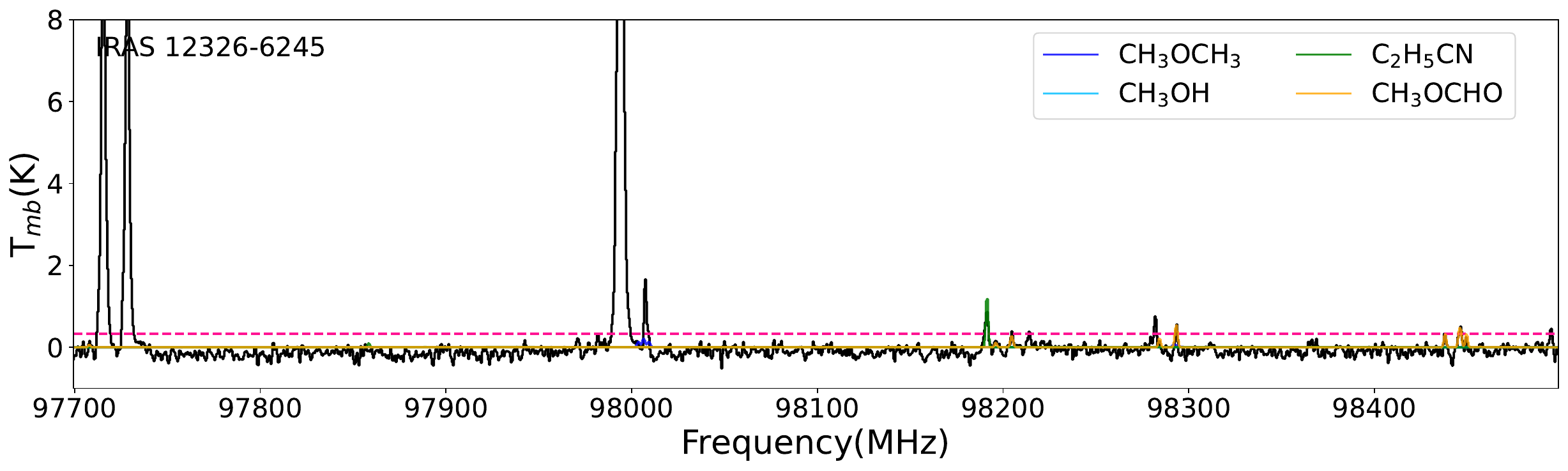}}
\quad
{\includegraphics[height=4.5cm,width=15.93cm]{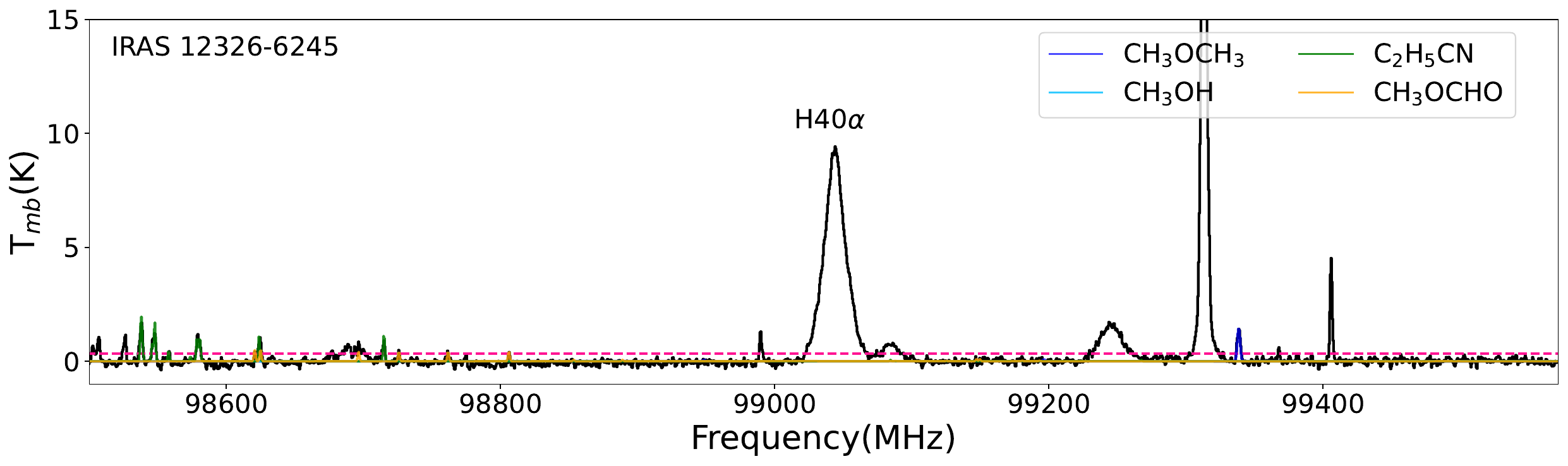}}
\quad
{\includegraphics[height=4.5cm,width=15.93cm]{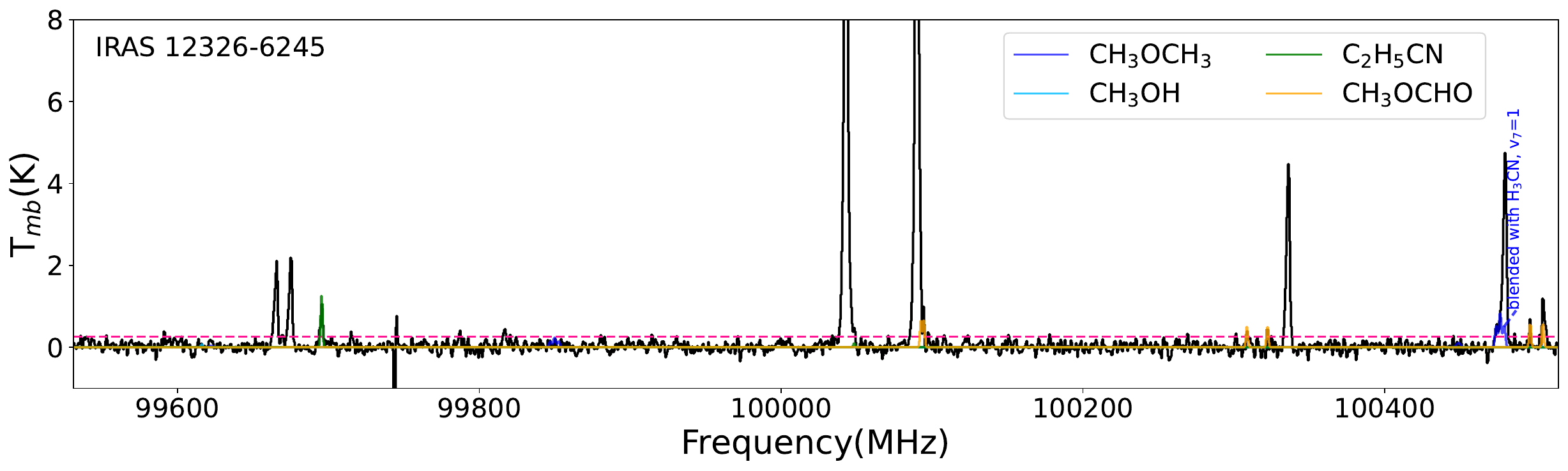}}
\quad
{\includegraphics[height=4.5cm,width=15.93cm]{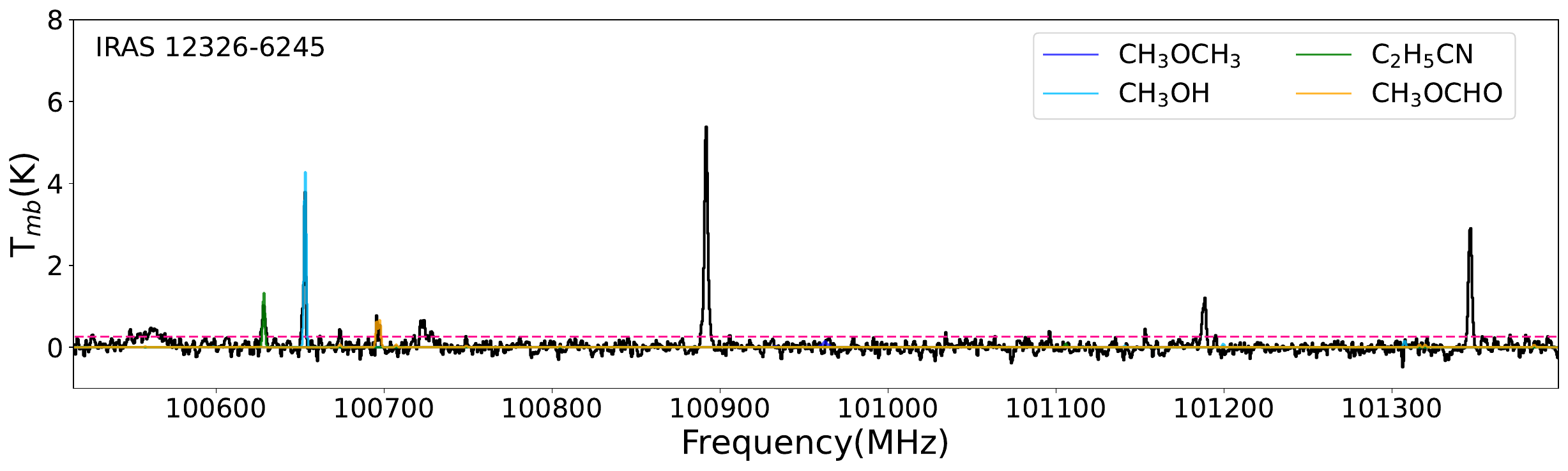}}
\caption{Continued.}
\end{figure}
\setcounter{figure}{\value{figure}-1}
\begin{figure}
  \centering 
{\includegraphics[height=4.5cm,width=15.93cm]{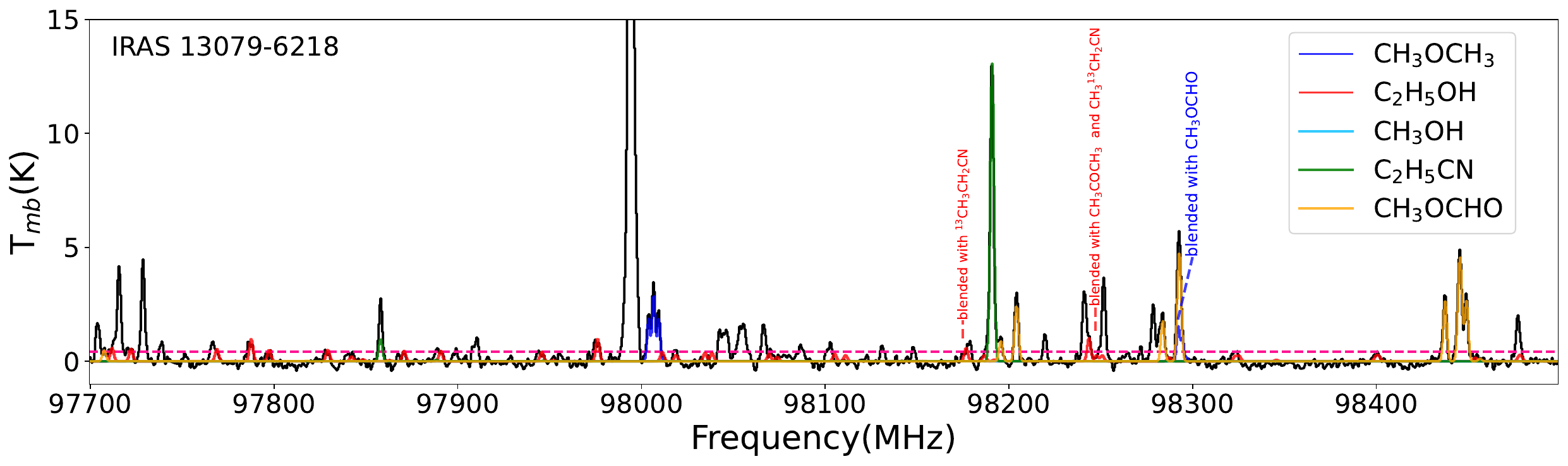}}
\quad
{\includegraphics[height=4.5cm,width=15.93cm]{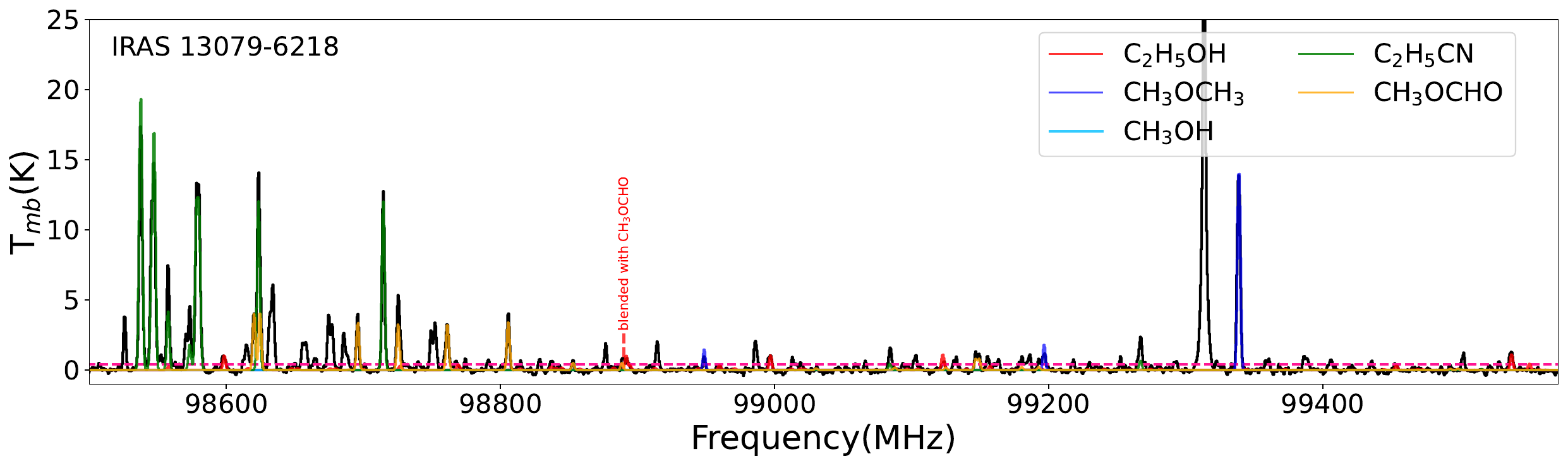}}
\quad
{\includegraphics[height=4.5cm,width=15.93cm]{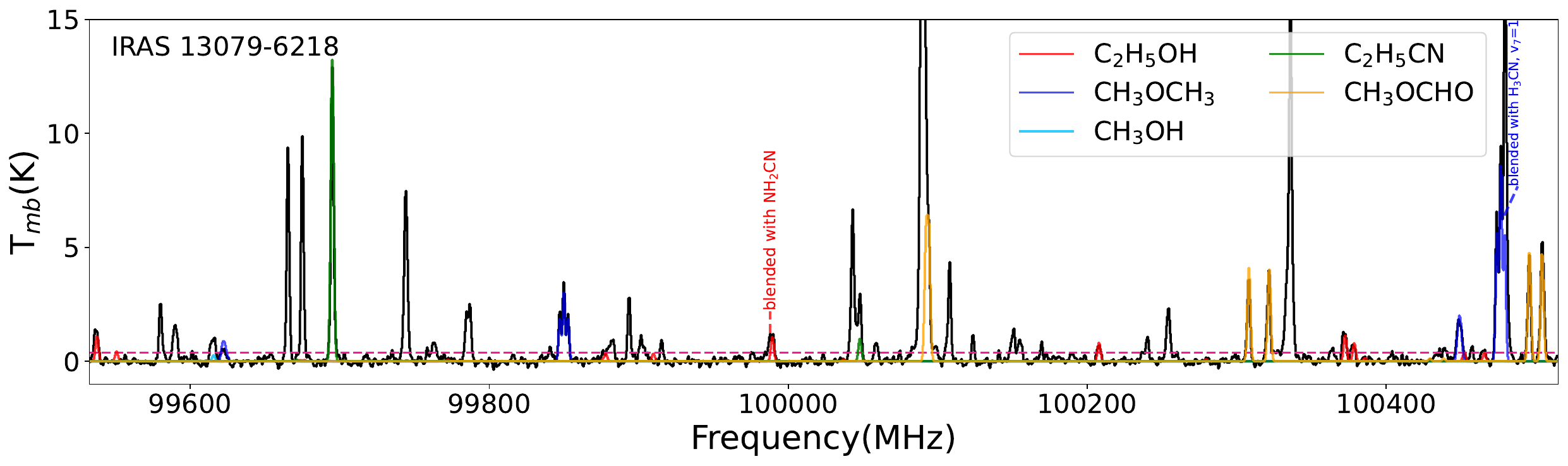}}
\quad
{\includegraphics[height=4.5cm,width=15.93cm]{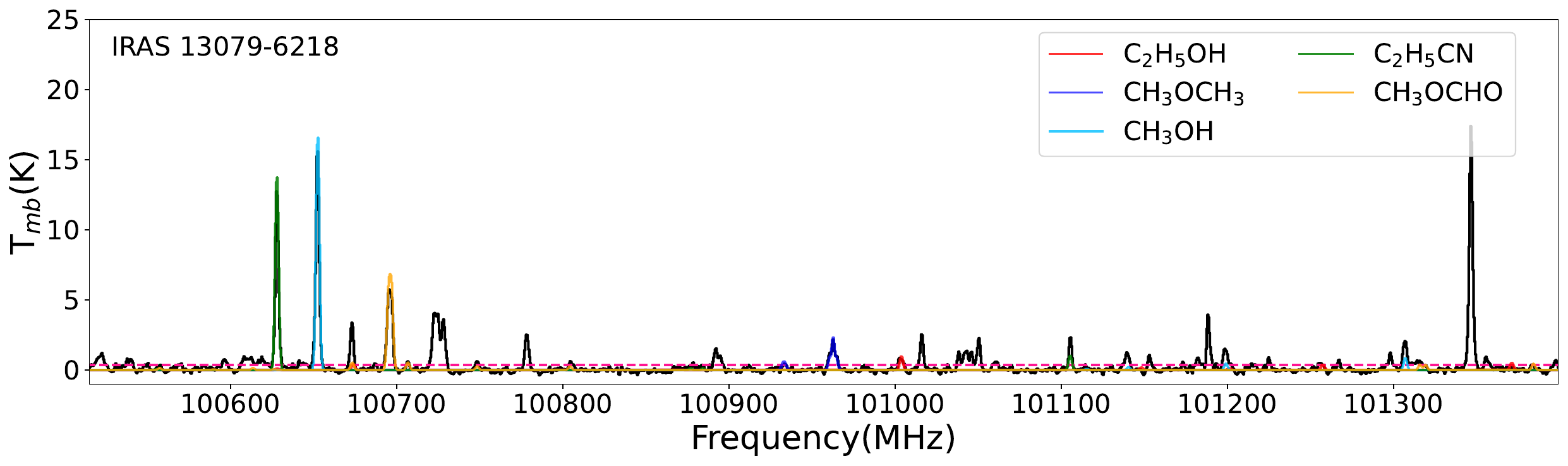}}
\quad
{\includegraphics[height=4.5cm,width=15.93cm]{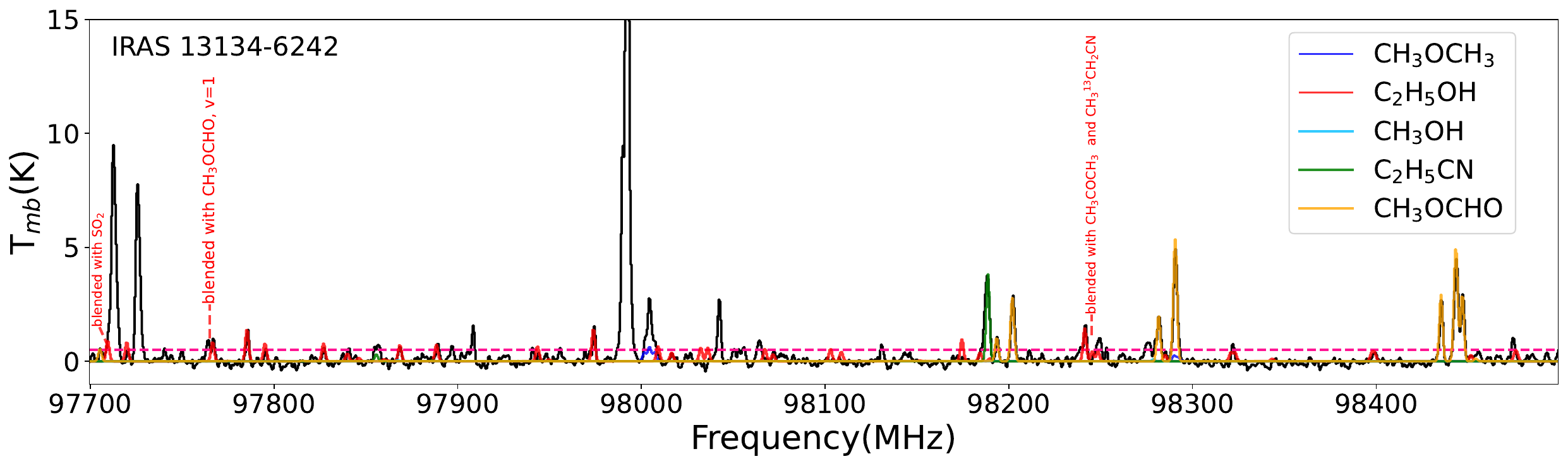}}
\caption{Continued.}
\end{figure}
\setcounter{figure}{\value{figure}-1}
\begin{figure}
  \centering 
{\includegraphics[height=4.5cm,width=15.93cm]{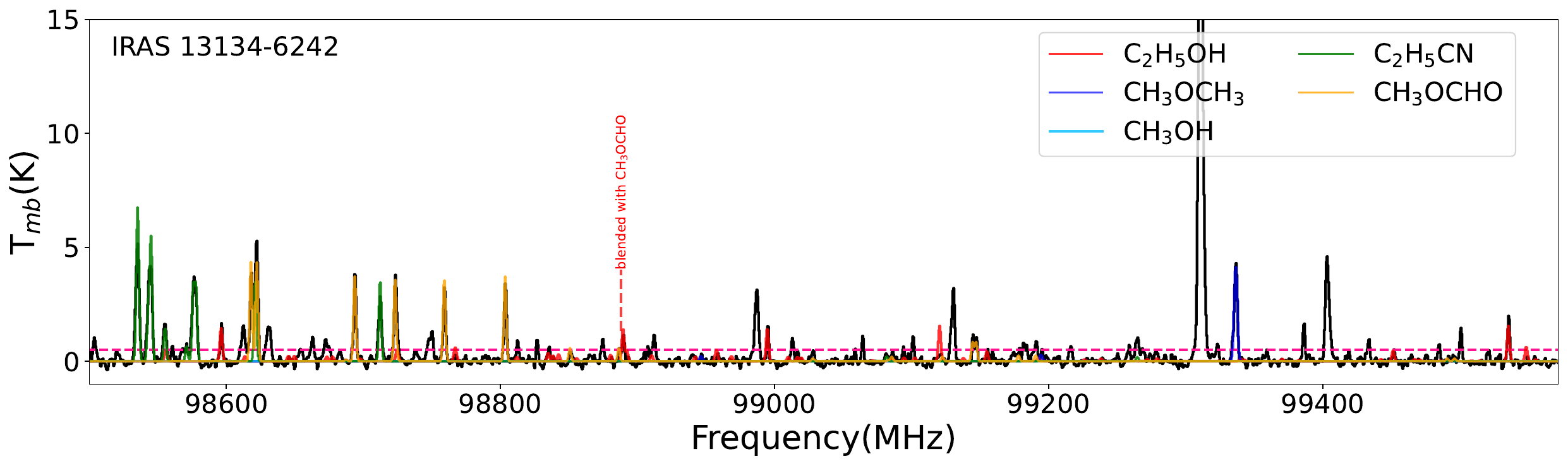}}
\quad
{\includegraphics[height=4.5cm,width=15.93cm]{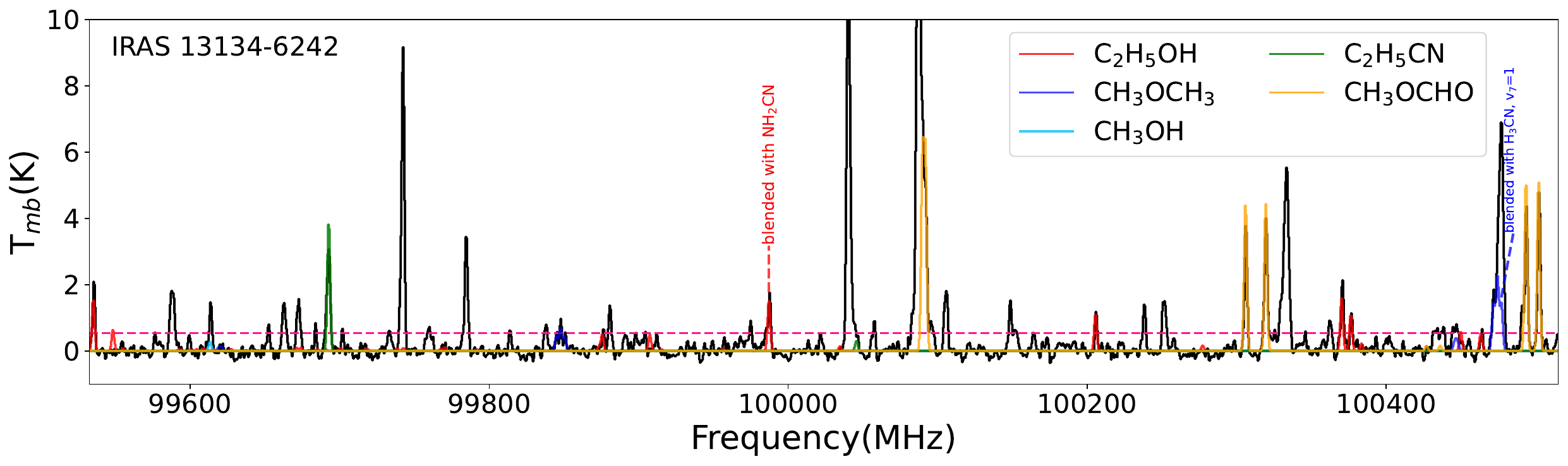}}
\quad
{\includegraphics[height=4.5cm,width=15.93cm]{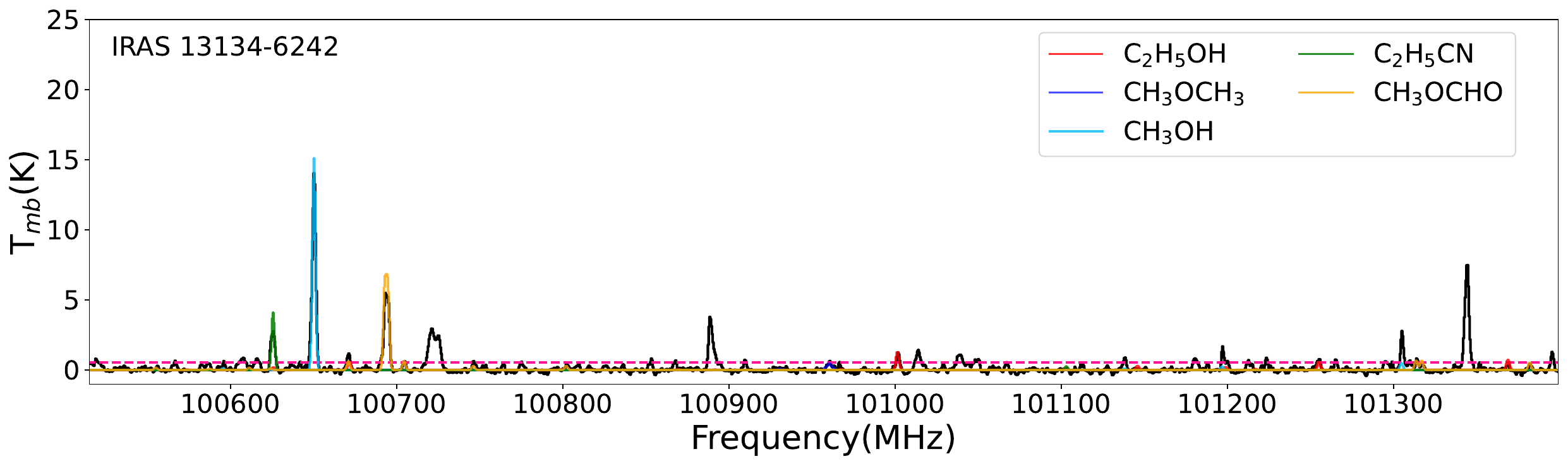}}
\quad
{\includegraphics[height=4.5cm,width=15.93cm]{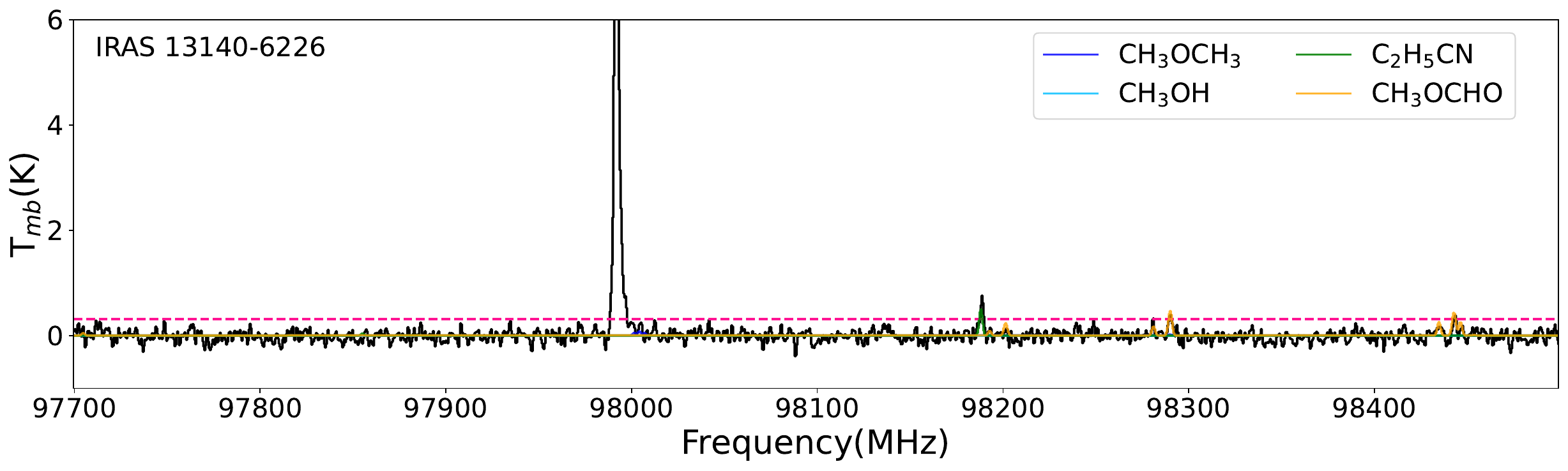}}
\quad
{\includegraphics[height=4.5cm,width=15.93cm]{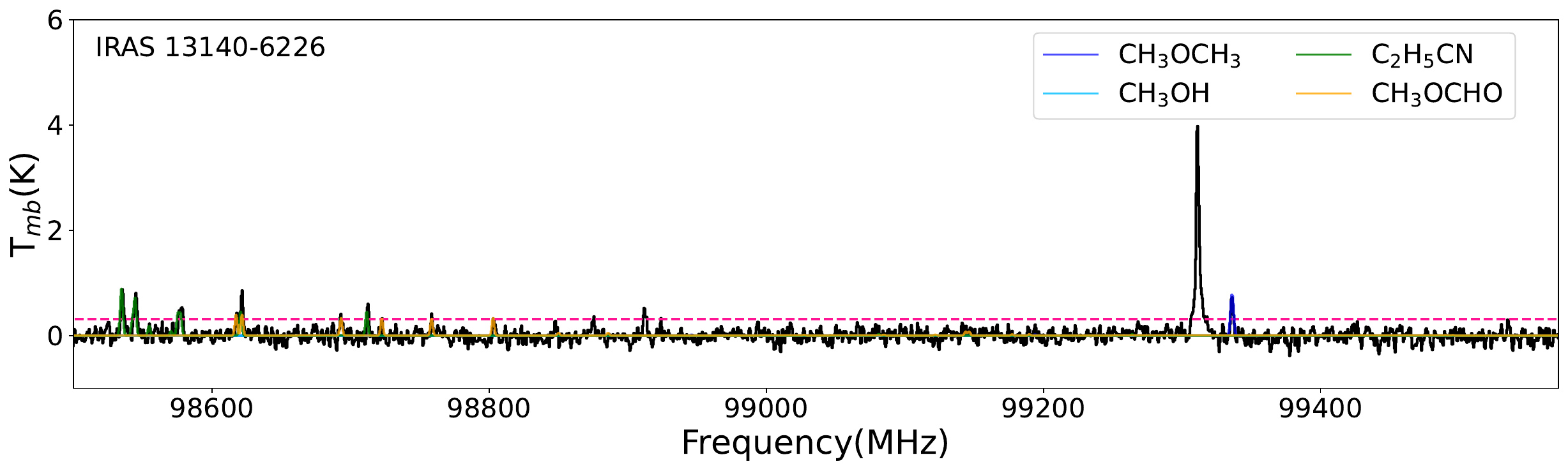}}
\caption{Continued.}
\end{figure}
\setcounter{figure}{\value{figure}-1}
\begin{figure}
  \centering 
{\includegraphics[height=4.5cm,width=15.93cm]{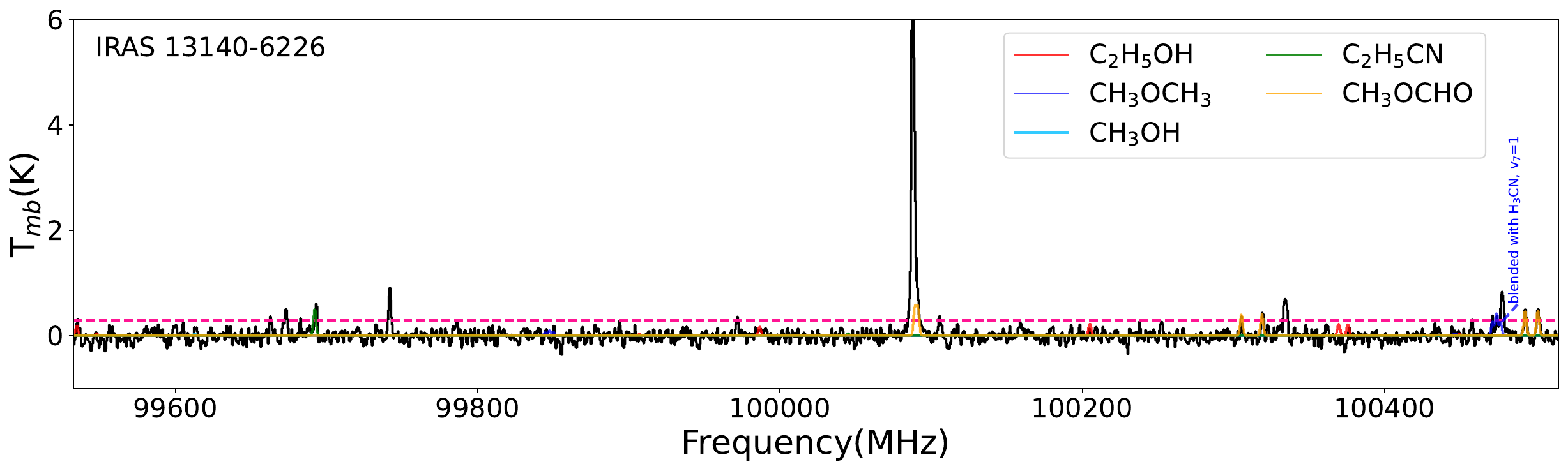}}
\quad
{\includegraphics[height=4.5cm,width=15.93cm]{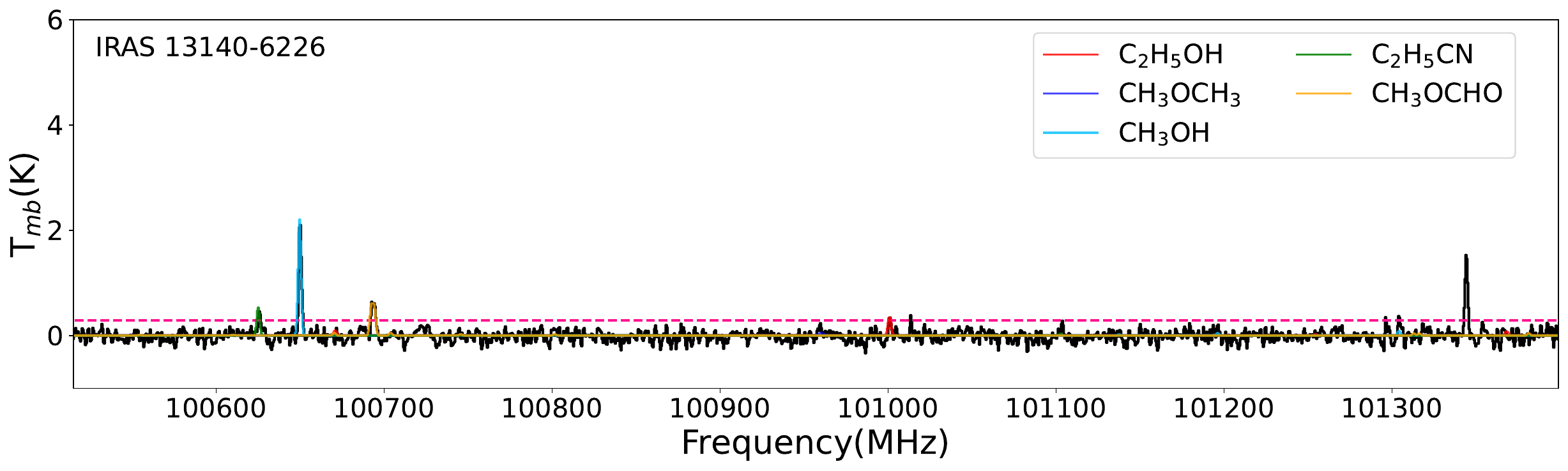}}
\quad
{\includegraphics[height=4.5cm,width=15.93cm]{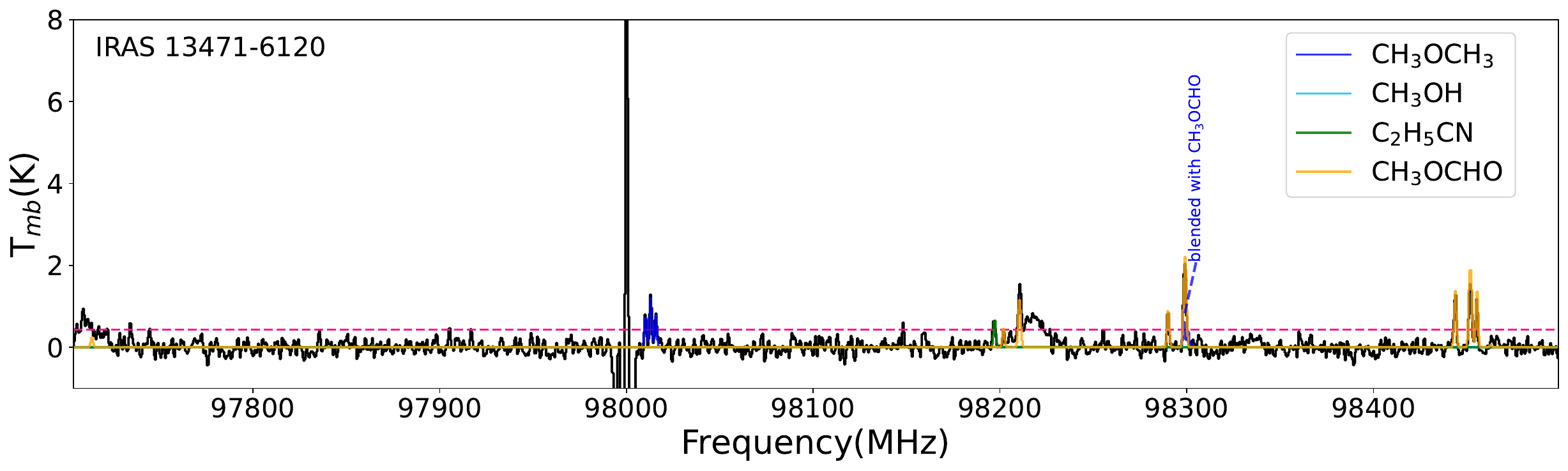}}
\quad
{\includegraphics[height=4.5cm,width=15.93cm]{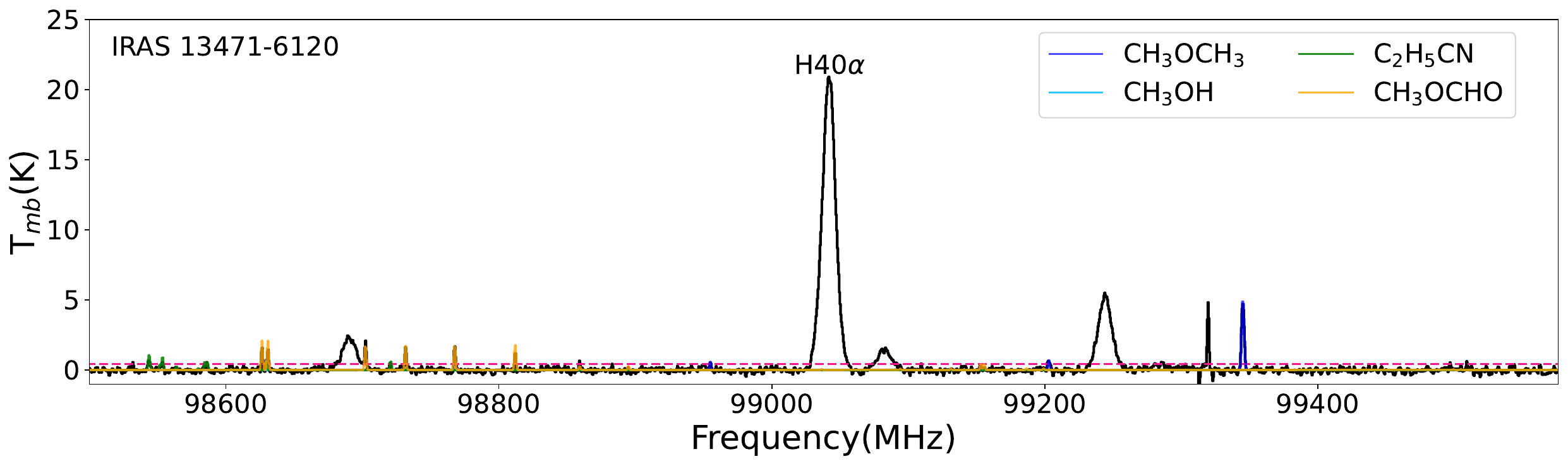}}
\quad
{\includegraphics[height=4.5cm,width=15.93cm]{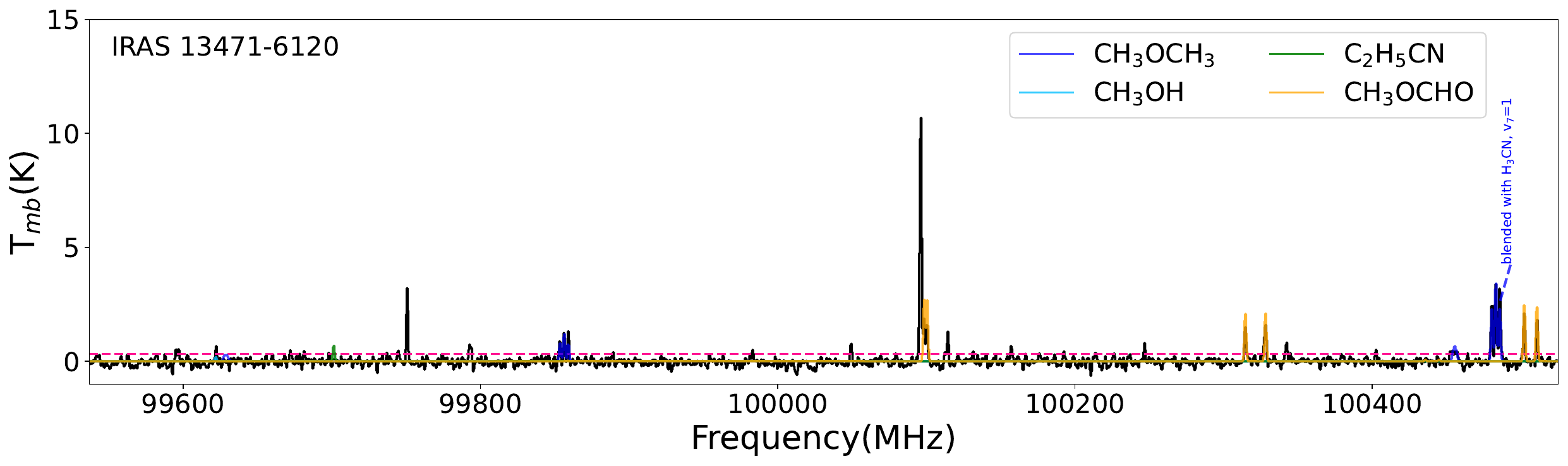}}
\caption{Continued.}
\end{figure}
\setcounter{figure}{\value{figure}-1}
\begin{figure}
  \centering 
{\includegraphics[height=4.5cm,width=15.93cm]{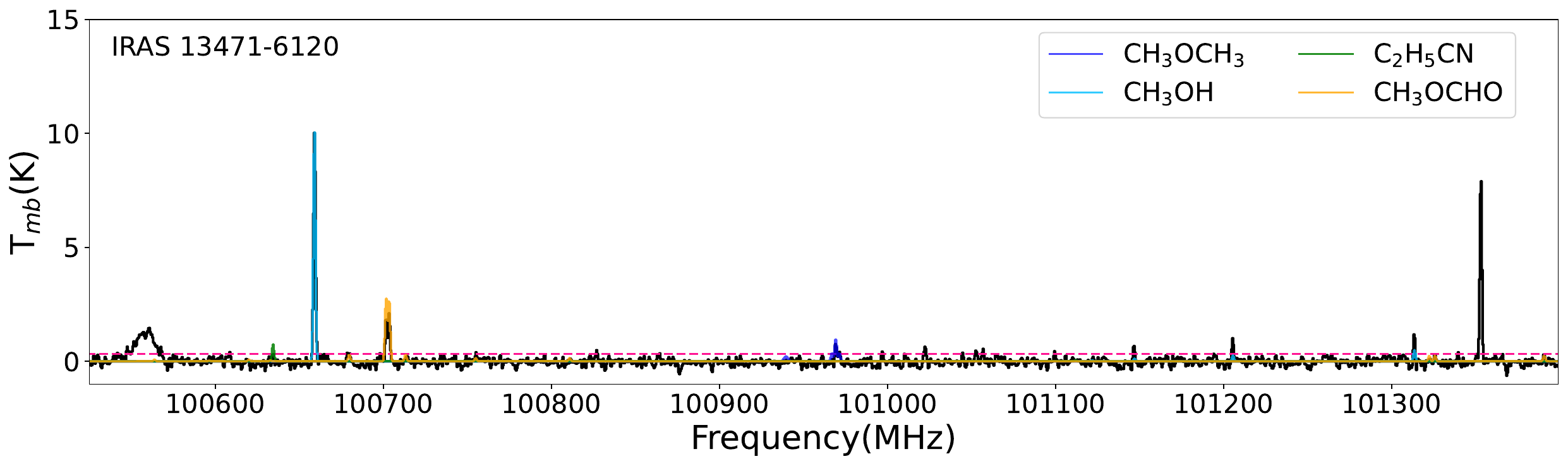}}
\quad
{\includegraphics[height=4.5cm,width=15.93cm]{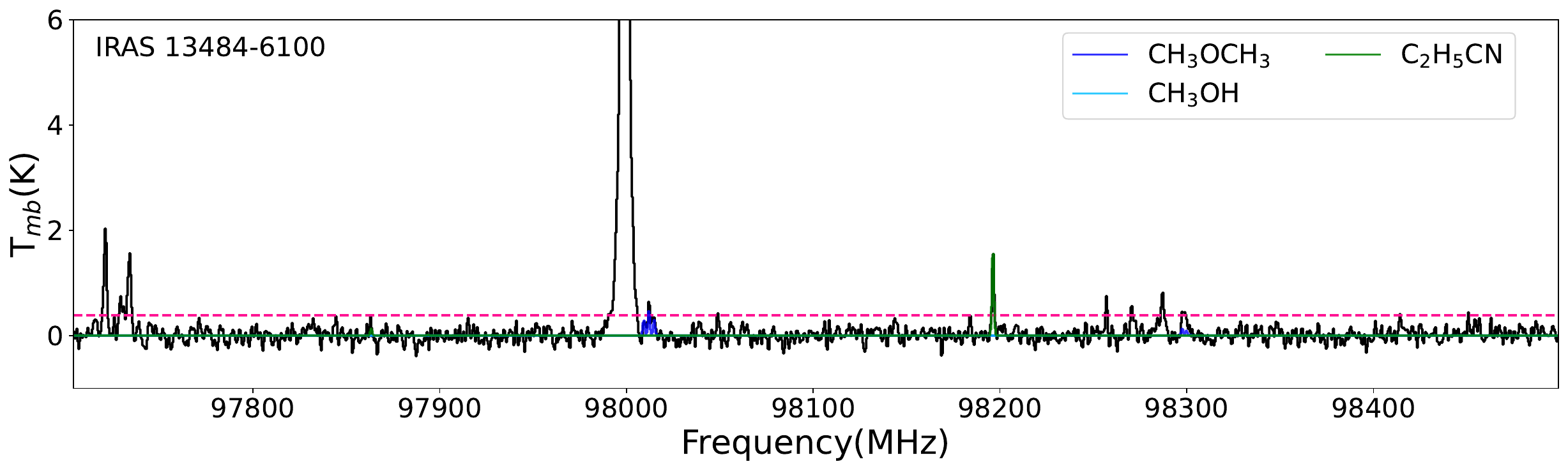}}
\quad
{\includegraphics[height=4.5cm,width=15.93cm]{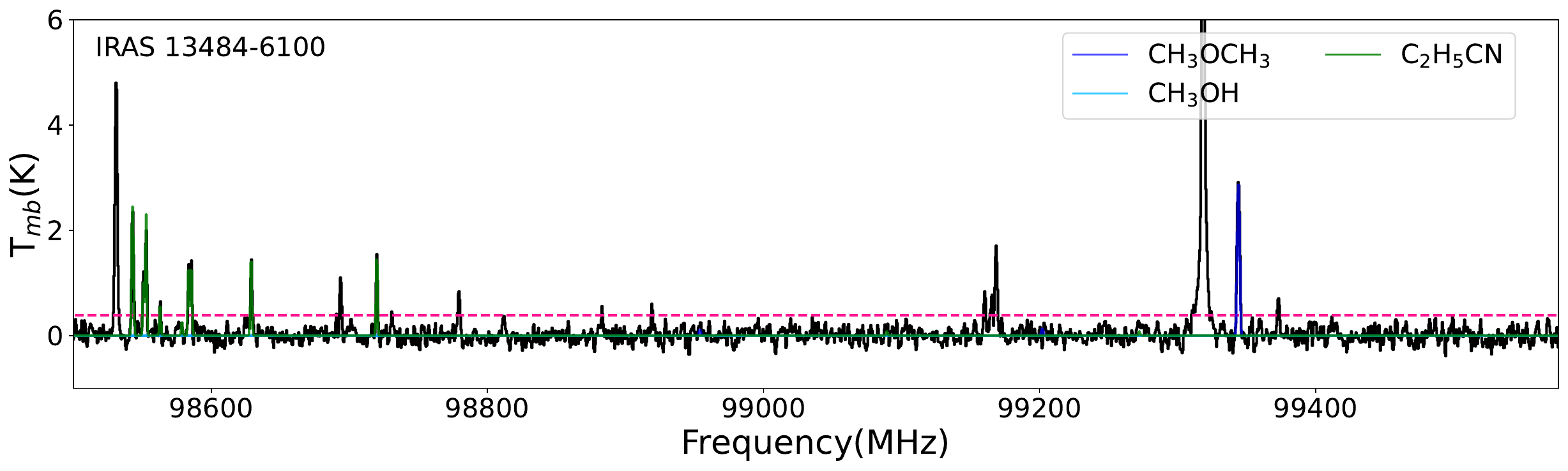}}
\quad
{\includegraphics[height=4.5cm,width=15.93cm]{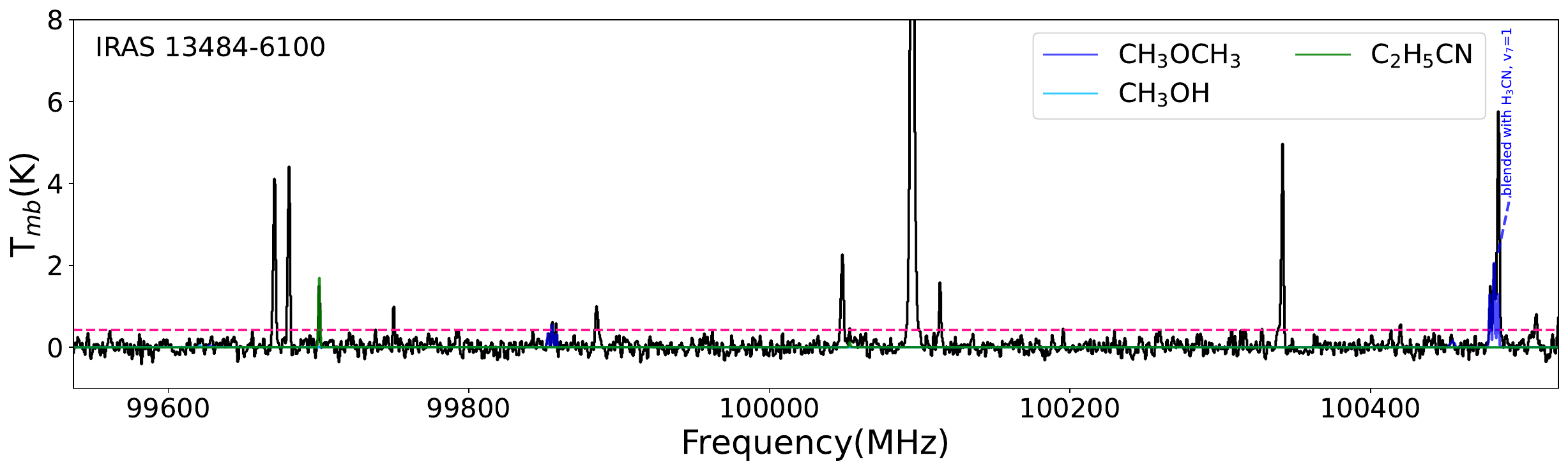}}
\quad
{\includegraphics[height=4.5cm,width=15.93cm]{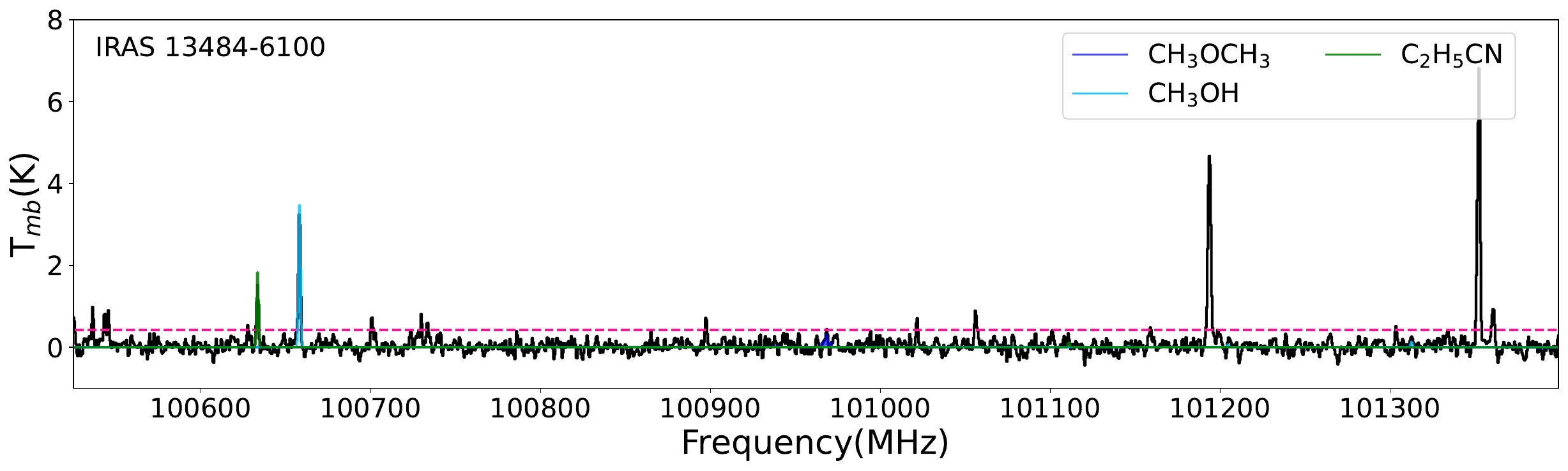}}
\caption{Continued.}
\end{figure}
\setcounter{figure}{\value{figure}-1}
\begin{figure}
  \centering 
{\includegraphics[height=4.5cm,width=15.93cm]{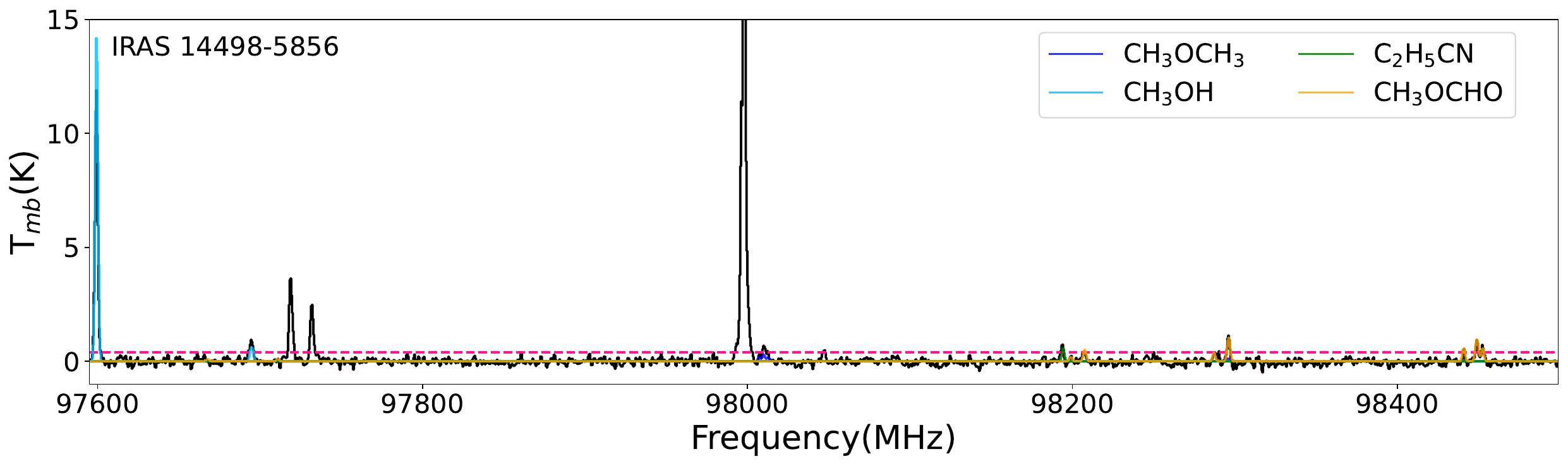}}
\quad
{\includegraphics[height=4.5cm,width=15.93cm]{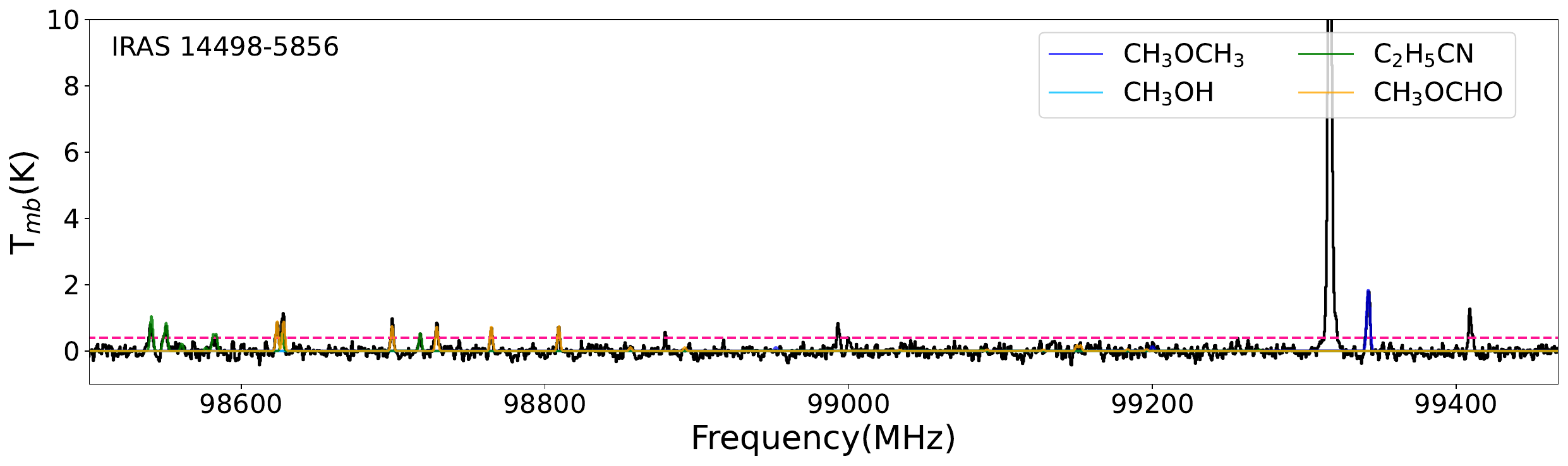}}
\quad
{\includegraphics[height=4.5cm,width=15.93cm]{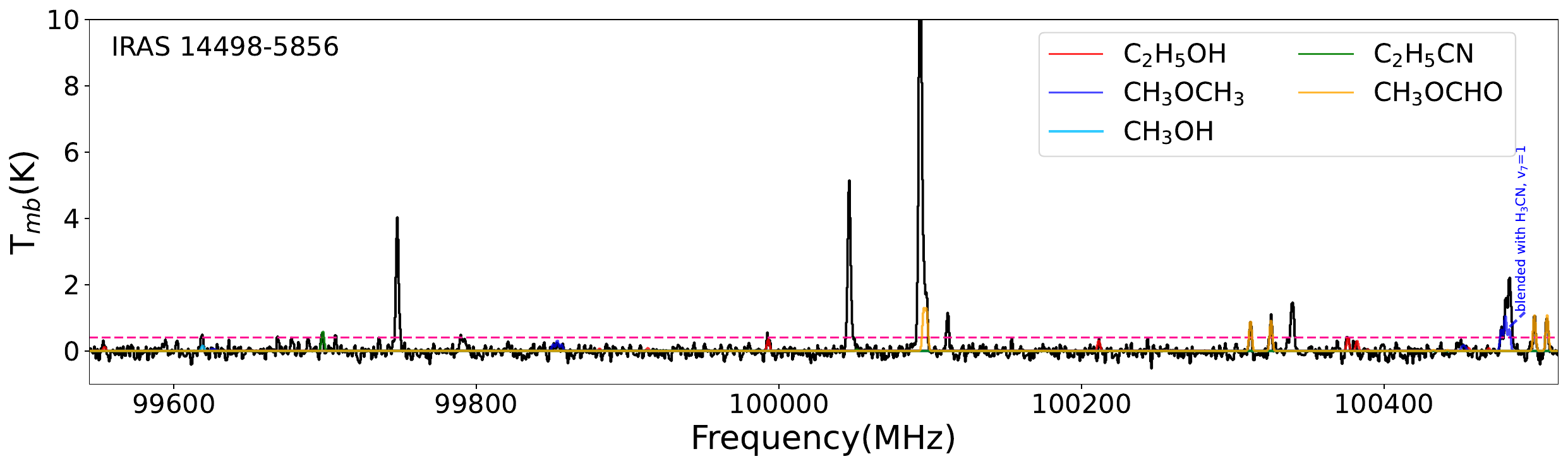}}
\quad
{\includegraphics[height=4.5cm,width=15.93cm]{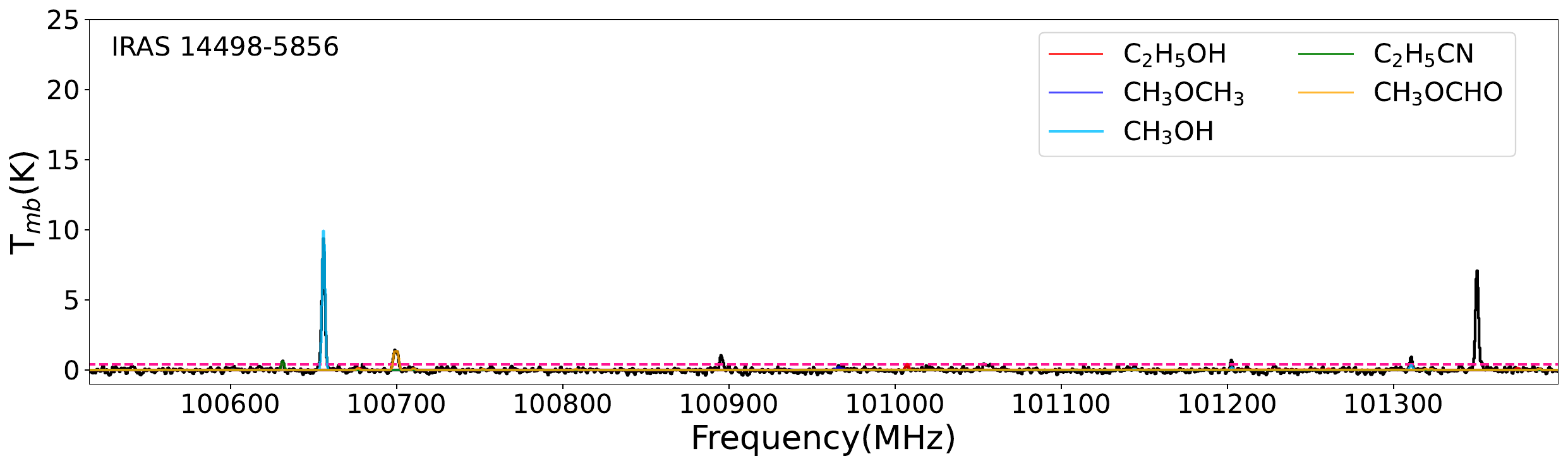}}
\quad
{\includegraphics[height=4.5cm,width=15.93cm]{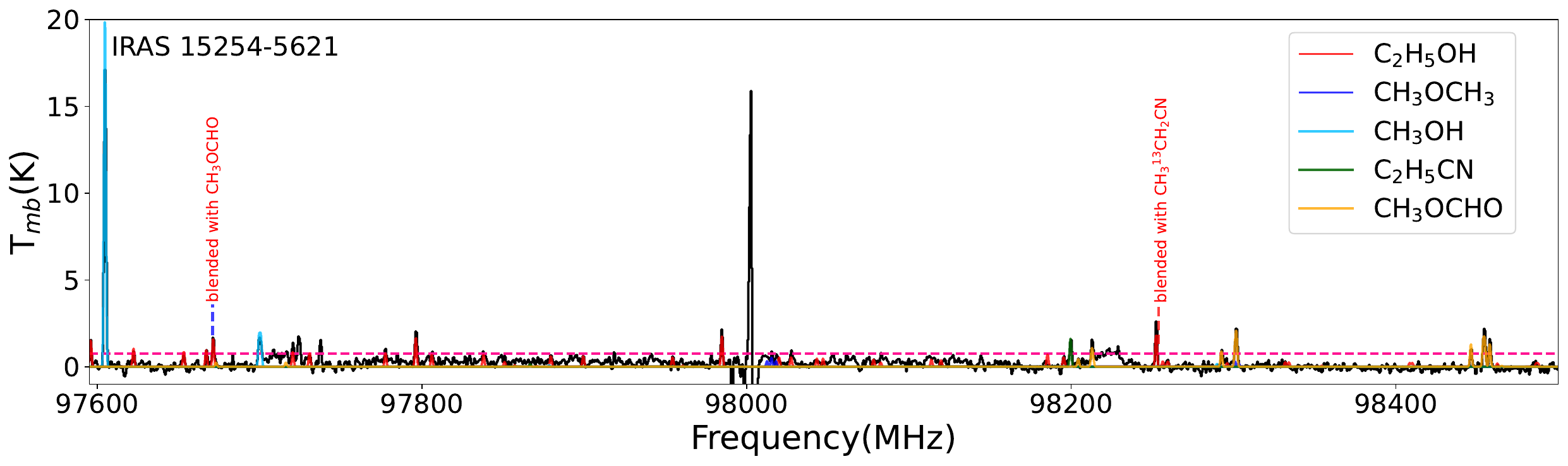}}
\caption{Continued.}
\end{figure}
\setcounter{figure}{\value{figure}-1}
\begin{figure}
  \centering 
{\includegraphics[height=4.5cm,width=15.93cm]{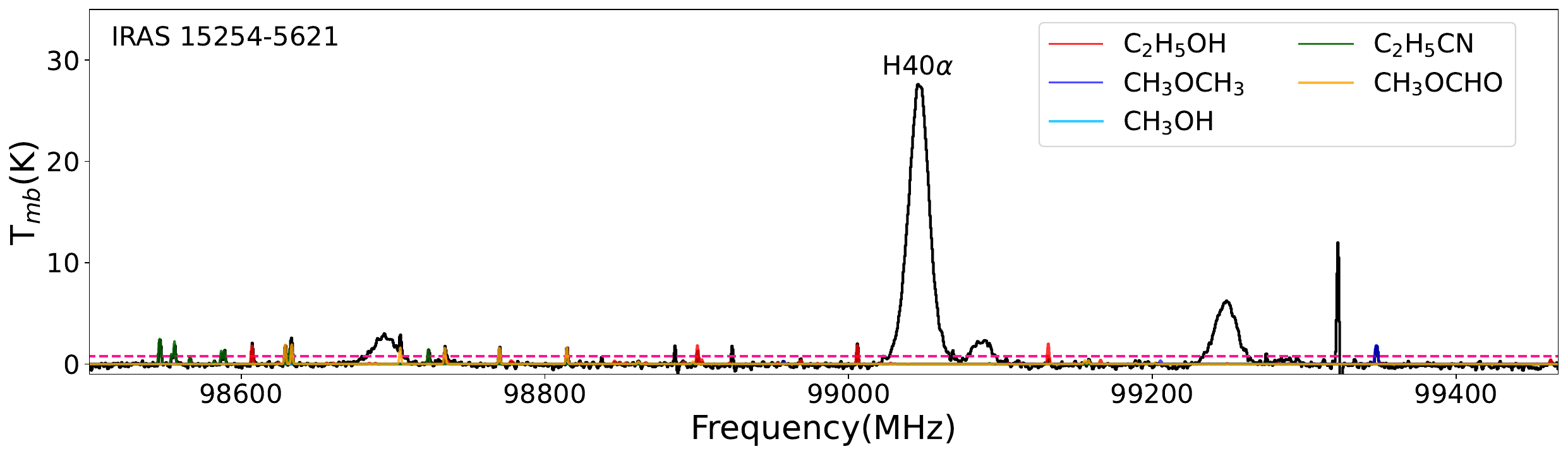}}
\quad
{\includegraphics[height=4.5cm,width=15.93cm]{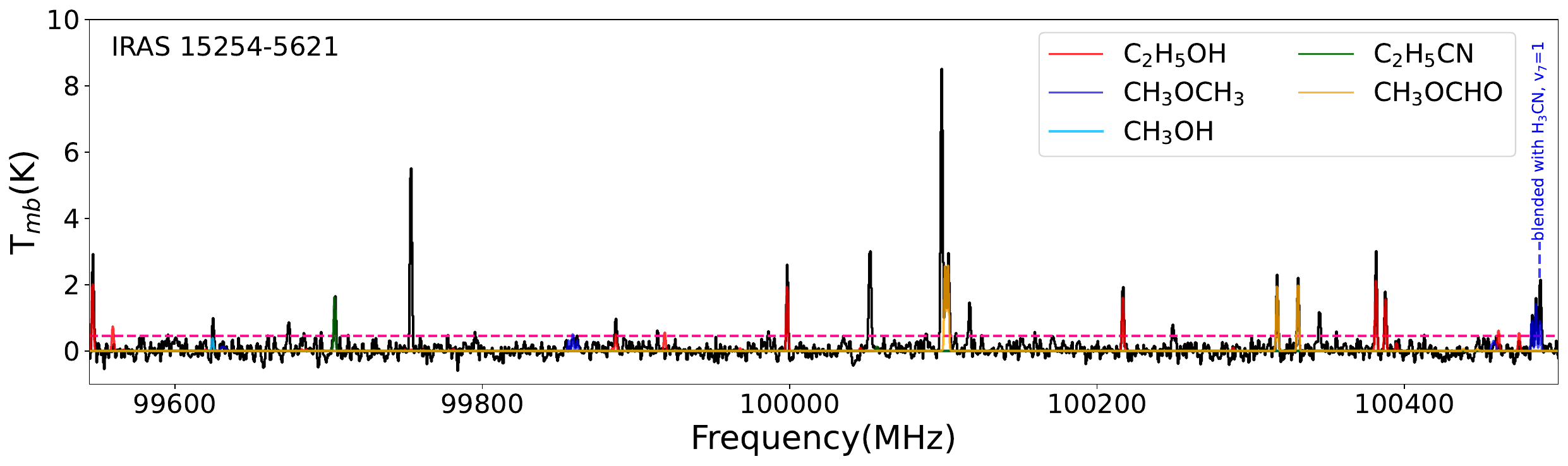}}
\quad
{\includegraphics[height=4.5cm,width=15.93cm]{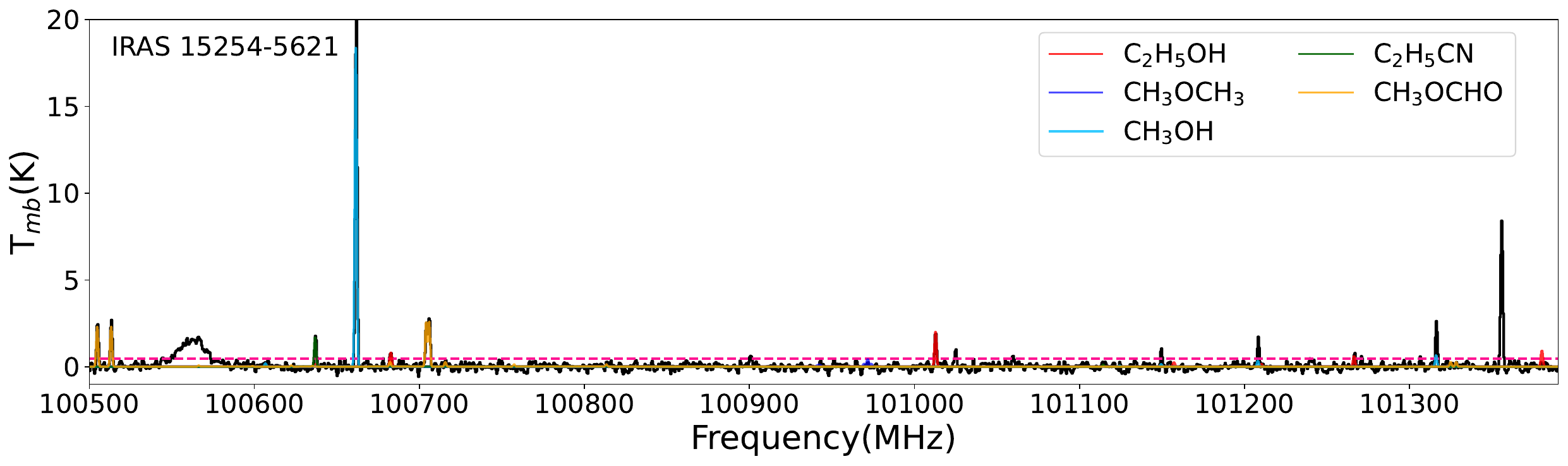}}
\quad
{\includegraphics[height=4.5cm,width=15.93cm]{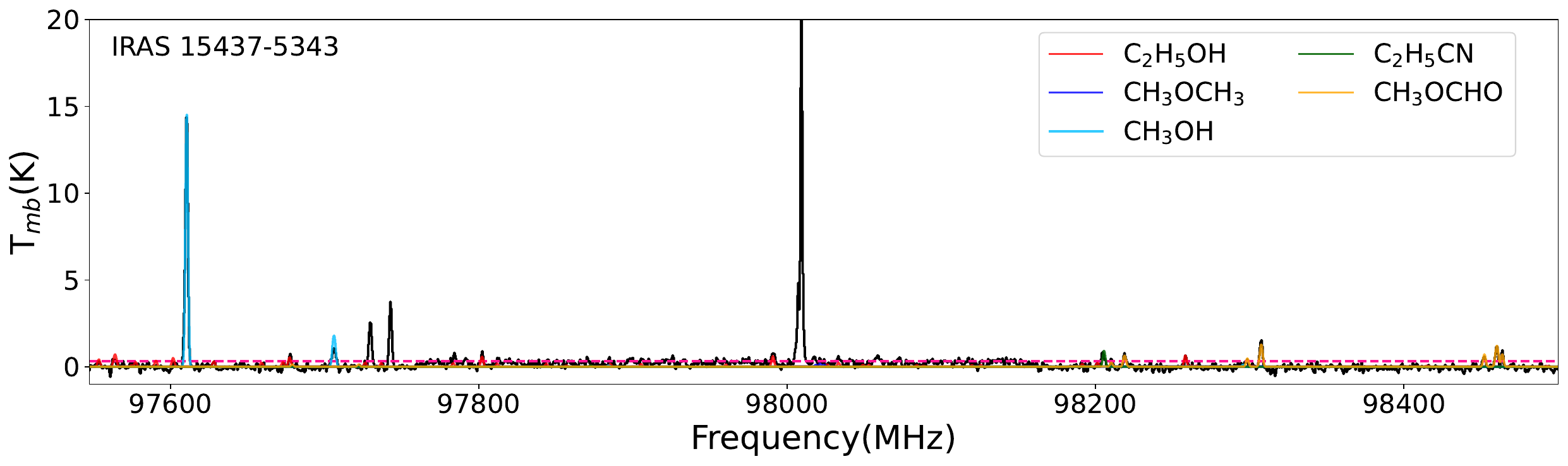}}
\quad
{\includegraphics[height=4.5cm,width=15.93cm]{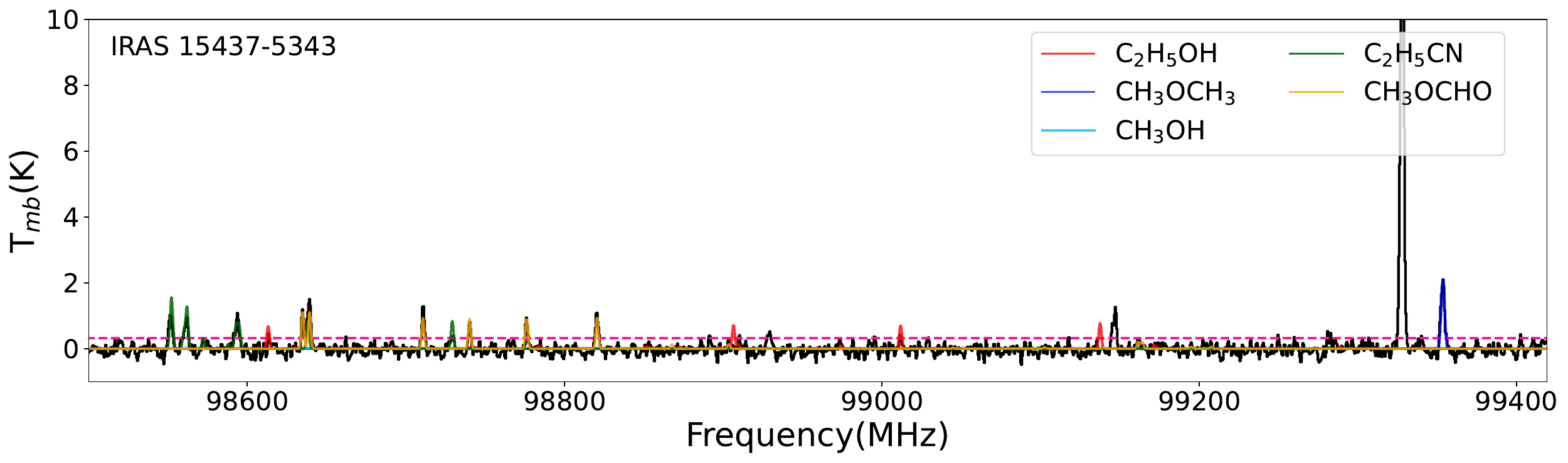}}
\caption{Continued.}
\end{figure}
\setcounter{figure}{\value{figure}-1}
\begin{figure}
  \centering 
{\includegraphics[height=4.5cm,width=15.93cm]{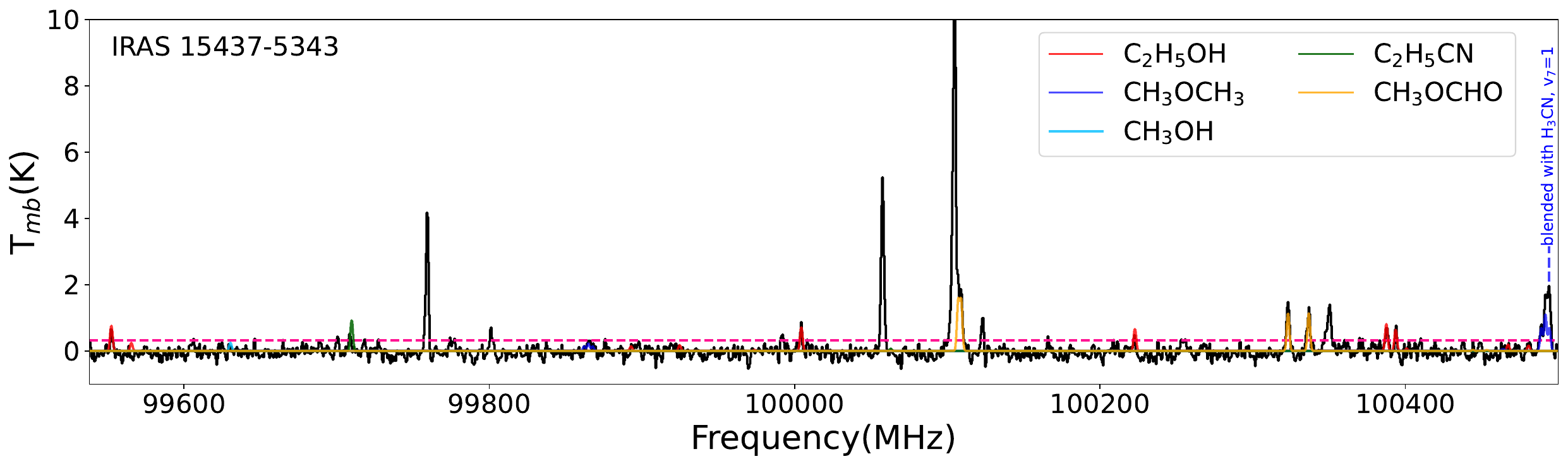}}
\quad
{\includegraphics[height=4.5cm,width=15.93cm]{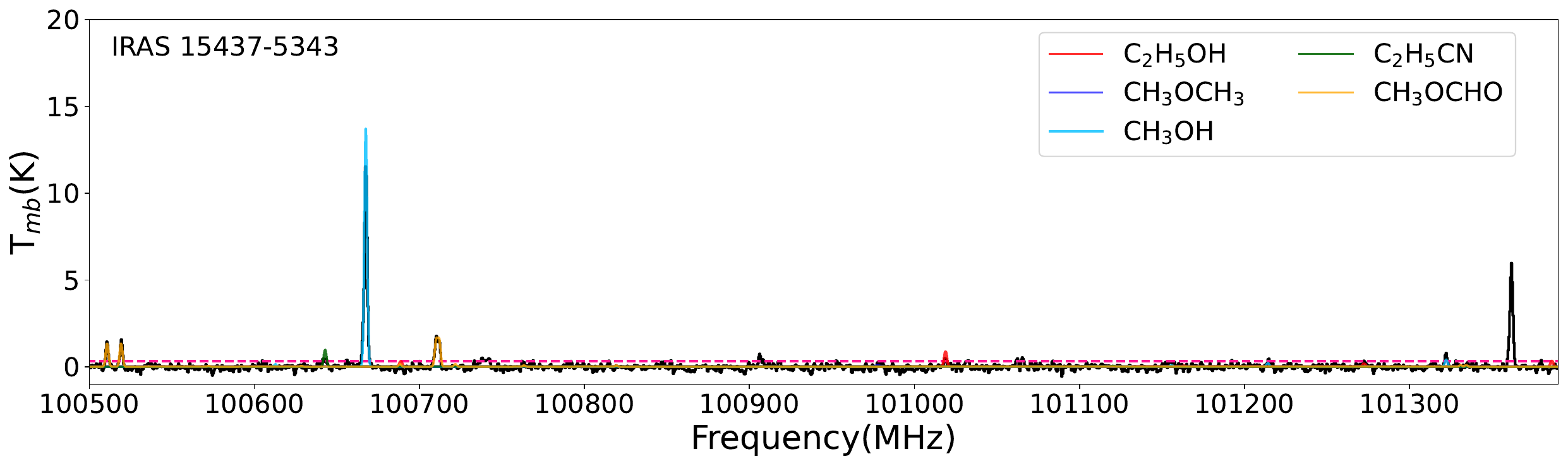}}
\quad
{\includegraphics[height=4.5cm,width=15.93cm]{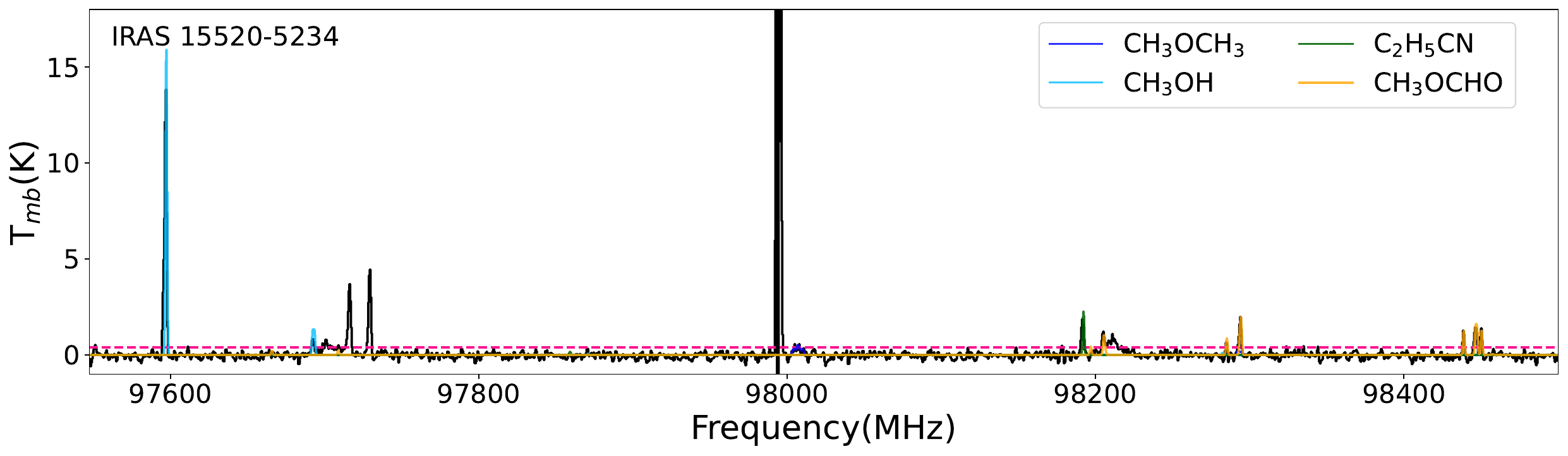}}
\quad
{\includegraphics[height=4.5cm,width=15.93cm]{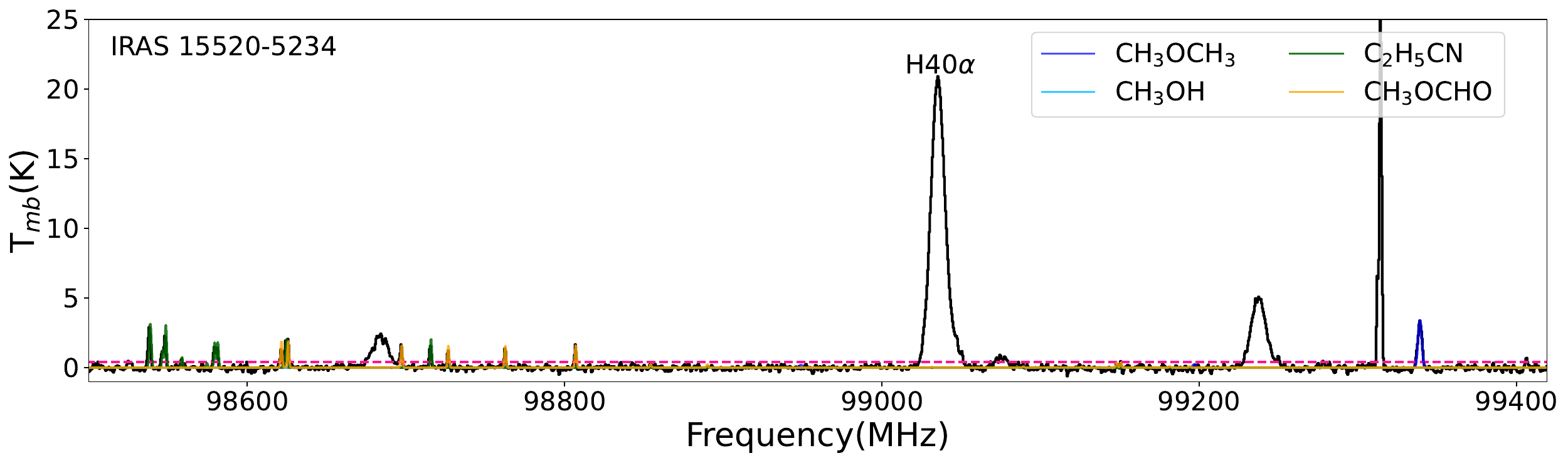}}
\quad
{\includegraphics[height=4.5cm,width=15.93cm]{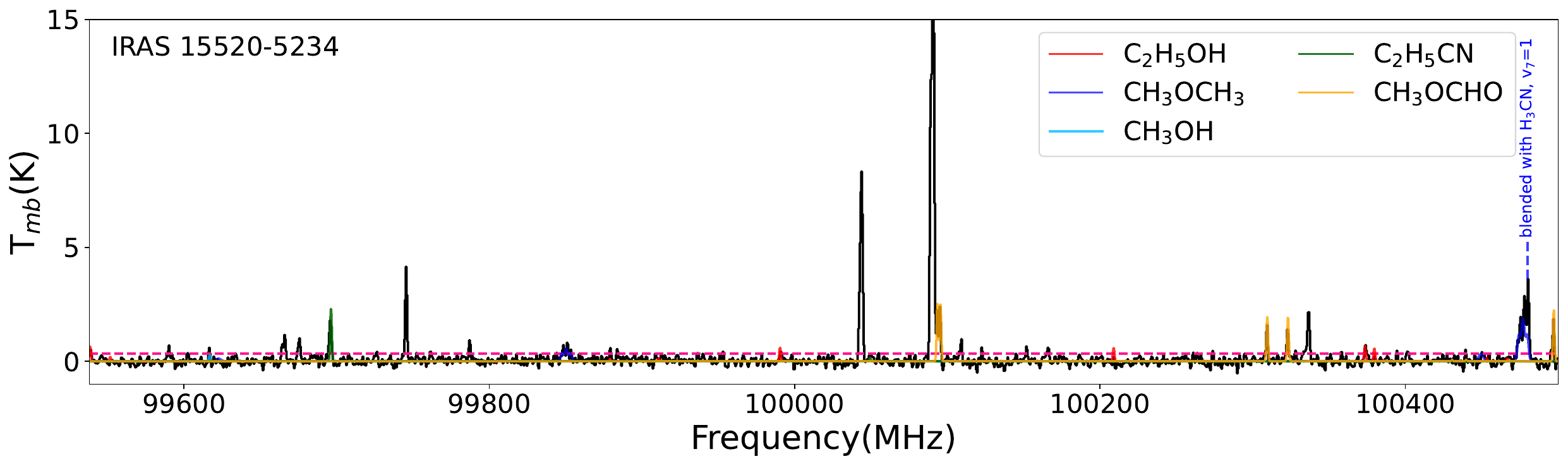}}
\caption{Continued.}
\end{figure}
\setcounter{figure}{\value{figure}-1}
\begin{figure}
  \centering 
{\includegraphics[height=4.5cm,width=15.93cm]{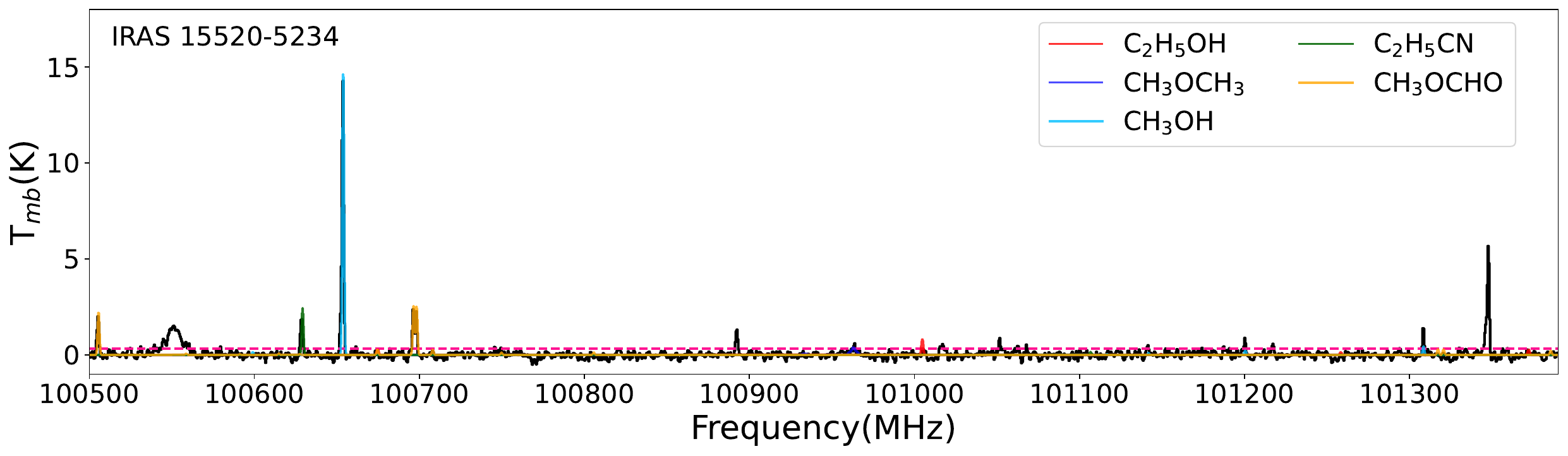}}
\quad
{\includegraphics[height=4.5cm,width=15.93cm]{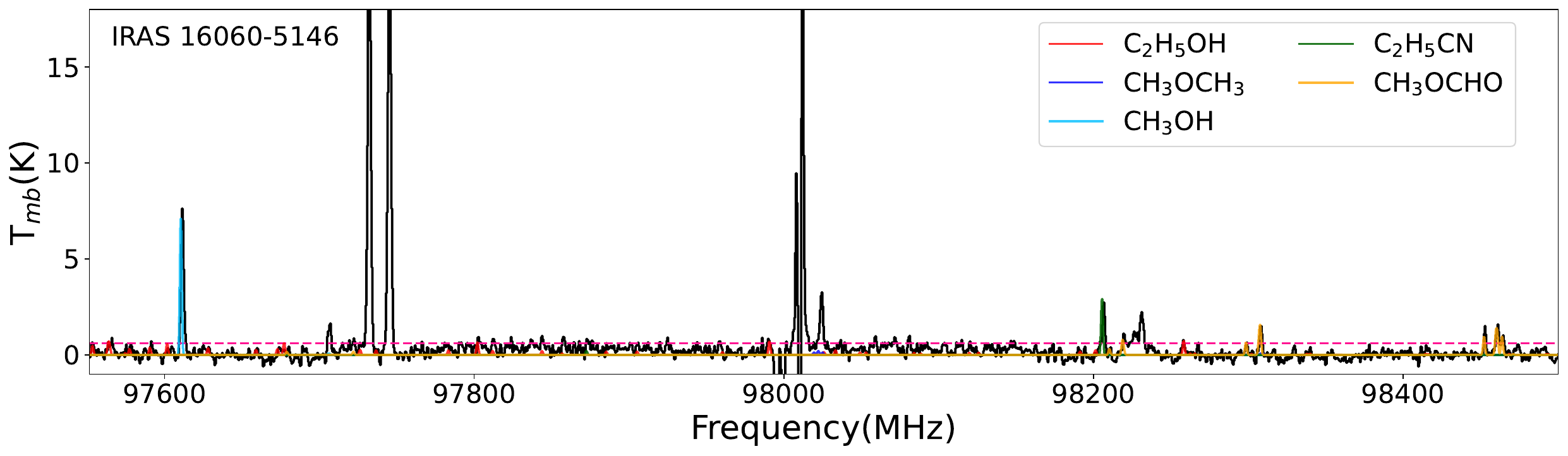}}
\quad
{\includegraphics[height=4.5cm,width=15.93cm]{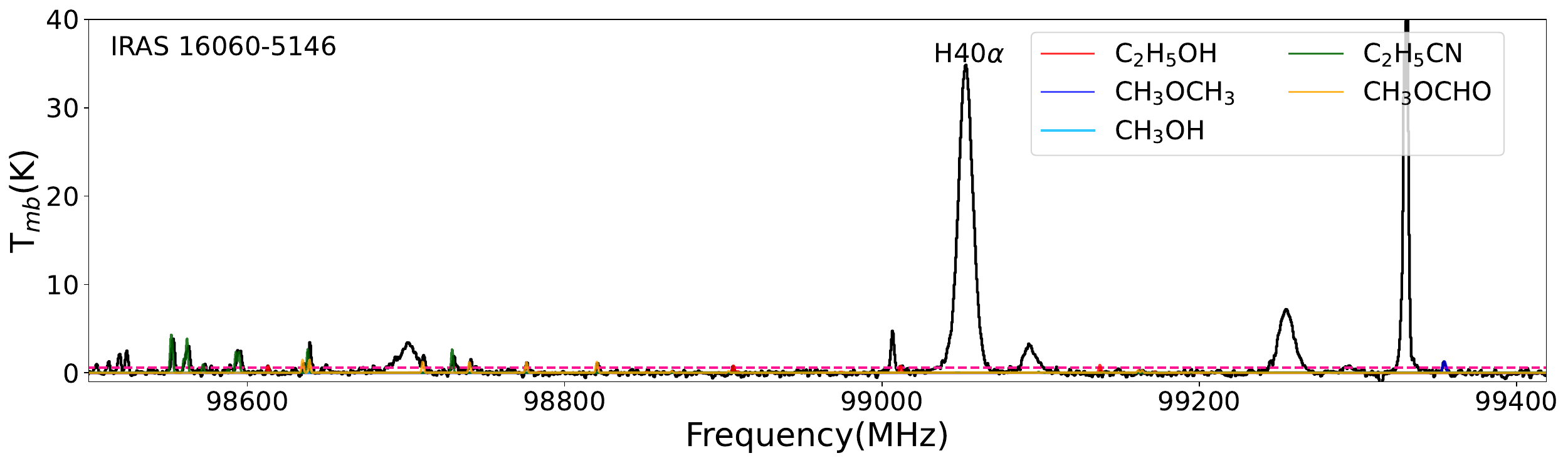}}
\quad
{\includegraphics[height=4.5cm,width=15.93cm]{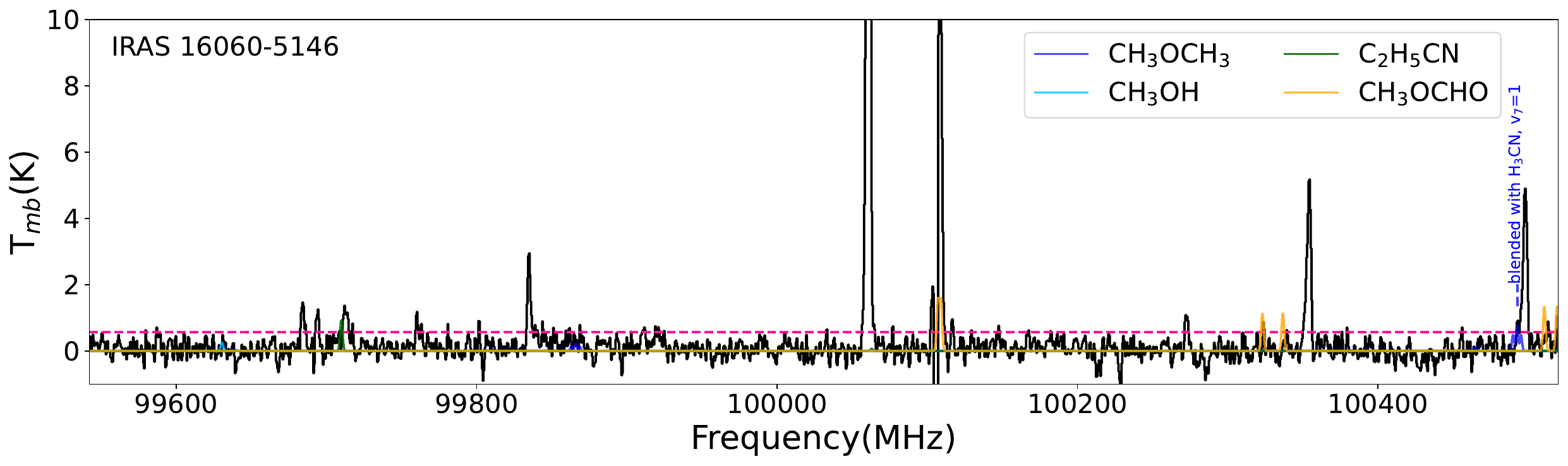}}
\quad
{\includegraphics[height=4.5cm,width=15.93cm]{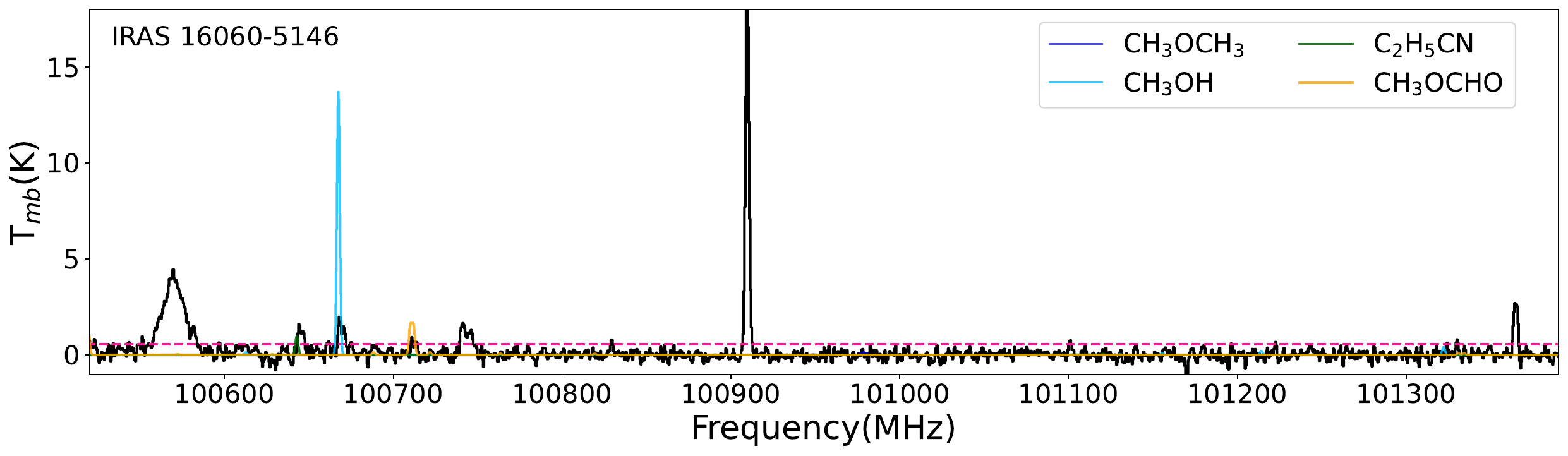}}
\caption{Continued.}
\end{figure}
\setcounter{figure}{\value{figure}-1}
\begin{figure}
  \centering 
{\includegraphics[height=4.5cm,width=15.93cm]{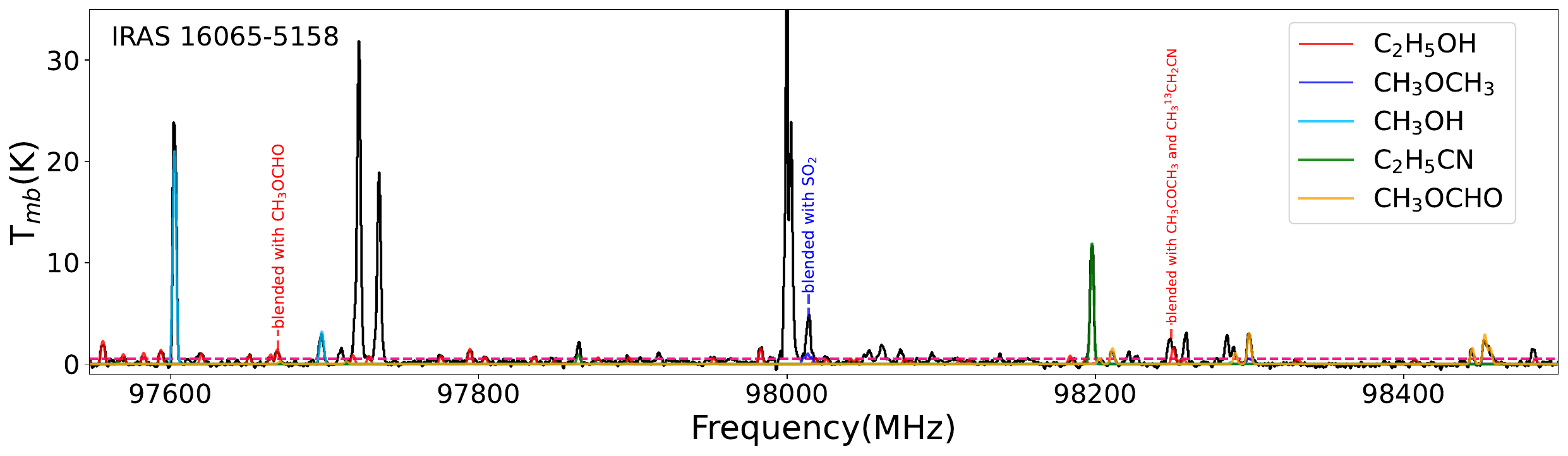}}
\quad
{\includegraphics[height=4.5cm,width=15.93cm]{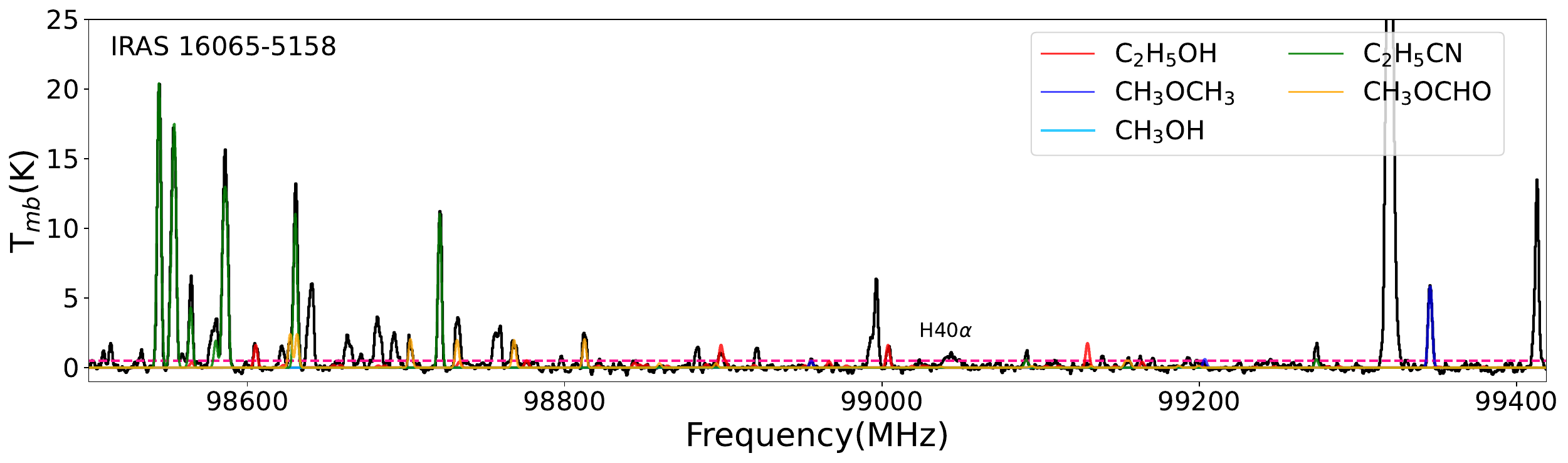}}
\quad
{\includegraphics[height=4.5cm,width=15.93cm]{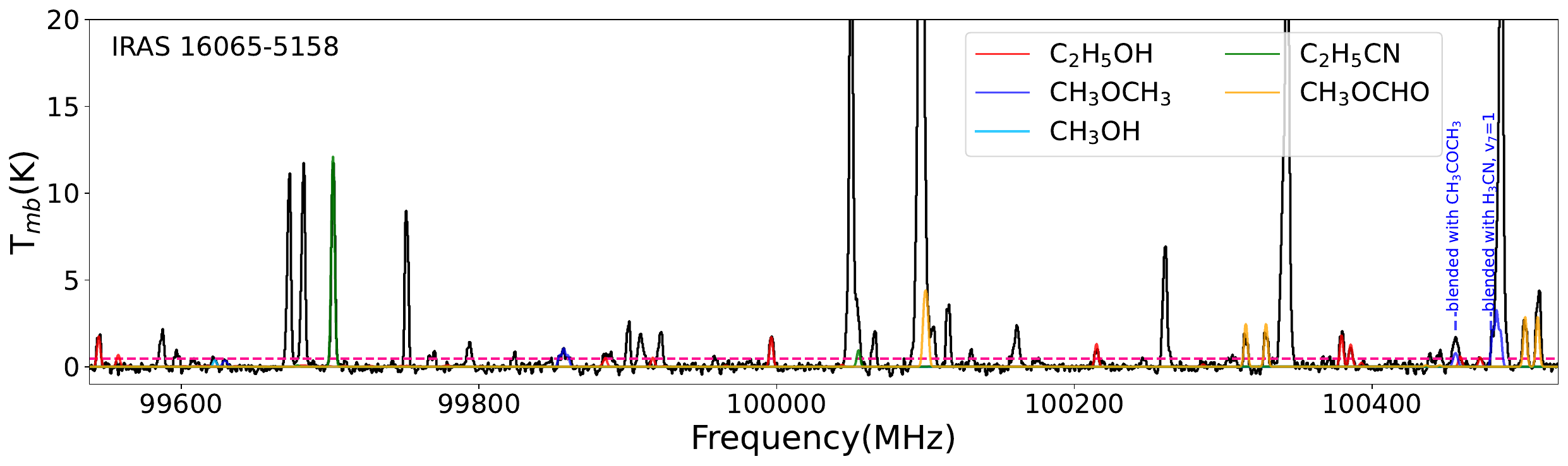}}
\quad
{\includegraphics[height=4.5cm,width=15.93cm]{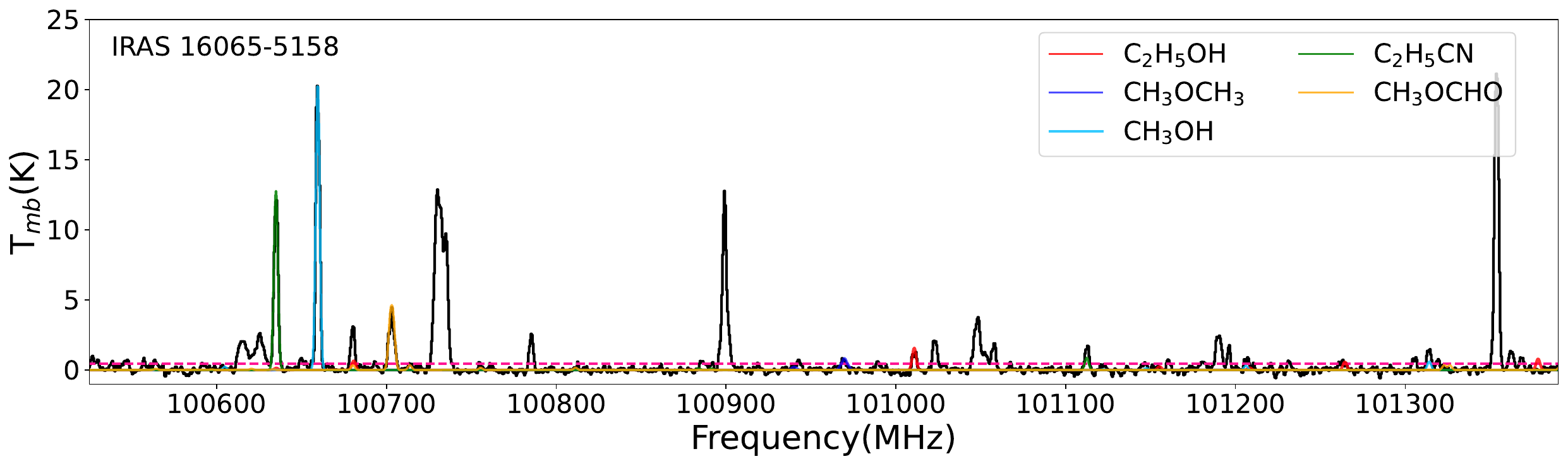}}
\quad
{\includegraphics[height=4.5cm,width=15.93cm]{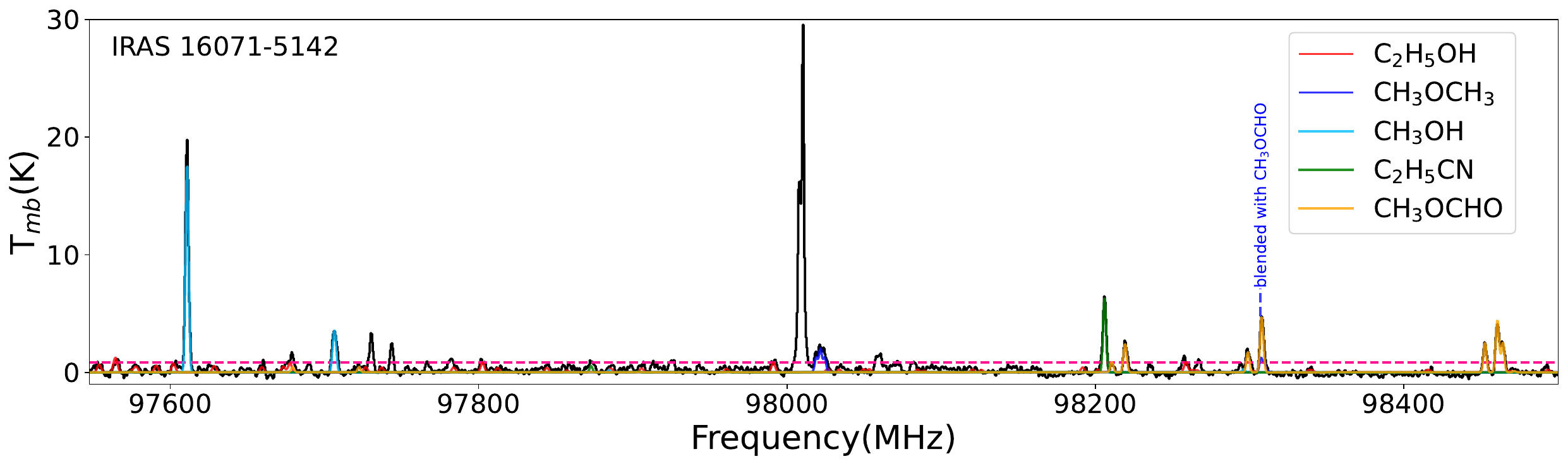}}
\caption{Continued.}
\end{figure}
\setcounter{figure}{\value{figure}-1}
\begin{figure}
  \centering 
{\includegraphics[height=4.5cm,width=15.93cm]{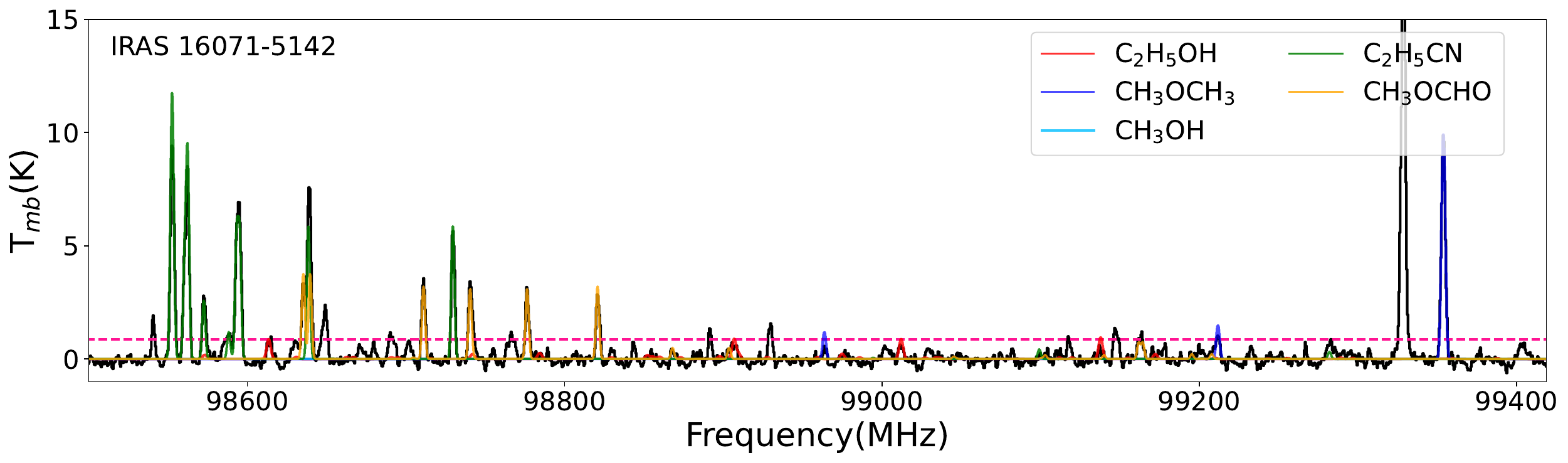}}
\quad
{\includegraphics[height=4.5cm,width=15.93cm]{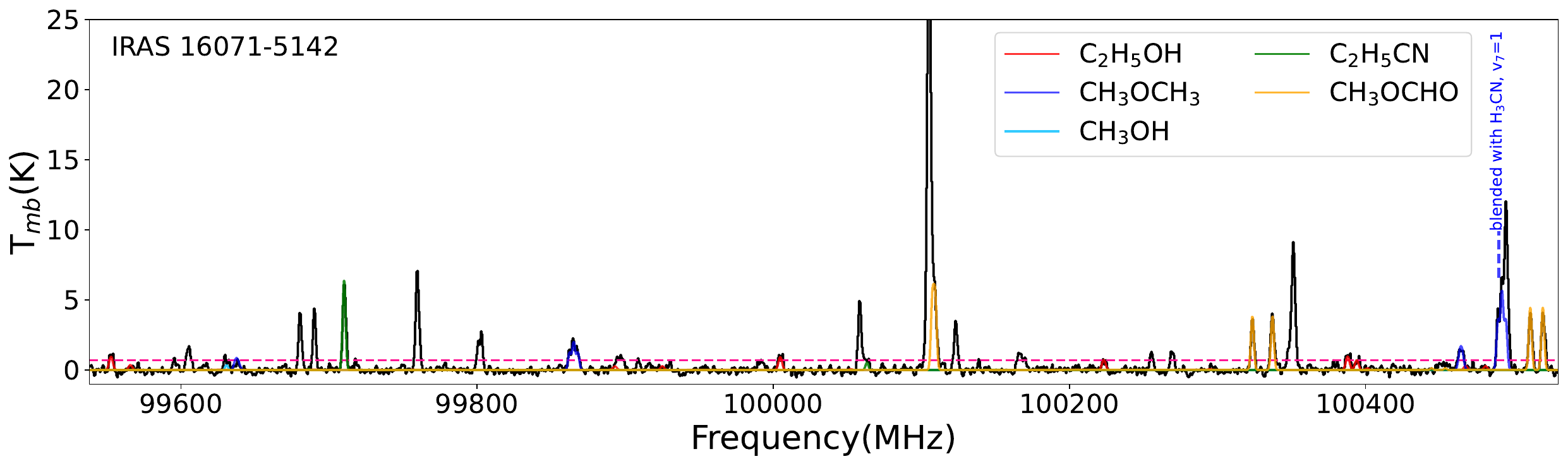}}
\quad
{\includegraphics[height=4.5cm,width=15.93cm]{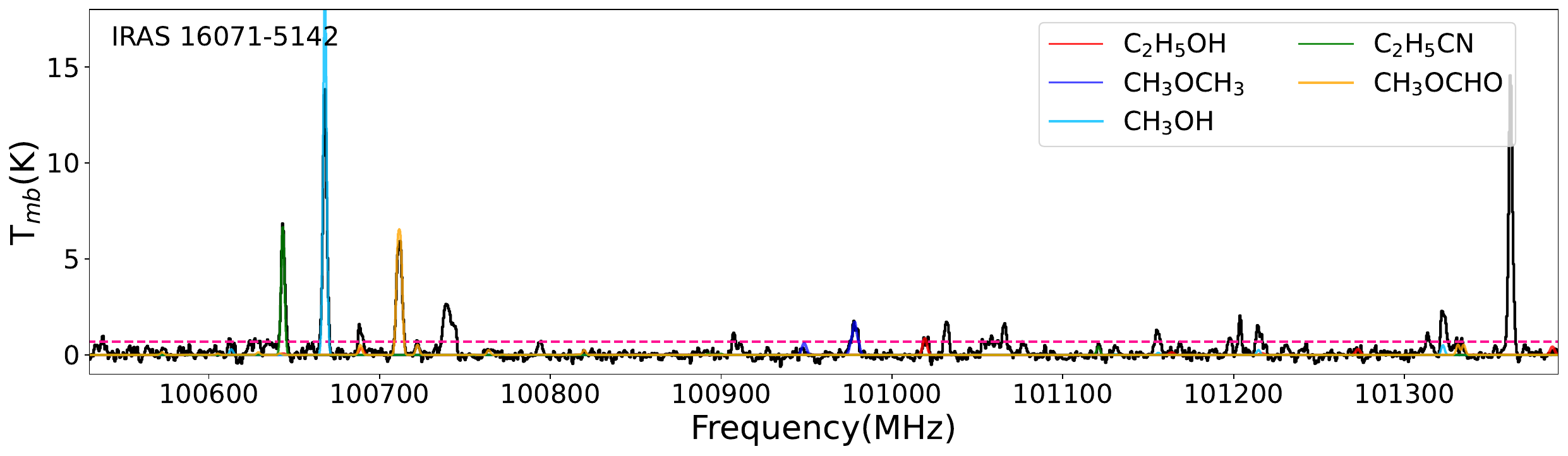}}
\quad
{\includegraphics[height=4.5cm,width=15.93cm]{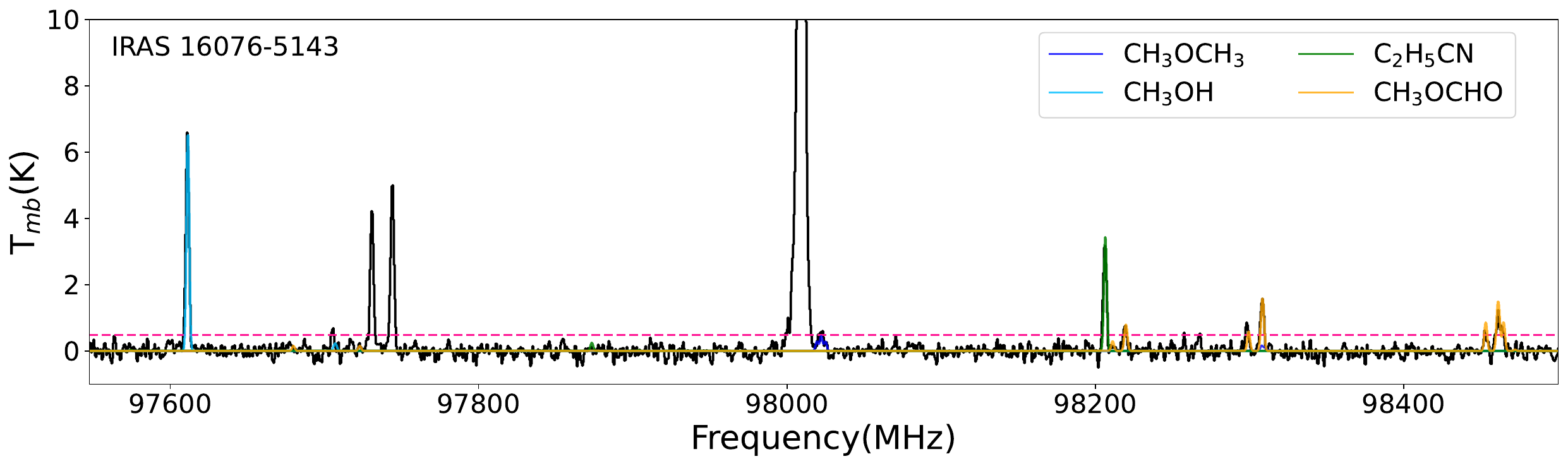}}
\quad
{\includegraphics[height=4.5cm,width=15.93cm]{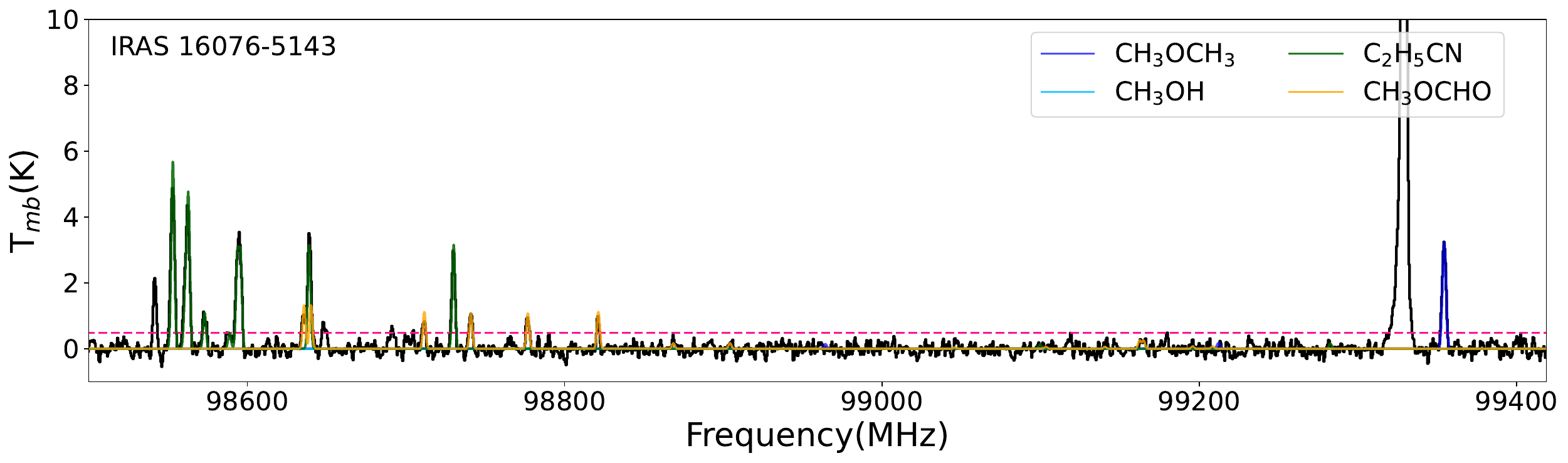}}
\caption{Continued.}
\end{figure}
\setcounter{figure}{\value{figure}-1}
\begin{figure}
  \centering 
{\includegraphics[height=4.5cm,width=15.93cm]{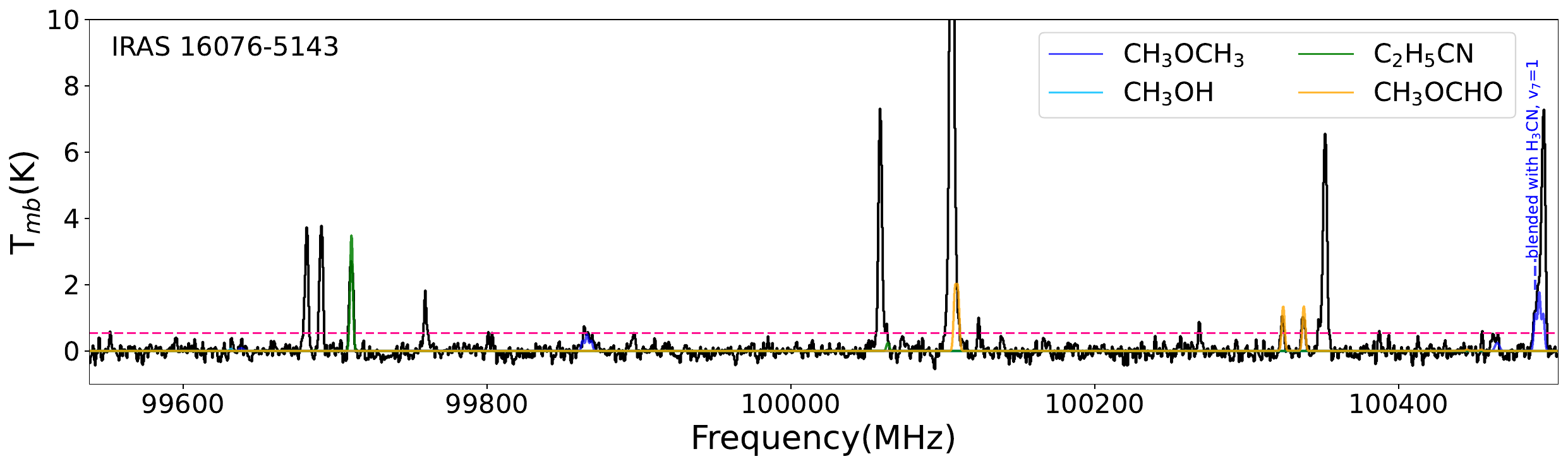}}
\quad
{\includegraphics[height=4.5cm,width=15.93cm]{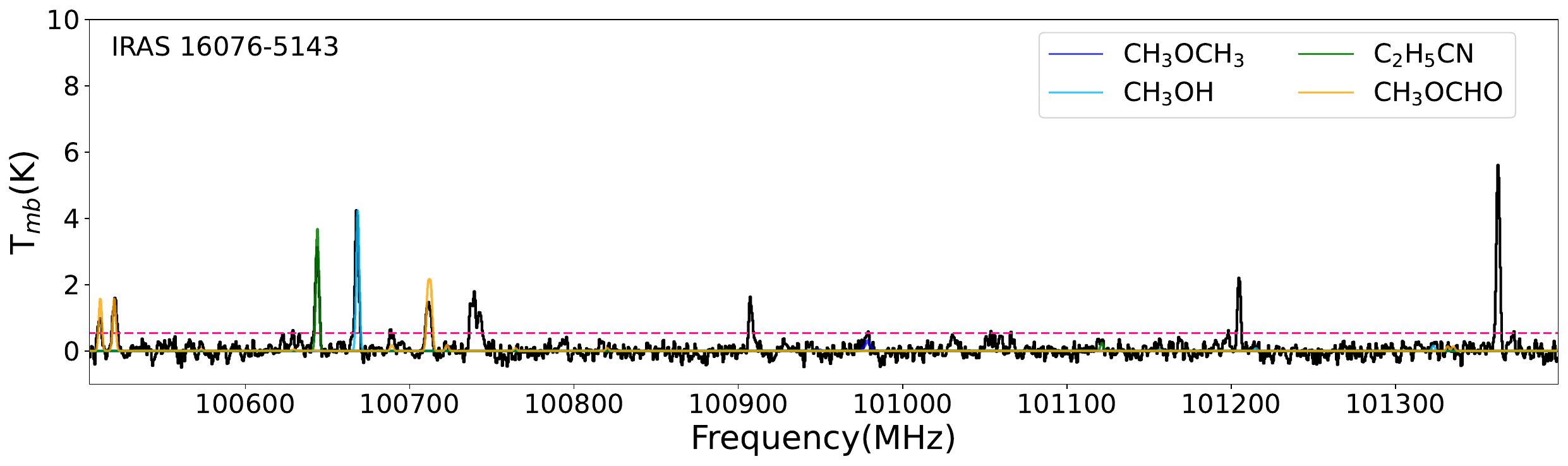}}
\quad
{\includegraphics[height=4.5cm,width=15.93cm]{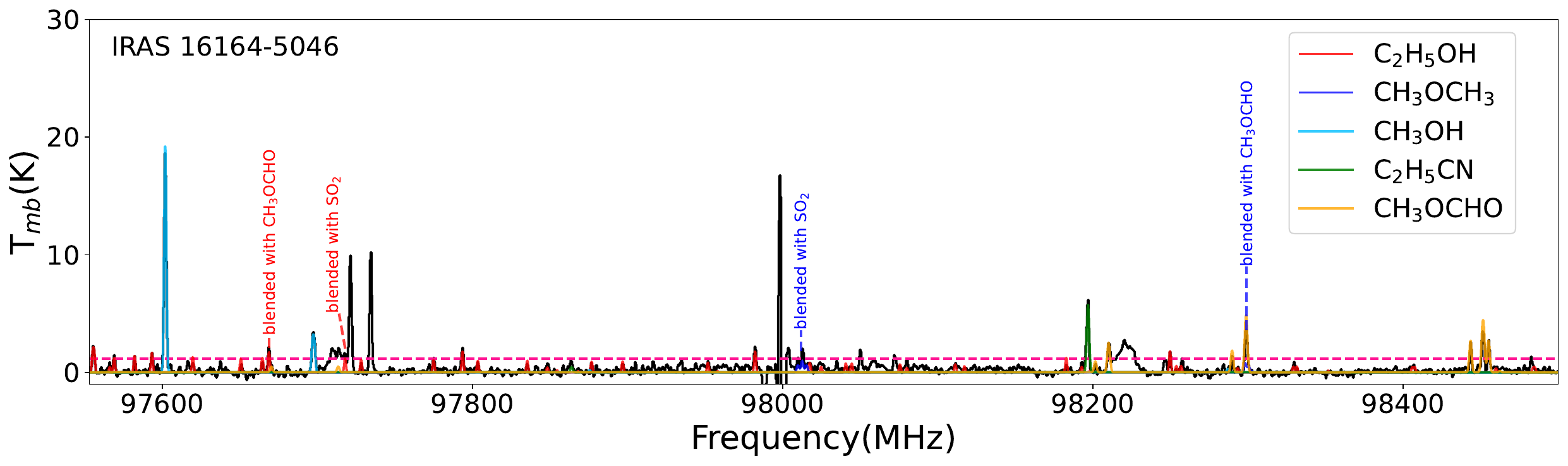}}
\quad
{\includegraphics[height=4.5cm,width=15.93cm]{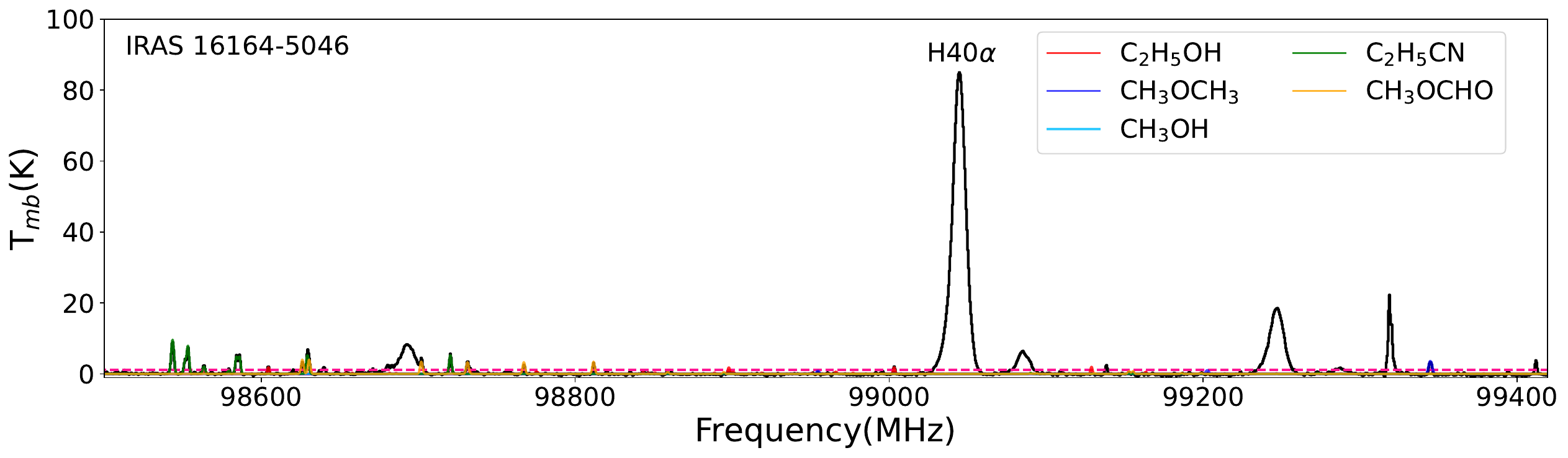}}
\quad
{\includegraphics[height=4.5cm,width=15.93cm]{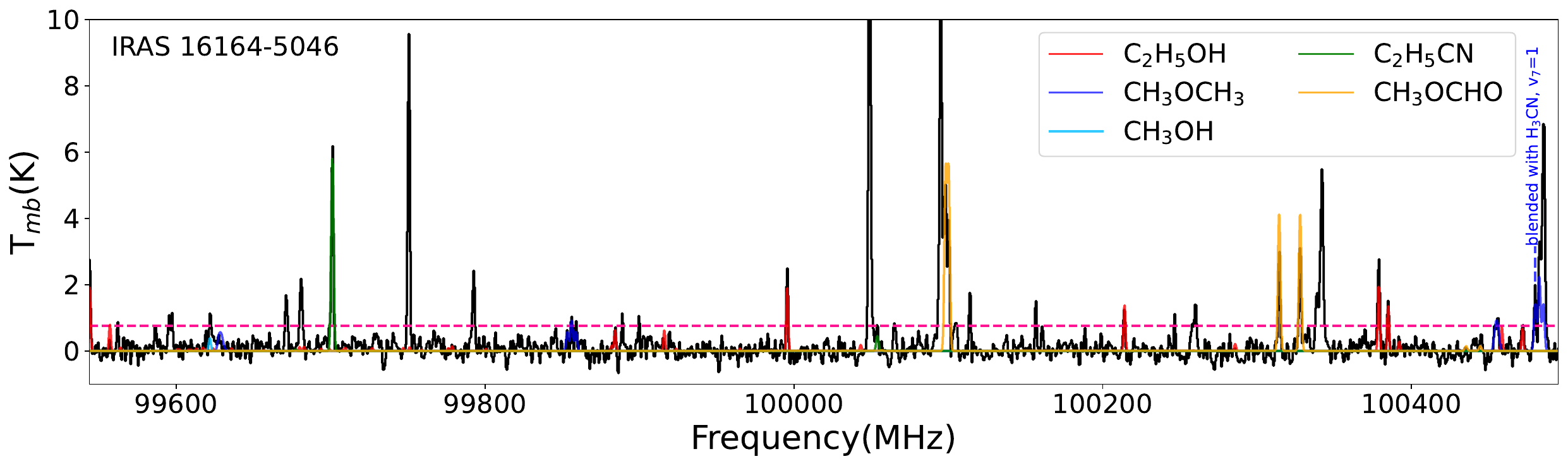}}
\caption{Continued.}
\end{figure}
\setcounter{figure}{\value{figure}-1}
\begin{figure}
  \centering 
{\includegraphics[height=4.5cm,width=15.93cm]{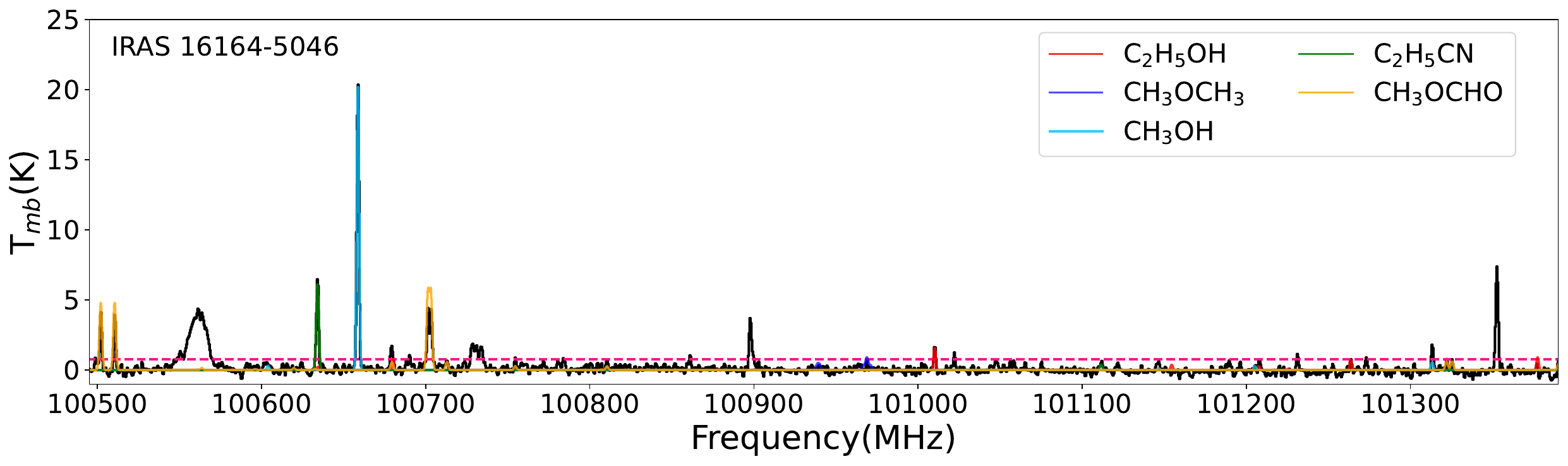}}
\quad
{\includegraphics[height=4.5cm,width=15.93cm]{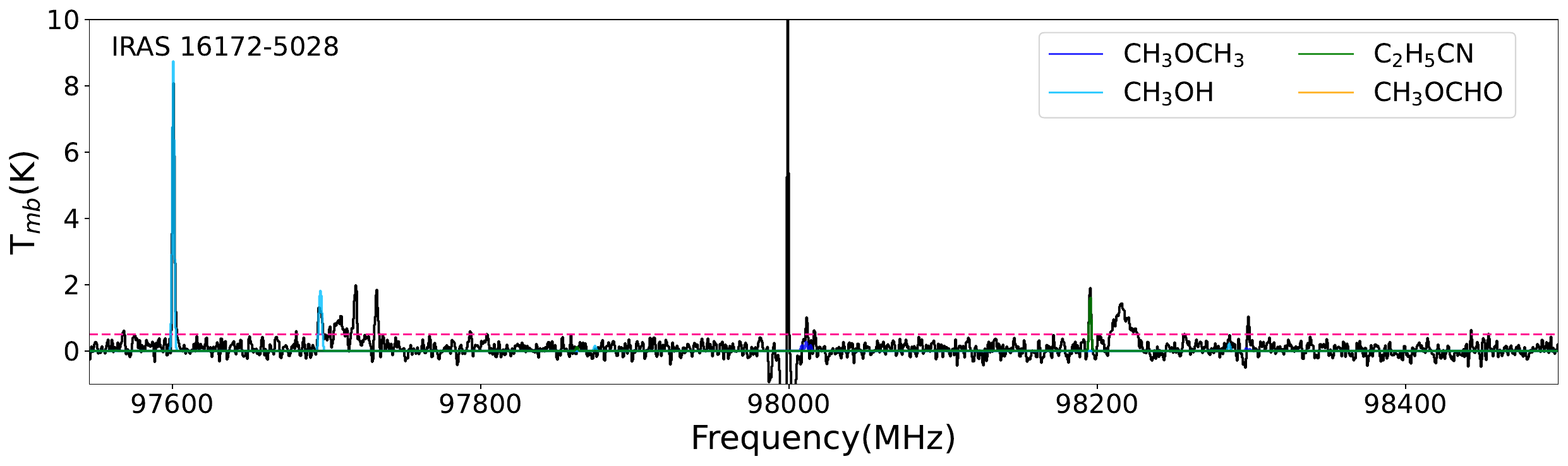}}
\quad
{\includegraphics[height=4.5cm,width=15.93cm]{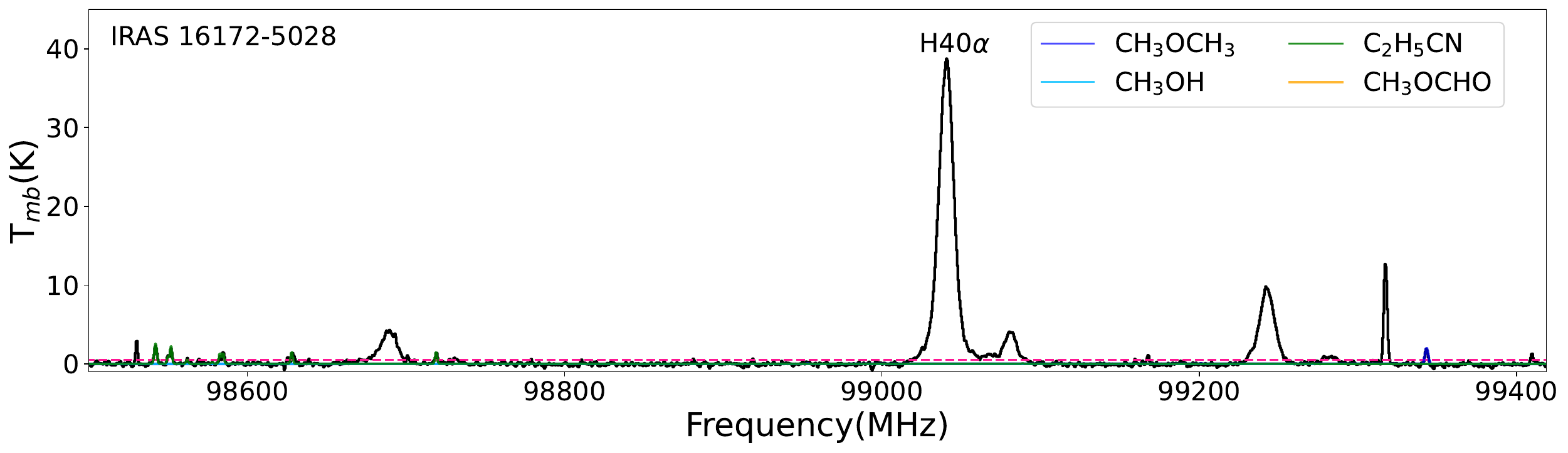}}
\quad
{\includegraphics[height=4.5cm,width=15.93cm]{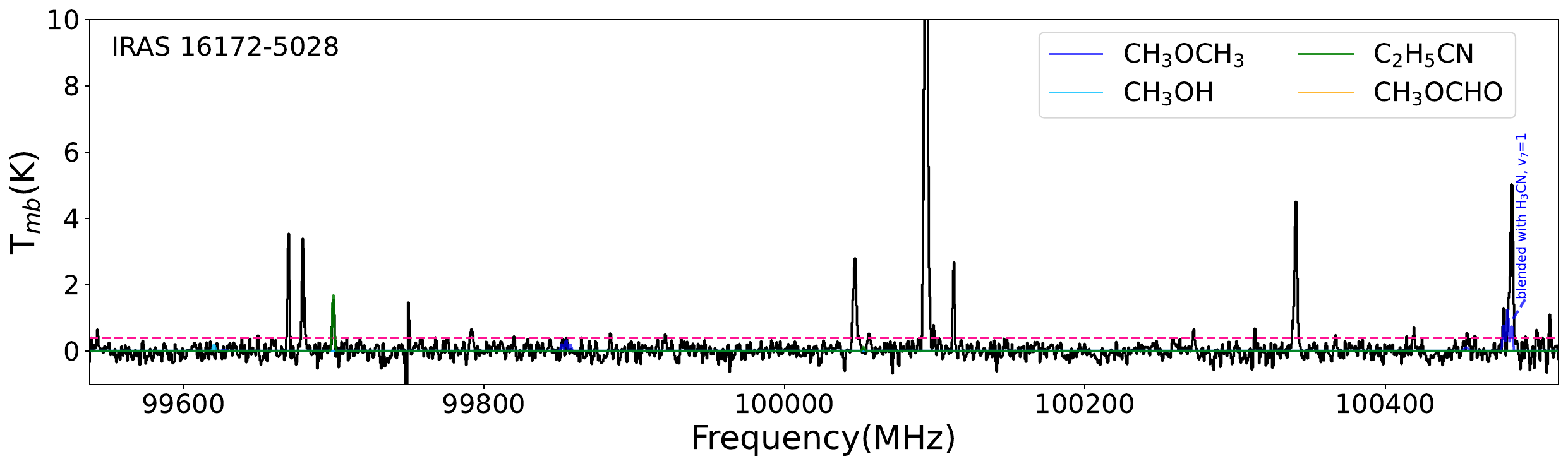}}
\quad
{\includegraphics[height=4.5cm,width=15.93cm]{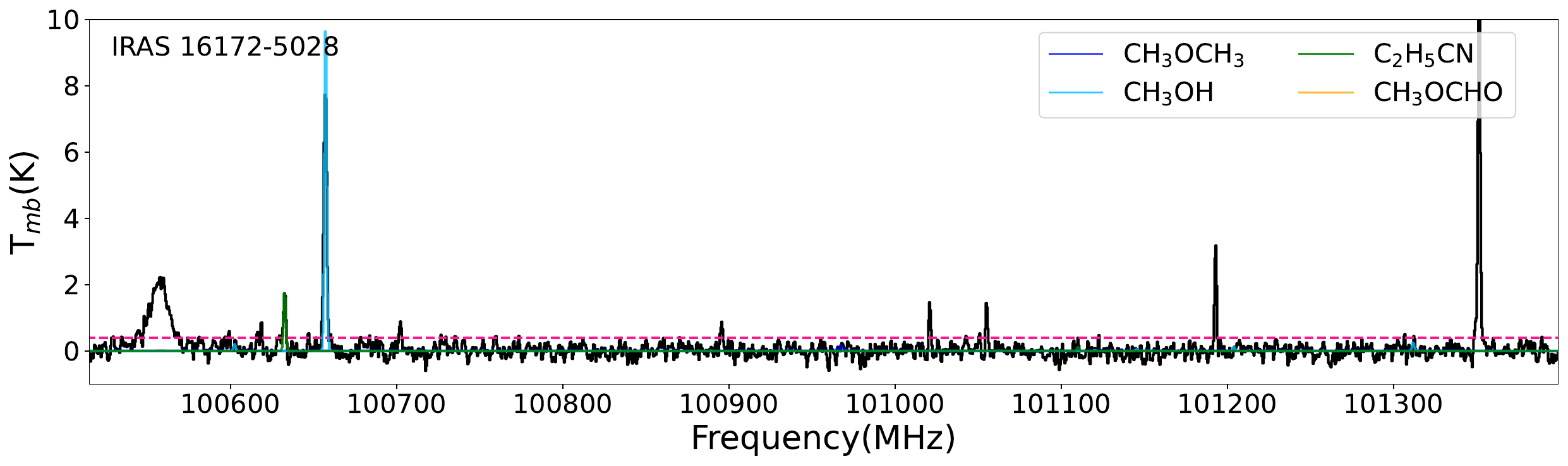}}
\caption{Continued.}
\end{figure}
\setcounter{figure}{\value{figure}-1}
\begin{figure}
  \centering 
{\includegraphics[height=4.5cm,width=15.93cm]{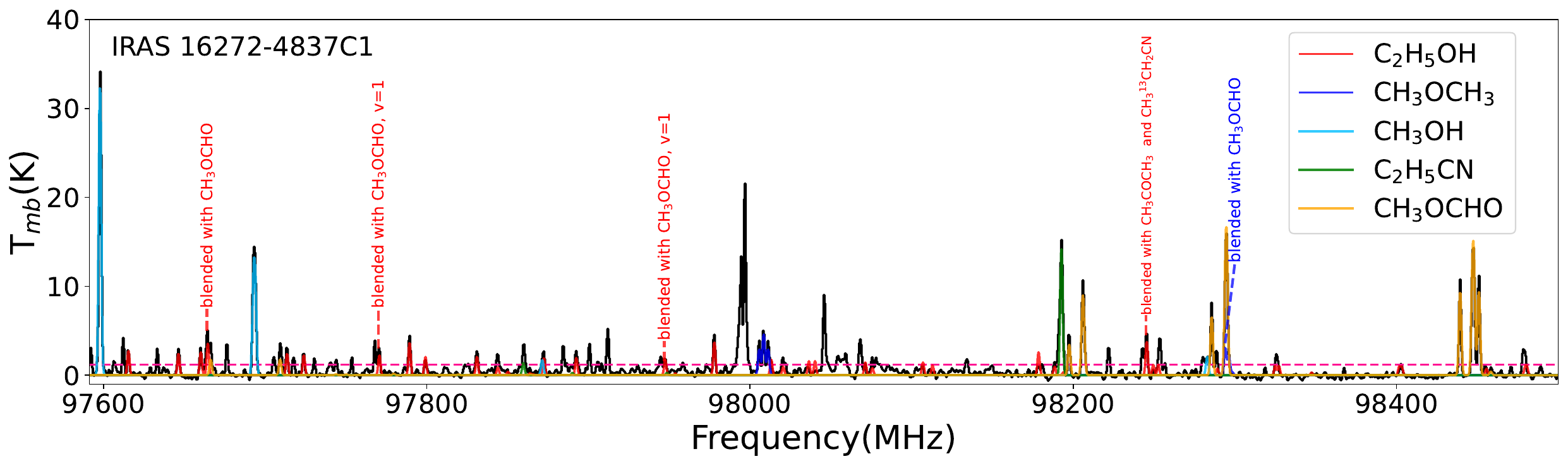}}
\quad
{\includegraphics[height=4.5cm,width=15.93cm]{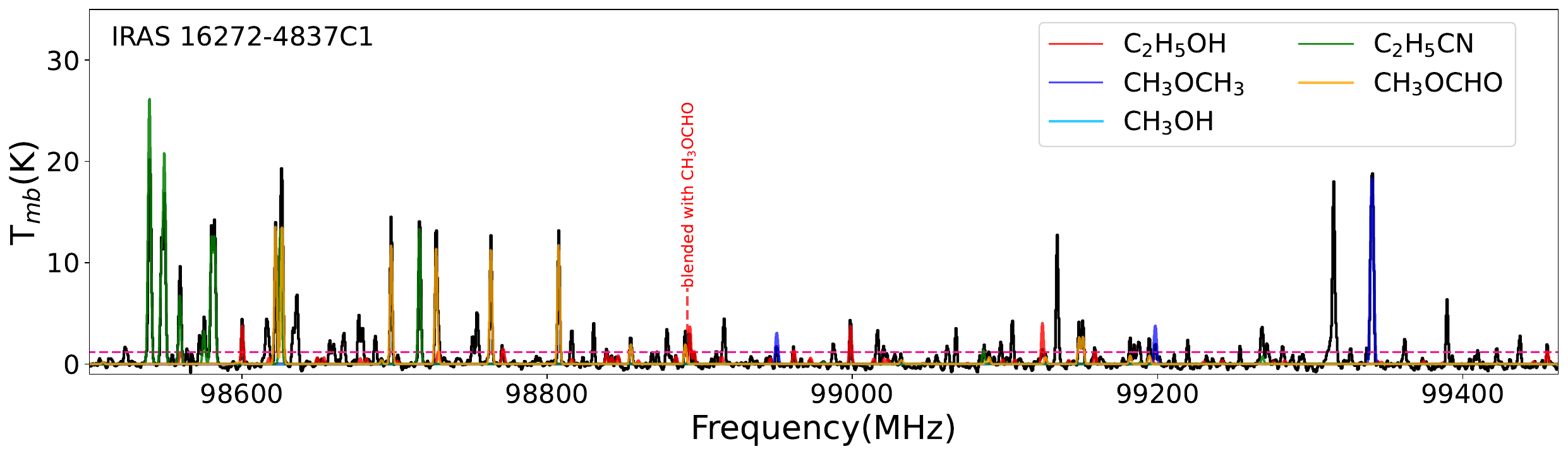}}
\quad
{\includegraphics[height=4.5cm,width=15.93cm]{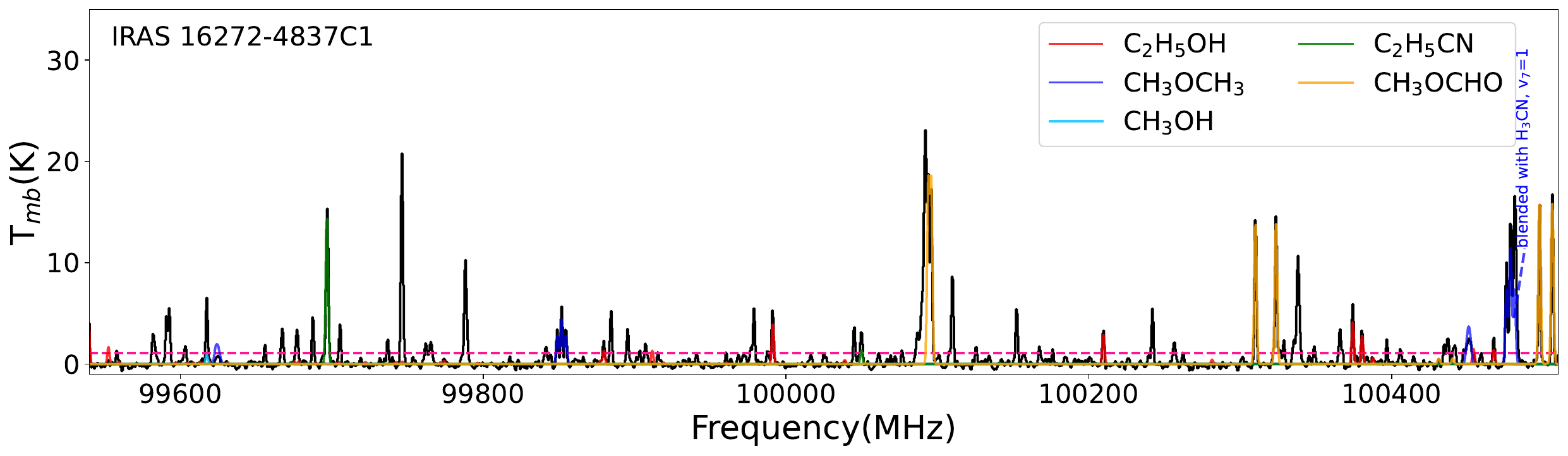}}
\quad
{\includegraphics[height=4.5cm,width=15.93cm]{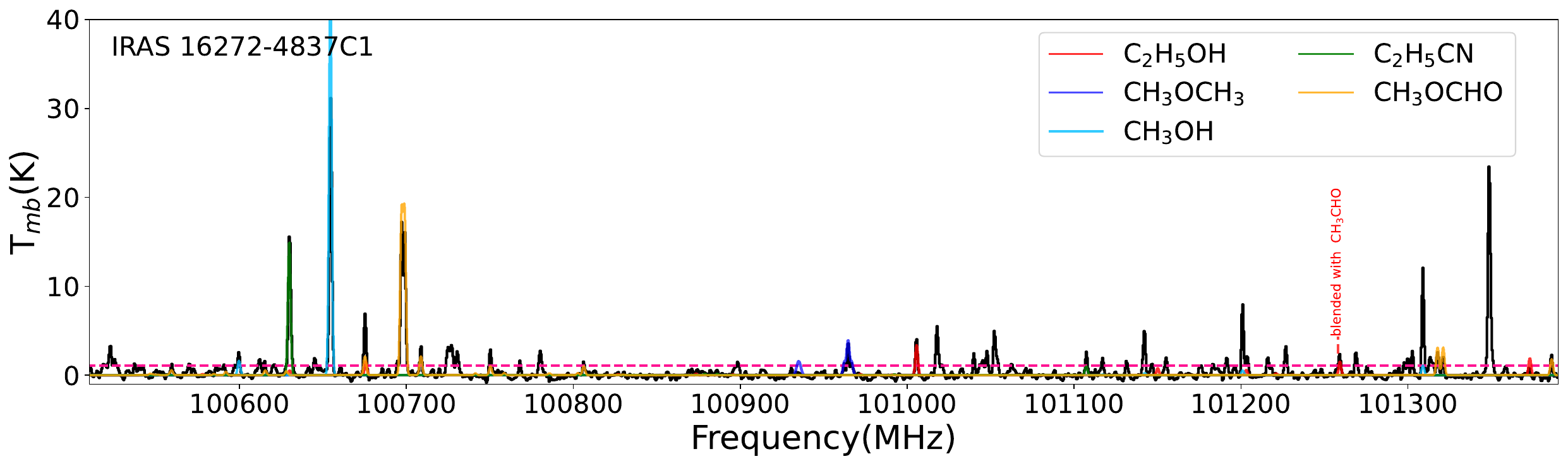}}
\quad
{\includegraphics[height=4.5cm,width=15.93cm]{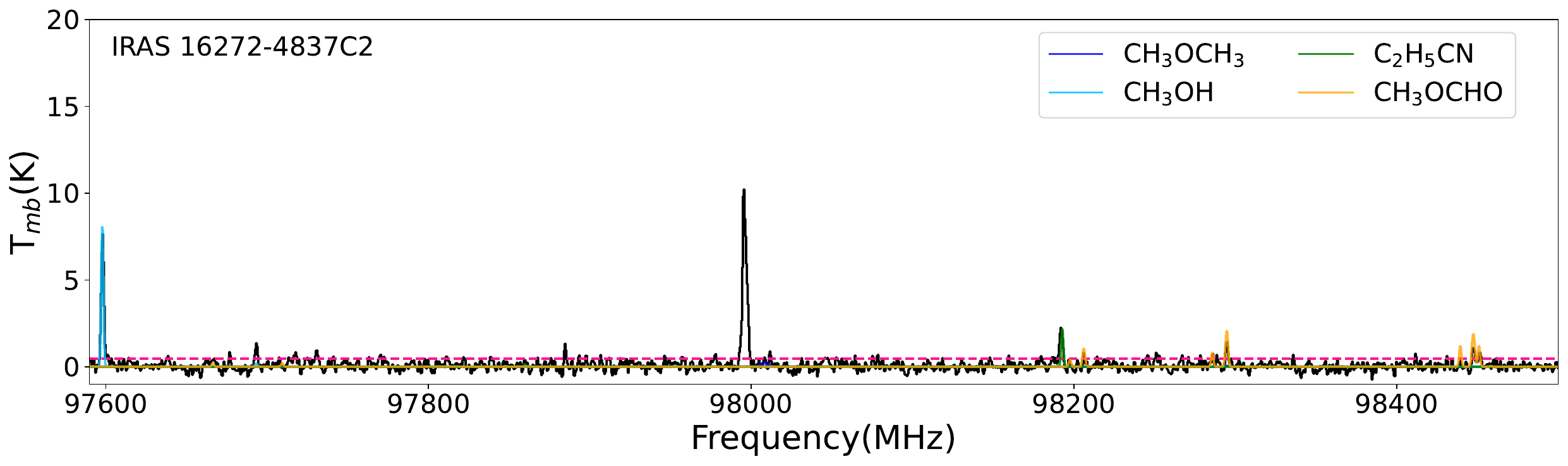}}
\caption{Continued.}
\end{figure}
\setcounter{figure}{\value{figure}-1}
\begin{figure}
  \centering 
{\includegraphics[height=4.5cm,width=15.93cm]{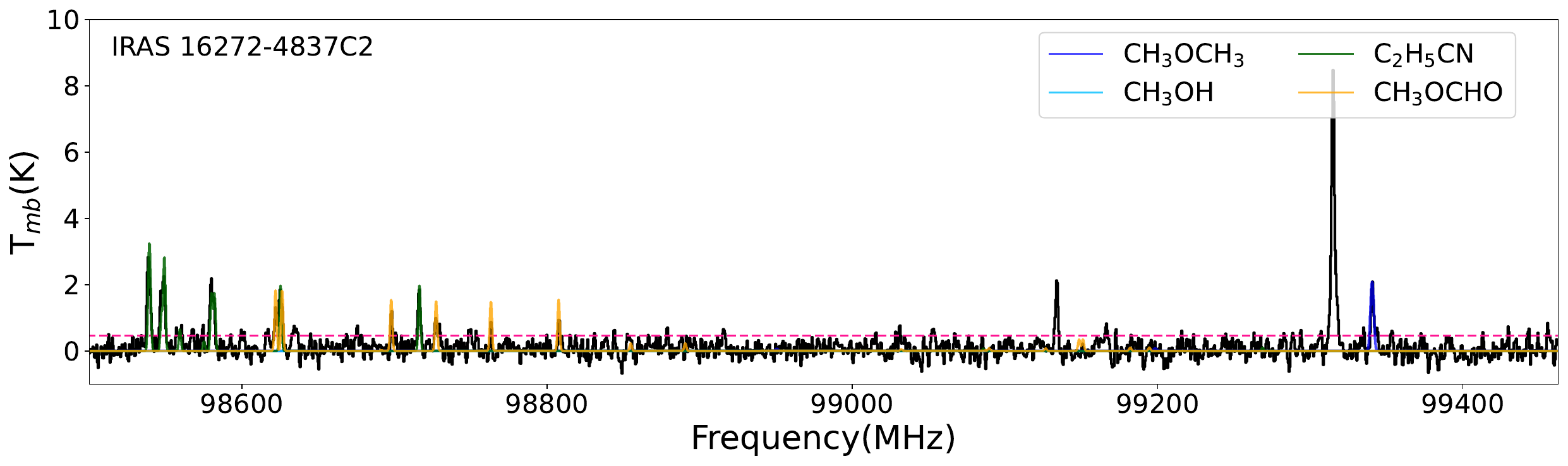}}
\quad
{\includegraphics[height=4.5cm,width=15.93cm]{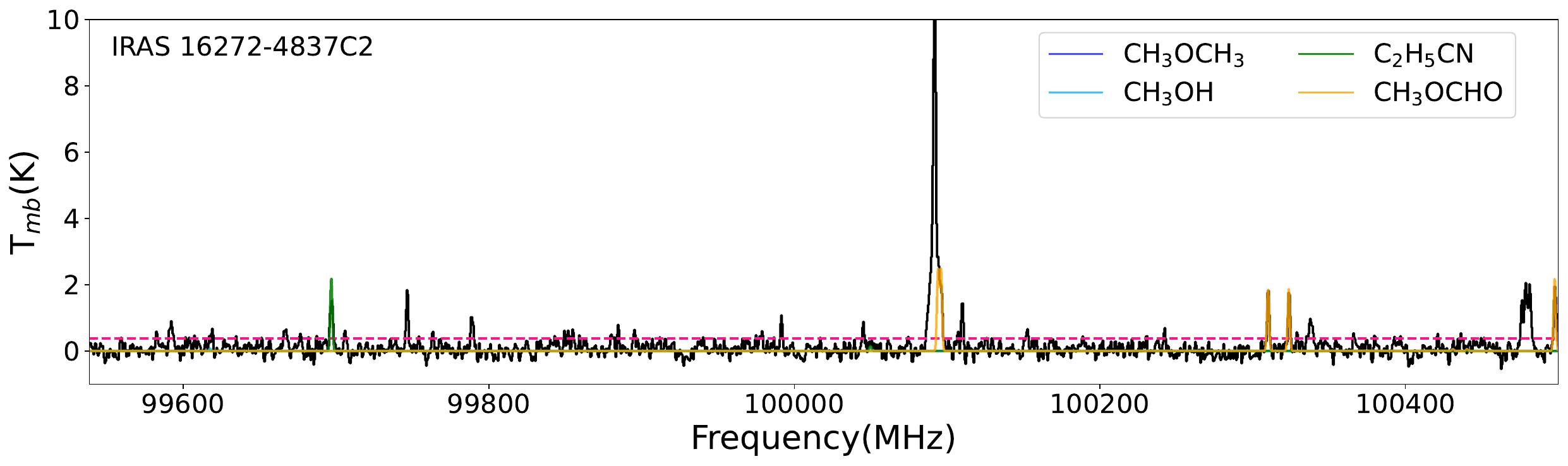}}
\quad
{\includegraphics[height=4.5cm,width=15.93cm]{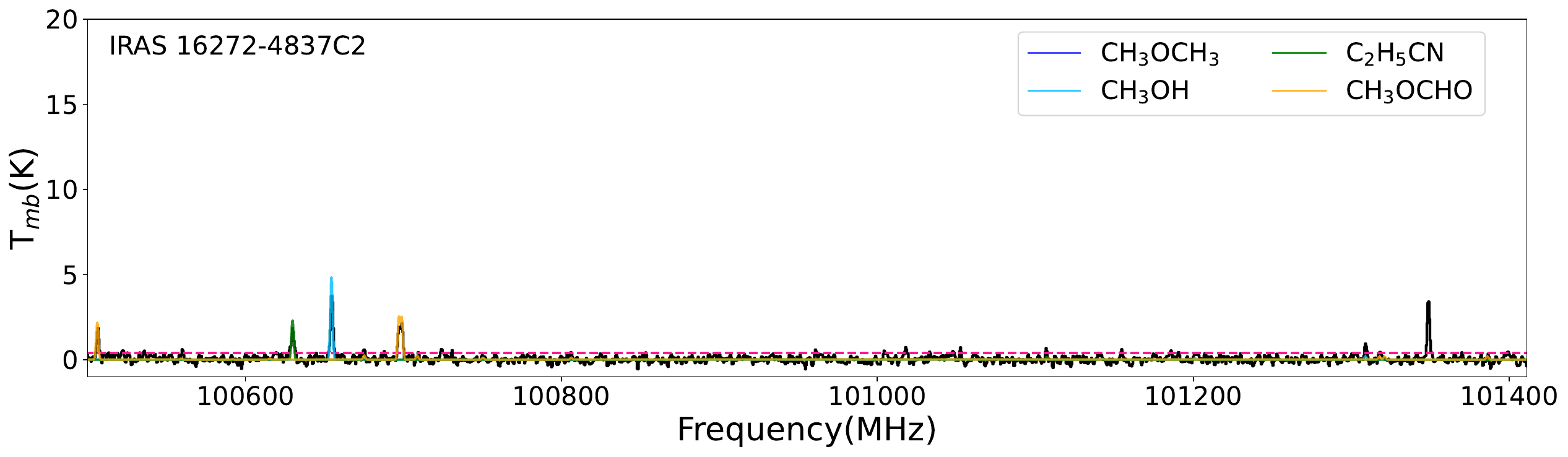}}
\quad
{\includegraphics[height=4.5cm,width=15.93cm]{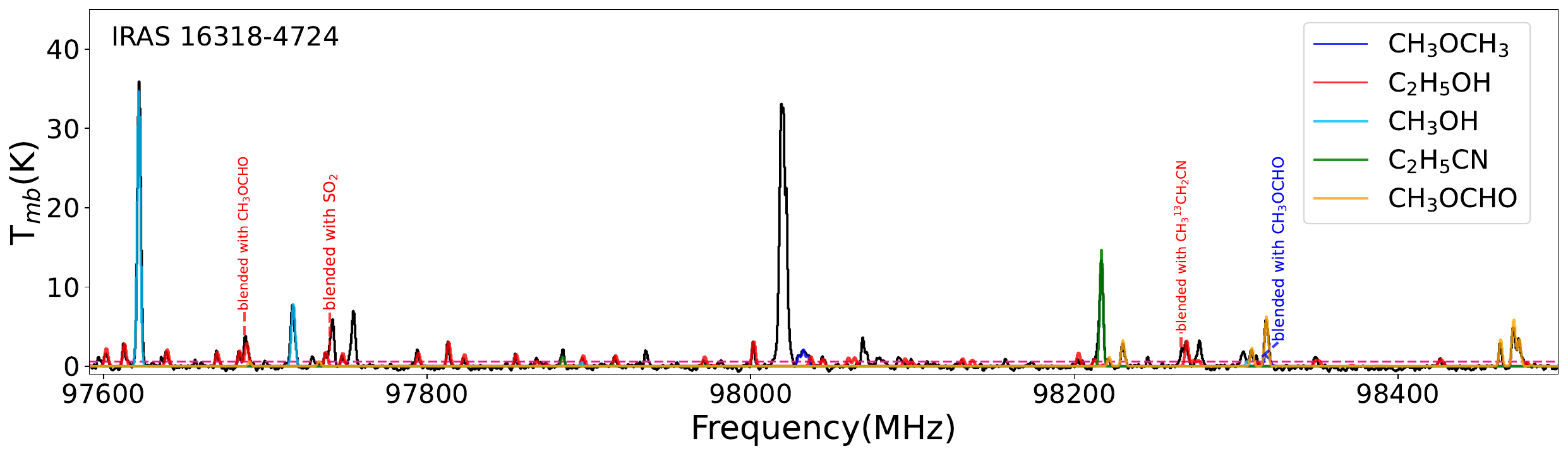}}
\quad
{\includegraphics[height=4.5cm,width=15.93cm]{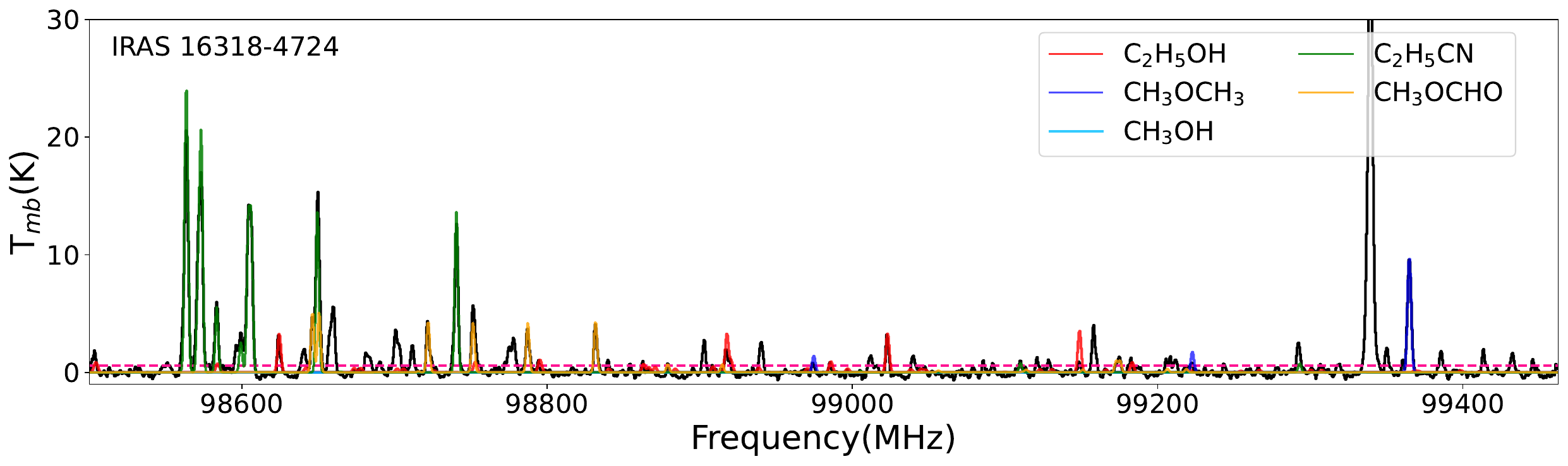}}
\caption{Continued.}
\end{figure}
\setcounter{figure}{\value{figure}-1}
\begin{figure}
  \centering 
{\includegraphics[height=4.5cm,width=15.93cm]{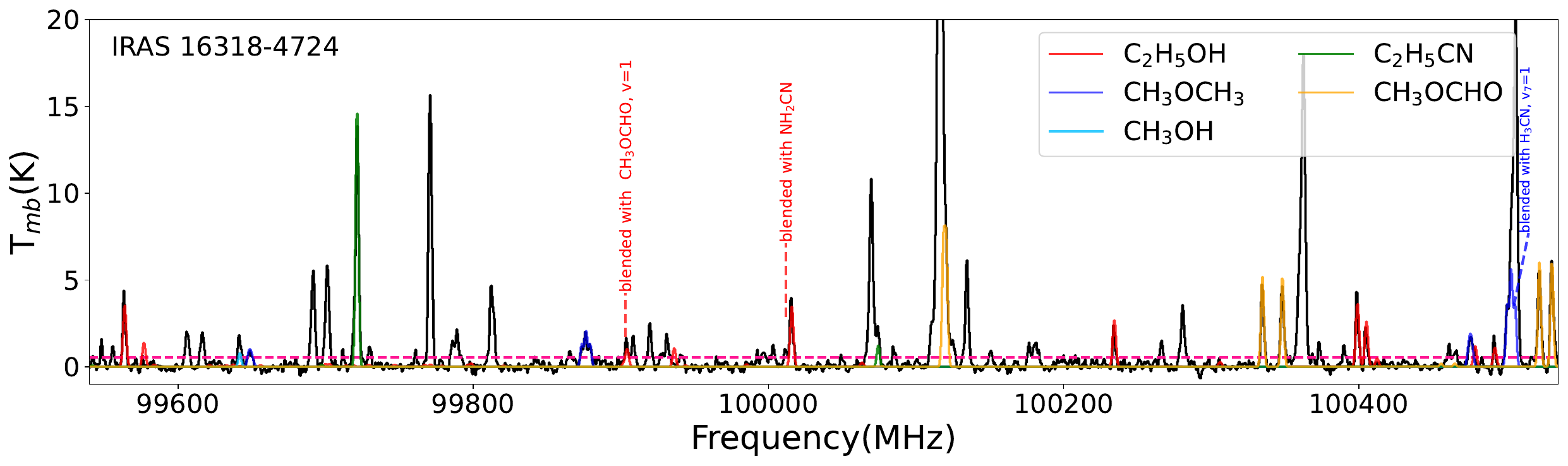}}
\quad
{\includegraphics[height=4.5cm,width=15.93cm]{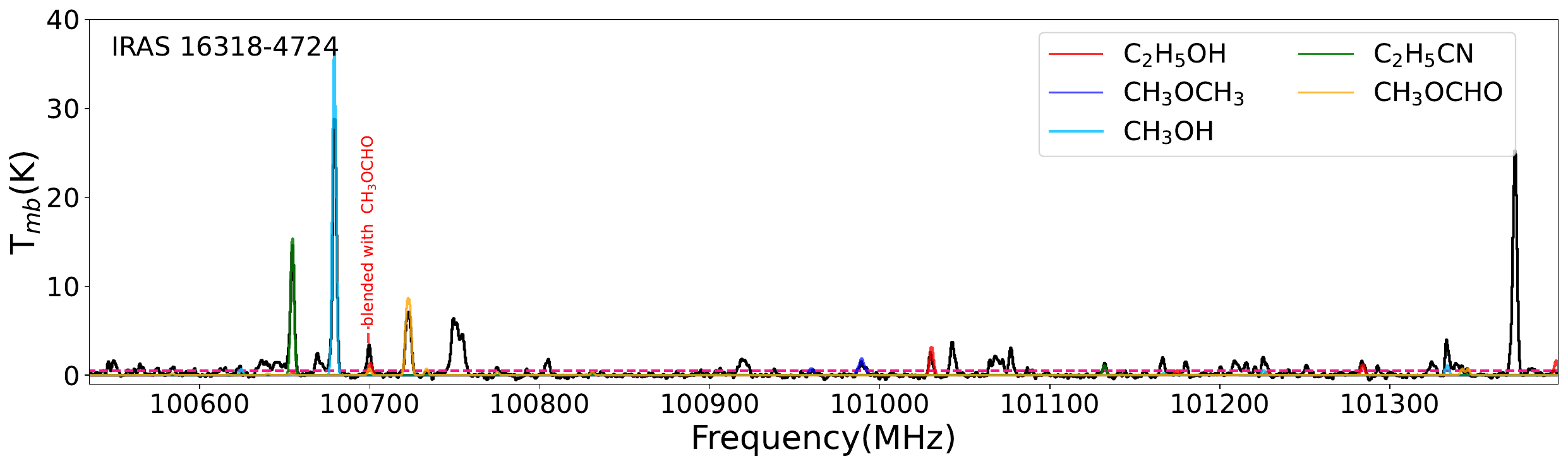}}
\quad
{\includegraphics[height=4.5cm,width=15.93cm]{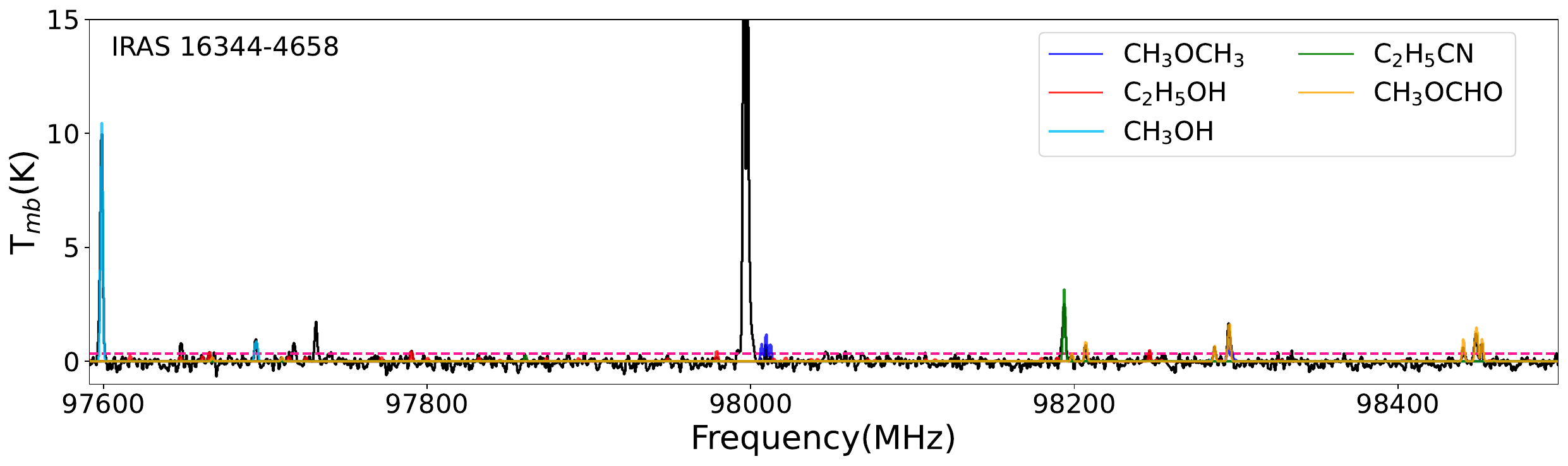}}
\quad
{\includegraphics[height=4.5cm,width=15.93cm]{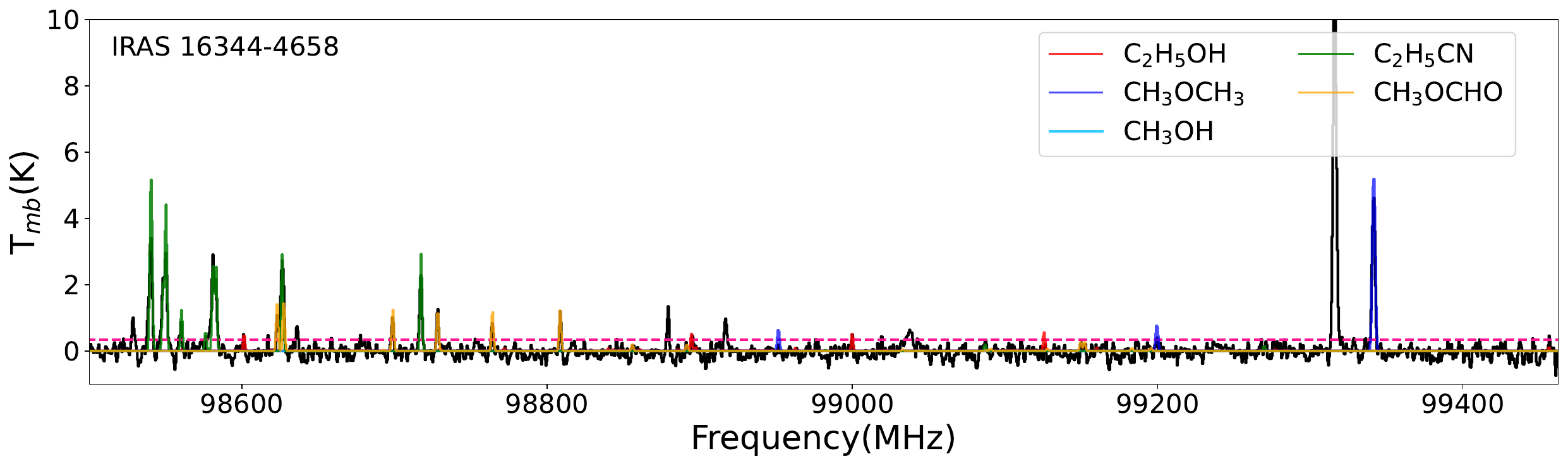}}
\quad
{\includegraphics[height=4.5cm,width=15.93cm]{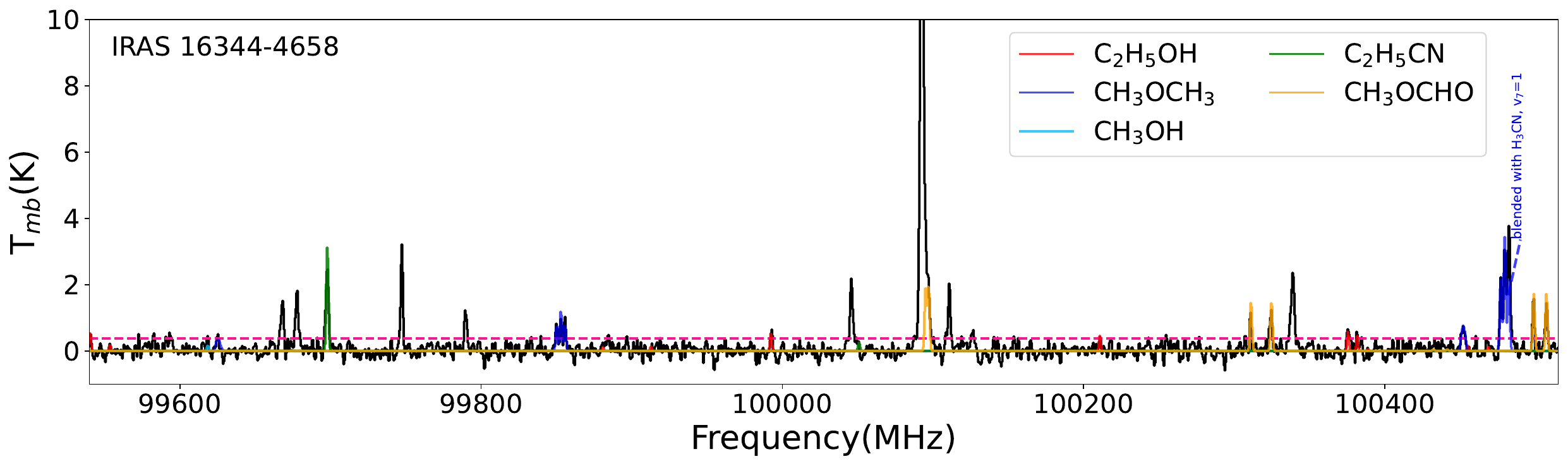}}
\caption{Continued.}
\end{figure}
\setcounter{figure}{\value{figure}-1}
\begin{figure}
  \centering 
{\includegraphics[height=4.5cm,width=15.93cm]{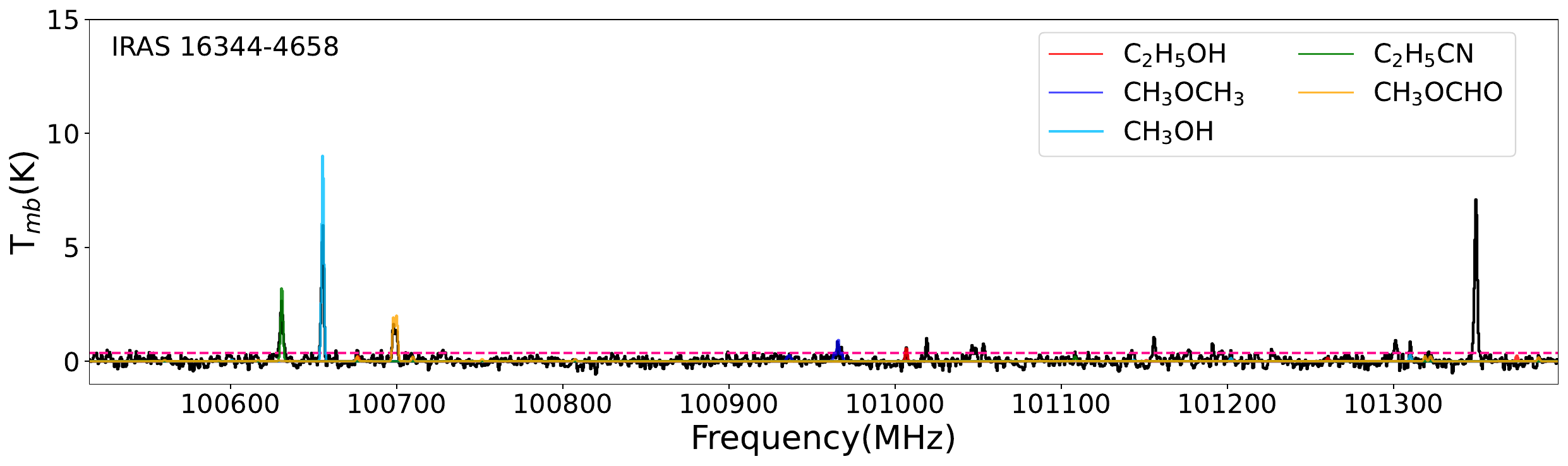}}
\quad
{\includegraphics[height=4.5cm,width=15.93cm]{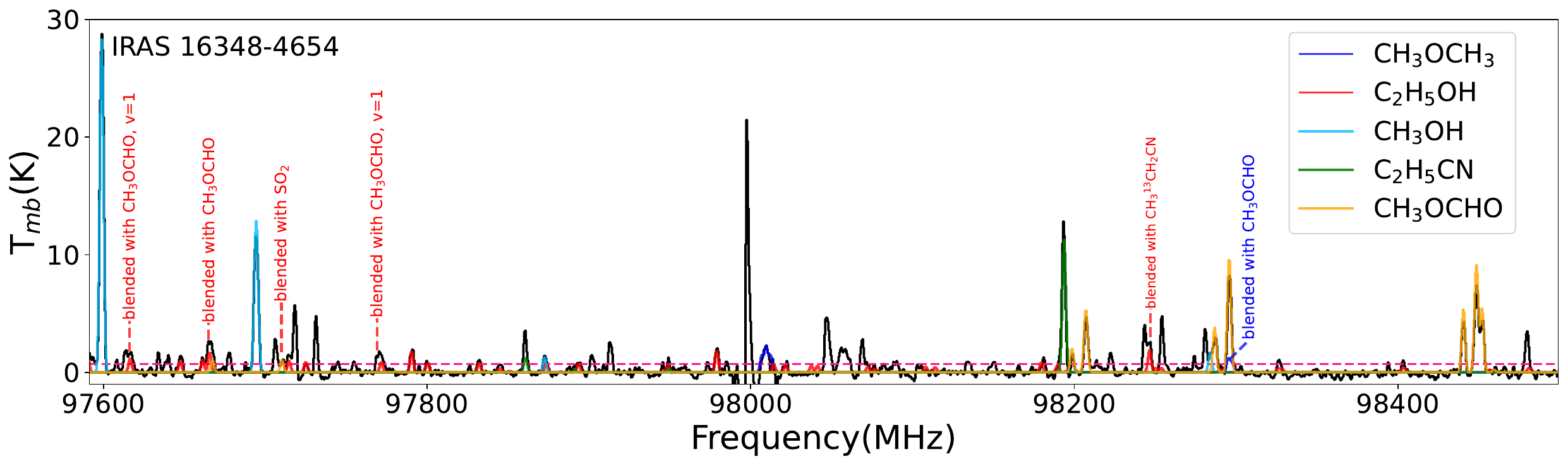}}
\quad
{\includegraphics[height=4.5cm,width=15.93cm]{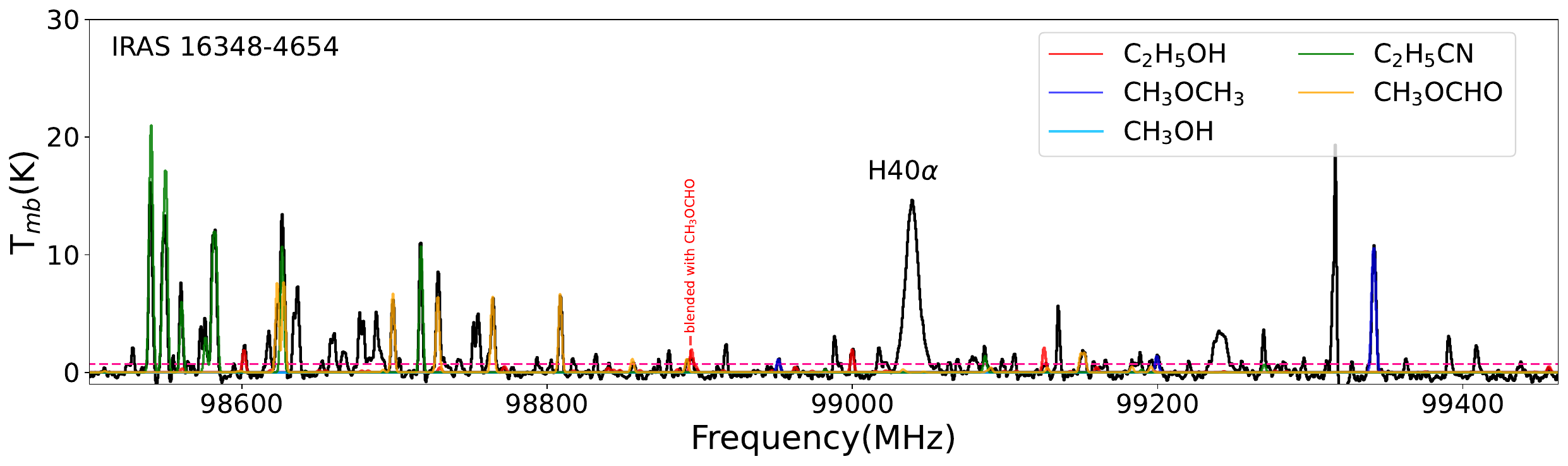}}
\quad
{\includegraphics[height=4.5cm,width=15.93cm]{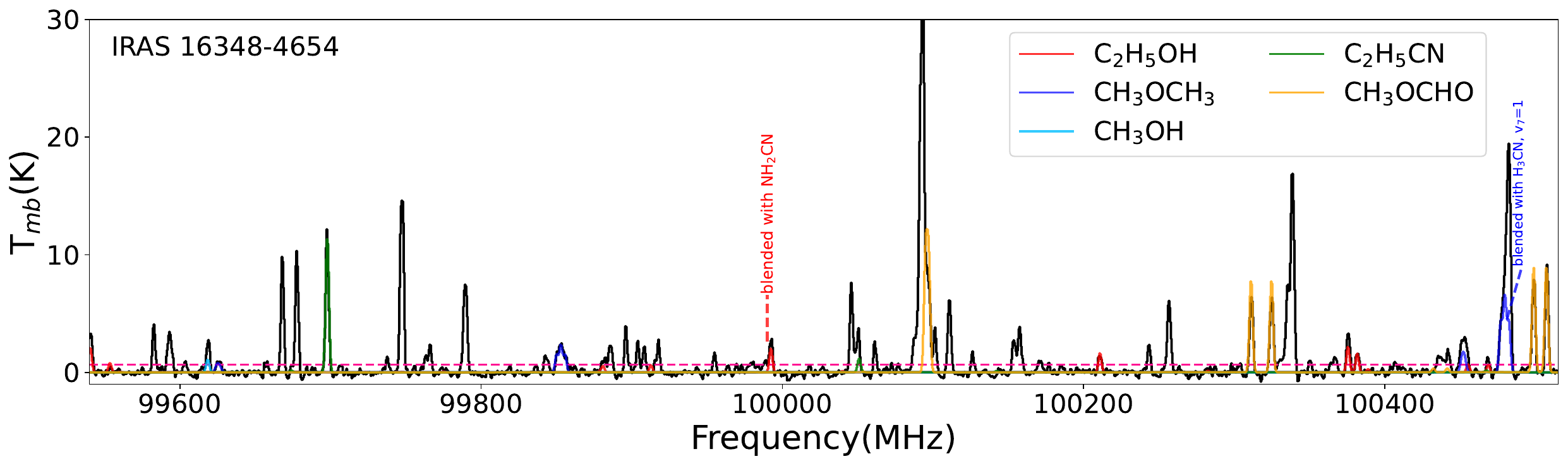}}
\quad
{\includegraphics[height=4.5cm,width=15.93cm]{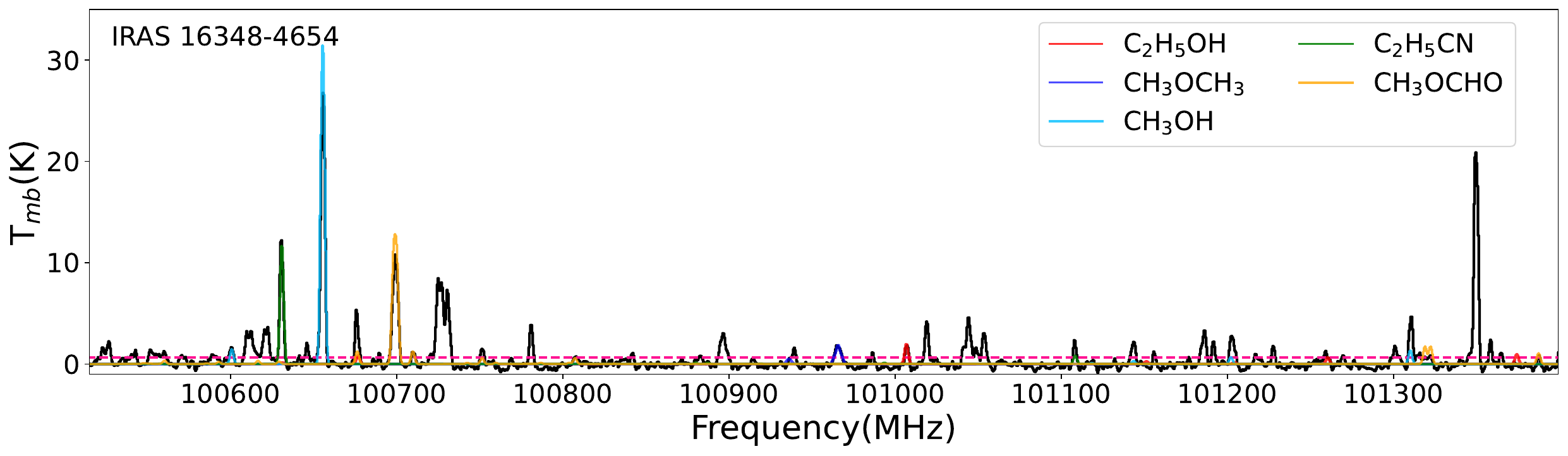}}
\caption{Continued.}
\end{figure}
\setcounter{figure}{\value{figure}-1}
\begin{figure}
  \centering 
{\includegraphics[height=4.5cm,width=15.93cm]{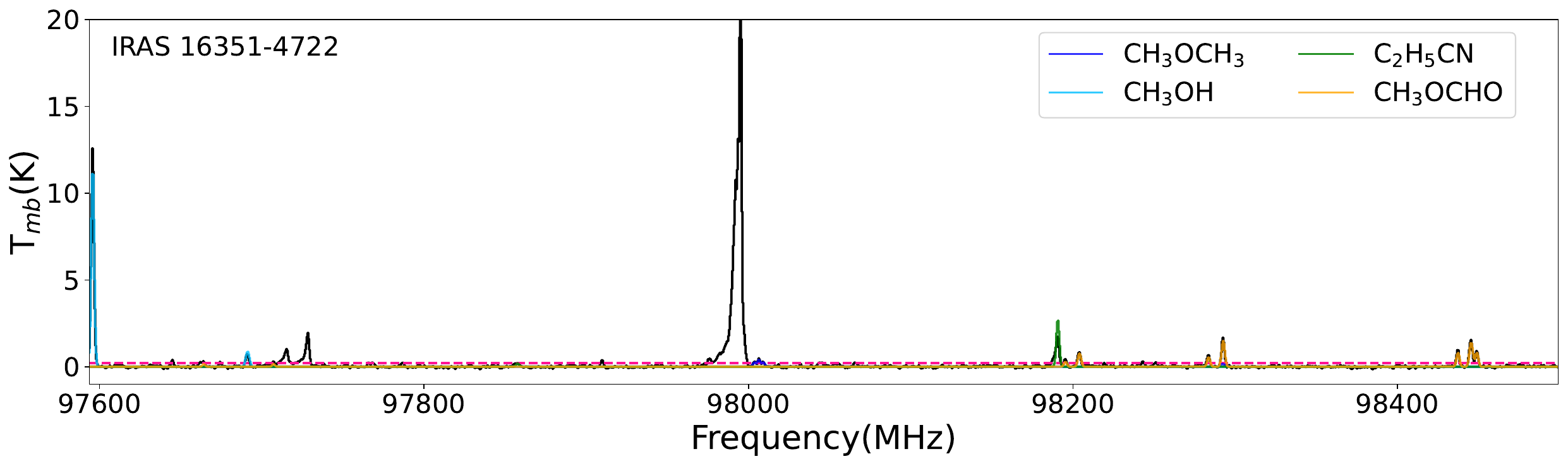}}
\quad
{\includegraphics[height=4.5cm,width=15.93cm]{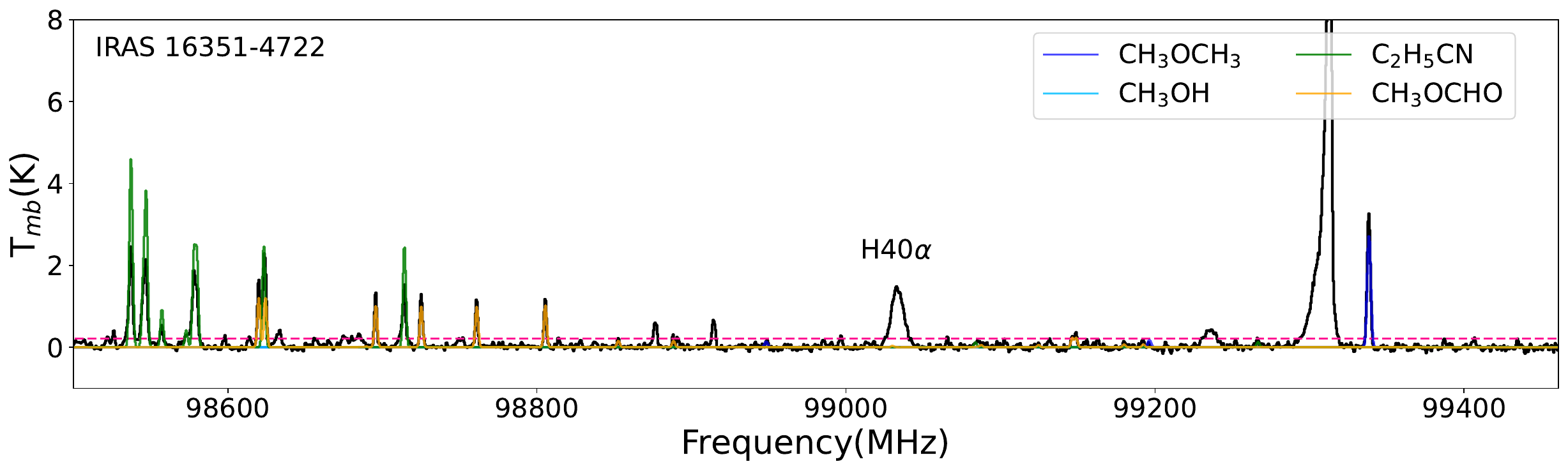}}
\quad
{\includegraphics[height=4.5cm,width=15.93cm]{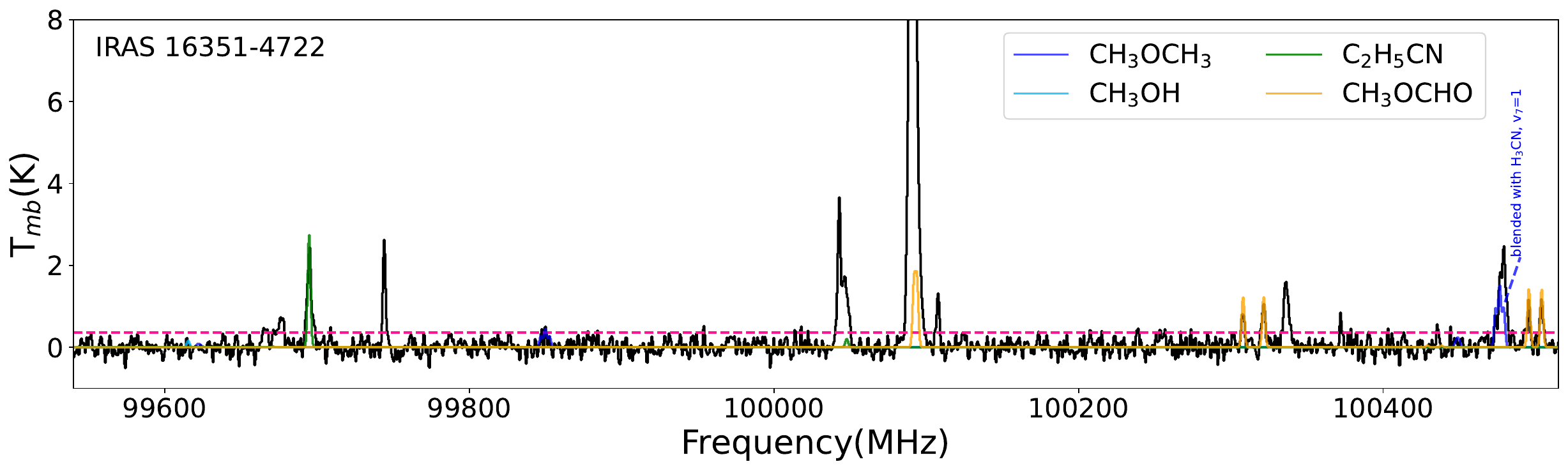}}
\quad
{\includegraphics[height=4.5cm,width=15.93cm]{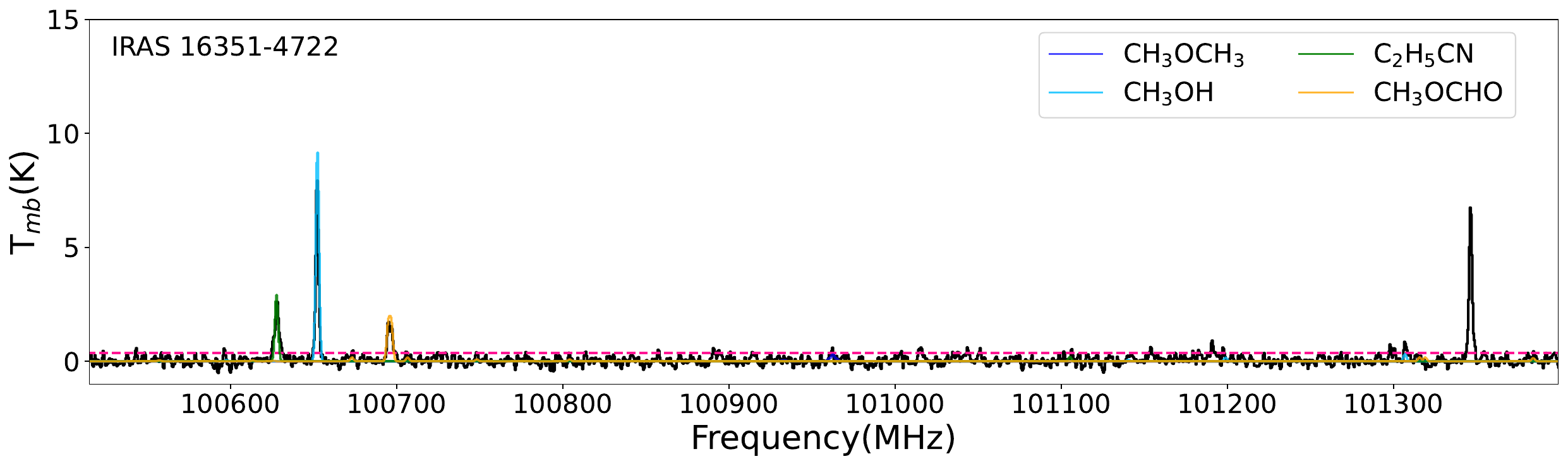}}
\quad
{\includegraphics[height=4.5cm,width=15.93cm]{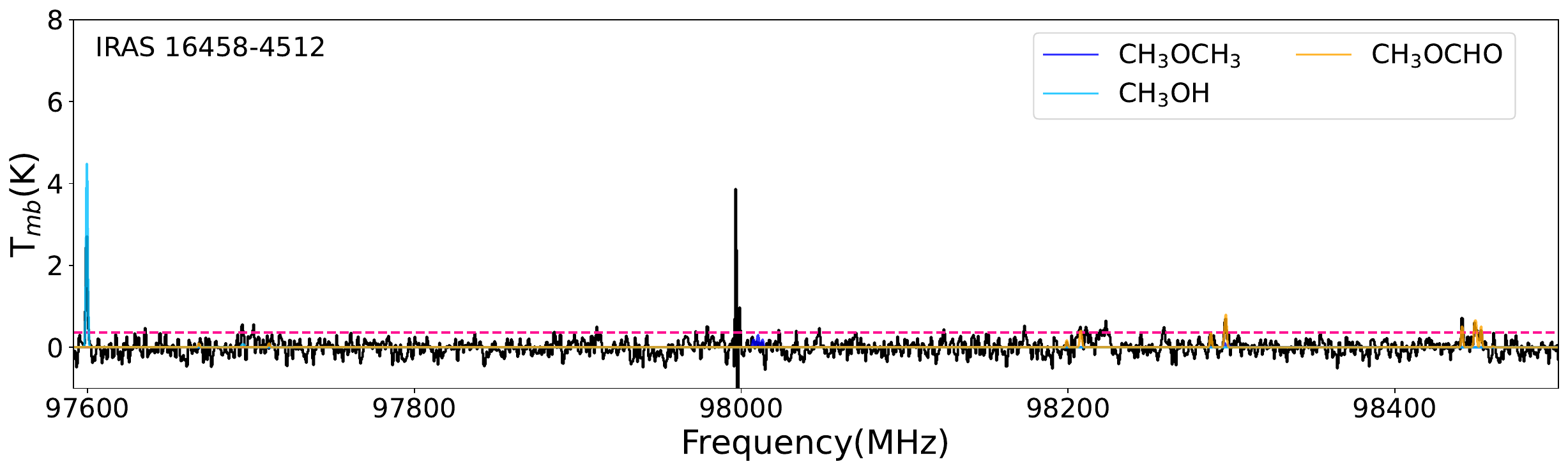}}
\caption{Continued.}
\end{figure}
\setcounter{figure}{\value{figure}-1}
\begin{figure}
  \centering 
{\includegraphics[height=4.5cm,width=15.93cm]{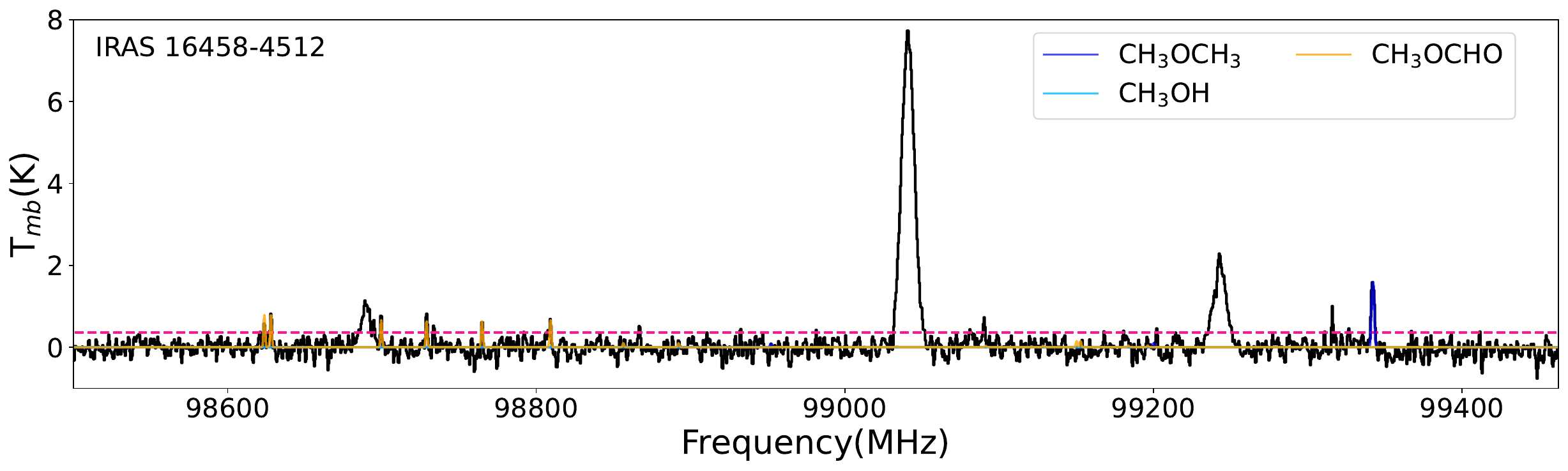}}
\quad
{\includegraphics[height=4.5cm,width=15.93cm]{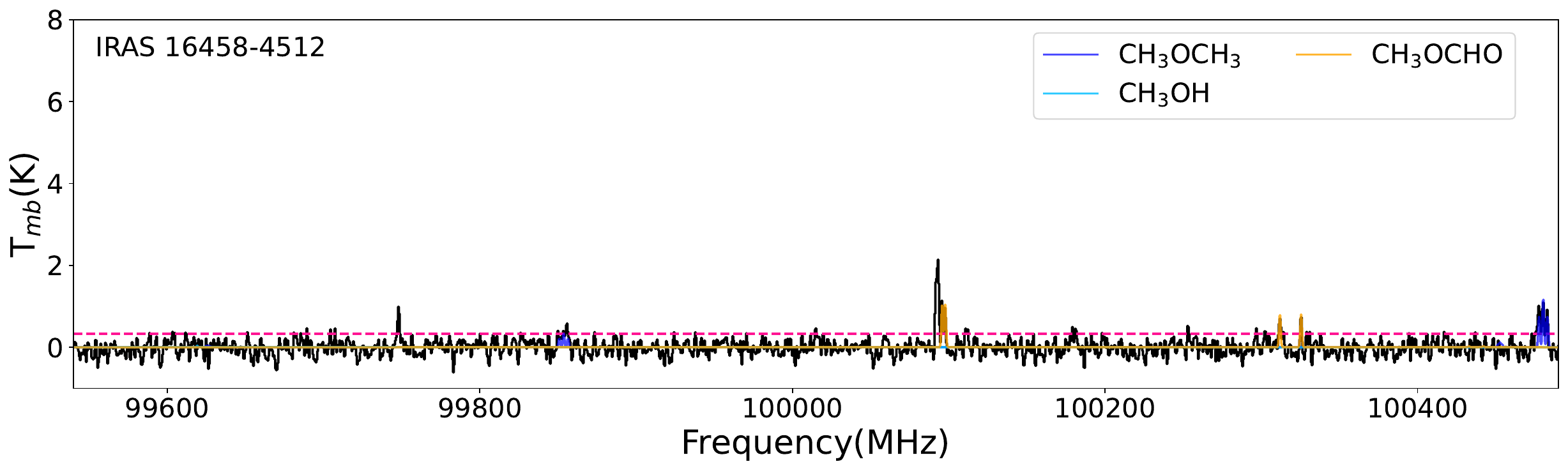}}
\quad
{\includegraphics[height=4.5cm,width=15.93cm]{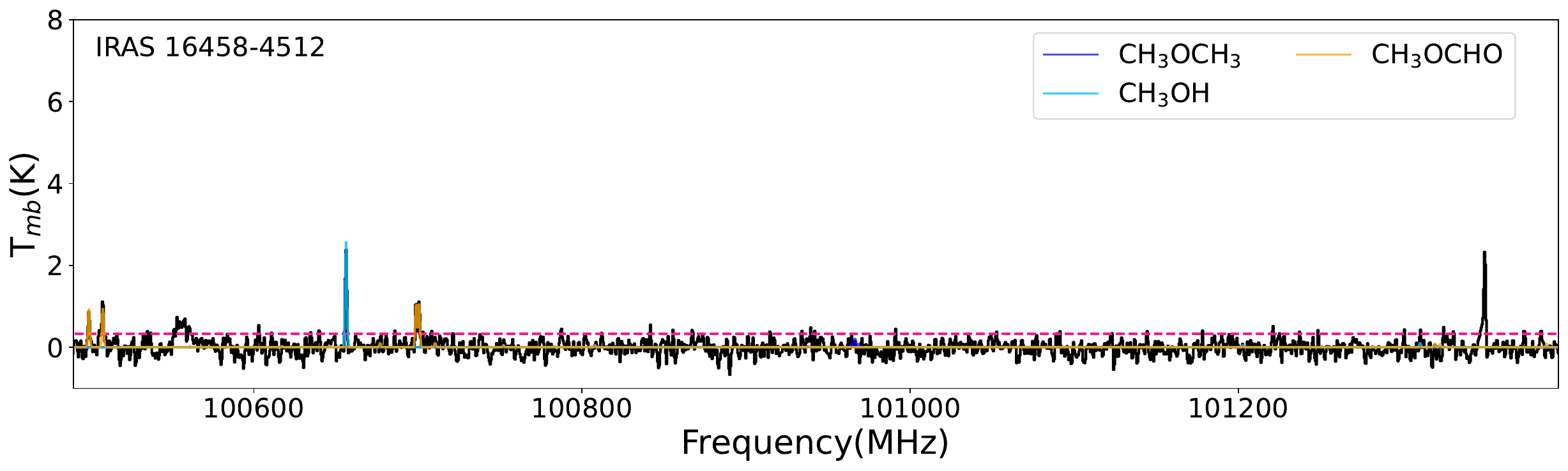}}
\quad
{\includegraphics[height=4.5cm,width=15.93cm]{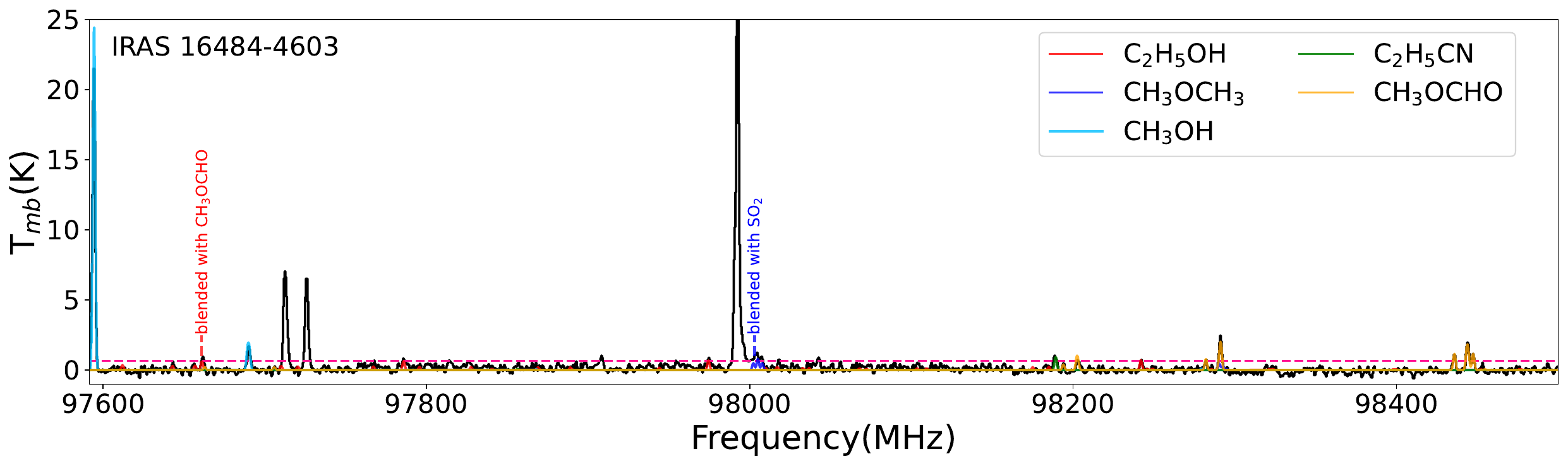}}
\quad
{\includegraphics[height=4.5cm,width=15.93cm]{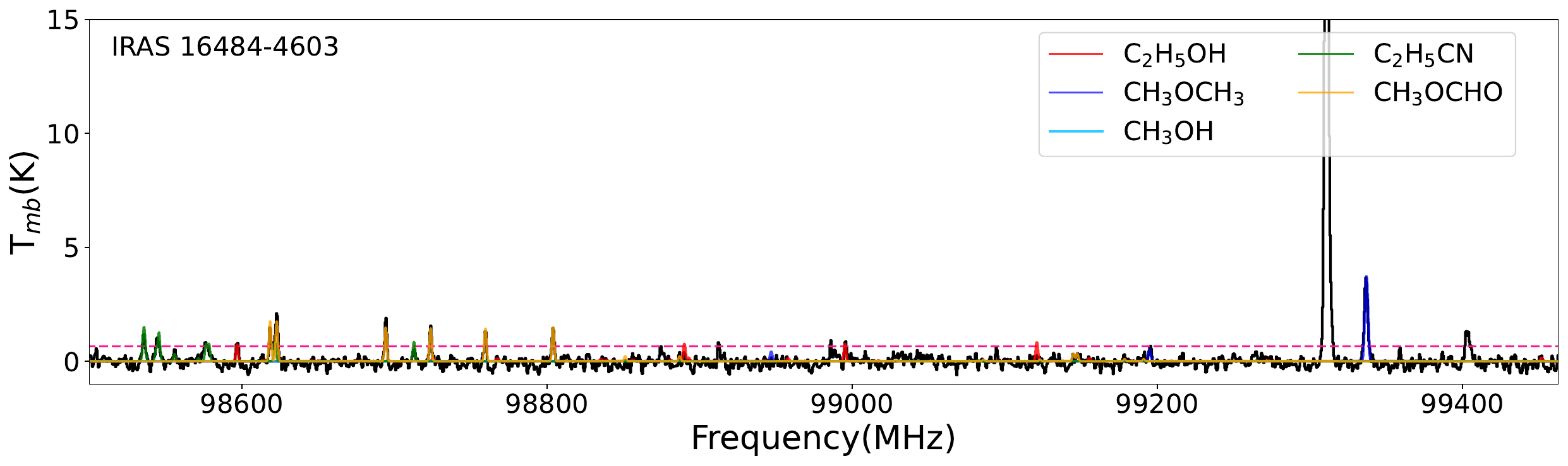}}
\caption{Continued.}
\end{figure}
\setcounter{figure}{\value{figure}-1}
\begin{figure}
\centering 
{\includegraphics[height=4.5cm,width=15.93cm]{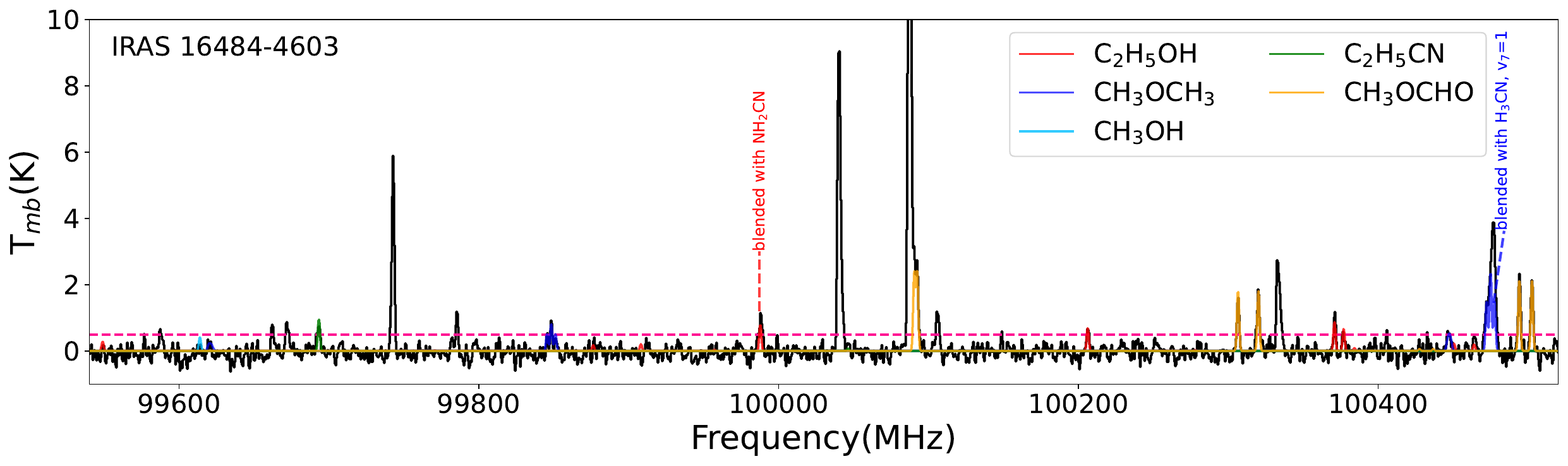}}
\quad
{\includegraphics[height=4.5cm,width=15.93cm]{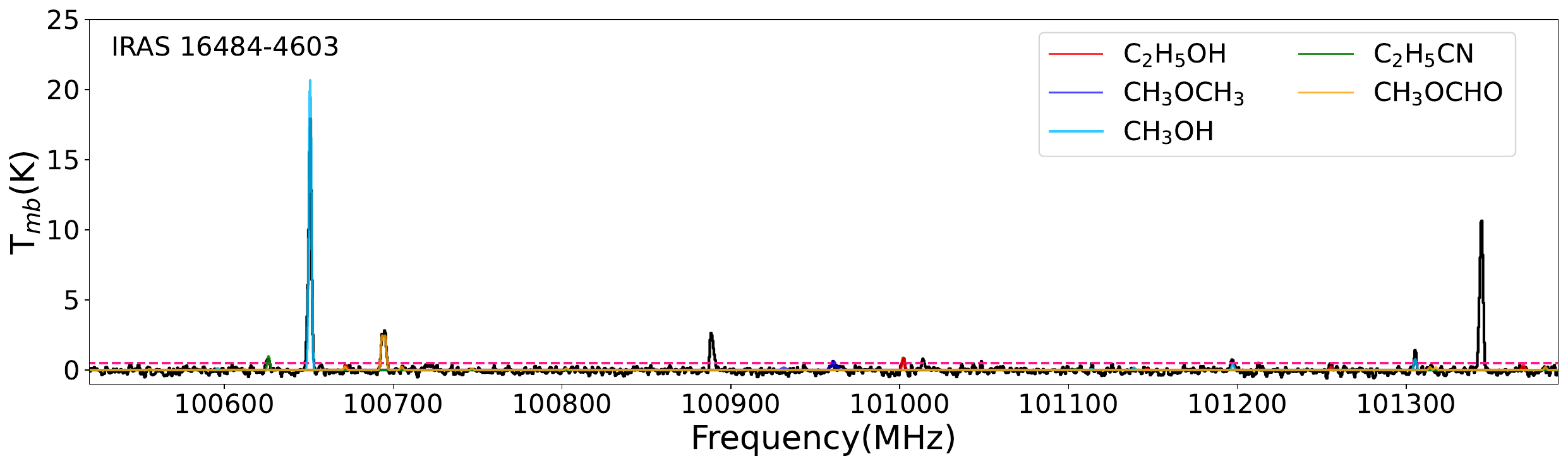}}
\quad
{\includegraphics[height=4.5cm,width=15.93cm]{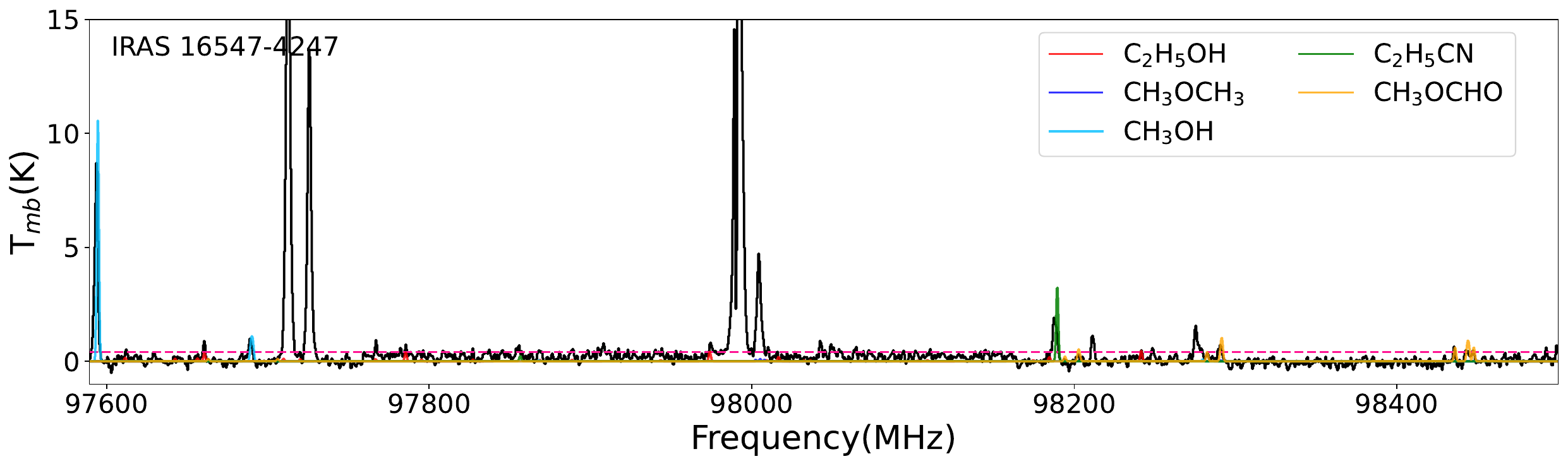}}
\quad
{\includegraphics[height=4.5cm,width=15.93cm]{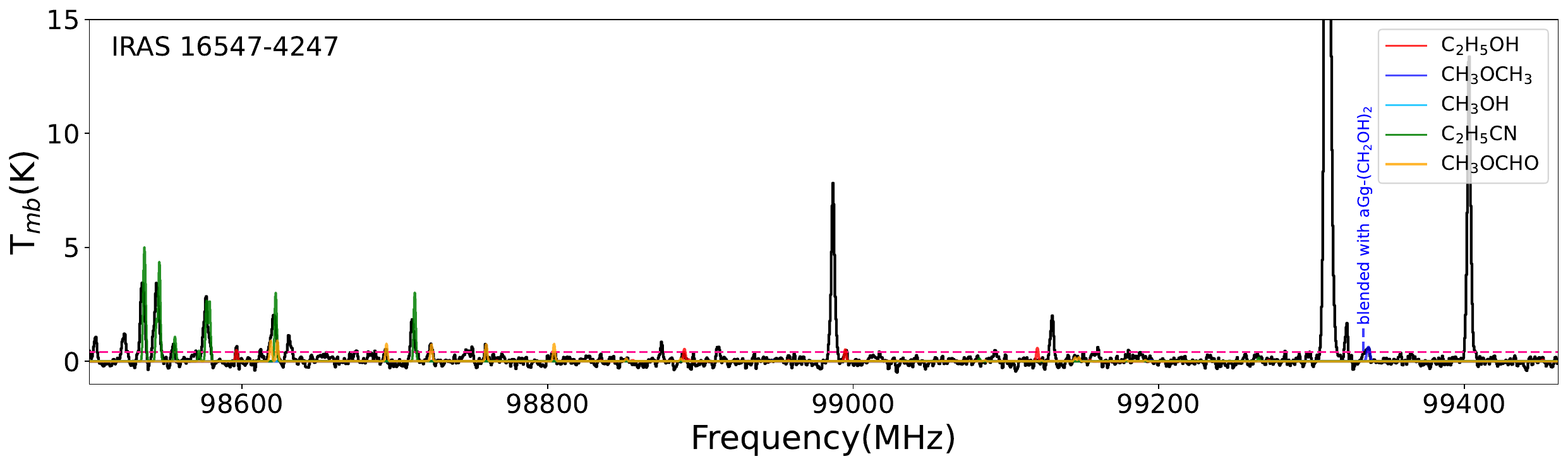}}
\quad
{\includegraphics[height=4.5cm,width=15.93cm]{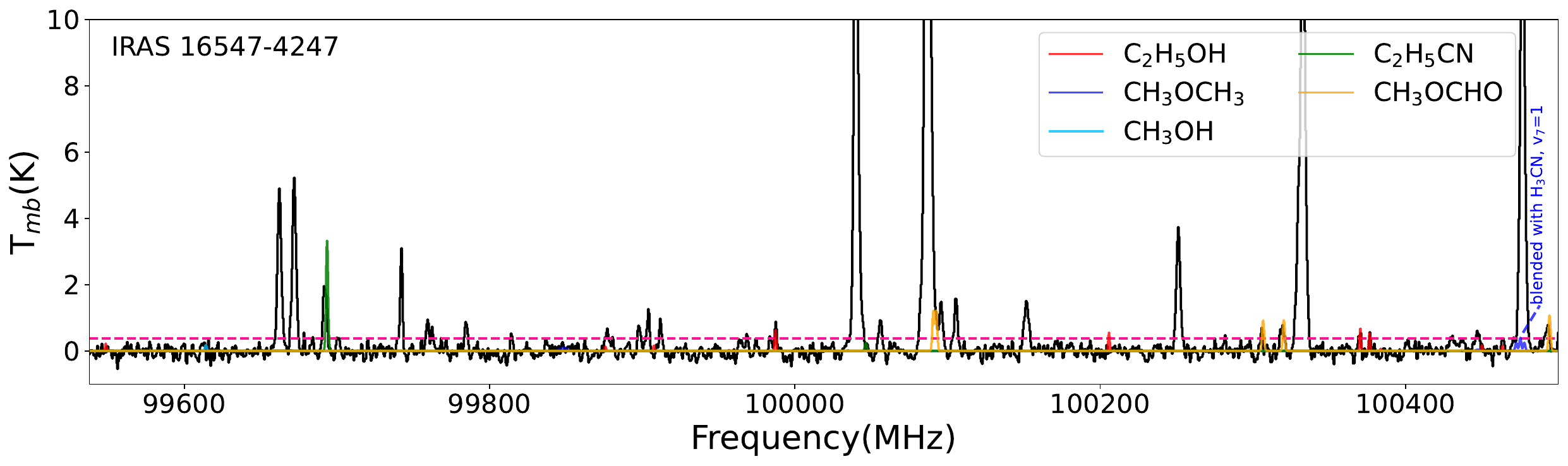}}
\caption{Continued.}
\end{figure}
\setcounter{figure}{\value{figure}-1}
\begin{figure}
\centering 
{\includegraphics[height=4.5cm,width=15.93cm]{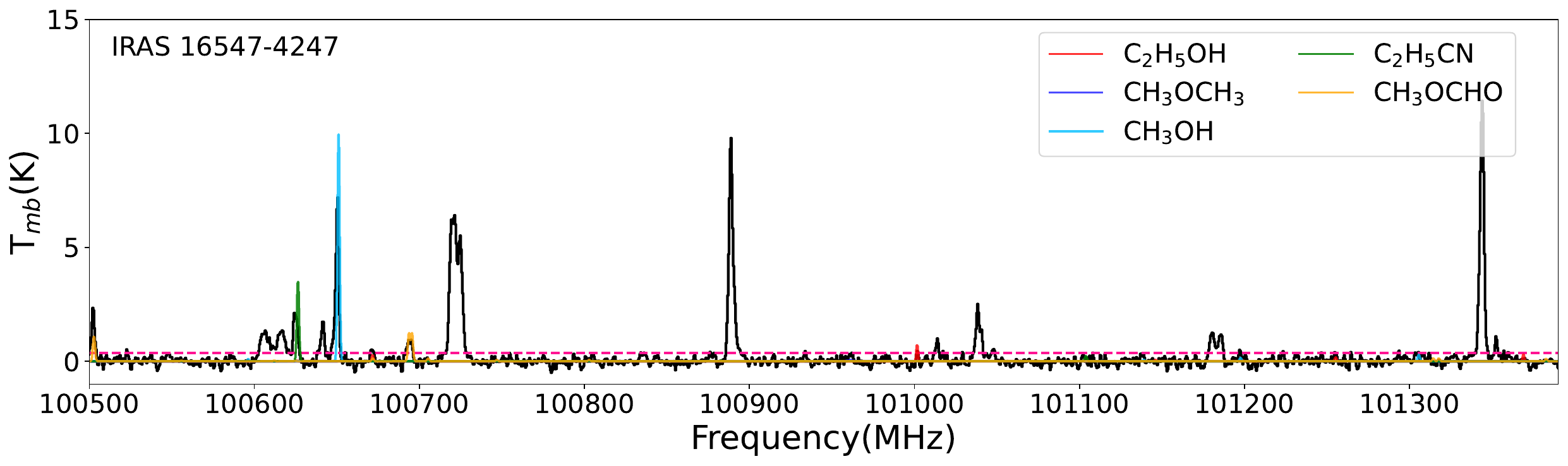}}
\quad
{\includegraphics[height=4.5cm,width=15.93cm]{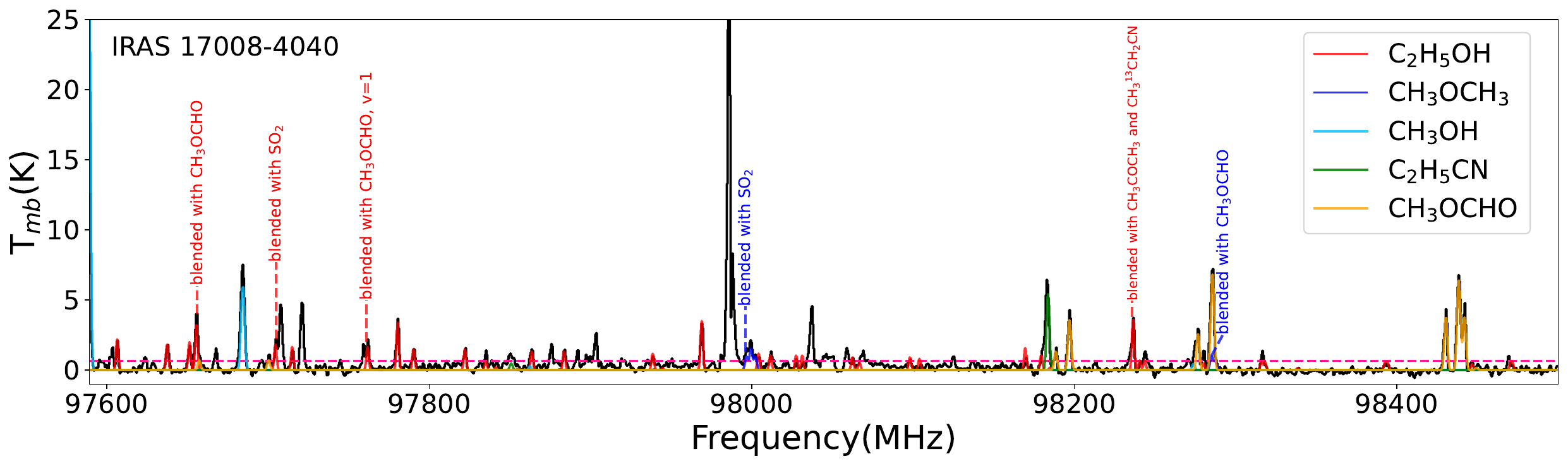}}
\quad
{\includegraphics[height=4.5cm,width=15.93cm]{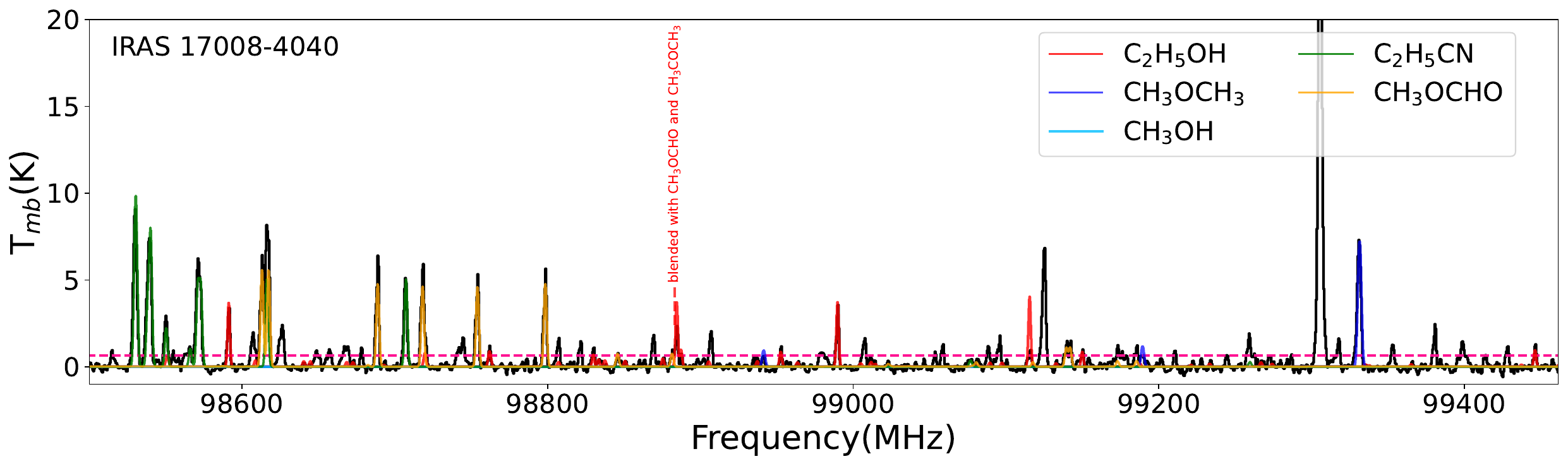}}
\quad
{\includegraphics[height=4.5cm,width=15.93cm]{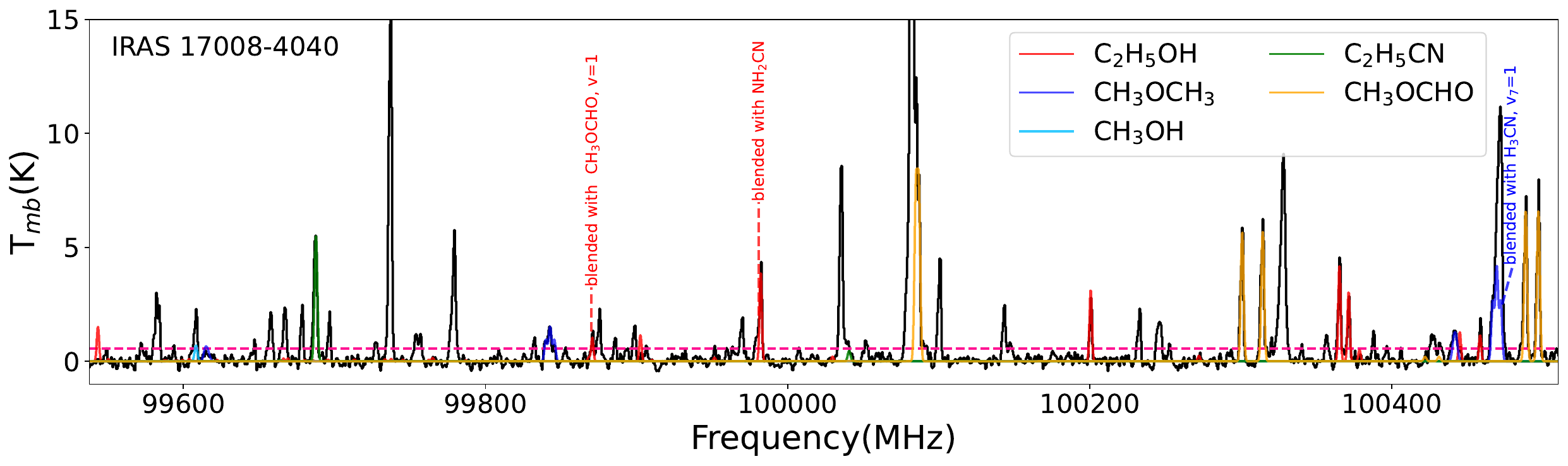}}
\quad
{\includegraphics[height=4.5cm,width=15.93cm]{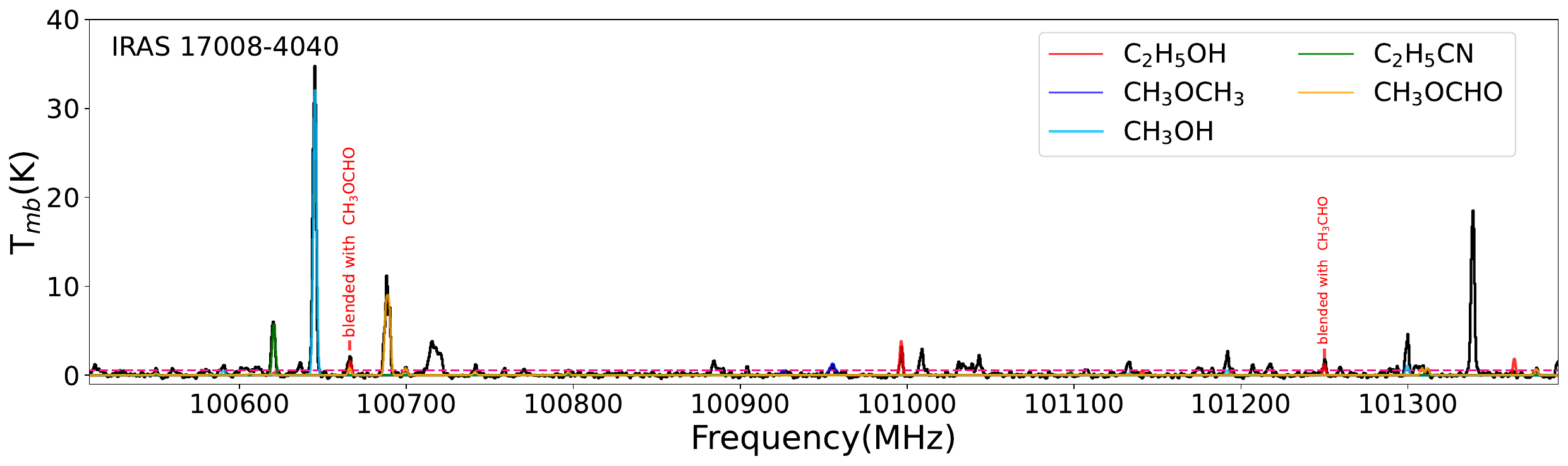}}
\caption{Continued.}
\end{figure}
\setcounter{figure}{\value{figure}-1}
\begin{figure}
\centering 
{\includegraphics[height=4.5cm,width=15.93cm]{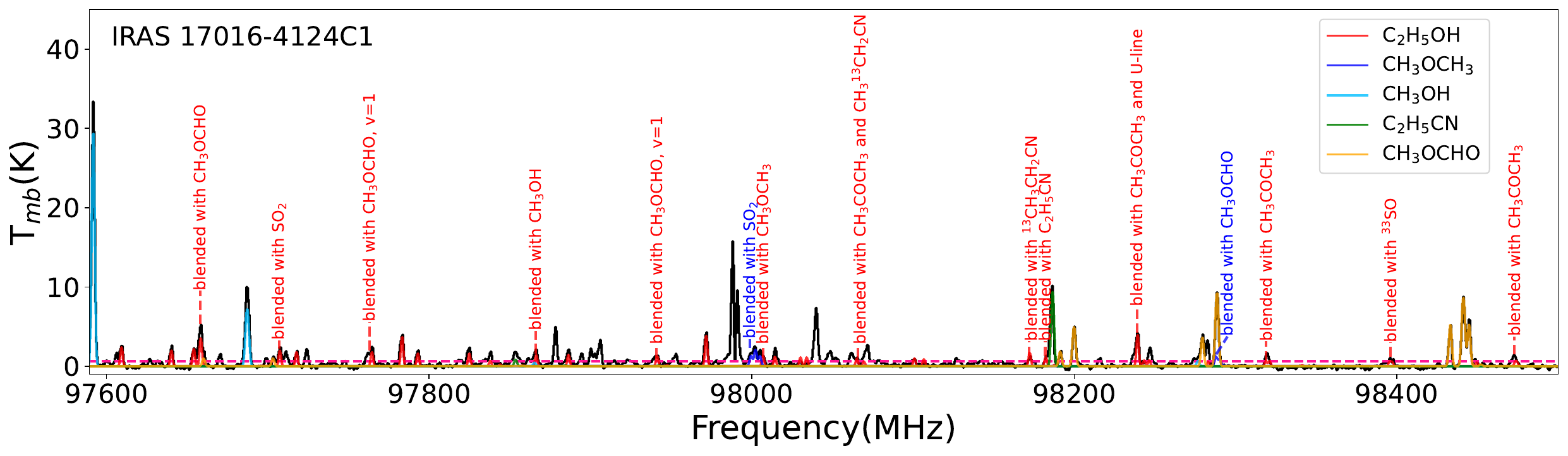}}
\quad
{\includegraphics[height=4.5cm,width=15.93cm]{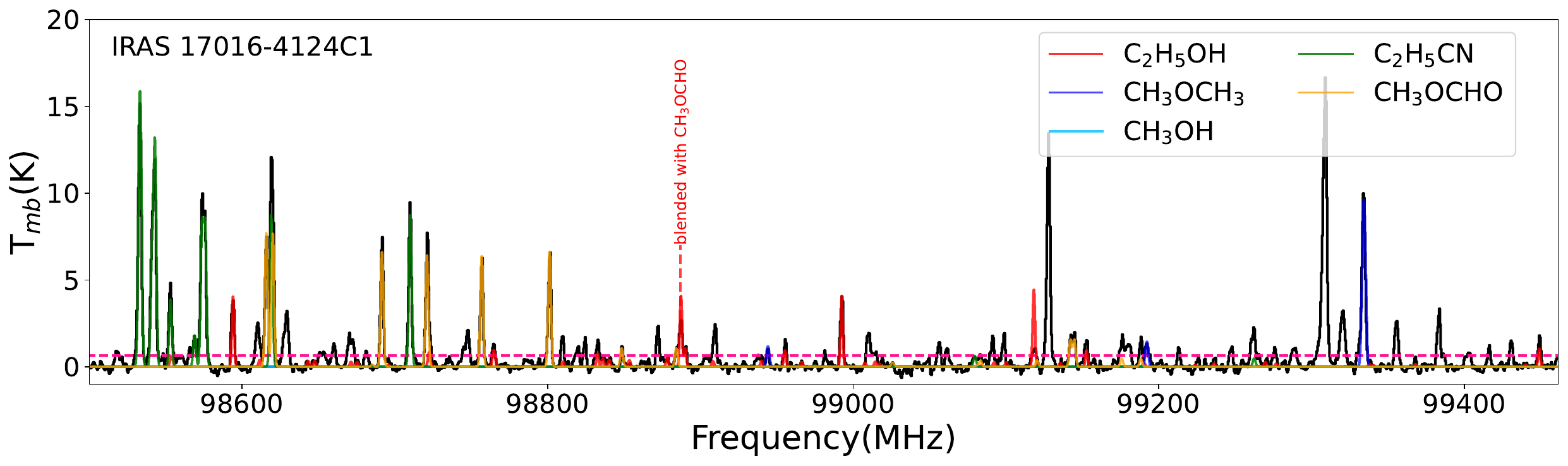}}
\quad
{\includegraphics[height=4.5cm,width=15.93cm]{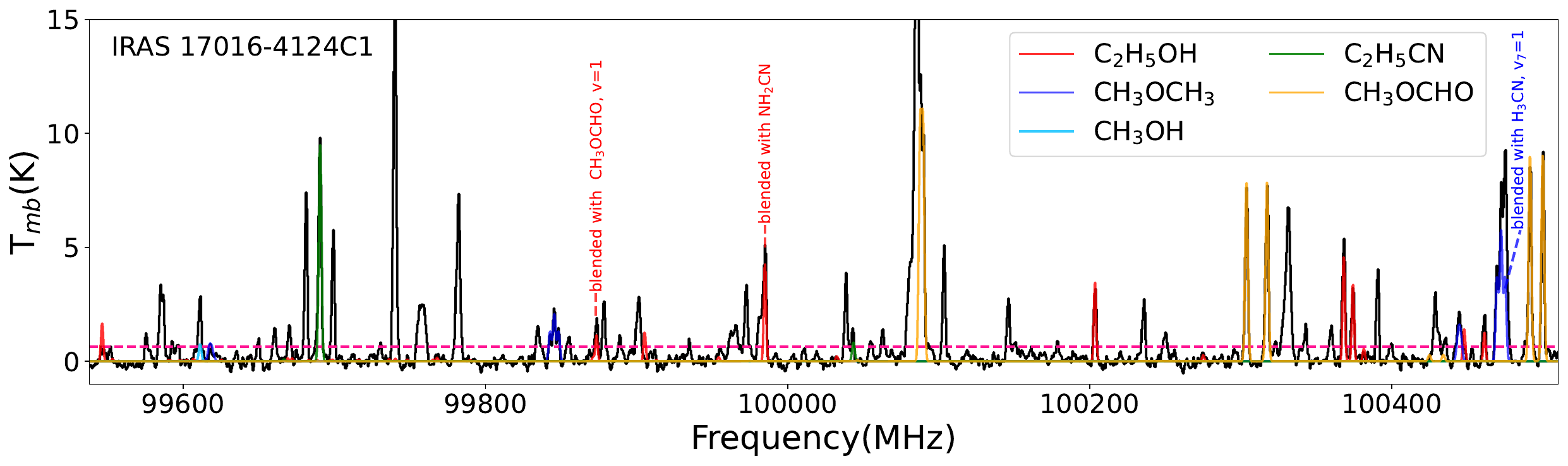}}
\quad
{\includegraphics[height=4.5cm,width=15.93cm]{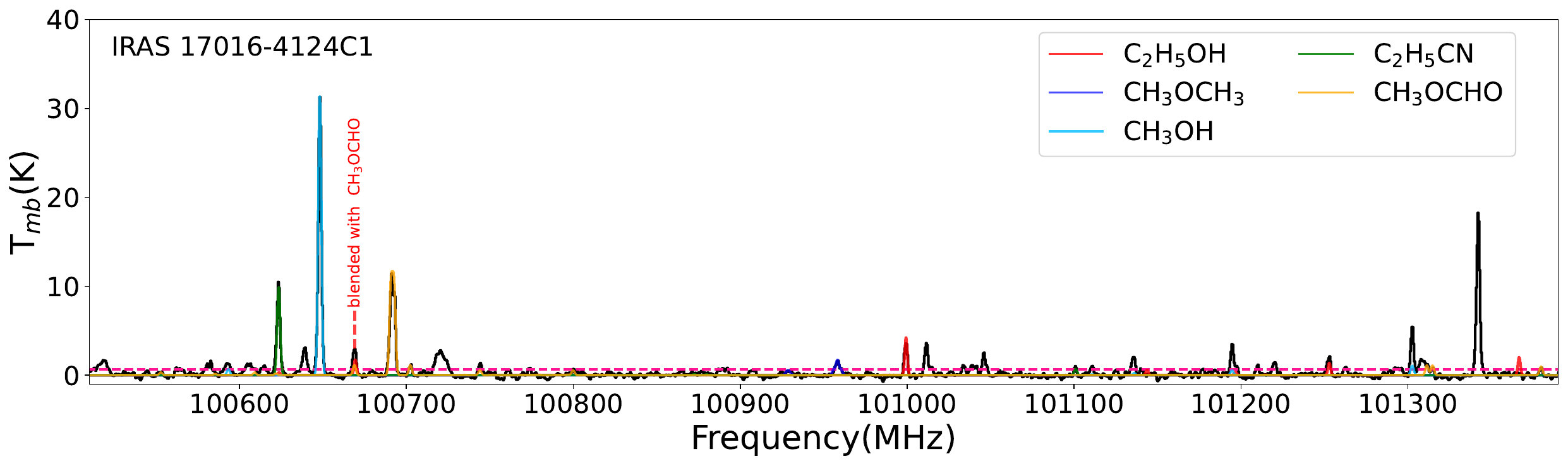}}
\quad
{\includegraphics[height=4.5cm,width=15.93cm]{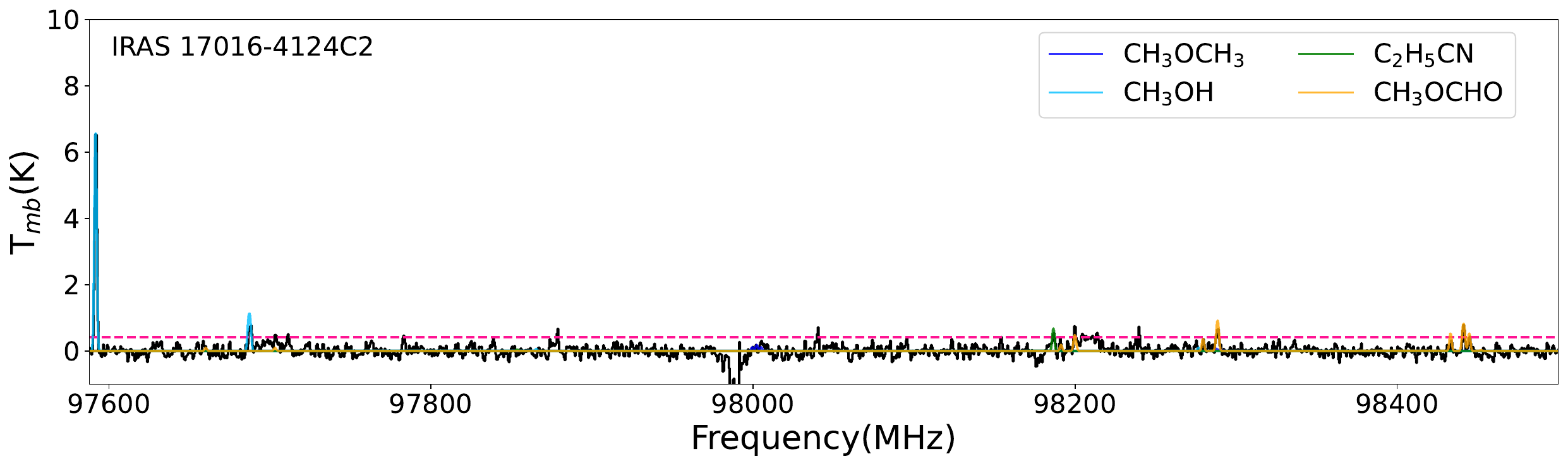}}
\caption{Continued.}
\end{figure}
\setcounter{figure}{\value{figure}-1}
\begin{figure}
\centering 
{\includegraphics[height=4.5cm,width=15.93cm]{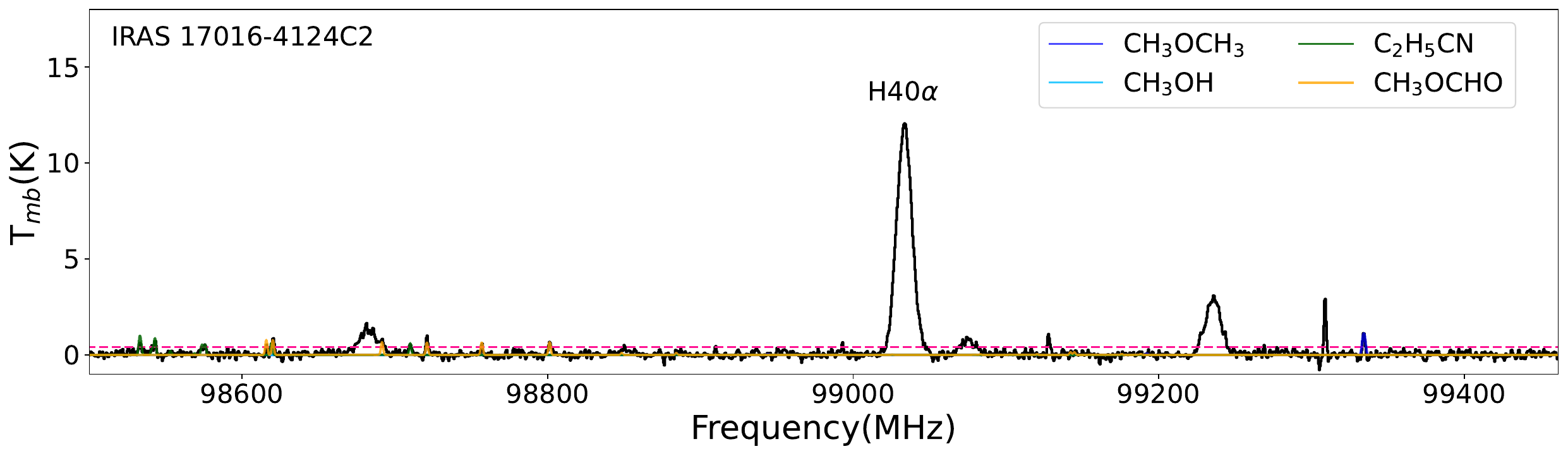}}
\quad
{\includegraphics[height=4.5cm,width=15.93cm]{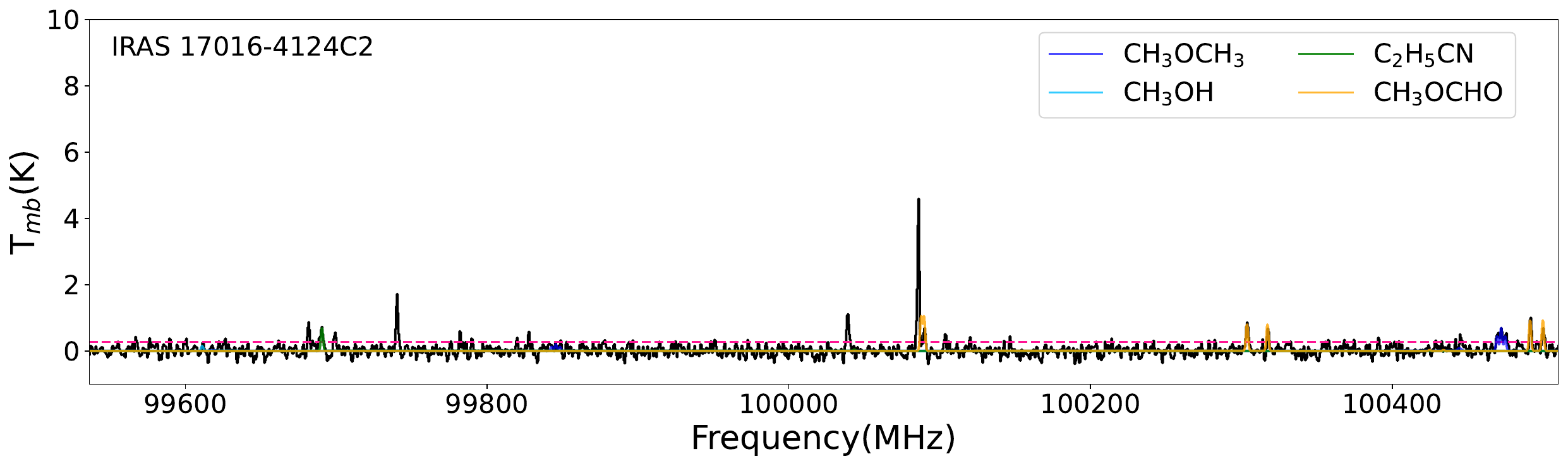}}
\quad
{\includegraphics[height=4.5cm,width=15.93cm]{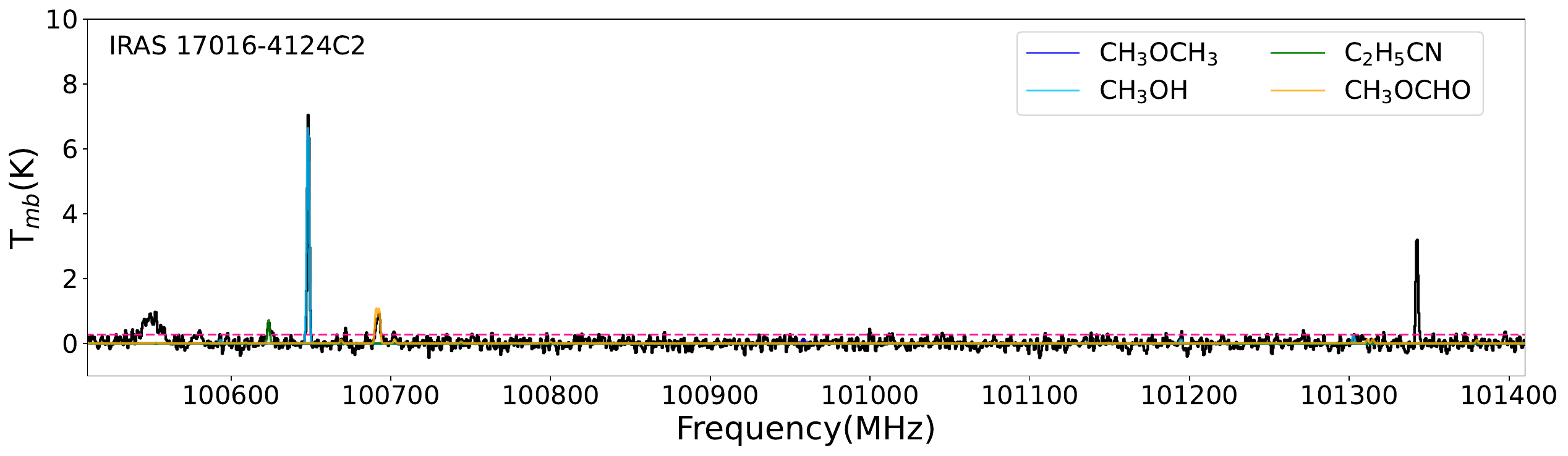}}
\quad
{\includegraphics[height=4.5cm,width=15.93cm]{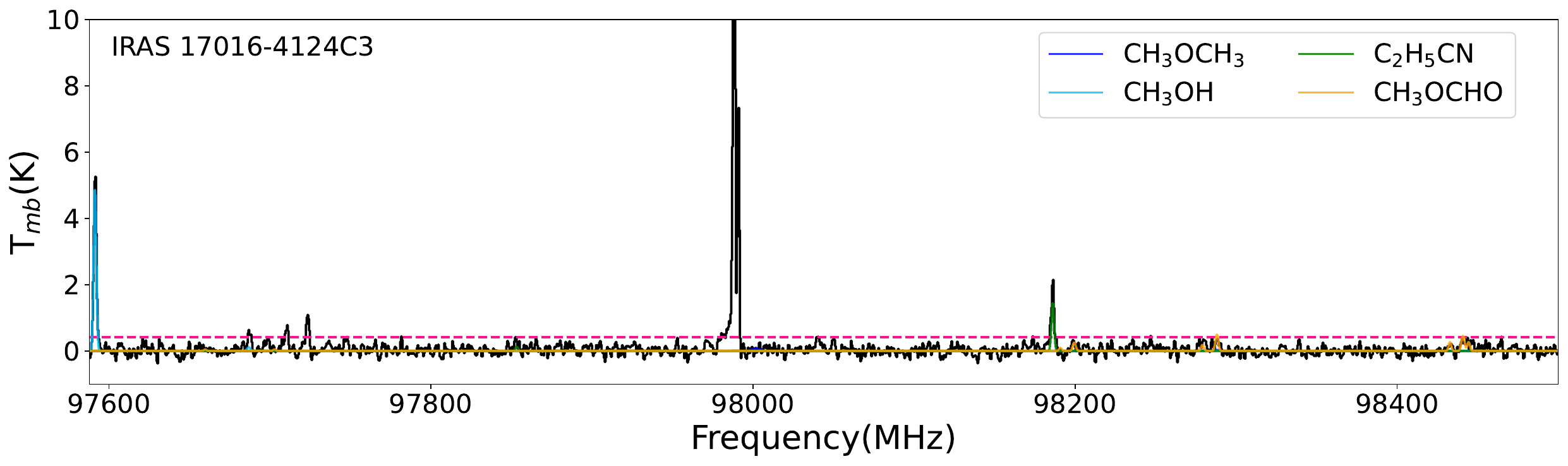}}
\quad
{\includegraphics[height=4.5cm,width=15.93cm]{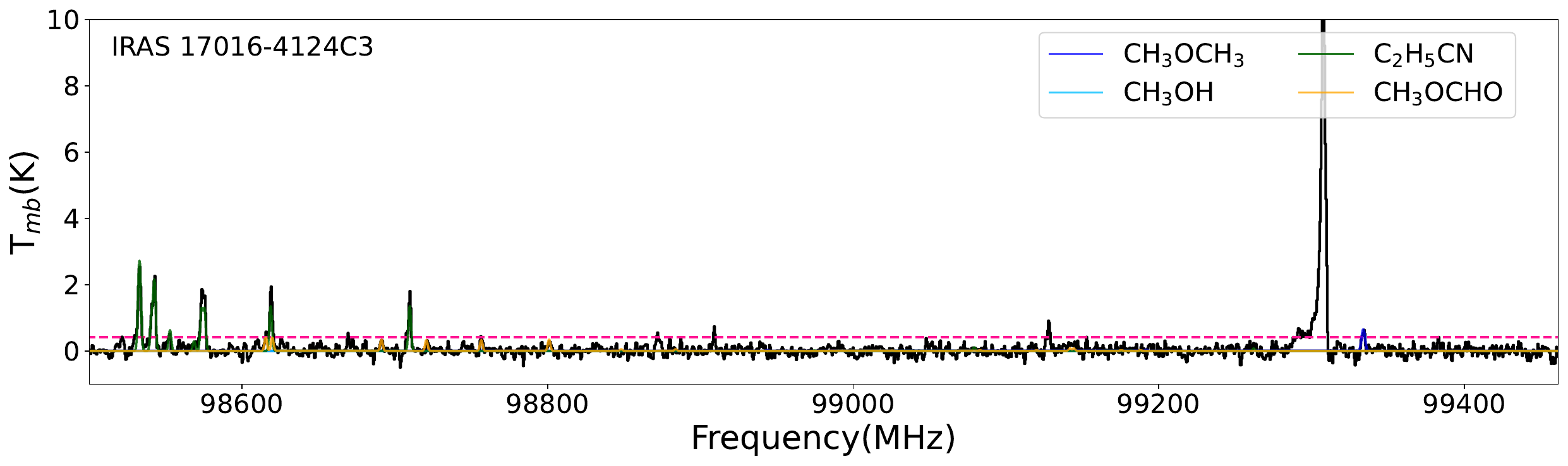}}
\caption{Continued.}
\end{figure}
\setcounter{figure}{\value{figure}-1}
\begin{figure}
  \centering 
{\includegraphics[height=4.5cm,width=15.93cm]{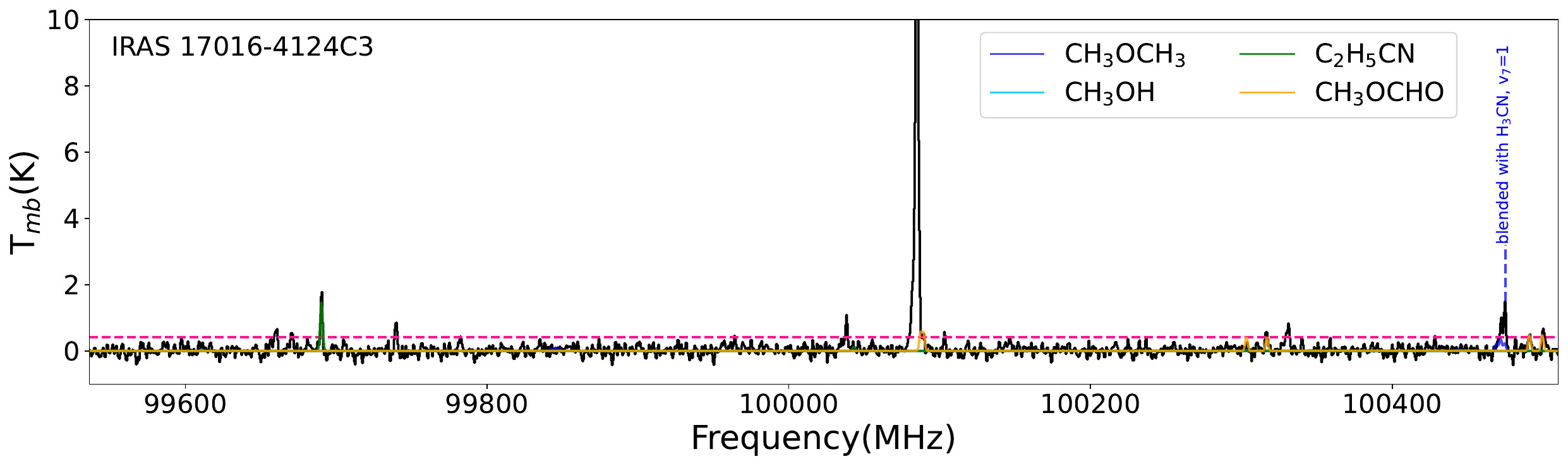}}
\quad
{\includegraphics[height=4.5cm,width=15.93cm]{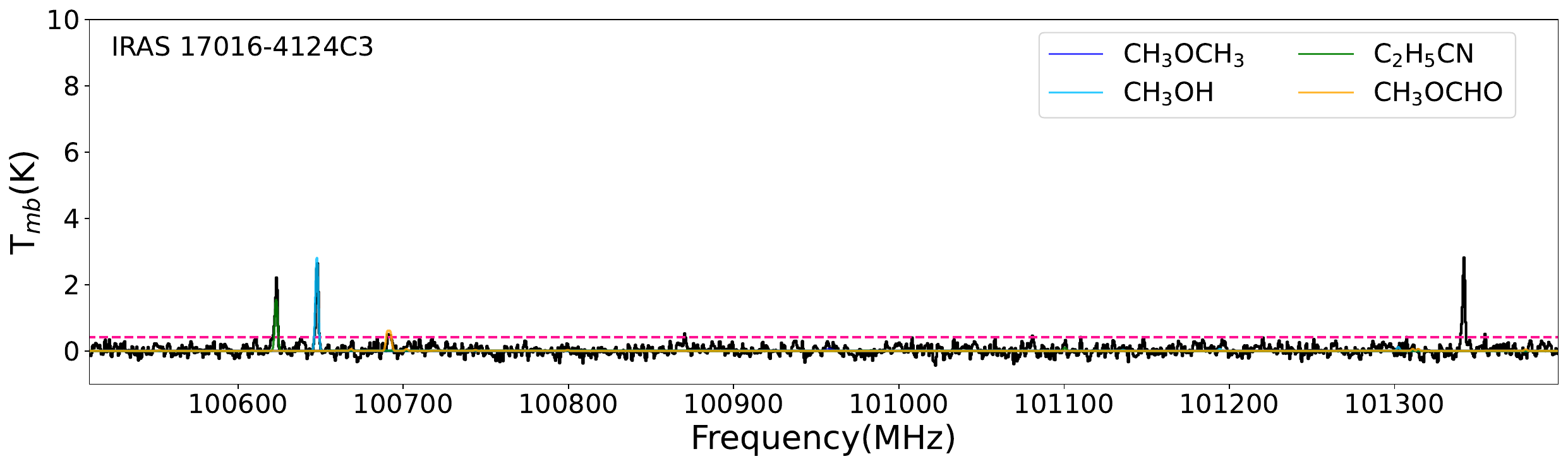}}
 \quad 
{\includegraphics[height=4.5cm,width=15.93cm]{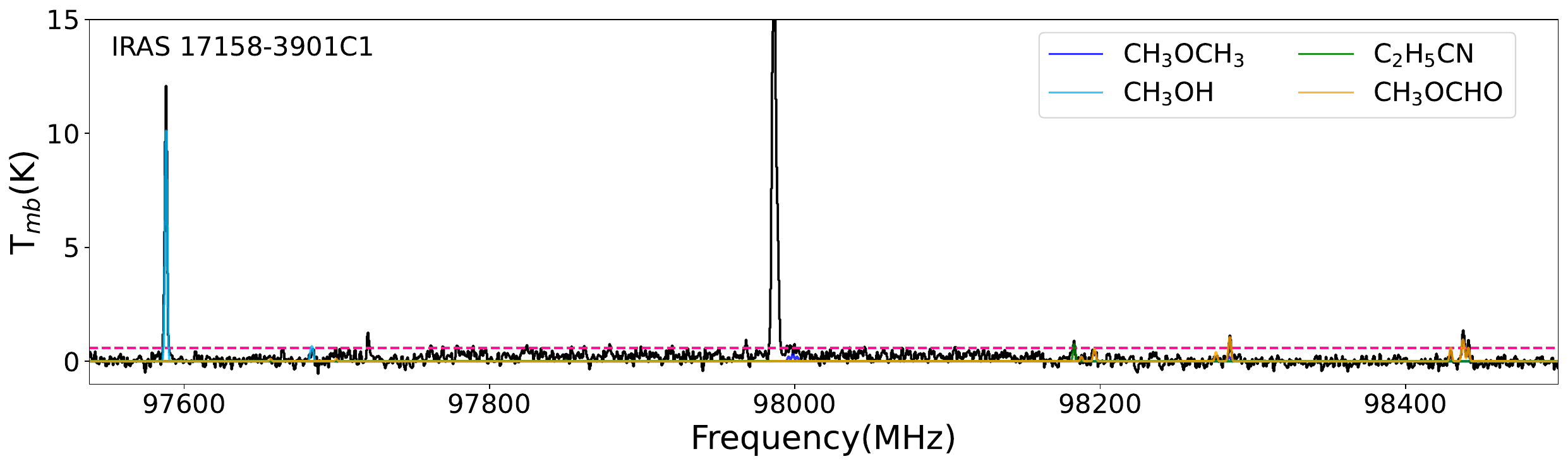}}
\quad
{\includegraphics[height=4.5cm,width=15.93cm]{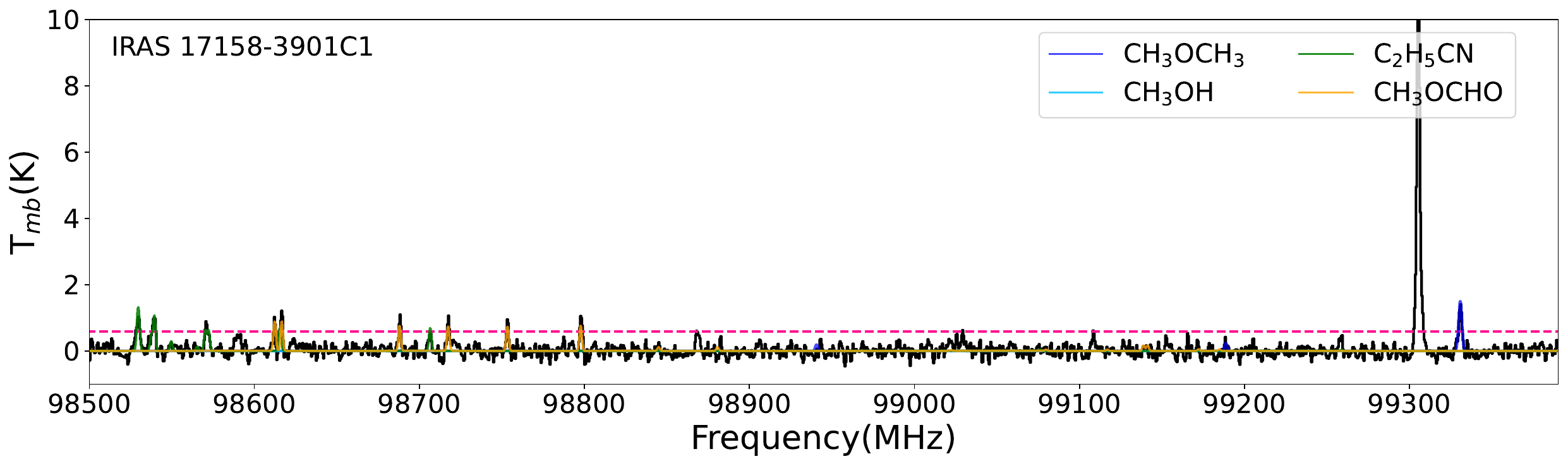}}
\quad
{\includegraphics[height=4.5cm,width=15.93cm]{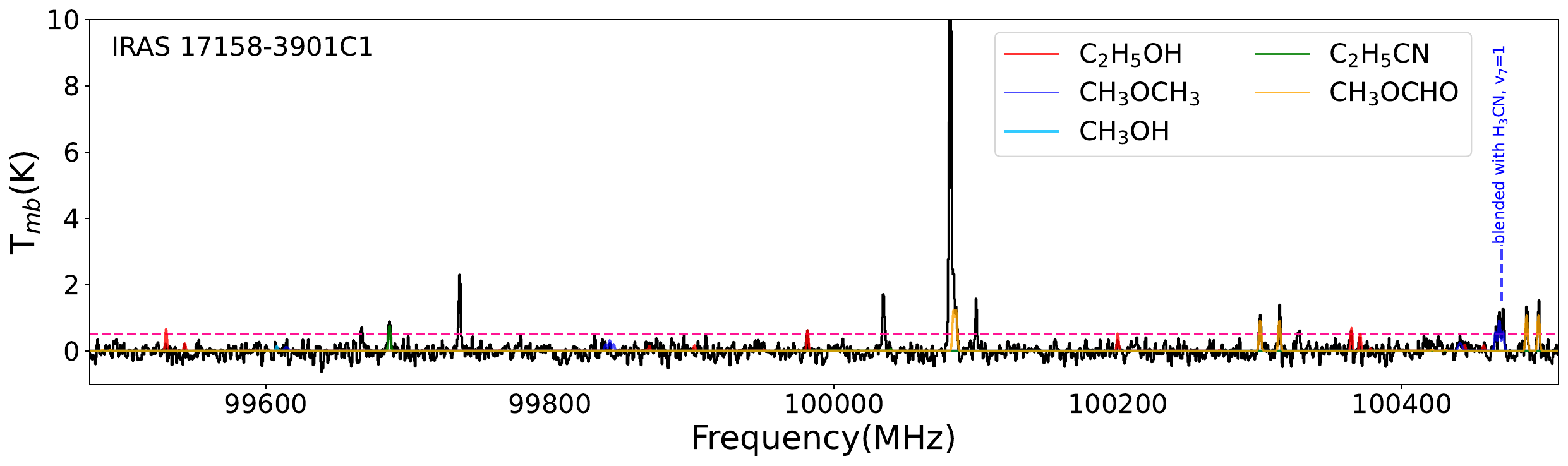}}
\caption{Continued.}
\end{figure}
\setcounter{figure}{\value{figure}-1}
\begin{figure}
  \centering 
{\includegraphics[height=4.5cm,width=15.93cm]{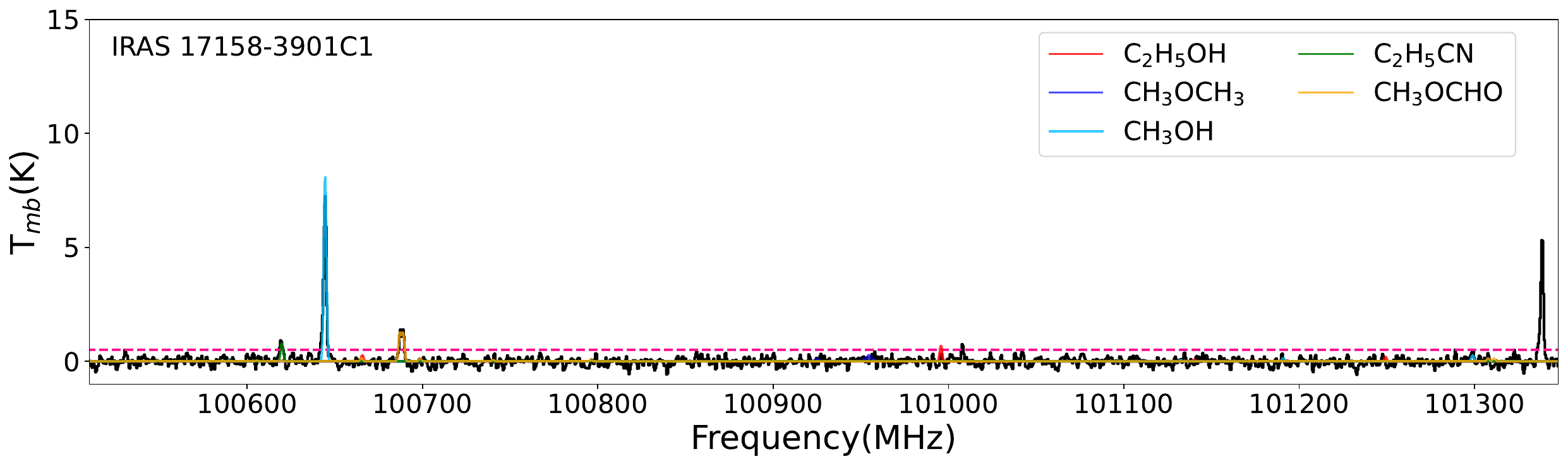}}
\quad
{\includegraphics[height=4.5cm,width=15.93cm]{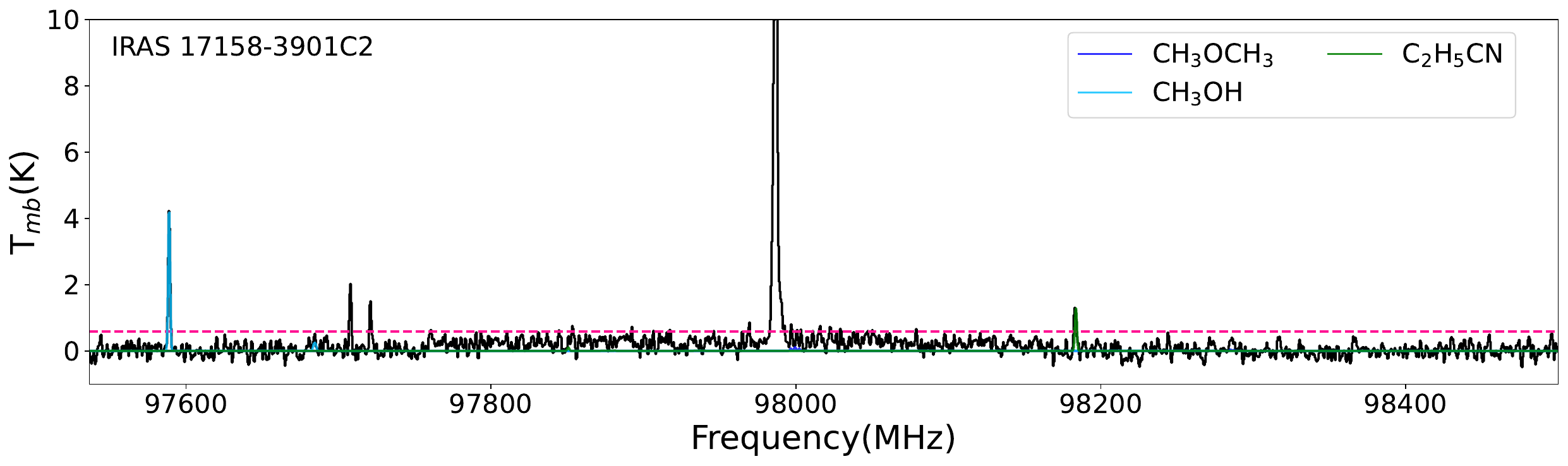}}
 \quad 
{\includegraphics[height=4.5cm,width=15.93cm]{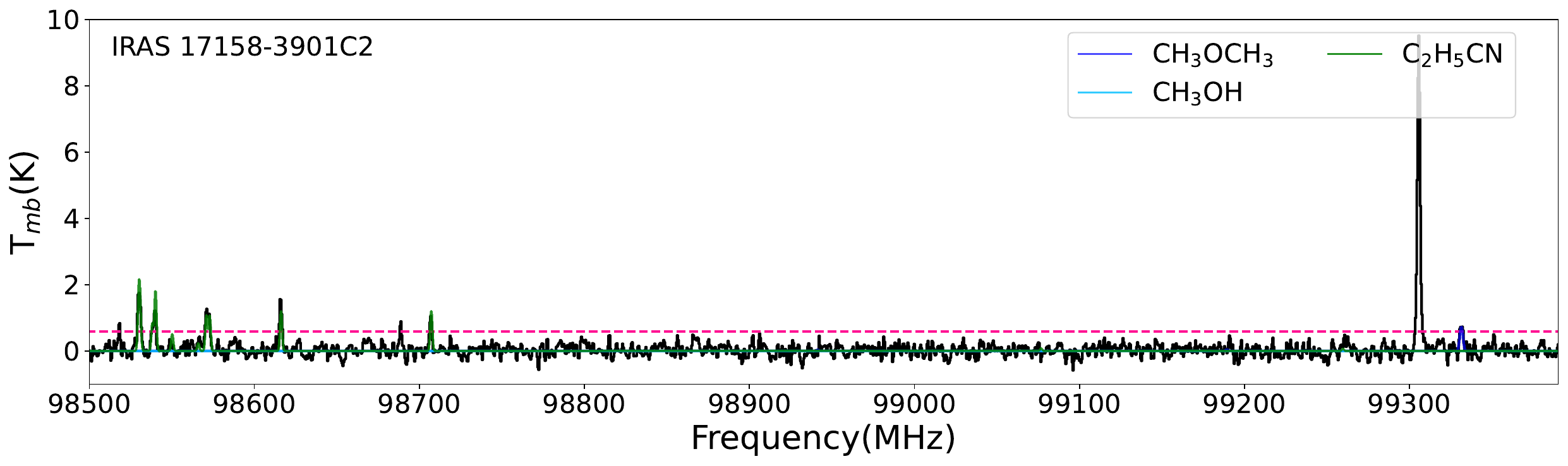}}
\quad
{\includegraphics[height=4.5cm,width=15.93cm]{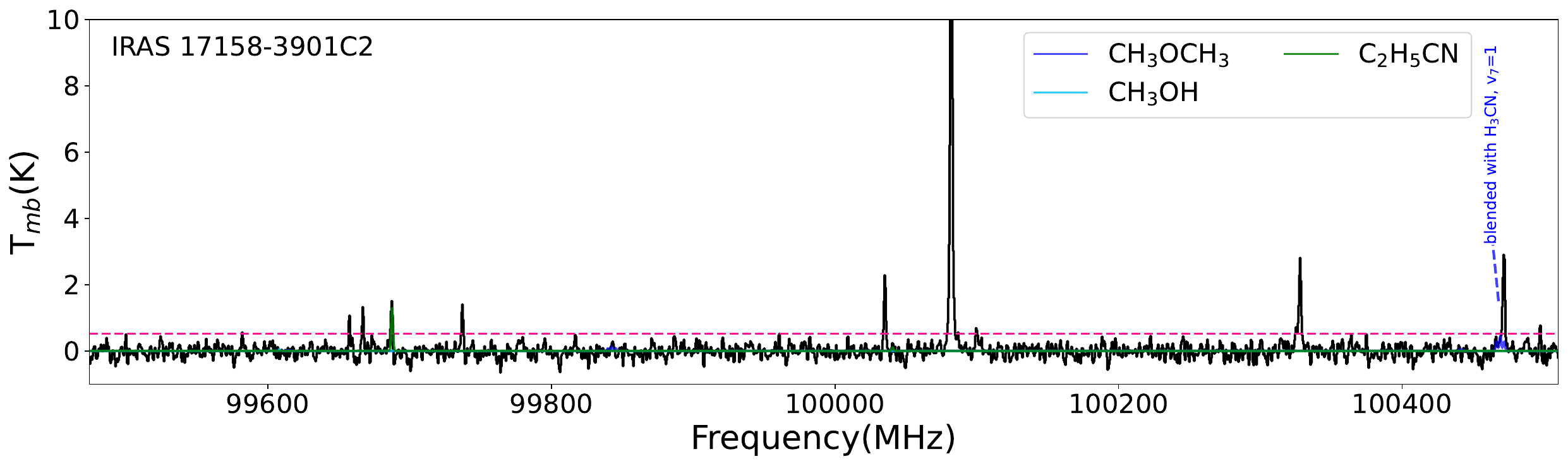}}
\quad
{\includegraphics[height=4.5cm,width=15.93cm]{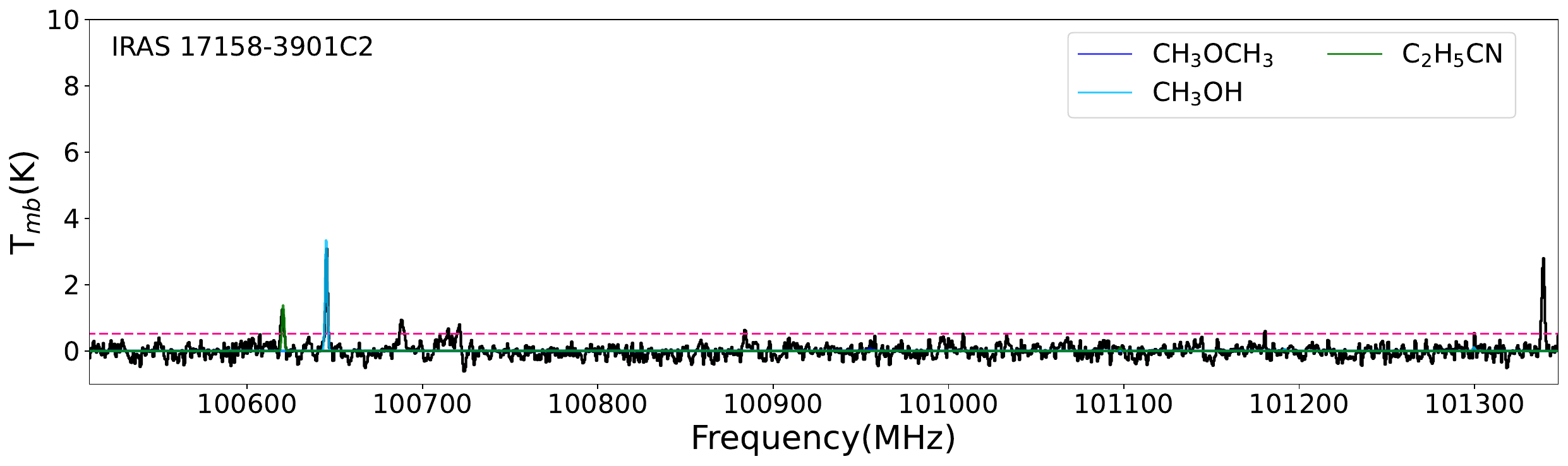}}
\caption{Continued.}
\end{figure}
\setcounter{figure}{\value{figure}-1}
\begin{figure}
  \centering 
{\includegraphics[height=4.5cm,width=15.93cm]{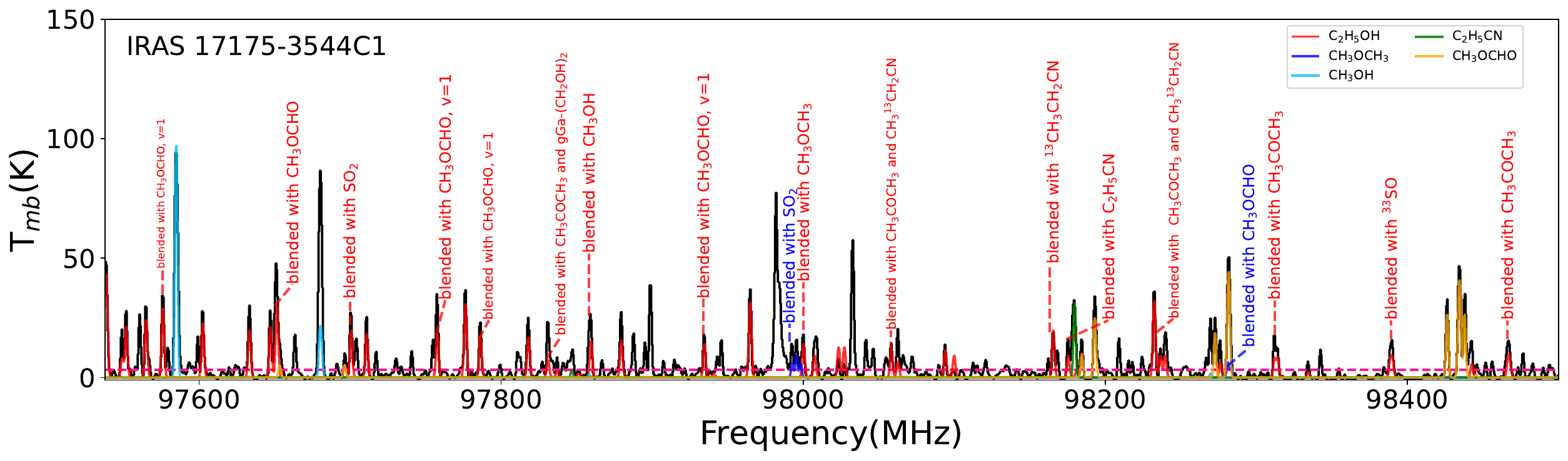}}
\quad
{\includegraphics[height=4.5cm,width=15.93cm]{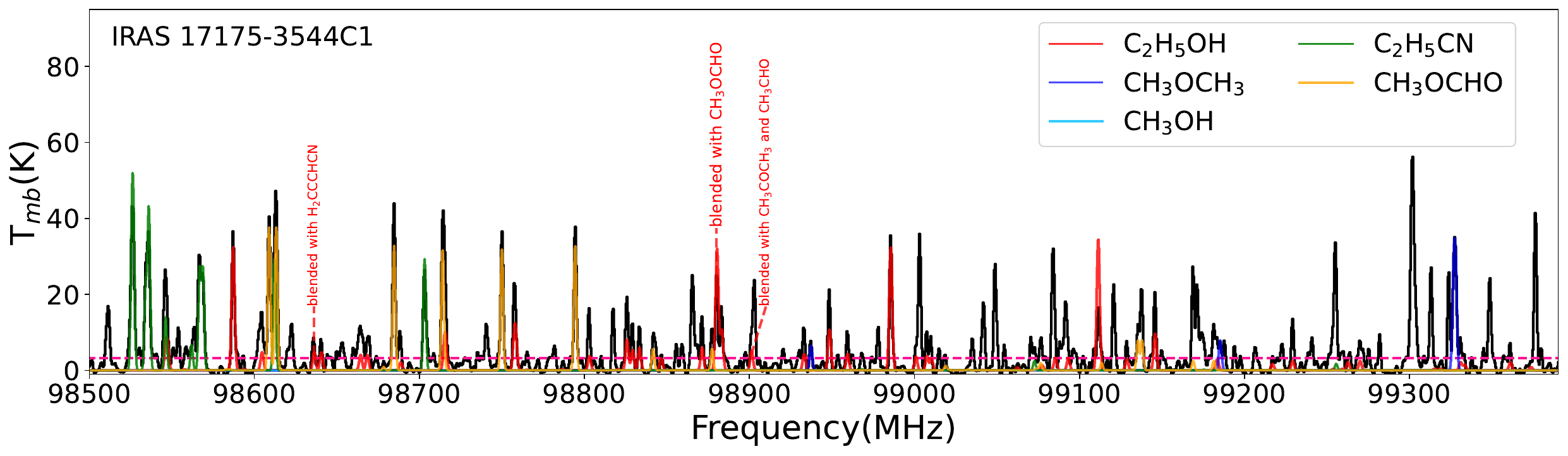}}
 \quad 
{\includegraphics[height=4.5cm,width=15.93cm]{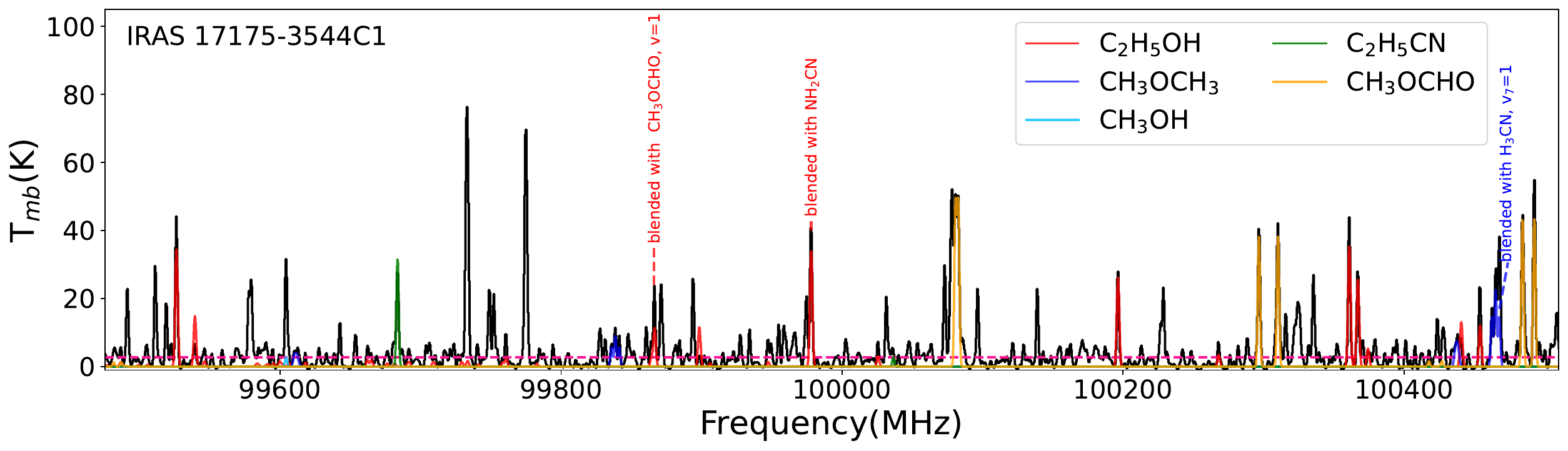}}
\quad
{\includegraphics[height=4.5cm,width=15.93cm]{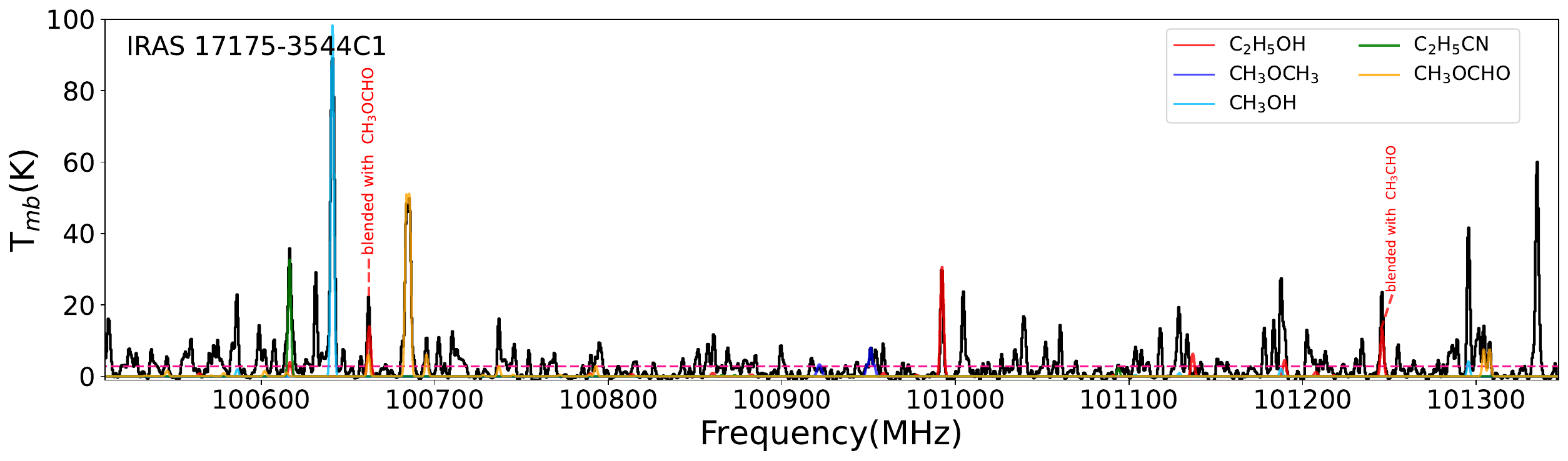}}
\quad
{\includegraphics[height=4.5cm,width=15.93cm]{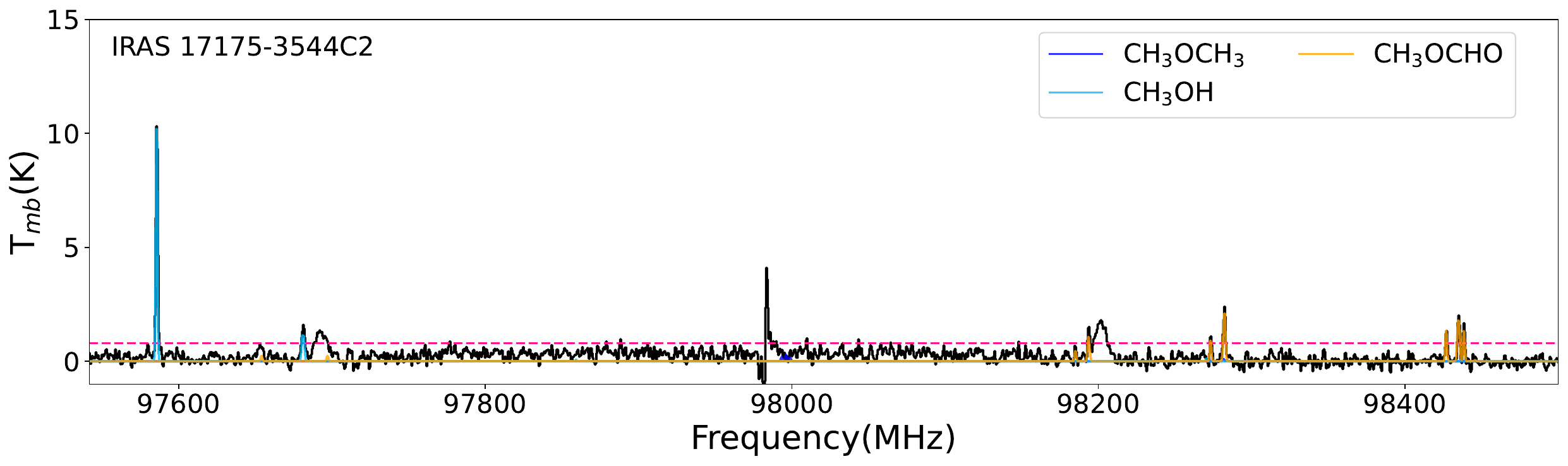}}
\caption{Continued.}
\end{figure}
\setcounter{figure}{\value{figure}-1}
\begin{figure}
  \centering 
{\includegraphics[height=4.5cm,width=15.93cm]{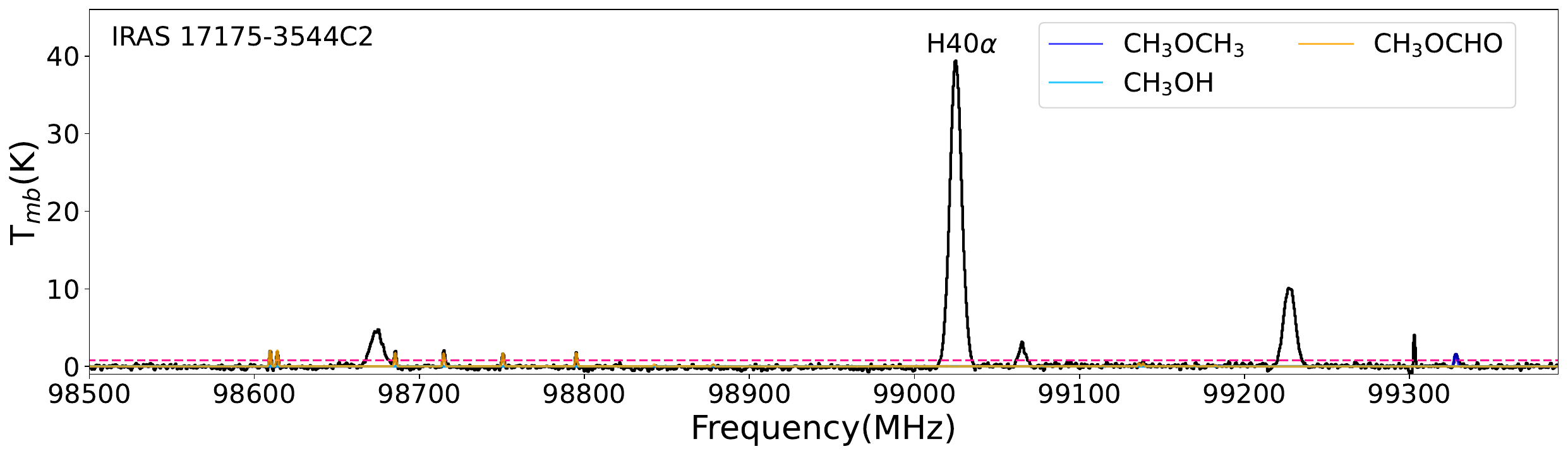}}
\quad
{\includegraphics[height=4.5cm,width=15.93cm]{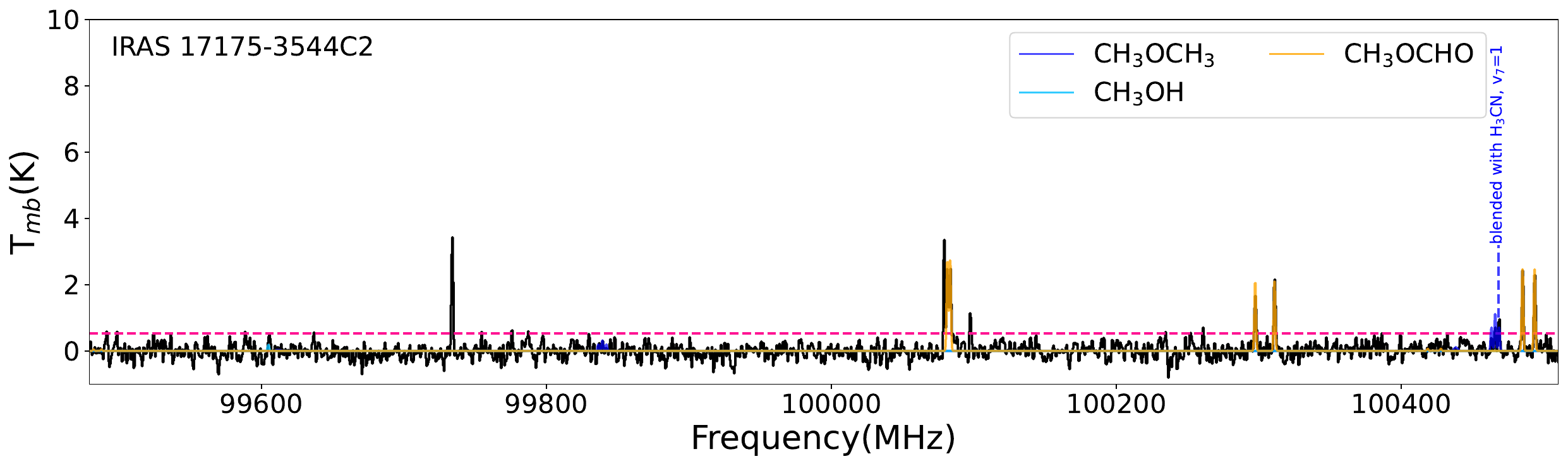}}
 \quad 
{\includegraphics[height=4.5cm,width=15.93cm]{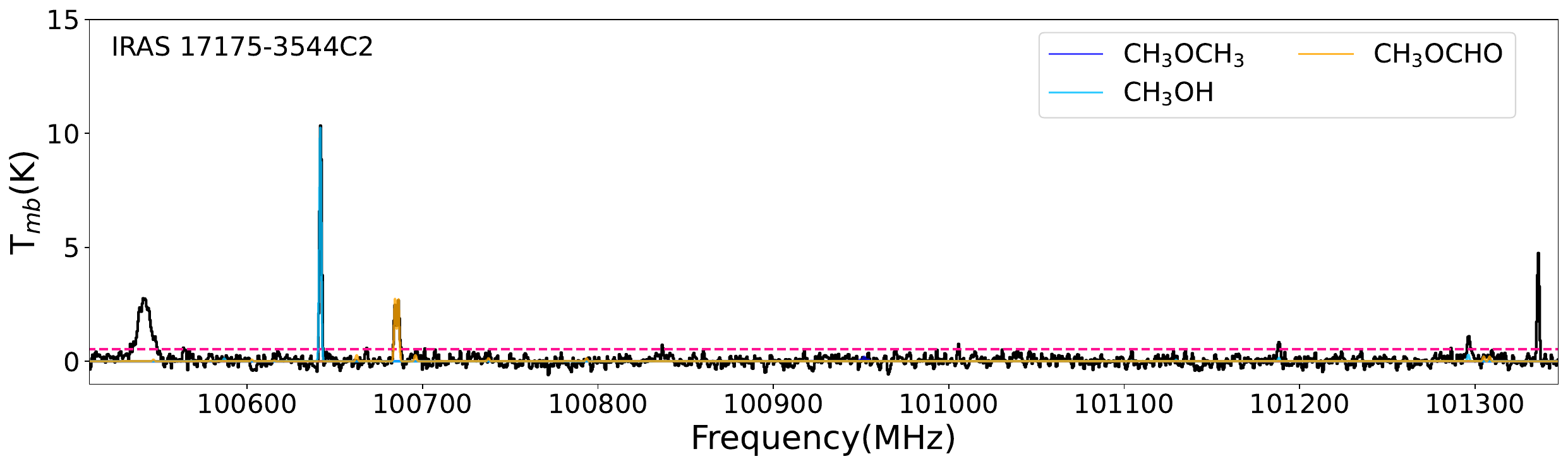}}
\quad
{\includegraphics[height=4.5cm,width=15.93cm]{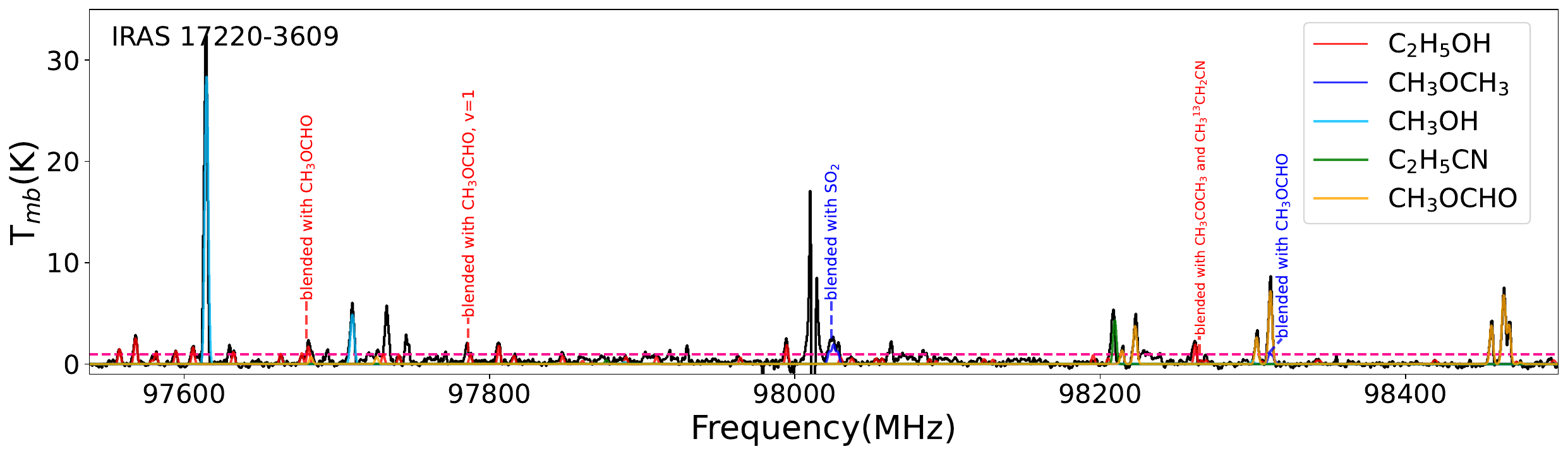}}
\quad
{\includegraphics[height=4.5cm,width=15.93cm]{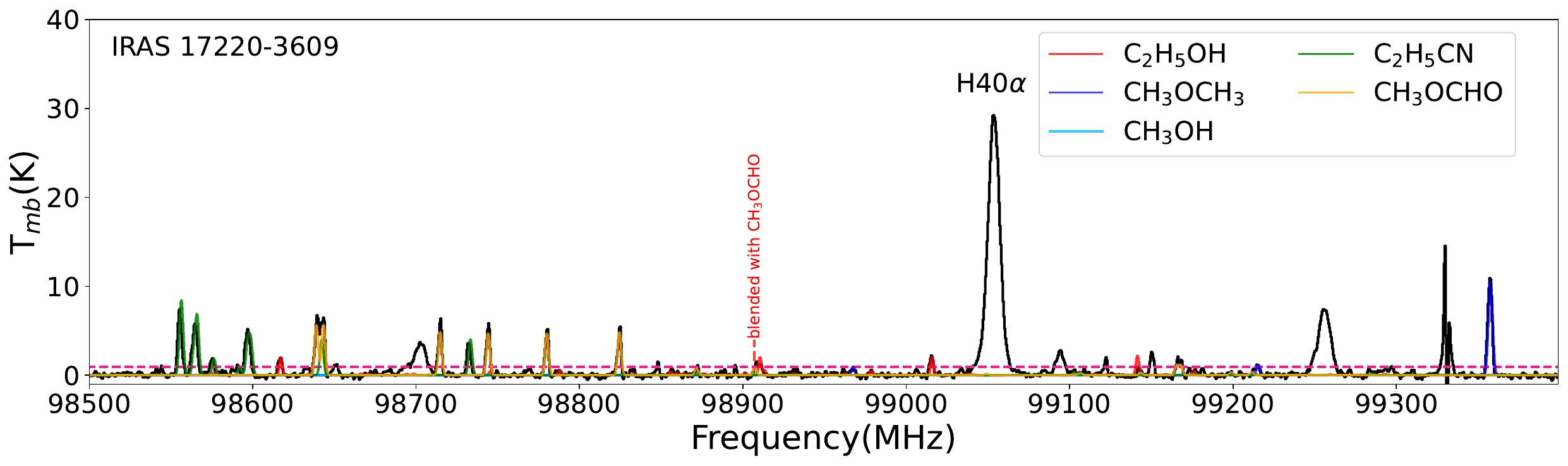}}
\caption{Continued.}
\end{figure}
\setcounter{figure}{\value{figure}-1}
\begin{figure}
  \centering 
{\includegraphics[height=4.5cm,width=15.93cm]{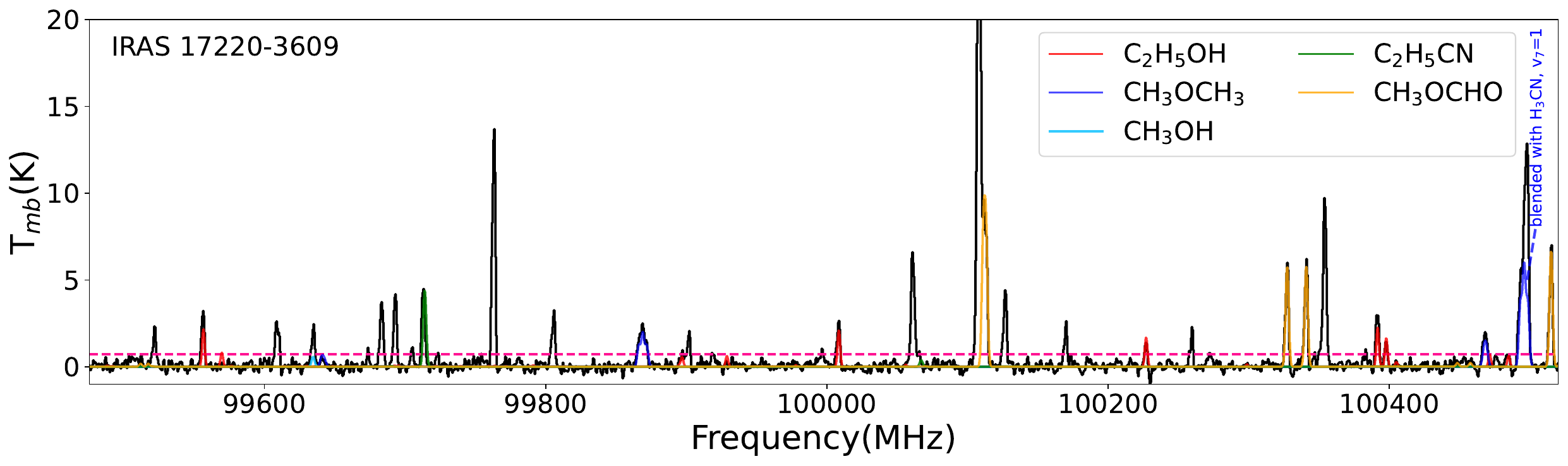}}
\quad
{\includegraphics[height=4.5cm,width=15.93cm]{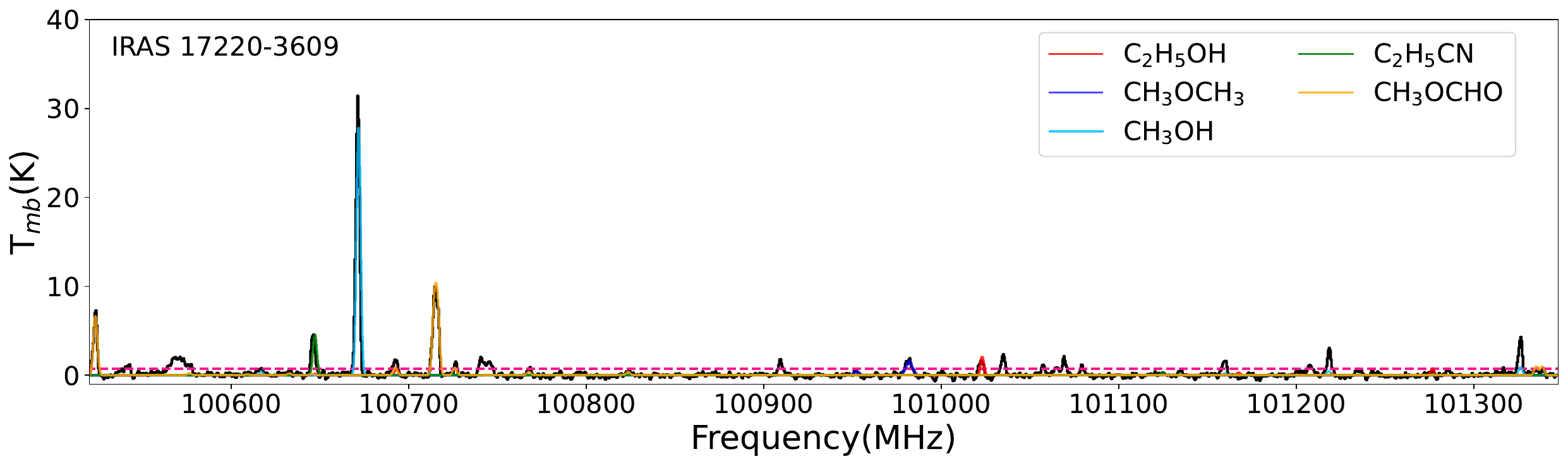}}
 \quad 
{\includegraphics[height=4.5cm,width=15.93cm]{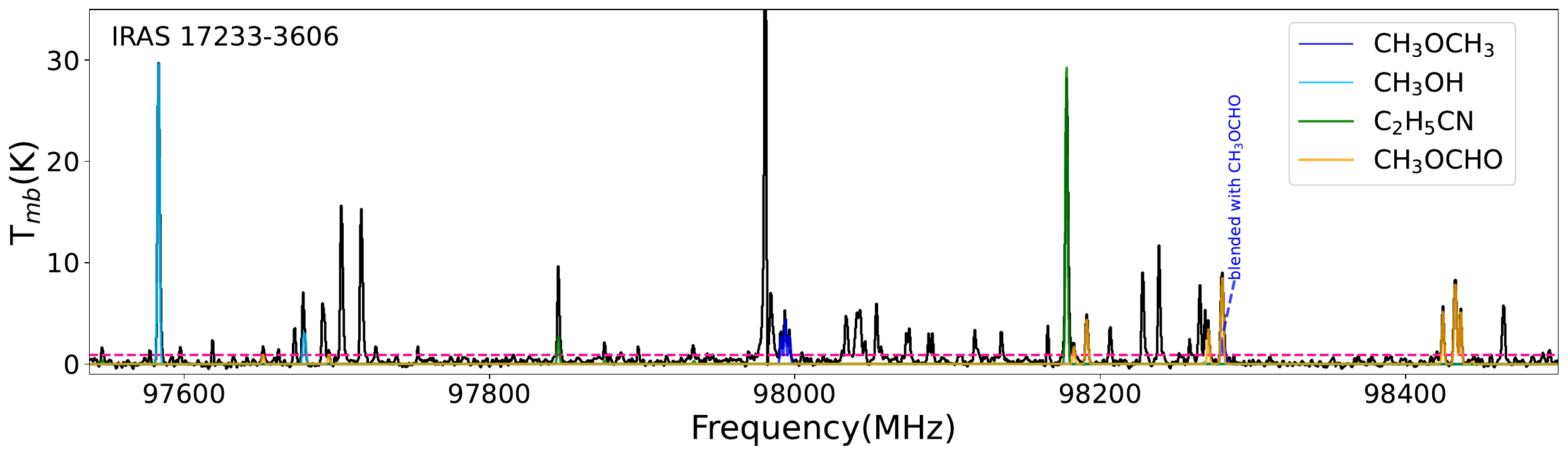}}
\quad
{\includegraphics[height=4.5cm,width=15.93cm]{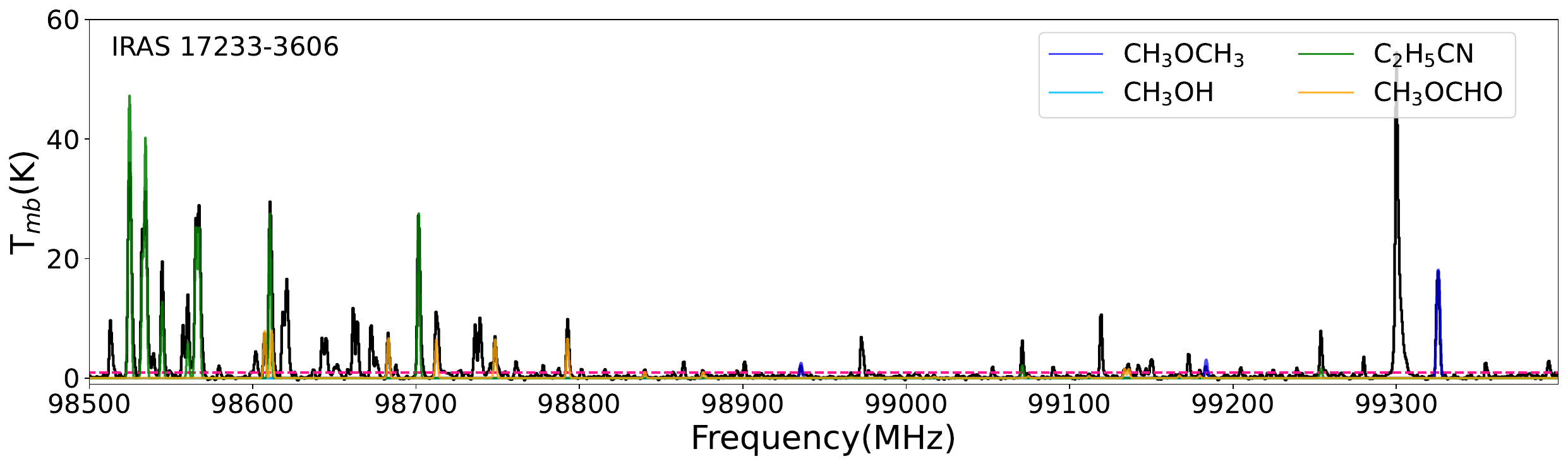}}
\quad
{\includegraphics[height=4.5cm,width=15.93cm]{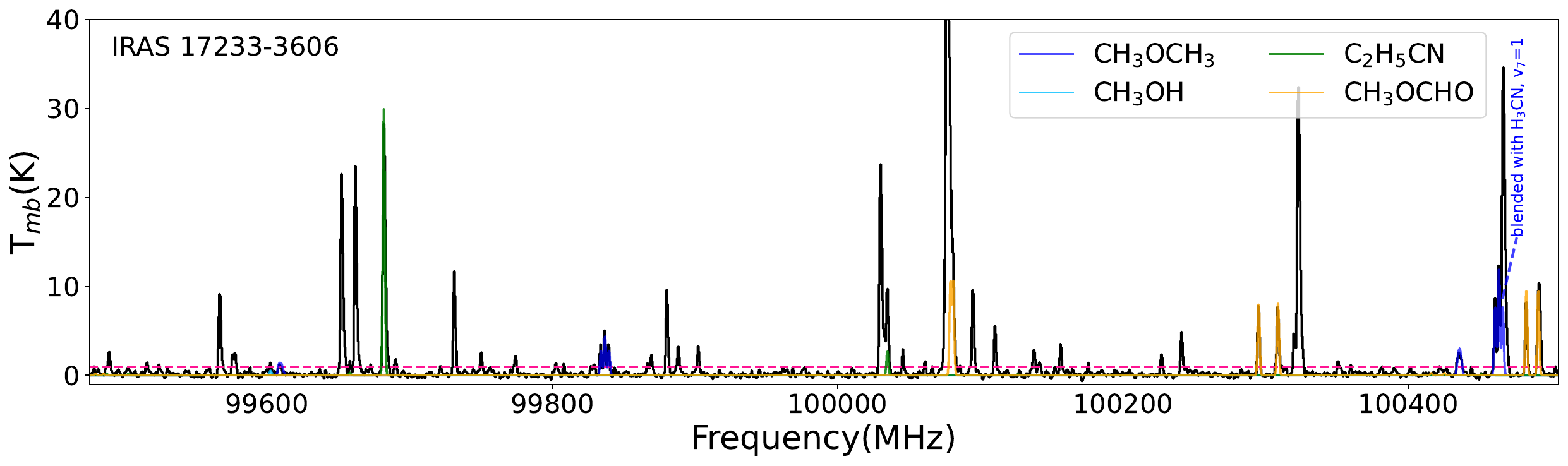}}
\caption{Continued.}
\end{figure}
\setcounter{figure}{\value{figure}-1}
\begin{figure}
  \centering 
{\includegraphics[height=4.5cm,width=15.93cm]{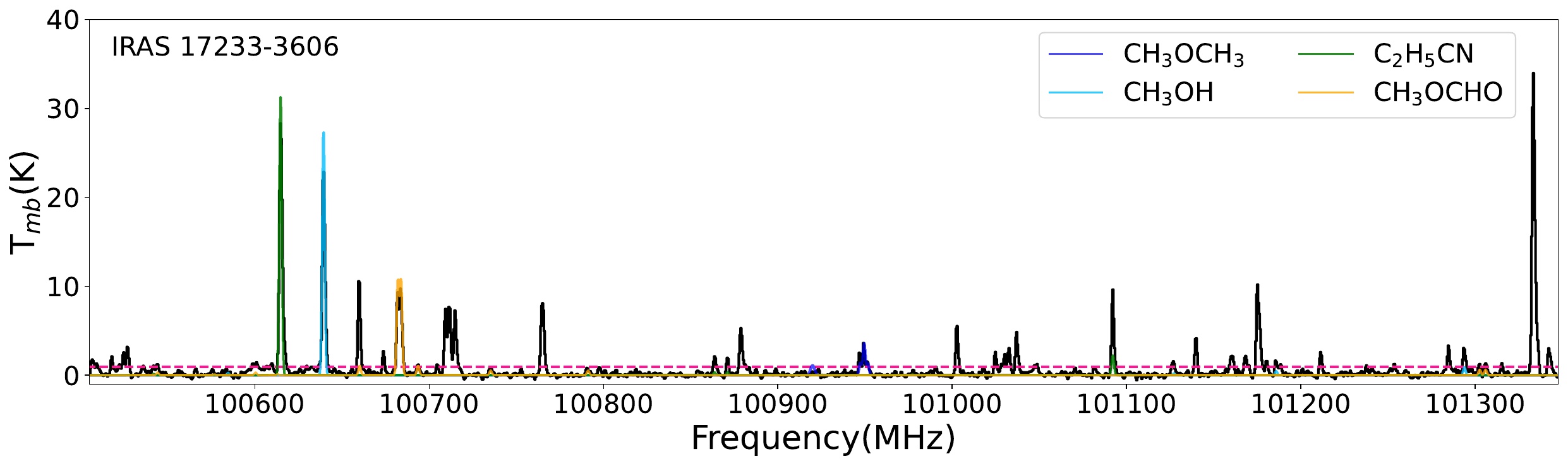}}
\quad
{\includegraphics[height=4.5cm,width=15.93cm]{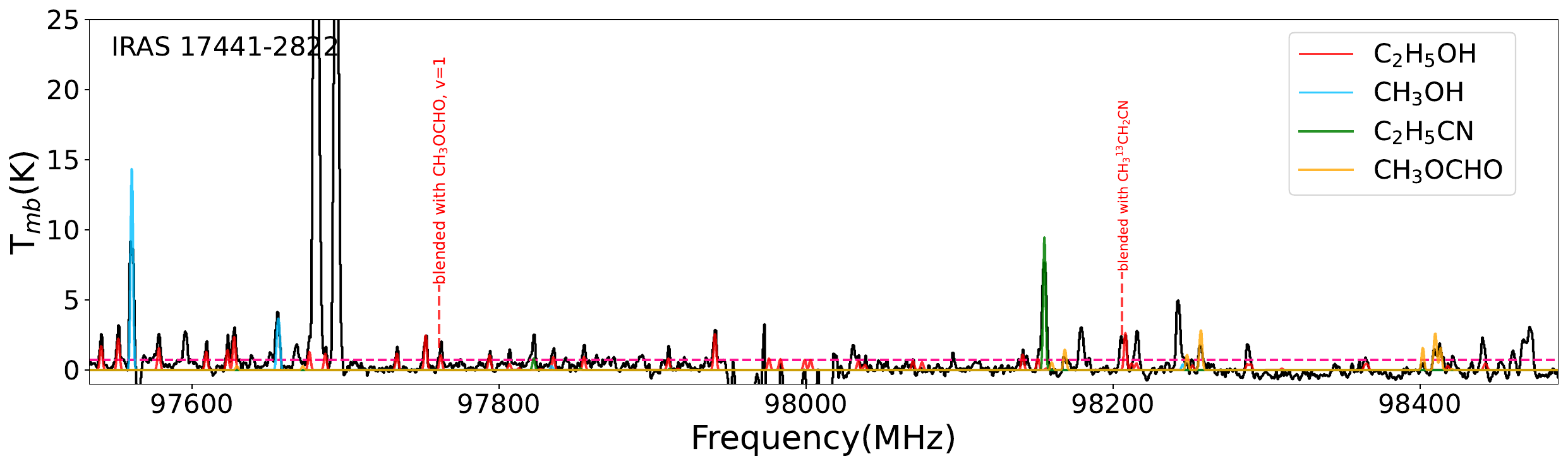}}
 \quad 
{\includegraphics[height=4.5cm,width=15.93cm]{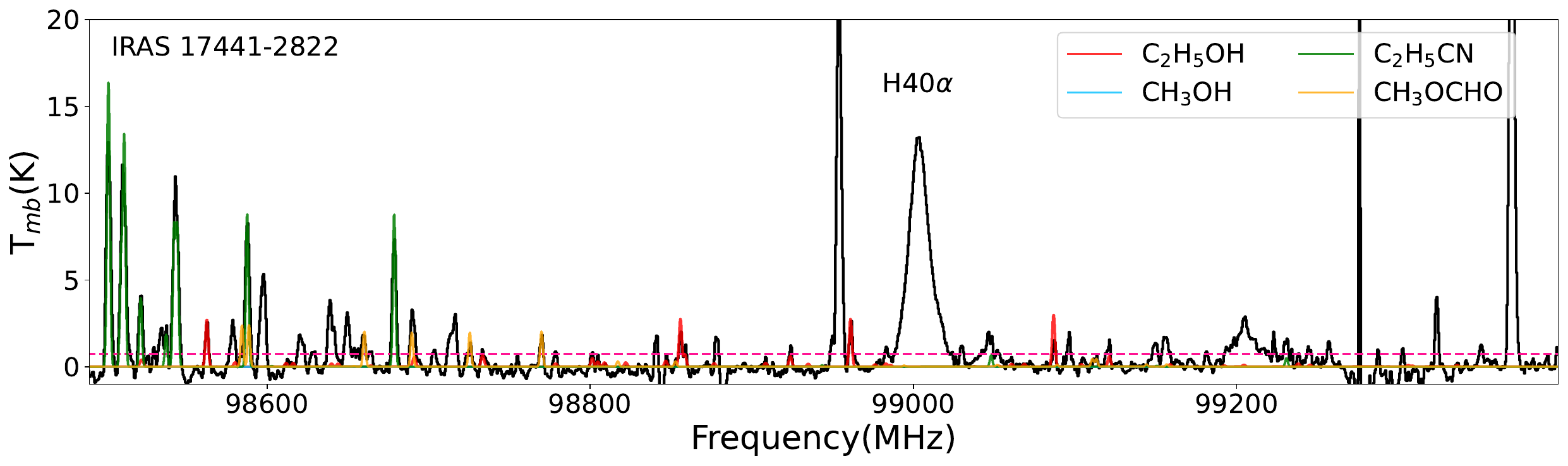}}
\quad
{\includegraphics[height=4.5cm,width=15.93cm]{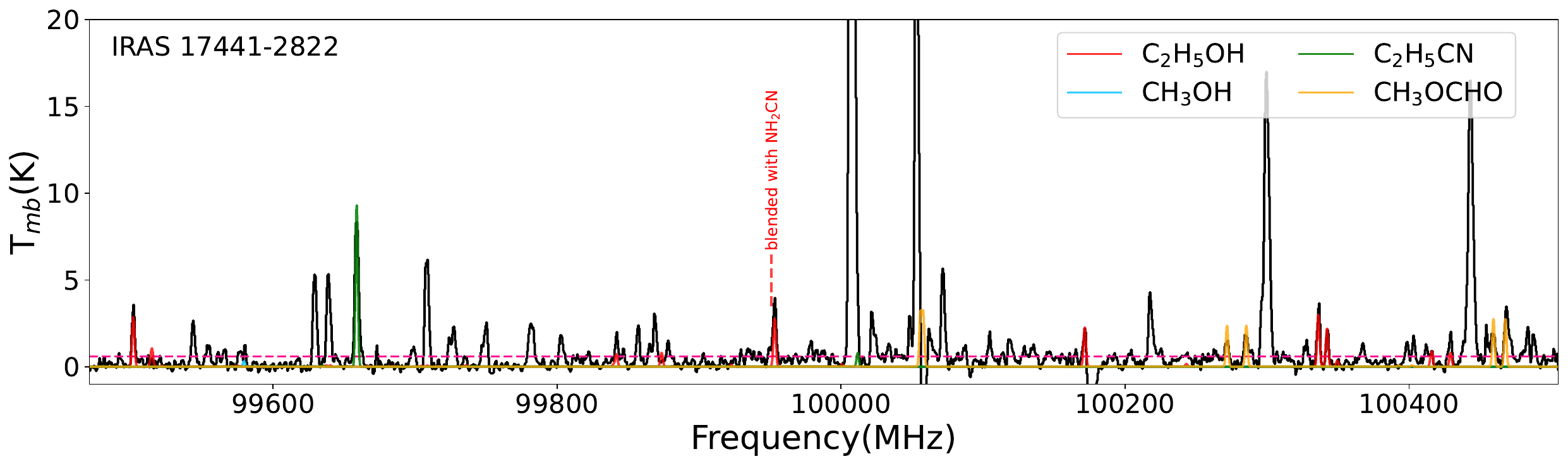}}
\quad
{\includegraphics[height=4.5cm,width=15.93cm]{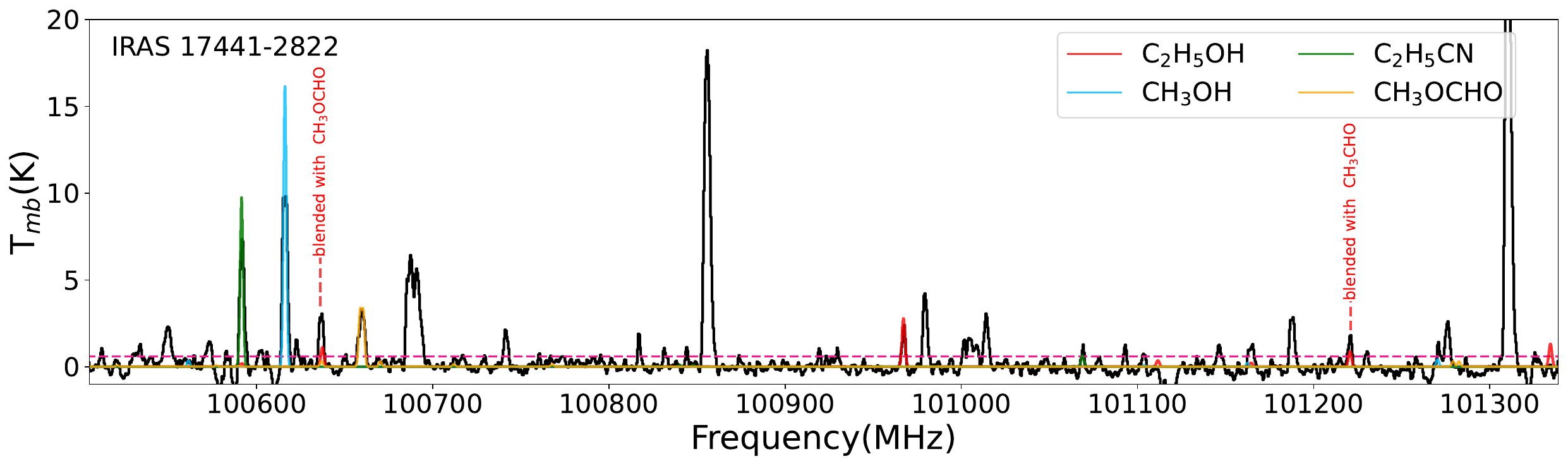}}
\caption{Continued.}
\end{figure}
\setcounter{figure}{\value{figure}-1}
\begin{figure}
  \centering 
{\includegraphics[height=4.5cm,width=15.93cm]{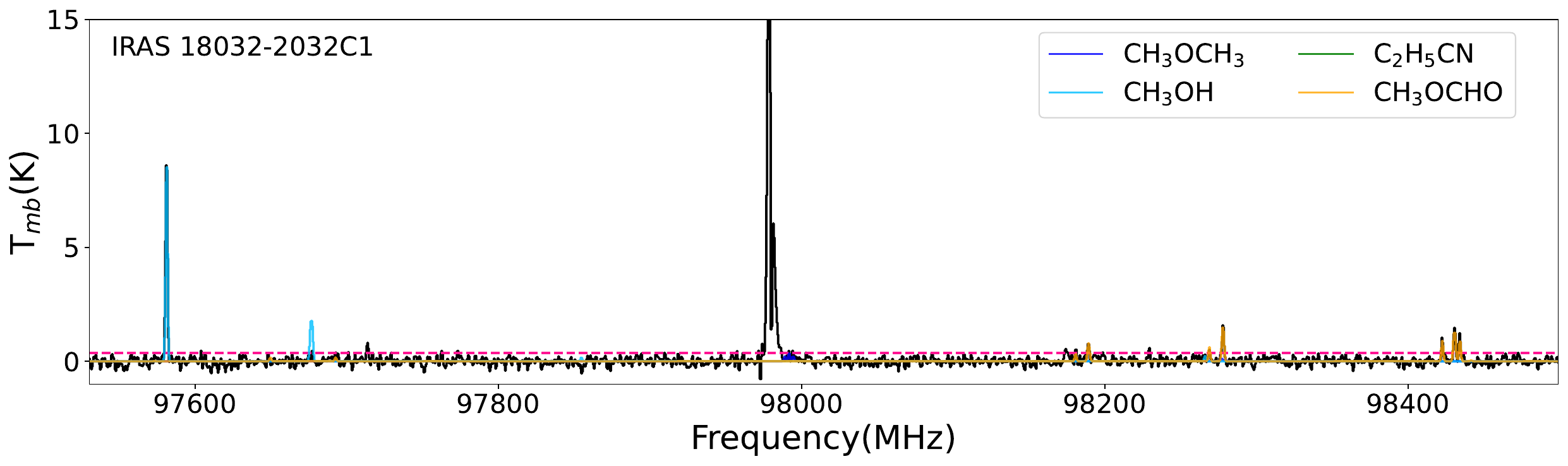}}
\quad
{\includegraphics[height=4.5cm,width=15.93cm]{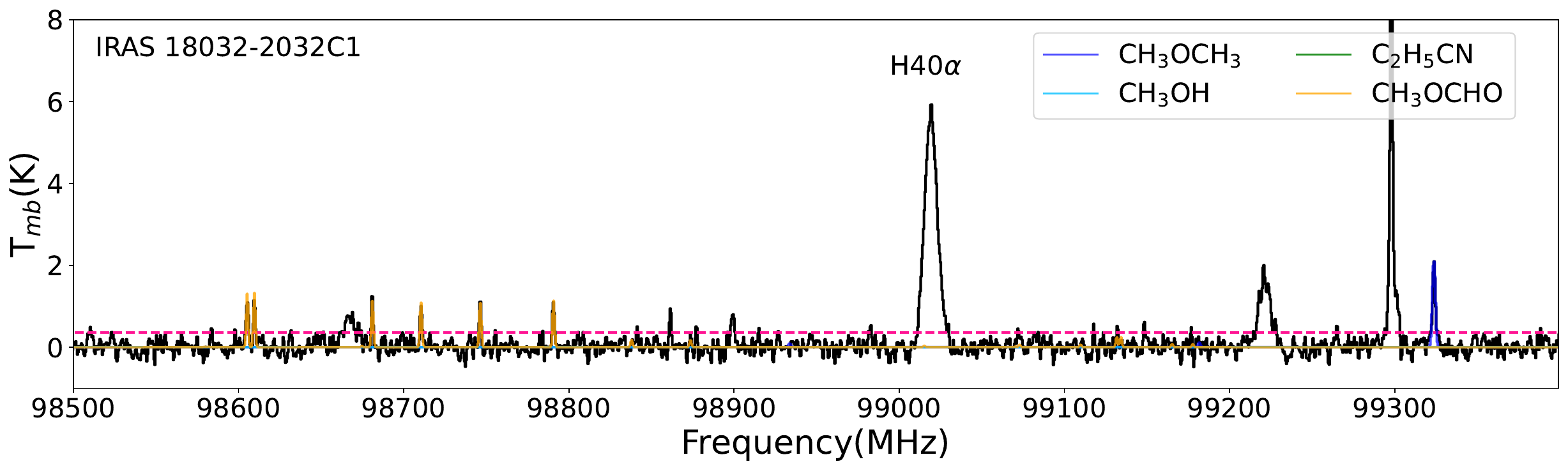}}
 \quad 
{\includegraphics[height=4.5cm,width=15.93cm]{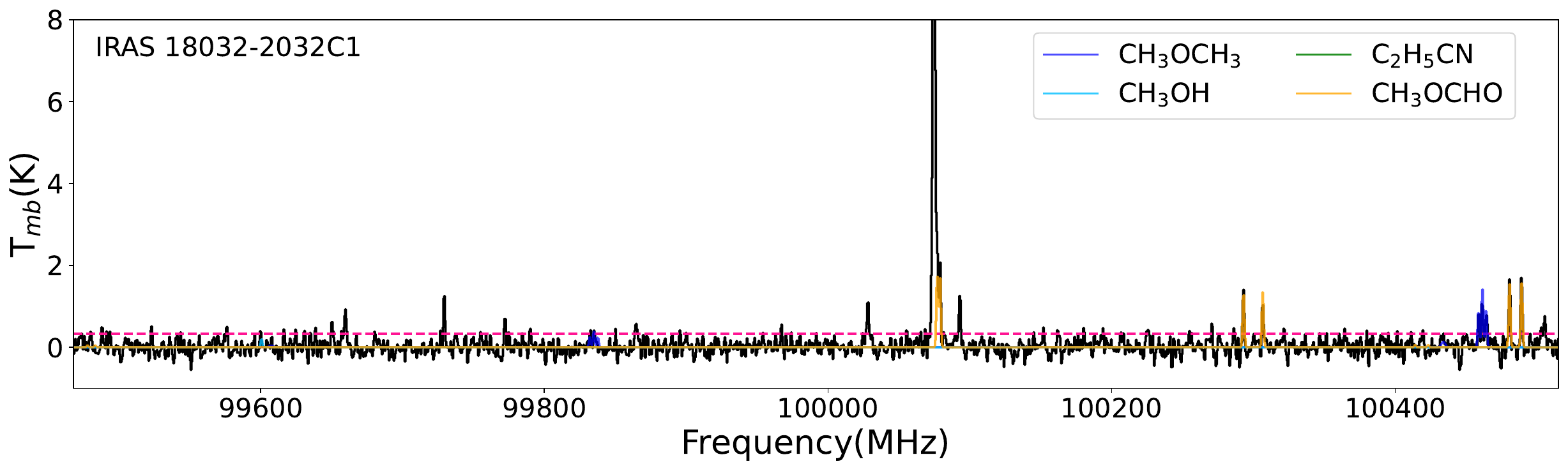}}
\quad
{\includegraphics[height=4.5cm,width=15.93cm]{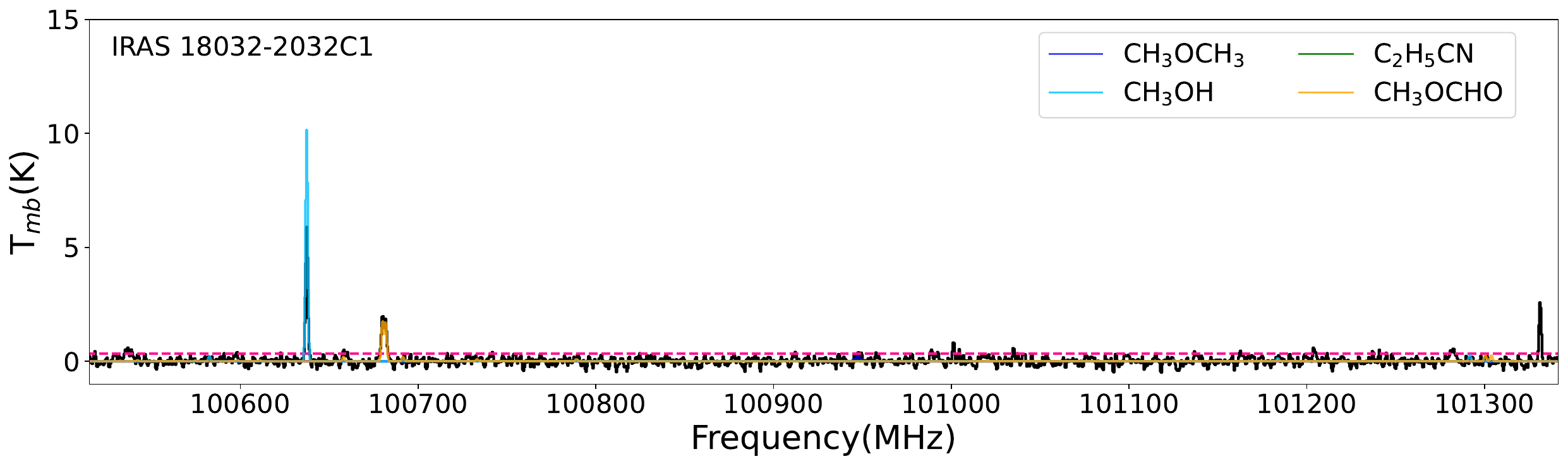}}
\quad
{\includegraphics[height=4.5cm,width=15.93cm]{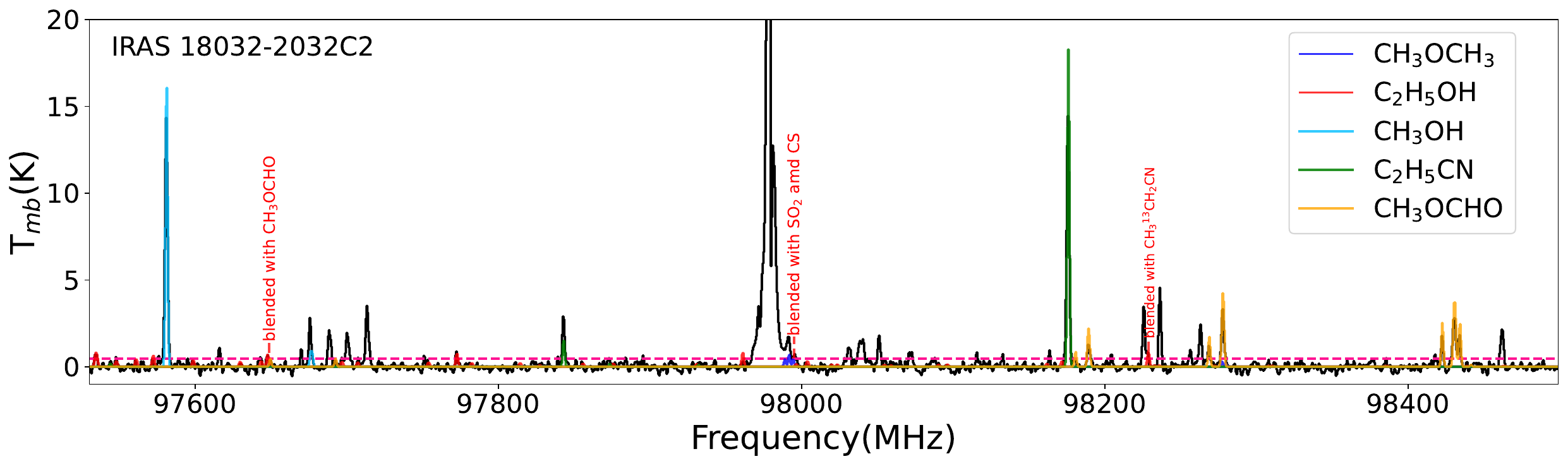}}
\caption{Continued.}
\end{figure}
\setcounter{figure}{\value{figure}-1}
\begin{figure}
  \centering 
{\includegraphics[height=4.5cm,width=15.93cm]{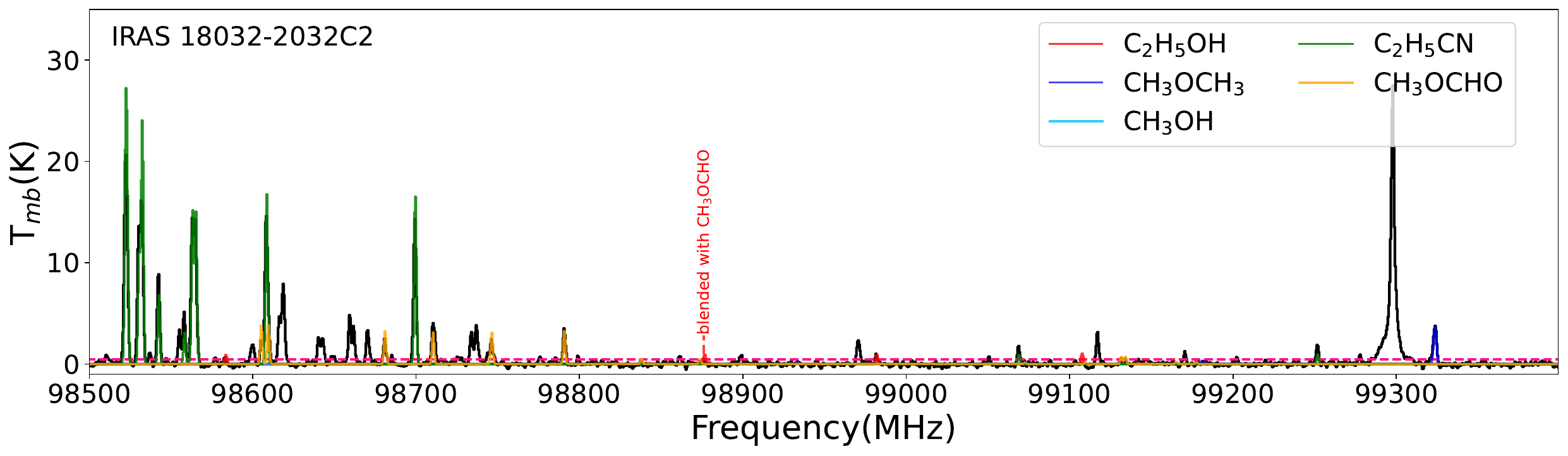}}
\quad
{\includegraphics[height=4.5cm,width=15.93cm]{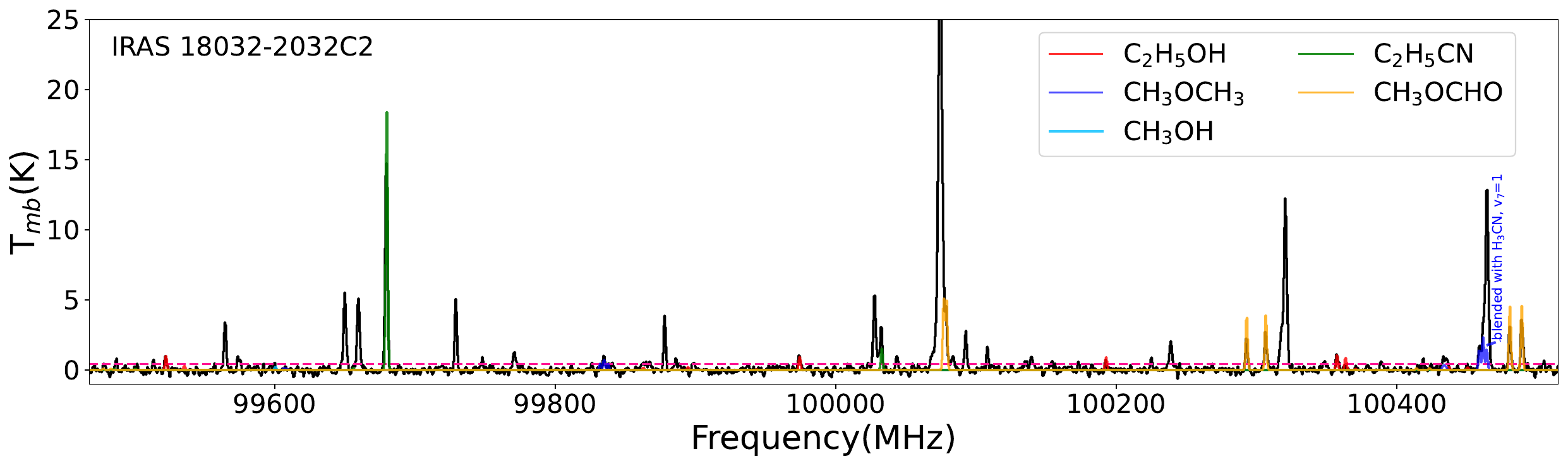}}
 \quad 
{\includegraphics[height=4.5cm,width=15.93cm]{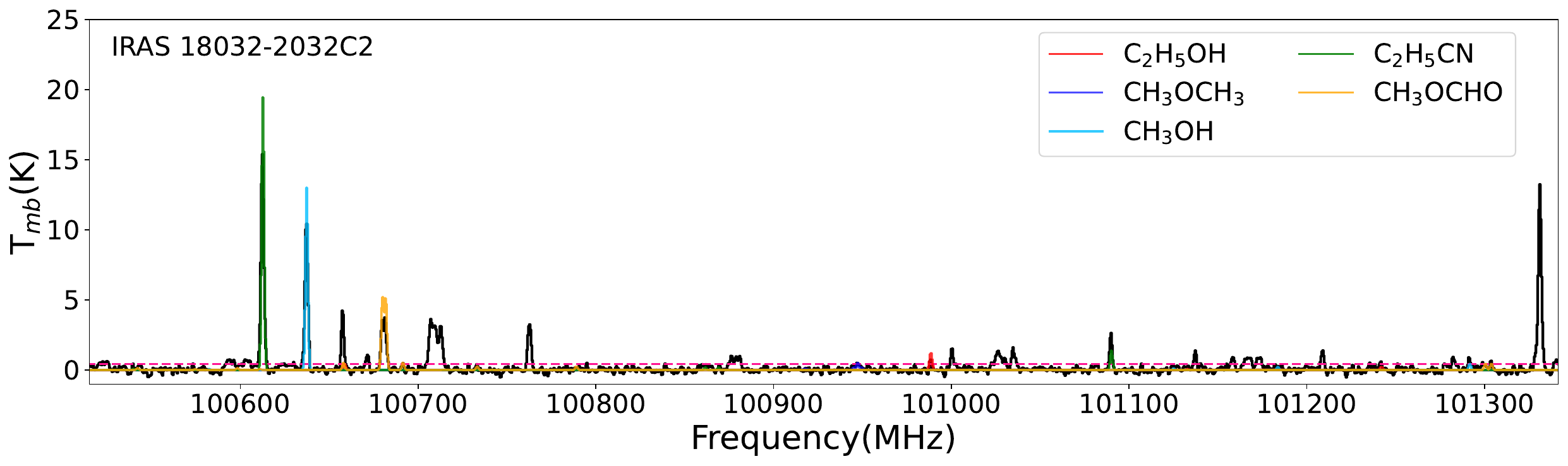}}
\quad
{\includegraphics[height=4.5cm,width=15.93cm]{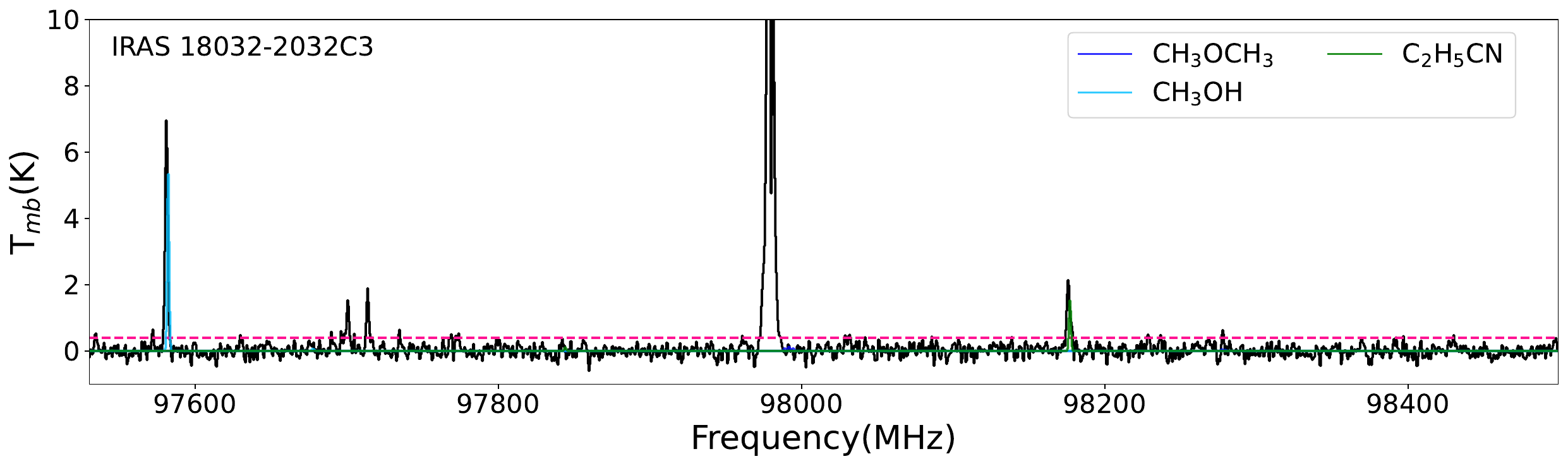}}
\quad
{\includegraphics[height=4.5cm,width=15.93cm]{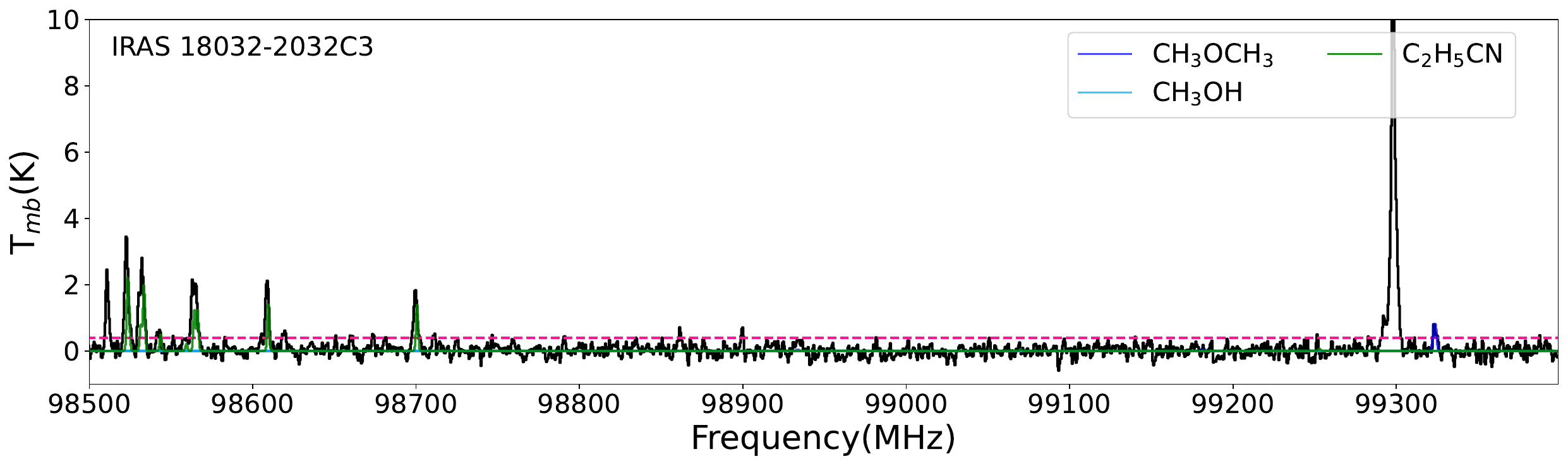}}
\caption{Continued.}
\end{figure}
\setcounter{figure}{\value{figure}-1}
\begin{figure}
  \centering 
{\includegraphics[height=4.5cm,width=15.93cm]{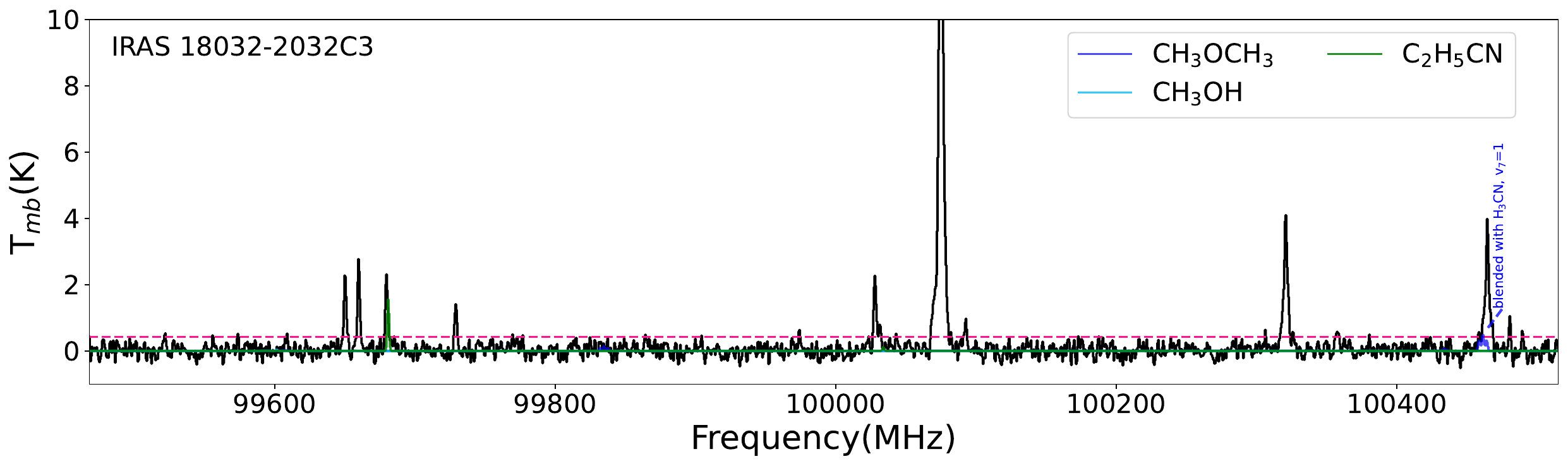}}
\quad
{\includegraphics[height=4.5cm,width=15.93cm]{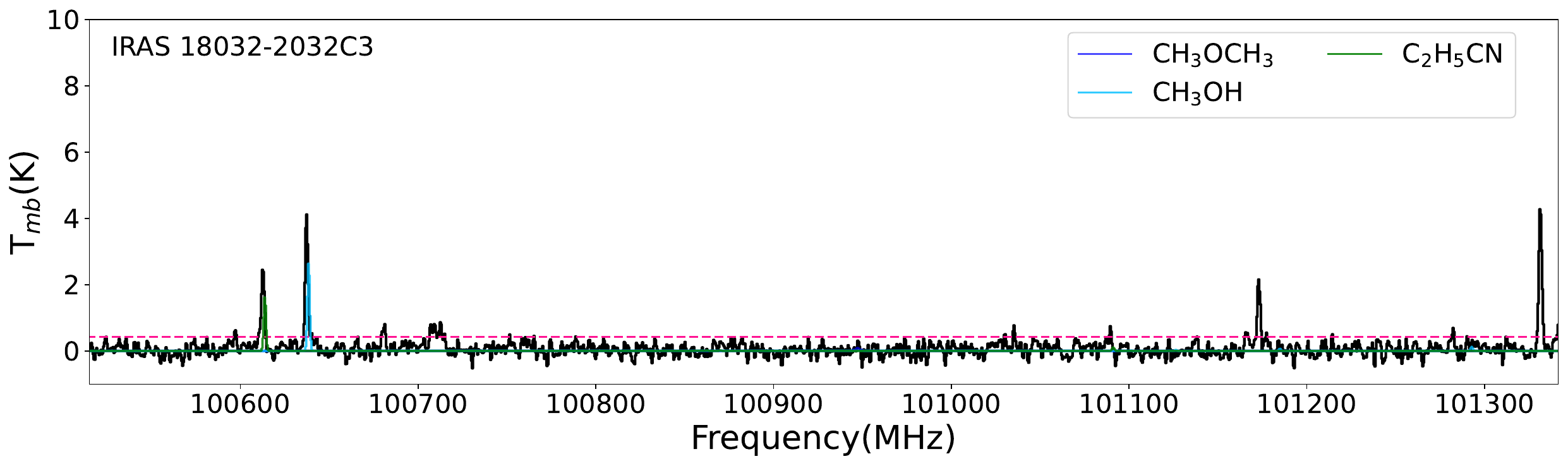}}
 \quad 
{\includegraphics[height=4.5cm,width=15.93cm]{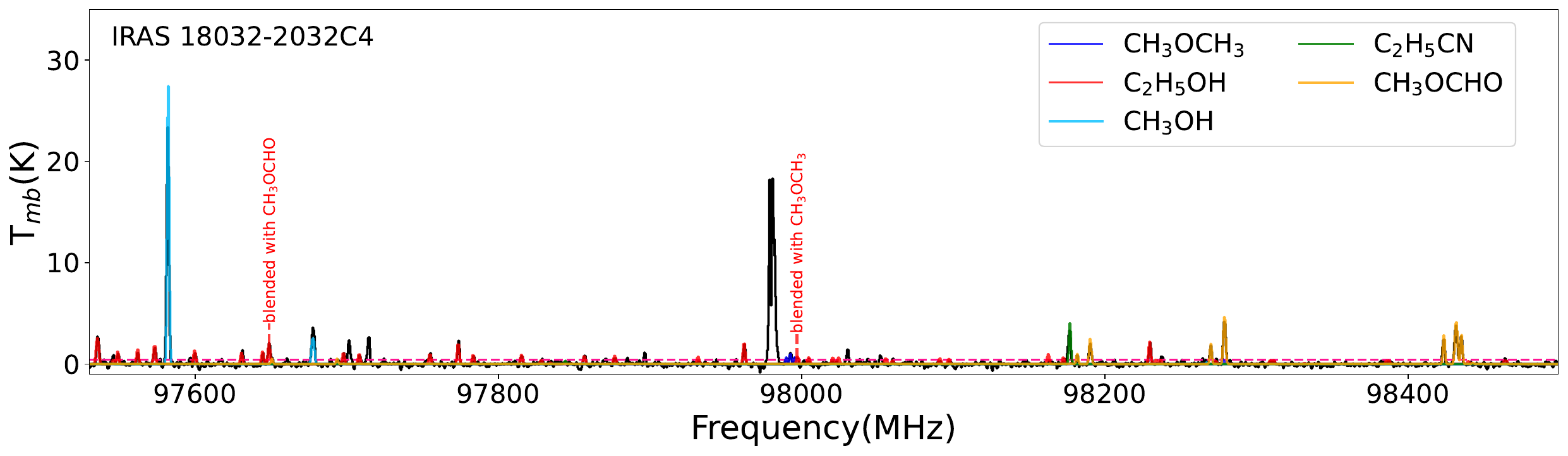}}
\quad
{\includegraphics[height=4.5cm,width=15.93cm]{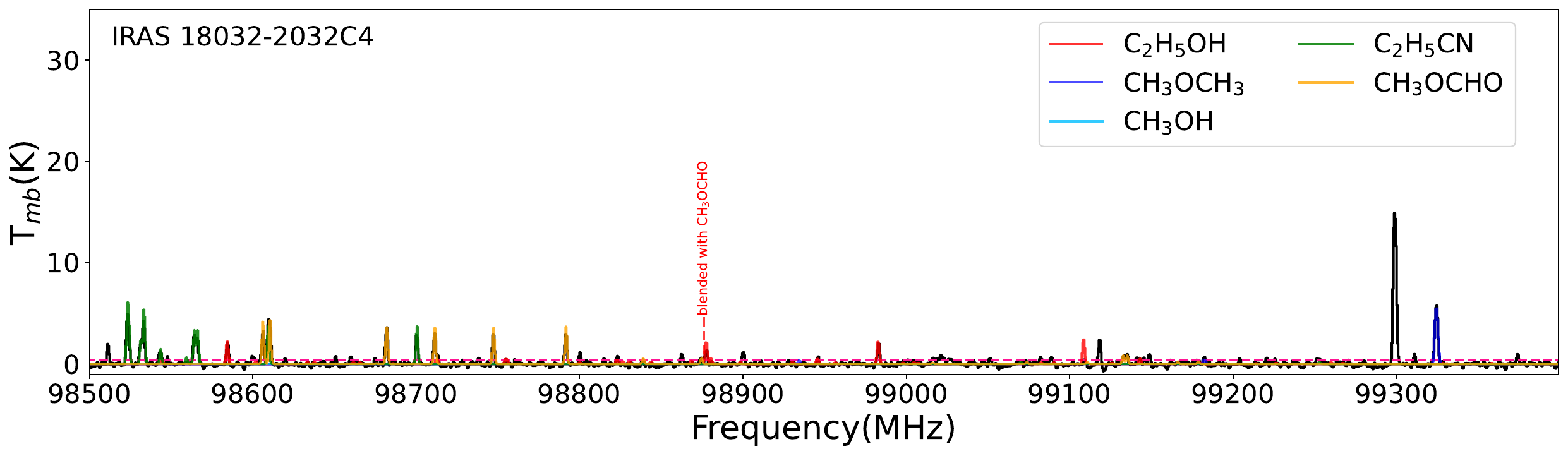}}
\quad
{\includegraphics[height=4.5cm,width=15.93cm]{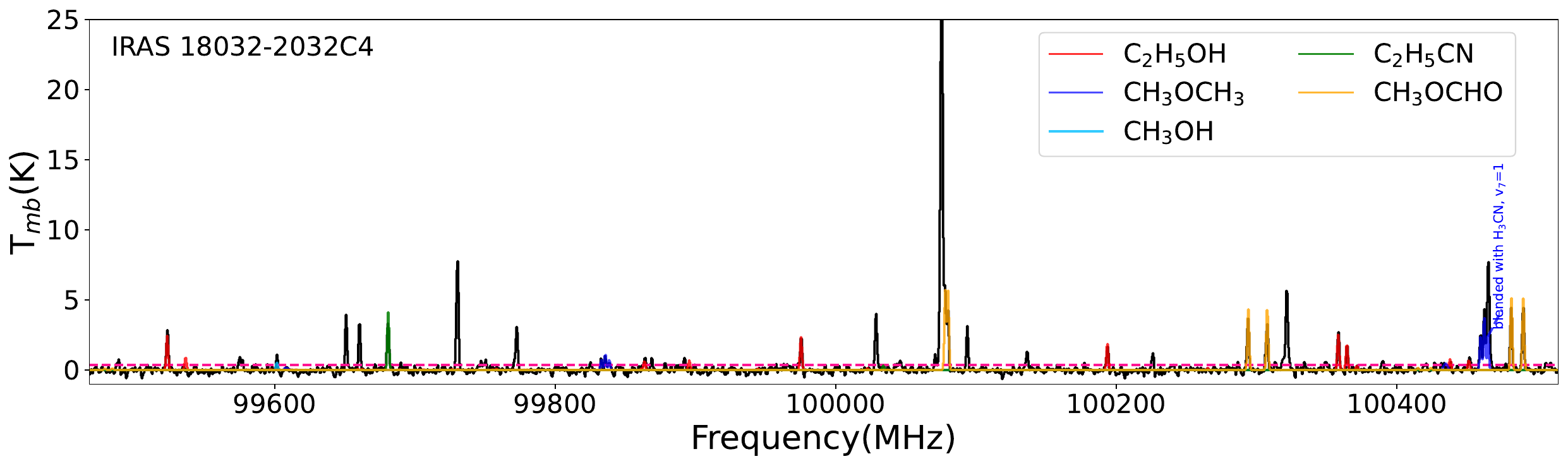}}
\caption{Continued.}
\end{figure}
\setcounter{figure}{\value{figure}-1}
\begin{figure}
  \centering 
{\includegraphics[height=4.5cm,width=15.93cm]{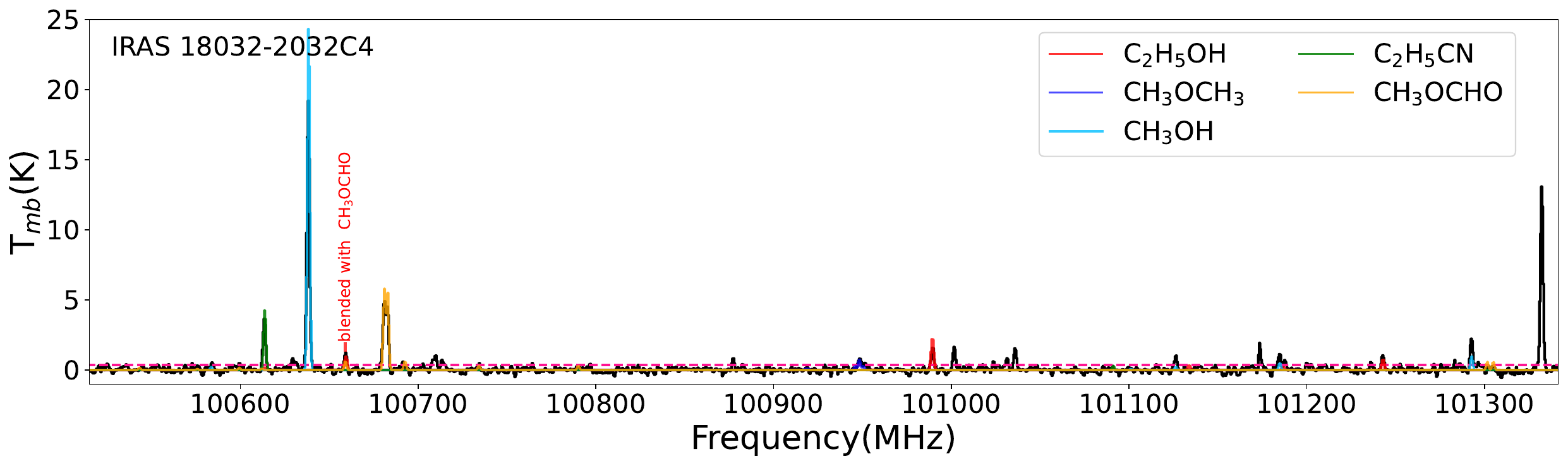}}
\quad
{\includegraphics[height=4.5cm,width=15.93cm]{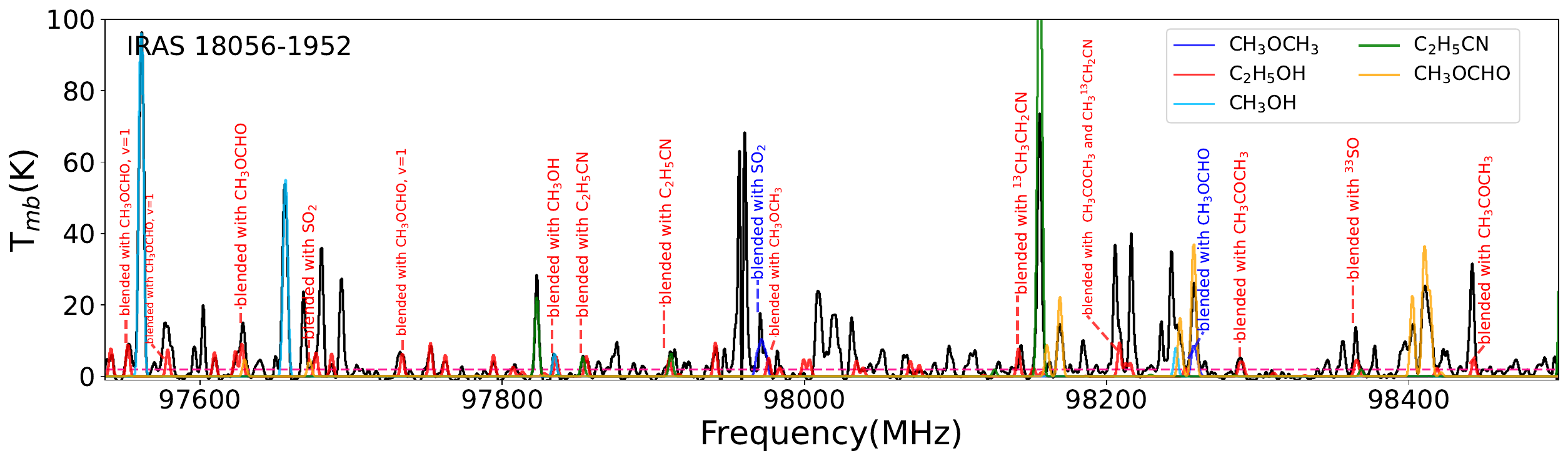}}
\quad
{\includegraphics[height=4.5cm,width=15.93cm]{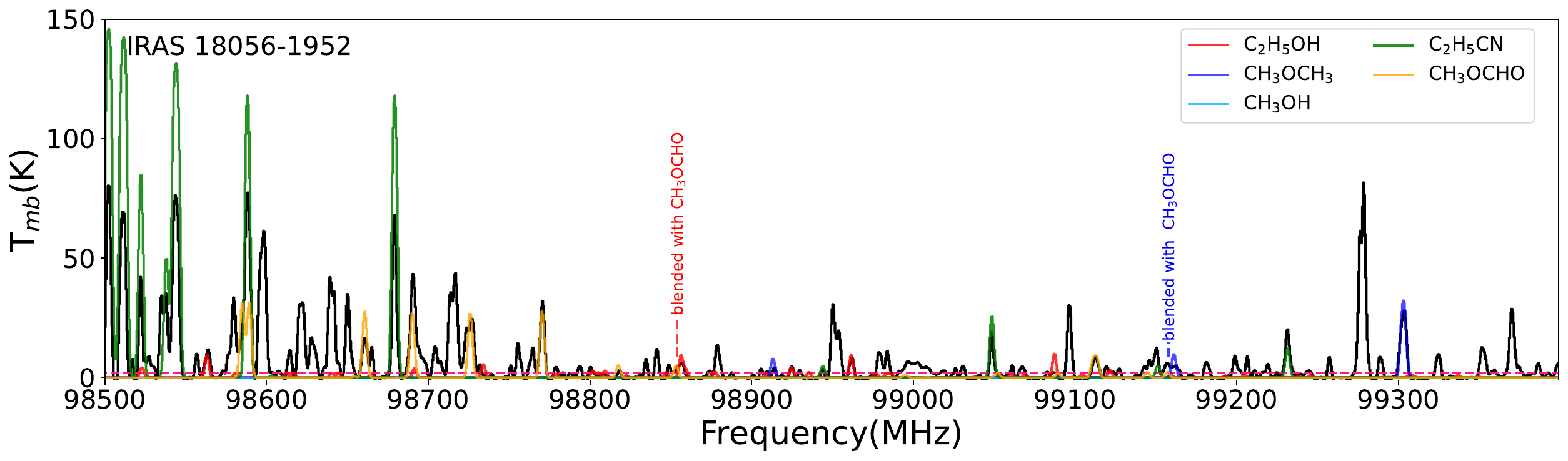}}
\quad
{\includegraphics[height=4.5cm,width=15.93cm]{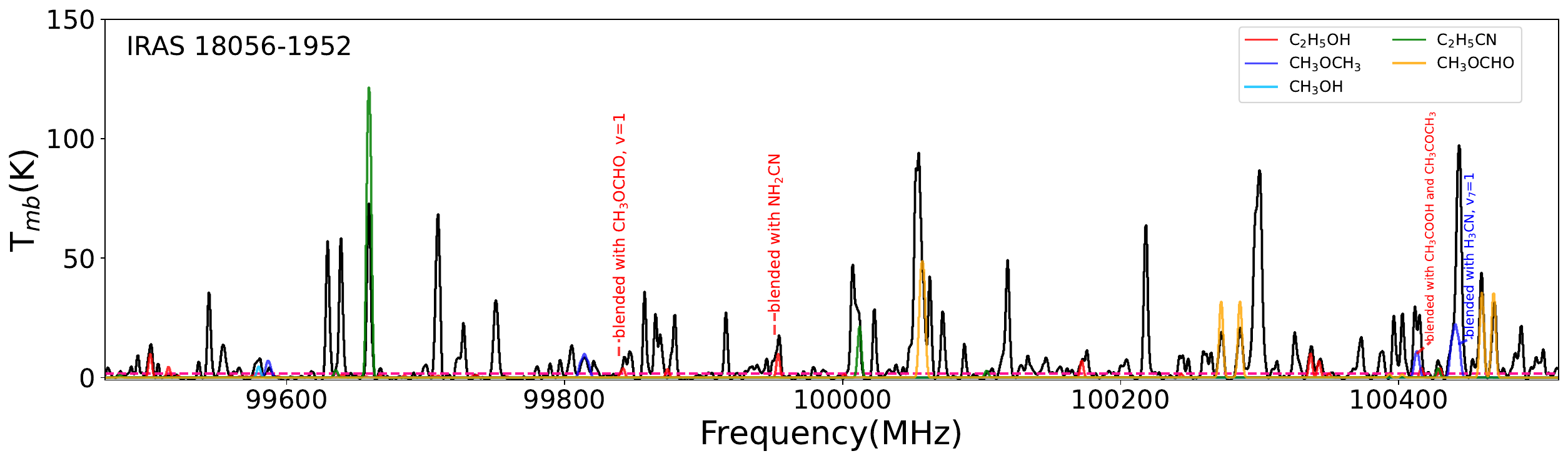}}
\quad
{\includegraphics[height=4.5cm,width=15.93cm]{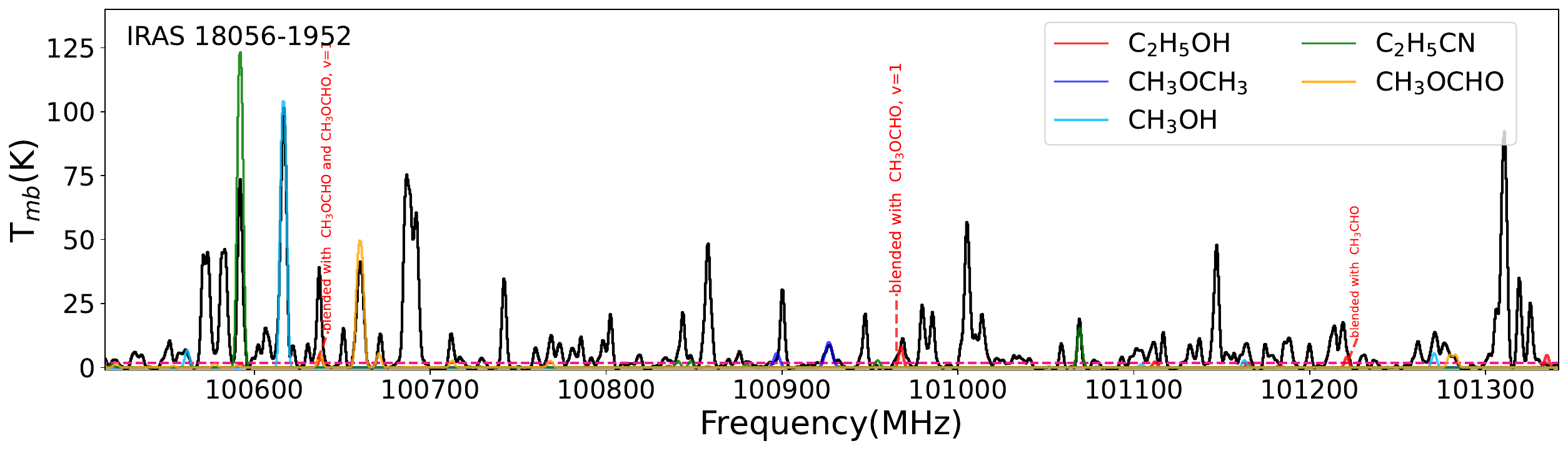}}
\caption{Continued.}
\end{figure}
\setcounter{figure}{\value{figure}-1}
\begin{figure}
  \centering 
{\includegraphics[height=4.5cm,width=15.93cm]{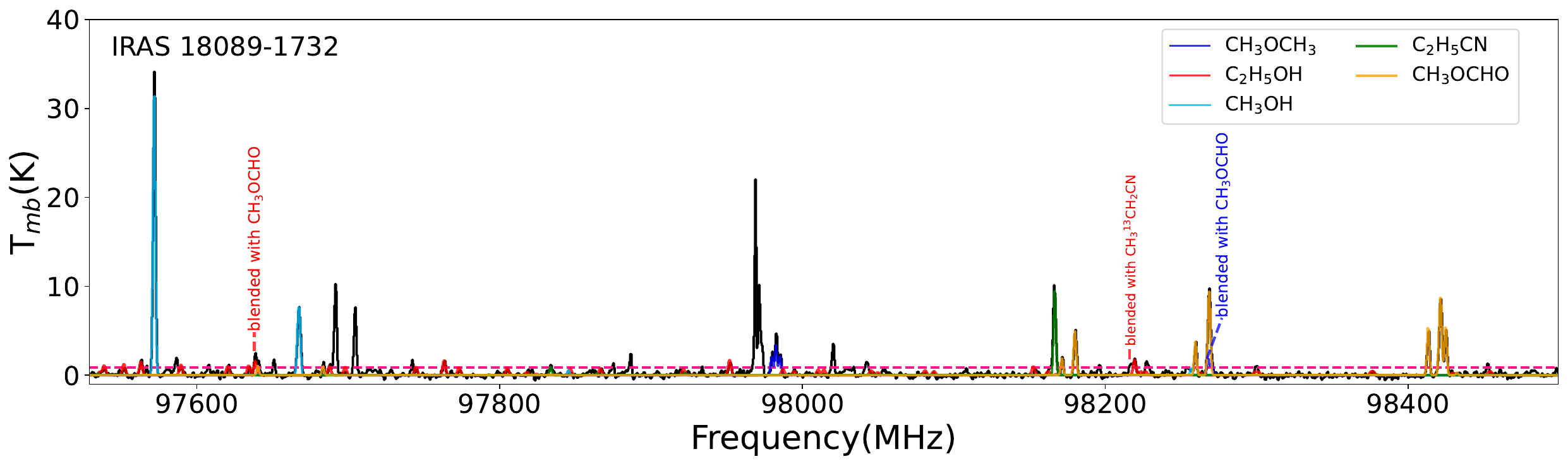}}
 \quad 
{\includegraphics[height=4.5cm,width=15.93cm]{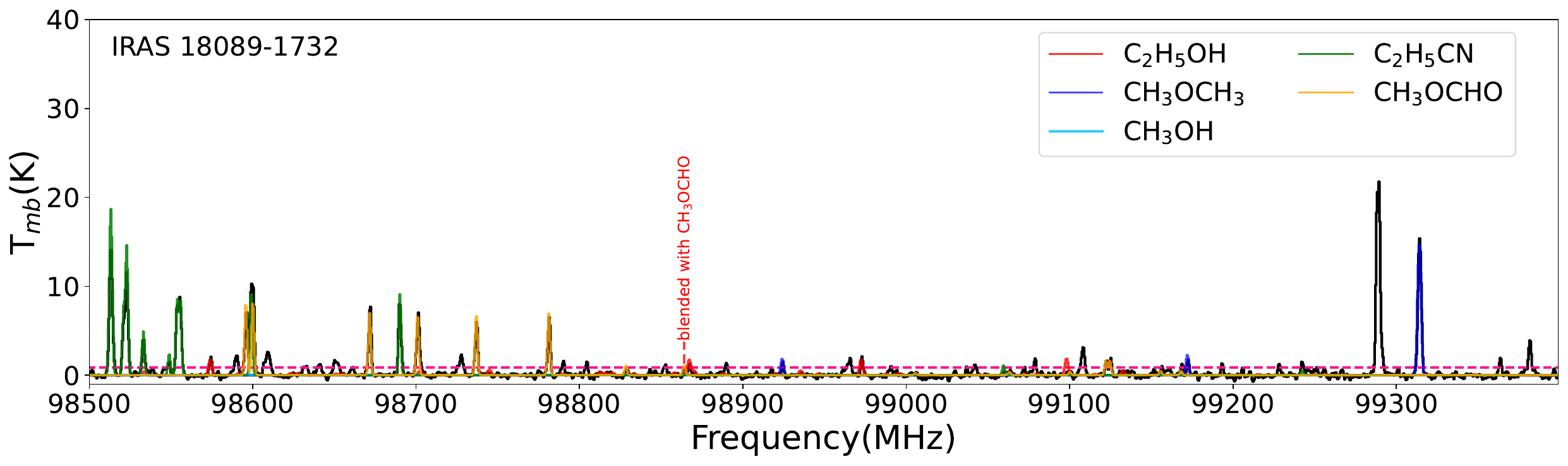}}
\quad
{\includegraphics[height=4.5cm,width=15.93cm]{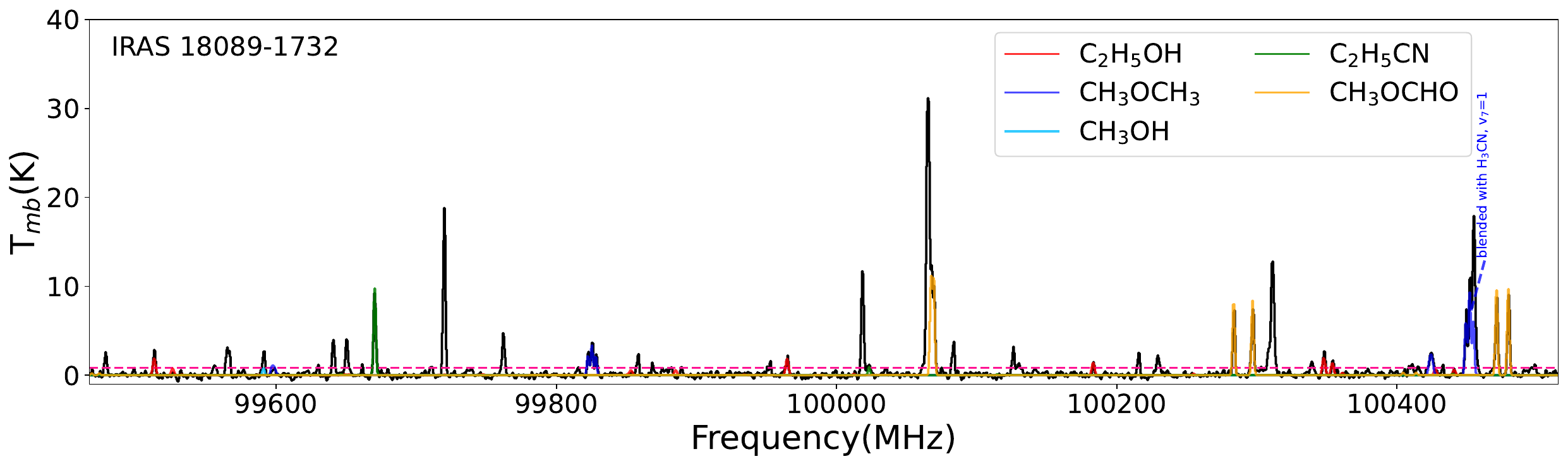}}
\quad
{\includegraphics[height=4.5cm,width=15.93cm]{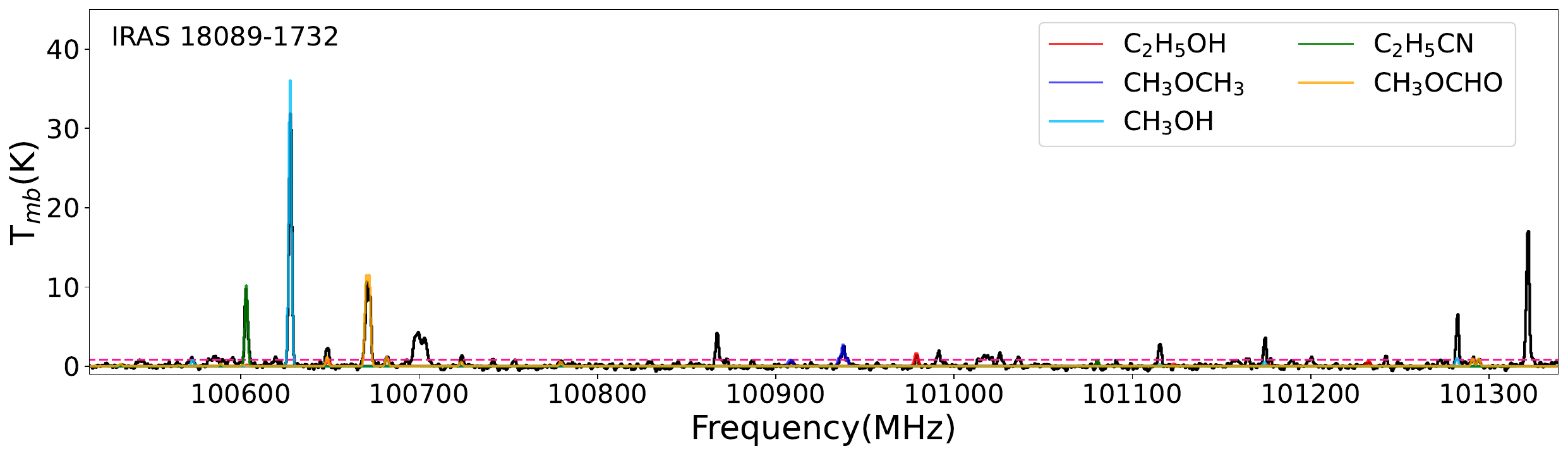}}
\quad
{\includegraphics[height=4.5cm,width=15.93cm]{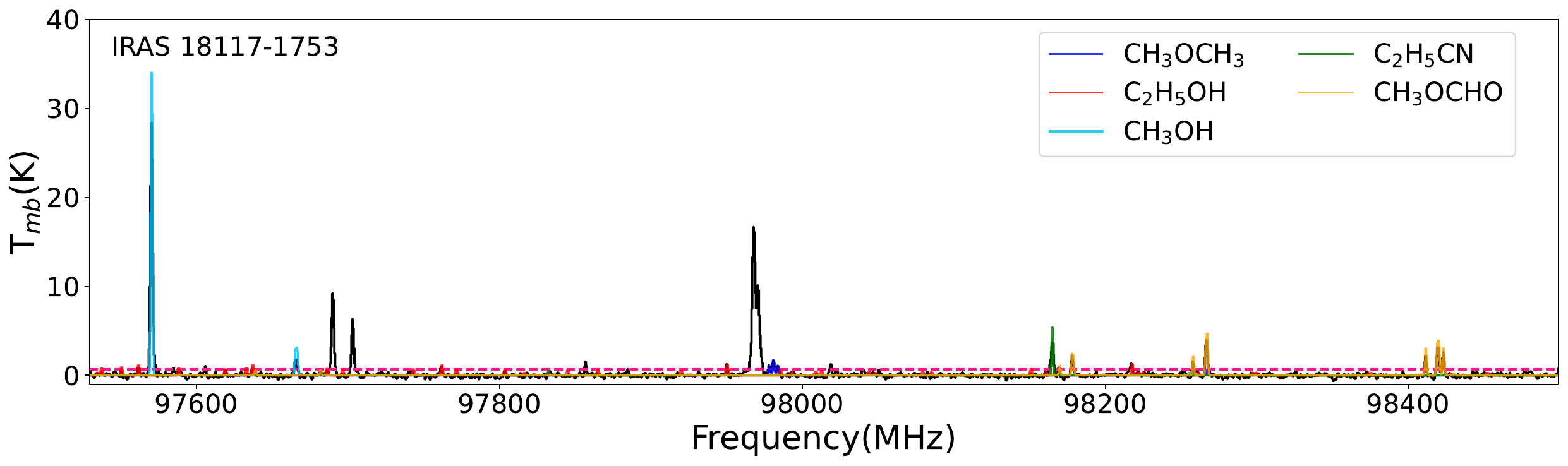}}
\caption{Continued.}
\end{figure}
\setcounter{figure}{\value{figure}-1}
\begin{figure}
  \centering 
{\includegraphics[height=4.5cm,width=15.93cm]{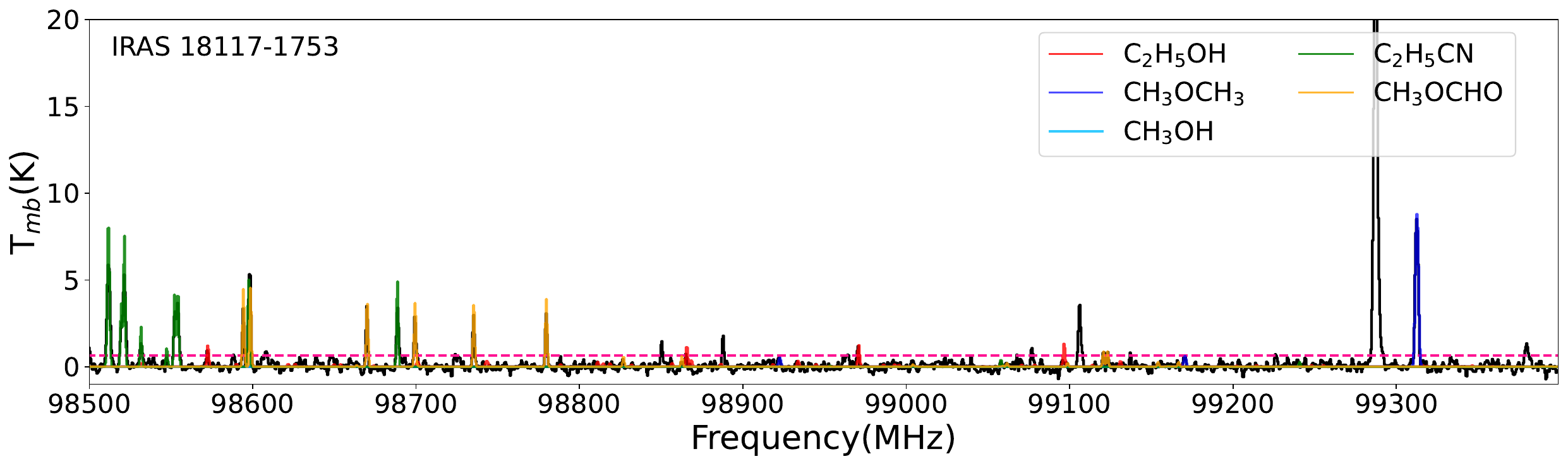}}
 \quad 
{\includegraphics[height=4.5cm,width=15.93cm]{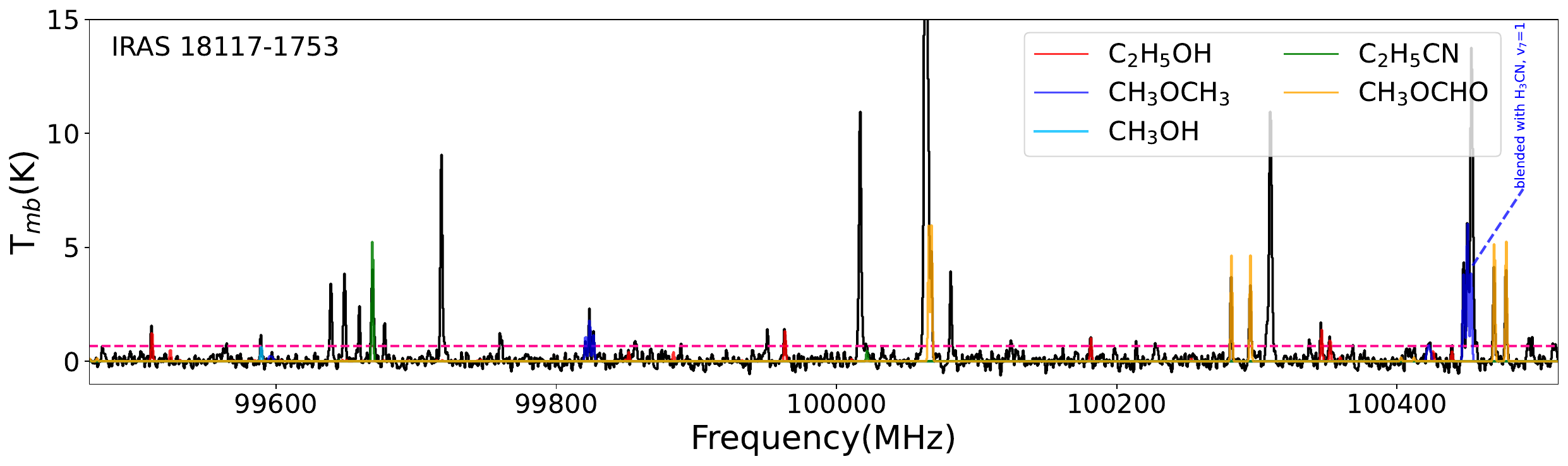}}
\quad
{\includegraphics[height=4.5cm,width=15.93cm]{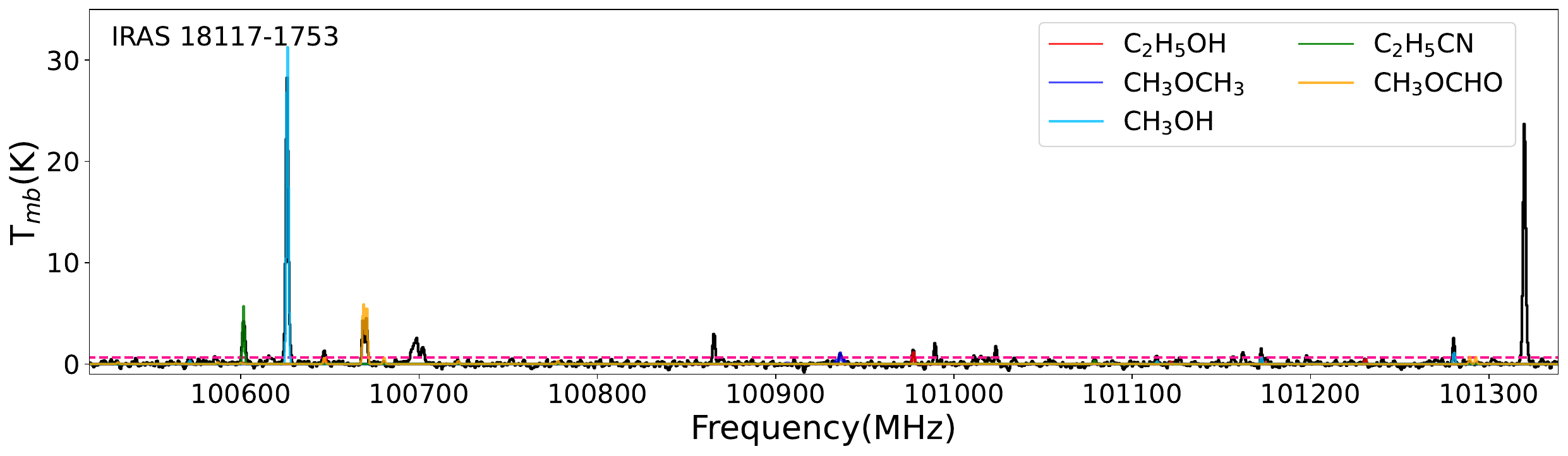}}
\quad
{\includegraphics[height=4.5cm,width=15.93cm]{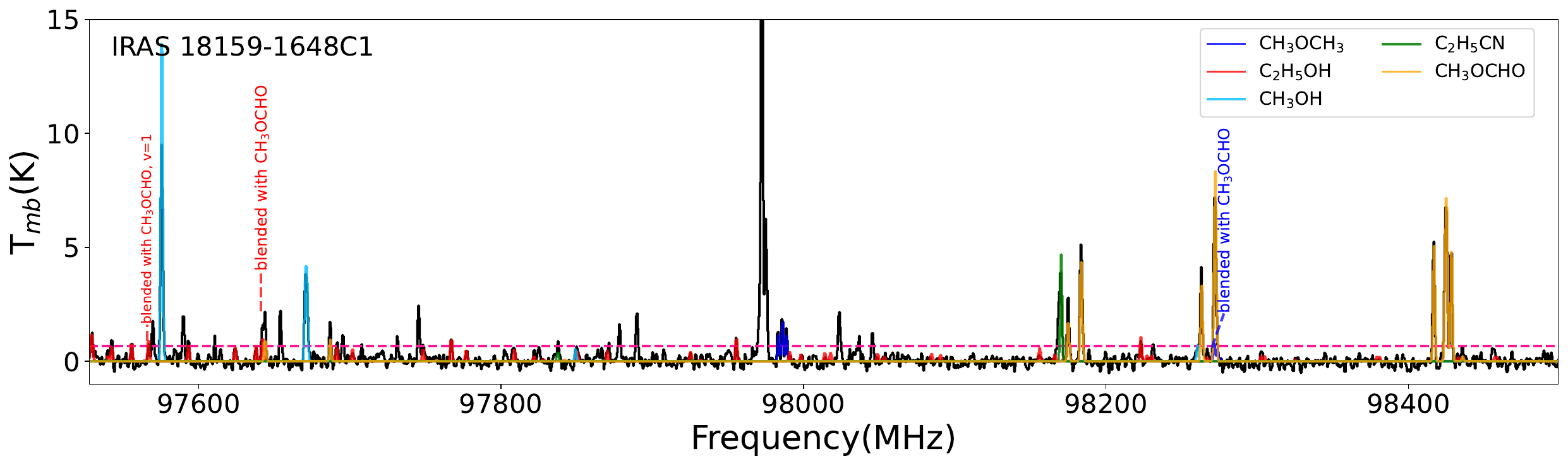}}
\quad
{\includegraphics[height=4.5cm,width=15.93cm]{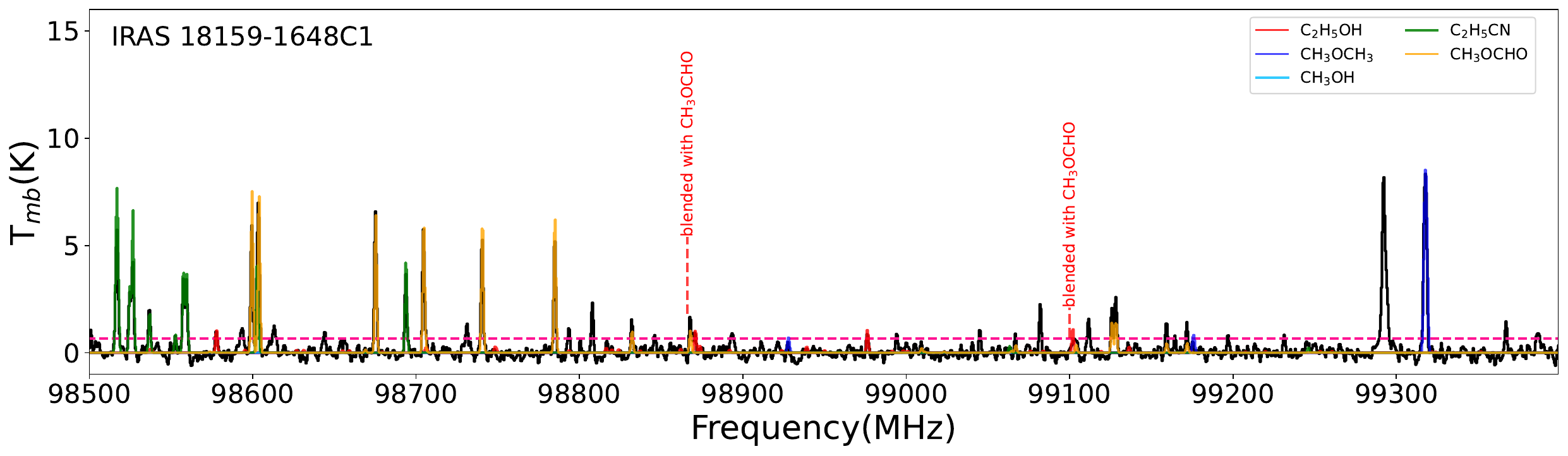}}
\caption{Continued.}
\end{figure}
\setcounter{figure}{\value{figure}-1}
\begin{figure}
  \centering 
{\includegraphics[height=4.5cm,width=15.93cm]{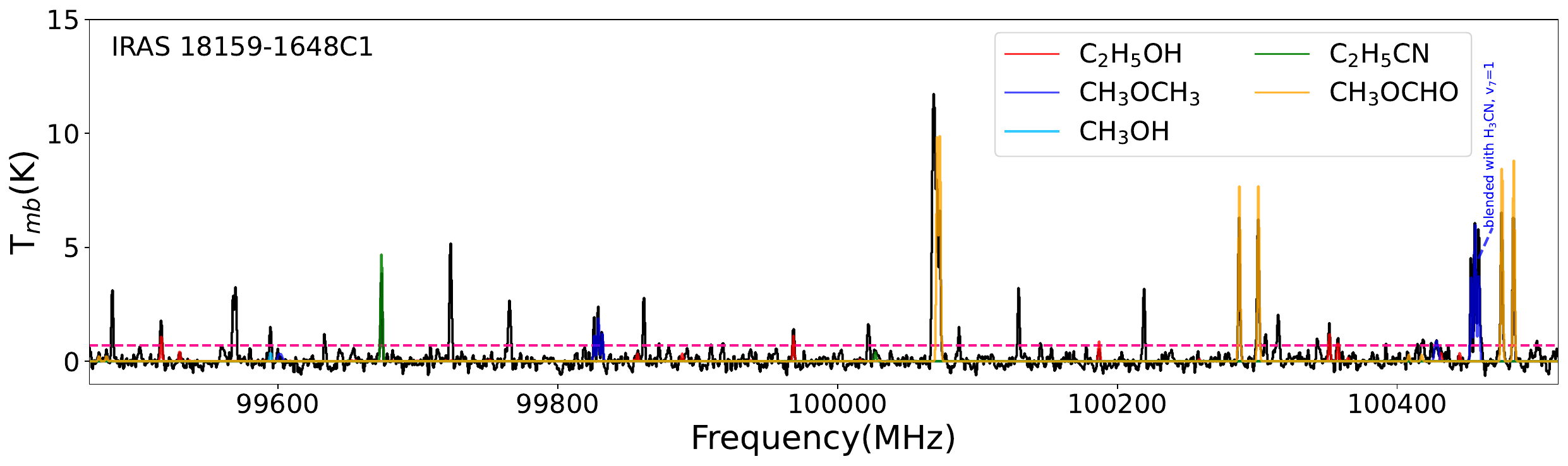}}
 \quad 
{\includegraphics[height=4.5cm,width=15.93cm]{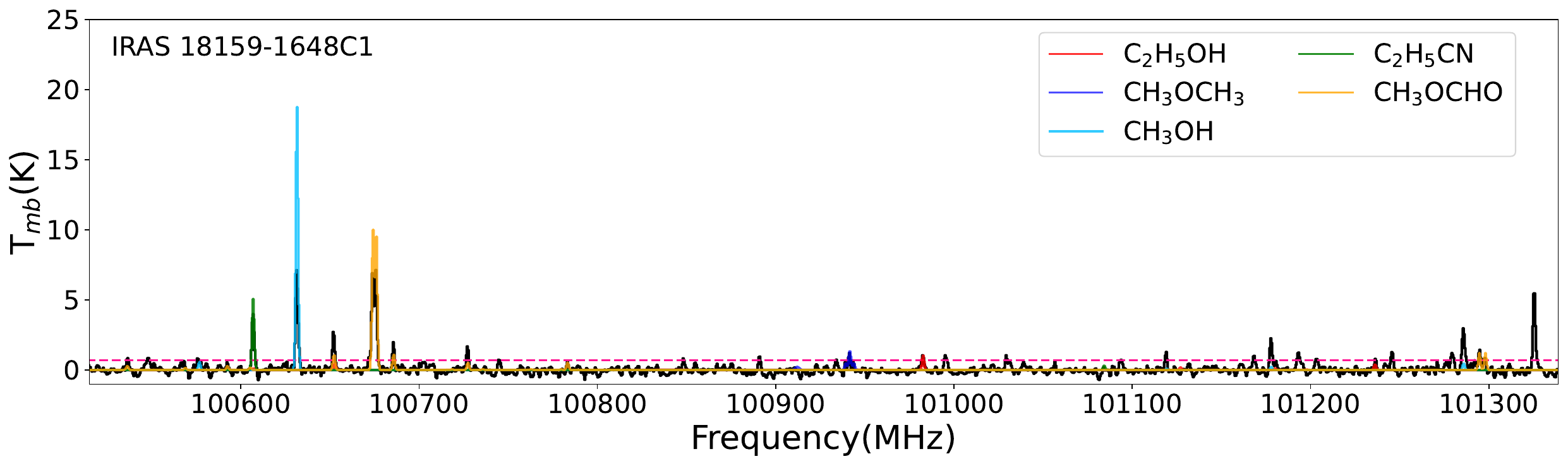}}
\quad
{\includegraphics[height=4.5cm,width=15.93cm]{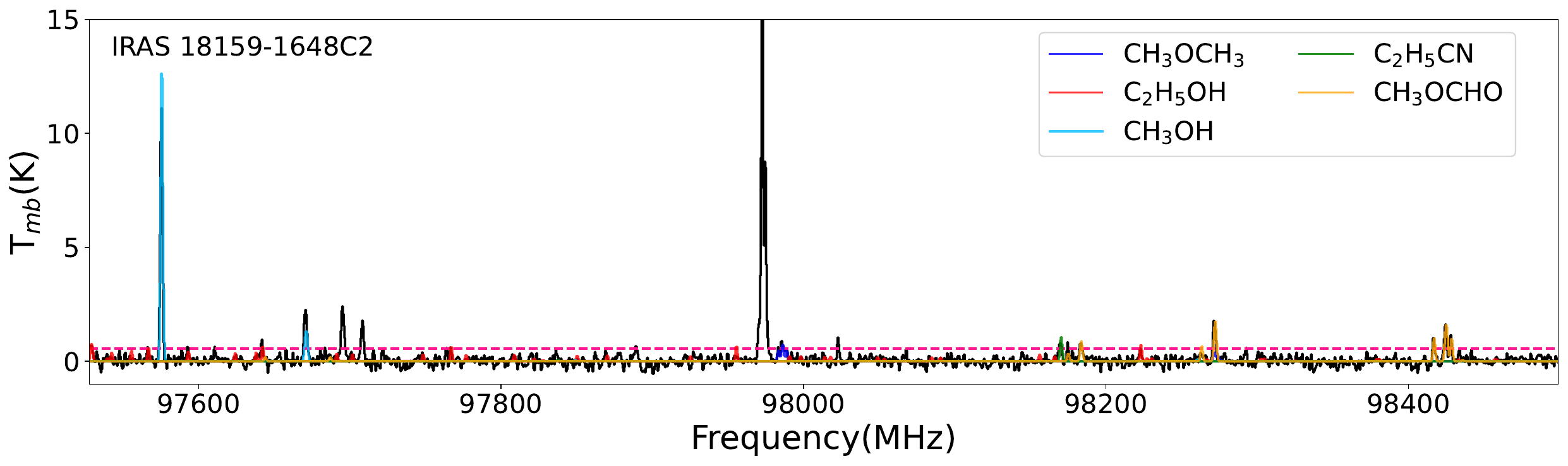}}
\quad
{\includegraphics[height=4.5cm,width=15.93cm]{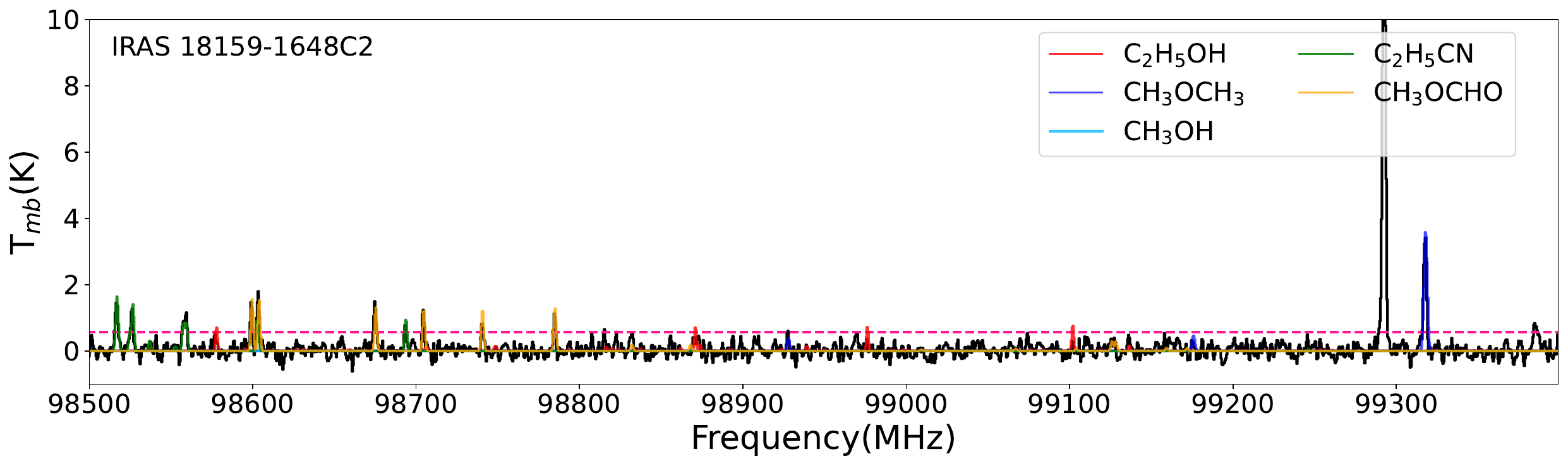}}
\quad
{\includegraphics[height=4.5cm,width=15.93cm]{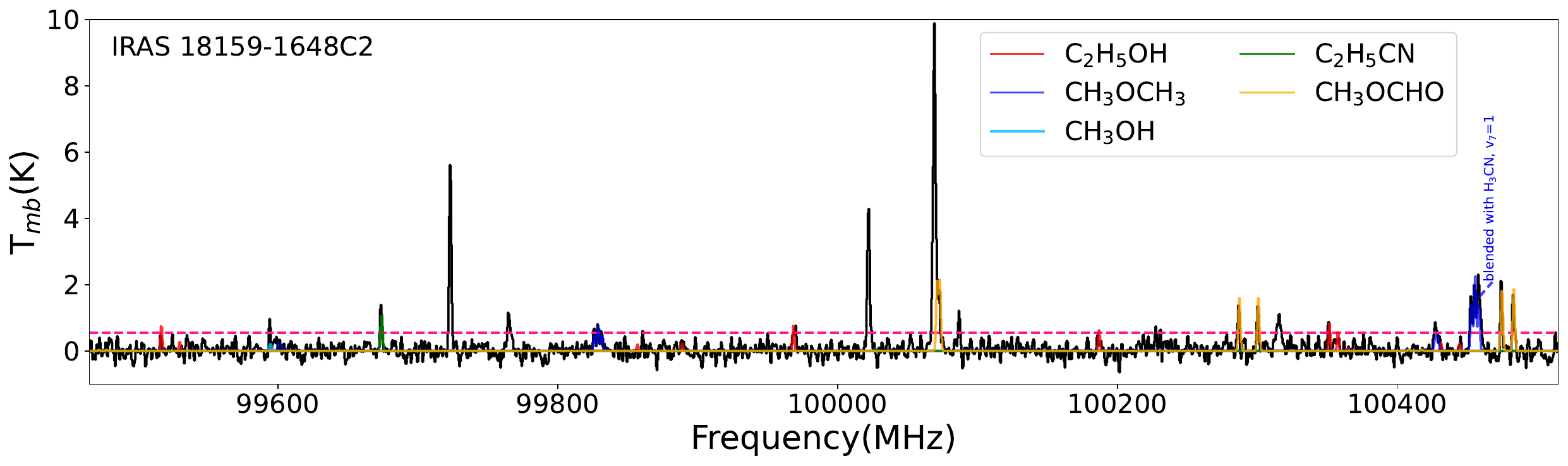}}
\caption{Continued.}
\end{figure}
\setcounter{figure}{\value{figure}-1}
\begin{figure}
  \centering 
{\includegraphics[height=4.5cm,width=15.93cm]{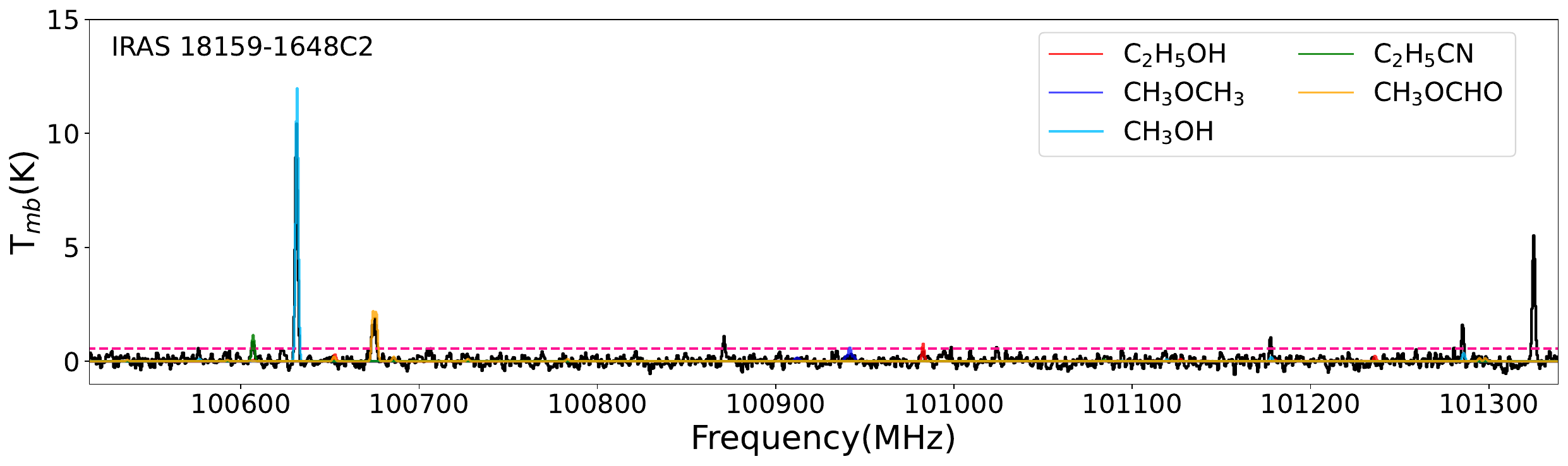}}
 \quad 
{\includegraphics[height=4.5cm,width=15.93cm]{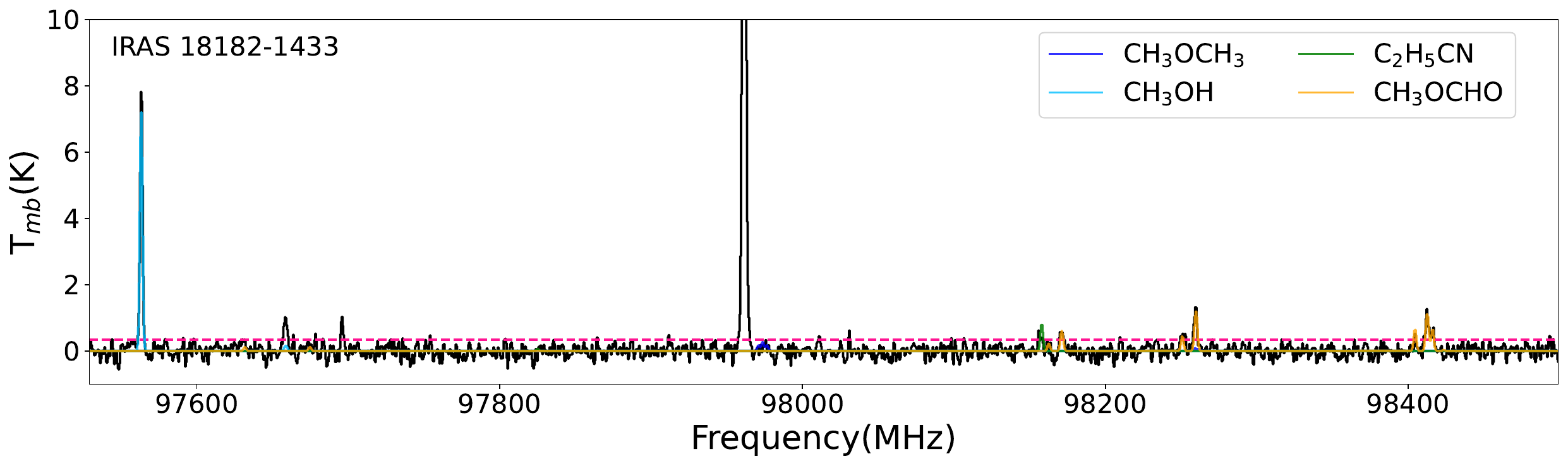}}
\quad
{\includegraphics[height=4.5cm,width=15.93cm]{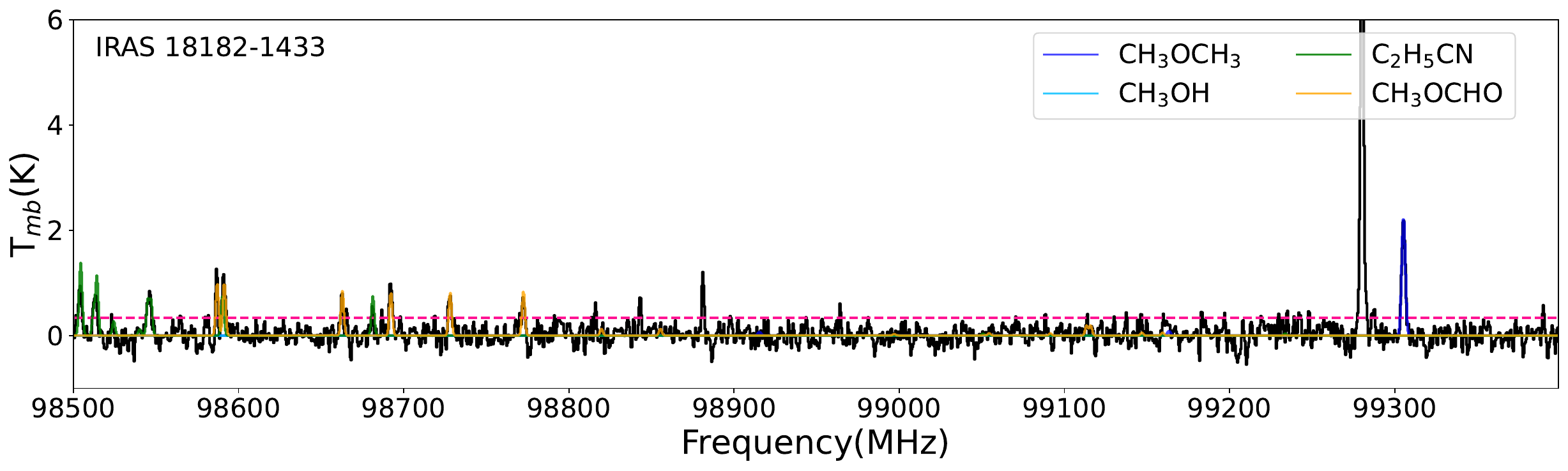}}
\quad
{\includegraphics[height=4.5cm,width=15.93cm]{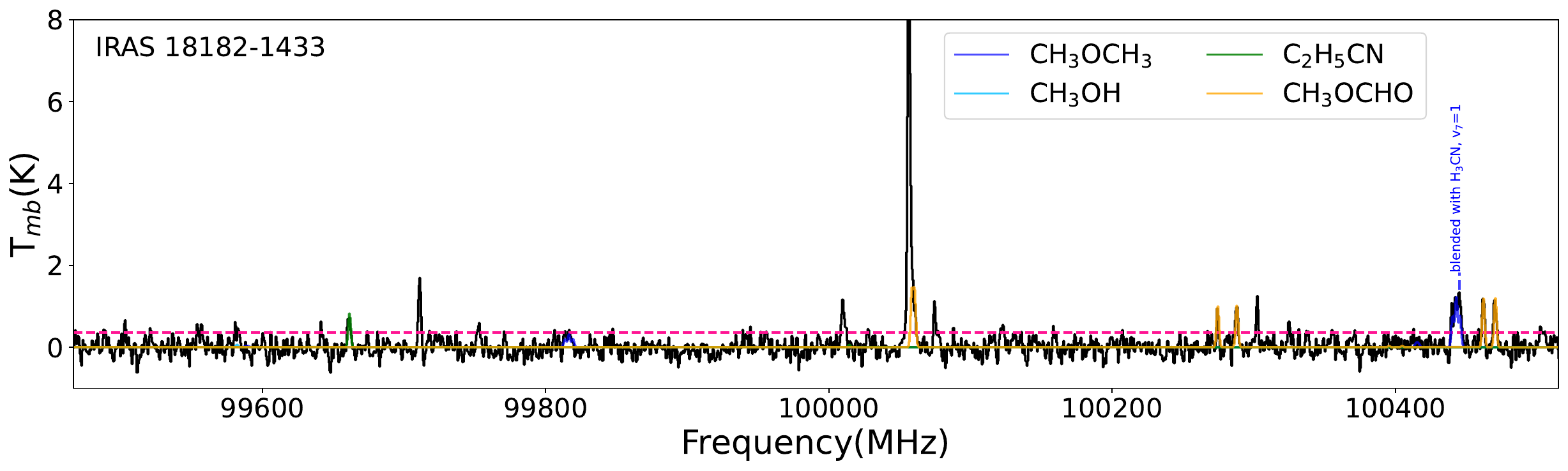}}
\quad
{\includegraphics[height=4.5cm,width=15.93cm]{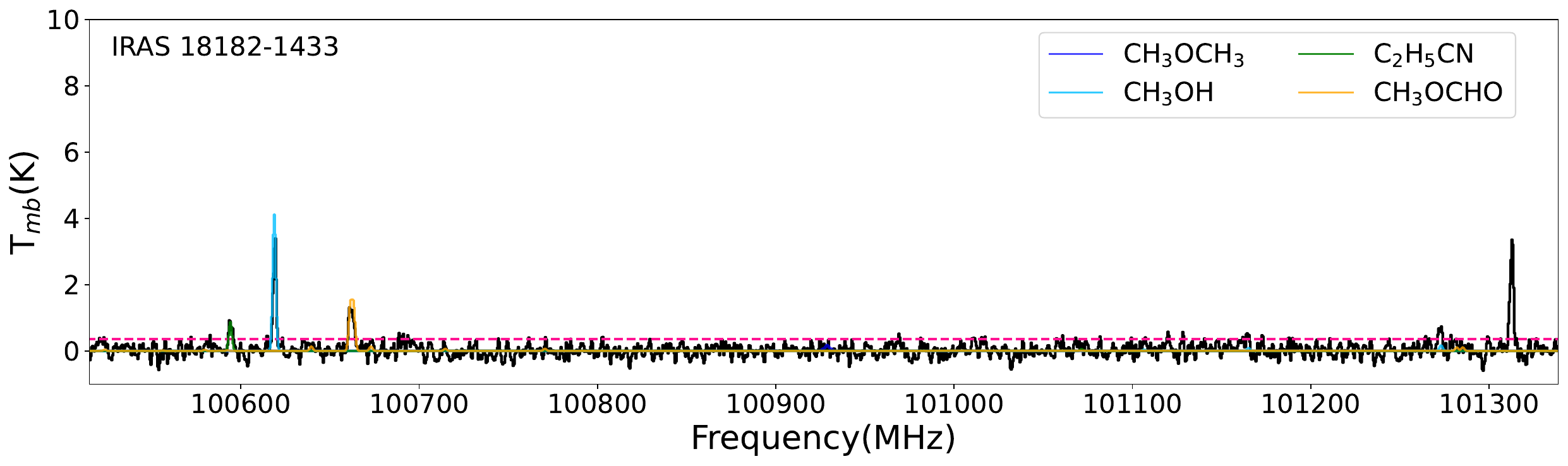}}
\caption{Continued.}
\end{figure}
\setcounter{figure}{\value{figure}-1}
\begin{figure}
\centering 
{\includegraphics[height=4.5cm,width=15.93cm]{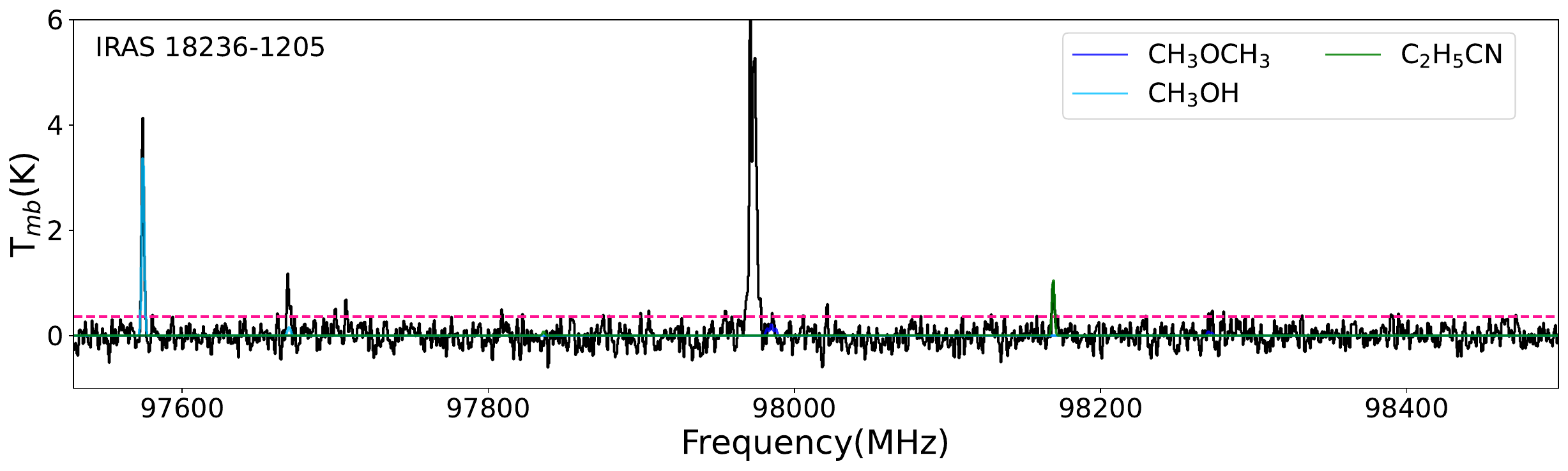}}
 \quad 
{\includegraphics[height=4.5cm,width=15.93cm]{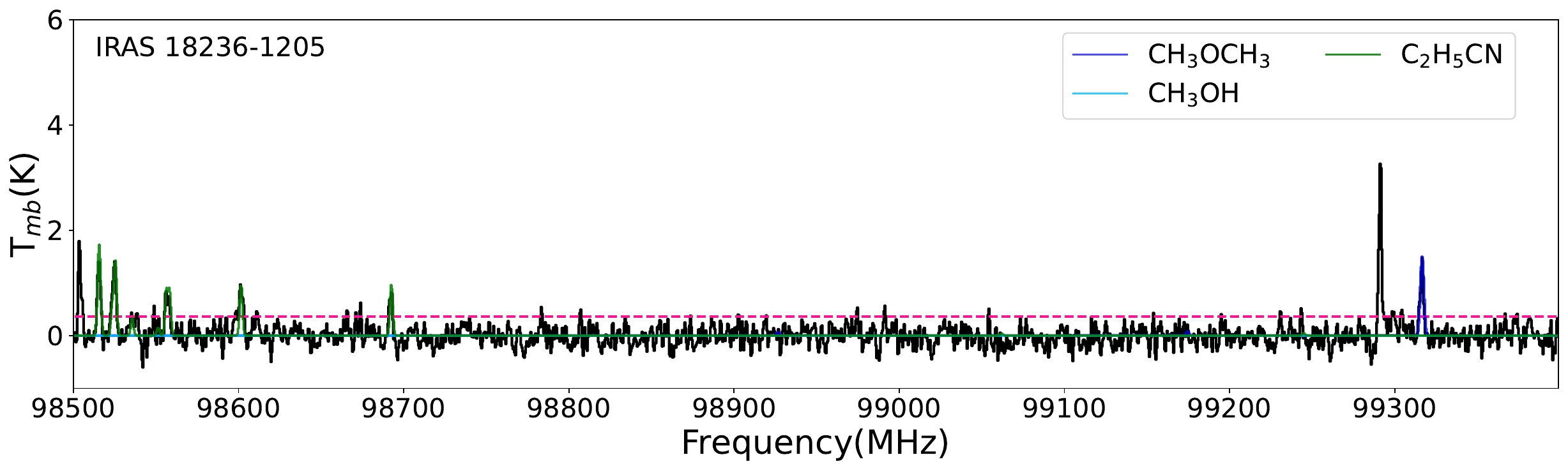}}
\quad
{\includegraphics[height=4.5cm,width=15.93cm]{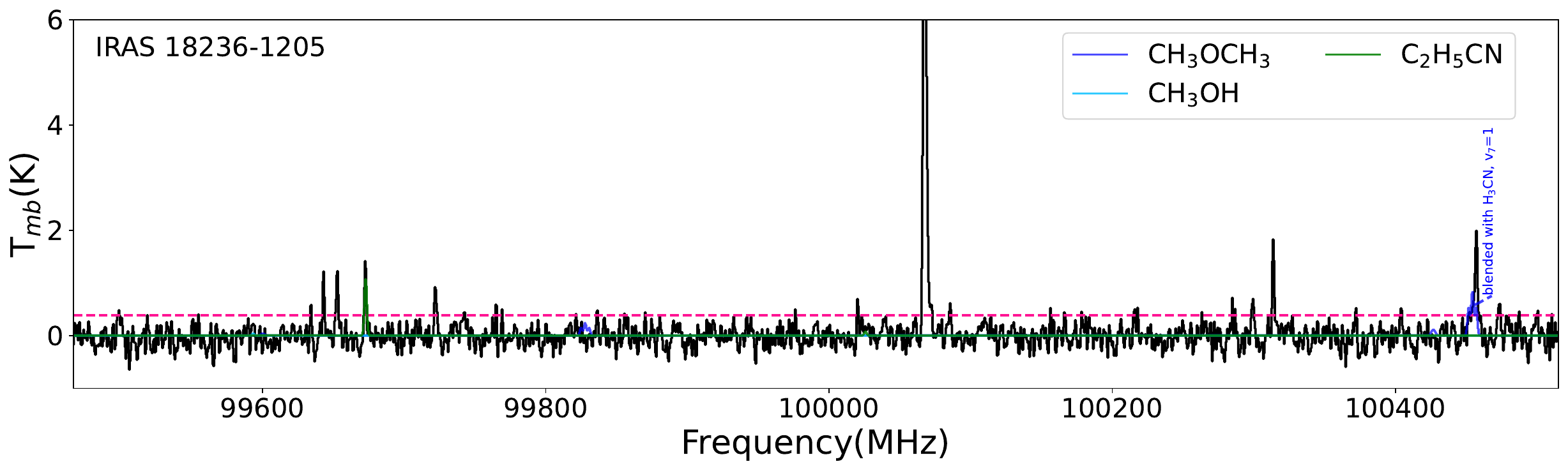}}
\quad
{\includegraphics[height=4.5cm,width=15.93cm]{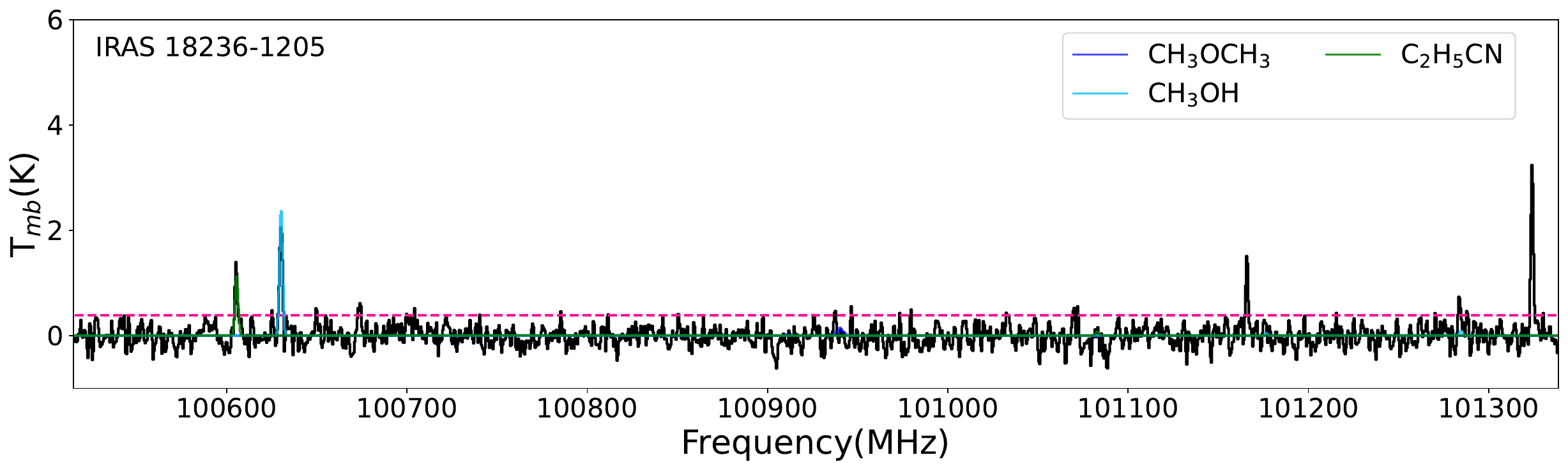}}
\quad
{\includegraphics[height=4.5cm,width=15.93cm]{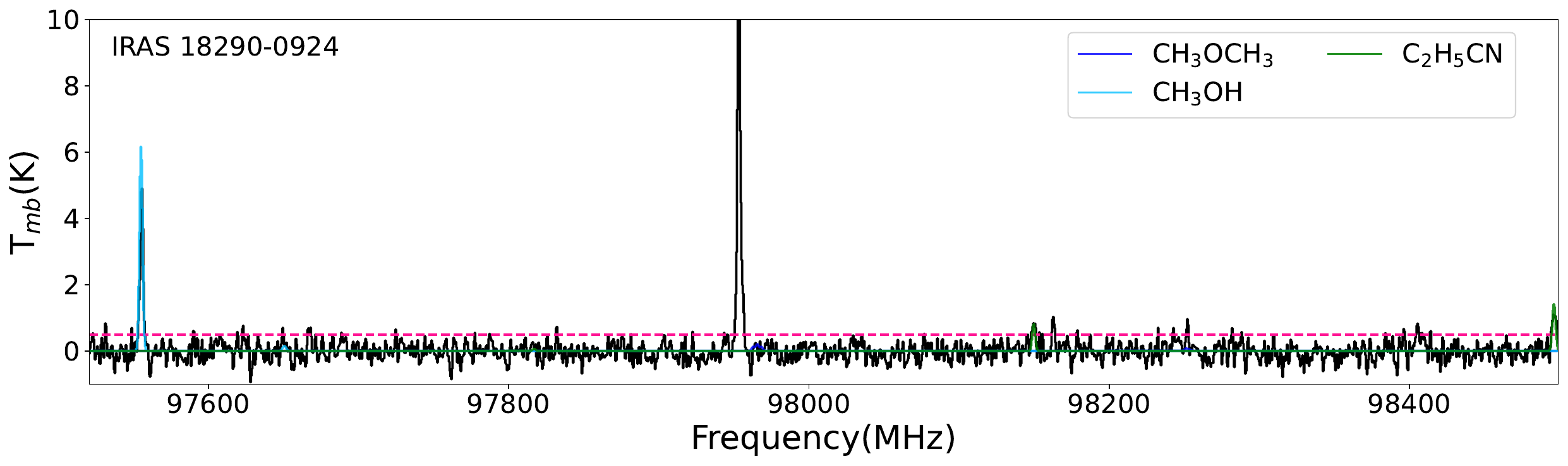}}
\caption{Continued.}
\end{figure}
\setcounter{figure}{\value{figure}-1}
\begin{figure}
  \centering 
{\includegraphics[height=4.5cm,width=15.93cm]{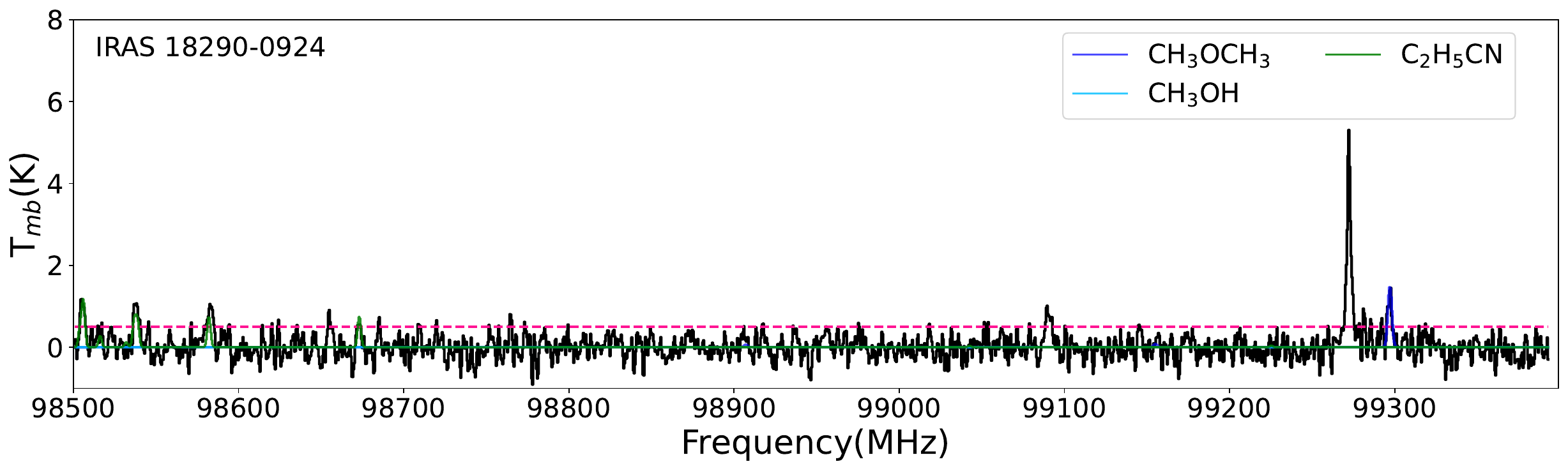}}
 \quad 
{\includegraphics[height=4.5cm,width=15.93cm]{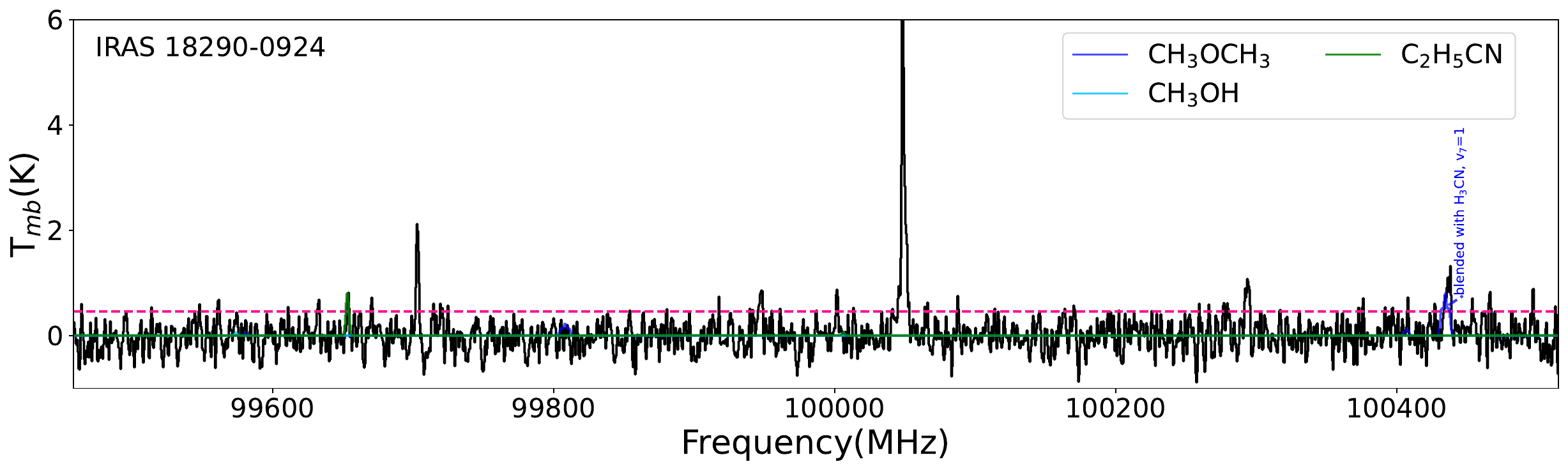}}
\quad
{\includegraphics[height=4.5cm,width=15.93cm]{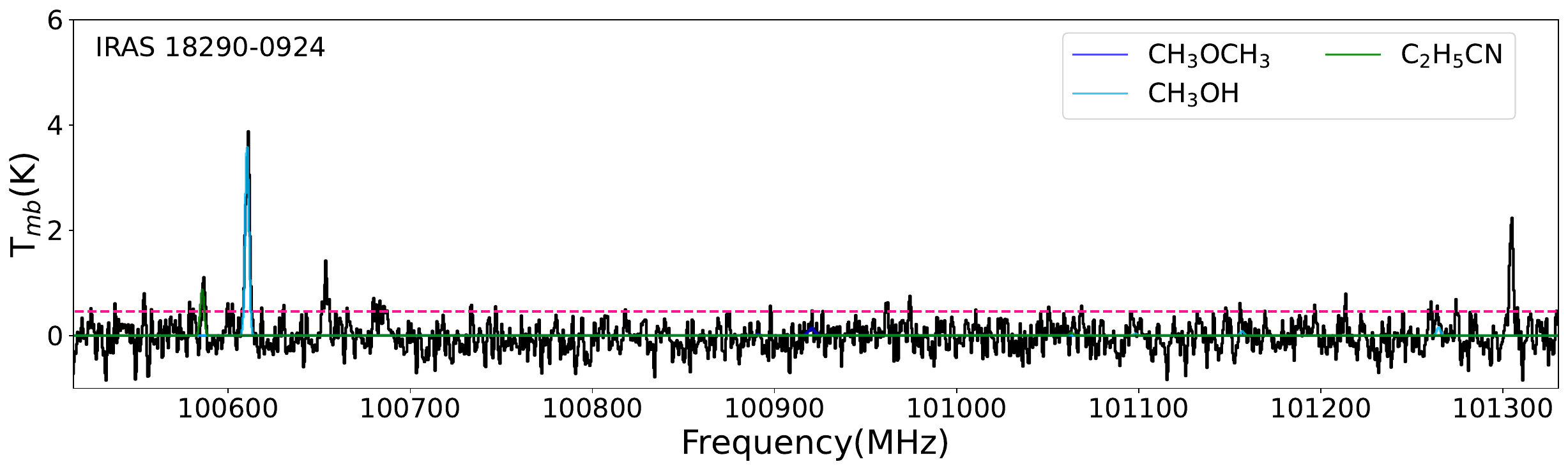}}
\quad
{\includegraphics[height=4.5cm,width=15.93cm]{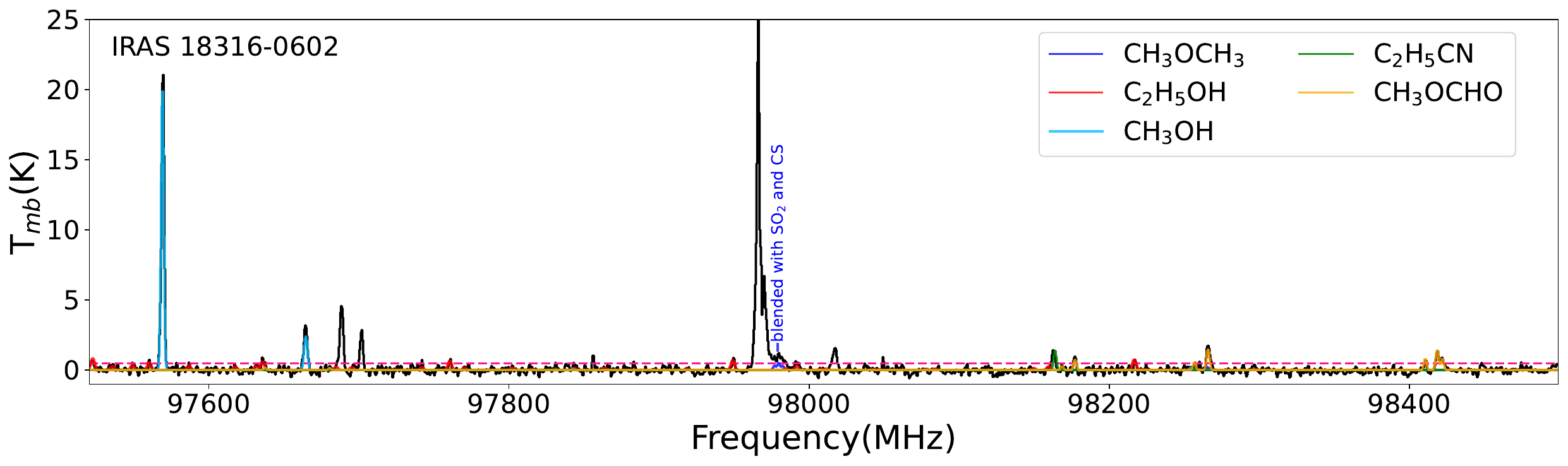}}
\quad
{\includegraphics[height=4.5cm,width=15.93cm]{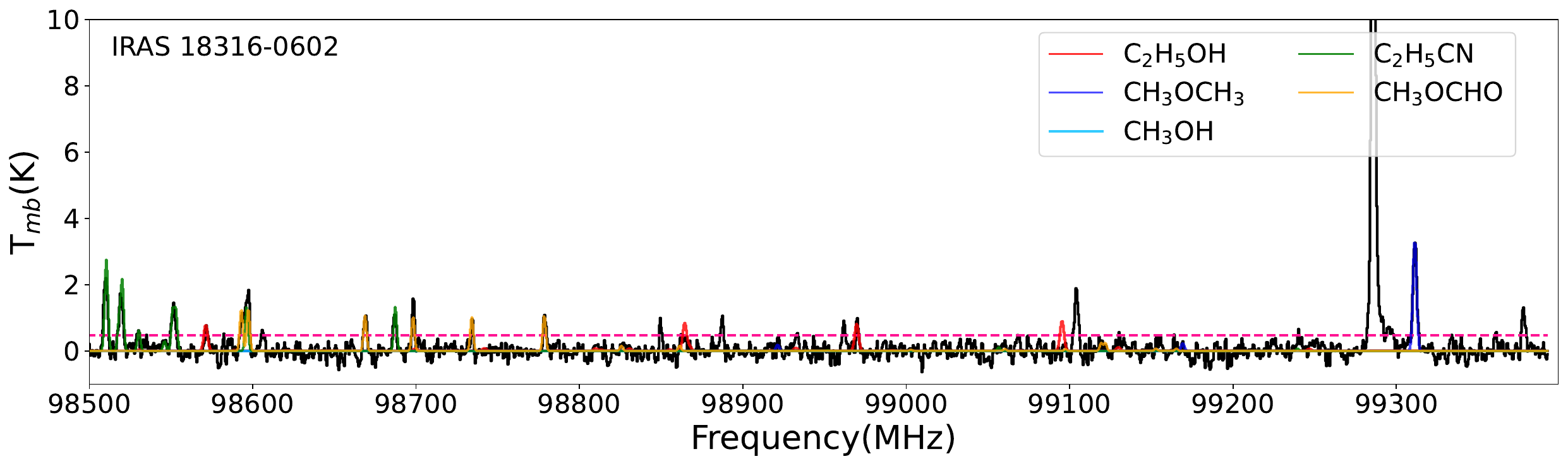}}
\caption{Continued.}
\end{figure}
\setcounter{figure}{\value{figure}-1}
\begin{figure}
\centering 
{\includegraphics[height=4.5cm,width=15.93cm]{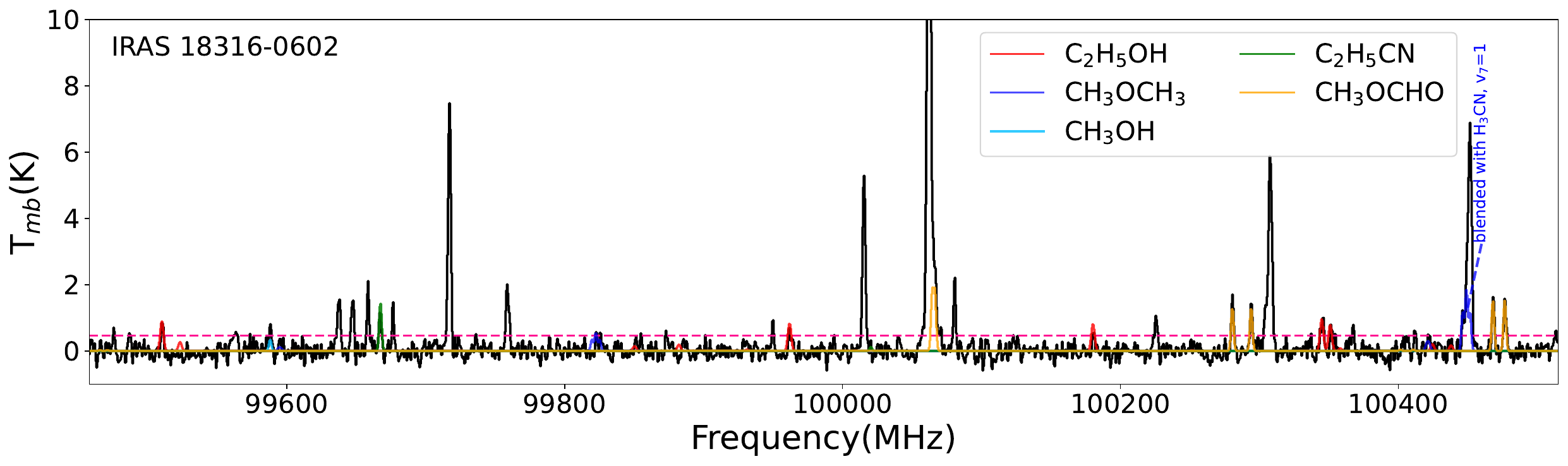}}
 \quad 
{\includegraphics[height=4.5cm,width=15.93cm]{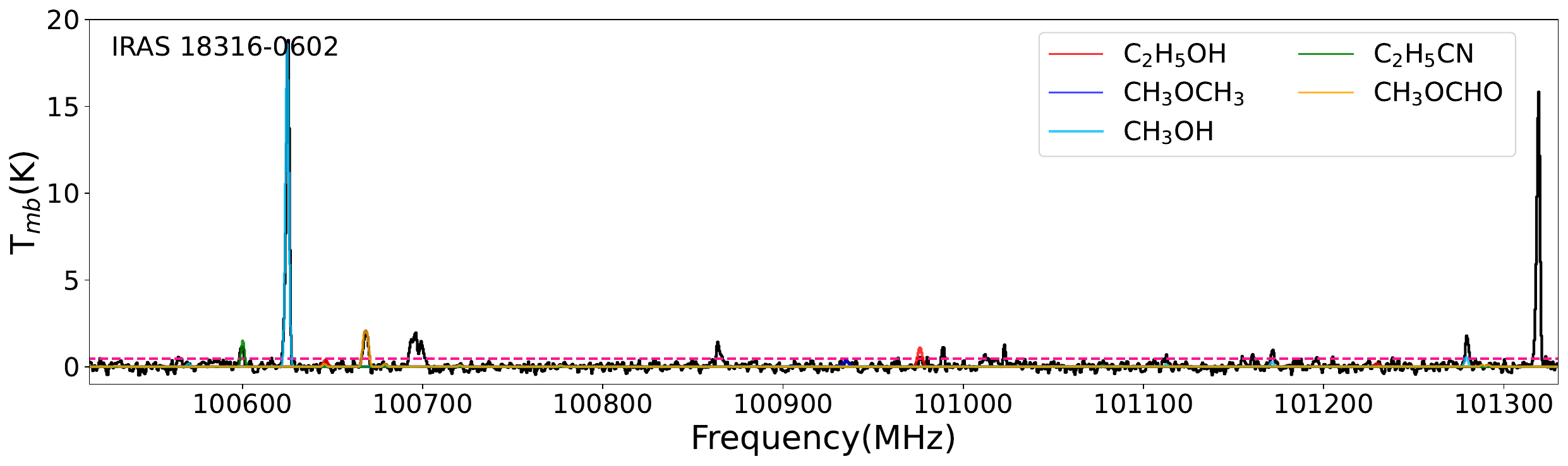}}
\quad
{\includegraphics[height=4.5cm,width=15.93cm]{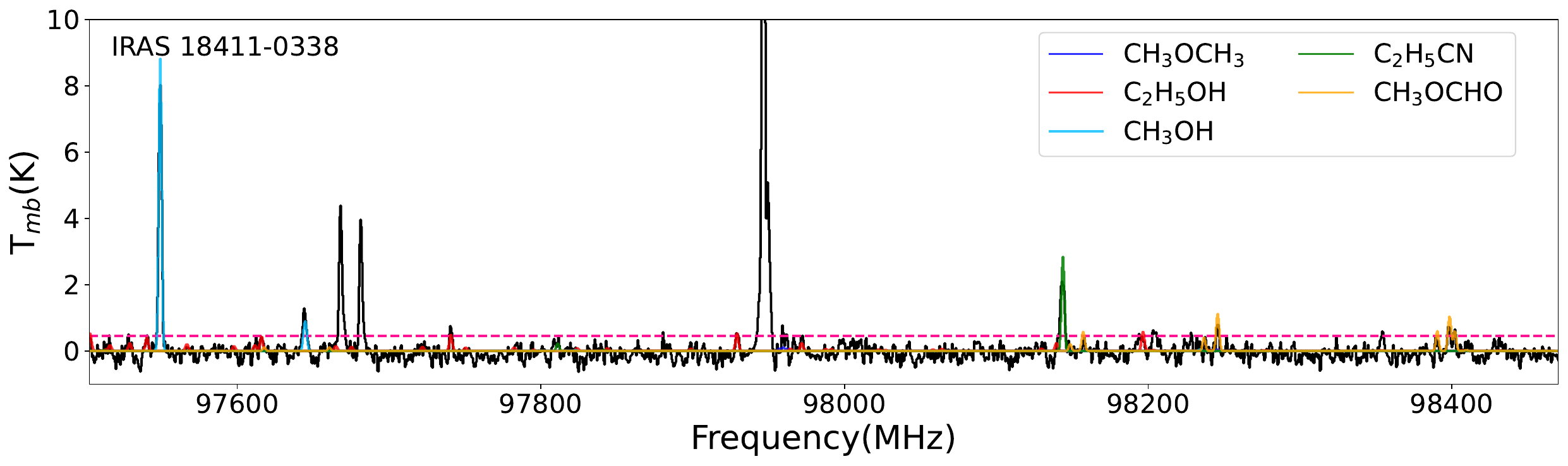}}
\quad
{\includegraphics[height=4.5cm,width=15.93cm]{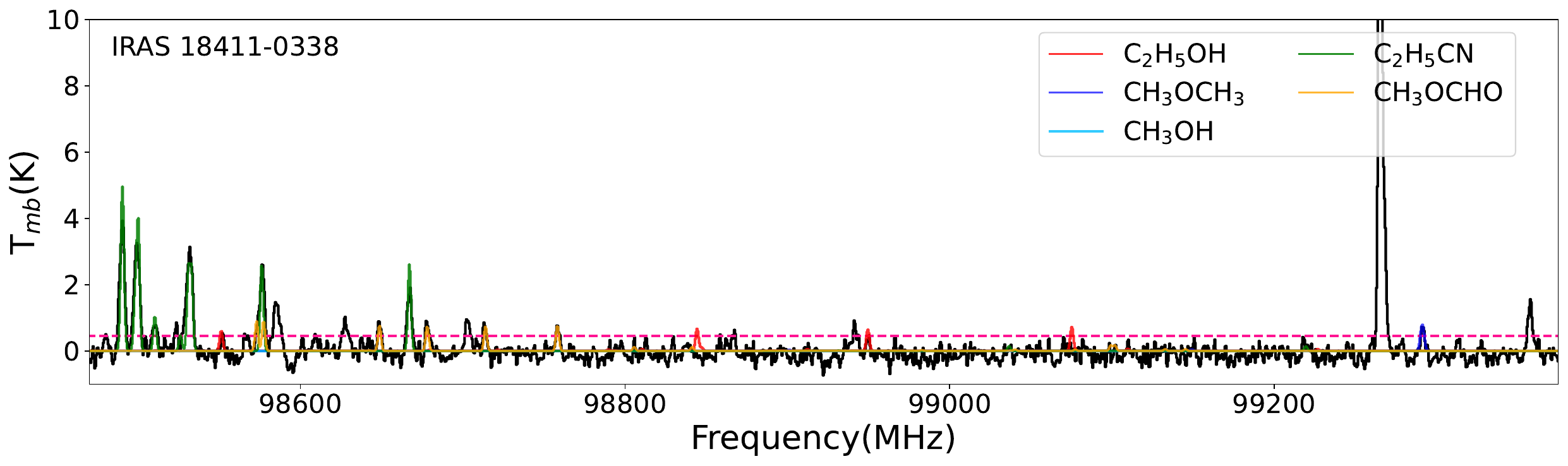}}
\quad
{\includegraphics[height=4.5cm,width=15.93cm]{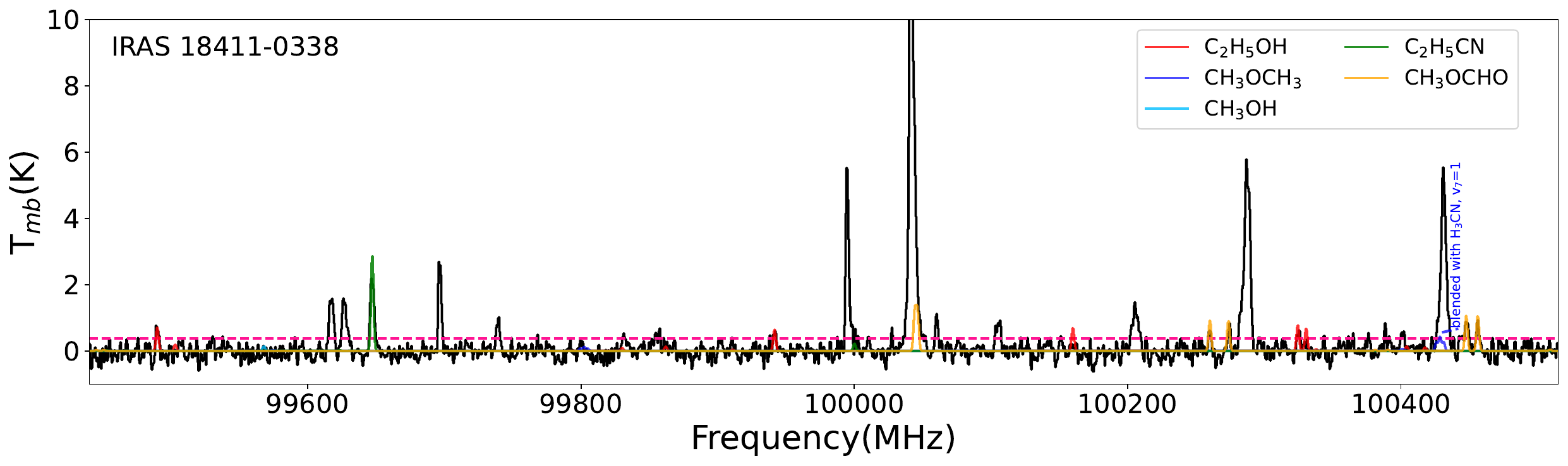}}
\caption{Continued.}
\end{figure}
\setcounter{figure}{\value{figure}-1}
\begin{figure}
  \centering 
{\includegraphics[height=4.5cm,width=15.93cm]{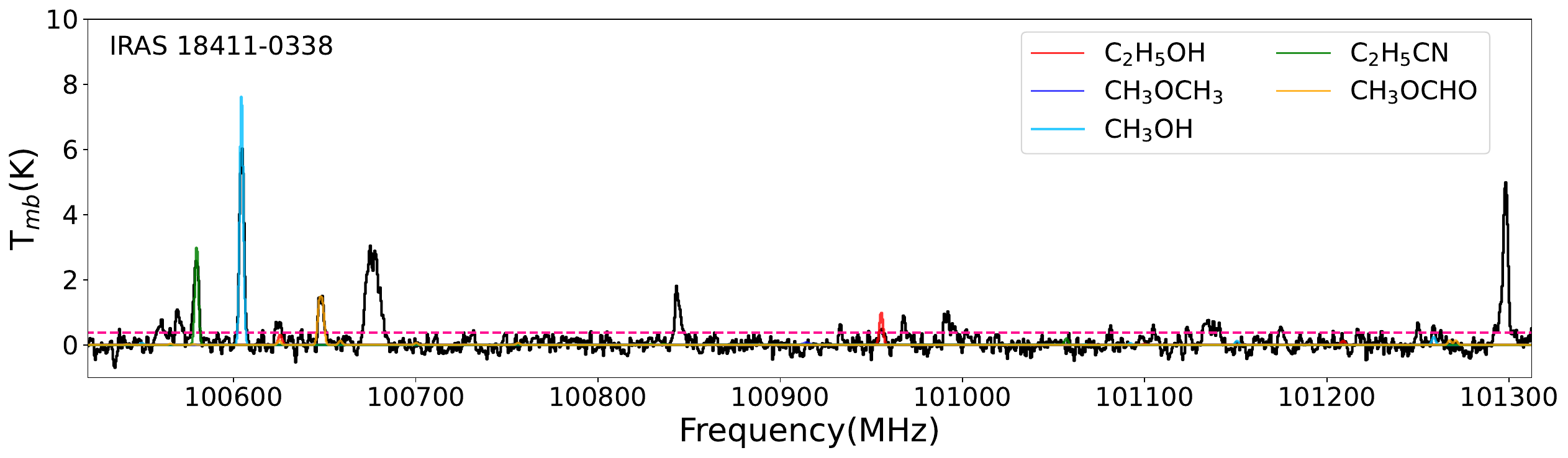}}
 \quad 
{\includegraphics[height=4.5cm,width=15.93cm]{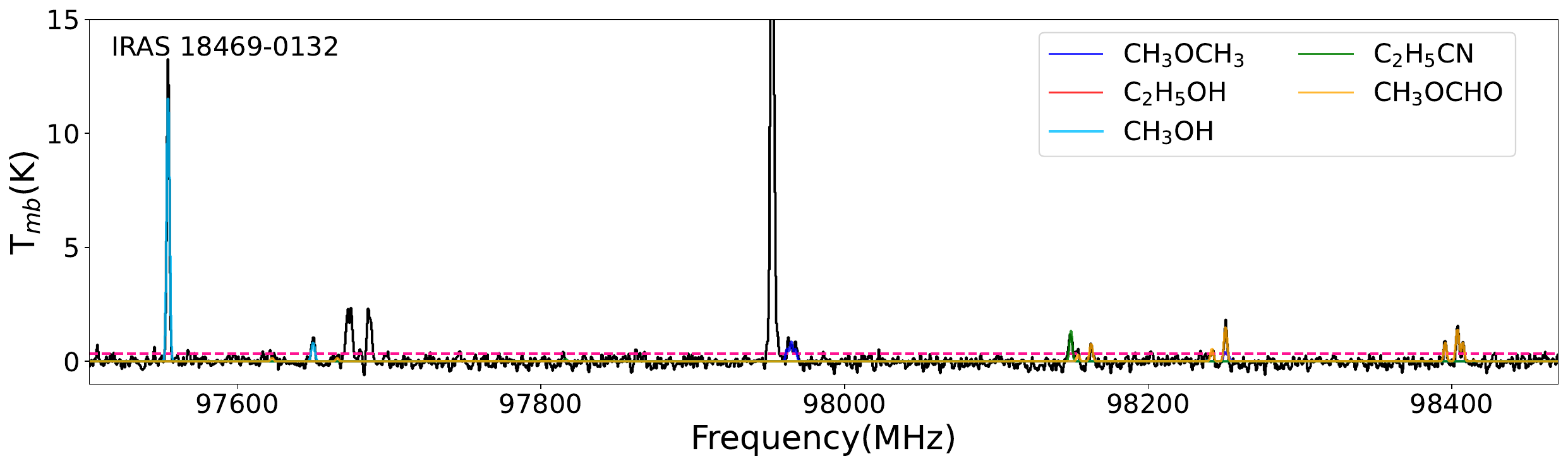}}
\quad
{\includegraphics[height=4.5cm,width=15.93cm]{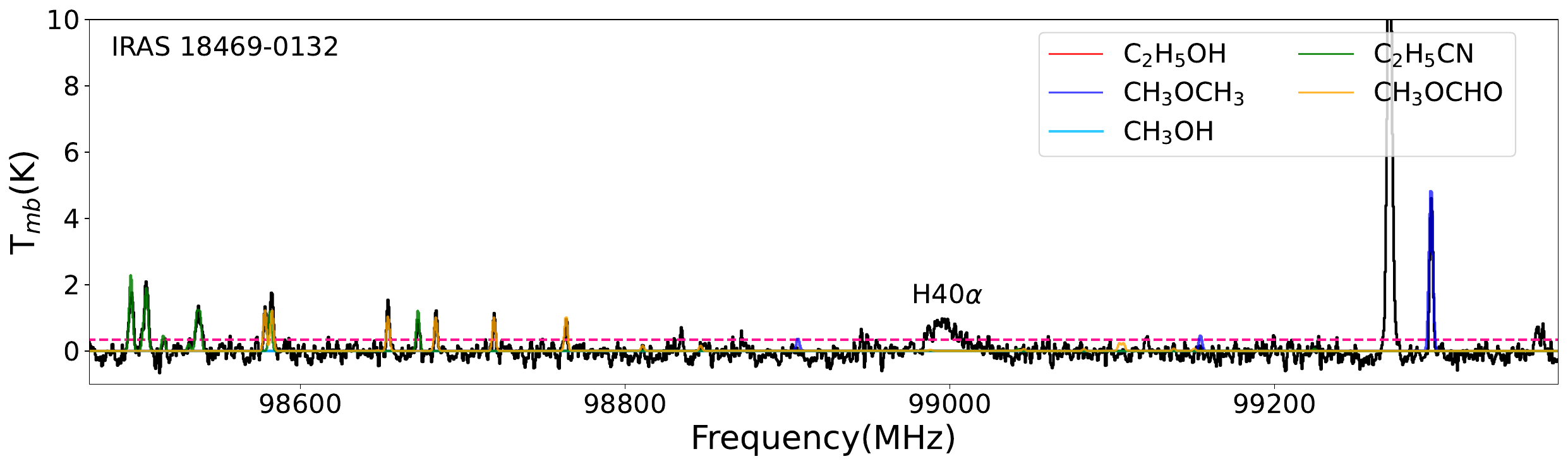}}
\quad
{\includegraphics[height=4.5cm,width=15.93cm]{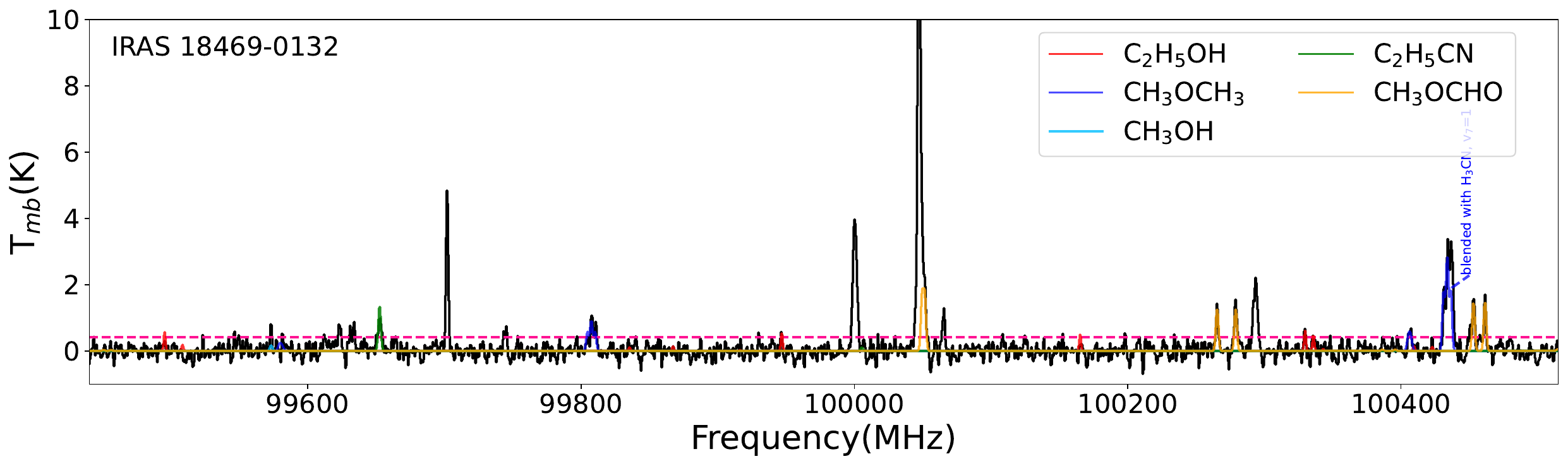}}
\quad
{\includegraphics[height=4.5cm,width=15.93cm]{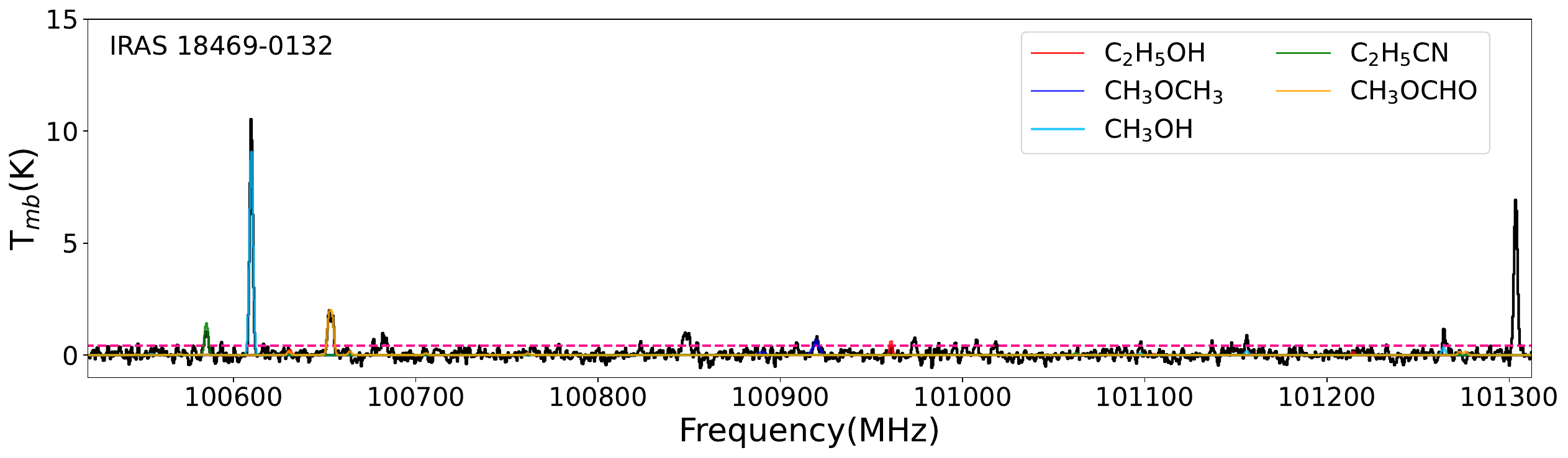}}
\caption{Continued.}
\end{figure}
\setcounter{figure}{\value{figure}-1}
\begin{figure}
  \centering 
{\includegraphics[height=4.5cm,width=15.93cm]{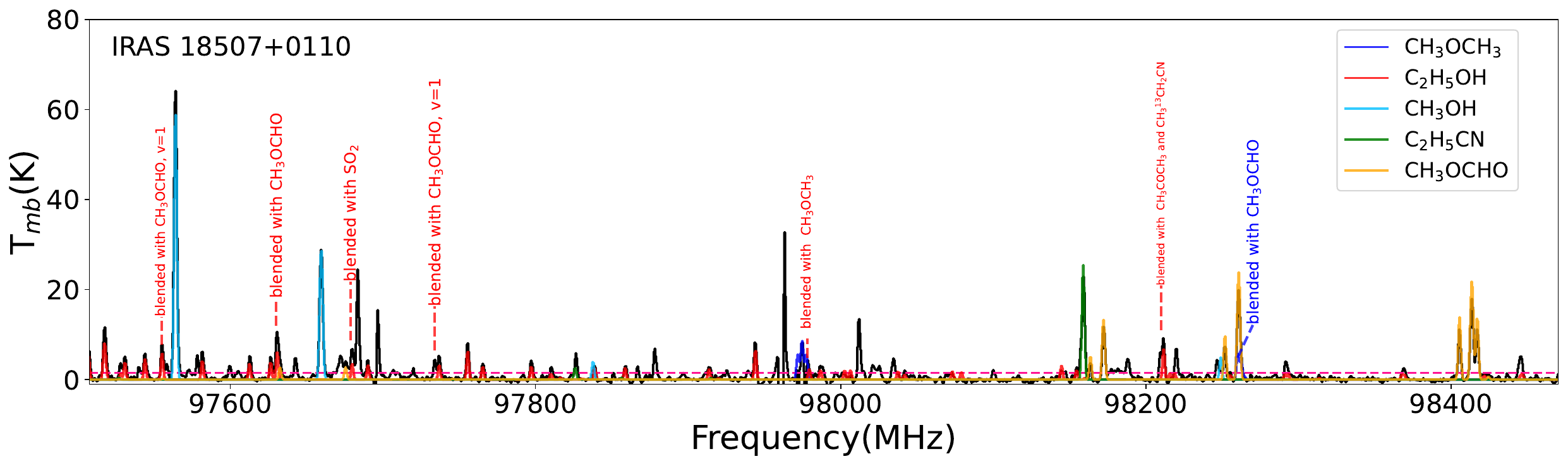}}
 \quad 
{\includegraphics[height=4.5cm,width=15.93cm]{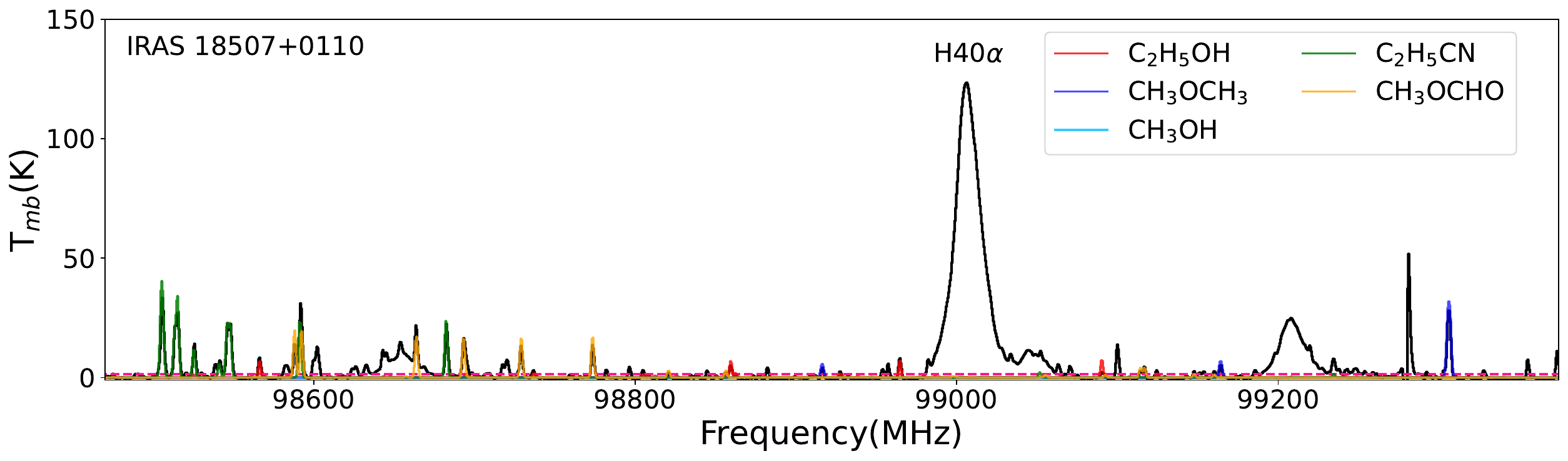}}
\quad
{\includegraphics[height=4.5cm,width=15.93cm]{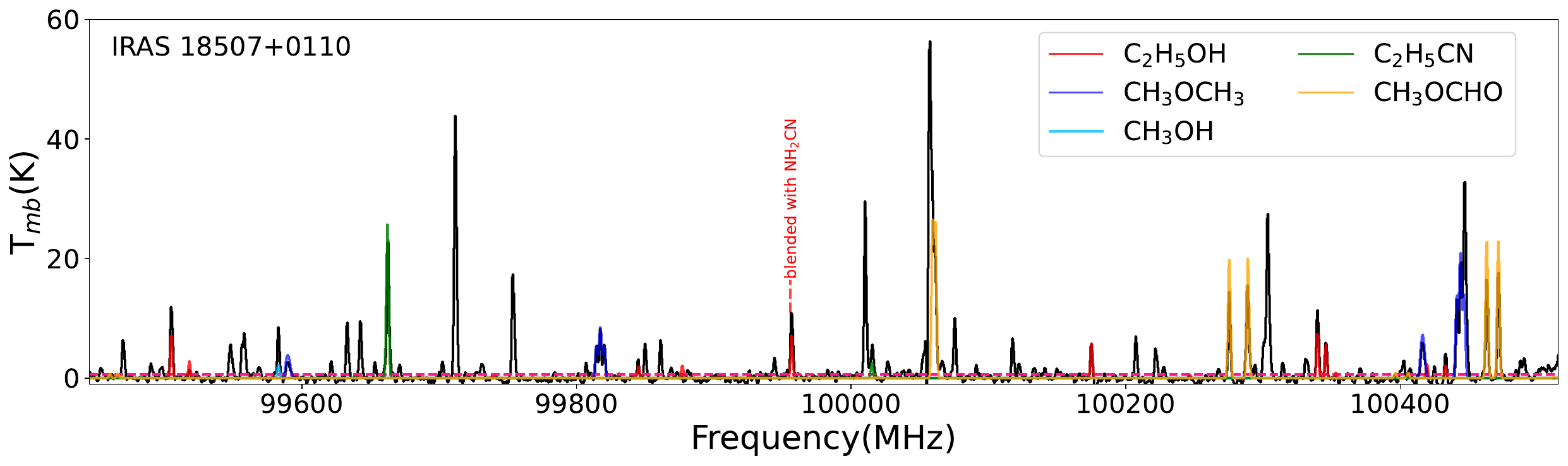}}
\quad
{\includegraphics[height=4.5cm,width=15.93cm]{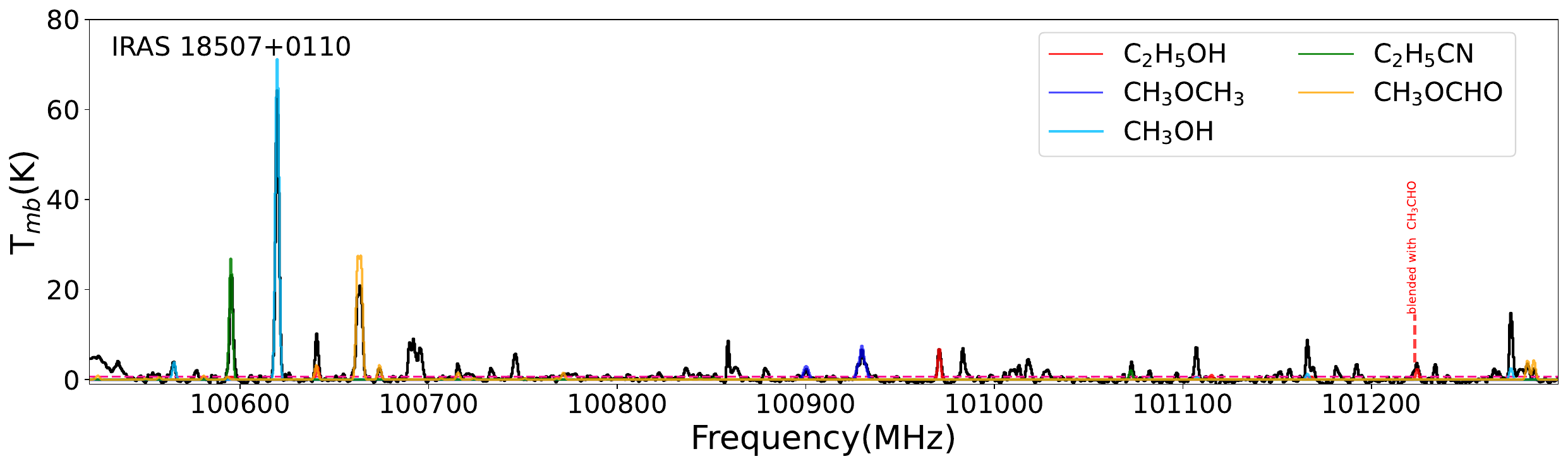}}
\quad
{\includegraphics[height=4.5cm,width=15.93cm]{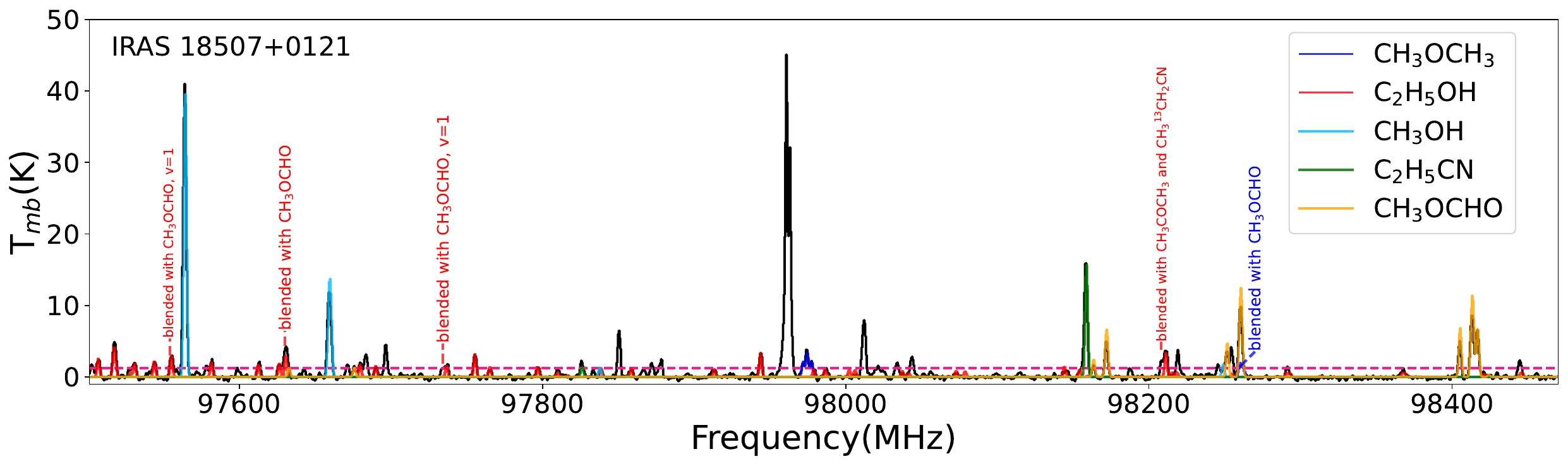}}
\caption{Continued.}
\end{figure}
\setcounter{figure}{\value{figure}-1}
\begin{figure}
  \centering 
{\includegraphics[height=4.5cm,width=15.93cm]{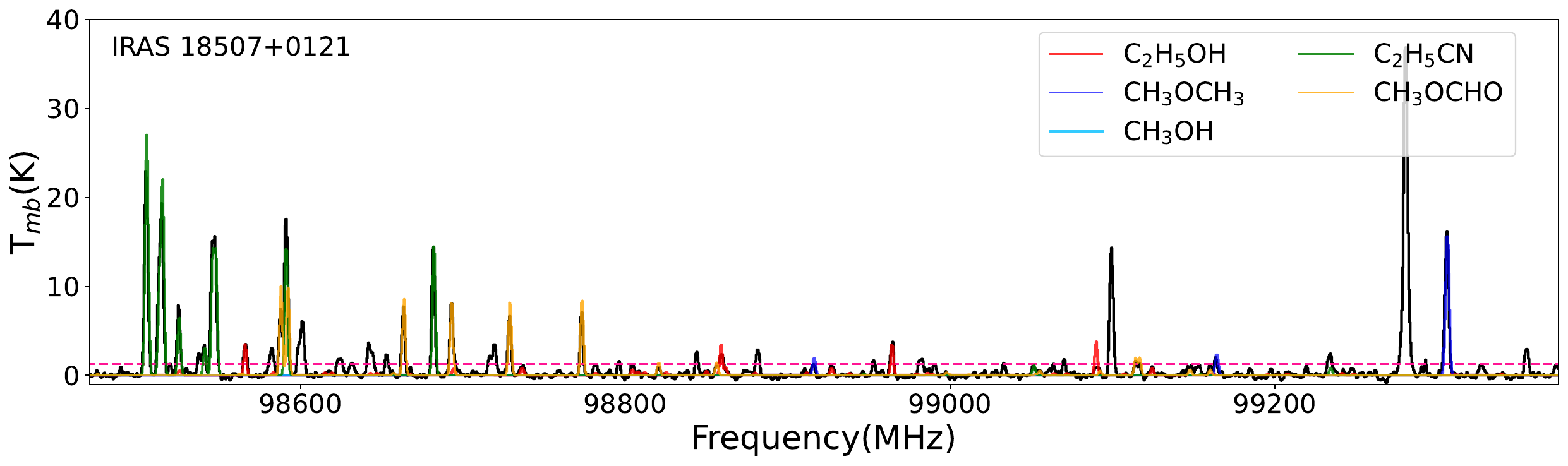}}
 \quad 
{\includegraphics[height=4.5cm,width=15.93cm]{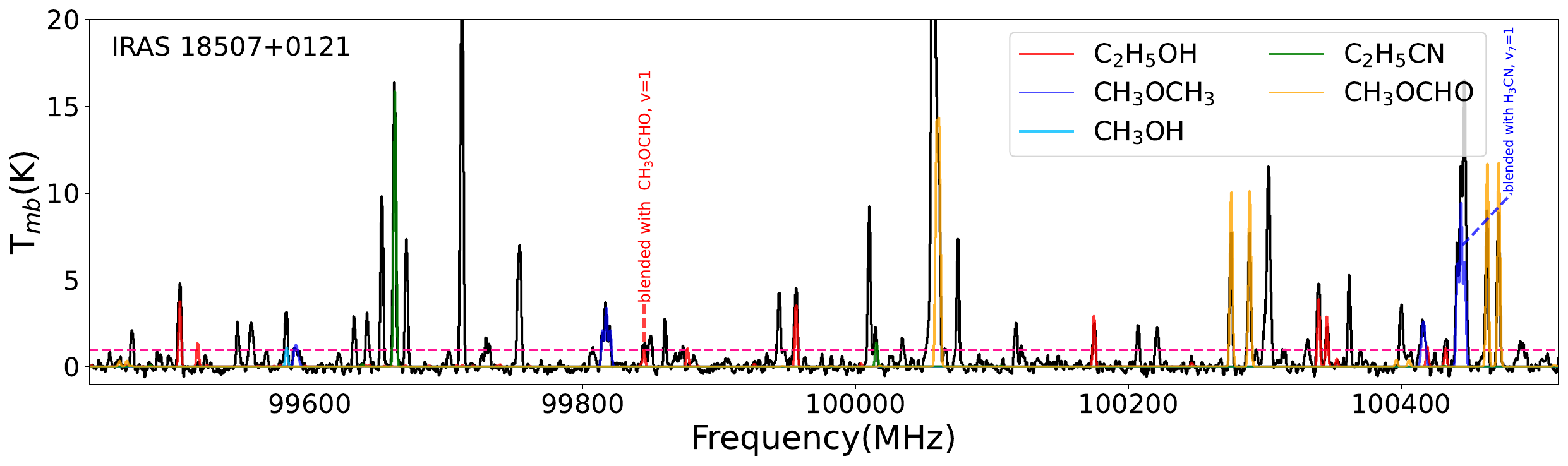}}
\quad
{\includegraphics[height=4.5cm,width=15.93cm]{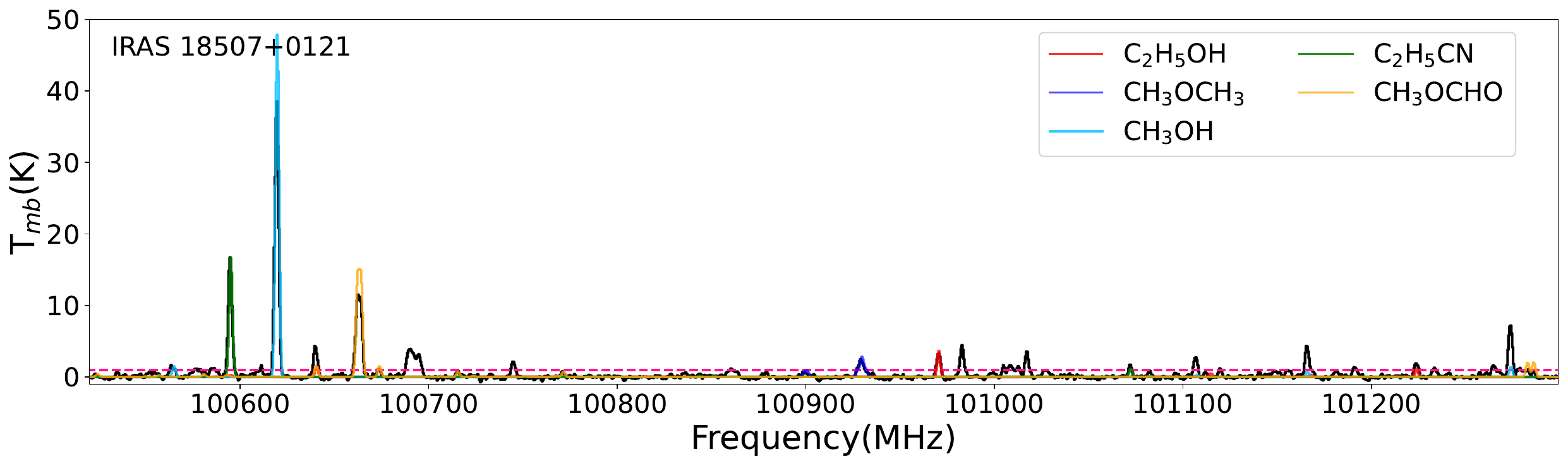}}
\quad
{\includegraphics[height=4.5cm,width=15.93cm]{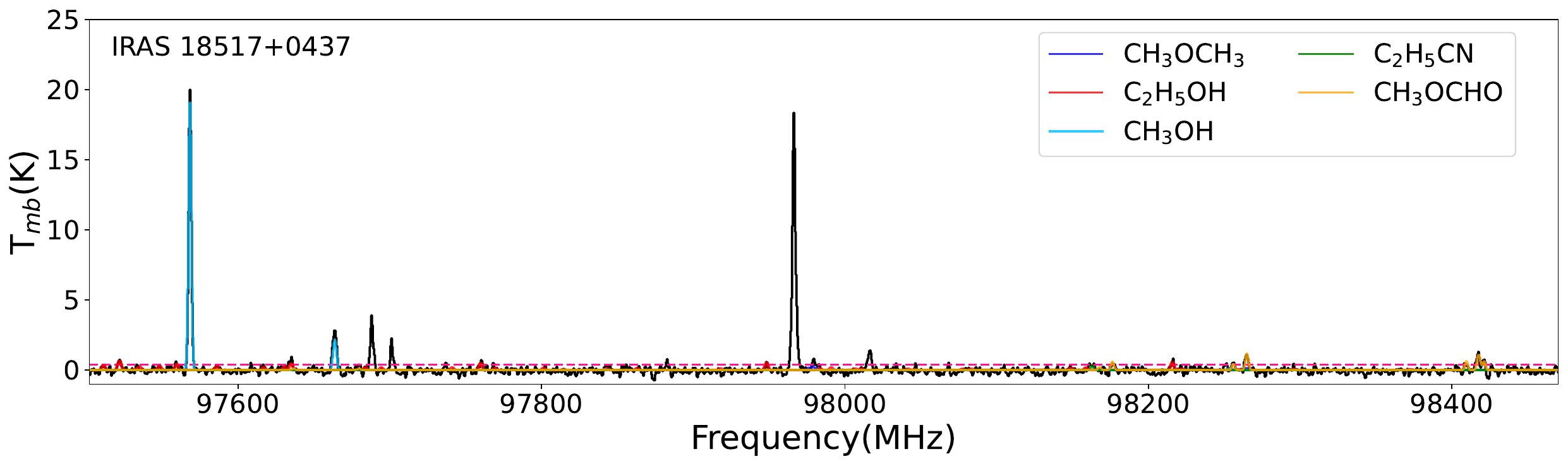}}
\quad
{\includegraphics[height=4.5cm,width=15.93cm]{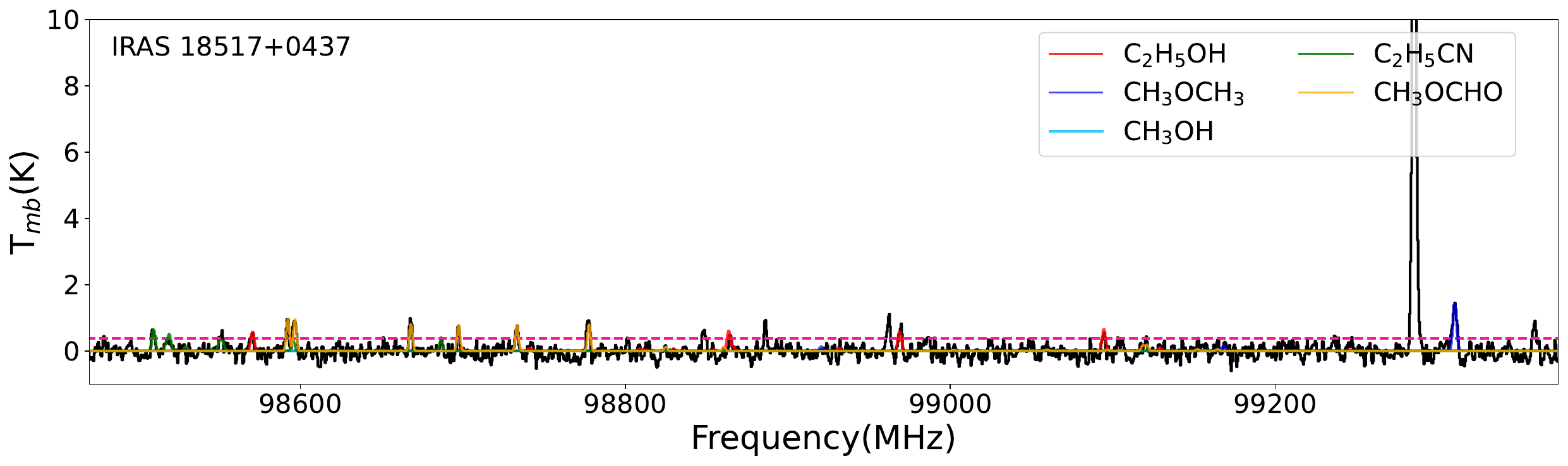}}
\caption{Continued.}
\end{figure}
\setcounter{figure}{\value{figure}-1}
\begin{figure}
  \centering 
{\includegraphics[height=4.5cm,width=15.93cm]{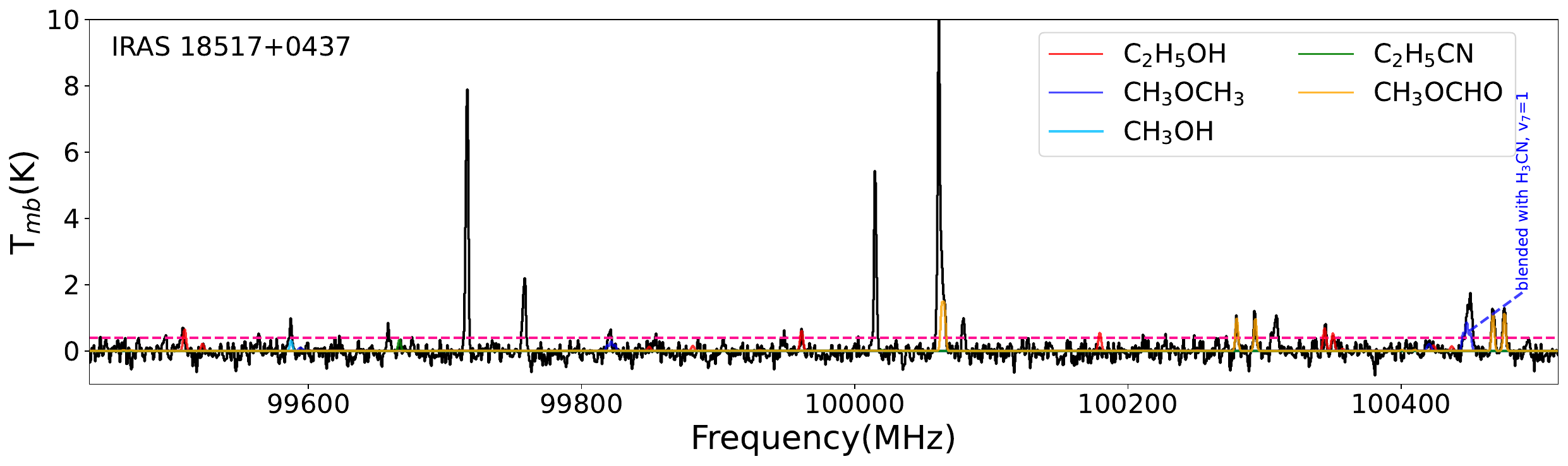}}
 \quad 
{\includegraphics[height=4.5cm,width=15.93cm]{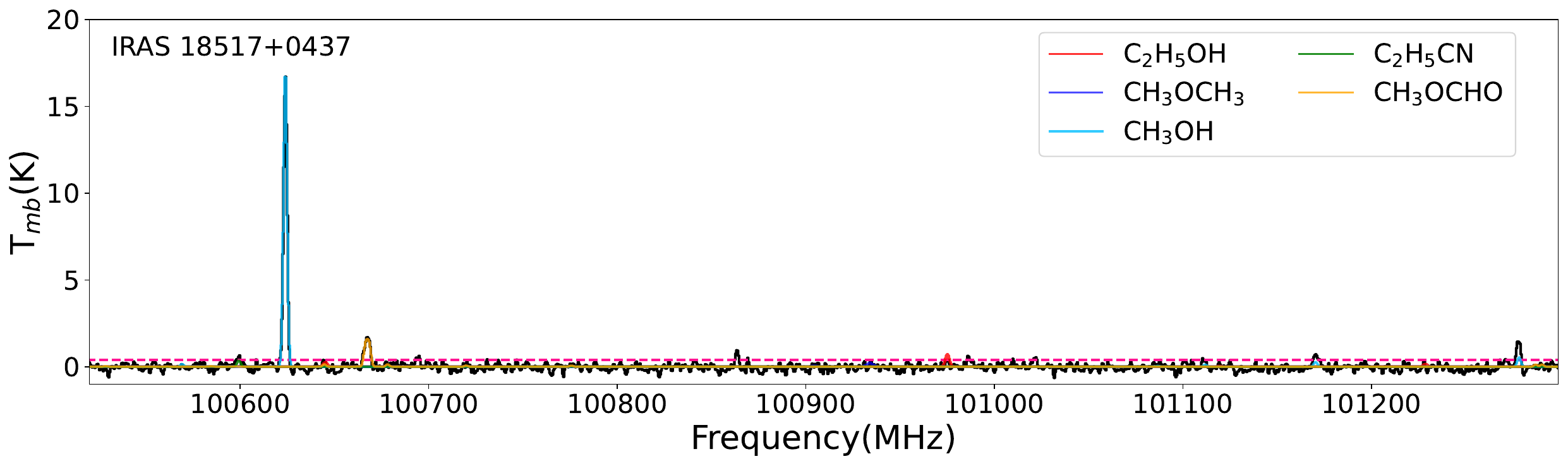}}
\quad
{\includegraphics[height=4.5cm,width=15.93cm]{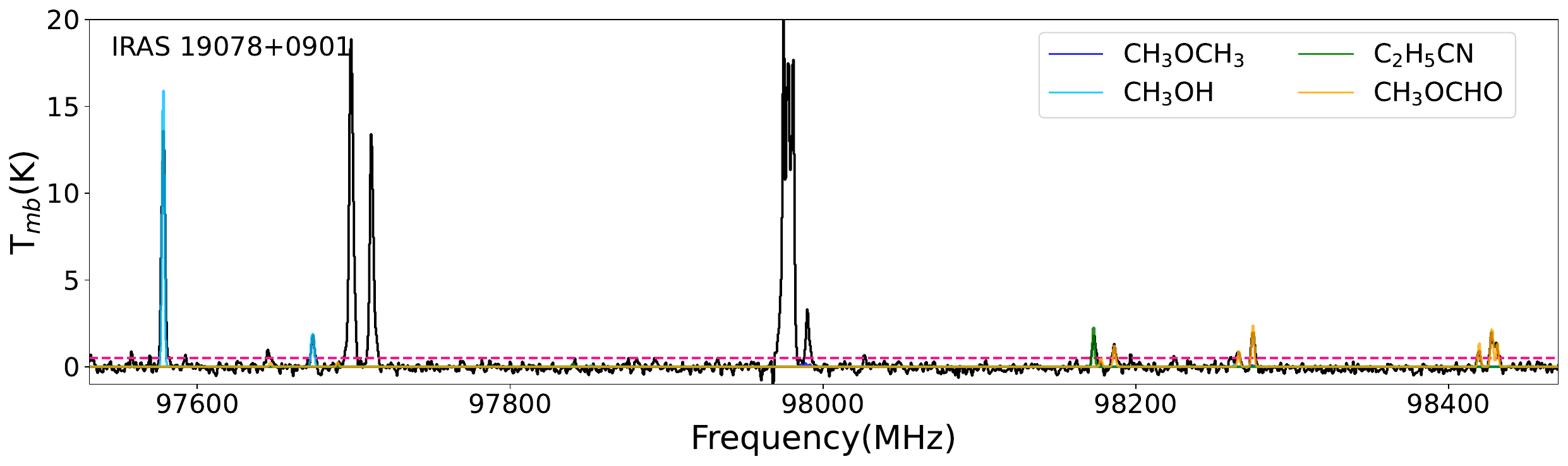}}
\quad
{\includegraphics[height=4.5cm,width=15.93cm]{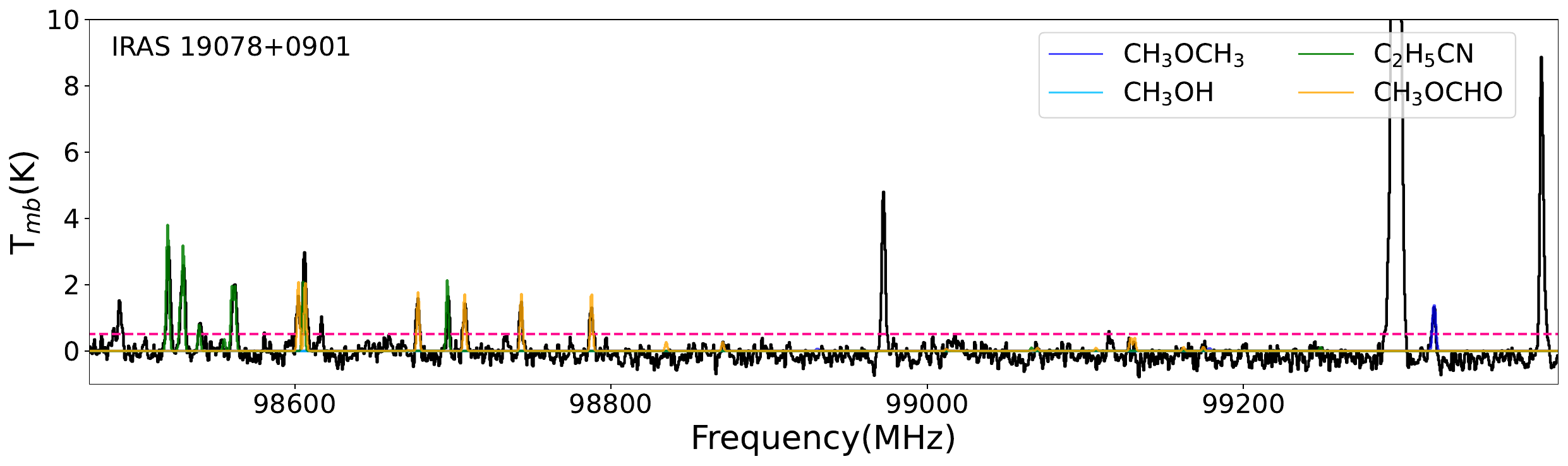}}
\quad
{\includegraphics[height=4.5cm,width=15.93cm]{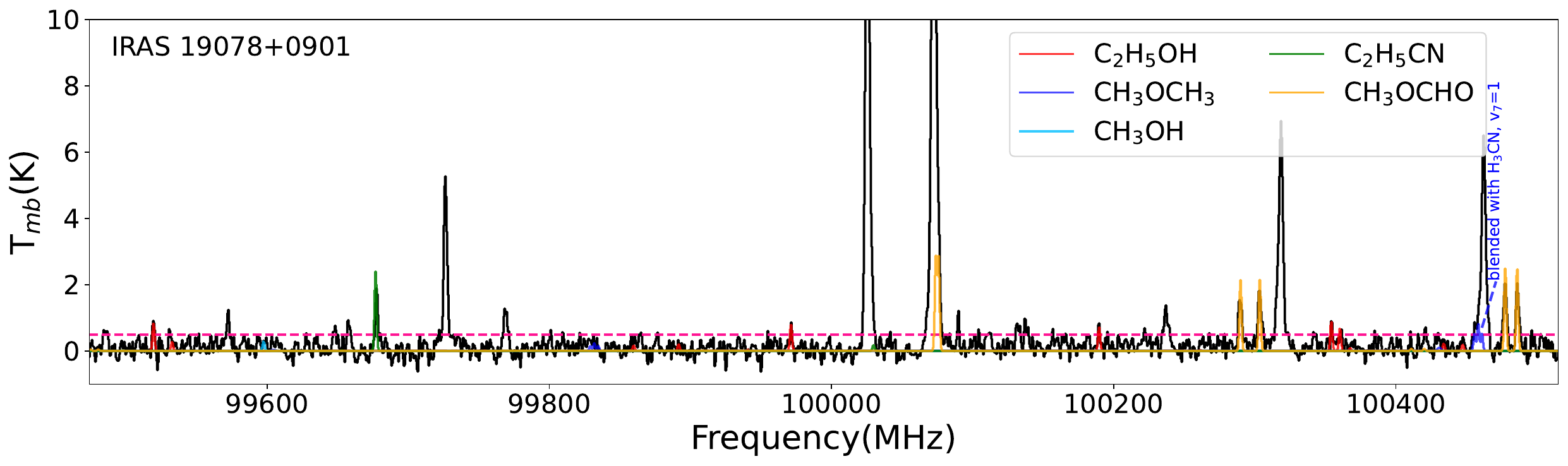}}
\caption{Continued.}
\end{figure}
\setcounter{figure}{\value{figure}-1}
\begin{figure}
  \centering 
{\includegraphics[height=4.5cm,width=15.93cm]{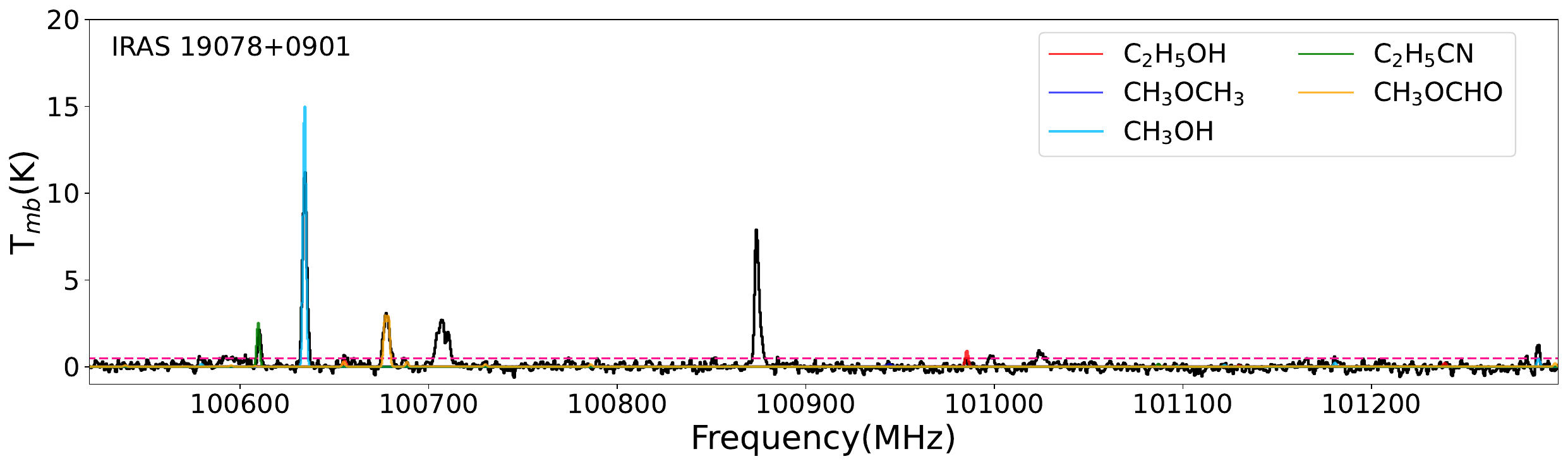}}
 \quad 
{\includegraphics[height=4.5cm,width=15.93cm]{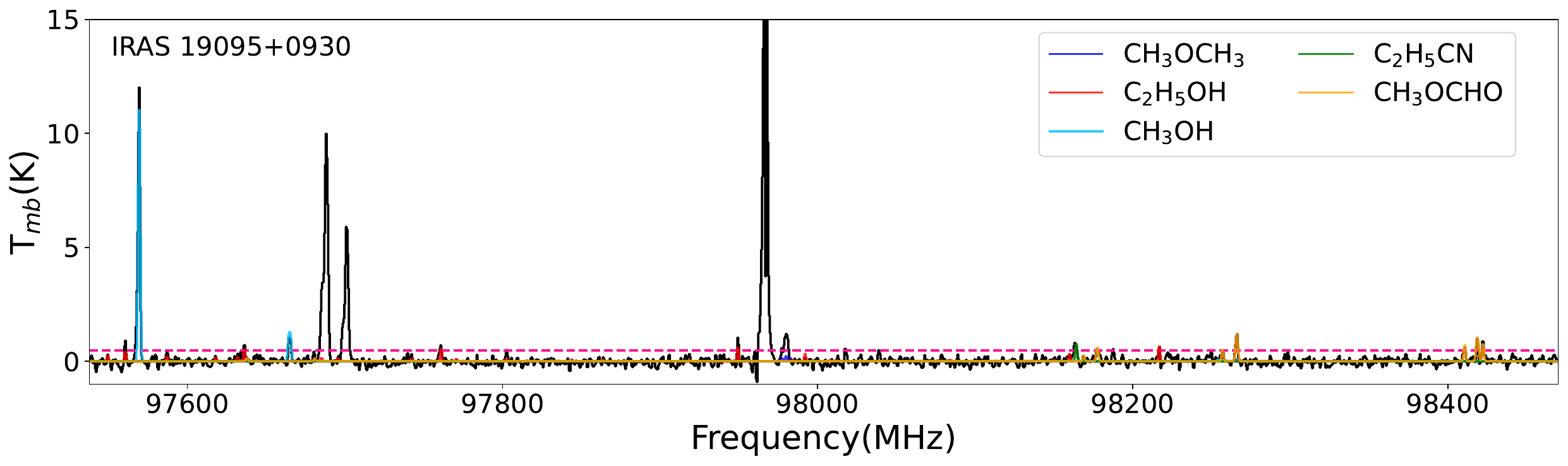}}
\quad
{\includegraphics[height=4.5cm,width=15.93cm]{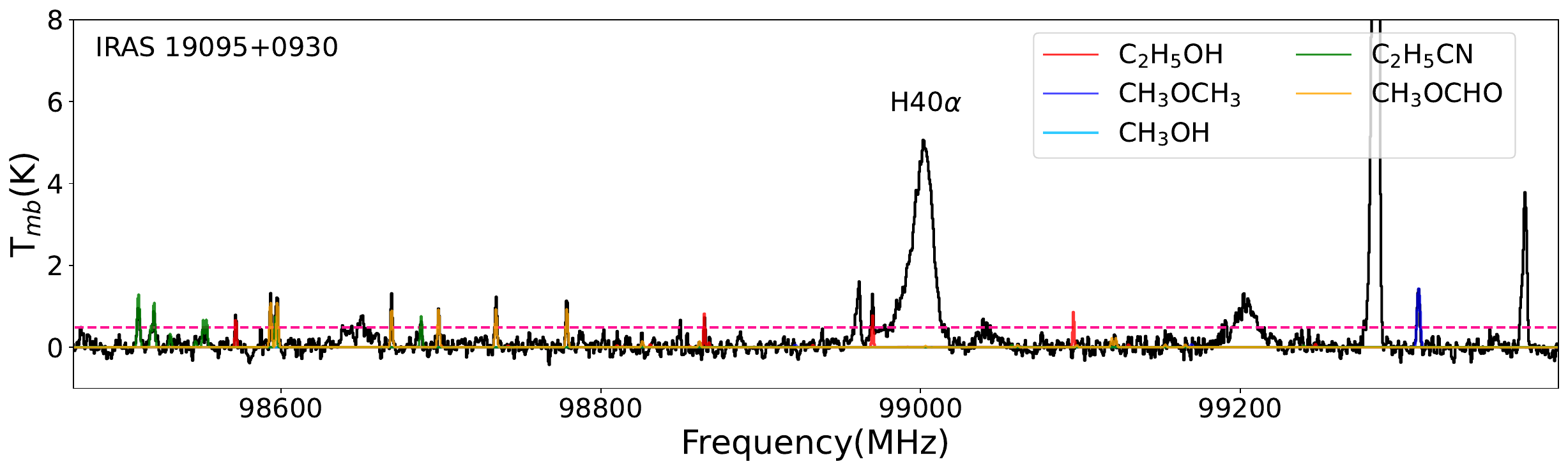}}
\quad
{\includegraphics[height=4.5cm,width=15.93cm]{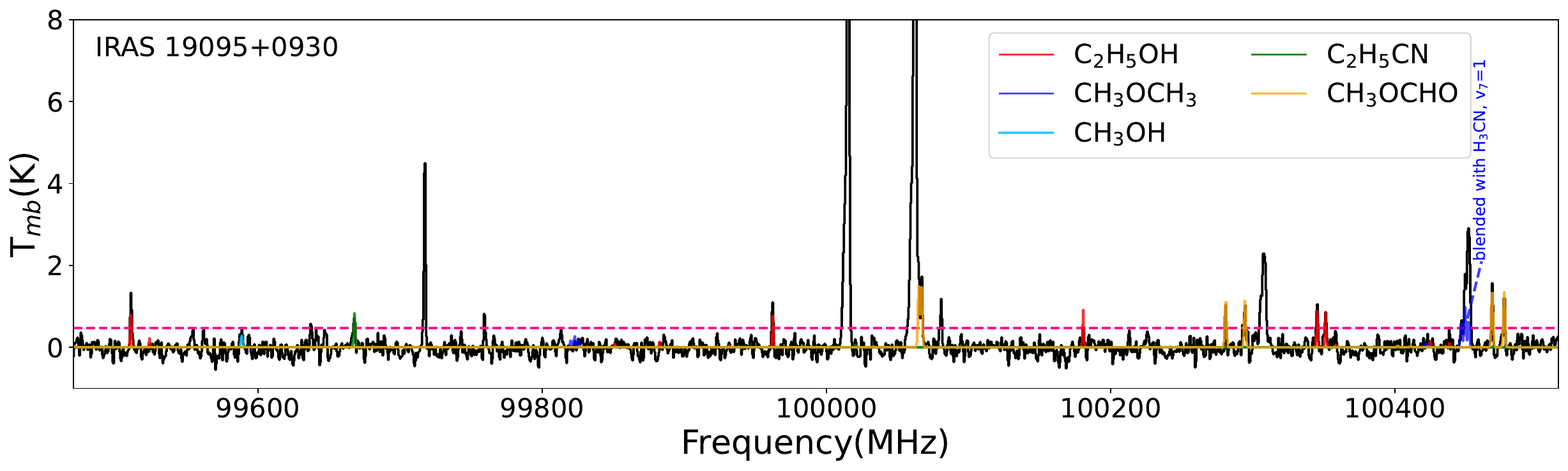}}
\quad
{\includegraphics[height=4.5cm,width=15.93cm]{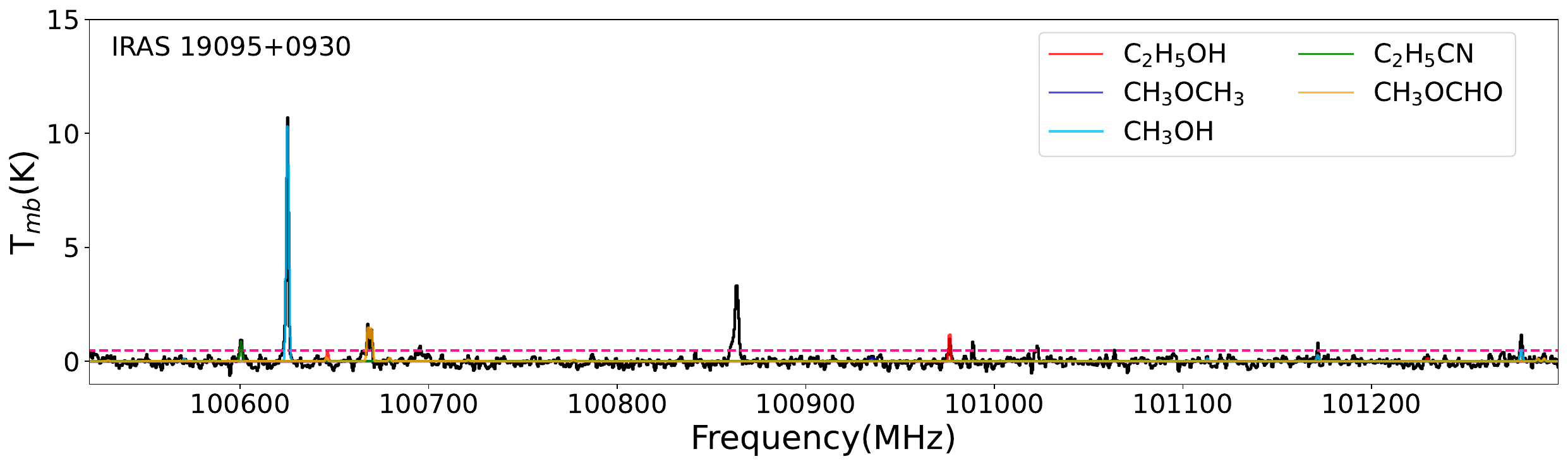}}
\caption{Continued.}
\end{figure}

\section{Moment 0 maps}  \label{app:C}
Figure \ref{fig:C1} shows continuum maps in color scale and moment 0 maps of C$_2$H$_5$OH and CH$_3$OCH$_3$. The white contours indicate the CH$_3$OCH$_3$ emission at the 3, 5, 10, 15, 25, 40, 60, 80, and 110$\sigma$ levels
($\sigma$ = 10-90 mJy beam$^{−1}\,{\rm km}\,{\rm s}^{-1}$). The red contours indicate the C$_2$H$_5$OH emission at the 3, 5, 7, 10, 15, 30, 50, 60, and 90$\sigma$ levels ($\sigma$ = 10-90 mJy beam$^{−1}\,{\rm km}\,{\rm s}^{-1}$). 
\begin{figure}
\centering
{\includegraphics[height=4.01cm,width=5.21cm]{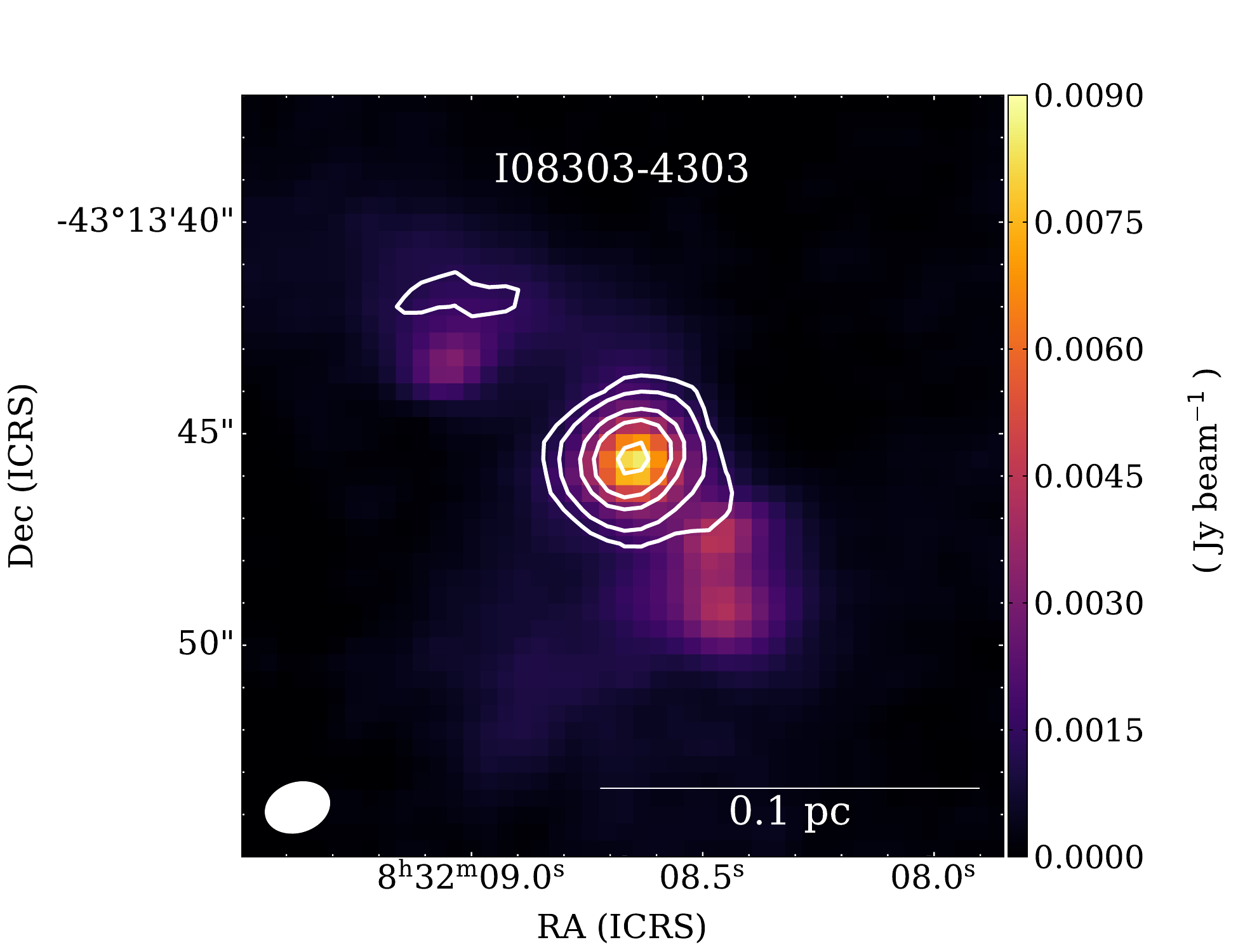}}
\quad
{\includegraphics[height=4.01cm,width=5.21cm]{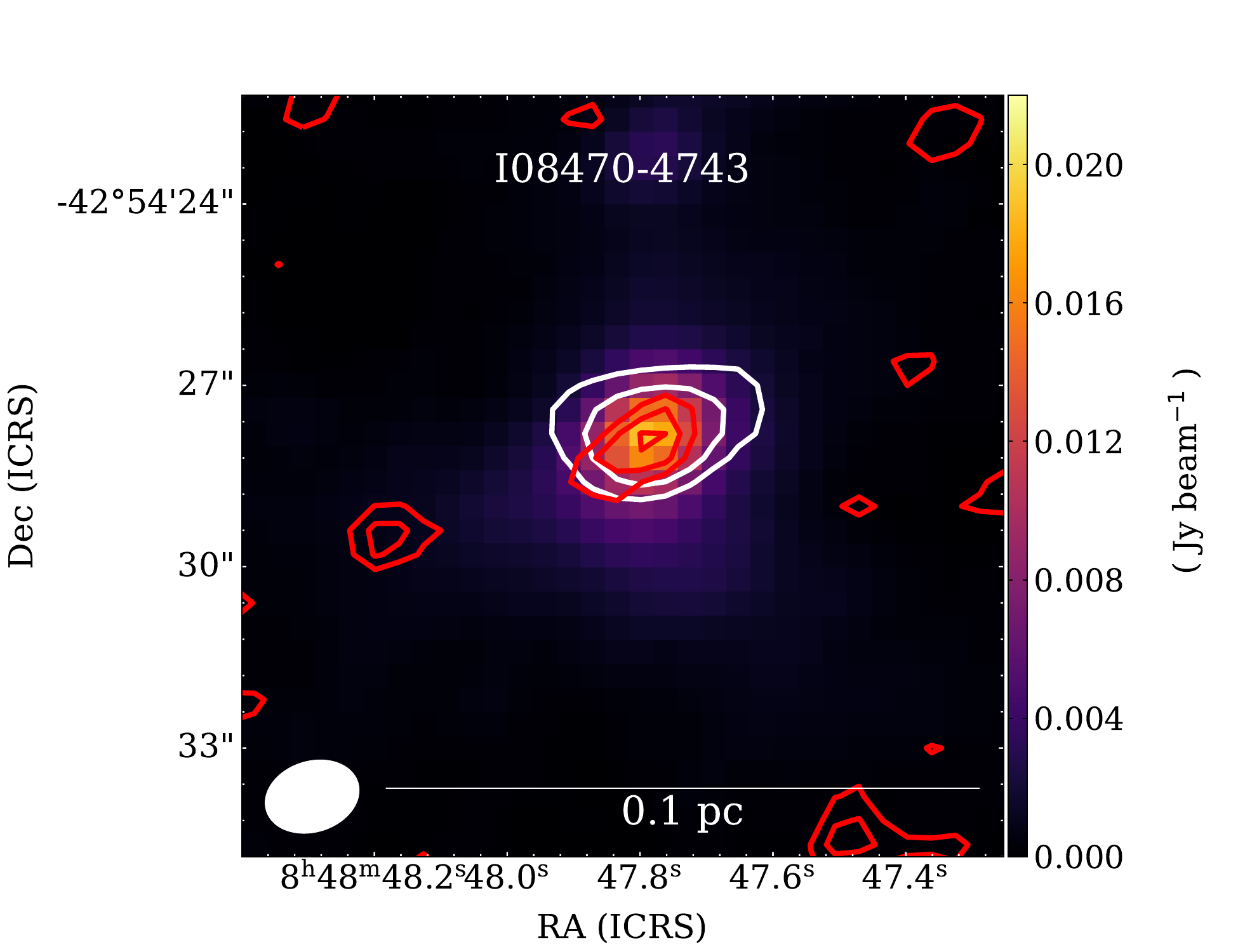}}
\quad
{\includegraphics[height=4.01cm,width=5.21cm]{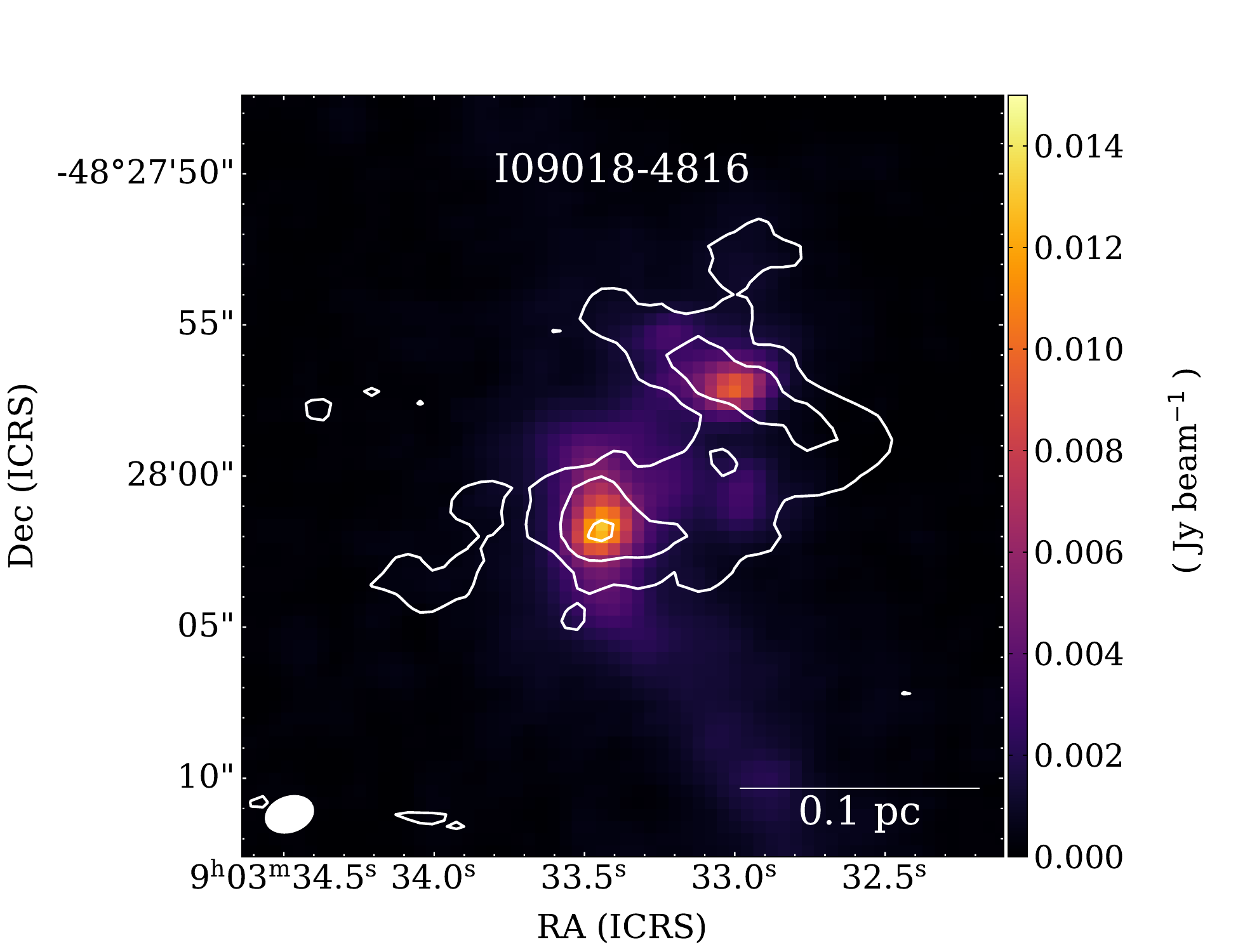}}
  \quad
{\includegraphics[height=4.01cm,width=5.21cm]{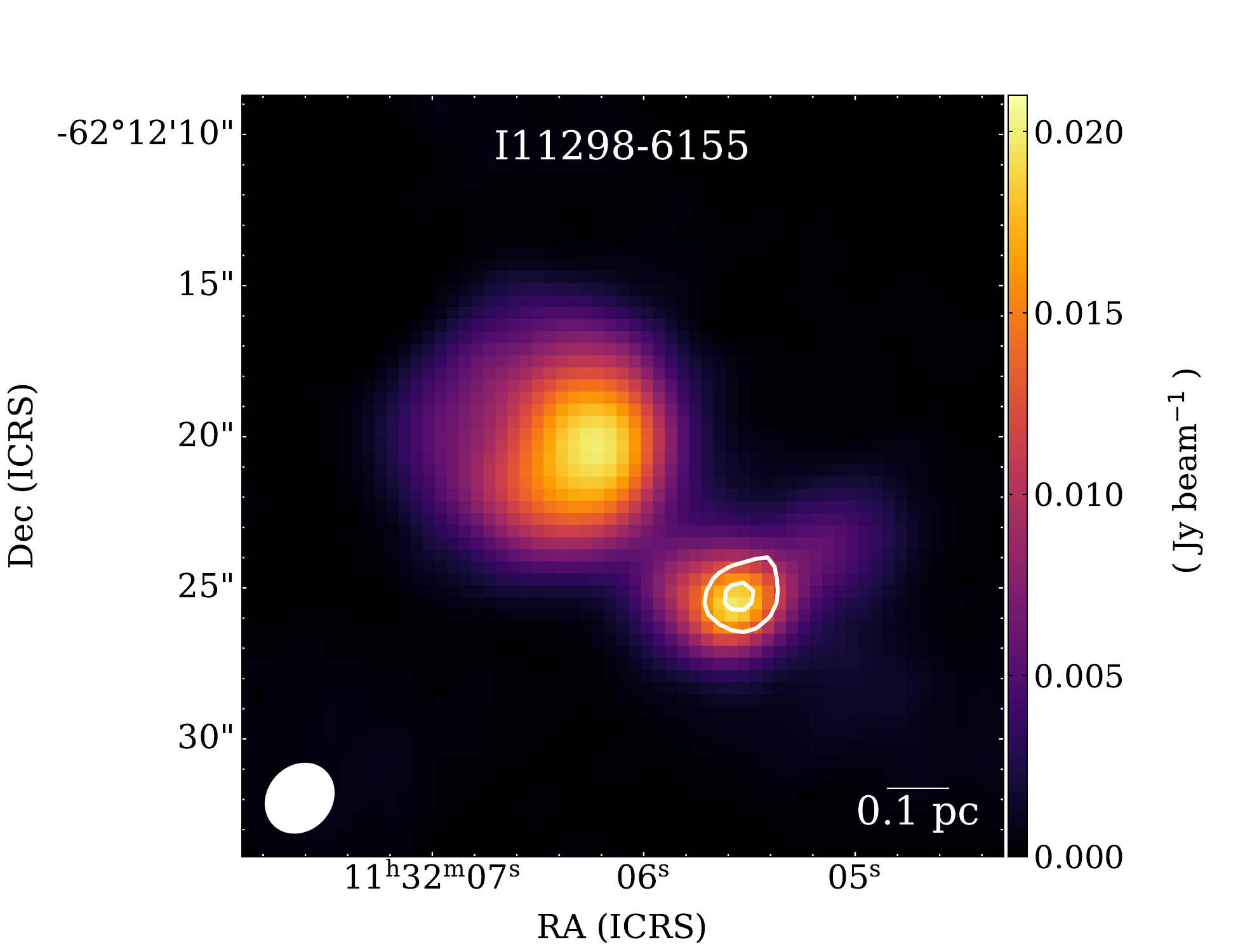}}
  \quad 
{\includegraphics[height=4.01cm,width=5.21cm]{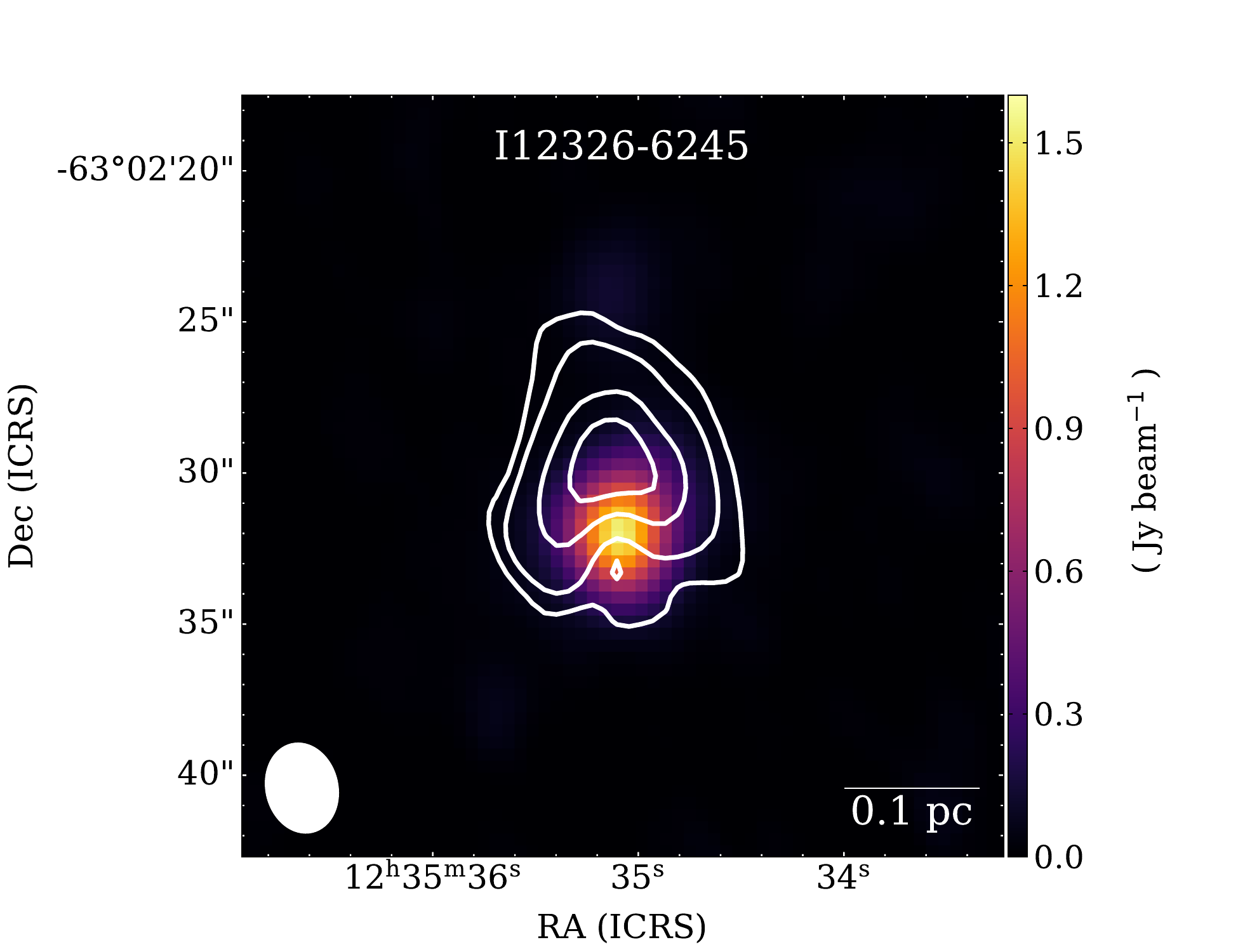}}
\quad
{\includegraphics[height=4.01cm,width=5.21cm]{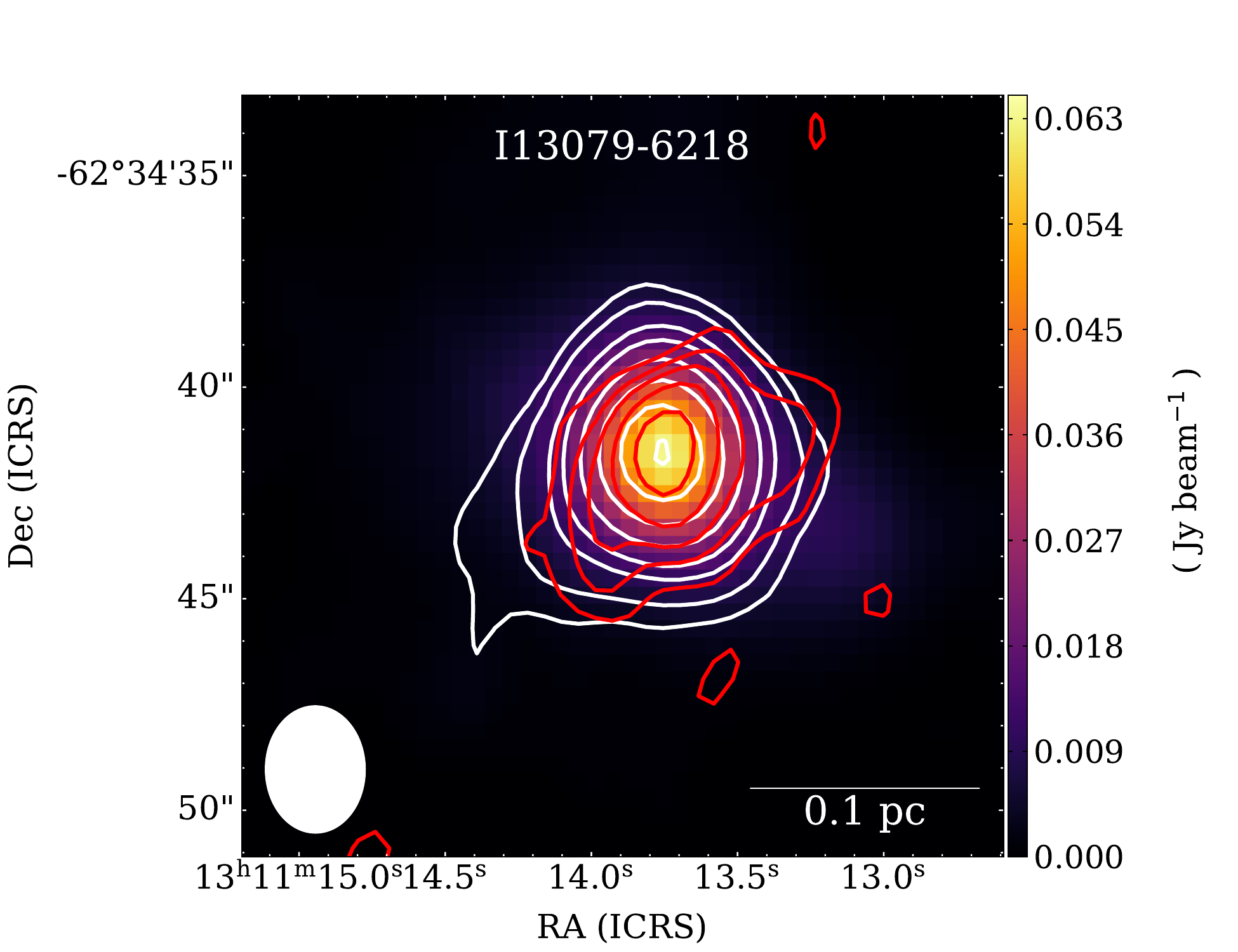}}
\quad
{\includegraphics[height=4.01cm,width=5.21cm]{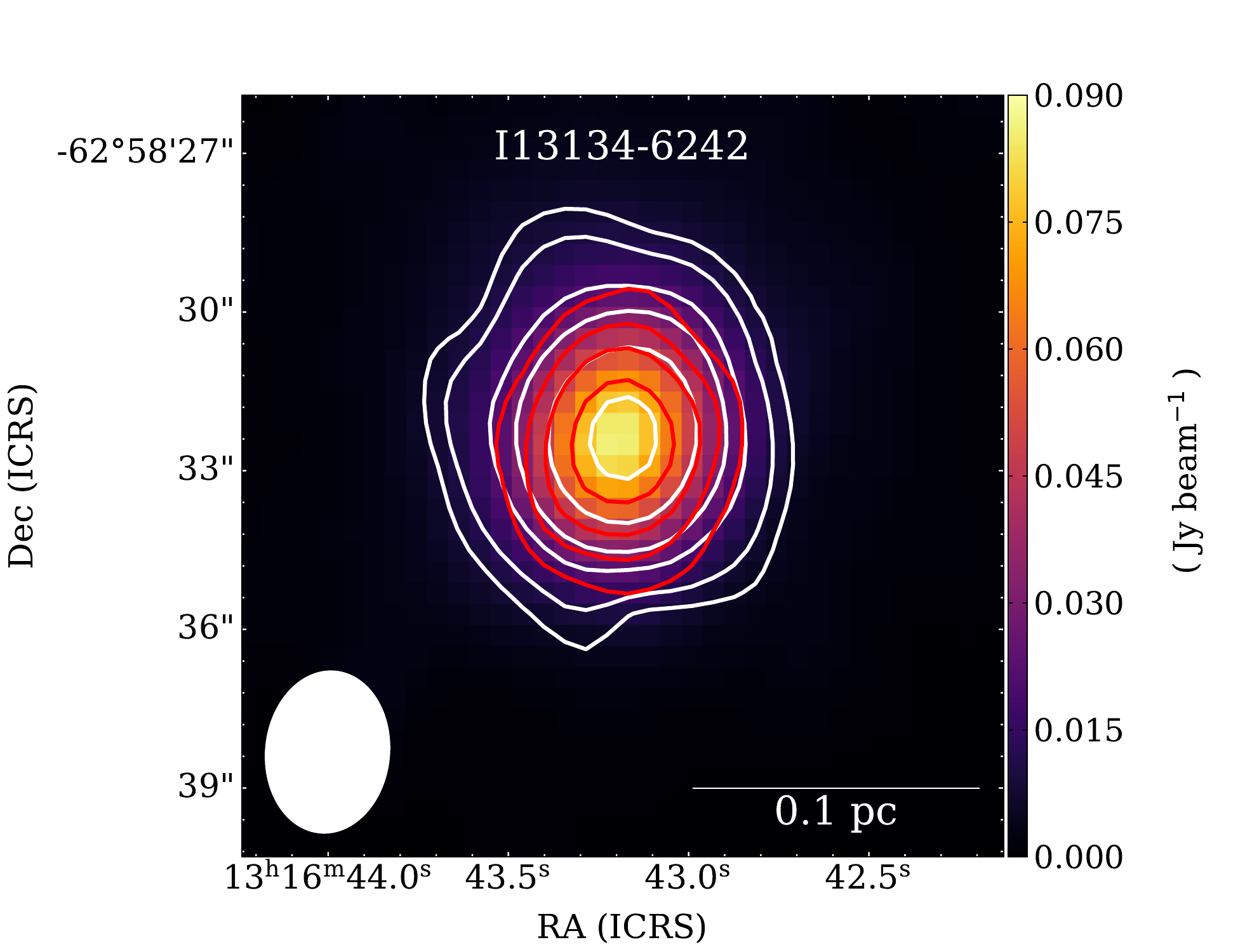}}
\quad
{\includegraphics[height=4.01cm,width=5.21cm]{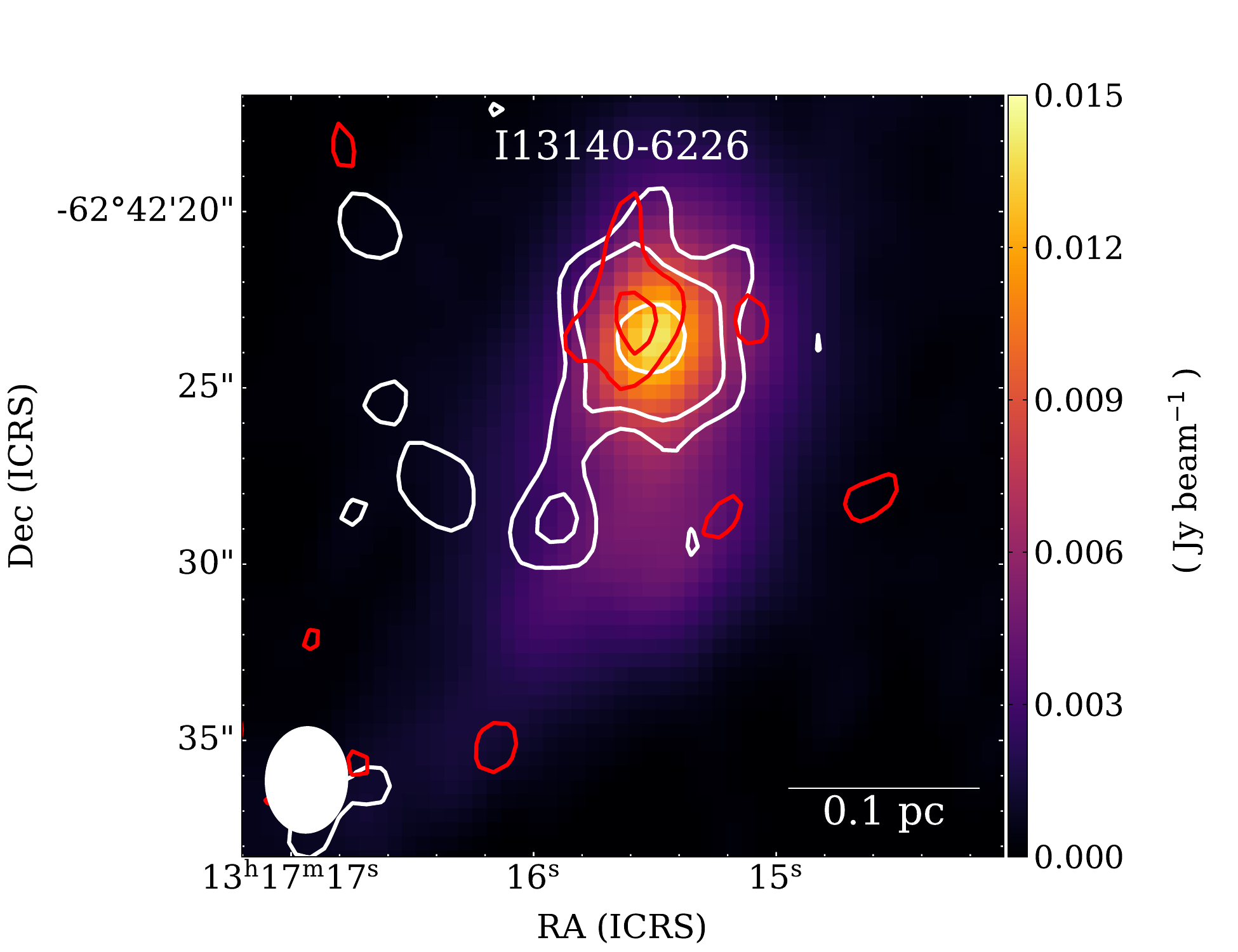}}
\quad
{\includegraphics[height=4.01cm,width=5.21cm]{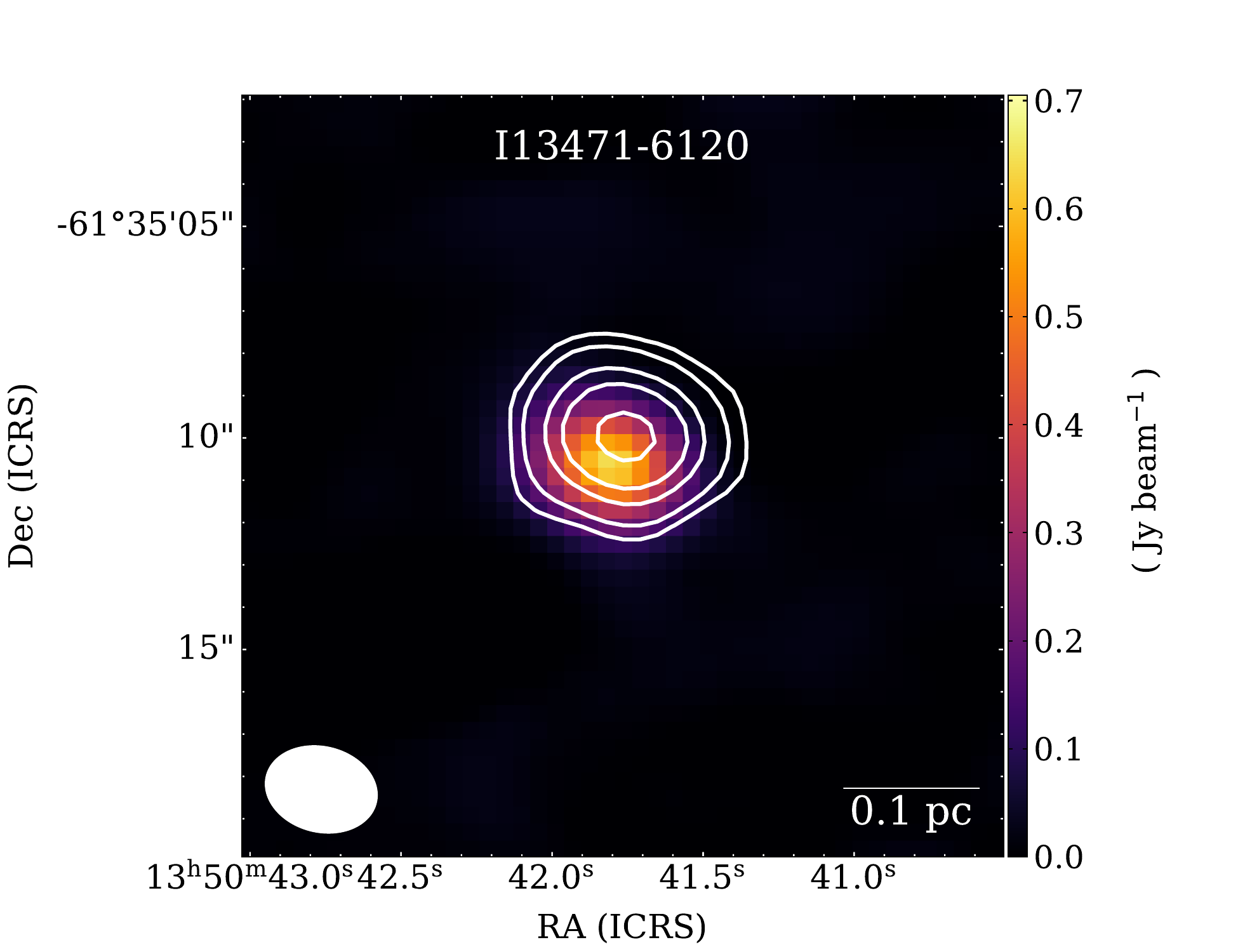}}
\quad
{\includegraphics[height=4.01cm,width=5.21cm]{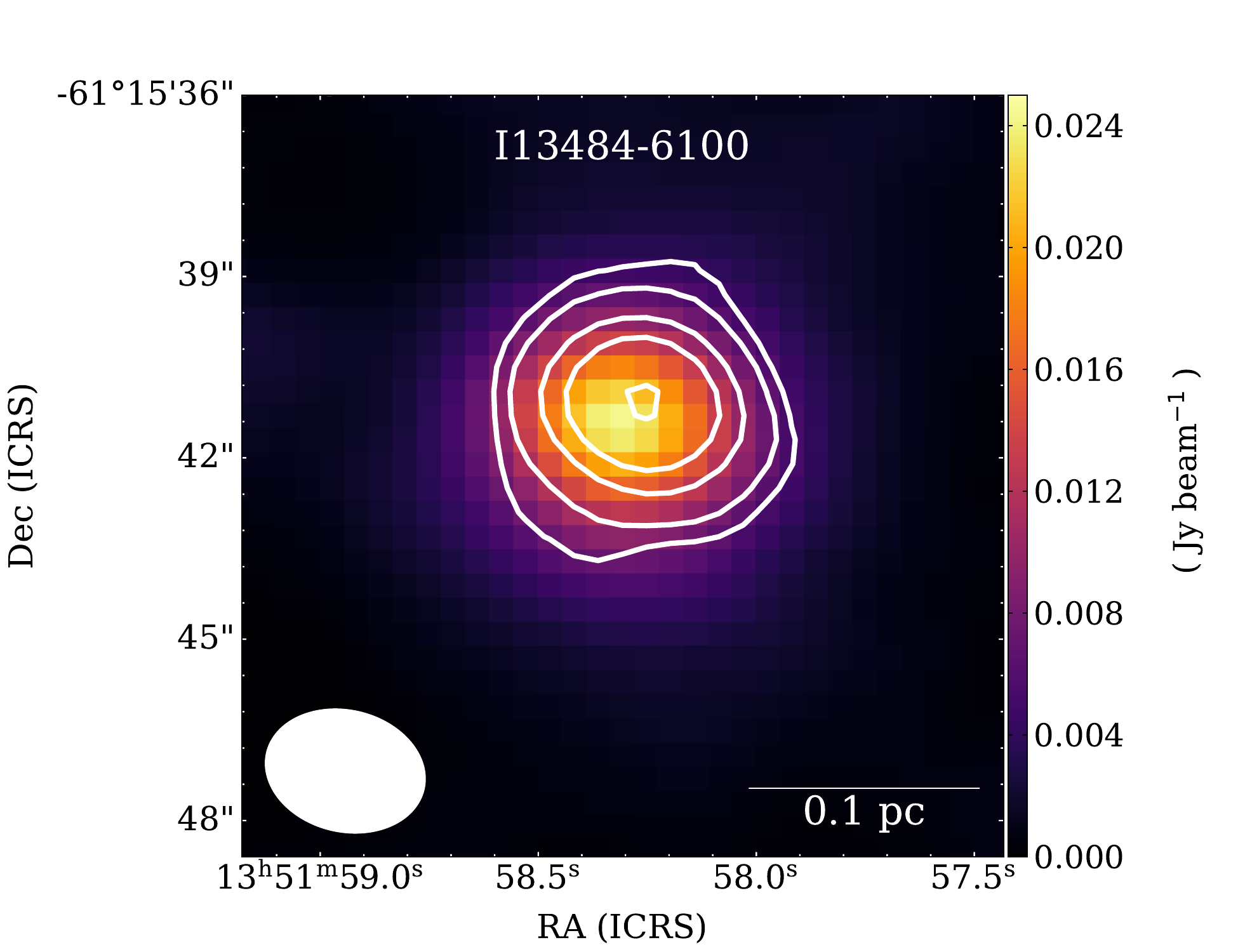}} 
\quad
{\includegraphics[height=4.01cm,width=5.21cm]{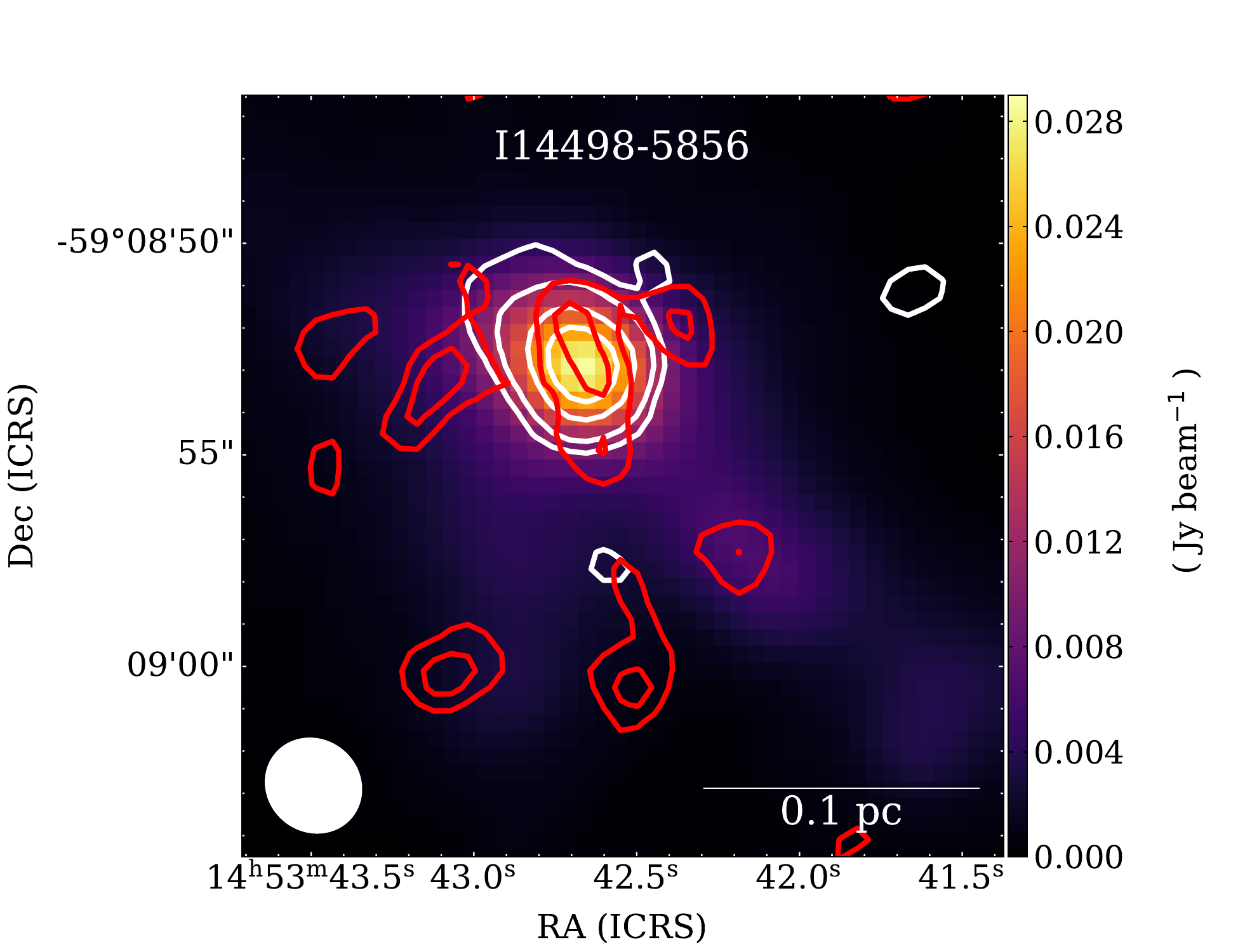}} 
\quad
{\includegraphics[height=4.01cm,width=5.21cm]{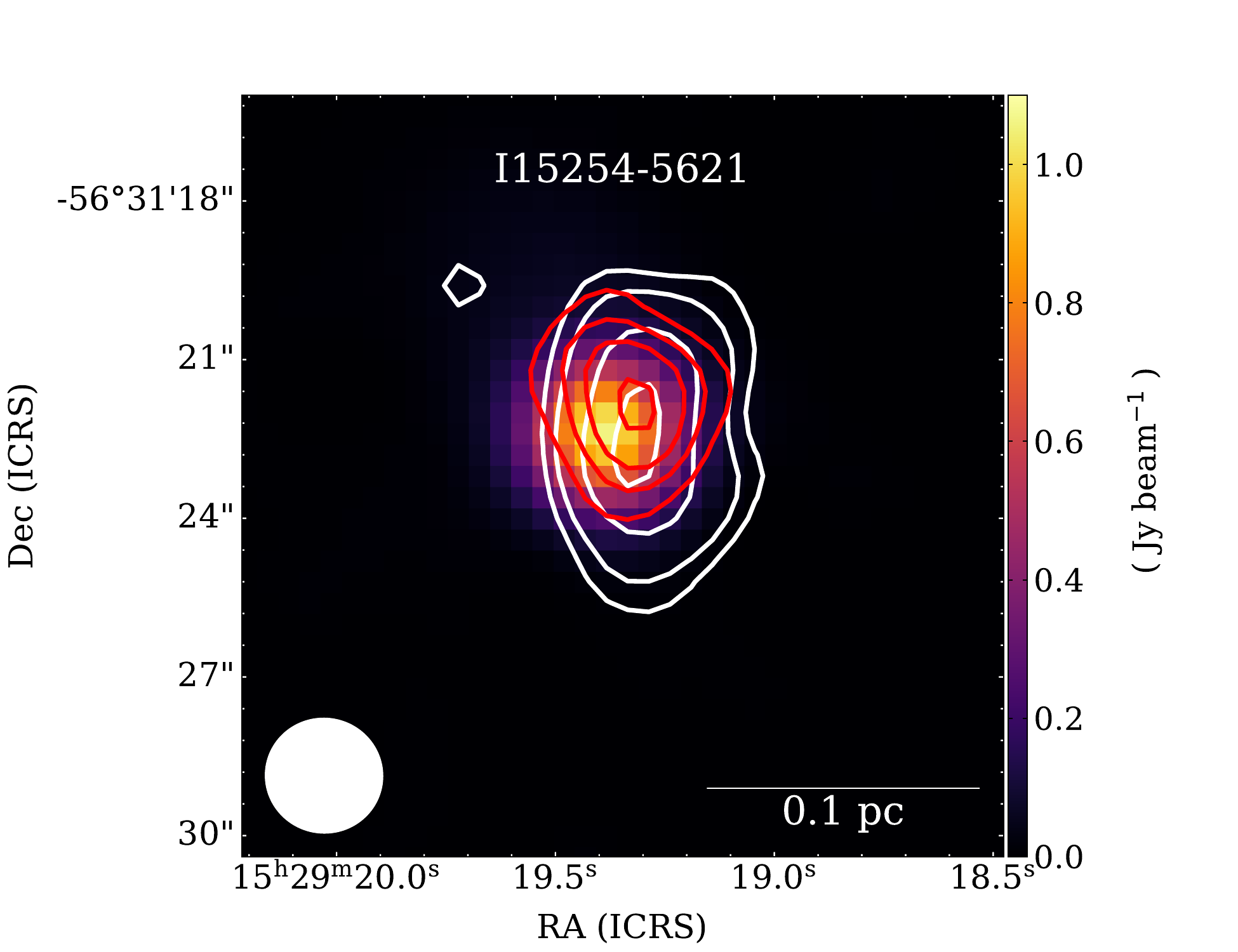}} 
\quad
{\includegraphics[height=4.01cm,width=5.21cm]{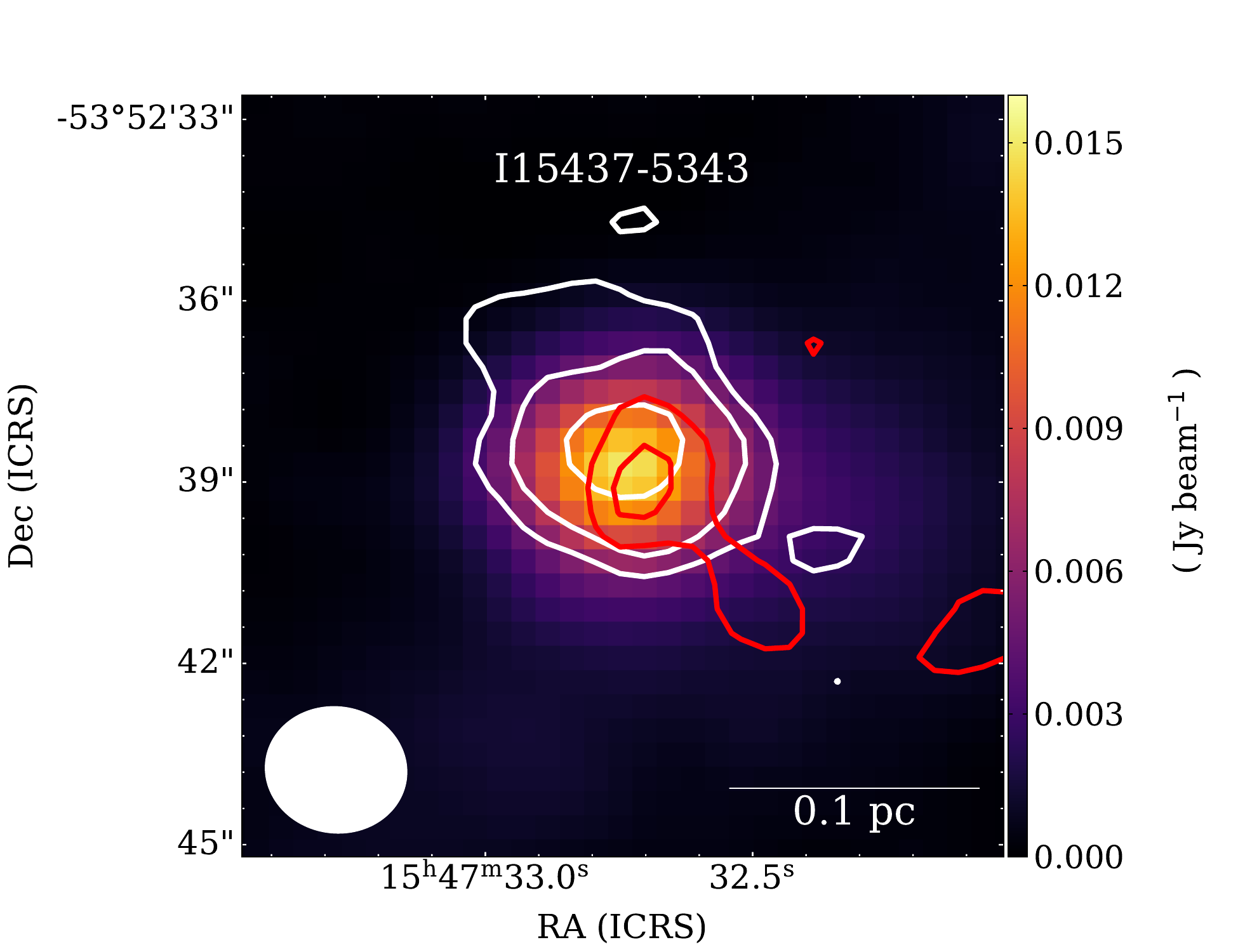}} 
\quad 
{\includegraphics[height=4.01cm,width=5.21cm]{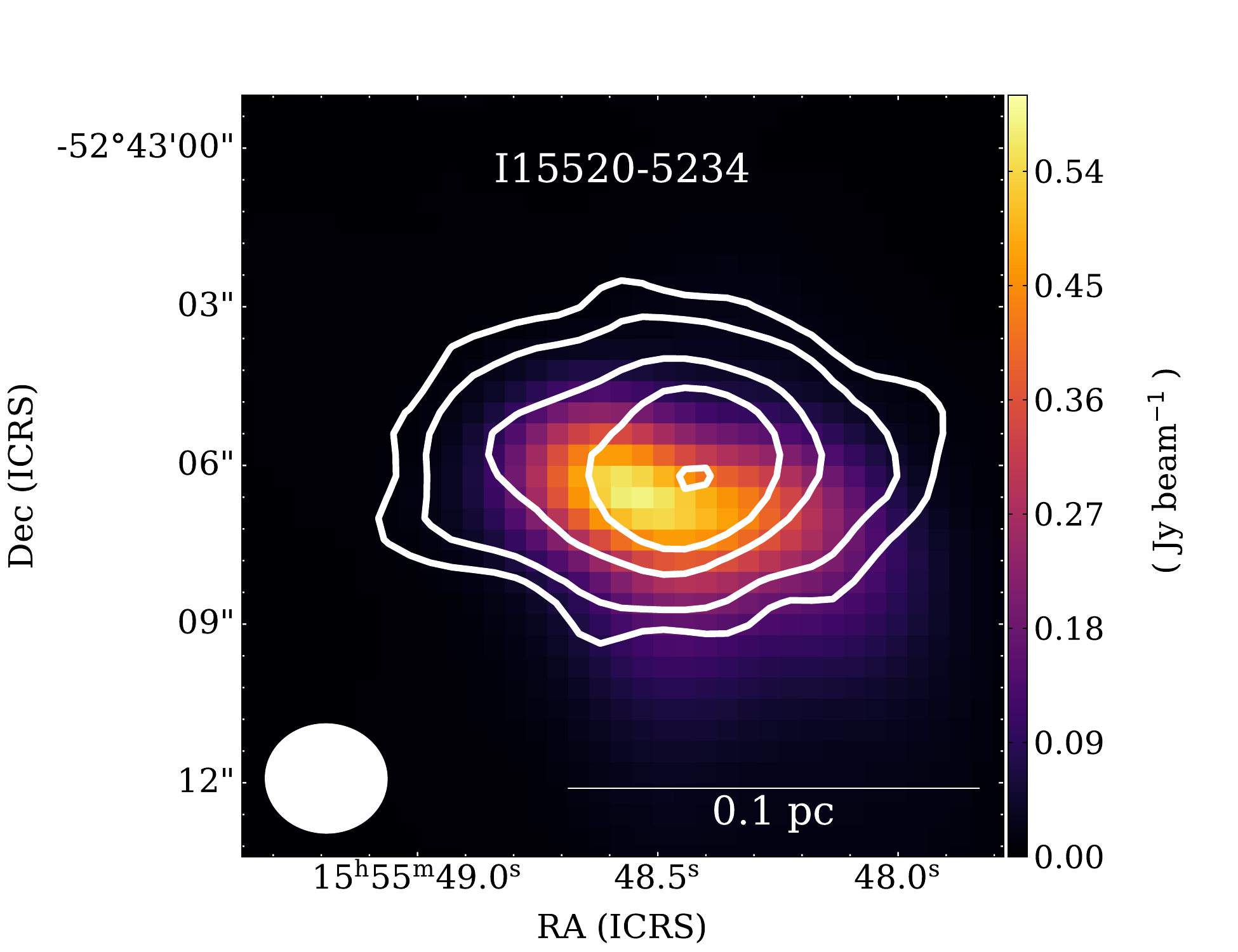}}
\quad 
{\includegraphics[height=4.01cm,width=5.21cm]{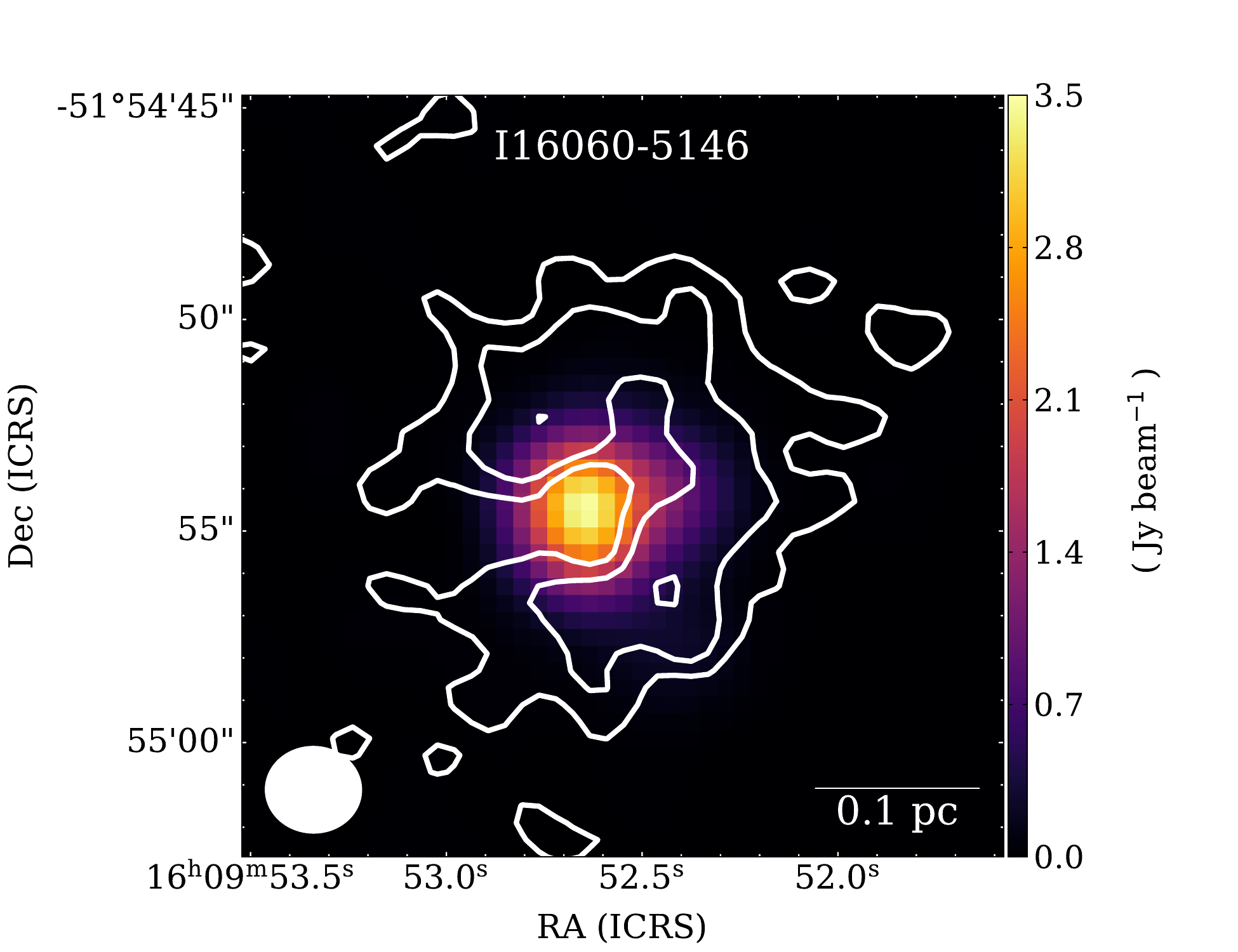}}
\caption{Moment 0 maps of the C$_2$H$_5$OH and CH$_3$OCH$_3$. The color grayscale background represents the continuum emission at a wavelength of 3 mm. The white contours delineate the flux levels of CH$_3$OCH$_3$ (E$_{\rm u}$ = 10.214 K). The red contours delineate the flux levels of C$_2$H$_5$OH (E$_{\rm u}$ = 35.173 K). The synthesized beam is shown in the lower left of each panel.}
\label{fig:C1}
\end{figure}
\setcounter{figure}{\value{figure}-1}
\begin{figure}
  \centering 

{\includegraphics[height=4.01cm,width=5.21cm]{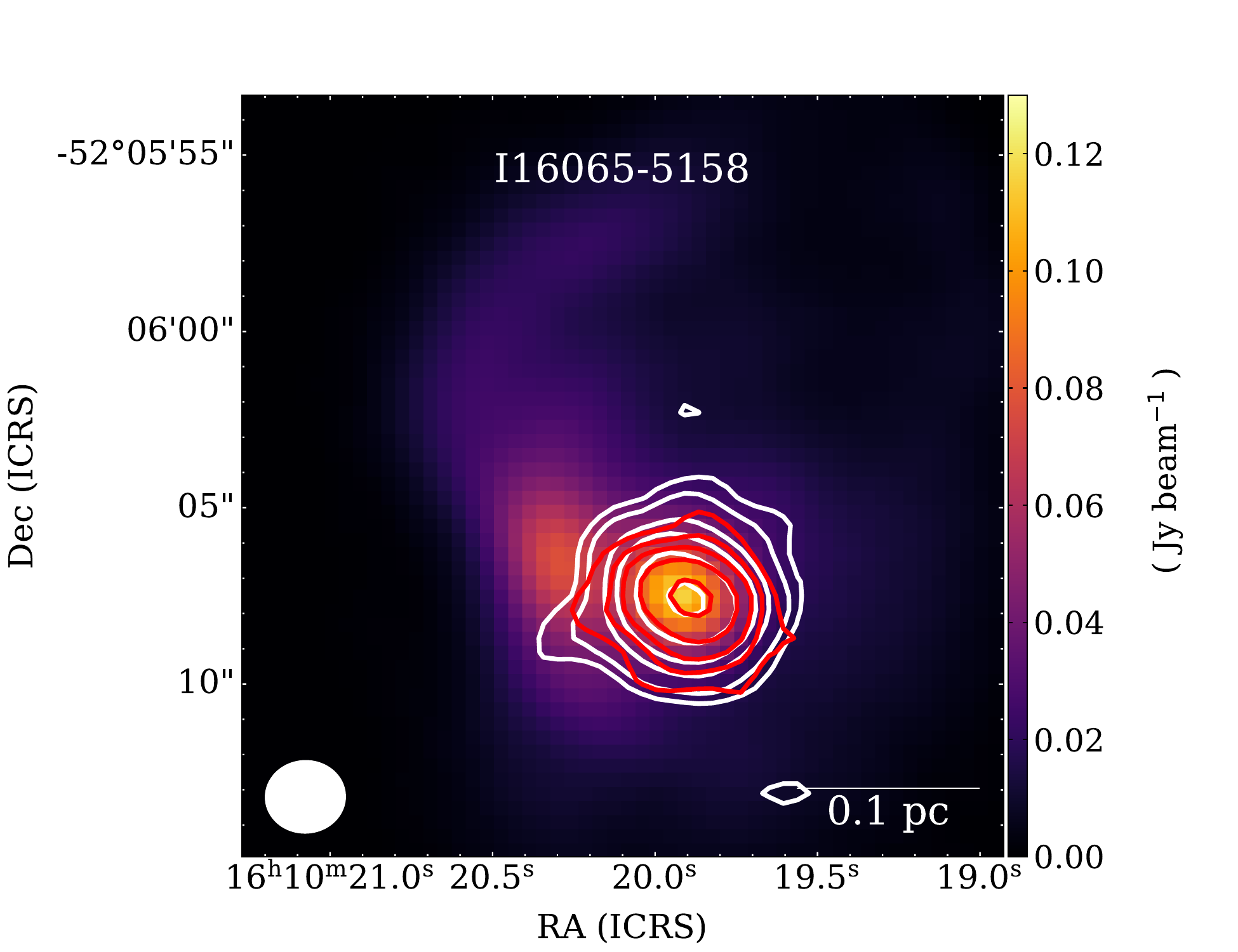}} 
\quad
{\includegraphics[height=4.01cm,width=5.21cm]{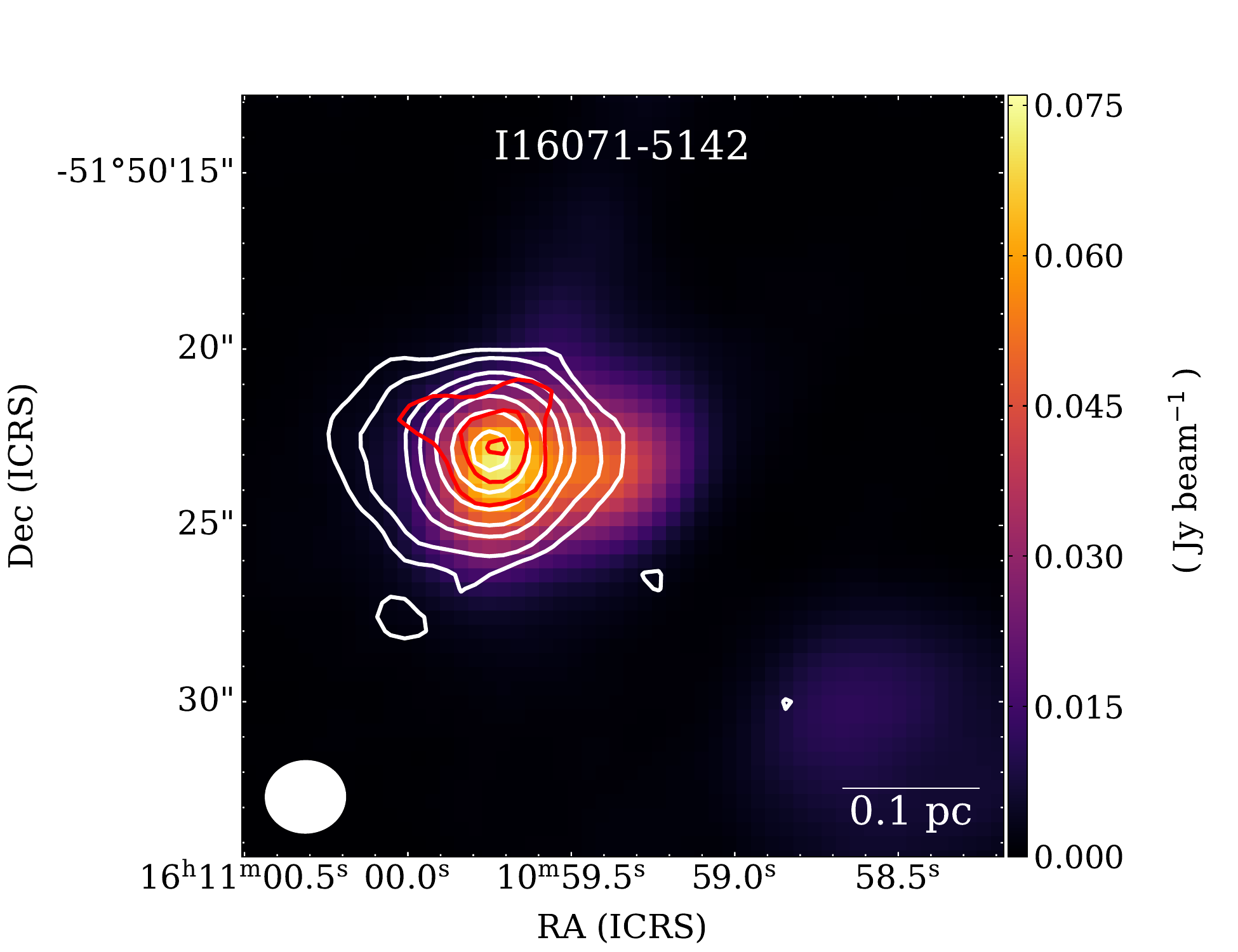}}
\quad
{\includegraphics[height=4.01cm,width=5.21cm]{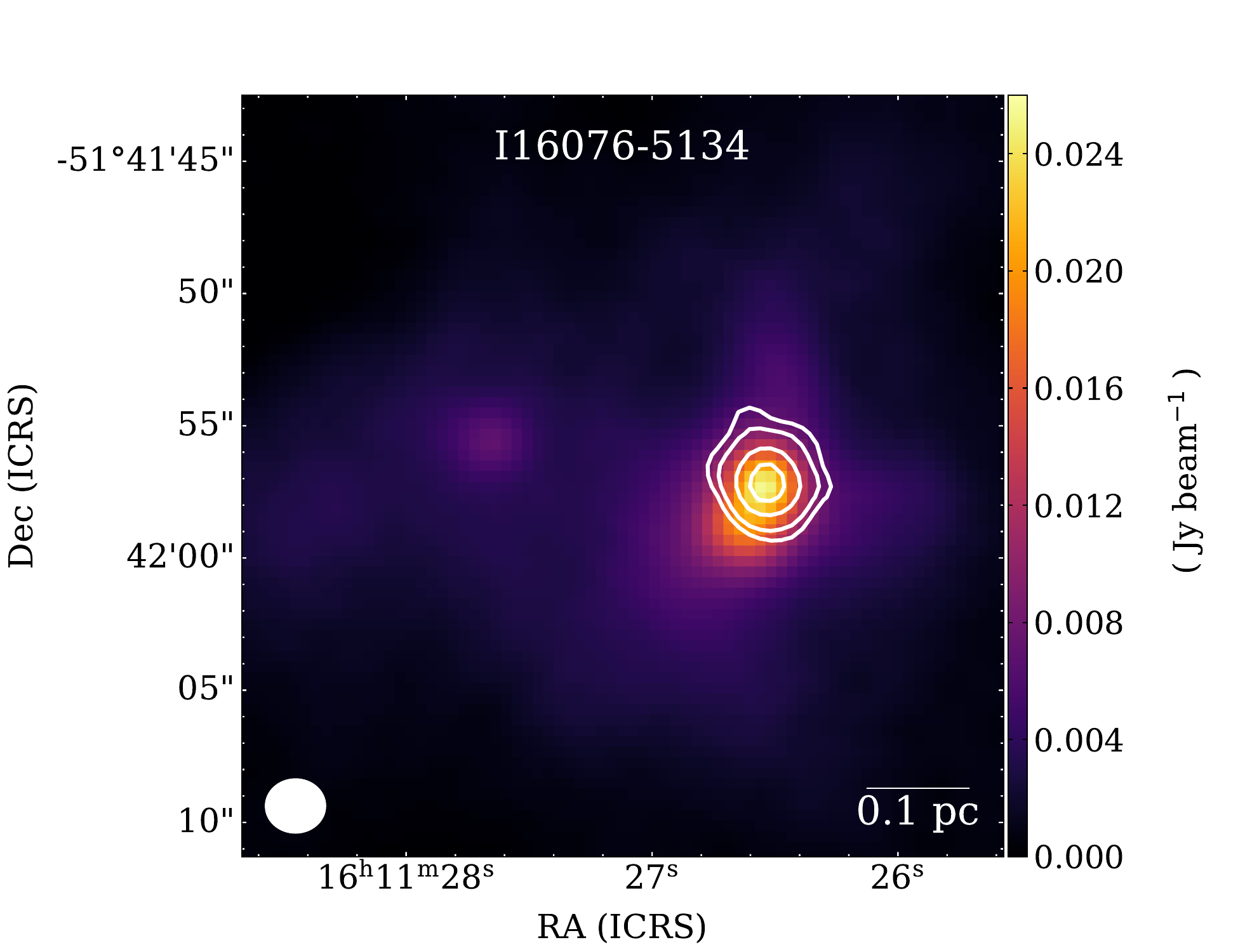}}
\quad 
{\includegraphics[height=4.01cm,width=5.21cm]{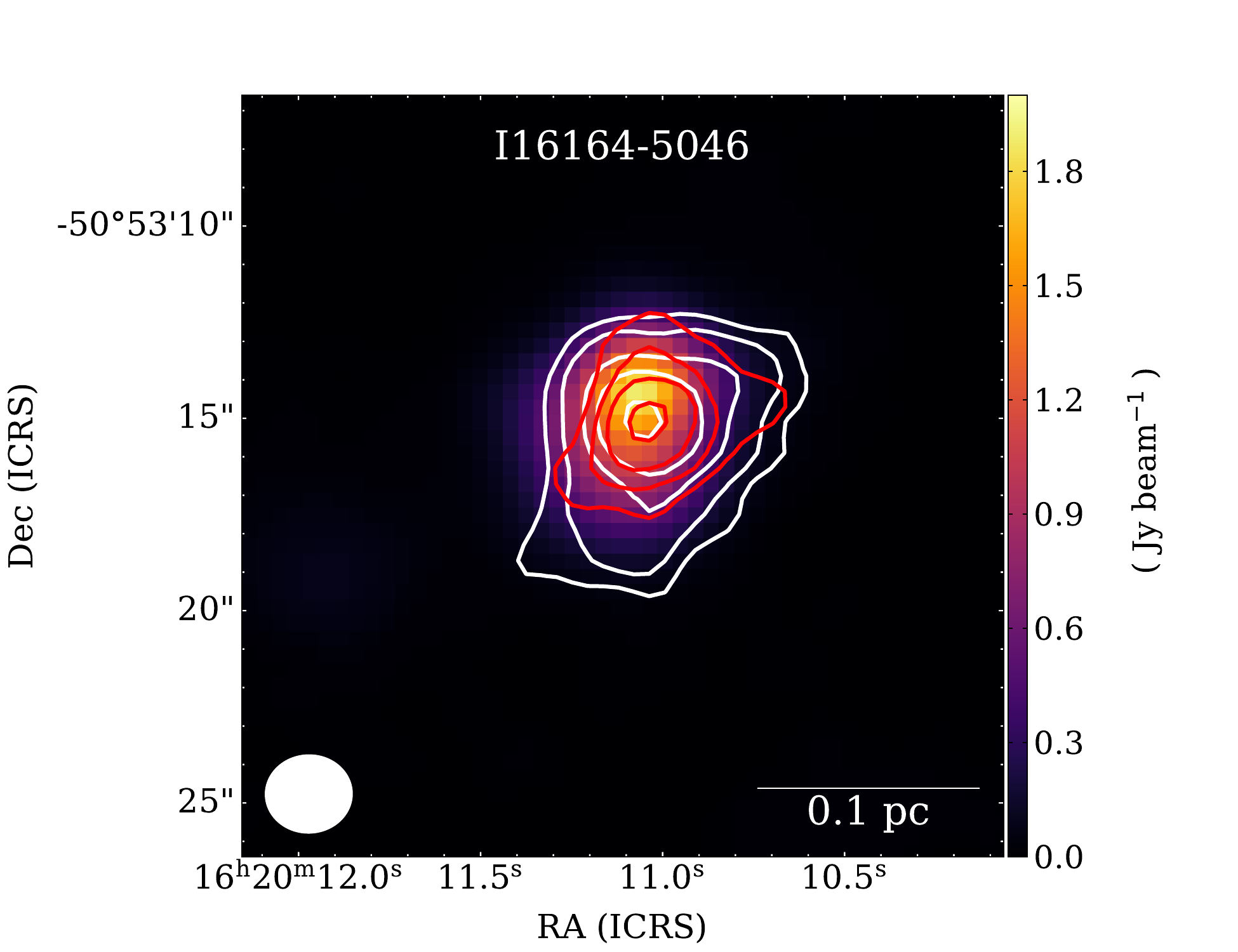}}
\quad 
{\includegraphics[height=4.01cm,width=5.21cm]{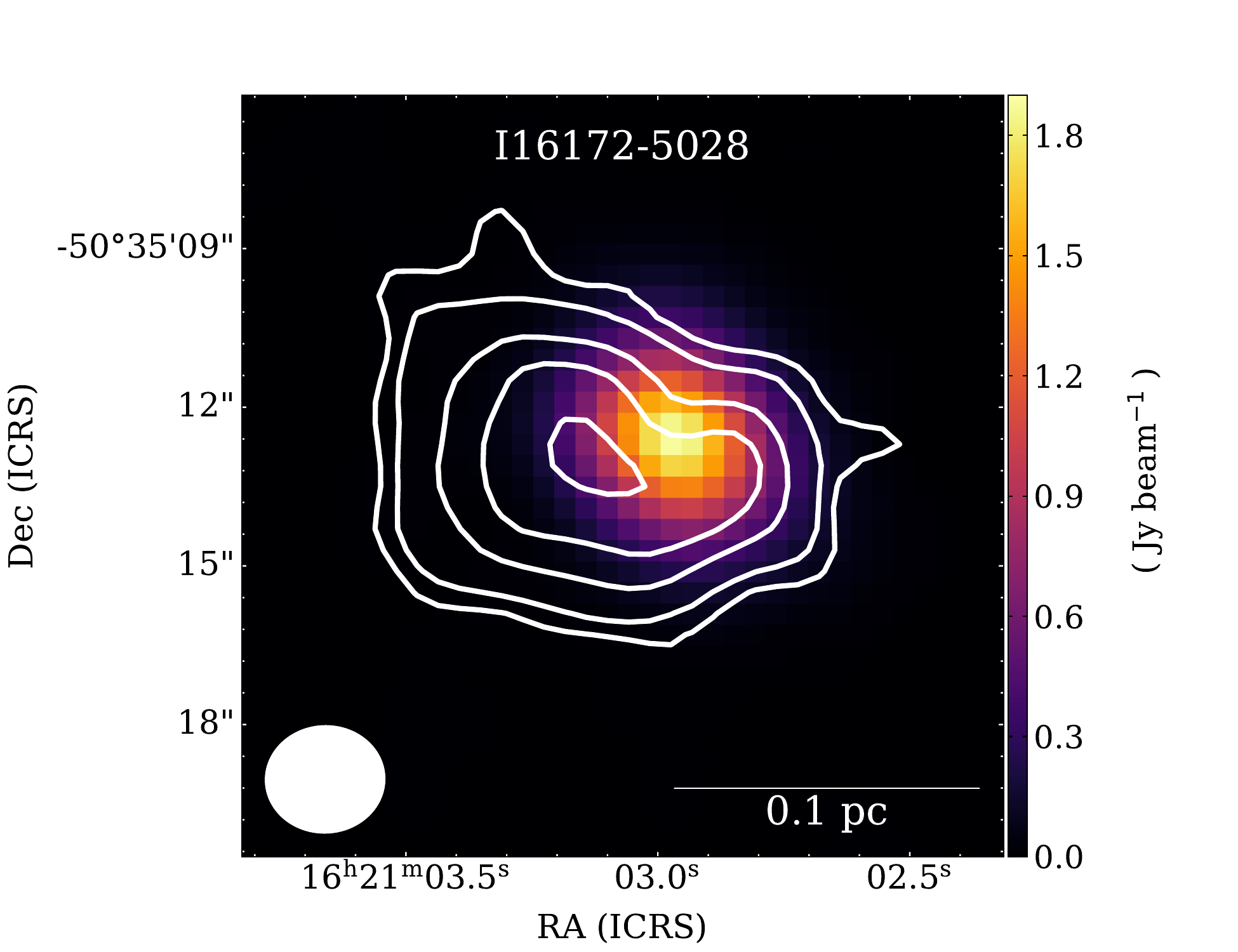}}
\quad
{\includegraphics[height=4.01cm,width=5.21cm]{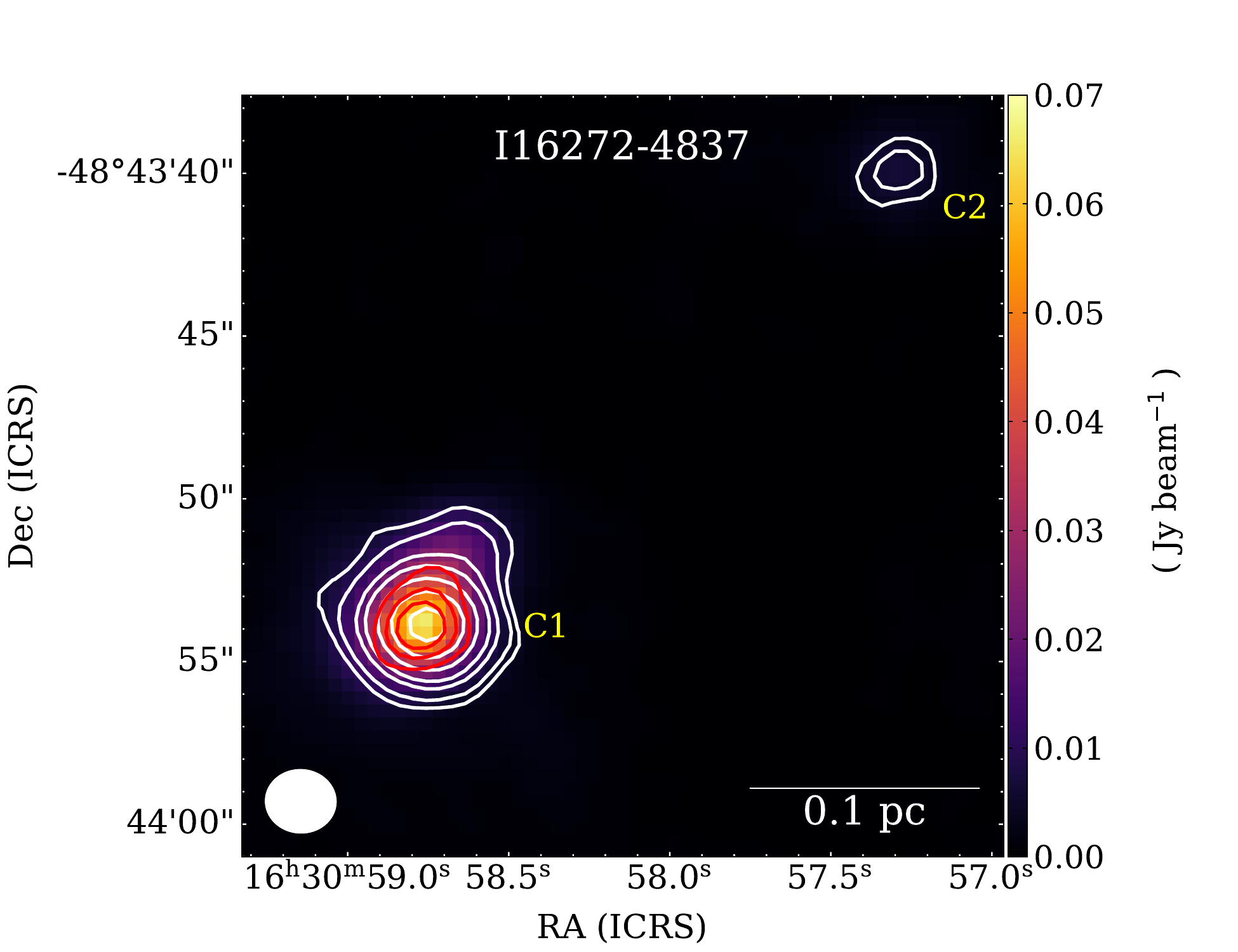}}
\quad
{\includegraphics[height=4.01cm,width=5.21cm]{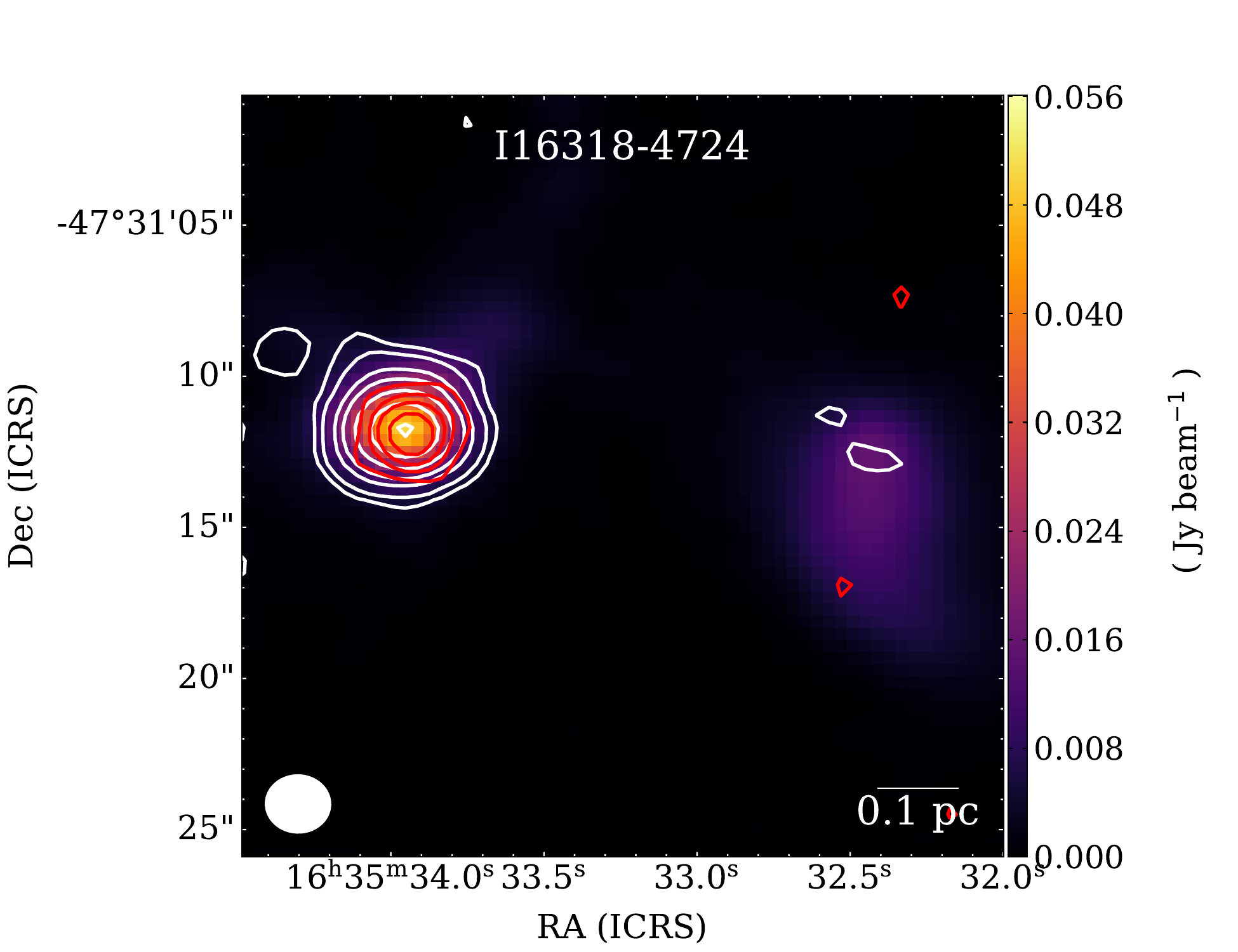}}
\quad
{\includegraphics[height=4.01cm,width=5.21cm]{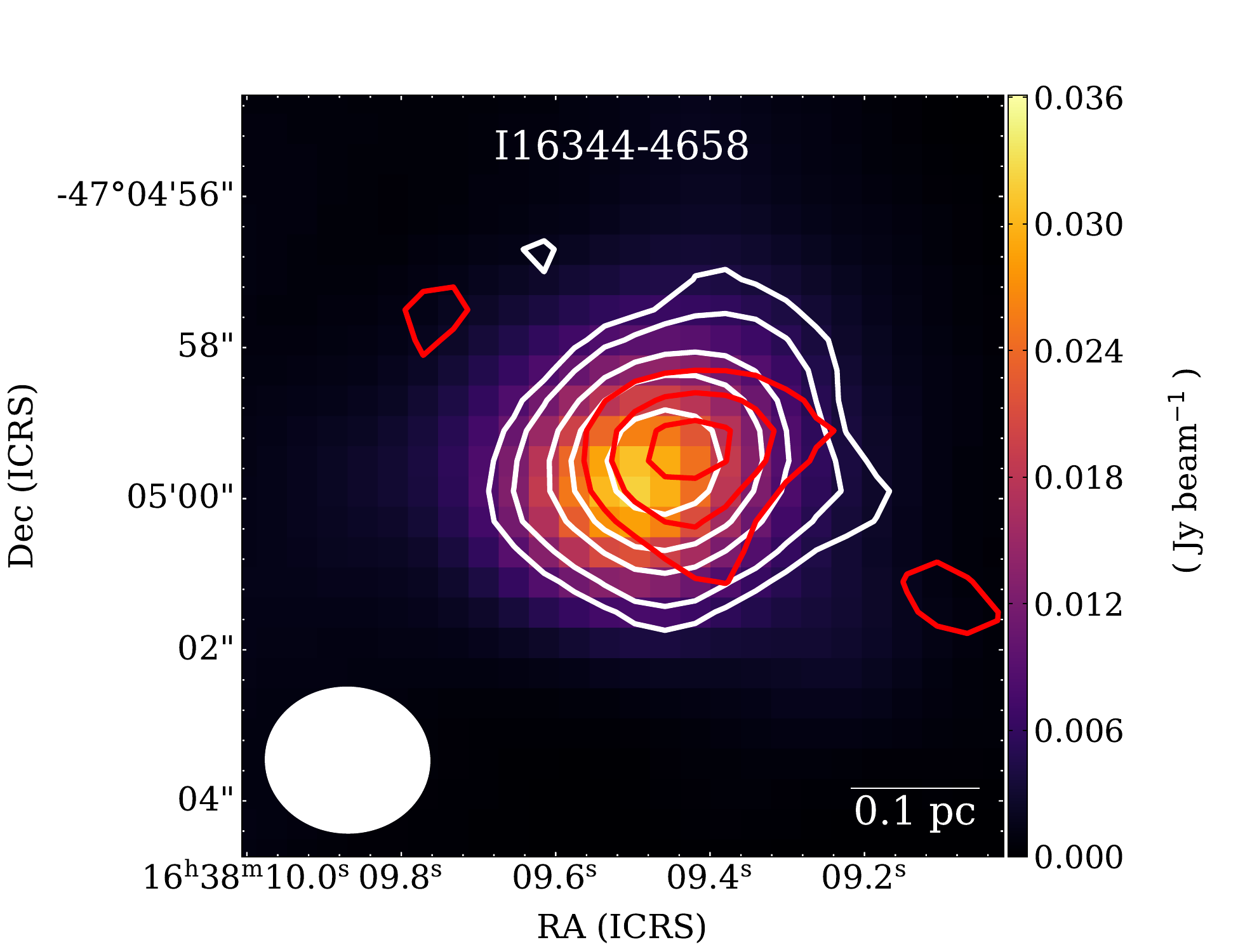}}
\quad
{\includegraphics[height=4.01cm,width=5.21cm]{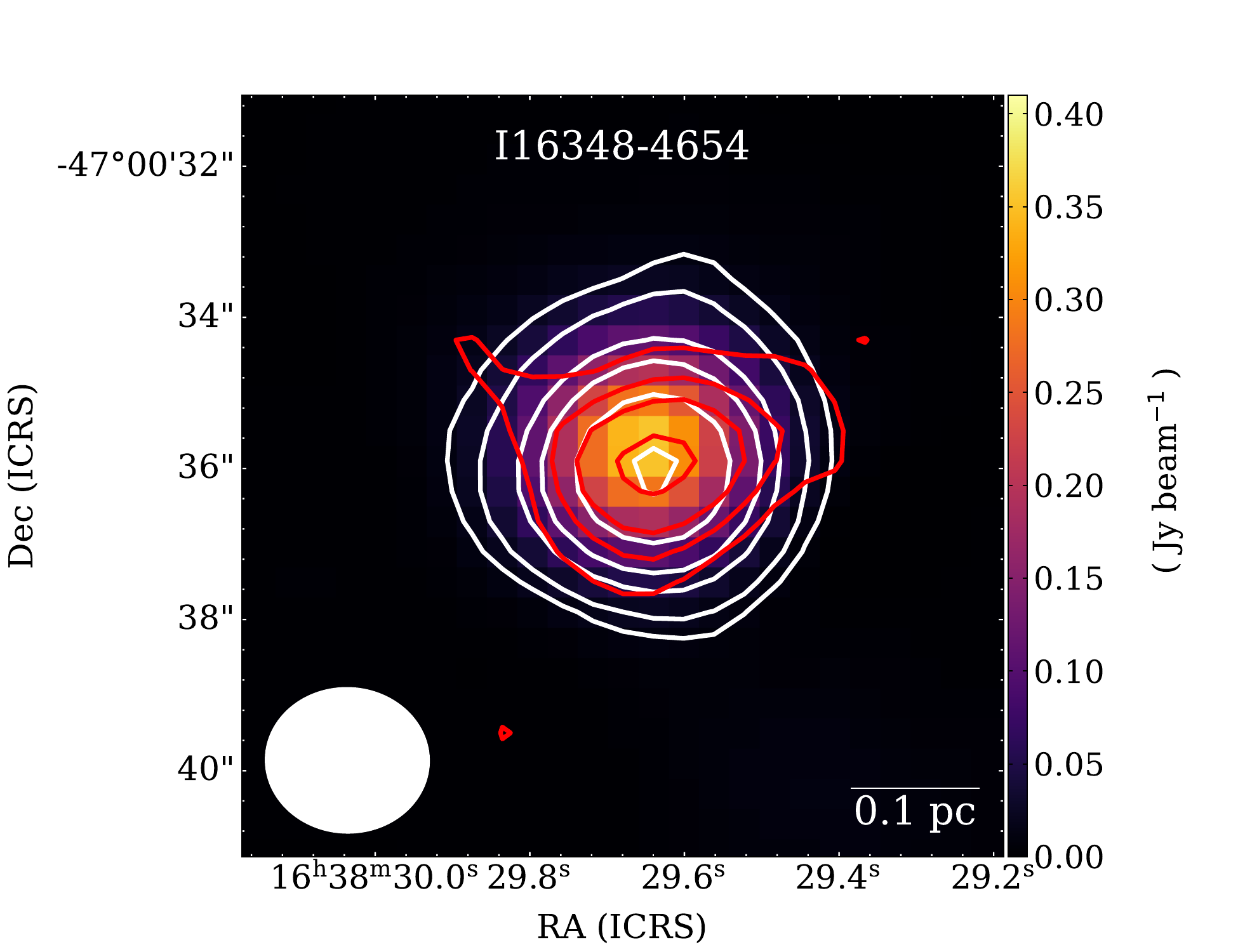}}
\quad 
{\includegraphics[height=4.01cm,width=5.21cm]{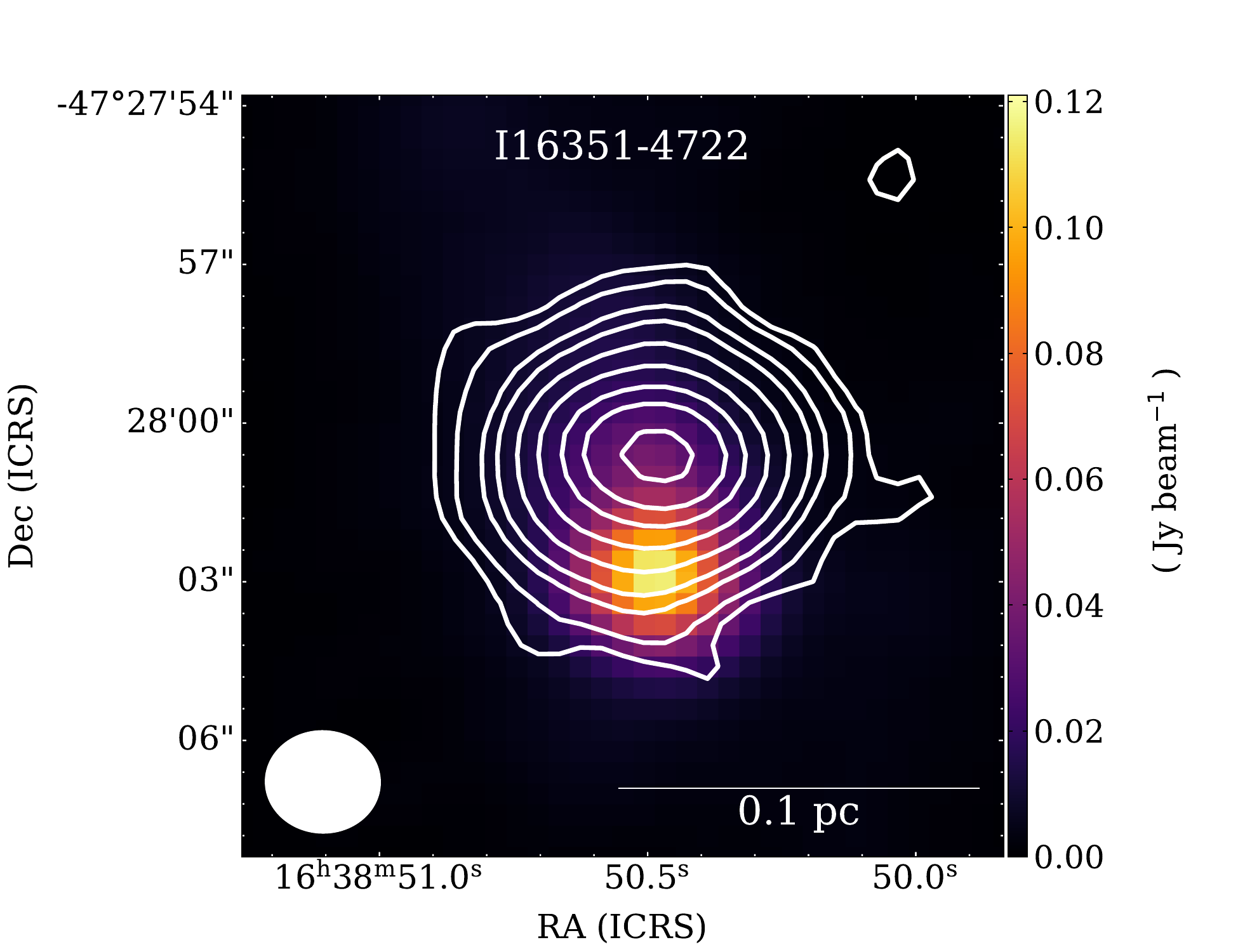}}
\quad
{\includegraphics[height=4.01cm,width=5.21cm]{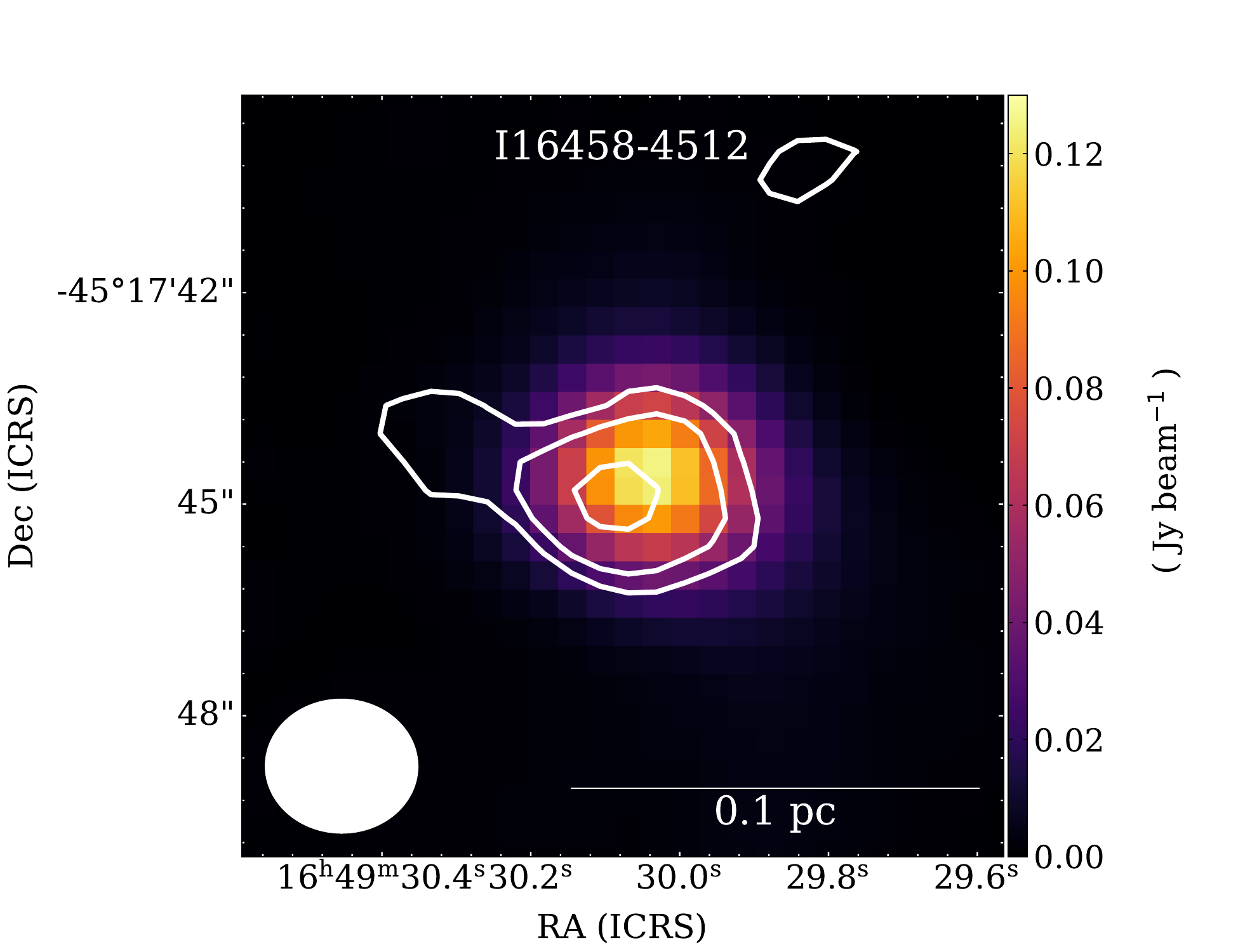}}
\quad
{\includegraphics[height=4.01cm,width=5.21cm]{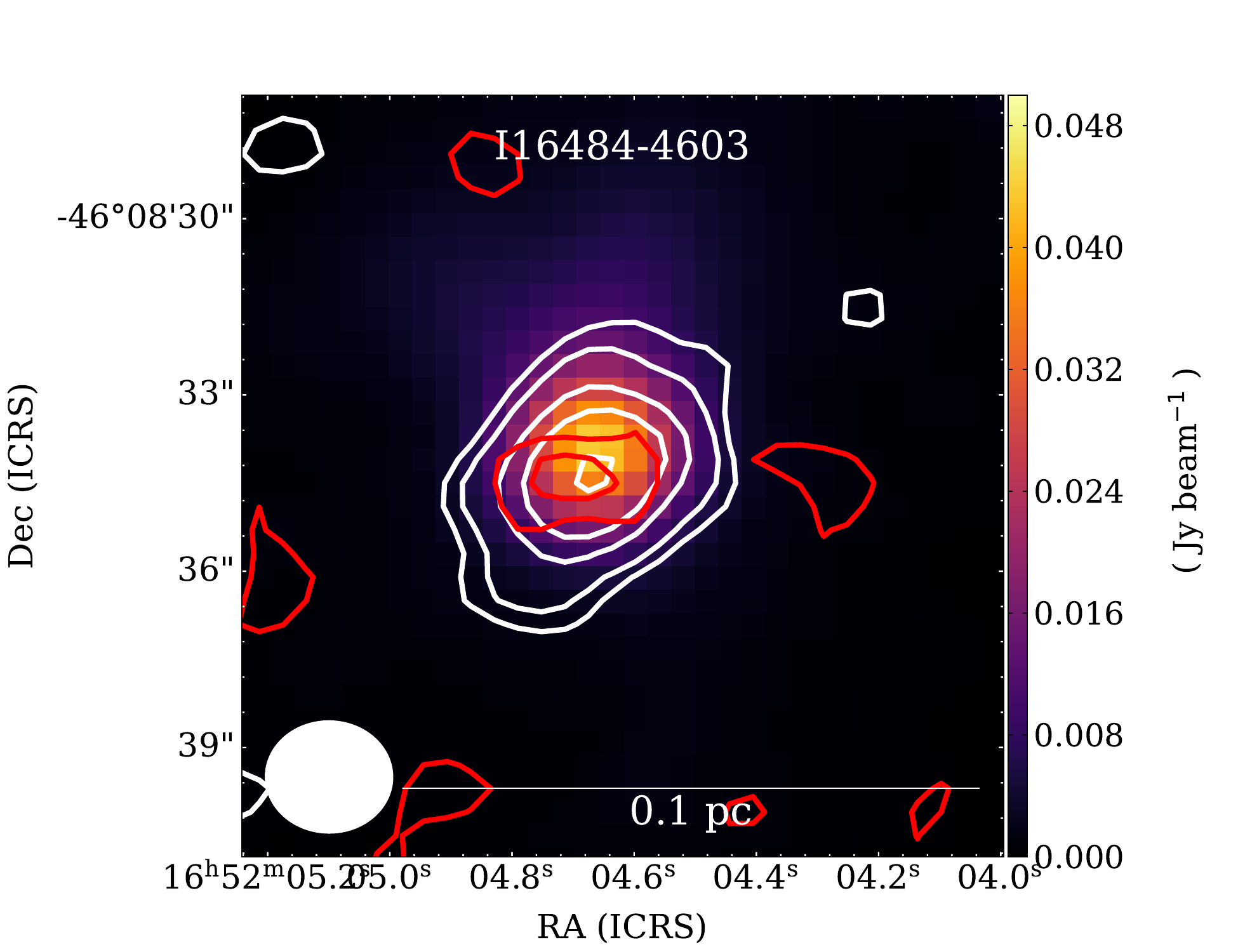}}
\quad
{\includegraphics[height=4.01cm,width=5.21cm]{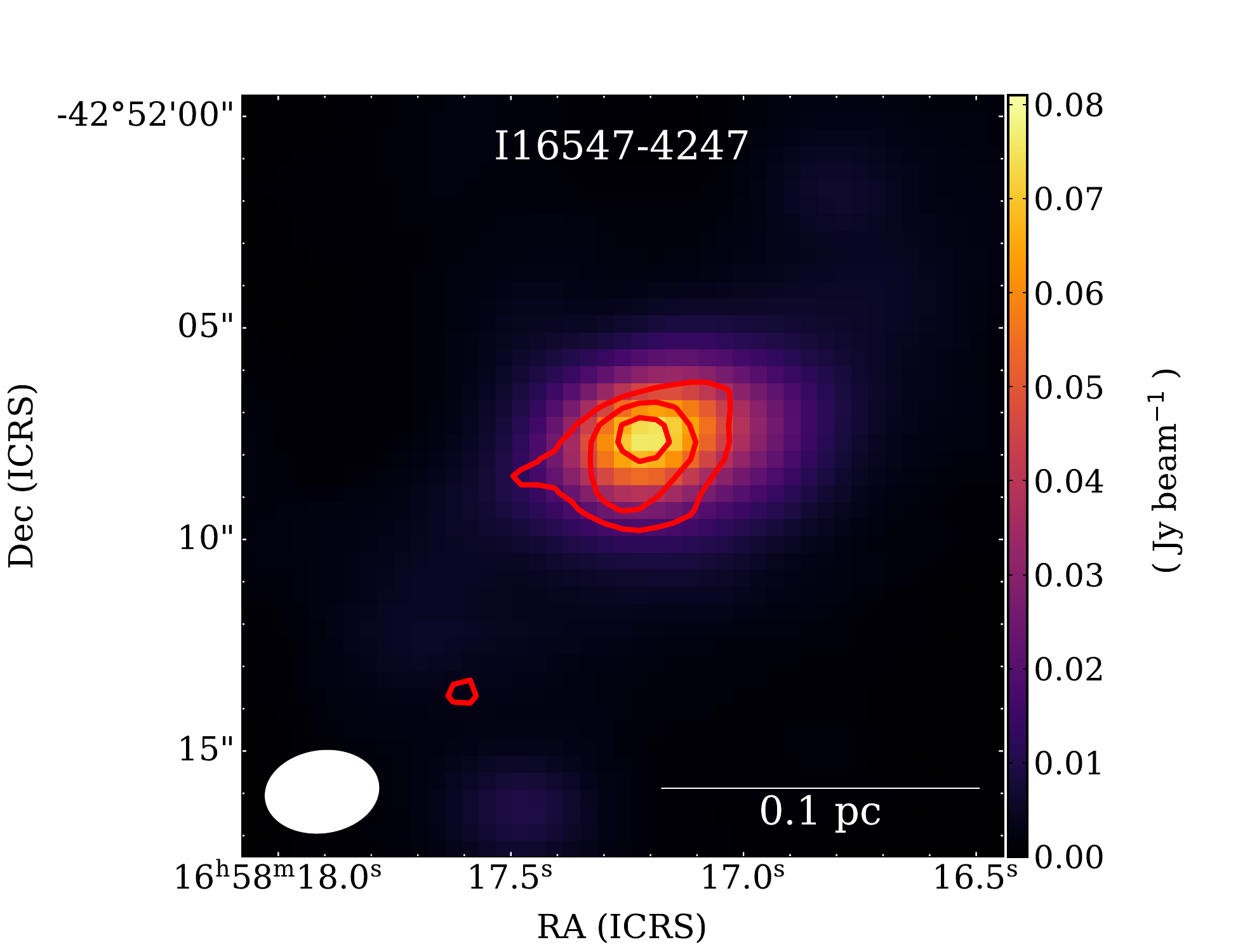}} 
\quad
{\includegraphics[height=4.01cm,width=5.21cm]{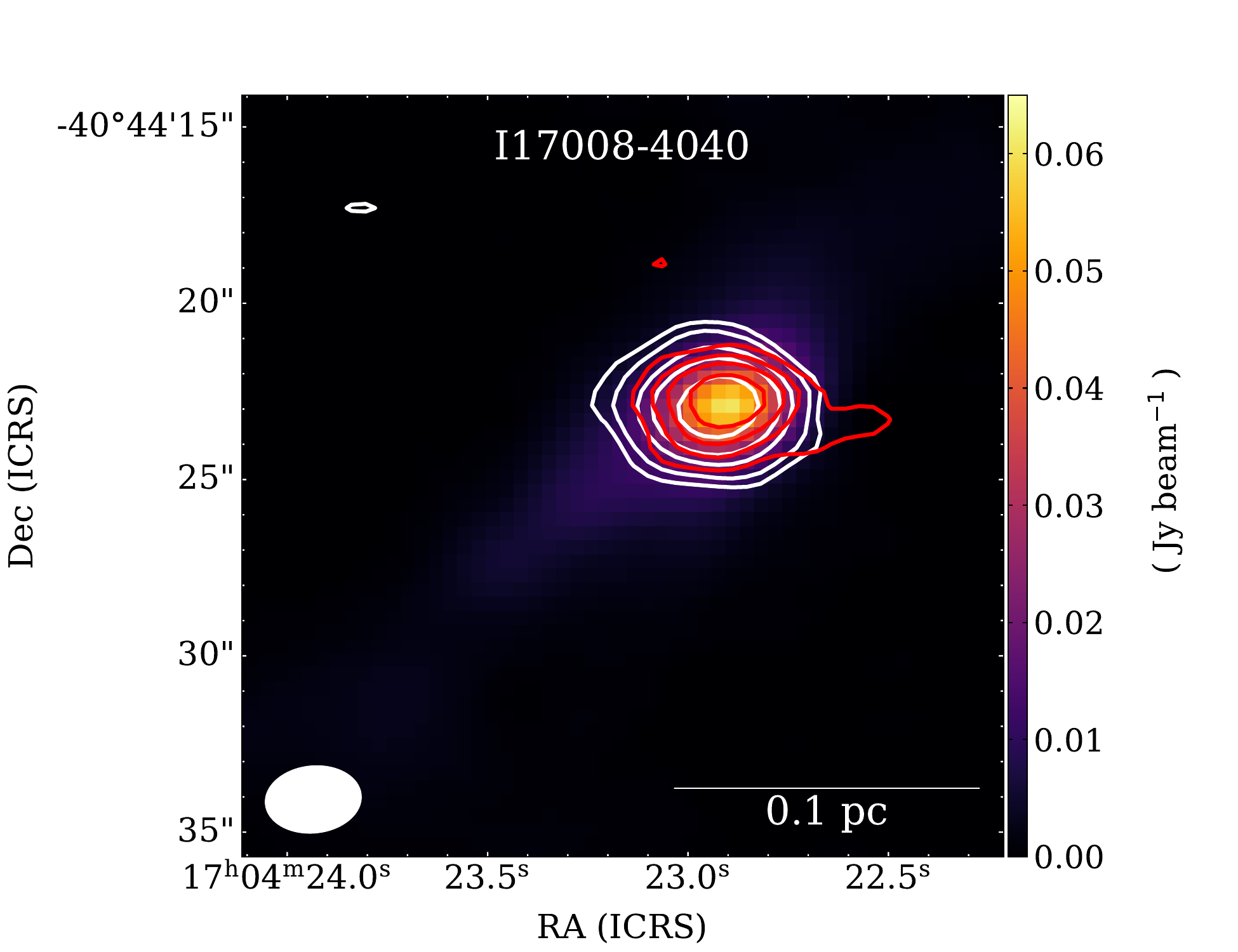}}
\quad
{\includegraphics[height=4.01cm,width=5.21cm]{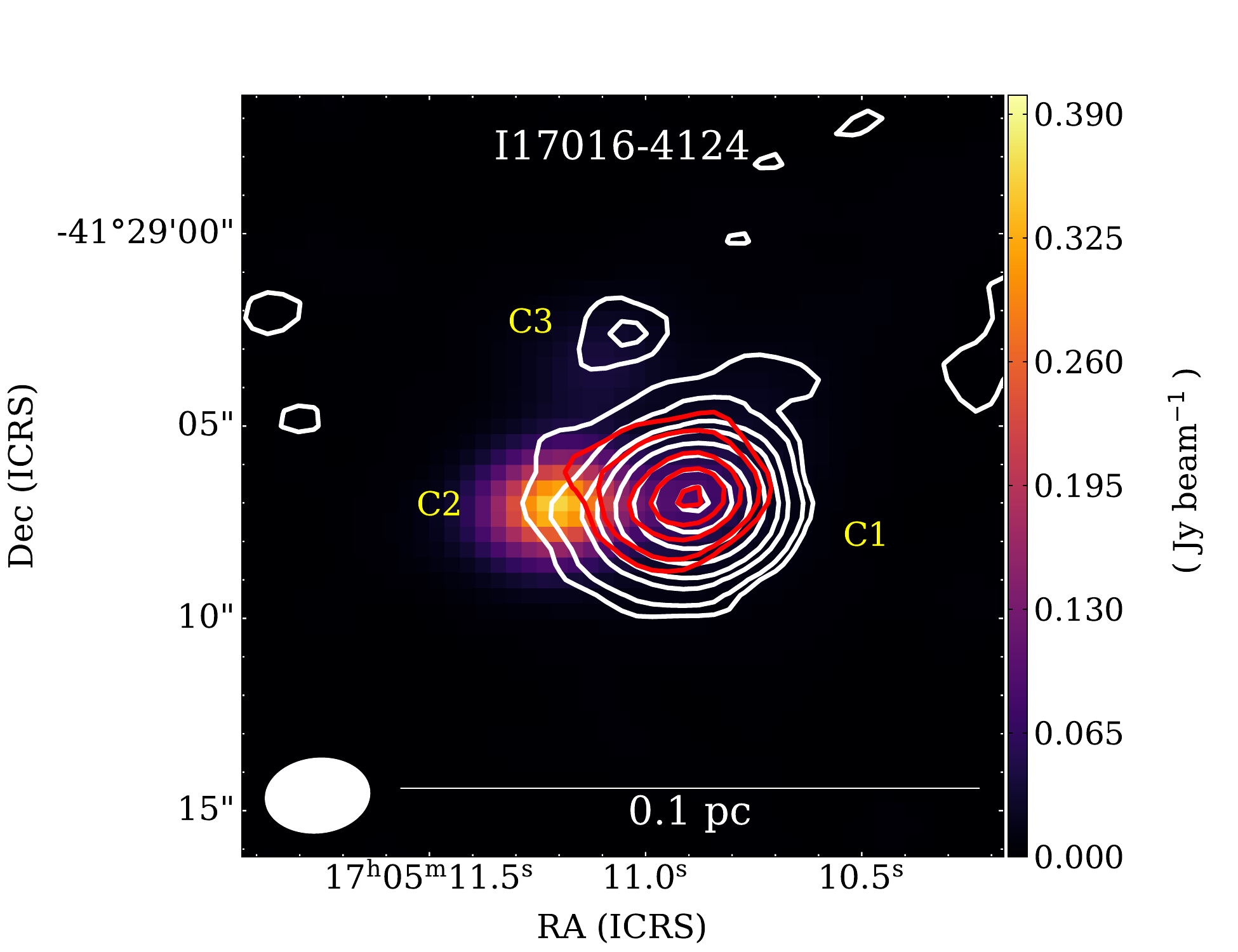}}
\caption{Continued.}
\end{figure}

\setcounter{figure}{\value{figure}-1}
\begin{figure}
  \centering 
{\includegraphics[height=4.01cm,width=5.21cm]{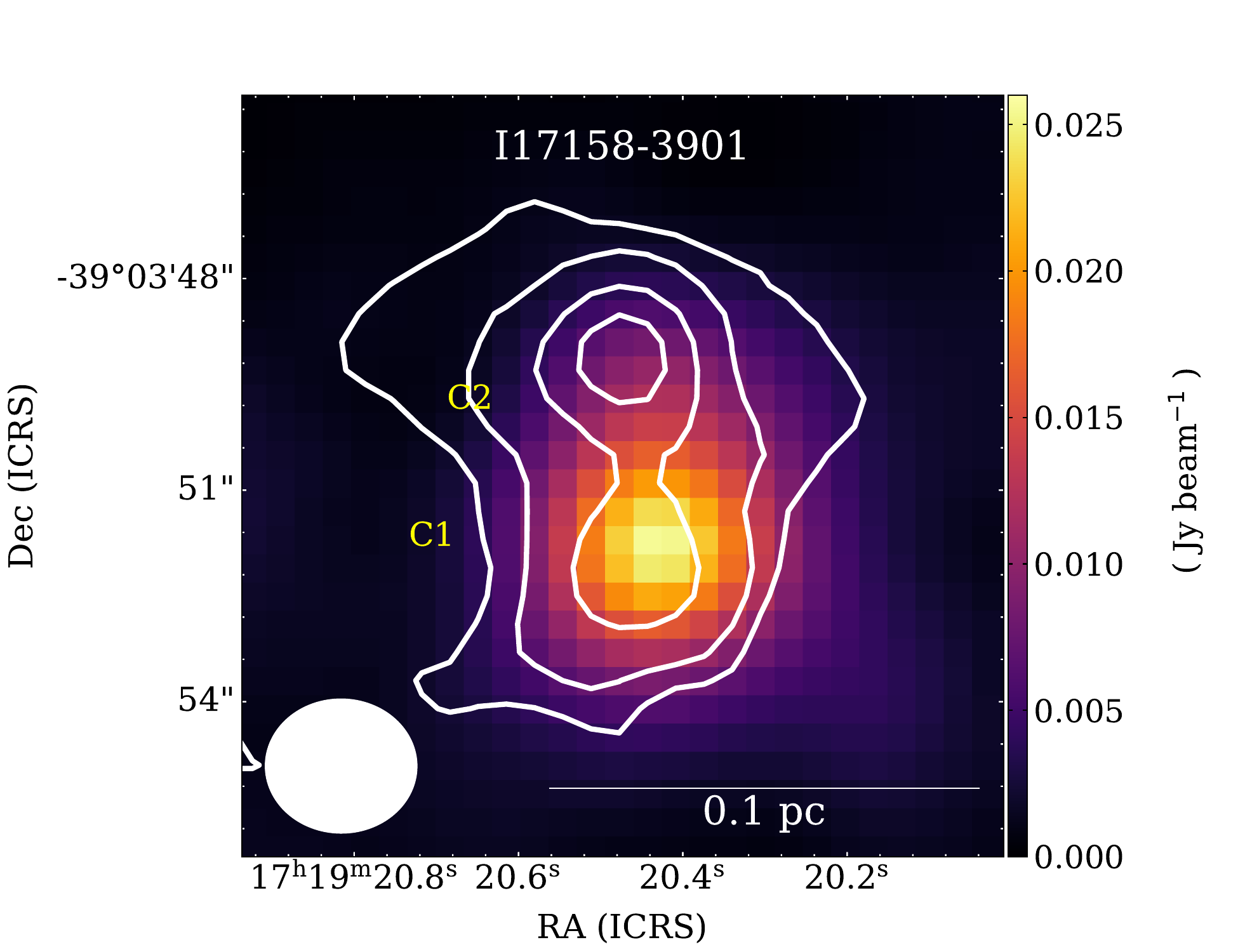}} 
\quad
{\includegraphics[height=4.01cm,width=5.21cm]{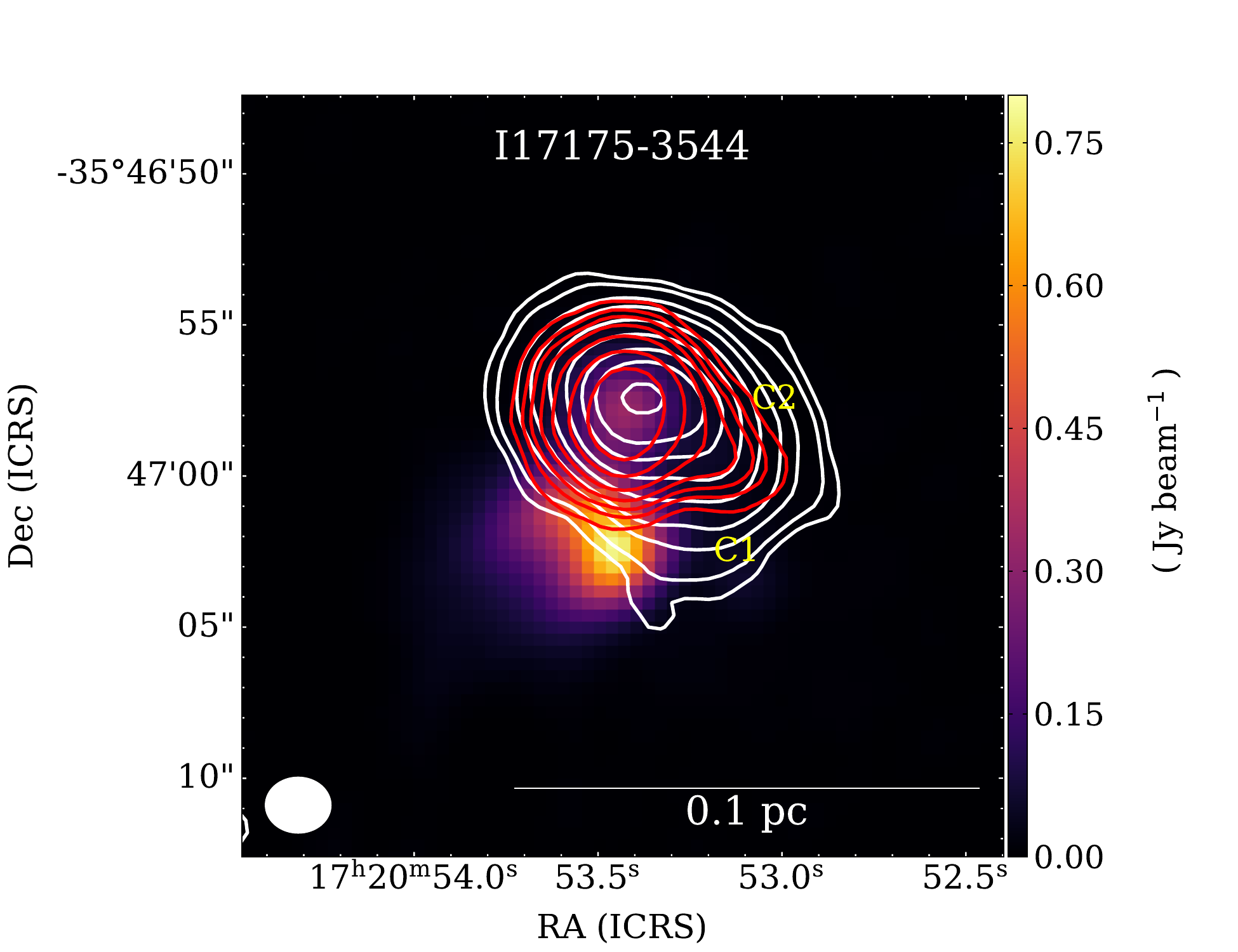}}
\quad
{\includegraphics[height=4.01cm,width=5.21cm]{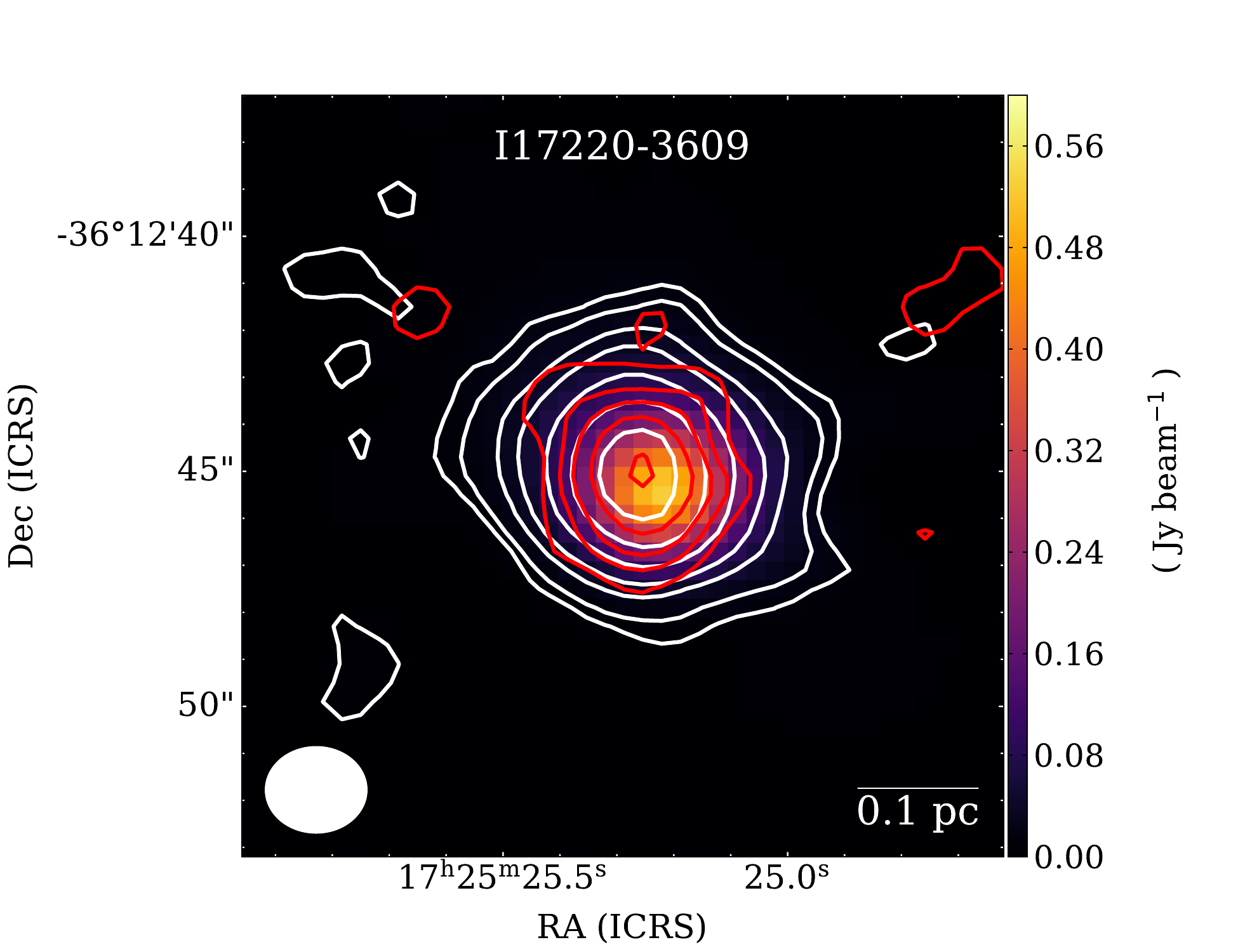}}
\quad 
{\includegraphics[height=4.01cm,width=5.21cm]{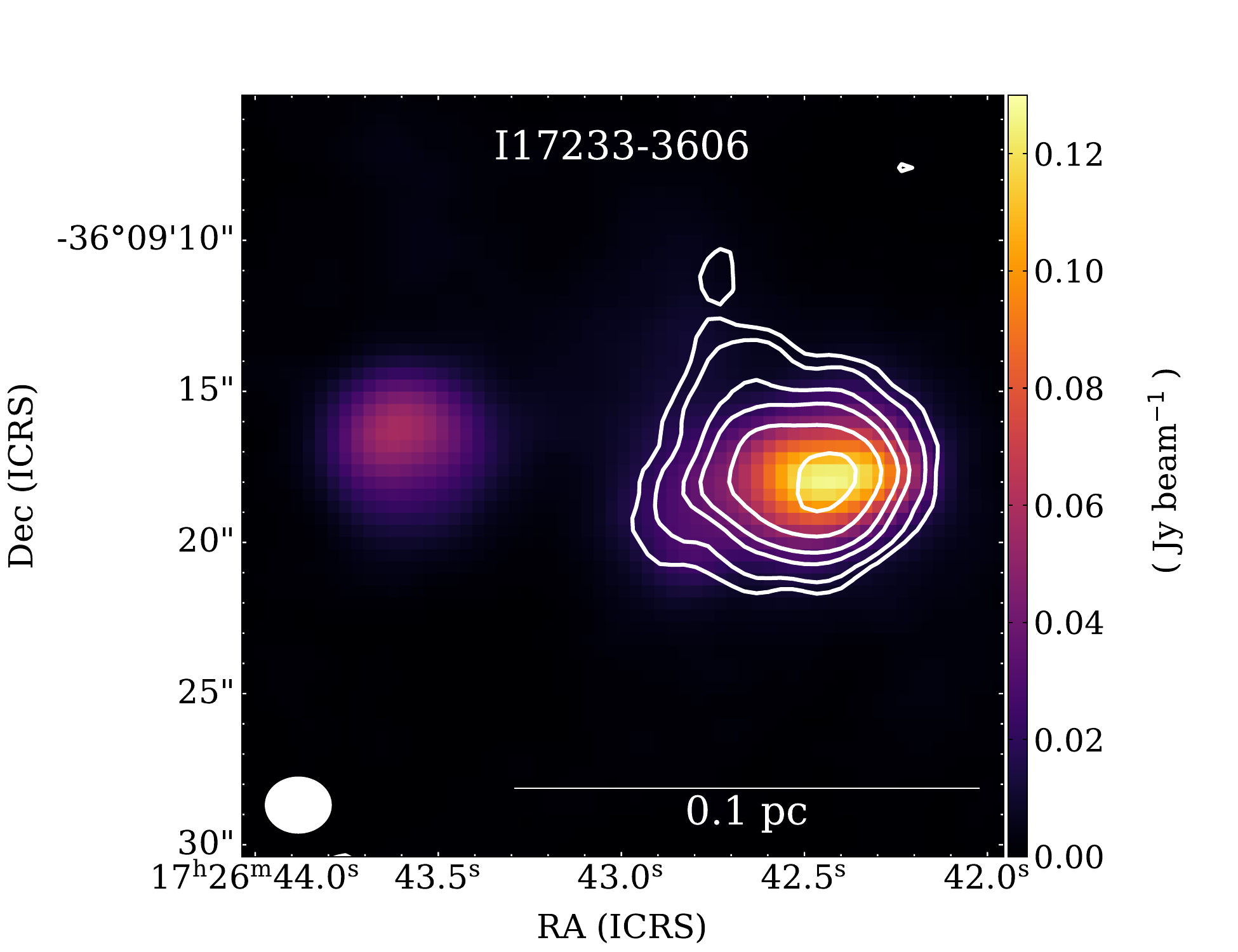}}
\quad 
{\includegraphics[height=4.01cm,width=5.21cm]{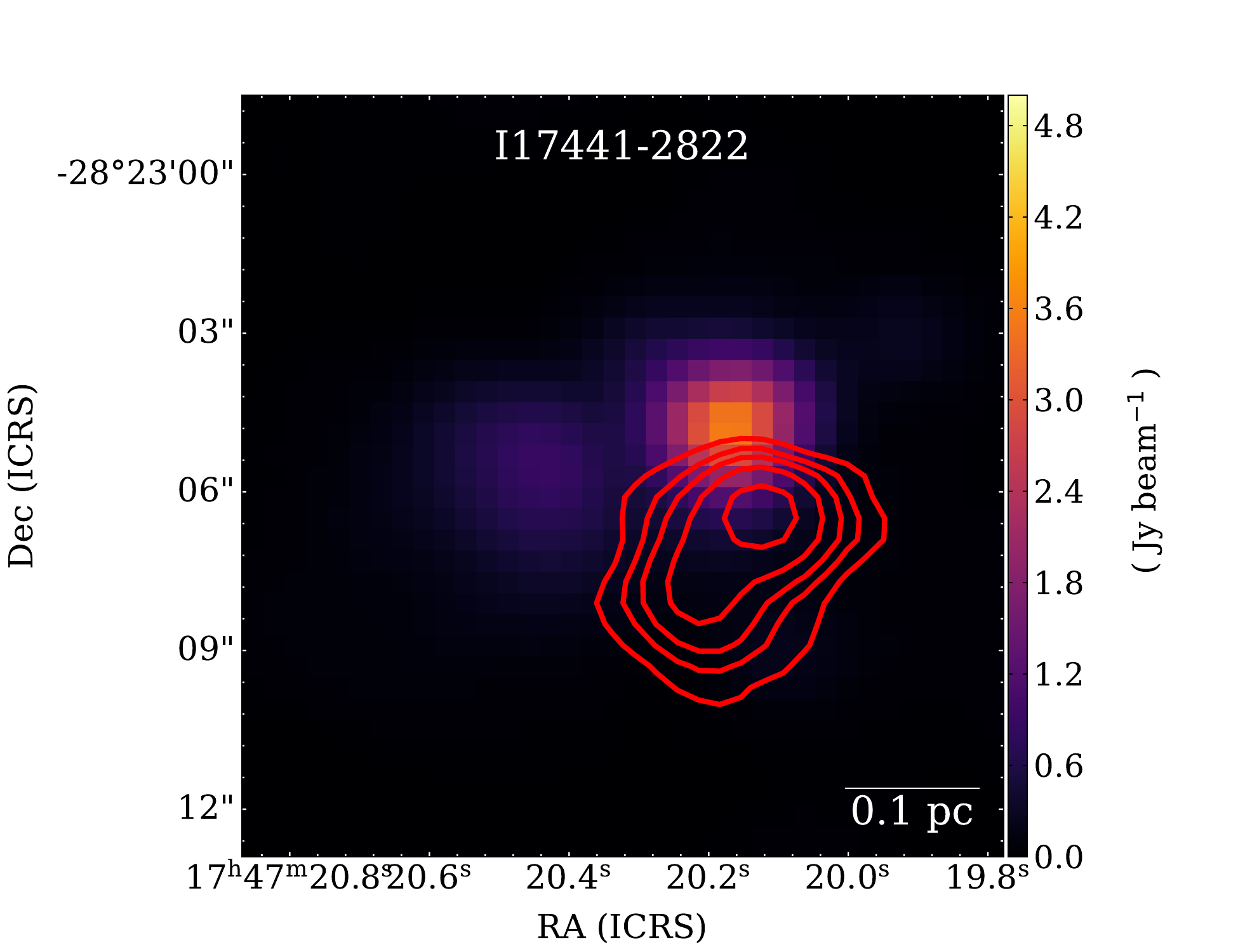}}
\quad
{\includegraphics[height=4.01cm,width=5.21cm]{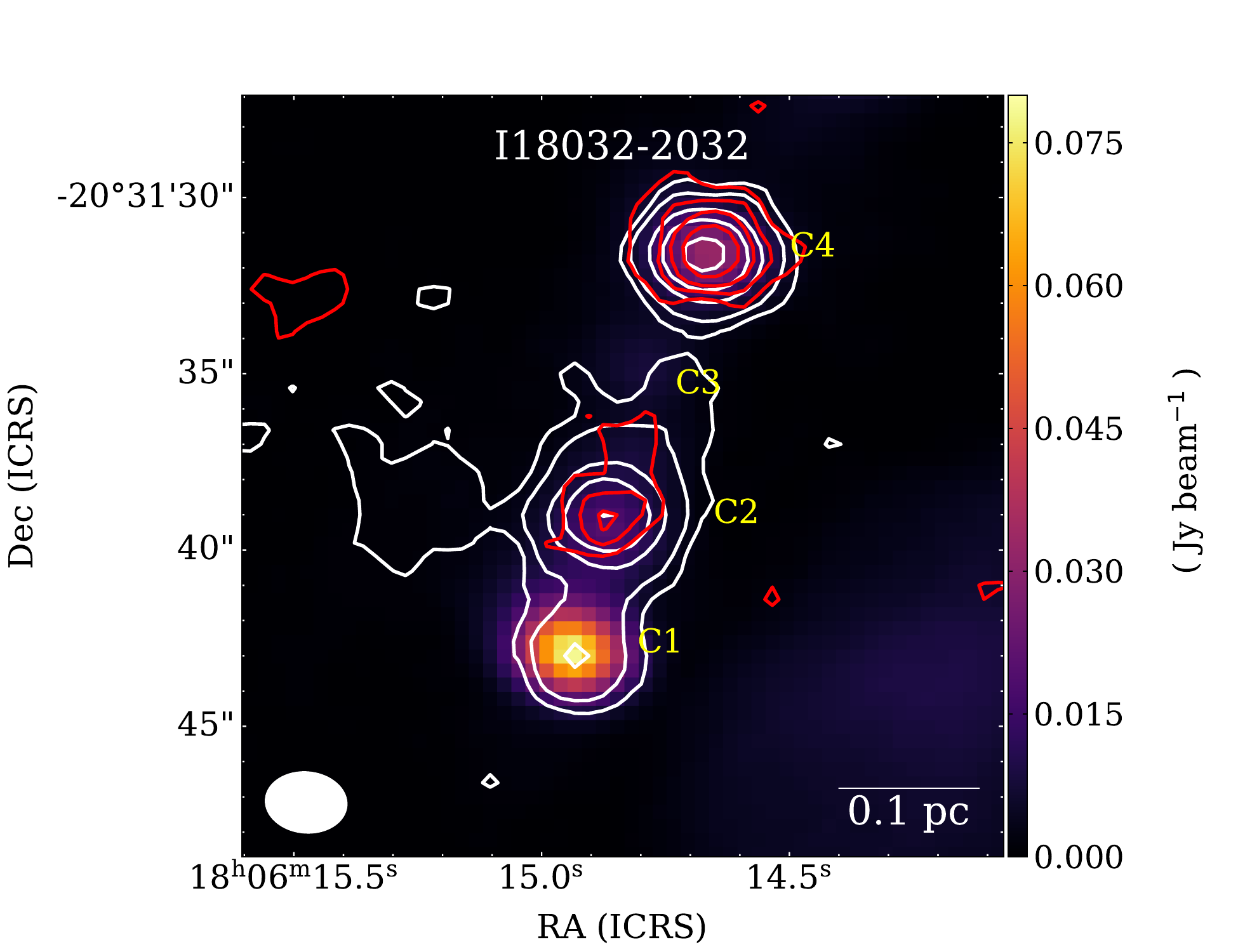}}
\quad
{\includegraphics[height=4.01cm,width=5.21cm]{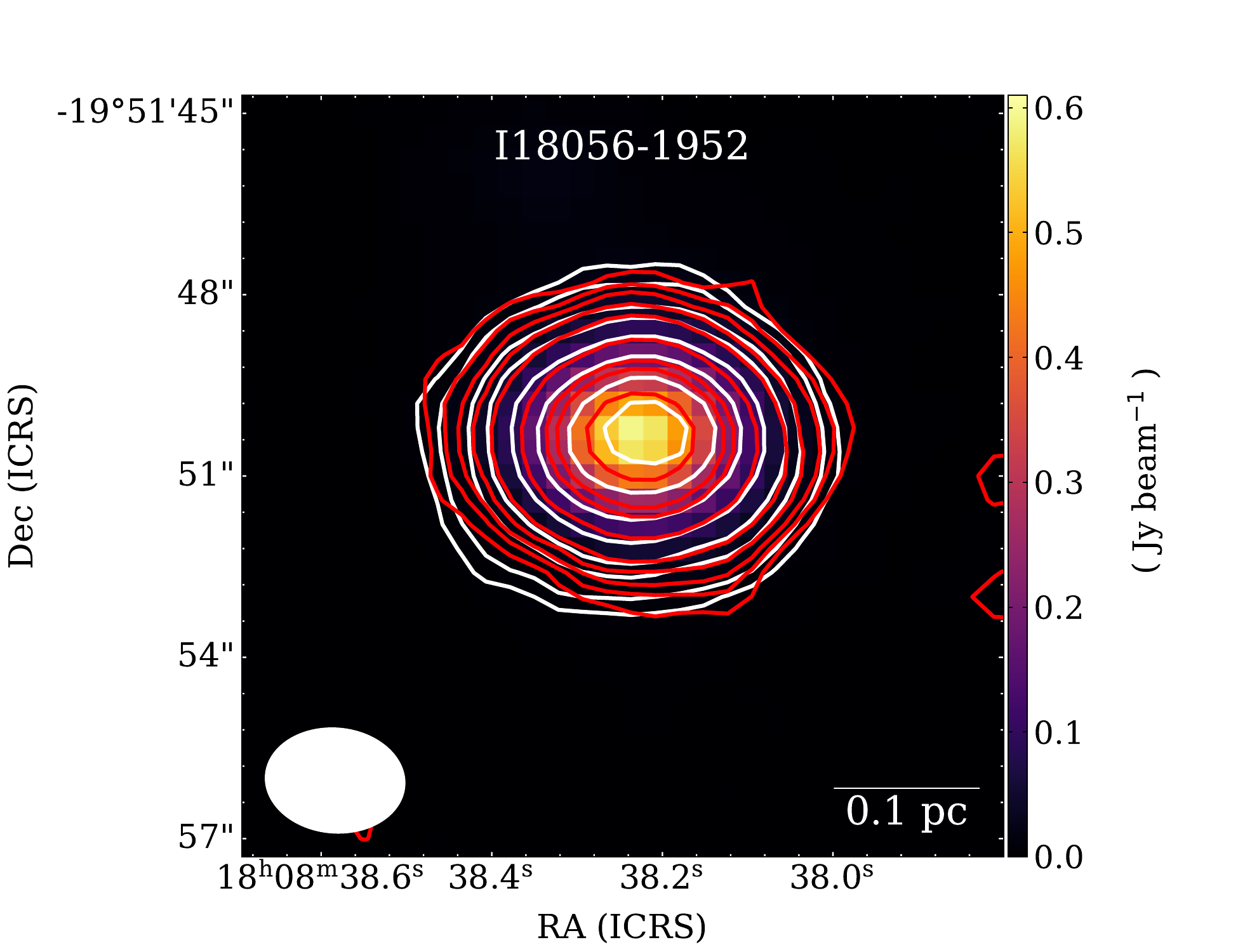}}
\quad
{\includegraphics[height=4.01cm,width=5.21cm]{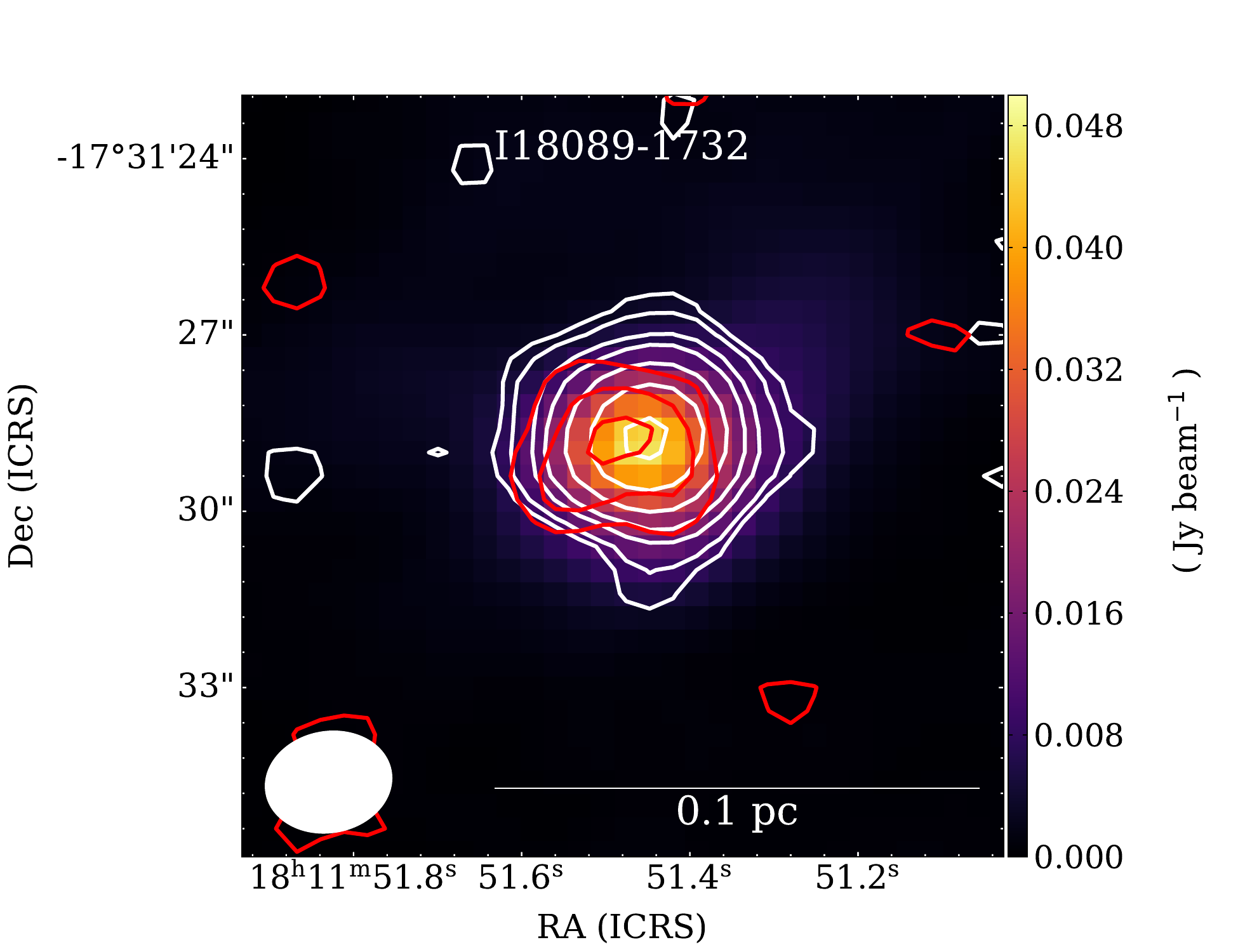}}
\quad
{\includegraphics[height=4.01cm,width=5.21cm]{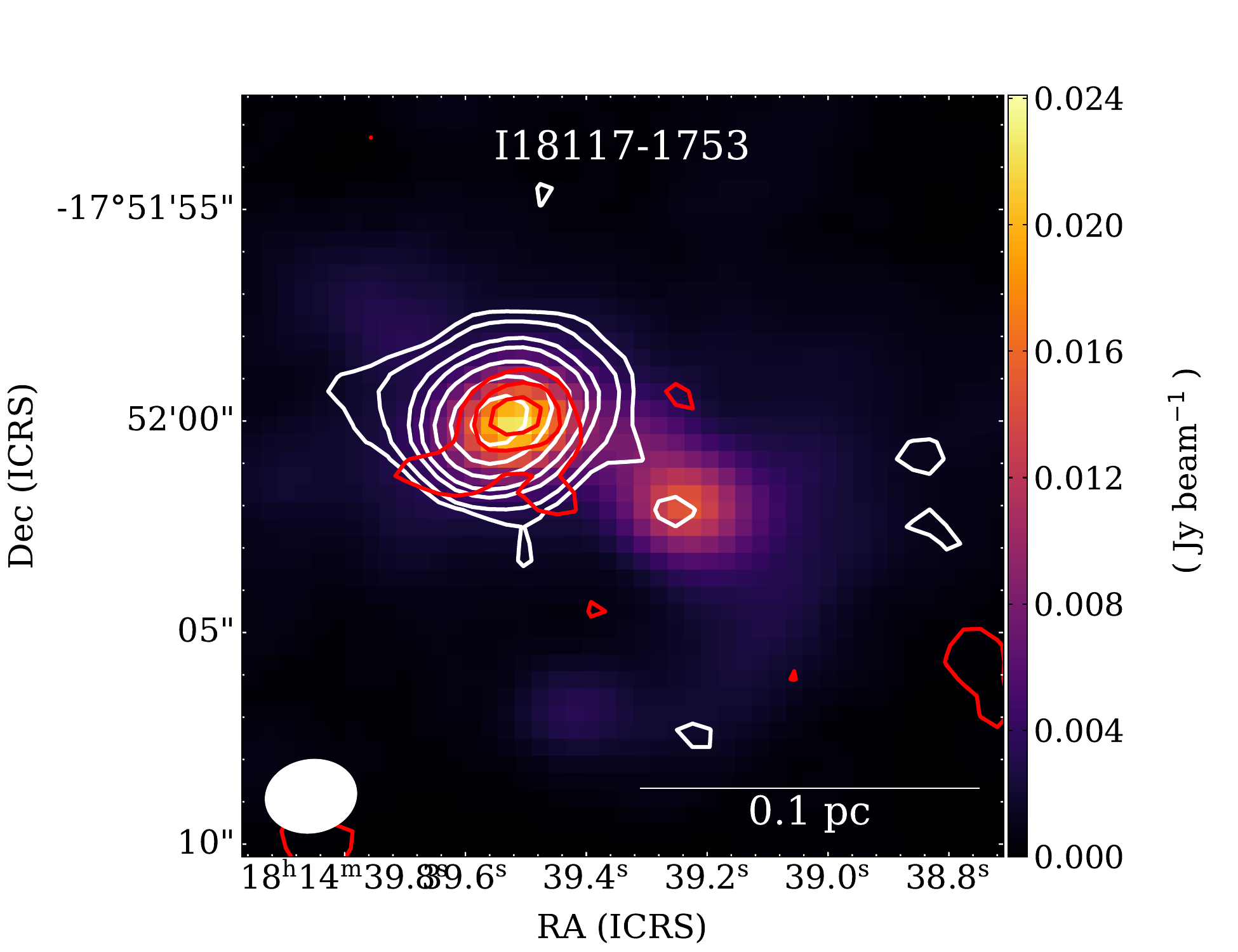}}
\quad 
{\includegraphics[height=4.01cm,width=5.21cm]{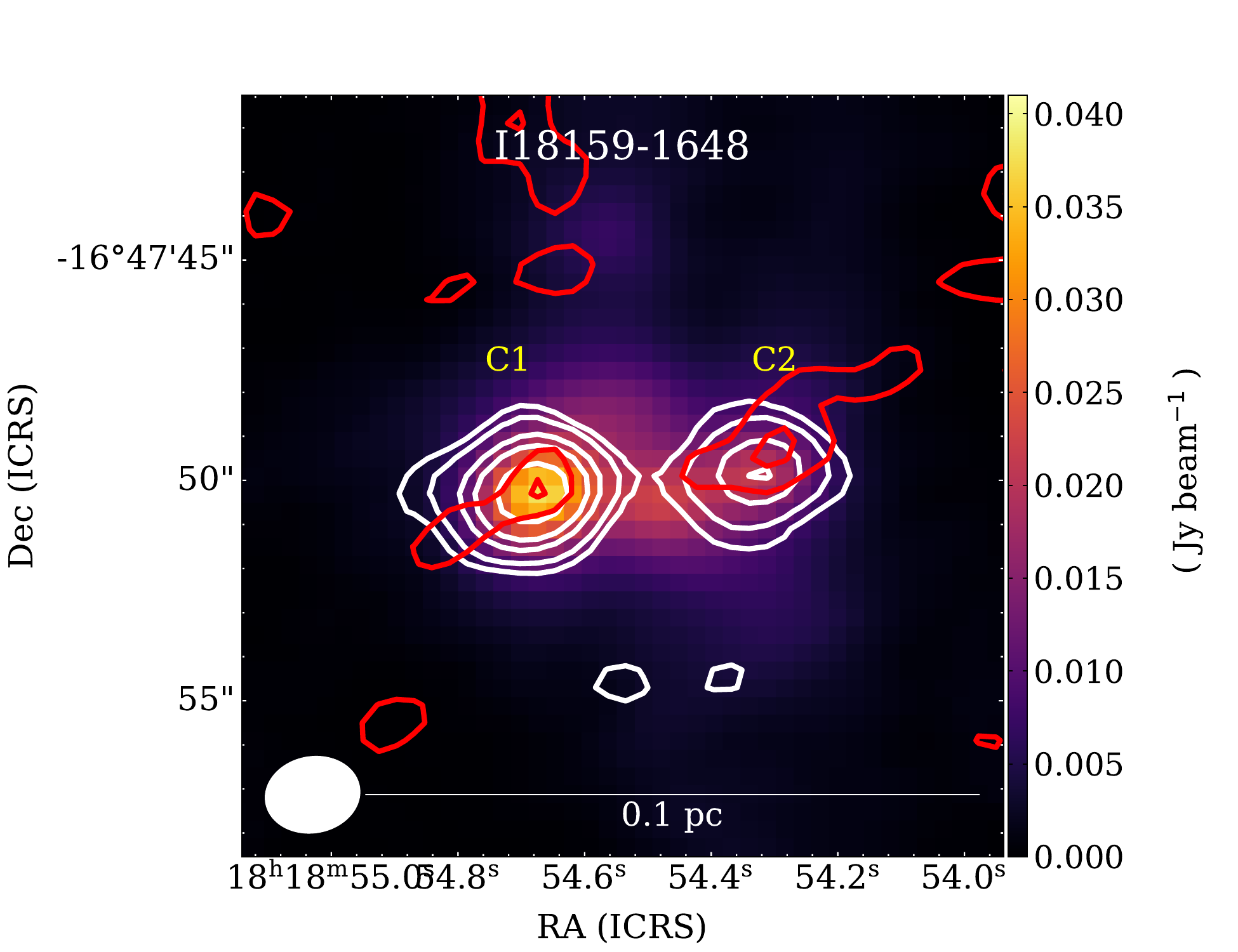}}
\quad
{\includegraphics[height=4.01cm,width=5.21cm]{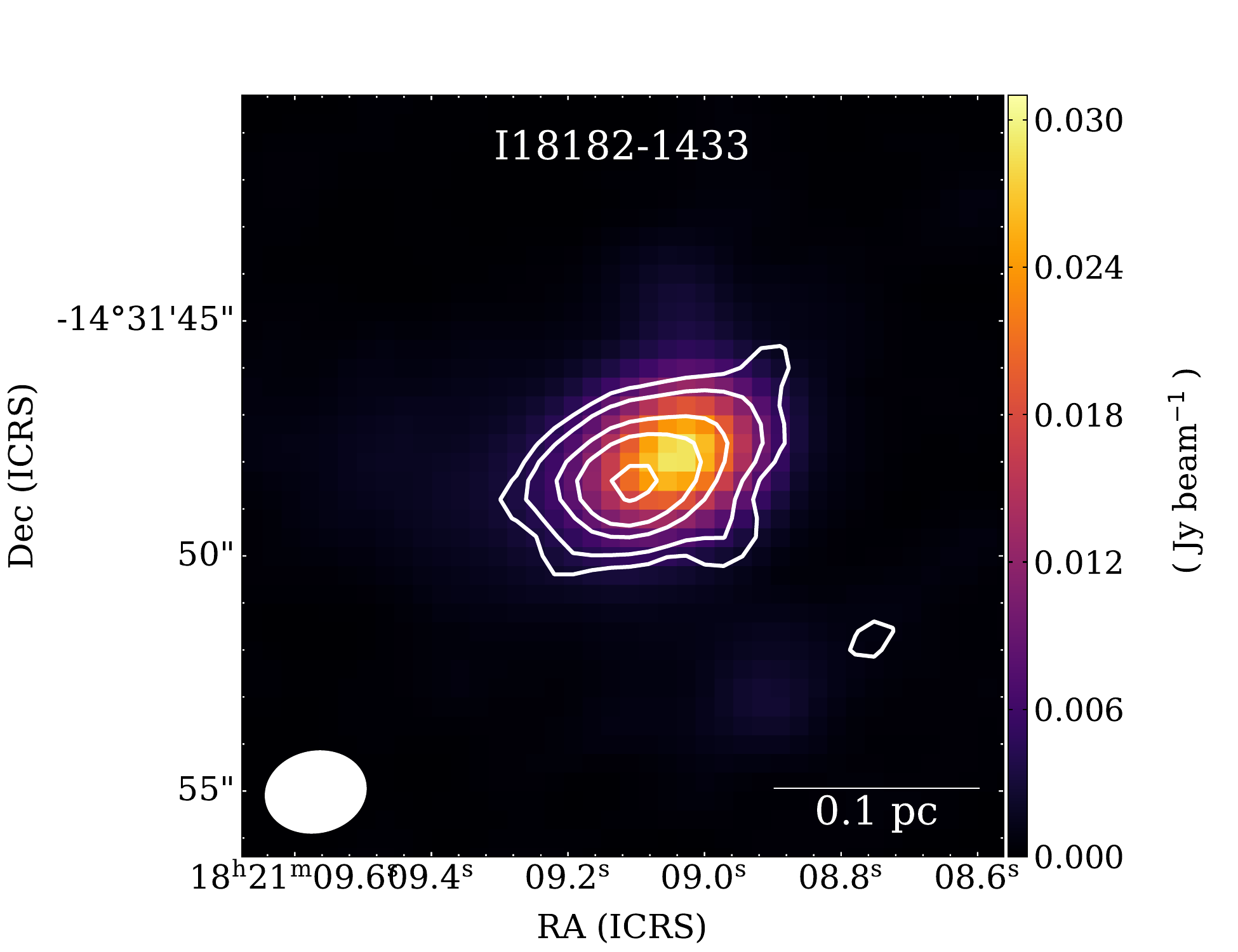}}
\quad
{\includegraphics[height=4.01cm,width=5.21cm]{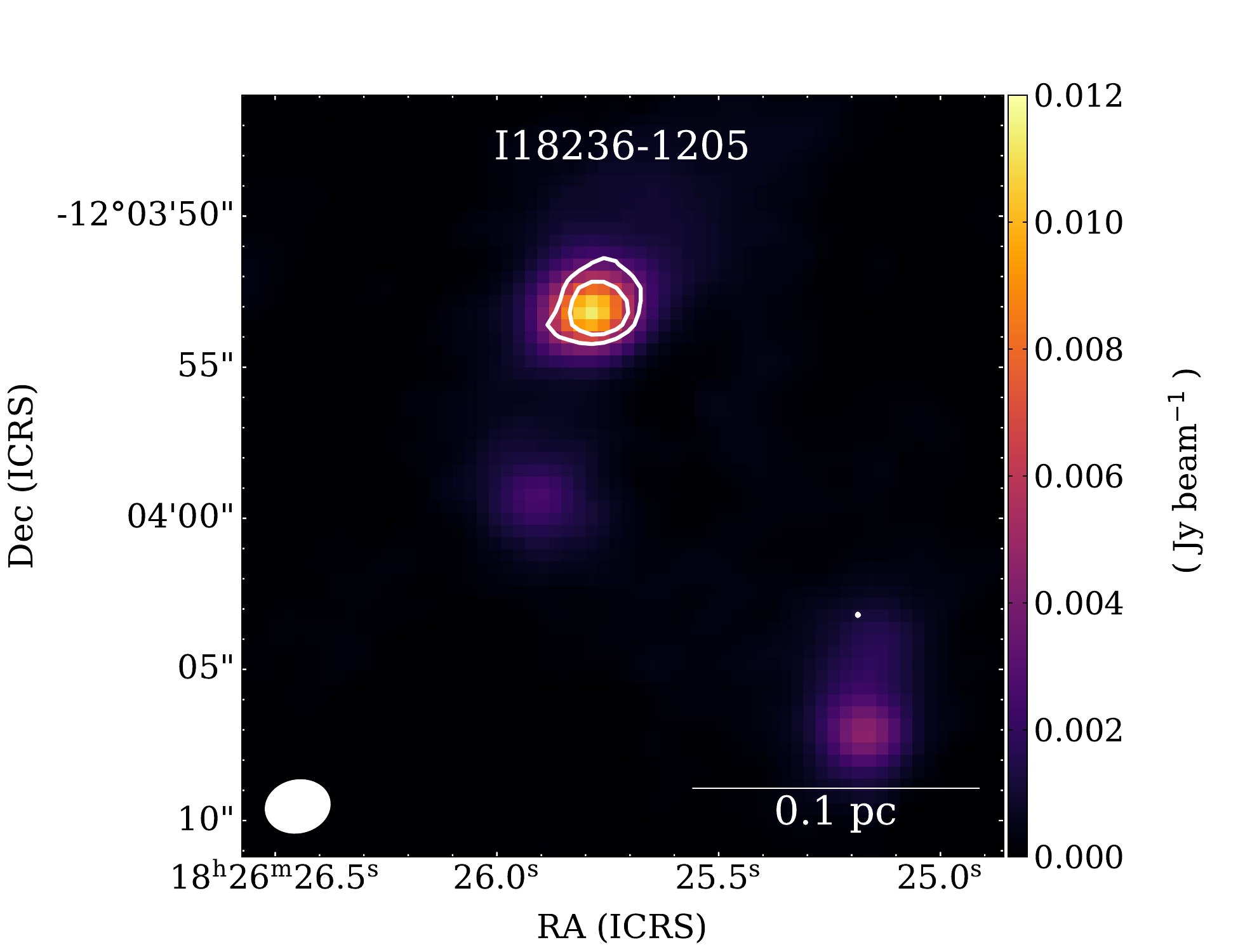}}
\quad
{\includegraphics[height=4.01cm,width=5.21cm]{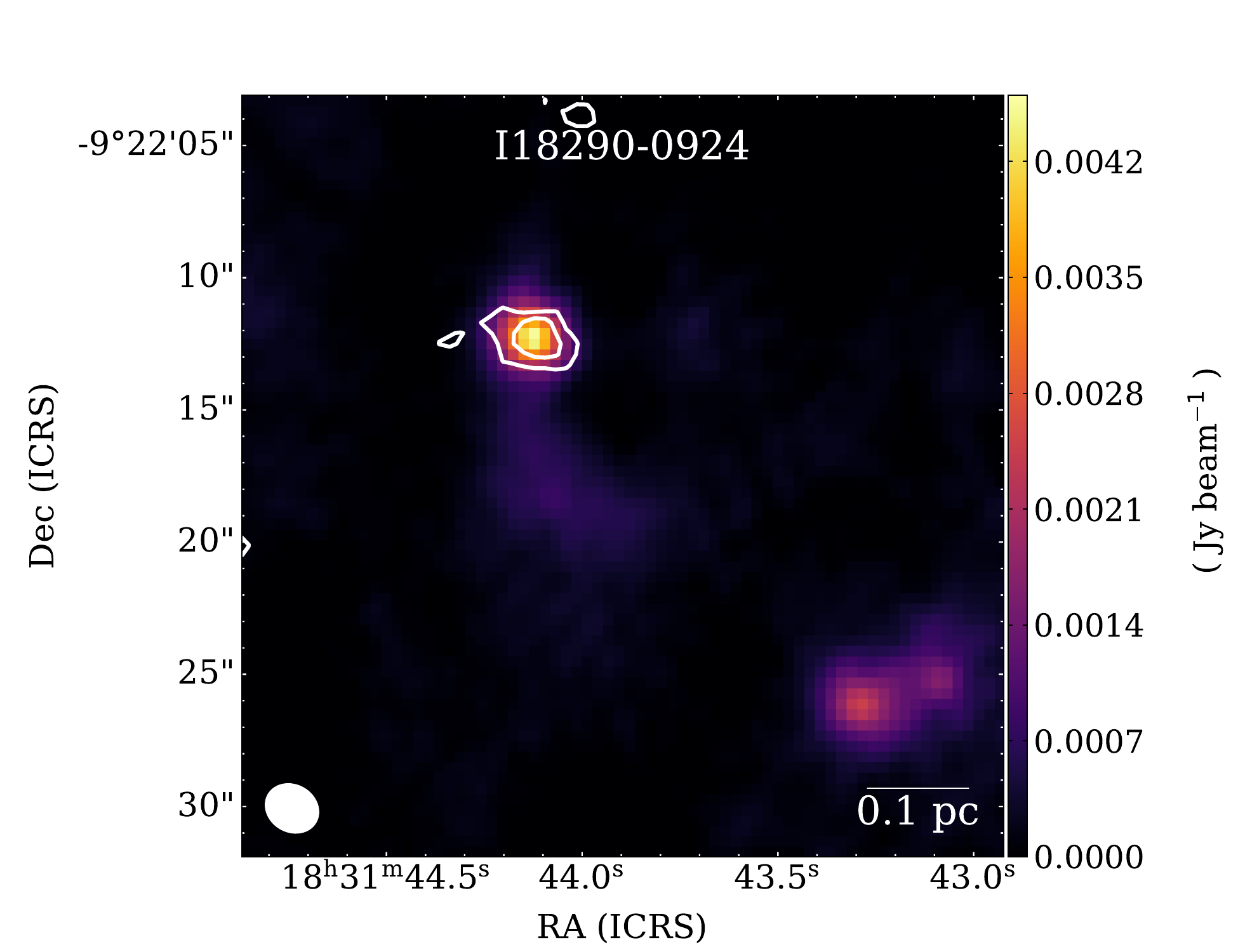}} 
\quad
{\includegraphics[height=4.01cm,width=5.21cm]{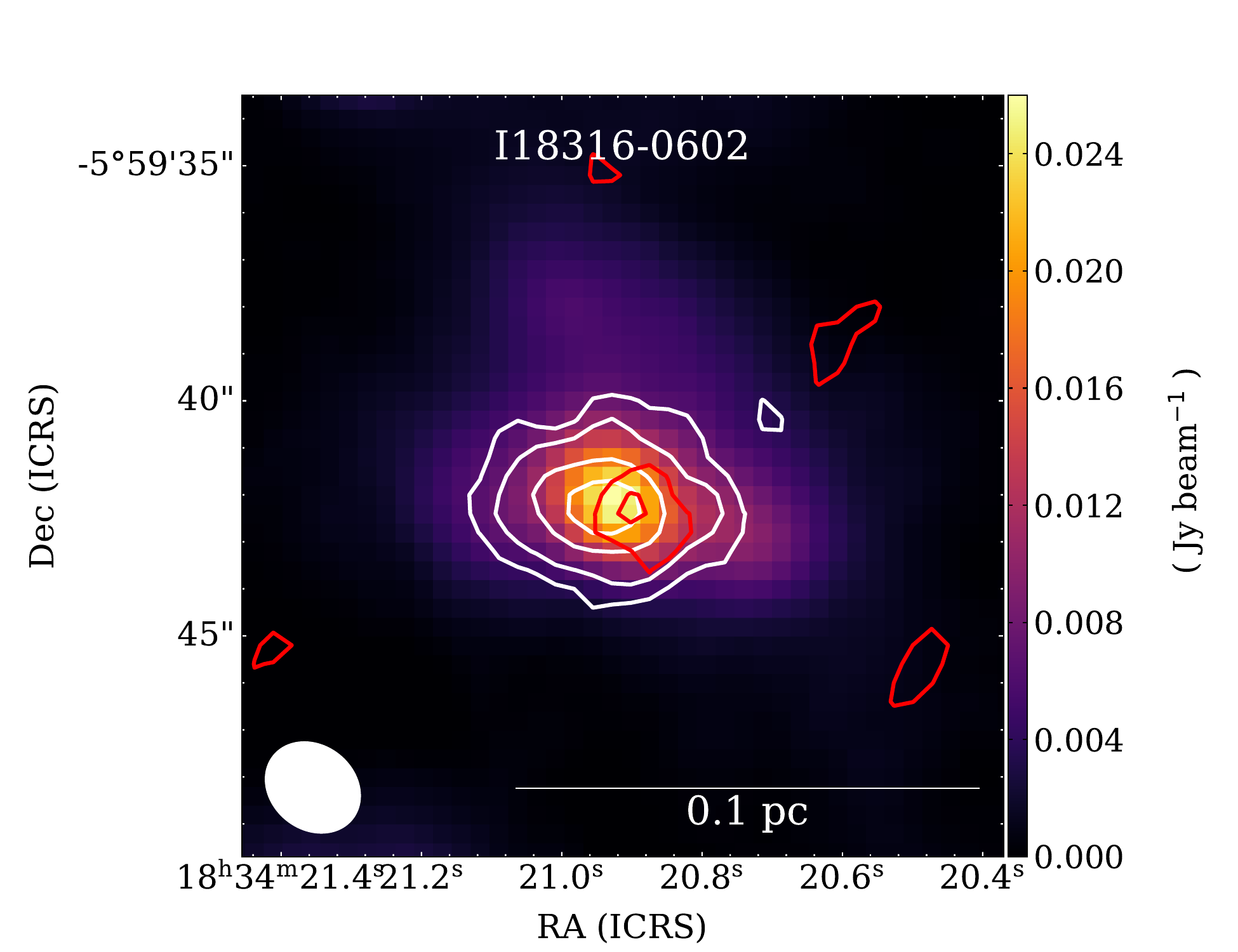}}
\quad
{\includegraphics[height=4.01cm,width=5.21cm]{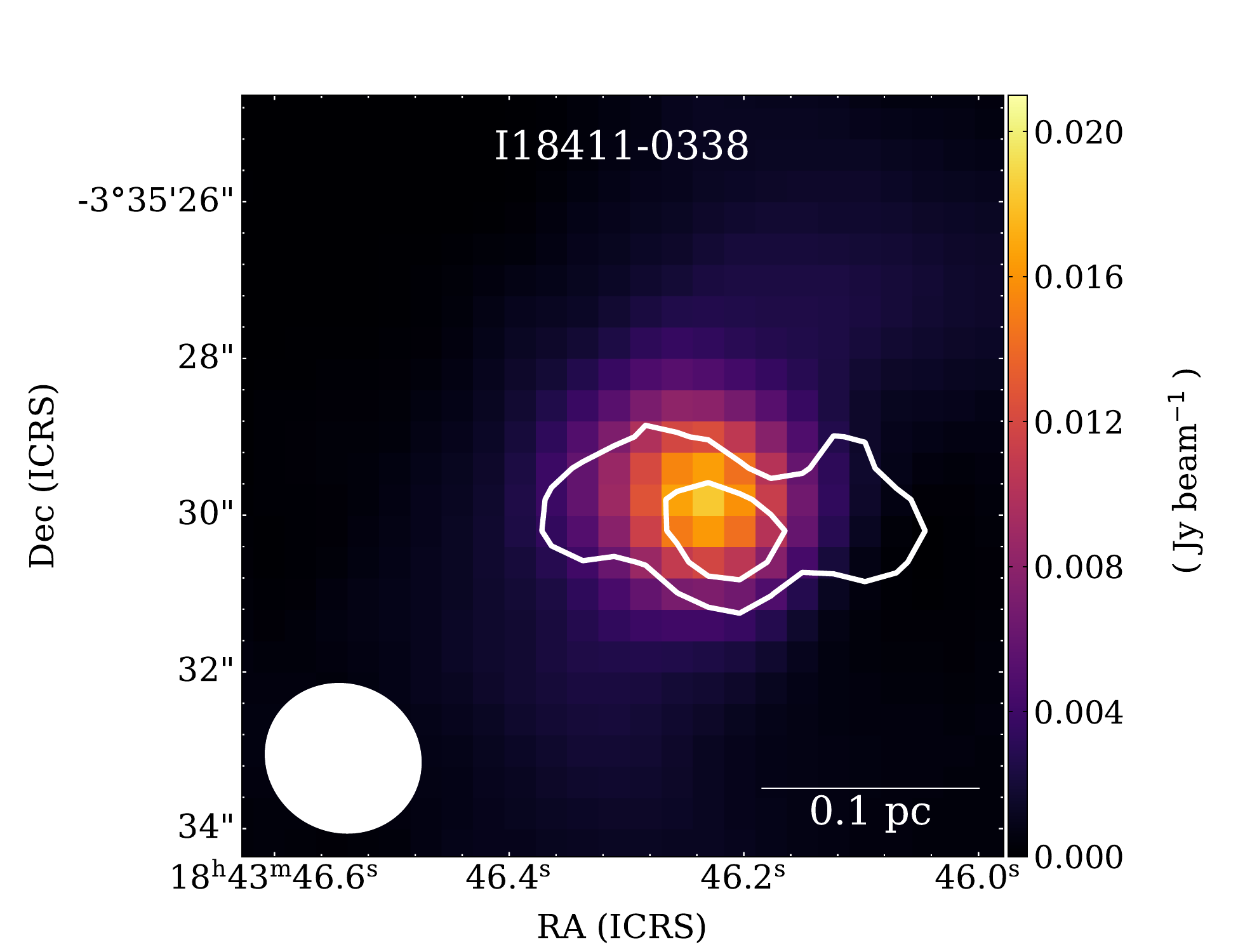}}
\caption{Continued.}
\end{figure}

\setcounter{figure}{\value{figure}-1}
\begin{figure}
  \centering 
{\includegraphics[height=4.01cm,width=5.21cm]{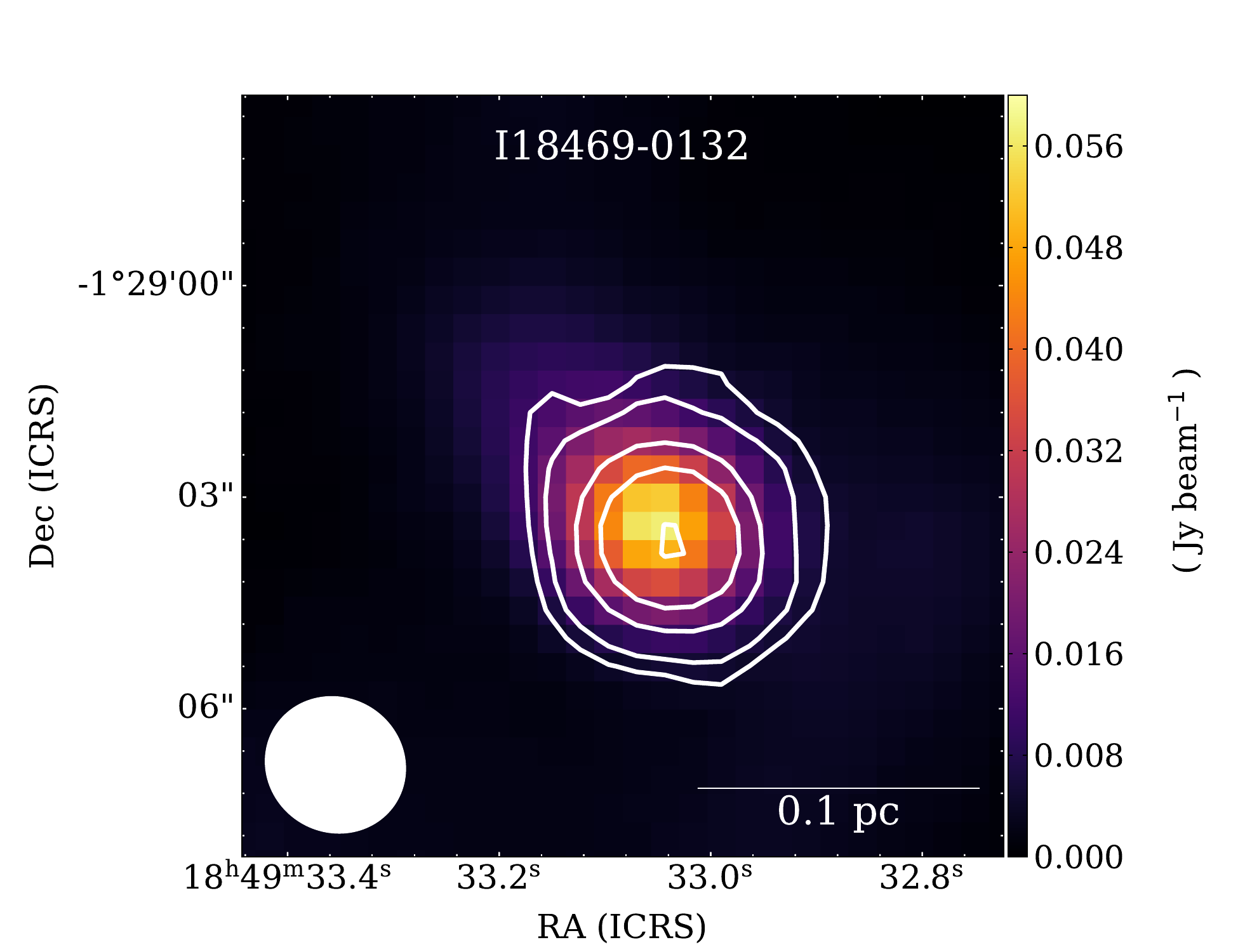}} 
\quad
{\includegraphics[height=4.01cm,width=5.21cm]{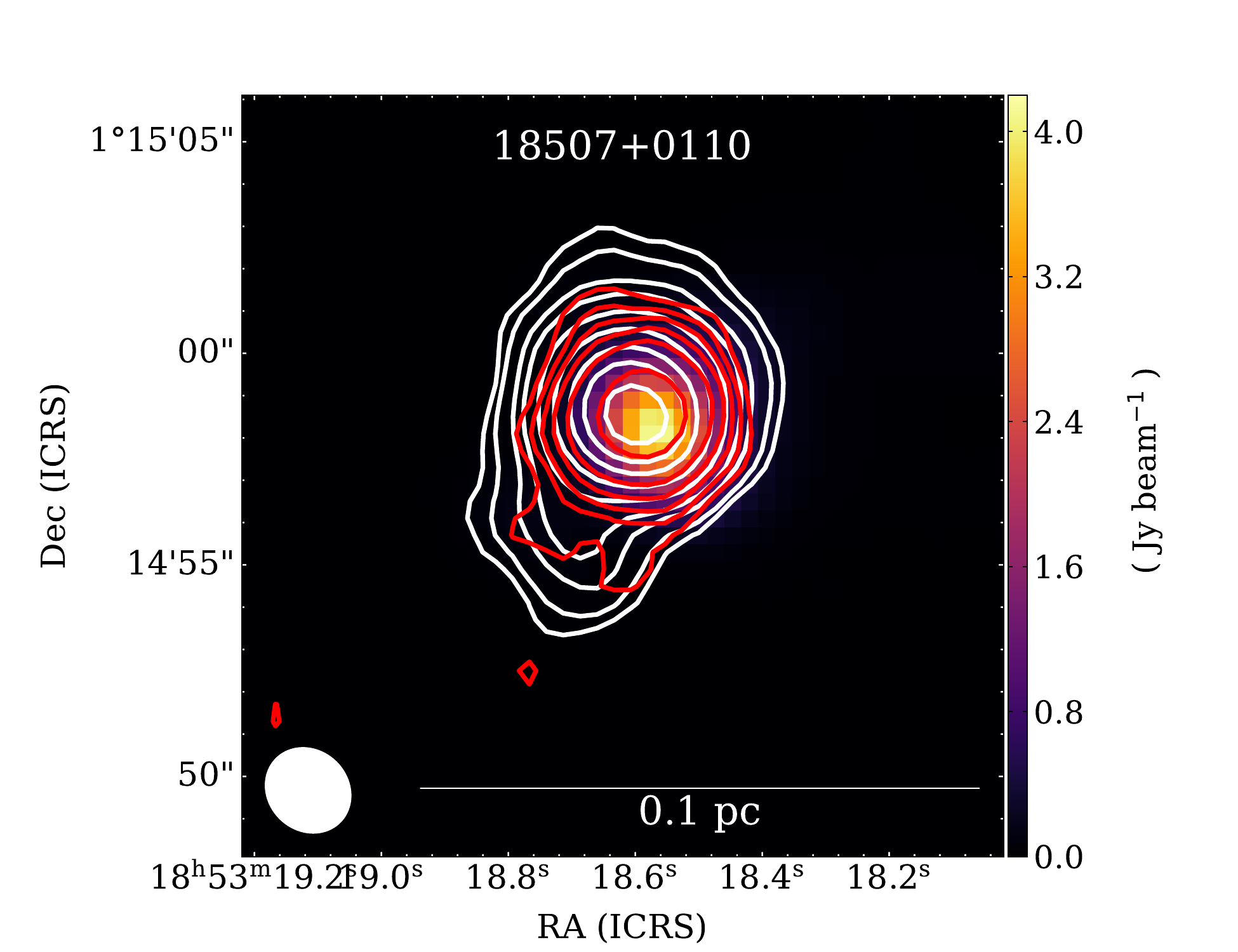}}
\quad
{\includegraphics[height=4.01cm,width=5.21cm]{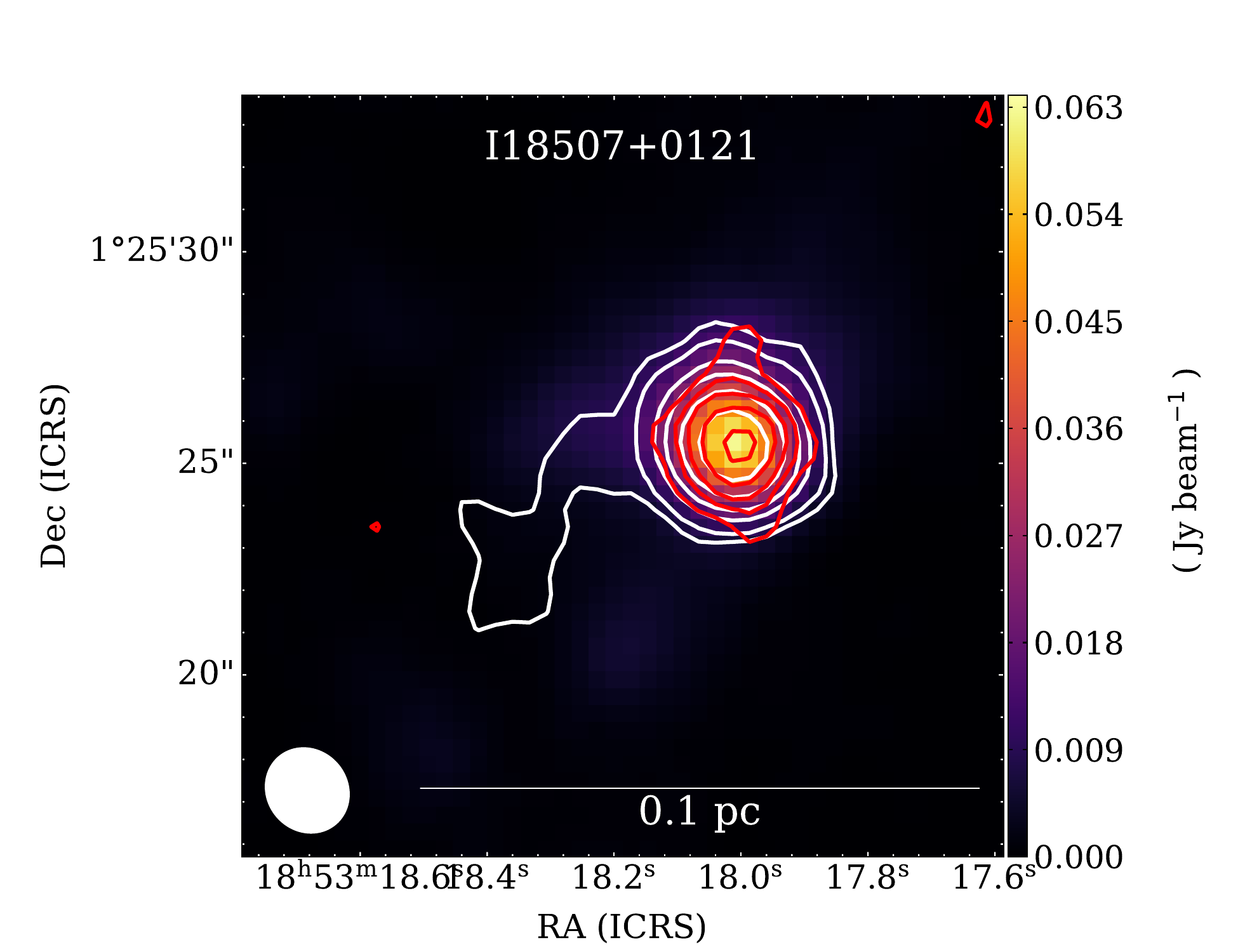}}
\quad 
{\includegraphics[height=4.01cm,width=5.21cm]{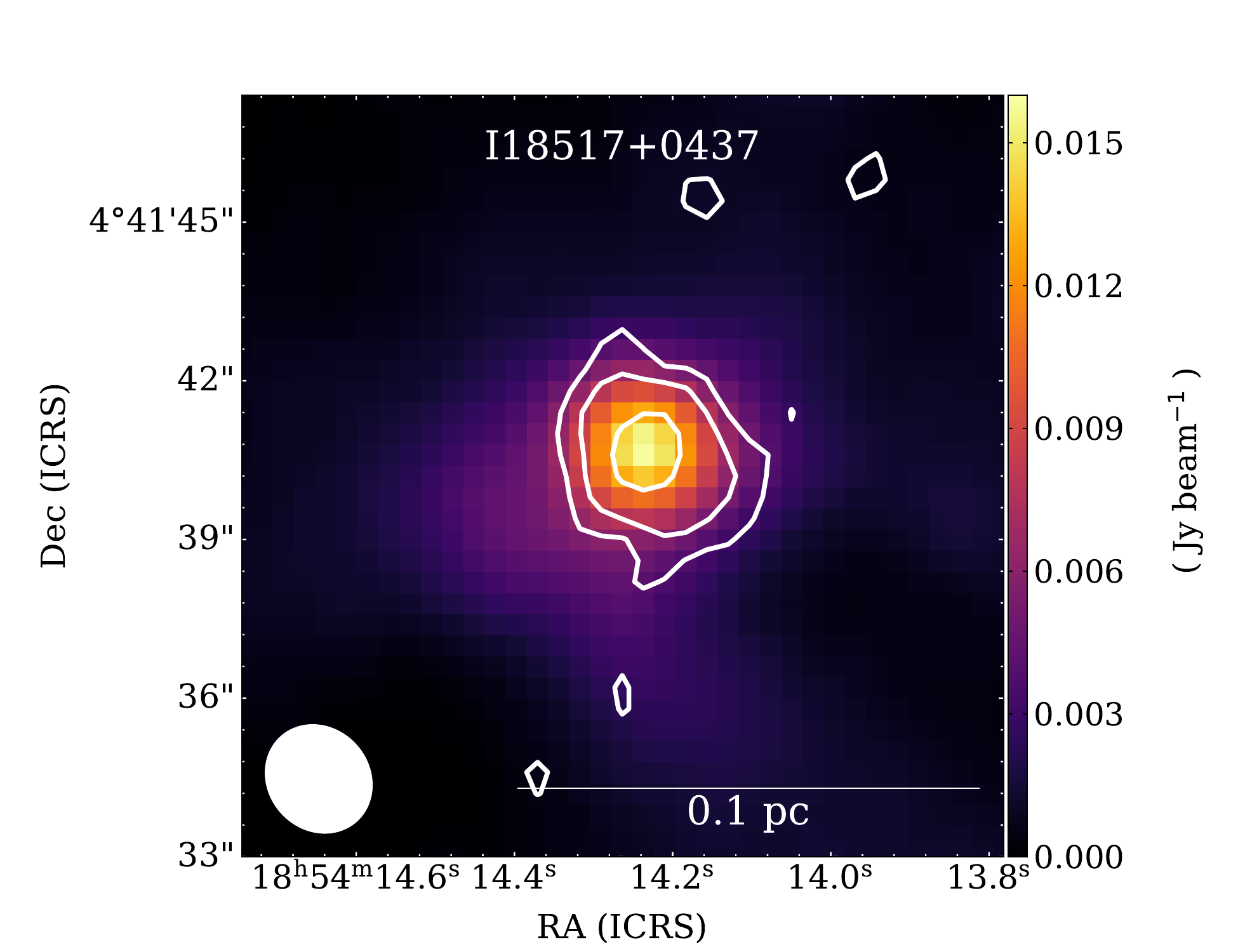}}
\quad 
{\includegraphics[height=4.01cm,width=5.21cm]{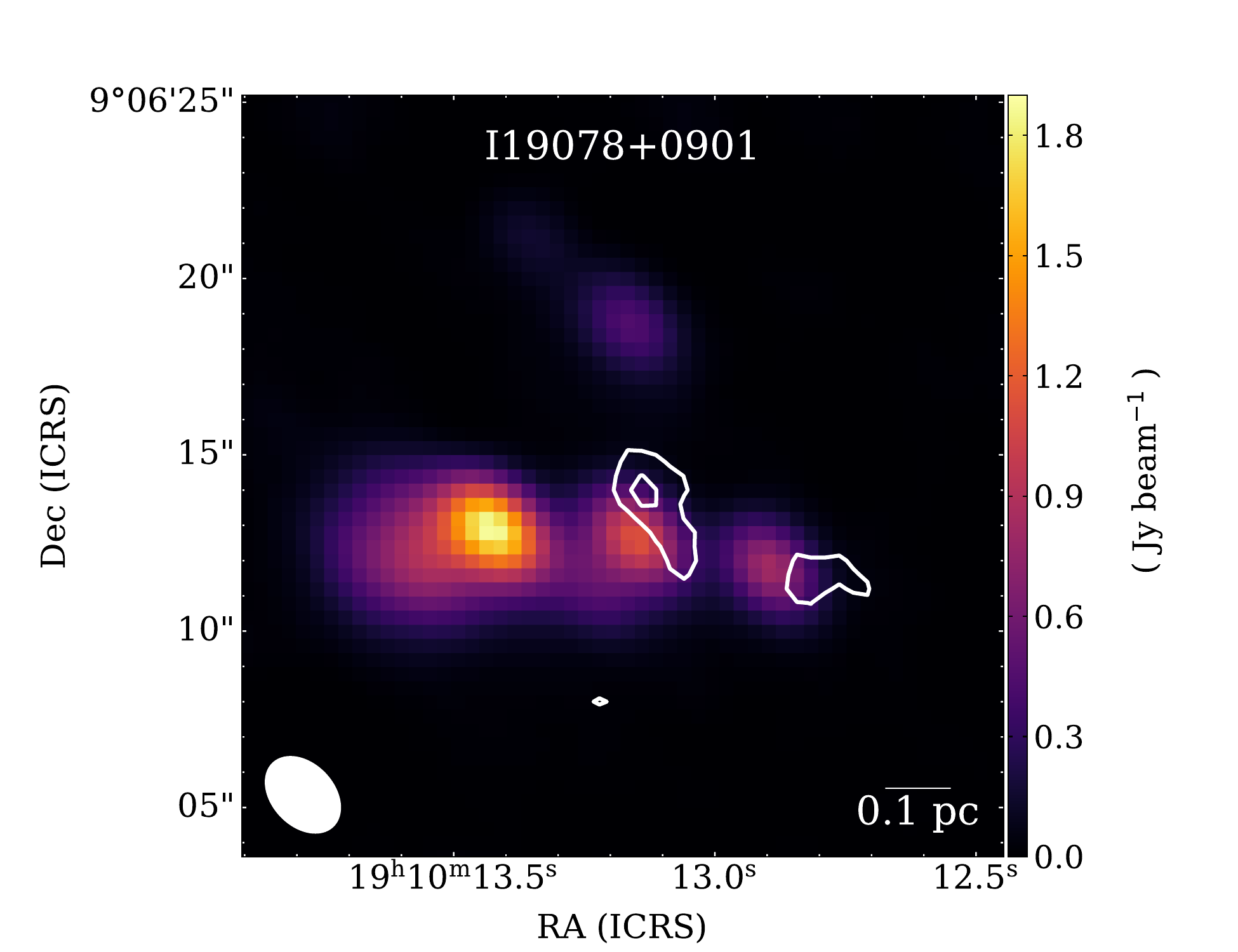}}
\quad
{\includegraphics[height=4.01cm,width=5.21cm]{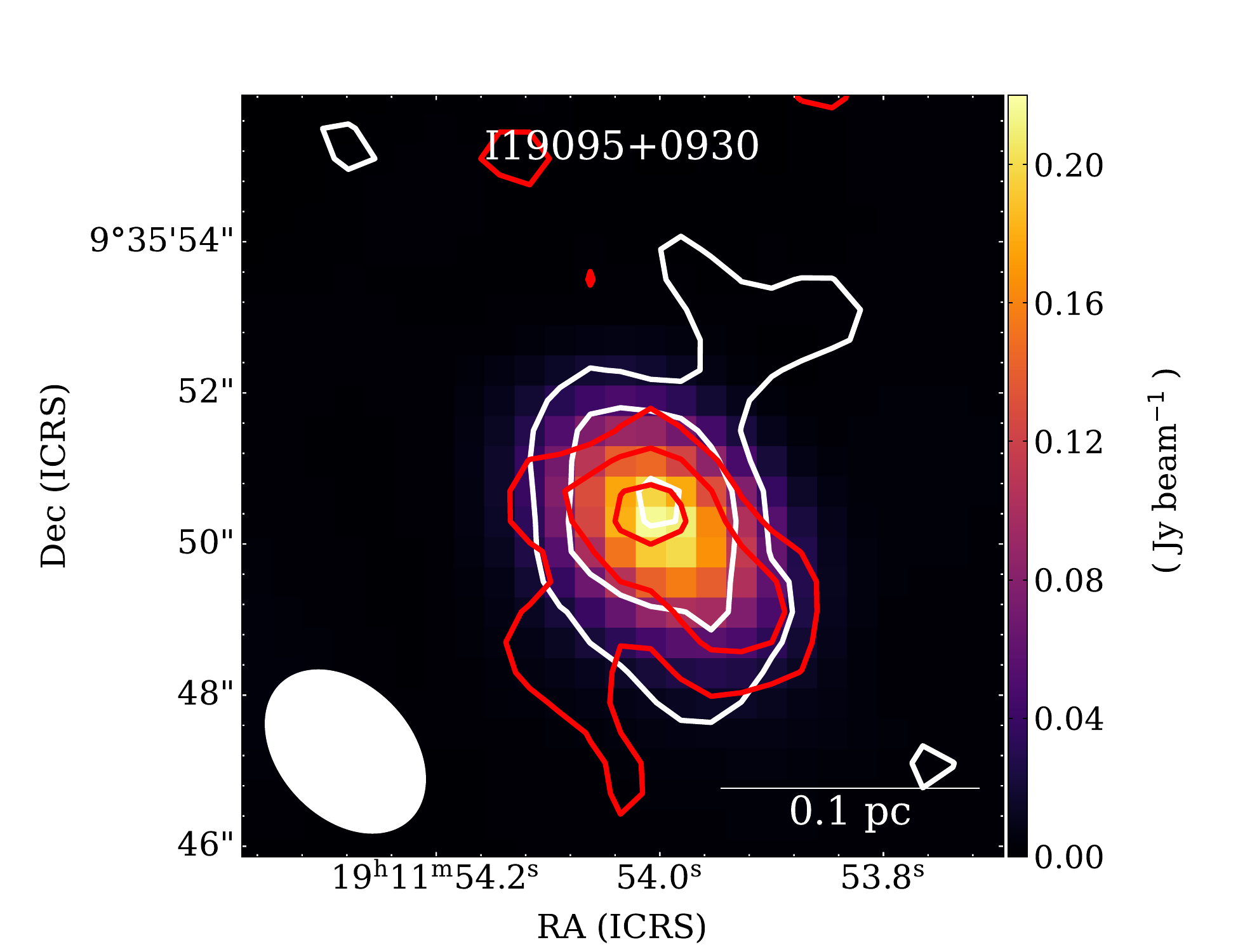}}
\caption{Continued.}
\end{figure}
\bsp
\label{lastpage}
\end{document}